\documentclass[%
 aip,
 amsmath,amssymb,
 reprint,%
]{revtex4-1}

\usepackage{braket}
\usepackage{graphicx}% Include figure files
\usepackage{dcolumn}% Align table columns on decimal point
\usepackage{bm}% bold math
\usepackage[utf8]{inputenc}
\usepackage[T1]{fontenc}
\usepackage{mathptmx}
\usepackage{multirow}
\usepackage{longtable}
\usepackage{enumerate}
\usepackage{soul, color, xcolor}

\renewcommand{\hl}[1]{#1}

\begin{document}

\preprint{AIP/123-QED}

\title[]{Continuous-variable quantum key distribution system: past, present, and future}
% Force line breaks with \\

\author{Yichen Zhang}
\affiliation{State Key Laboratory of Information Photonics and Optical Communications, School of Electronic Engineering, Beijing University of Posts and Telecommunications, Beijing 100876, China.}
 %\altaffiliation[]{State Key Laboratory of Information Photonics and Optical Communications, Beijing University of Posts and Telecommunications, Beijing 100876, China.}%Lines break automatically or can be forced with \\
\author{Yiming Bian}
\affiliation{State Key Laboratory of Information Photonics and Optical Communications, School of Electronic Engineering, Beijing University of Posts and Telecommunications, Beijing 100876, China.}

\author{Zhengyu Li}%
\email{lizhengyu2@huawei.com}
\affiliation{Central Research Institute, 2012 Labs, Huawei Technologies Co., Ltd, Shenzhen 518129, Guangdong, China.}%

\author{Song Yu}
\email{yusong@bupt.edu.cn}
\affiliation{State Key Laboratory of Information Photonics and Optical Communications, School of Electronic Engineering, Beijing University of Posts and Telecommunications, Beijing 100876, China.}%

\author{Hong Guo}
\email{hongguo@pku.edu.cn}
\affiliation{State Key Laboratory of Advanced Optical Communication Systems and Networks, Department of Electronics, and Center for Quantum Information Technology, Peking University, Beijing 100871, China.}

\date{\today}% It is always \today, today,

\begin{abstract}
\noindent\rule[0.15\baselineskip]{16.5cm}{0.1pt} \\
\textsf{\textbf{ABSTRACT}}
\\
\\
Quantum key distribution provides secure keys with information-theoretic security ensured by the principle of quantum mechanics. The continuous-variable version of quantum key distribution using coherent states offers the advantages of its compatibility with telecom industry, e.g., using commercial laser and homodyne detector, is now going through a booming period. In this review article, we describe the principle of continuous-variable quantum key distribution system, focus on protocols based on coherent states, whose systems are gradually moving from proof-of-principle lab demonstrations to in-field implementations and technological prototypes. We start by reviewing the theoretical protocols and the current security status of these protocols. Then, we discuss the system structure, the key module, and the mainstream system implementations. The advanced progresses for future applications are discussed, including the digital techniques, system on chip and point-to-multipoint system. Finally, we discuss the practical security of the system and conclude with promising perspectives in this research field. 
\noindent\rule[0.15\baselineskip]{16.5cm}{0.1pt} \\
\end{abstract}

\maketitle

%\begin{quotation}
%The ``lead paragraph'' is encapsulated with the \LaTeX\
%\verb+quotation+ environment and is formatted as a single paragraph before the first section heading.
%(The \verb+quotation+ environment reverts to its usual meaning after the first sectioning command.)
%Note that numbered references are allowed in the lead paragraph.
%%
%The lead paragraph will only be found in an article being prepared for the journal \textit{Chaos}.
%\end{quotation}
\tableofcontents

\section{INTRODUCTION}\label{sec:1}
Since 1984, quantum key distribution (QKD) \cite{Bennett_BB84_1984} has ushered in an era of secure communications using quantum methods by providing information-theoretic secure key distribution \cite{Ekert_PhysRevLett_1991,Gisin_RevModPhys_2002,Scarani_RevModPhys_2009,Pirandola_Advances_2020,Xu_RevModPhys_2020,cao_IEEECommunSurvTutor_2022,portmann_RevModPhys_2022}. 
The combination of this method with one-time-pad encryption provides the ultimate protection for the transmission of confidential messages. In general, for \hl{simplified} implementations, QKD protocols are formulated in a prepare-and-measure fashion, where \hl{the} classical information is encoded on non-orthogonal quantum states: \hl{they} are randomly prepared by Alice (the sender) and then transmitted to Bob (the receiver) through an insecure quantum channel. At the output of the channel, the states are measured by Bob to retrieve the encoded classical information. The quantum no-cloning theorem dictates that an unknown quantum state cannot be reliably cloned \cite{Wootters1982ASQ}, ensuring long-term security based on physical principles \cite{diamanti_npjQuantumInf_2016} against unlimited computational power.

% Since 1984, quantum key distribution (QKD)  has opened an era of secure communication using quantum methods, by providing information-theoretical secure key distribution~\cite{Ekert_PhysRevLett_1991,Gisin_RevModPhys_2002,Scarani_RevModPhys_2009,Pirandola_Advances_2020,Xu_RevModPhys_2020,cao_IEEECommunSurvTutor_2022,portmann_RevModPhys_2022}. Combining this method with one-time pad encryption provides ultimate protection to the transmission of confidential messages. In general, for cost-effective implementations, QKD protocols are formulated in a prepare-and-measure fashion, where classical information is encoded in nonorthogonal quantum states: these are randomly prepared by Alice (the sender) and then transmitted to Bob (the receiver) through an insecure quantum channel. At the output of the channel, the states will be measured by Bob, so as to retrieve the encoded classical information. The quantum no-cloning theorem dictates that an unknown quantum state cannot be cloned reliably \cite{Wootters1982ASQ}, which ensures the long-term security based on physical principles \cite{diamanti_npjQuantumInf_2016}, agianst un-limited computation power.
% An important advantage of QKD is that, once a QKD session is over, there is no classical transcript for Eve to keep since the communication is quantum . Therefore, an eavesdropper has to break a QKD session in real time or it will be secure forever. This differs significantly from conventional key distribution schemes.

So far, various QKD protocols with discrete variables have been proposed to support the long-distance and practical-secure system implementations \cite{Bennett_BB84_1984, Ekert_PhysRevLett_1991, lutkenhaus2000security, inoue2002differential, hwang2003quantum, scarani2004quantum, barrett2005no, stucki2005fast, wang2005beating, lo2005decoy, acin2007device, inamori2007unconditional, pironio2009device, masanes2011secure, braunstein2012side, lo2012measurement, sasaki2014practical, lucamarini2018overcoming, arnon2018practical, vazirani2019fully}, including the decoy states experiments \cite{zhao2006experimental,peng2007experimental,rosenberg2007long,schmitt2007experimental,yuan2007unconditionally,yin2007experimental,wang2008experimental,dixon2008gigahertz,rosenberg2009practical,yuan2009practical,liu2010decoy,chen2010metropolitan,wang2013direct,boaron2018secure}, 
the measurement-device-independent (MDI) experiments \cite{rubenok2013real,liu2013experimental,da2013proof,tang2014experimental,tang2014measurement,tang2014field,wang2015phase,valivarthi2015measurement,yin2016measurement,tang2016measurement,tang2016experimental,comandar2016quantum,kaneda2017quantum,wang2017measurement,valivarthi2017cost,liu2018polarization,wei2020high}, 
the twin-field experiments \cite{minder2019experimental,wang2019beating,liu2019experimental,zhong2019proof,fang2020implementation,chen2020sending, Pittaluga2020600kmRQ, chen2021twin, wang2022twin, liu2023experimental}, 
the system on chip \cite{ma2016silicon,sibson2017chip,sibson2017integrated, bunandar2018metropolitan, ding2017high, paraiso2019modulator, wei2020high}, 
and so on \cite{lee2014entanglement,guan2015experimental,takesue2015experimental,wang2015experimental,zhong2015photon, korzh2015provably,mirhosseini2015high, sit2017high, li2016experimental,yuan2016directly,islam2017provably}.
\hl{Specifically, the twin-field QKD with a 3-station scheme has significantly promoted the development of the long-distance QKD, where the total distance can break the PLOB bound} \cite{pirandola2017fundamental}.
These achievements have resulted in the long-haul point-to-point connection up to 1000 km \cite{liu2023experimental}, high-speed metropolitan system \cite{Tanaka2012High, Lucamarini2013Efficient, Frohlich2017Long, Yuan201810Mbps,li2023high}, and field deployed QKD network \cite{elliott2005current, Peev_NewJPhys_2009, chen2009field, sasaki2011field,  chen2021implementation}, even with satellite-to-ground links \cite{Yin2017SatellitebasedED,liao2017satellite, Ren2017GroundtosatelliteQT, Yin2017SatellitetoGroundEQ, Liao2018SatelliteRelayedIQ, chen2021integrated}. 
Furthermore, it is worth noting a distinct category of protocols where the information is encoded on the quadrature of light that is continuous-variable.
The use of such continuous-variable quantum information carriers, instead of qubits, constitutes a potent and alternative approach for QKD~\cite{ralph1999continuous, Cerf_PhysRevA_2001, Grosshans_PhysRevLett_2002, Weedbrook_PhysRevLett_2004, Braunstein_RevModPhys_2005,Wang_PhysRep_2007,Weedbrook_RevModPhys_2012,Lam_NatPhotonics_2013,Diamanti_Entropy_2015} and more broadly, for quantum information processing \cite{Braunstein_RevModPhys_2005,cerf2007quantum, furusawa2007quantum, andersen2010continuous, Weedbrook_RevModPhys_2012, van2014quantum, adesso2014continuous, andersen2015hybrid, kurizki2015quantum, serafini2017quantum}.  

\begin{figure*}
\includegraphics[width=1\textwidth]{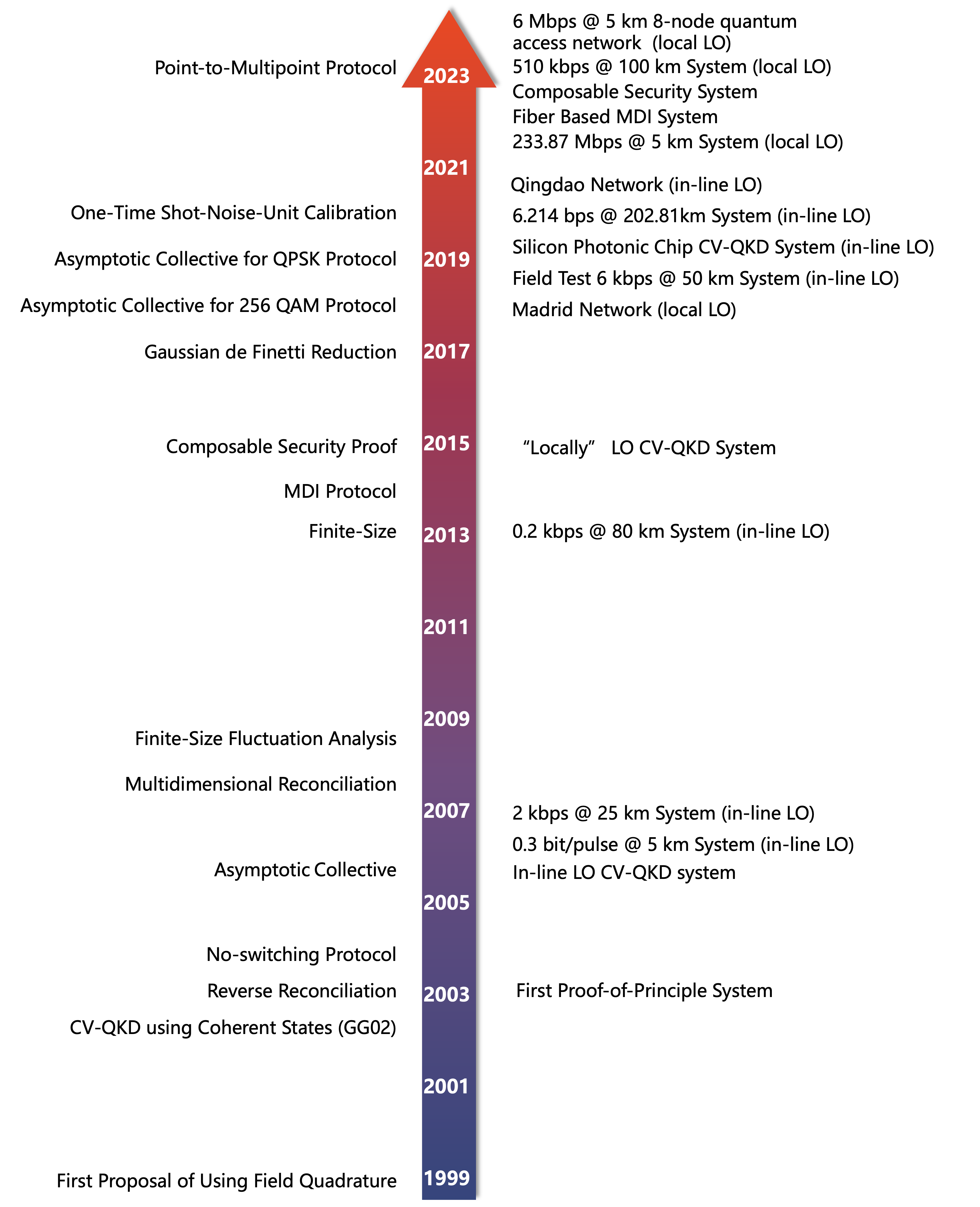}% Here is how to import EPS art
\caption{\label{fig:fig1-History} Selected key developments in CV-QKD. On both sides of the arrow are theoretical research \hl{progresses} (left) and experimental research \hl{progresses} (right). Here, the theoretical research \hl{progresses} includes the protocol proposal, security proof and  optimization of protocol steps. The experimental \hl{progresses} includes the development of in-line LO and local LO systems.}
\end{figure*}

Continuous-variable QKD (CV-QKD) using coherent states \cite{Grosshans_PhysRevLett_2002, Weedbrook_PhysRevLett_2004} is now currently experiencing a booming period due to its compatibility with telecom industry, e.g., using commercial continuous-wave laser and coherent receiver. 
This potential has led to significant advancements in CV-QKD, including \hl{the} protocol design, security analysis, and system implementation (See Fig.~\ref{fig:fig1-History}).
The security of CV-QKD protocol using Gaussian-modulated coherent states was initially proved under asymptotic conditions~\cite{Garcia_PhysRevLett_2006,Navascues_PhysRevLett_2006}, and later extended to the finite-size regime with universal composability against collective attack~\cite{Leverrier_PhysRevLett_2015}, and general attacks by exploiting Gaussian de Finetti theorem~\cite{Leverrier_PhysRevLett_2017}. 
In addition to continuous modulation, \hl{the discrete modulation of coherent states has also been well investigated}~\cite{Li_Arxiv_2018,Ghorai_PhysRevX_2019,Lin_PhysRevX_2019, Denys2021explicitasymptotic, matsuura2021finite}. 
Along with the improving security proof, the implementation of CV-QKD system has progressed from the initial proof-of-principle demonstration to the second stage with swift advancements in high performance and system robustness. 
Currently, it has entered the third stage where a modern architecture is evolving with the benefit of being fully \hl{compatible with classical optical coherent communication}.

% the system implementation of CV-QKD has gone through the early stage of proof-of-principle demonstration to the second stage with rapid development of high performance and system robustness, and now entered the third stage in which a modern architecture is evolving with the advantage of fully compatible with coherent communication.

\begin{figure*}
  \includegraphics[width=0.8\textwidth]{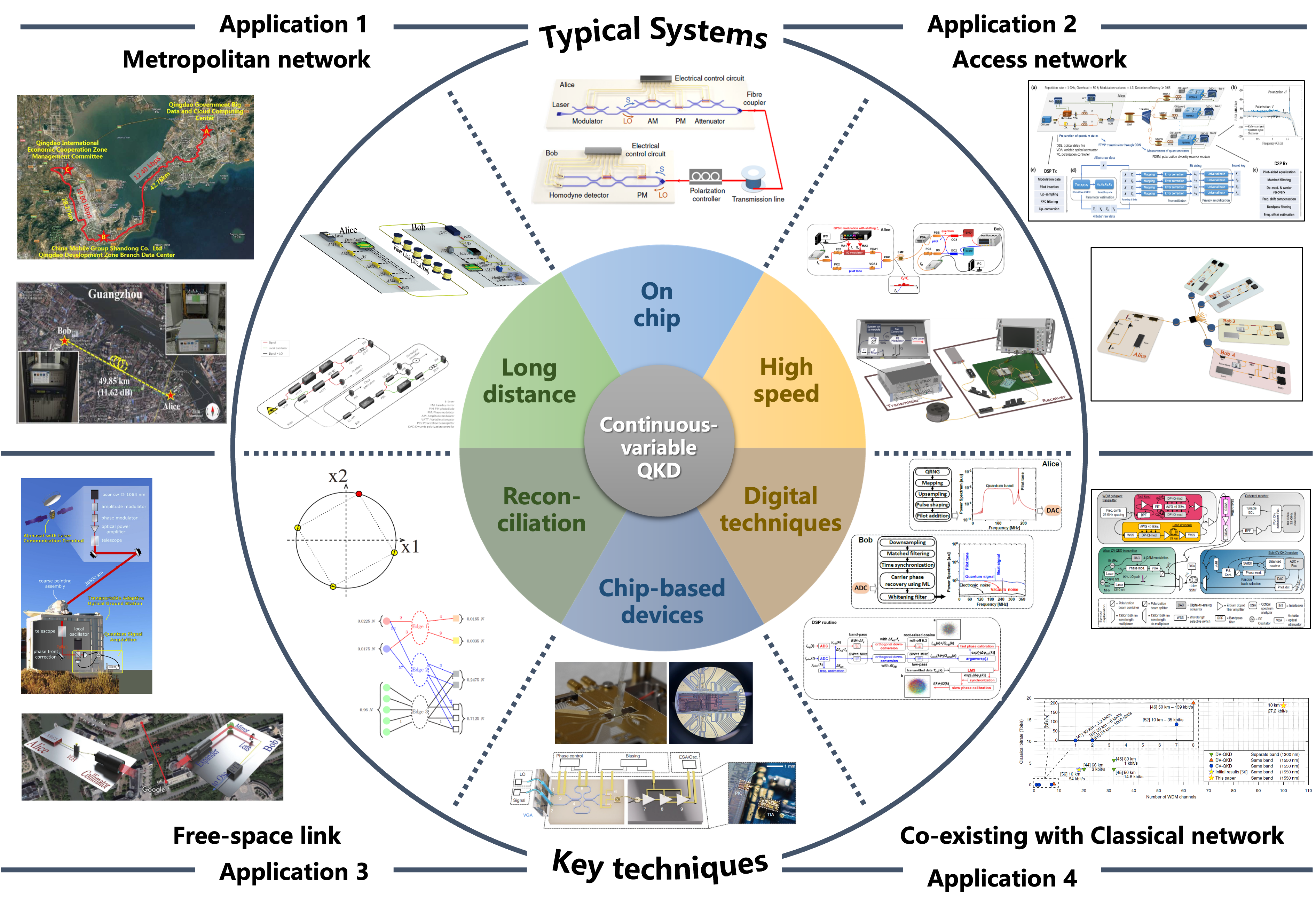}% Here is how to import EPS art
  \caption{\label{fig:OverviewIntro} The overview of the CV-QKD system, including the typical systems, key techniques and the applications. Here we mainly show the long-distance, chip-based and high-speed system, as well as the key techniques including the reconciliation, chip-based devices and digital techniques.}
\end{figure*}

During the initial \hl{phase} of the CV-QKD system, the primary challenge was to overcome the 3 dB limit, which was solved by the reverse reconciliation \cite{Grosshans_Nature_2003}.
Subsequently, the CV-QKD system progressed towards allowing long-distance transmission to facilitate two-user interconnection across a wide range without a trusted relay. 
Developments of the reconciliation resulted in enhanced error correction capability even with an extremely low signal-to-noise ratio (SNR) \cite{leverrier_PhysRevA_2008}, which played a significant role in the long-distance system covering a distance from 25 km \cite{Lodewyck_PhysRevA_2007} up to 80 km \cite{Jouguet_NatPhotonics_2013}.
% the CV-QKD system has evolved towards long transmission distance to enable relay-less two-user interconnection over a wide range. 
% The enhancement of the reconciliation \cite{leverrier_PhysRevA_2008} enables the error correction with extremely low signal-to-noise ratio (SNR), which significantly promoted the development of the long-distance system, from 25 km \cite{Lodewyck_PhysRevA_2007} to 80 km \cite{Jouguet_NatPhotonics_2013}. 
At this stage, both the quantum signal and local oscillator (LO) are generated by the same laser of transmitter and co-propagated in the quantum channel, known as the in-line LO system, which contributes to the suppression of phase noise when the quantum \hl{signal} and LO interfere for \hl{coherent} detection.
However, a significant challenge towards achieving long-distance and stable transmission is to reduce crosstalk between the strong LO and the weak quantum signal. 
The most effective current approach is using pulsed signals with high extinction ratio, then  combining polarization multiplexing and time-division multiplexing \cite{Jouguet_OptExpress_2012, Jouguet_NatPhotonics_2013, Huang_OptLett_2016, Huang_SciRep_2016, Zhang_QuantumSciTechnol_2019, Zhang_NatPhotonics_2019, Zhang_PhysRevLett_2020}. 
This methodology has resulted in the longest lab experiment over 202 km \cite{Zhang_PhysRevLett_2020}, the longest field test of 50 km \cite{Zhang_QuantumSciTechnol_2019}, the long-term test of a 3-node CV-QKD network in Qingdao, China \cite{zhang2020continuous}, and the first chip-based system \cite{Zhang_NatPhotonics_2019}.

In 2015, an alternative scheme was proposed, which relaxed the requirement of the extremely high isolation by  generating LO inside the receiver and using a pilot assisted phase recovery to suppress the phase noise introduced by the different laser source \cite{Qi_PhysRevX_2015, Soh_PhysRevX_2015}. 
As the pilot signal has significantly lower power compared to the LO, the high-extinction pulse generation for \hl{time-division} multiplexing is no longer required. 
Instead, frequency-division multiplexing is widely used, which simplifies the signal generation, and sometimes can be combined with polarization multiplexing for better isolation.
For higher secret key rate and better phase recovery, the repetition rate of the system is gradually enhanced, where digital techniques in classical optical communication systems, such as the pulse shaping and matched filter, are introduced to overcome the limited bandwidth of devices. 
% Typically, pulse shaping and matched filter in digital domain are respectively processed before modulation and after detection for generating quantum signal pulses, instead of using amplitude modulation with high extinction ratio. 
Further, more and more digital algorithms for a CV-QKD system are developed, including the de-multiplexing, impairment compensation, synchronization etc., which then drive the innovation in system architecture.
To date, the local LO scheme contributes to the high-speed system with the repetition rate of 5 GBaud, resulting in 190 Mbps secret key rate at 5 km \cite{SubGbps}, as well as a flexible network deployment, which has been demonstrated by the software-defined CV-QKD network in Madrid, Spain \cite{aguado2019engineering}.

The use of digital techniques from classical communications has significantly advanced the development of CV-QKD systems, allowing for the completion of most operations in digital domain, and resulting in a system compatible with classical optical communication in the aspect of both architectures and algorithms \cite{roumestan2022experimental}.
Meanwhile, the advanced progresses of the homodyne detector integrated on chip have shown the potential of a compact system with high-performance, where the baud rate of the system using chip-based homodyne detector can reach 10 GBaud \cite{hajomer2023continuous}.
Additionally, the implementation of a high-rate downstream point-to-multipoint CV-QKD network can facilitate large-scale deployments, enabling multi-user access with low-cost devices and simplified network structures \cite{bian2023high}.

An overview of the typical CV-QKD systems, key techniques and application scenarios is presented in Fig. \ref{fig:OverviewIntro}.
These typical system achievements, supported by the advanced reconciliation, digital signal processing (DSP) and chip-based devices, have proved that the CV-QKD system is suitable for the metropolitan network and access network, as well as the free space communication and co-existing with the classical optical networks.
% In addition, since the LO also acts as a filter, which enhances the out-of-band noise immunity, CV-QKD can also be used in free space link and co-existing with the classical optical networks, where the resistance of the background noise is crucial. 
With these advanced techniques, a large-scale cost-effective QKD network supported by advanced CV-QKD systems is on the way. 

In all this panorama, the present review aims at providing an overview of the most important results and the most recent advances in the field of CV-QKD system. After a brief introduction of the general notions, we review the main CV-QKD protocols and security analysis in Sec. II. The system structure and key modules are reviewed in Sec. III, and the typical currently-achievable implementations are detailed in Sec. IV, including the in-line LO systems, local LO systems, systems co-existing with classical networks, and so on. We then discuss the advanced progress for future applications, such as \hl{the} digital continuous-variable system and point-to-multipoint network, in Sec. V. Finally, we will discuss the practical security of the system in Sec. VI and conclude with promising perspectives in this research field in Sec. VII.

\section{CV-QKD PROTOCOL AND SECURITY PROOF}\label{sec:2}
The CV-QKD system relies on a protocol to establish the system's operating procedures, where the security of the protocol is determined by security proof.
In this section, we introduce the basic notions, the CV-QKD protocols and security analysis.

\subsection{Basic notions of continuous-variable systems}
The continuous-variable system is a typical infinite dimensional quantum system. Suppose that a CV system consists of a sequence of $n$ modes in the Hilbert space $\mathcal{H}=\otimes_{i=1}^{n}\mathcal{H}_i$, there are $n$ pairs of annihilation and creation operators $\left\{\hat{a}_i,\hat{a}^{\dag}_i\right\}$ with $i=1,2,...,n$, which satisfies

\begin{equation}\label{Eq.1}
\begin{split}
&\left[\hat{a}_i,\hat{a}^{\dag}_k\right]=\delta_{ik}, \\
&\left[\hat{a}_i,\hat{a}_k\right]=0,\\
&\left[\hat{a}^{\dag}_i,\hat{a}^{\dag}_k\right]=0.
\end{split}
\end{equation}

For each mode, we can correspondingly define the operators $\hat{x}$ and $\hat{p}$ as
\begin{equation}\label{Eq.2}
\hat{x}=\hat{a}^{\dagger}+\hat{a}, \hat{p}=i(\hat{a}^\dagger-\hat{a}),
\end{equation}
which are so-called quadratures of electromagnetic field, and the quadratures can be described by a $n$-mode vector $\hat{r}=\left(\hat{x}_1,\hat{p}_1,\hat{x}_2,\hat{p}_2,...,\hat{x}_n,\hat{p}_n\right)^T$. Using the standard bosonic canonical commutation relations as well as Eq.~(\ref{Eq.1}) and (\ref{Eq.2}) we can easily get

\begin{equation}
\begin{split}
&\left[\hat{x}_i,\hat{x}_k\right]=0, \\
&\left[\hat{p}_i,\hat{p}_k\right]=0,\\
&\left[\hat{x}_i,\hat{p}_k\right]=2i\delta_{ik},
\end{split}
\end{equation}
which gives the well-known Heisenberg uncertainty relation 

\begin{equation}
  \Delta\hat{x}\Delta\hat{p}\geq |\langle [x, p] \rangle| = 1.
\end{equation}
Here,  $\Delta \hat{A} = (\langle \hat{A}^2 \rangle- \langle \hat{A} \rangle ^2)^{1/2}.$ Now it is straightforward to derive the following relation

\begin{equation}
\left[\hat{r}_i,\hat{r}_k\right]=2i\Omega_{ik},
\end{equation}
where $\Omega=\bigoplus_{i=1}^{n}\left[\begin{matrix}0 & 1\\-1 &0\end{matrix}\right]$, and $\Omega_{ik}$ is the generic element of $\Omega$.

Gaussian states are the states that their characteristic function is a Gaussian function in phase space.
% \begin{equation}
%   \chi_{\rho}(\xi)=e^{-\frac{1}{4}\xi^T\Gamma\xi+iD^T\xi}.
% \end{equation}
Its displacement operator can be defined as $d=\langle \hat{r} \rangle =Tr\left[\rho \hat{r} \right],$ while the positive-semidefinite symmetric covariance matrix is defined as 
\begin{equation}
  \gamma_{ij}=\frac{1}{2}Tr[\rho\{(\hat{r}_i-d_i),(\hat{r}_j-d_j)\}],
\end{equation}
where $\{\}$ denotes the anticommutator, and $\rho$ is a general density operator. Since a state is Gaussian if its Wigner function is Gaussian, it is completely characterized by the first two statistical moments, the mean value $d$, and the covariance matrix $\gamma$.

Since all physical existed Gaussian states should obey the Heisenberg uncertainty relation, therefore the covariance matrix is generally \cite{Simon_1994_PRA}. 

\begin{equation}
  \label{eq:Uncertainty}
  \gamma+i\Omega\geq0.
\end{equation}

% The states of a CV system can be expressed by the set of density operator $\hat{\rho}$ on the Hilbert space $\mathcal{H}$ or a characteristic function $\chi(\xi)$ defined on the phase space. When carrying out Fourier transform on $\chi(\xi)$, one can obtain the Wigner function $W(r)$ of the state. Particularly, Gaussian states are attained only when their characteristic function is Gaussian in phase space. Owing to the characters of Gaussian function, these Gaussian states can be represented by the first two statistical moments displacement vectors $\textbf{d}$ and covariance matrices $\gamma$. For a general Gaussian state $\rho$ made of $n$ modes, we define the displacement vector $\textbf{d}=\mathrm{Tr}\left[\rho\hat{r}\right]$, where $\textbf{d}\in\mathbb{R}^{2n}$. The $2n\times 2n$ covariance matrix is defined as

% \begin{equation}\label{Eq.3}
% \gamma_{ik}=\mathrm{Tr}\left[\rho\left\{\left(\hat{r}_i-\textbf{d}_i\right),\left(\hat{r}_k-\textbf{d}_k\right)\right\}\right],
% \end{equation}
% where $\left\{\cdot,\cdot\right\}$ represents for the anti-commutator. When the displacement vector of the state is 0, Eq.~(\ref{Eq.3}) becomes

% \begin{equation}\label{Eq.4}
% \gamma_{ik}=\mathrm{Tr}\left[\rho\left\{\hat{r}_i,\hat{r}_k\right\}\right]=\left\langle \hat{r}_i\hat{r}_k\right\rangle.
% \end{equation}

% Notice that a legitimate covariance matrix of a physical Gaussian state must obey the Heisenberg uncertainty relation, thus we have

% \begin{equation}
% \gamma + i\Omega \geq 0.
% \end{equation}

The most common single-mode Gaussian states include vacuum state, coherent states and squeezed states. The vacuum state is centered at the origin of the phase space and the covariance matrix is an identity matrix. The coherent states are the displaced vacuum state with non-zero displacement vectors $\textbf{d}=\left(d_x,d_p\right)$. Thus, the covariance matrices of the coherent states are also identity matrices. The squeezed states can be obtained by squeezing coherent states at one of the two quadratures. Suppose the states are squeezed on $x$ quadrature, the conjugate $p$ quadrature is anti-squeezed.  

\begin{figure}[t]
  \includegraphics[width=0.45\textwidth]{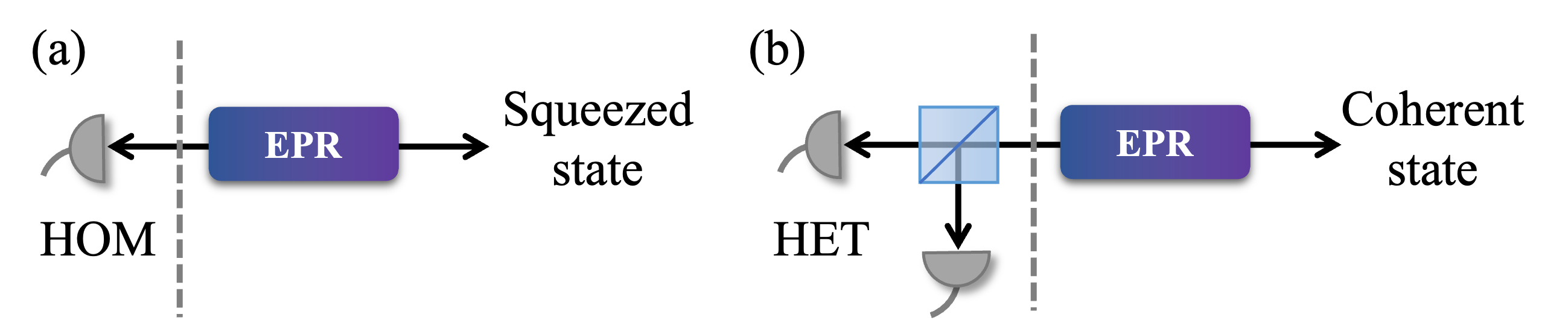}% Here is how to import EPS art
  \caption{\label{fig:EPR} The preparation of a squeezed (a) or coherent state (b) using homodyne or heterodyne detection on one mode of the EPR state. HOM: homodyne detection, HET: heterodyne detection.}
\end{figure}

For two-mode Gaussian states, the commonly used states in CV system are two mode squeezed vacuum states, which can also be noted as the Einstein-Podolsky-Rosen (EPR) states in CV system \cite{Einstein_PhysRev_Can}, of which the covariance matrix reads
\begin{equation}
  \gamma_{EPR}=\left[ \begin{aligned}cosh~2r~I_{2}~~~ sinh~2r~\sigma_z\\
    sinh~2r~\sigma_{z} ~~~ cosh~2r~I_{2}
    \end{aligned} \right]\\
    = \left[ \begin{aligned}V~I_{2}~~~ \sqrt{V^2-1}~\sigma_z\\
      \sqrt{V^2-1}~\sigma_{z} ~~~ V~I_{2}
      \end{aligned} \right],
\end{equation}
where
\begin{equation}
  I_2=\left[\begin{aligned} 1 ~~~0\\0~~~1 \end{aligned}\right],~\sigma_{z} = \left[\begin{aligned} &1 ~~~~~~0\\&0~~-1 \end{aligned}\right],
\end{equation}
$r$ is the squeezed ratio, and $V$ is called the variance of the EPR state.
In particular, performing homodyne detection on one of the modes of an EPR state results in the other mode being projected on a squeezed state, while performing heterodyne detection on one of the modes of an EPR state results in the other mode being projected on a coherent state \cite{Grosshans2003VirtualEA}, as shown in Fig. \ref{fig:EPR}.

\newcommand{\tabincell}[2]{\begin{tabular}{@{}#1@{}}#2\end{tabular}}
\begin{table*}[t]
\renewcommand{\arraystretch}{1.4}

\caption{\label{tab:table1}Current security status of the main one-way CV-QKD protocols.}
\begin{ruledtabular}
\begin{tabular}{c c c c c}

Protocol & State & Modulation & Measurement & Best Current-Available Security Proof
\\
\hline
F. Grosshans et al. \cite{Grosshans_PhysRevLett_2002}& Coherent & Gaussian & Homodyne & \tabincell{c}{Asymptotic Collective~\cite{Renner_PhysRevLett_2009}\\ \hl{Finite-Size Collective}~\cite{pirandola2021composable, pirandola2021limits, pirandola2023improved}}
\\
\hline
&  &  &  & Finite-Size~\cite{Leverrier_PhysRevLett_2015,Leverrier_PhysRevLett_2017, pirandola2021composable, pirandola2021limits, pirandola2023improved}
\\
C. Weedbrook et al. \cite{Weedbrook_PhysRevLett_2004} & Coherent & Gaussian & Heterodyne & $ K_{\text {coll }}^{\epsilon}(N) \approx K_{\text {coll}}^{\text {asympt }} \text { for practical } N $
\\
&  &  &  & $K^{\epsilon}(N) =0 \text { for practical } N $~\cite{Leverrier_PhysRevLett_2013}
\\
\hline
&  &  &  & Finite-Size~\cite{Furrer_PhysRevA_2014,Furrer_PhysRevLett_2012}
\\
N. J. Cerf et al. \cite{Cerf_PhysRevA_2001}& Squeezed & Gaussian & Homodyne &$K^{\epsilon}(N)>0$ for practical $N$
\\
&  &  &  & $\lim _{N \rightarrow \infty} K^{\epsilon}(N)<K_{\text {coll }}^{\text {asympt }}$
\\
\hline
A. Leverrier et al. \cite{Leverrier_PhysRevLett_2009}& Coherent & QPSK & Homodyne & Asymptotic Collective with linear assumption ~\cite{Leverrier_PhysRevLett_2009}
\\
Z. Li et al. \cite{Li_Arxiv_2018}& Coherent & QPSK, Arbitrary & Homo/Heterodyne & Asymptotic Collective~\cite{Li_Arxiv_2018}
\\
S. Ghorai et al. \cite{Ghorai_PhysRevX_2019}& Coherent & QPSK, Arbitrary \cite{Denys2021explicitasymptotic} & Homo/Heterodyne & Asymptotic Collective~\cite{Ghorai_PhysRevX_2019,Denys2021explicitasymptotic}
\\
J. Lin et al. \cite{Lin_PhysRevX_2019} & Coherent & QPSK & Homo/Heterodyne & Asymptotic Collective~\cite{Lin_PhysRevX_2019}
\\
\hline
V. Usenko et al. \cite{Usenko_PhysRevA_2015}& Coherent & Gaussian 1D & Homodyne & Finite-size \hl{Collective}~\cite{Liao_QuantumInfProc_2018}
\\
R. Garc\'{\i}a-Patr\'on et al. \cite{Garcia-Patron_PhysRevLett_2009}& Squeezed & Gaussian & Heterodyne & Asymptotic Collective~\cite{Garcia-Patron_PhysRevLett_2009}
\\
R. Filip \cite{vidiella2006continuous,Filip_PhysRevA_2008}& Thermal & Gaussian & Homo/Heterodyne & Asymptotic Collective~\cite{Papanastasiou_PhysRevA_2018}
\\
\hline
\tabincell{c}{J. Fiur\'a\ifmmode \check{s}\else \v{s}\fi{}ek  et al. \cite{Fiurasek_PhysRevA_2012} \\ N. Walk et al. \cite{Walk_PhysRevA_2013}} & Coherent & Gaussian & \tabincell{c}{Homo/Heterodyne + \\ Gaussian Post-selection} & Asymptotic Collective ~\cite{Fiurasek_PhysRevA_2012,Walk_PhysRevA_2013,Hosseinidehaj_PhysRevA_2020}
\\
Z. Li et al. \cite{Li_PhysRevA_2016}& Coherent & \tabincell{c}{Gaussian + Non-\\Gaussian Post-selection}& Homo/Heterodyne & Asymptotic Collective~\cite{Li_PhysRevA_2016} \\
L. S. Madsen et al. \cite{Madsen_NatCommun_2012}& Squeezed & \tabincell{c}{Gaussian + \\Additional Gaussian} & Homodyne & Asymptotic Collective~\cite{Madsen_NatCommun_2012}

\end{tabular}
\end{ruledtabular}
\end{table*}

\begin{table*}[t]
\renewcommand{\arraystretch}{1.4}

\caption{\label{tab:table2}Current security status of the main two-way and MDI CV-QKD protocols.}
\begin{ruledtabular}
\begin{center}
\begin{tabular}{ccccccc}
\multicolumn{1}{c}{Protocol} & \multicolumn{2}{c} { Alice's side } & \multicolumn{2}{c} { Bob's side } & \multicolumn{1}{c}{Measurement} & \multicolumn{1}{c}{Best Currently-Available} \\
\cline {2 -5} & State & Modulation & State & Modulation & & Security Proofs \\
\hline
S. Pirandola et al. \cite{Pirandola_NatPhys_2008} & Coherent & Gaussian & Coherent & Gaussian & Homo/Heterodyne & Finite-size ~\cite{Ghorai_PhysRevA_2019} \\
M. Sun et al. \cite{Sun_IntJQuantumInform_2012} & Coherent & Gaussian & Coherent & Gaussian & Homo/Heterodyne & Asymptotic ~\cite{Sun_IntJQuantumInform_2012} \\
\hline
Y. Zhao et al. \cite{Zhao_QuantumInfProc_2017} & Coherent & \tabincell{c}{Gaussian + Non- \\ Gaussian Post-selection} & Coherent & \tabincell{c}{Gaussian + Non- \\ Gaussian Post-selection} & Homodyne & Asymptotic collective ~\cite{Zhao_QuantumInfProc_2017} \\
\hline
C. Li et al. \cite{li2016performance} & Coherent & Gaussian & Coherent & Gaussian & \tabincell{c}{Homo/Heterodyne + \\ Gaussian Post-selection} & Asymptotic collective \\
\hline
Y. Bian et al. \cite{bian2021unidimensional}& Coherent & Gaussian $1D$ & Coherent & Gaussian & Homodyne &  Asymptotic collective~\cite{bian2021unidimensional} \\
\hline
\tabincell{c}{Z. Li et al. \cite{Li_PhysRevA_2014} \\ S. Pirandola et al. \cite{Pirandola_NatPhoton_2015}}
& Coherent & Gaussian & Coherent & Gaussian & Bell-state Measurement & \tabincell{c}{Finite-size ~\cite{Zhang_PhysRevA_2017,Lupo_PhysRevA_2018}\\$K^{\epsilon}(N)>0$ for practical $N$ \\ $\lim _{N \rightarrow \infty} K^{\epsilon}(N)<K_{\text {coll }}^{\text {asympt }}$}\\
\hline
Y. Zhang et al. \cite{Zhang_PhysRevA_2014} & Squeezed & Gaussian & Squeezed & Gaussian & Bell-state Measurement & \tabincell{c}{Finite-size~\cite{Chen_PhysRevA_2018}\\$K^{\epsilon}(N)>0$ for practical $N$ \\ $\lim _{N \rightarrow \infty} K^{\epsilon}(N)<K_{\text {coll }}^{\text {asympt }}$} \\
\hline
L. Huang et al. \cite{Huang_Entropy_2019} & Coherent & Gaussian $1D$ & Coherent & Gaussian $1D$ & Bell-state Measurement & Asymptotic collective~\cite{Huang_Entropy_2019,Bai_QuantumInfProc_2020} \\
\hline
H. Ma et al. \cite{Ma_PhysRevA_2019} & Coherent & QPSK & Coherent & QPSK & Bell-state Measurement & Asymptotic collective~\cite{Ma_PhysRevA_2019}\\
\hline
Y. Zhao et al. \cite{Zhao_PhysRevA_2018} & Coherent & \tabincell{c}{Gaussian + Non- \\ Gaussian Post-selection} & Coherent & \tabincell{c}{Gaussian + Non- \\ Gaussian Post-selection} & Bell-state Measurement & Asymptotic collective~\cite{Zhao_PhysRevA_2018}
\end{tabular}
\end{center}
\end{ruledtabular}
\end{table*}

%\subsection{CV-QKD protocols}

\subsection{A historical outline of CV-QKD protocols and the current security status}
% Implemented using off-the-shelf standard telecom components, CV protocols become another choice to accomplish QKD. Different from the discrete-variable protocols based on single-photon detection technology, CV protocols utilize interferometry to perform measurement, including homodyne detection and heterodyne detection. 
% In homodyne detection, the signal to be measured is mixed with a known reference signal, which is often called a local oscillator (LO), and the carrier frequency of the signal to be measured and the reference signal are the same, so that the interference light field obtained can eliminate the frequency noise of the electromagnetic wave itself. 

The first CV-QKD protocol was proposed in 1999, using squeezed states to achieve secret key distribution \cite{ralph1999continuous}. However, due to the challenges in preparing squeezed states, a protocol of using coherent states and homodyne detection to distribute secret key was proposed in 2002~\cite{Grosshans_PhysRevLett_2002}, namely the GG02 protocol. Because the coherent states can be easily generated by a laser, this protocol has received an increasing attention in recent years. Subsequently the protocol based on Gaussian modulated coherent states got further developments. The no-switching protocol was reported in 2004~\cite{Weedbrook_PhysRevLett_2004}, in which \hl{the} heterodyne detection instead of homodyne detection was used. 
In 2009, heterodyne detection was also utilized in the Gaussian modulated squeezed-state protocol and an improvement of performance was found~\cite{Garcia-Patron_PhysRevLett_2009}.

Although Gaussian modulated CV-QKD protocols have undergone extensive \hl{studies and developments}, there remain technical challenges to implementing ideal Gaussian modulation in experiments.
Therefore, discrete modulated CV-QKD protocol~\cite{Leverrier_PhysRevLett_2009} was proposed. The initial discrete modulated CV protocol is the four-state modulation protocol, where a modulation method similar to QPSK in classical communications was used. Soon afterwards, A. Leverrier~\cite{Leverrier_PhysRevA_2011} highlighted that CV protocols can be used for secret key distribution with multi-dimensional discrete modulation.  In addition to discrete modulation, other modulation methods are also considered, including unidimensional modulation~\cite{Usenko_PhysRevA_2015}. It simplifies the modulation process at the Alice side can be compared with the GG02 protocol when the excess noise is small. 
\hl{Recently, a phase-sensitive multimode protocol is proposed, which achieves higher secret key rate and better excess noise tolerance} \cite{su2023experimental}.
\hl{Most of the theoretical analysis of the CV-QKD protocols are based on fiber channels, and recently, the study is extended to the free space scenario for considerations of satellite quantum communications.} \cite{pirandola2021satellite}

Generally, the classification of standard one-way CV-QKD protocols, in which the quantum state passes through a single channel, can follow the type of the used quantum states  (coherent or squeezed), the methods of modulation (Gaussian modulation, unidimensional modulation or discrete modulation) or the type of measurement (homodyne or heterodyne). In addition to the one-way protocols, various other protocols correspond to different application scenarios are proposed, such as two-way protocols, source-device-independent protocols, \hl{MDI} protocols, and so forth \cite{goncharov2022rationale}.

In 2008, the original two-way protocol was proposed \cite{Pirandola_NatPhys_2008}, in which an optical switch was used to randomly switch between two working statuses ``ON'' or ``OFF''. The tolerable noise on ON mode is higher than that of one-way protocol. However, the use of optical switches cannot meet the demand for high-speed quantum key distribution. Subsequently in 2012, an improved two-way protocol was reported~\cite{Sun_IntJQuantumInform_2012}, in which Alice uses a Gaussian modulated coherent state and a beamsplitter to replace the ON-OFF switch and translation operation in the original two-way protocol. Very recently, the protocol is proved to be secure in the finite-size regime ~\cite{Ghorai_PhysRevA_2019}. In addition, the unidimensional two-way CV-QKD protocol is proposed to simplify the system realization and is proved to be secure against collective attack ~\cite{Bian_FiO_2020}.

It should be noted, however, that the one-way and two-way protocols can be considered theoretically secure only with the trustworthy equipments.
The issue of practical security remains a major concern due to the possible mismatch between practical devices and theoretical assumptions. 
An optimal solution would be a device-independent protocol, in which system security is not influenced by the trustworthiness of the devices.
However, since achieving comprehensive device independence is challenging, semi-device-independence protocols have been developed and well studied, including the MDI protocols~\cite{Pirandola_NatPhoton_2015,Li_PhysRevA_2014} and the source-device-independent (SDI) protocols~\cite{Weedbrook_PhysRevA_2013}. These protocols eliminate certain assumptions regarding the reliability of the devices and, thus, close the corresponding security loopholes.

The CV-MDI QKD was independently proposed by Z. Li et al.~\cite{Li_PhysRevA_2014} and S. Pirandola et al.~\cite{Pirandola_NatPhoton_2015}, in which Alice and Bob are both senders, preparing coherent states and sending them through two independent channels to the untrusted party, Charlie, to perform Bell-state measurement. Note that there are no assumptions about the trustworthiness on Charlie, which implies that Eve can have complete control over Charlie, and it can withstand all attacks that are based on detector's loopholes. 
However, the CV-MDI QKD protocol has a limited transmission distance which restricts the long-haul deployment. 
% Although the CV-MDI QKD protocol using squeezed states is proposed and is proved to get some improvement, the original protocols are still limited by transmission distance. 
In 2019, the discrete modulation is used for CV-MDI QKD~\cite{Ma_PhysRevA_2019}, whose secret key rate correspondingly decreases but can still guarantee its security against collective attack. The unidimensional CV-MDI QKD protocol was also proposed \cite{Huang_Entropy_2019,Bai_QuantumInfProc_2020}, in which more cases are considered in~\cite{Huang_Entropy_2019}, while the finite-size effect is involved in~\cite{Bai_QuantumInfProc_2020}. 
\hl{The security of the CV-MDI protocol under the source intensity errors is also investigated} \cite{wang2020continuous}.
Here, the current security proofs status of the two-way protocols and the MDI protocols are revealed in Table~\ref{tab:table2}.

\begin{figure}[b]
  \includegraphics[width=0.45\textwidth]{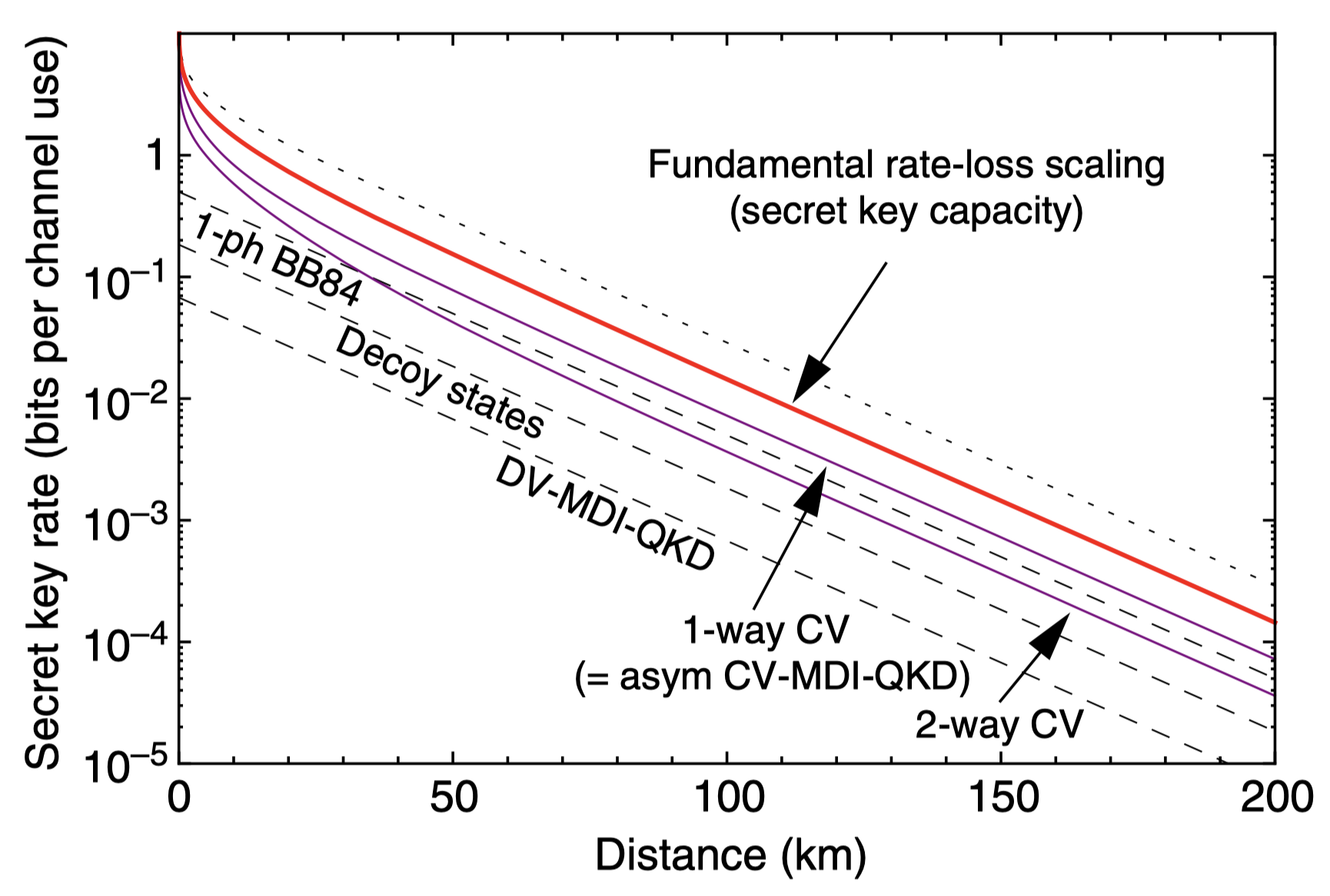}% Here is how to import EPS art
  \caption{\label{fig:PLOB} The PLOB bound and the distance between the PLOB bound and various QKD protocols in ideal case. From S. Pirandola et al. \cite{pirandola2017fundamental}. S. Pirandola et al., Nat. Commun., 8, 15043, 2017; licensed under a Creative Commons Attribution (CC BY) license.}
\end{figure}

Another semi-device-independent protocol, known as the SDI protocol, has also undergone development in addition to the MDI protocol. The CV-SDI QKD protocol was initially proposed in 2013 and has been demonstrated to be resistant to collective attacks \cite{Weedbrook_PhysRevA_2013}. Furthermore, Y. Zhang et al. supplemented the security proof in 2020 \cite{Zhang_SciRep_2020}. Similar to the CV-MDI QKD, the CV-SDI QKD protocol does not make any assumptions about the source's credibility. In this protocol, an entanglement source is positioned between Alice and Bob as both Alice and Bob are receivers.
In addition, there are also some special CV-QKD protocols such as using a thermal state as the source, so called the passive protocol \cite{qi2018passive}. 
% The passive protocol provides simpler system structure and can reduce the security loopholes caused by modulation.

\hl{As} shown in Fig. \ref{fig:PLOB}, ideal CV-QKD protocols are closer to the theoretical limit, known as the PLOB bound \cite{pirandola2017fundamental}, which shows the potential of achieving high secret key rate and long transmission distance using CV-QKD. 
But we have to remark that, there is still a gap between the optimal parameters of the existing CV-QKD protocols for practical and ideal situations, where the reconciliation efficiency is normally less than 100 \% and the optimal modulation variance is limited to less than 5. Therefore the performance in a practical CV-QKD system still has room to be improved, and more CV-QKD protocols are expected to be proposed for further approaching the theoretical limit.

\subsection{Security analysis}
\begin{figure}[t]
  \includegraphics[width=0.45\textwidth]{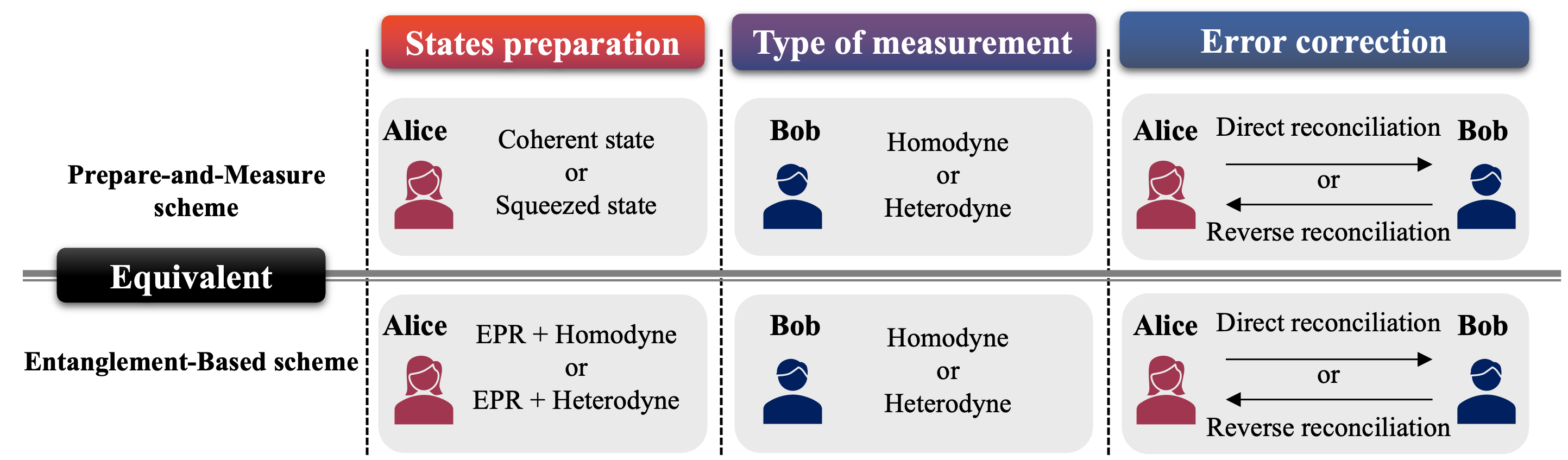}% Here is how to import EPS art
  \caption{\label{fig:PMandEB} The PM protocol and the equivalent EB protocol in CV-QKD. The most important steps in a CV-QKD protocol are the state preparation, the measurement and the error correction. The main difference between these two types of the protocols are the state preparation, where the coherent state preparation is equal to the heterodyne detection on one of the modes of the EPR state, and the squeezed state preparation is equal to the homodyne detection on one of the modes of the EPR state.}
\end{figure}

Before starting the security analysis, it is essential to introduce two equivalent schemes, namely the prepare-and-measurement (PM) scheme and the entanglement-based (EB) scheme, as shown in Fig. \ref{fig:PMandEB}. 
These schemes differ mainly in the method of preparing quantum states. In the PM scheme, Alice generates the states with a light resource, usually a laser, whereas in the EB scheme, Alice generates EPR states. As homodyne (heterodyne) detection performed on one mode of the EPR state has the effect of projecting the other mode onto a squeezed (coherent) state, the transmitter's outputs of both methods are identical for the third party. Therefore, both schemes are equivalent to Bob and Eve, which is an important property. 

It is important to note that although both schemes are equivalent, they are used in different application scenarios. For instance, the PM scheme is used for experimental implementation, whereas the EB scheme is used for theoretical analysis due to its ease of calculation. Our analysis is based on the EB scheme in the following part.

Reconciliation is the key technique to make Alice and Bob share a same bit string, which can be categorized into two types, \hl{the direct reconciliation and the reverse reconciliation} \cite{pirandola2009direct}.
In direct reconciliation, the detection data is corrected to the modulation data, however, the tolerable channel loss is limited under 3 dB. Reverse reconciliation is proposed to solve this issue, by correcting the modulation data to the detection data.

In the aspect of an eavesdropper, the attack strategy can be categorized into 3 types, individual attack, collective attack and coherent attack. In individual attack, Eve manipulates and measures the transmitted states independently and identically. In collective attack, Eve manipulates the transmitted states independently and identically, but measures them jointly. While for coherent attack, Eve manipulates and measures the transmitted states jointly. Usually, coherent attack is the strongest one, while in asymptotic case the collective attack is normally the most powerful coherent attack. 

% As we know, in the CV protocols, the squeezed or coherent states are prepared by Alice points out that the states can be equivalently prepared by a pair of quantum entangled beams with measurement on one (or both) quadratures of one beam. Now let us assume that Alice prepares a pair of EPR beams, and use $\left(X_1,P_1\right)$ to represent for the quadratures of the beam sent to Bob and $\left(X_0,P_0\right)$ to represent for the quadratures of the beam kept by Alice. Owing to ease to analyze, the beams are set to be $\left\langle X_0^{2}\right\rangle=\left\langle X_1^{2}\right\rangle=\left\langle P_0^{2}\right\rangle=\left\langle P_1^{2}\right\rangle=V N_{0}$, where $N_0$ is the shot-noise variance.

\subsubsection{Gaussian-modulated protocol }

\begin{table*}[t]
\renewcommand{\arraystretch}{1.8}

\caption{\label{tab:table7} Main calculation equations of secret key rate of Gaussian-modulated CV-QKD protocols. }
\begin{ruledtabular}
\begin{scriptsize}
\begin{center}
\begin{tabular}{ccccccc}
States & Measurement &  $I_{AB}$ & \tabincell{c}{Eigenvalues of the \\Covariance Matrix\\ before Measurement} &\tabincell{c}{Type of \\ Reconciliation}&\tabincell{c}{Eigenvalues of the \\Covariance Matrix\\ after Measurement}&$\chi_{B E}$\\
\hline
\multirow{2}{*}{Squeezed} & \multirow{2}{*}{Homodyne} &  \multirow{2}{*}{$\frac{1}{2}\log_2{\frac{V+\chi}{\chi+1/V}}$}  &  \multirow{8}{*}{~}  & Direct  & $\lambda_3=\sqrt{\frac{T^2\left(V+\chi\right)}{\left(\chi+1/V\right)}}$ & $\sum\limits_{i=1}^{2}G\left(\lambda_i\right)-G\left(\lambda_3\right)$ \\
\cline{5-7}
& & & & Reverse &$\lambda_4=\sqrt{\frac{V\left(V\chi+1\right)}{\left(\chi+V\right)}}$ &$\sum\limits_{i=1}^{2}G\left(\lambda_i\right)-G\left(\lambda_4\right)$\\
\cline{1-3}\cline{5-7}
\multirow{2}{*}{Coherent} & \multirow{2}{*}{Homodyne} &  \multirow{2}{*}{$\frac{1}{2}\log_2{\frac{V+\chi}{\chi+1}}$}  &  &  Direct & $\lambda_{5,6}=\sqrt{\frac{1}{2}\left[A\pm\sqrt{A^2-4B}\right]}$ & $\sum\limits_{i=1}^{2}G\left(\lambda_i\right)-\sum\limits_{i=5}^{6}G\left(\lambda_i\right)$ \\
\cline{5-7}
& & &$\lambda_{1,2}=\sqrt{\frac{1}{2}\left[\mathrm{\Delta}\pm\sqrt{\mathrm{\Delta}^2-4D^2}\right]}$ & Reverse &${\lambda_7=\lambda}_4=\sqrt{\frac{V\left(V\chi+1\right)}{\left(\chi+V\right)}}$ &$\sum\limits_{i=1}^{2}G\left(\lambda_i\right)-G\left(\lambda_7\right)$ \\
\cline{1-3}\cline{5-7}
\multirow{2}{*}{Squeezed} & \multirow{2}{*}{Heterodyne} &  \multirow{2}{*}{$\frac{1}{2}\log_2{\frac{T(V+\chi)+1}{T(\chi+1/V)+1}}$}  &  &  Direct & $\lambda_8=\lambda_3=\sqrt{\frac{T^2\left(V+\chi\right)}{\left(\chi+1/V\right)}}$ & $\sum\limits_{i=1}^{2}G\left(\lambda_i\right)-G\left(\lambda_8\right)$ \\
\cline{5-7}
& & & & Reverse&$\lambda_{9,10}=\sqrt{\frac{1}{2}\left[A'\pm\sqrt{{A'}^2-4B'}\right]}$ &$\sum\limits_{i=1}^{2}G\left(\lambda_i\right)-\sum\limits_{i=9}^{10}G\left(\lambda_i\right)$\\
\cline{1-3}\cline{5-7}
\multirow{2}{*}{Coherent} & \multirow{2}{*}{Heterodyne} &  \multirow{2}{*}{$\log_2{\frac{T(V+\chi)+1}{T(\chi+1)+1}}$}  &  &  Direct & $\lambda_{11}=T\left(\chi+1\right)$ & $\sum\limits_{i=1}^{2}G\left(\lambda_i\right)-G\left(\lambda_{11}\right)$ \\
\cline{5-7}
& & & & Reverse &$\lambda_{12}=\frac{T\left(V\chi+1\right)+1}{T\left(V+\chi\right)+1}$ &$\sum\limits_{i=1}^{2}G\left(\lambda_i\right)-G\left(\lambda_{12}\right)$\\
\hline Notice&\multicolumn{6}{l}{\tabincell{l}{$\left(\begin{array}{cc}
a ~ {I_2} & c ~ \sigma_{z} \\
c ~ \sigma_{z} & b ~ {I_2}
\end{array}\right)=\left(\begin{array}{cc}
V ~ {I_2} & \sqrt{T\left(V^{2}-1\right)} ~ \sigma_{z} \\
\sqrt{T\left(V^{2}-1\right)} ~ \sigma_{z} & T(V+\chi) ~ {I_2}
\end{array}\right),\Delta=a^{2}+b^{2}-2 c^{2},D=ab-c^2,A=\frac{1}{a+1} \left(a+bD+\Delta\right), A'=\frac{1}{b+1} \left(b+aD+\Delta\right),B=\frac{D}{a+1} \left(b+D\right)$,\\$B'=\frac{D}{b+1} \left(a+D\right),
G\left(x\right)= \frac{x+1}{2}\text{log}_2\frac{x+1}{2}-\frac{x-1}{2}\text{log}_2 \frac{x-1}{2}$, $V$ is the variance of the EPR state which is also written as $V_A$ when representing the variance of mode $V$.}}\\
\end{tabular}
\end{center}
\end{scriptsize}
\end{ruledtabular}
\end{table*}

% Among all the CV protocols, the protocols based on Gaussian modulation is undoubtedly the most basic but important because the characters of Gaussian states make these protocols easy to be theoretically analyzed. From the foregoing, it can be easily known that the CV protocols based on the Gaussian modulated states can be divided into four types according to the types of quantum states (coherent states or squeezed states) and detection methods (homodyne detection or heterodyne detection). 
% Although in the theoretical simulation the protocols based on Gaussian modulated squeezed states perform better than those based on the coherent states (probably because the thermal noise in the squeezed quadrature is declined), there are still many technical challenges to prepare ideal squeezed states. 
% In this section, the coherent-state CV protocols and these security analysis method are focused. Since the main meaning of the theoretical security of the protocols is that the security of the protocols is guaranteed under the collective attack, in this section we will introduce the security proof method of the coherent-state CV protocols under the attack in detail, and the cases with squeezed states can be analyzed in parallel.
The protocols based on Gaussian modulation are the most fundamental among all CV-QKD protocols, which are usually categorized into four types based on types of quantum states (coherent or squeezed) and detection methods (homodyne or heterodyne).
So far, the protocols based on Gaussian modulated coherent states which can be generated by the off-the-shelf laser have received much attentions. Therefore, we detail the security analysis of the Gaussian-modulated coherent state protocol, and briefly introduce the squeezed-state protocol.

The security indicator is the secret key rate. If the secret key rate is greater than zero while considering attacks, then Alice and Bob can distill a secret key under the attack. Otherwise, the protocol will not be effective.
The asymptotic secret key rate with reverse reconciliation can be given by \cite{Devetak2003DistillationOS}
\begin{equation}\label{Eq.5}
R=\beta I_{AB}-S_{BE},
\end{equation}
where $I_{AB}$ is the classical mutual information between Alice and Bob, that can be easily estimated, normally written as
\begin{equation}
  \label{IAB}
  \begin{split}
  &I_{AB}^{hom}=\frac{1}{2}\text{log}_2\frac{V_B}{V_{B|A}}=\frac{1}{2}\text{log}_2\frac{V+\chi_{line}}{1+\chi_{line}},\\
  &I_{AB}^{het}=\text{log}_2\frac{V_B+1}{V_{B|A}+1}=\text{log}_2\frac{T\left(V+\chi_{line}\right)+1}{T\left(1+\chi_{line}\right)+1},
  \end{split}
\end{equation}
for \hl{perfect} homodyne and heterodyne detection. 
Here $V$ is the variance of EPR state owned by Alice, $V_B$ is the variance of the state Bob receives, $V_{A|B}$ is the conditional variance given Bob's measurement result. $\chi_{line}=1/T-1+\varepsilon$ is the total channel-added noise expressed in shot noise units (SNUs), relevant to the channel parameters, including the transmittance ($T$) and \hl{excess noise} ($\varepsilon$) \cite{pirandola2018theory}. 
Since mainstream CV-QKD implementations use reverse reconciliation, where the modulation data is corrected to the detection data, the quantum mutual information between Bob and Eve, $S_{BE}$, is the concern.

It is pointed out that the maximum information available to Eve is bounded by Holevo quantity \cite{Holevo1973BoundsFT}:

\begin{equation}\label{Eq.6}
  S_{BE}=\chi_{BE}=S\left(E\right)-S\left(E|m_B\right),\\
\end{equation}
where $S\left(E\right)$ is the von Neumann entropy of the eavesdropper's state $\rho_E$, and $S\left(E|m_B\right)$ is von Neumann entropy conditional on Bob's measurement result and is determined by the detection method. Based on the fact that the eavesdropper Eve is able to purify the binary quantum system $\rho_{AB}$, Eq. (\ref{Eq.6}) becomes

\begin{equation}
\begin{split}
\chi_{BE}=S\left(\rho_{AB}\right)-S\left(\rho_{AB|m_B}\right).\\
\end{split}
\end{equation}

As for $\chi_{BE}$, usually we use the Gaussian extremity theorem to find its upper bound \cite{Wolf_2006_PRL}, corresponding to the lower bound of the secret key rate. This means there always exists a Gaussian state $\rho^G_{AB}$ with the same covariance matrix as $\rho_{AB}$ who makes  
\begin{equation}
  \chi_{BE} \leq \chi_{BE}^G=S\left(\rho^{G}_{AB}\right)-S\left(\rho^{G}_{AB|m_B}\right).
\end{equation}

Specifically, calculations of $S\left(\rho^{G}_{AB}\right)$ and $S\left(\rho^{G}_{AB|m_B}\right)$ can be simplified using the symplectic eigenvalues of the covariance matrix $\gamma_{AB}$ and $\gamma_{AB|m_B}$ respectively, corresponding to the states $\rho_{AB}$ and $\rho_{AB|m_B}$. 
Covariance matrix $\gamma_{AB}$ can be estimated from the modulation and detection data 
\begin{equation}
  \label{Cov}
  \gamma_{AB} = 
  \begin{pmatrix}
    \gamma_A & \sigma_{AB} \\
    \sigma_{AB} & \gamma_B
   \end{pmatrix}
   = \begin{pmatrix}
    V_A~I_2 & C_{AB} ~ \sigma_z \\
    C_{AB} ~ \sigma_z & V_B~I_2
   \end{pmatrix}
   = \begin{pmatrix}
    a ~ I_2 & c ~ \sigma_z \\
    c ~ \sigma_z & b ~ I_2
   \end{pmatrix}.
\end{equation}
then $\gamma_{AB|m_B}$ is given by
\begin{equation}
\gamma_{AB|m_B}=\gamma_A-\sigma_{AB}^{T}H\sigma_{AB}.
\end{equation}
Here, $H$ is the symplectic matrix on behalf of the measurement operation on mode $B$, and can be written respectively:
\begin{equation}
\begin{split}
&H^{hom}=\left(X\gamma_B X\right)^{MP},\\
&H^{het}=\left(\gamma_B+I_2\right)^{-1},
\end{split}
\end{equation}
where $X=\left(\begin{array}{cc}1 &0 \\0&0\end{array}\right)$ (for homodyne detection on $p$ quadrature, we use $P=\left(\begin{array}{cc}0 &0 \\0&1\end{array}\right)$), and $MP$ represents for the Moore-Penrose pseudo-inverse of a matrix. 
Notice that $\gamma_A$, $\gamma_B$ and $\sigma_{AB}$ are the subsmatrices of the covariance matrix $\gamma_{AB}$ as detailed in Eq. (\ref{Cov}).

Generally, the symplectic eigenvalues of a covariance matrix $\gamma$ can be achieved by calculating the eigenvalues of $i\Omega \gamma$. For the two-mode system, one can easily achieve the analytical solution, as detailed in Table. \ref{tab:table7}, where $S\left(\rho^G_{AB}\right)$ can be calculated with $\lambda_{1,2}$, and $S\left(\rho^G_{AB|m_B}\right)$ can be calculated with $\lambda_7$ for homodyne detection and $\lambda_{12}$ for heterodyne detection.
\begin{equation}
\begin{split}
&\chi_{BE}^{hom}=\sum_{i=1}^{2}G\left(\lambda_i \right)-G\left(\lambda_7 \right),\\
&\chi_{BE}^{het}=\sum_{i=1}^{2}G\left(\lambda_i \right)-G\left(\lambda_{12} \right),
\end{split}
\end{equation}
where 
\begin{equation}
  G\left(x\right)= \frac{x+1}{2}\text{log}_2\frac{x+1}{2}-\frac{x-1}{2}\text{log}_2 \frac{x-1}{2}.
\end{equation}
% $\lambda_{1,2}$ are the symplectic eigenvalues of the covariance matrix $\gamma_{AB}$ corresponding to the states $\rho_{AB}$, 
% which can be written by:
% \begin{equation}
% \gamma_{AB}=\left[\begin{array}{cc} aI_2 & c\sigma_z\\c\sigma_z& bI_2\end{array}\right]=\left[\begin{array}{cc} \gamma_A & \sigma_{AB}^{T}\\\sigma_{AB}& \gamma_B\end{array}\right],
% \end{equation}
% with $\sigma_z=\left[\begin{array}{cc}1 &0\\0& -1\end{array}\right]$. 
% $\lambda_{3,4}$ is the symplectic eigenvalues of the covariance matrix $\gamma_{AB|m_B}$ corresponding to the states $\rho_{AB|m_B}$, which is given by

% For the two-mode system, it is possible to achieve the analytical solutions of the symplectic eigenvalue:

% \begin{equation}\label{Eq.7}
% \begin{split}
% \lambda_{1,2}^2&=\frac{1}{2}\left[\Delta\pm\sqrt{\Delta^2-4D}\right],\\
% \lambda_3&=\sqrt{a\left(a-c^2/b\right)},\\
% \lambda_4&=a-c^2/\left(b+1\right).
% \end{split}
% \end{equation}
% with
% \begin{equation} \label{eq:Delta}
% \Delta=a^2+b^2-2c^2, D=ab-c^2.
% \end{equation}

Similarly, the case with direct reconciliation or the security analysis of squeezed states protocols can be analyzed, for simplicity, the analytical solutions are concluded in Table. \ref{tab:table7}.
% where Bob's data is corrected to Alice's data. The secret key rate can be given by
% \begin{equation}
% R=\beta I_{AB}-\chi_{AE},
% \end{equation}
% and the calculation for $\chi_{AE}$ can be derived by
% \begin{equation}
% \begin{split}
% \chi_{AE}^{hom}=\sum_{i=1}^{2}G\left(\lambda_i \right)-G\left(\lambda_5 \right),\\
% \chi_{AE}^{het}=\sum_{i=1}^{2}G\left(\lambda_i \right)-G\left(\lambda_6 \right),
% \end{split}
% \end{equation}
% where $\lambda_{1,2}$ keep unchanged, and $\lambda_{5,6}$ are
% \begin{equation}
% \begin{split}
% \lambda_5=\sqrt{b\left(b-c^2/a\right)},\\
% \lambda_6=b-c^2/\left(a+1\right).
% \end{split}
% \end{equation}

\subsubsection{Discrete-modulated protocol}

The earliest investigation on discrete-modulated CV-QKD is in 2009, when the transmission distance of a CV-QKD system is limited to less than 30 km due to the lack of the \hl{efficient error correction strategy of Gaussian variable at low SNR}, and the proposed discrete-modulated protocol reduces the requirement of error correction, which can be a promising way to enhance the system transmission distance \cite{Leverrier_PhysRevLett_2009}. 

The early security proof requires the linear channel assumption, where the input-output relations of the quadrature operators in Heisenberg representation are given by
\begin{equation}
  X_{out} = g_X X_{in} + B_X, P_{out} = g_P P_{in} + B_P.
\end{equation}
Here the added noises $B_X,$ $B_P$ are uncorrelated with the input quadratures $X_{in}$, $P_{in}$.
Only based on this assumption, the covariance matrix describing the system after the quantum state transmission can be written as 
\begin{equation}
  \label{eq:MatSecurity}
  \gamma = \begin{pmatrix}
    (V_M+1)~I_2 & \sqrt{T}Z~\sigma_z \\
     \sqrt{T}Z~\sigma_z & (TV_M+1+T \epsilon )~I_2
   \end{pmatrix},
\end{equation}
where $V_M$, $T$, and $\epsilon$ correspond, respectively, to modulation variance ($V_M = V - 1$, $V$ is the variance of mode $A$), the channel transmittance, and the excess noise estimated experimentally in the prepare and measure scenario. Here, $Z$ is a function of $V_M$.
Once the covariance matrix is achieved, the Gaussian extremity theorem can be used to calculate the upper bound of the eavesdropper's knowledge, which is similar to the security analysis method in Gaussian-modulated protocol. Then, the lower bound of secret key rate can be derived. 

The linear channel assumption is adopted since the quadrature measurement of one mode of an EPR state cannot directly map the other mode to a group of discrete-modulated coherent states.
This results in the lack of covariance item $Z$ in the covariance matrix, leading to the need for an additional assumption.  
If considering general case without linear channel assumption, the unknown of $Z$ will lead to the worst-case estimation of the secret key rate, which is zero. 
Since the security proof of the discrete-modulated CV-QKD is not perfect, 
% while the proposal of multidimensional reconciliation significantly enhances the reconciliation efficiency of a Gaussian modulated system with low signal-to-noise ratio, and successfully supports the demonstration of a 80 km Gaussian modulated CV-QKD system \cite{Jouguet_NatPhotonics_2013}, 
the discrete-modulated CV-QKD system is less concerned for a period of time.

This situation did not change until 2018, as the security proof of discrete-modulated protocols gradually improved and no longer required linear channel assumptions \cite{Li_Arxiv_2018}. 
The eavesdropping behavior can be bounded through methods such as the uncertainty principle \cite{Li_Arxiv_2018}, the semidefine programming \cite{Ghorai_PhysRevX_2019,Lin_PhysRevX_2019}, and the entropy uncertainty \cite{matsuura2021finite}, then the secret key rate of the protocol can be obtained by searching for the lower bound.
% Moreover, the structure of the CV-QKD system has also undergone several development, which has attracted widespread attention to discrete-modulated systems that are closer to coherent optical communication, and has developed rapidly in recent years.

Introducing an auxiliary mode into the EB scheme is a feasible strategy to continuously use the security analysis method based on covariance matrix \cite{Li_Arxiv_2018}.
Before sending the mode into the unsecured quantum channel, the mode is divided by a beamsplitter, and one of the output mode is preserved by the sender. 
This preserved mode can be used to construct correlations between Alice and Bob.
Therefore, the covariance matrix describing the system contains 3 modes, which weakens the negative influence of the undefined covariance on secret key rate. 
This works for arbitrary discrete modulation formats, specifically, 256 QAM with Gaussian probability shaping performs close to the Gaussian modulated protocol, and the 64 QAM system can still support the transmission of more than 100 km. 
% The security framework can approach the Gaussian modulation, and can support any modulation format. 

Specifically, after the transmission of the quantum state, the covariance matrix of the quantum state shared by Alice and Bob can be written as
\begin{equation}
\gamma_{\mathbf{A C} B}=\left(\begin{array}{ccc}
\gamma_{\mathbf{A}} & \phi_{\mathbf{A C}} & \kappa_{\mathbf{A} B} \\
\phi_{\mathbf{A C}}^{T} & \gamma_{\mathbf{C}} & \phi_{\mathbf{C} B} \\
\kappa_{\mathbf{A} B}^{T} & \phi_{\mathbf{C} B}^{T} & \gamma_{B}
\end{array}\right),
\end{equation}
where $\gamma_\mathbf{A}$, $\gamma_\mathbf{C}$, and $\phi_\mathbf{AC}$ can be theoretically calculated, $\phi_{\mathbf{C}B}$, $\gamma_B$ can be estimated through the measured data, and only the covariance term $\kappa_{\mathbf{A}B}$ is unknown. Since the covariance matrix $\gamma_N$ for a $N$-mode state is constrained by the uncertainty principle (Eq. (\ref{eq:Uncertainty})).
The constraint set $S_\kappa$ which limits the possible value of $\kappa_{\mathbf{A}B}$ can be obtained as
\begin{equation}
S_{\kappa}=\left\{\phi_{\mathbf{A} B} \mid \gamma_{\mathbf{A C} B}\left[\kappa_{\mathbf{A} B}=\phi_{\mathbf{A} B}\right]+i \Omega_{N} \geq 0\right\}
\end{equation}
and the maximum value of $S_{B E}^{G}$ can be obtained by searching in the set.

Overall, the secret key rate at the asymptotic limit can be calculated as follow:
\begin{equation}
K=\beta I\left(M_{\mathbf{A}}: H_{B}\right)-\sup _{\kappa_{\mathbf{A} B} \in S_{\kappa}} S_{B E}^{G}\left(\kappa_{\mathbf{A} B}\right),
\end{equation}
where $I(M_\mathbf{A}:H_B)$is the classical mutual information, and $\beta$ is the reconciliation efficiency.

In 2019, S. Ghorai et al. proposed another security framework based on formulating the above problem as a semidefinite program (SDP) and focused on analyzing the security of a QPSK protocol \cite{Ghorai_PhysRevX_2019}. In this framework, SDP is used to search a state bounded by specially designed statistics between modulation and detection data.  It can be extended to a high-order modulation format and can be derived with an analytical solution of secret key rate in a symmetric modulation case \cite{Denys2021explicitasymptotic}. 

Soon later in 2019, J. Lin et al. proposed another strategy to analyze the security of a discrete-modulated CV-QKD protocol with SDP \cite{Lin_PhysRevX_2019}. The security of two QPSK protocols, with homodyne and heterodyne detection respectively are analyzed. 
It adopts a mapping and postselection based on the detection data to construct the positive operator valued measurement of the receiver.
In principle, this security framework can also be extended to high-order modulation format, but practically, it is limited by the demand of high computational resource. 

In the method proposed by J. Lin et al. \cite{Lin_PhysRevX_2019}, the secret key rate under collective attacks in the asymptotic limit can be can rewritten as
\begin{equation}
K=\min _{\rho_{A B} \in \mathbf{S}} D\left(\mathcal{G}\left(\rho_{A B}\right) \| \mathcal{Z}\left[\mathcal{G}\left(\rho_{A B}\right)\right]\right)-p_{\text {pass }} \delta_{\mathrm{EC}}
\end{equation}
where $D(\sigma\|\tau):=\operatorname{tr}(\sigma log\sigma)-\operatorname{tr}(\sigma log\tau)$ is the quantum relative entropy, $\mathcal{G}$ is a completely positive map related to the postselection in terms of actions on the bipartite state $\rho_{AB}$ and $\mathcal{Z}$ is a completely positive trace preserving map related to the key map, $\delta_{\mathrm{EC}}$ stands for the actual amount of information leakage per signal and $p_{\text {pass }}$ is the sifting probability.

In the transformed formula, only the first term is unknown, so the rate calculation problem has become a convex optimization problem for solving the minimum value of the first term in the formula. The set $\mathbf{S}$ is the feasible set of the optimization problem, which contains all bipartite density operators $\rho_{AB}$ that are compatible with experimental observations.

From the measurement of Bob, the expectation values of the first and second moments of the quadrature operators $\langle\hat{x}\rangle$, $\langle\hat{x}^2\rangle$, $\langle\hat{p}\rangle$, $\langle\hat{p}^2\rangle$ can be obtained. In addition, the operators $\hat{n}=\hat{a}^{\dagger}\hat{a}$ and $\hat{d}=\hat{a}^{2}+(\hat{a}^{\dagger})^2$ to constrain $\rho_{AB}$.

\begin{table}[t]
  \renewcommand{\arraystretch}{1.8}
  \caption{\label{tab:DMCVQKD} The security framework of discrete-modulated CV-QKD.}
  \begin{ruledtabular}
  \begin{scriptsize}
  \begin{center}
    \begin{tabular}{ccc}
      Security analysis  methods     & Years & Representative results                               \\ \hline
      Uncertainty principle \cite{Li_Arxiv_2018}              & 2018  & Arbitrary formats, 256 QAM, QPSK et al.   \\ \hline
      Linear semidefinite program \cite{Ghorai_PhysRevX_2019} & 2019  & Arbitrary formats, 256 QAM, QPSK et al.   \\ \hline
      Nonlinear semidefinite program \cite{Lin_PhysRevX_2019} & 2019  & 12-state double ring, QPSK et al.    \\ \hline
      Entangled photon pairs  \cite{matsuura2021finite}       & 2021  & 2-state  
      \end{tabular}
  \end{center}
  \end{scriptsize}
  \end{ruledtabular}
\end{table}

The relevant optimization problem is 
\begin{equation}
  \operatorname{minimize} D\left(\mathcal{G}\left(\rho_{A B}\right) \| \mathcal{Z}\left[\mathcal{G}\left(\rho_{A B}\right)\right]\right),
\end{equation}
subject to
\begin{equation}
\left\{
\begin{aligned}\label{eq:1}
&\operatorname{Tr}\left[\rho_{A B}\left(|k\rangle\left\langle\left. k\right|_{A} \otimes \hat{x}\right)\right]=p_{k}\langle\hat{q}\rangle_{k},\right. \\
&\operatorname{Tr}\left[\rho_{A B}\left(|k\rangle\left\langle\left. k\right|_{A} \otimes \hat{p}\right)\right]=p_{k}\langle\hat{p}\rangle_{k},\right. \\
&\operatorname{Tr}\left[\rho_{A B}\left(|k\rangle\left\langle\left. k\right|_{A} \otimes \hat{n}\right)\right]=p_{k}\langle\hat{n}\rangle_{k},\right. \\
&\operatorname{Tr}\left[\rho_{A B}\left(|k\rangle\left\langle\left. k\right|_{A} \otimes \hat{d}\right)\right]=p_{k}\langle\hat{d}\rangle_{k},\right. \\
&\operatorname{Tr}\left[\rho_{A B}\right]=1 \text {, } \\
&\operatorname{Tr}_{B}\left[\rho_{A B}\right]=\sum_{i, j=0}^{3} \sqrt{p_{i} p_{j}}\left\langle\varphi_{j} \mid \varphi_{i}\right\rangle|i\rangle\left\langle\left. j\right|_{A},\right. \\
&\rho_{A B} \geq 0, \\
\end{aligned}
\right.
\end{equation}
where $k\in \left\{0,1,...M \right\}$, $p_k$ is the probability of sending the corresponding state, $M$ is the modulation order, and $\langle\hat{q}\rangle_{k}$, $\langle\hat{p}\rangle_{k}$, $\langle\hat{n}\rangle_{k}$, $\langle\hat{d}\rangle_{k}$ denote the corresponding expectation values of operators $\hat{x}$, $\hat{q}$, $\hat{n}$, $\hat{d}$ for the conditional state $\rho_B^x$. 
This security analysis framework is further developed to higher modulation level with various formats \cite{almeida2021secret,lupo2022quantum,lin2020trusted,liu2021homodyne,upadhyaya2021dimension,kanitschar2022optimizing,wang2023discrete,kanitschar2023finite}.

\begin{figure}[t]
  \includegraphics[width=0.4\textwidth]{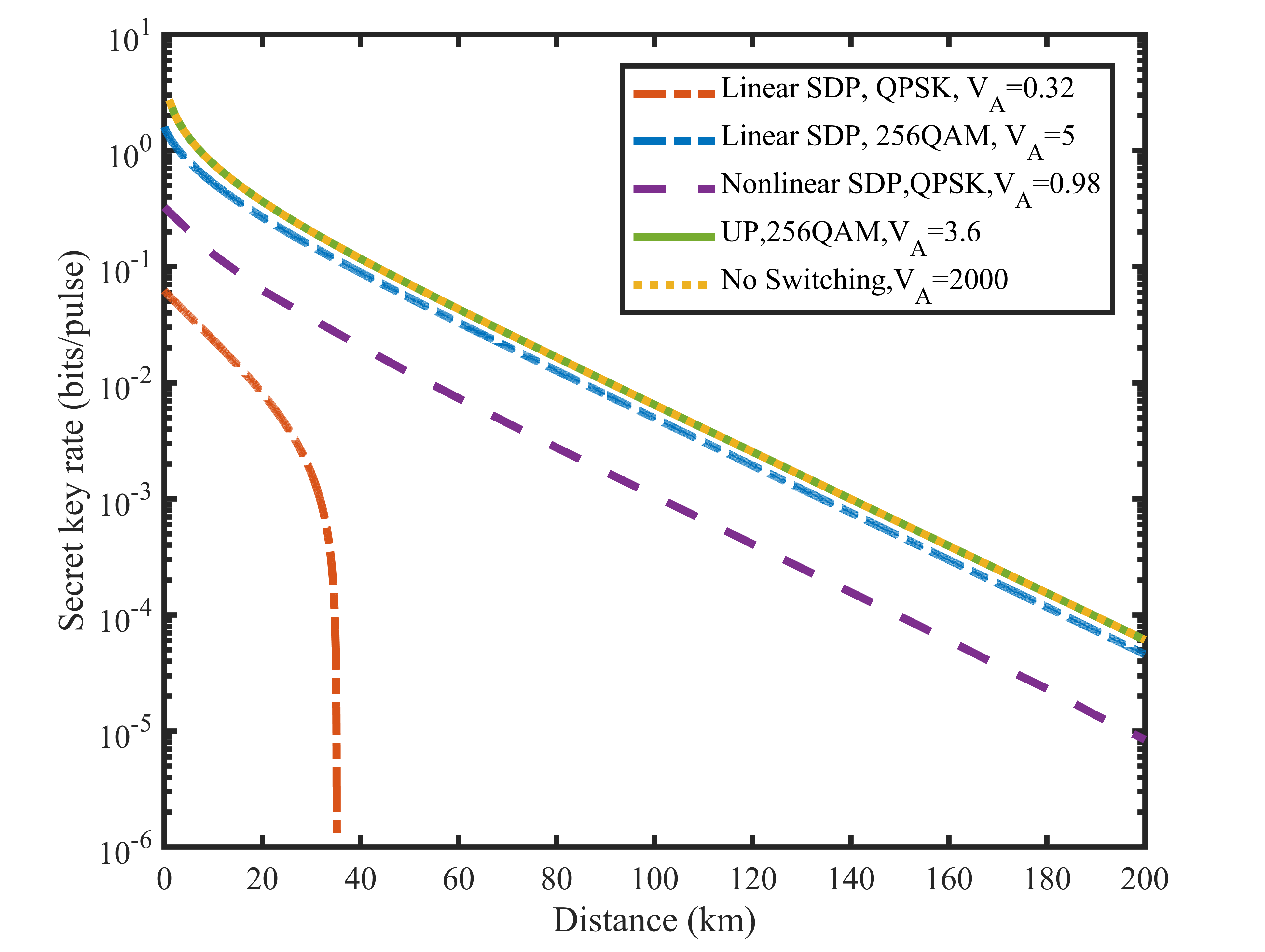}% Here is how to import EPS art
  \caption{\label{fig:DM} {The performance of different heterodyne detected discrete-modulated CV-QKD security analysis, including the linear SDP (red and blue line), the nolinear SDP (purple line) and the security analysis using uncertainty principle (UP) (green line). The modulation variance are the optimal for different security analysis. The performance of the no-switching protocol is also presented (yellow line).}}
\end{figure}

In addition to the security analysis methods we mentioned above, there also exists some other strategies that can reach security under finite-size effect, \hl{such as the binary modulated protocol} \cite{yamano2022finite}, the QPSK protocol using the entropy accumulation \cite{bauml2023security} and the \hl{discrete alphabet CV-QKD} \cite{papanastasiou2021continuous}.
The recent progress of the security analysis of discrete-modulated protocol is shown in Table. \ref{tab:DMCVQKD} and the performance is shown in Fig. \ref{fig:DM}. In conclusion, the mainstream strategy to provide a general security analysis of the discrete-modulated protocol is to introduce more correlation between Alice and Bob, which contributes to a tighter bound on the possible eavesdropping behavior. 

\subsubsection{Practical protocol with trusted noise}

\begin{figure}[tbp]
  \includegraphics[width= 0.45 \textwidth]{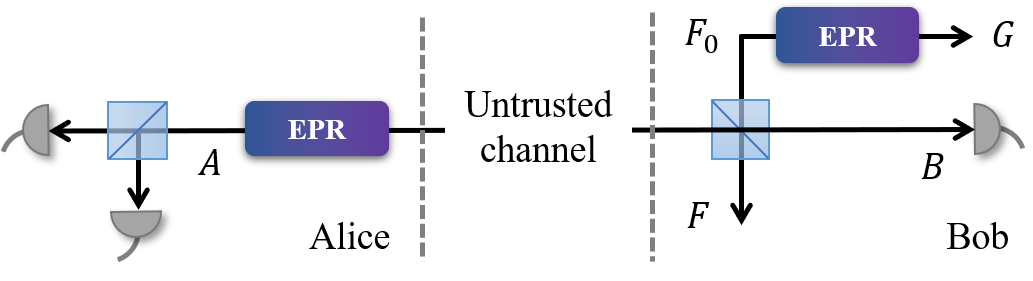}% Here is how to import EPS art
  \caption{\label{fig:TruDet} The EB scheme of the security analysis with trusted detector module. The structure of Alice is the same as preparing coherent state with EPR states. The beamsplitter at the receiver site has the transmittance of $\eta$, and the variance of the EPR state is $V_D = 1+\nu_{ele}/(1-\eta)$. Here, $\nu_{ele}$ is the electronic noise of  the homodyne detector.}
\end{figure}

The practical measurement of a CV-QKD system usually suffers from the imperfect detection efficiency and electronic noise, leading to the increase of channel loss and excess noise. While in device-dependent system, \hl{the trusted device with specific noise model can relax the degradation caused by device imperfections} \cite{Lodewyck_PhysRevA_2007,pirandola2014quantum, usenko_Entropy_2016}.
% However, since the devices inside the receiver can be trusted, a tighter estimation to the potential eavesdropping behavior is achievable through an accurate calibration of the imperfections.
Normally, the coherent receiver of a CV-QKD system can be trusted, where the loss and noise inside the detector can be modeled by  an EPR state with one mode (denoted as $F_0$) coupled into the signal path by a beamsplitter. 
The trusted detector module significantly enhances the transmission distance and the secret key rate of the system, which is demonstrated in the early experiment \cite{Lodewyck_PhysRevA_2007}. Further, the feasibility of enhancing the system performance using optical amplifiers is proved \cite{Fossier_JPhysBAtMolOptPhys_2009,Zhang2013ImprovementOT,Zhang2015NoiselessLA,Zhang2015ApplicationOP,Wu2022PerformanceAO}.
As shown in Fig. \ref{fig:TruDet}, here we denote the output mode after the coupling as $F$, and the other mode of the EPR state as $G$. The transmittance of the beamsplitter, $\eta$, reflects the detection efficiency. The variance of the EPR state is written as,
\begin{equation}\label{eq:TruDet}
  V_{D} = 1+\nu_{ele}/(1-\eta),
\end{equation} 
where $\nu_{ele}$ reflects the variance of the electronic noise of a homodyne detector. For heterodyne detection, $\nu_{ele}$ should be replaced by $2 \nu_{ele}$.
The security analysis is based on the covariance matrix $\gamma_{AFGB}$

% \begin{equation}
%   \gamma_{AFGB} = \begin{pmatrix}
%     \gamma_A    & \sigma_{AF} & \sigma_{AG} & \sigma_{AB} \\
%     \sigma_{AF} & \gamma_F    & \gamma_{FG} & \gamma_{BF}\\
%     \sigma_{AG} & \gamma_{FG} & \gamma_G    & \gamma_{BG}\\
%     \sigma_{AB} & \gamma_{BF} & \gamma_{BG} & \gamma_B
%   \end{pmatrix}
%   =
%   \begin{pmatrix}
%     \gamma_{AFG} & \sigma^T_{AFGB} \\
%     \sigma_{AFGB} & \gamma_B
%    \end{pmatrix}.
% \end{equation}

\begin{equation}
  \gamma_{AFGB} = 
  \begin{pmatrix}
    \gamma_{AFG} & \sigma^T_{AFGB} \\
    \sigma_{AFGB} & \gamma_B
   \end{pmatrix}.
\end{equation}

% In the security analysis based on the entanglement-based scheme, the system including the trusted modes is analyzed, written as $\rho_{AFGB}$.
% Compared with the security analysis without trusted modes, where the system $\rho_{AB}$ is analyzed, Eve's ability is weakened, therefore a tighter lower bound of the secret key rate can be achieved, contributes to higher secret key rate.

For convenience, we will discuss both homodyne and heterodyne scenarios simultaneously in this section.
A practical detector is characterized by an efficiency $\eta$ and a noise $\nu_{ele}$ due to detector electronics. As we did for the channel, we can define a detection-added noise referred to Bob's input and expressed in shot-noise units that we devote in general as $\chi_h$, and is given by the expressions 
\begin{equation}
  \chi_{hom}=[(1-\eta)+\nu_{ele}]/\eta,
\end{equation}
and 
\begin{equation}
  \chi_{het}=[1+(1-\eta)+2\nu_{ele}]/\eta,
\end{equation}
for homodyne and heterodyne detection, respectively. The total noise refered to the channel input can then be expressed as $\chi_{tot}=\chi_{line}+\chi_h/T$.

The mutual information between Alice and Bob is given in the case
of practical protocol with trusted noise by the same expressions as Eq. (\ref{IAB}) by replacing $\chi_{line}$ to $\chi_{tot}$.
The upper bound of Eve's knowledge, $\chi_{BE}$ becomes
\begin{equation}\label{eq:chiAE_trused}
\chi_{BE}=\sum_{i=1}^{2}G(\lambda'_i)-\sum_{i=3}^{5}G(\lambda'_i),
\end{equation}
where $\lambda^{\prime}_{1,2}$ represent the symplectic eigenvalues of covariance matrix $\gamma_{AB}$. $\lambda^{\prime}_{3,4,5}$ represent the symplectic eigenvalues of covariance matrix $\gamma_{AFGB|m_A}$ which is given by
\begin{equation}\label{eq:gamma_{AFGB|m_A}}
\gamma_{AFGB|m_A}=\gamma_{AFG}-\sigma_{AFGB}^{T}H\sigma_{AFGB}.
\end{equation}
Here, $\gamma_{AFG}$ and $\sigma_{AFGB}$ are the sub-matrix of {$\gamma_{AFGB}$}.
% In the above equation, $H$ is the symplectic matrix that represents the homodyne or heterodyne measurement on mode $B$. In the former case, $H_{hom}=(X\gamma_{B}X)^{MP}$, where $X=\left[\begin{aligned}
%   1 ~~~0\\0~~~ 0
% \end{aligned}\right]$, 
% and $MP$ stands for the Moore-Penrose pseudo-inverse of a matrix, while in the latter case, $H_{het}=(\gamma_B+I_2)^{-1}$. 
The symplectic eigenvalues can be given by expressions of the form,
\begin{equation}\label{eq:lambda1,2_trused}
{ \lambda^{\prime }_{1,2} }^2 =\frac{1}{2}[\Delta\pm\sqrt{\Delta^2-4D}],
\end{equation}

\begin{equation}\label{eq:lambda3,4_trusted}
{\lambda^{\prime}_{3,4}}^2 =\frac{1}{2}[E\pm\sqrt{E^2-4F}],
\end{equation}
and $\lambda^{\prime}_{5}=1$.
For the homodyne case,
\begin{equation}\label{eq:E_hom}
E_{hom}=\frac{\Delta\chi_{hom}+V\sqrt{D}+T(V+\chi_{line})}{T(V+\chi_{tot})},
\end{equation}
\begin{equation}\label{eq:F_hom}
F_{hom}=\sqrt{D}\frac{V+\sqrt{D}\chi_{hom}}{T(V+\chi_{tot})},
\end{equation}
and for the heterodyne case,
\begin{equation}\label{eq:E_het}
  \begin{aligned}
    &E_{het}=\\
    &\frac{\Delta\chi_{het}^2+D+1+2\chi_{het}(V\sqrt{D}+T(V+\chi_{line}))+2T(V^2-1)}{(T(V+\chi_{tot}))^2},
  \end{aligned}
\end{equation}
\begin{equation}\label{eq:F_het}
F_{het}=(\frac{V+\sqrt{D}\chi_{het}}{T(V+\chi_{tot})})^2,
\end{equation}
where $\Delta$, $D$ are given in Table. \ref{tab:table7}.

\begin{table*}[htpb]
  \renewcommand{\arraystretch}{1.4}
  
  \caption{\label{tab:KeyModule}Current implementation status of the key module technology of CV-QKD system using coherent states. TDM: Time-Division Multiplexing, Pol.M: Polarization Multiplexing, FDM: Frequency-Division Multiplexing, MIMO: Multiple-Input Multiple-Output, LDPC: Low Density Parity Check.}
  \begin{ruledtabular}
  \begin{scriptsize}
  \begin{center}

  \begin{tabular}{m{1cm}m{2cm}m{2cm}m{5cm}m{4cm}}
  \multicolumn{3}{c}{Key Module}                                                                            & Function                                                                         & Notes                                                     \\ \hline
  \multirow{8}{*}{Transmitter}               & \multirow{2}{*}{Source}    & Pulsed                    & Pulsed light generation                                                                & High extinction ratio of $\geq$ 60 dB with narrow linewidth                                    \\ \cline{3-5} 
                                             &                                  & Continuous-wave           & Continuous-wave light generation                                                 & Narrow linewidth $\leq 20$ kHz                                                              \\ \cline{2-5} 
                                             & \multirow{4}{*}{Modulation}      & Gaussian modulation       & Preparation of coherent states obeying Gaussian distribution                     & 1 GHz repetition frequency with IQ modulator                                            \\ \cline{3-5} 
                                             &                                  & Discrete modulation       & Preparation of coherent states obeying discrete distribution                   & 1 GBaud 256 QAM Gaussian probability shaping with IQ modulator                          \\ \cline{3-5} 
                                             &                                  & Attenuation               & Reducing the power of signals to quantum level for security                      & Variable optical attenuator or amplitued modulator                                      \\ \cline{3-5} 
                                             &                                  & Multiplexing              & Co-propagation of quantum signal and LO  / pilot-tone                            & TDM, Pol.M, TDM+Pol.M, FDM, FDM+Pol.M, Dual Pol.                                        \\ \cline{2-5} 
                                             & \multirow{2}{*}{Transmitter Monitoring}      & Source monitoring         & Source modeling                                                      & Detecting a fraction of the modulated signal                                            \\ \cline{3-5} 
                                             &                                  & Injected light monitoring & Closing potential security loopholes                                             & Detecting the injected light                                                                               \\ \hline
  \multirow{7}{*}{Receiver}                  & \multirow{2}{*}{Receiver Monitoring}      & LO monitoring             & \multirow{2}{*}{Closing potential security loopholes}                   & Detecting a fraction of the LO                                                          \\ \cline{3-3} \cline{5-5} 
                                             &                                  & Injected light monitoring &                                                                                  & Detecting the injected light                                                                                      \\ \cline{2-5} 
                                             & \multirow{3}{*}{De-modulation}   & Sychronization            & Keeping the receiver’s clock consistent with the sender’s clock                  & Co-propagation or a separately transmission of the clock signal                         \\ \cline{3-5} 
                                             &                                  & De-multiplexing           & Separating the quantum signal and LO (or pilot signal if necessary) before detection    & Optical delay line and / or polarization beamsplitter before detection                  \\ \cline{3-5} 
                                             &                                  & Optical compensation      & Suppressing the noise from polarization or phase mismatch                        & Dynamic polarization controller and / or phase shifter                                     \\ \cline{2-5} 
                                             & \multirow{2}{*}{Detection}       & Detection balance         & Compensation of the imbalance from different optical paths                       & Variable optical attenuator before photodiodes                                          \\ \cline{3-5} 
                                             &                                  & Hom / Het detection       & Measurement of the quadrature information                                        & Shot noise limited homodyne detection, heterodyne detection                             \\ \hline
  \multirow{2}{*}{QRNG}                      & \multirow{2}{*}{-}               & Continuous                & Getting continuous-distributed random numbers to support Gaussian modulation     & Transmitting from the discrete-distributed random number, or QRNG with continuous ouput \\ \cline{3-5} 
                                             &                                  & Discrete                  & Getting 0\&1 random numbers for discrete modulation, sifting and postprocessing & High-speed QRNG, normally based on continous-variable system                            \\ \hline
  \multirow{2}{*}{SNU Calibration}           & \multirow{2}{*}{-}               & Two time                  & \multirow{2}{*}{Achieving the SNU for data normoalization}                       & Calibrating the total noise and electronic noise                                        \\ \cline{3-3} \cline{5-5} 
                                             &                                  & One time                  &                                                                                  & Calibrating the total noise                                                             \\ \hline
  \multirow{6}{*}{DSP} & Pulse shaping                    & -                         & Generating signal pulses and raise the spectrum efficiency                                             & RRC filter                                                                              \\ \cline{2-5} 
                                             & Framing  & -                         & Quantum signal, LO / pilot tone and frame signal                                                         & Frequency division multiplexing                                                         \\ \cline{2-5} 
                                             & \multirow{2}{*}{Synchronization} & Clock                     & Keeping the receiver's clock consistent with the sender’s clock                  & Co-propagation or a separately transmission of the clock signal                         \\ \cline{3-5} 
                                             &                                  & Frame                     & Definition of the starting and ending positions                                  & Data frame in LO or pilot tone                                                          \\ \cline{2-5} 
                                             & \multirow{2}{*}{Equalization}    & Static                    & Compensation of device imperfections                                             & S21 compensation, matched filter, skew compensation                                                                          \\ \cline{3-5} 
                                             &                                  & Dynamic                   & Compensation of  transmission impairments                                        & Phase and polarization mismatch with MIMO algorithms                                                                                    \\ \hline
  \multirow{4}{*}{Post-processing}           & Sifting                          & -                         & Preseving the modulation data with the same detection basis as the detector      & Detection basis announcement                                                            \\ \cline{2-5} 
                                             & Parameter estimation             & -                         & Estimation of the secret key rate                                                & Estimating the covariance matrix or system parameters                                   \\ \cline{2-5} 
                                             & \multirow{2}{*}{Information reconciliation} & Reconciliation & Mapping the continous data to discrete form                                      & Slice reconciliation and multidimensional reconciliation                                   \\ \cline{3-5}
                                             &                                  & Error Correction          & Making the two parties share a same bit string                                   & LDPC and Polar code                              \\ \cline{2-5}  
                                             & Privacy amplification            & -                         & Distilling the final secret key bits                                             & Toepliz matrix                                                                         
  \end{tabular}

\end{center}
\end{scriptsize}
\end{ruledtabular}

\end{table*}

\section{CV-QKD SYSTEM AND KEY MODULE}\label{sec:3}

\begin{figure}[t]
  \includegraphics[width=0.45\textwidth]{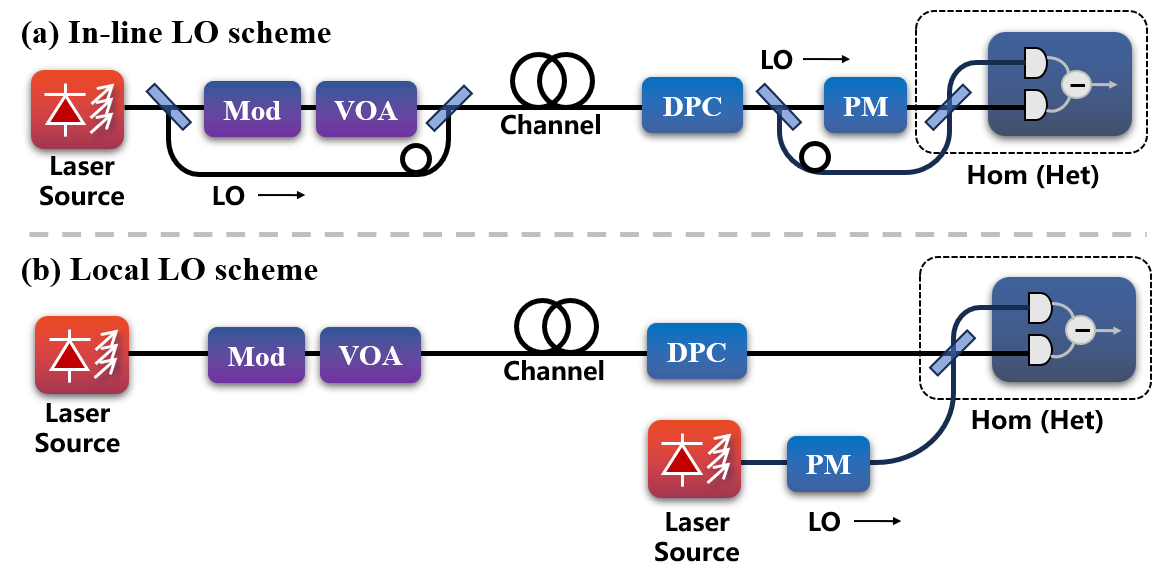}% Here is how to import EPS art
  \caption{\label{fig3:Structure} Two mainstream structures of CV-QKD system implementation: (a) In-line LO scheme and (b) Local LO scheme. The phase modulator at the receiver's side, is used to choose the quadrature of measurement for homodyne detection, which is not needed for heterodyne detection. Mod: Gaussian or discrete modulation, VOA: variable optical attenuator, DPC: dynamic polarization controller, PM: phase modulator, Hom: homodyne detector, Het: heterodyne detector. 
  % (c) Scheme of a practical CV-QKD system using off-the-shelf telecom components. At the transmitter's side, the signal is generated by a dual-polarization I/Q-modulator. At the receiver's side, the received signal is mixed with the LO in an optical hybrid and is detected by four balanced homodyne detectors. Then the signals are sent to the digital signal processing (DSP) for the phase and frequency compensation.
  }
\end{figure}

\begin{figure*}[tbp]
  \includegraphics[width= 1 \textwidth]{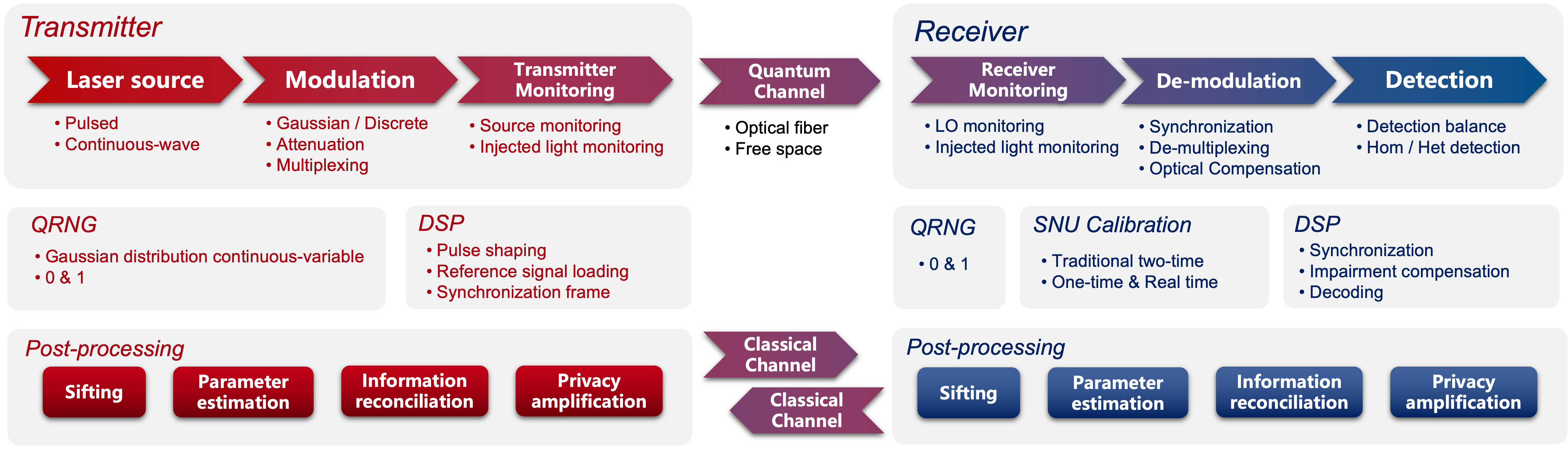}% Here is how to import EPS art
  \caption{\label{fig:KeyModule} The key modules in a CV-QKD system. The transmitter and receiver form the optical part. Optical fiber and free space are the two kinds of quantum channel, where the optical fiber is the mainstream way to transmit the quantum signals. System control is normally used to realize the synchronization between the transmitter and the receiver. QRNG are deployed at both sides to provide the random numbers for modulation, detection and postprocessing. Shot noise unit (SNU) calibration is the security module which provides the normalized unit of the detection data. Post-processing uses classical channel to exchange the information on modulation and detection data, and getting the final secret key bits.}
\end{figure*}

In this section, we will provide an overview of the architecture and the details of the main modules in a CV-QKD system. 
A summary of the current status of the key modules in a CV-QKD system using coherent states can be found in Table \ref{tab:KeyModule}.
Generally, CV-QKD system can be categorized into two types, the in-line LO and the local LO. 
As shown in Fig. \ref{fig3:Structure}, of all the differences between these two types of systems, the greatest one is whether the LO is generated inside the receiver. 
For the in-line LO system, as shown in Fig. \ref{fig3:Structure} (a), the light generated by the laser source is divided, where part of the light is modulated and attenuated to quantum level, while the other part of the light is used as the strong LO.
The quantum signal and LO are then multiplexed and transmitted to the receiver simultaneously. 
% Normally, the receiver firstly compensates the polarization, and then de-modulates the multiplexed signals and performs coherent detection. 
For the local LO system, displayed in Fig. \ref{fig3:Structure} (b), in principal the transmitter only need to modulate and send the quantum signal, and the receiver uses the locally generated LO to perform coherent detection, which simplifies the system structure and prevents the access of LO by Eve.
These two different system schemes result in different main noise source, therefore various system architectures are adopted for suppressing the excess noise and raise the secret key rate.

Generally, the structure of a CV-QKD system can be concluded in Fig.~\ref{fig:KeyModule}.
The transmitter and receiver are the optical modules of the system, which are responsible for the information encoding and decoding.

The transmitter consists of the laser source, the modulation module and the monitoring module.
The laser source can be pulsed or continuous-wave light. The mostly commonly employed modulation formats are Gaussian and discrete modulation. Additionally, the modulator must modulate and multiplex a classical auxiliary signal with the quantum signal, which is the LO in an in-line LO system and the pilot tone in a local LO system.
The quantum signal is multiplexed with LO or pilot tone using time-division, frequency-division and polarization multiplexing techniques.
For security reasons, monitoring is typically employed at the output stage.

At the receiver end, the co-transmitted quantum and classical signals are demodulated and detected by shot-noise-limited balanced homodyne detectors. 
A seperate monitoring module is deployed to ensure the practical security.
Several automatic feedback methods are used to calibrate sampling time, polarization, and phase of the quantum states in order to overcome perturbations in the channel due to changing environmental conditions. These calibrations can be realized using either hardware or digital techniques.

To map the output of the detector to the quadrature information, DSP is employed with the aim of achieving maximal correlation between the transmitter and receiver.
With the assist of DSP, SNU calibration enables the establishment of a connection between theoretical security analysis and the practical system, a crucial aspect of system security.
Lastly, postprocessing is processed to extract the secret key bits from the correlation established by the aforementioned procedures. It comprises sifting, parameter estimation, information reconciliation, and privacy amplification.
The randomness of the overall system is supported by the quantum random number generator (QRNG), which provides the information of modulation and controls the postprocessing.

\subsection{Quantum random number generator}

Randomness determines the security of a CV-QKD system. 
In fact, all of the random numbers used in a CV-QKD system should satisfy the true randomness, which is essential to the unconditional security~\cite{Gisin_RevModPhys_2002,Scarani_RevModPhys_2009,Pirandola_Advances_2020,Xu_RevModPhys_2020}. 
An outstanding alternative is a QRNG, which exploits the intrinsic random nature of quantum mechanics, acts as a promising method in generating truly random numbers \cite{bera_RepProgPhys_2017, ma_npjQuantumInf_2016, herrero_RevModPhys_2017}. 

The random numbers are used for controlling the modulators so that the sender can prepare random coherent states following a certain modulation format, determining the detection basis during homodyne detection, and constructing mappings or universal hash functions in postprocessing.
Specifically, Gaussian distributed random numbers, Rayleigh distributed or uniformly distributed random numbers in continuous-variable form are required in Gaussian modulation of coherent states. Discrete-variable random numbers are required in discrete-modulation, basis selection and postprocessing.

\begin{table}[t]
  \renewcommand{\arraystretch}{1.4}
  
  \caption{\label{tab:QRNG} Current QRNG implementations. The bandwidth with $^*$ represents the repetition rate of the laser pulse. }
  \begin{ruledtabular}
  \begin{scriptsize}
  \begin{center}
  
  \begin{tabular}{llll}
  Scheme                                & Year & Bandwidth     & Generation Rate                                                        \\ \hline
  \multirow{5}{*}{Vacuum   fluctuation} & 2011 & 120 MHz       & 2 Gbps (real-time)\cite{symul_ApplPhysLett_2011}    \\
                                        & 2019 & 1 GHz         & 6.83 Gbps (real-time)\cite{zheng20196}                \\
                                        & 2021 & 3.5 GHz       & 18.8 Gbps   (real-time)\cite{bai_ApplPhysLett_2021} \\
                                        & 2021 & 400 MHz       & 2.9 Gbps   (real-time) \cite{gehring_NatComm_2021}   \\
                                        & 2023 & 10 GHz        & 100 Gbps \cite{bruynsteen_PRXQuantum_2023}         \\ \hline
  \multirow{6}{*}{Laser phase noise}    & 2010 & 1 GHz         & 500 Mbps  \cite{qi_OptLett_2010}                \\
                                        & 2016 & 1 GHz         & 5.4 Gbps   (real-time) \cite{yang_OptExp_2016}      \\
                                        & 2014 & 2.5 GHz  $^*$ & 20 Gbps \cite{yuan_ApplPhysLett_2014}              \\
                                        & 2015 & 12 GHz        & 68 Gbps \cite{nie_RevSciInstr_2015}                 \\
                                        & 2019 & 1 GHz  $^*$   & 8 Gbps (real-time) \cite{roger_JOSAB_2019}          \\
                                        & 2021 & 2.5 GHz  $^*$ & 10 Gbps  \cite{imran_OptComm_2021}                   \\ 
                                        & 2023 & 20 GHz        & 218 Gbps  \cite{yang2023ultra}                   \\  \hline
  \multirow{3}{*}{ASE}                  & 2010 & 7.5 GHz       & 12.5 Gbps \cite{williams_OptExp_2010}               \\
                                        & 2011 & 15 GHz        & 20 Gbps \cite{li_OptLett_2011}                      \\
                                        & 2020 & 5 GHz         & 118.375 Gbps \cite{yang_QuanSciTech_2020}          
  \end{tabular}

  \end{center}
  \end{scriptsize}
  \end{ruledtabular}

\end{table}

Based on different security levels, the QRNG can be divided into the practical type \cite{jennewein_RevSciInstr_2000, stefanov_JourModOpt_2000, dynes_ApplPhysLett_2008, wayne_JourModOpt_2009, gabriel_NatPhot_2010, symul_ApplPhysLett_2011}, the semi-device independent type \cite{nie_PhysRevA_2016, cao_NewJourPhys_2015, avesani_NatComm_2018, marangon_PhysRevLett_2017, cao_PhysRevX_2016, brask_PhysRevAppl_2017, van_Quantum_2017, xu_QuanSciTech_2019, tebyanian_PhysRevA_2021, avesani_PhysRevAppl_2021} and the device independent type \cite{liu_Nature_2018, liu_NatPhys_2021, li_PhysRevLett_2021, zhangyb_PhysRevLett_2020}.
The QRNG can also be divided into the discrete-variable and continuous-variable types corresponding to different types of entropy source.
Normally, the continuous-variable QRNG can support high generation rate with simple structure, where the randomness can come from the amplifier spontaneous emission (ASE) noise \cite{williams_OptExp_2010,li_OptLett_2011, wei_IEEEPhotTechLett_2011, argyris_JourLigTech_2012, li_IEEEPhotJour_2014, zhou_PhysRevA_2015, yang_QuanSciTech_2020}, the phase noise \cite{qi_OptLett_2010, guo_PhysRevE_2010, xu_OptExp_2012, abellan_OptExp_2014, yuan_ApplPhysLett_2014, nie_RevSciInstr_2015, zhang_RevSciInstr_2016, abellan_Optica_2016,liu_IEEEPhotTechLett_2016, yang_OptExp_2016, sun_PhysRevA_2017,alvarez_OptExp_2020, raffaelli_OptExp_2018, imran_OptComm_2021, roger_JOSAB_2019, chrysostomidis_EPJQuanTech_2023} 
and the vacuum noise \cite{jofre_OptExp_2011,haw_PhysRevAppl_2015,raffaelli_QuanSciTech_2018,gehring_NatComm_2021, bai_ApplPhysLett_2021,bruynsteen_PRXQuantum_2023}. 
The QRNG based on vacuum noise can achieve high-speed and simple structure, using only a laser source and a homodyne detector \cite{gehring_NatComm_2021, bai_ApplPhysLett_2021, raffaelli_QuanSciTech_2018}.
The state of the art QRNG based on the vacuum noise can achieve a generation rate of 100 Gbps using photonic integrated circuits \cite{bruynsteen_Optica_2021}, where the speed, size and scalability can well satisfy the need of a CV-QKD system \cite{bruynsteen_PRXQuantum_2023}.
Here we summarize the most commonly used QRNG implementations as shown in Table \ref{tab:QRNG}.

However, the most of the QRNG generates random numbers that are uniform distributed. Therefore, a transformation from the uniform distributed random numbers to the Gaussian distributed random numbers would be a mandatory step. There are several ways of transforming the uniform distributed random numbers into the Gaussian distributed random numbers. For example, the CDF Inversion Method \cite{muller1958inverse}, Box-Muller Transform \cite{box1958note}, Central Limit Theorem \cite{teichroew1953distribution}, Piecewise Linear Approximation using Triangular Distributions \cite{kabal2000generating}, Rejection methods \cite{knuth1981art} and so on.
Typically, the Box-Muller Transform is one of the most widely used methods. It is based on the independent samples, $U_1$ and $U_2$, chosen from the uniform distribution on the unit interval $(0, 1)$. 
Let 
\begin{equation}
  \begin{aligned}
    Z_0 &= \sqrt{-2ln U_1} ~ cos(2\pi U_2), \\
    Z_1 &= \sqrt{-2ln U_1} ~ sin(2\pi U_2).
  \end{aligned}
\end{equation}
Then $Z_0$ and $Z_1$ are independent random variables with a standard normal distribution.
A discrete-modulated CV-QKD system can naturally avoid this issue, where the only step using continuous-variable random numbers, the Gaussian modulation, is replaced.

\subsection{Transmitter}

\begin{figure*}[t]
  \includegraphics[width=0.9\textwidth]{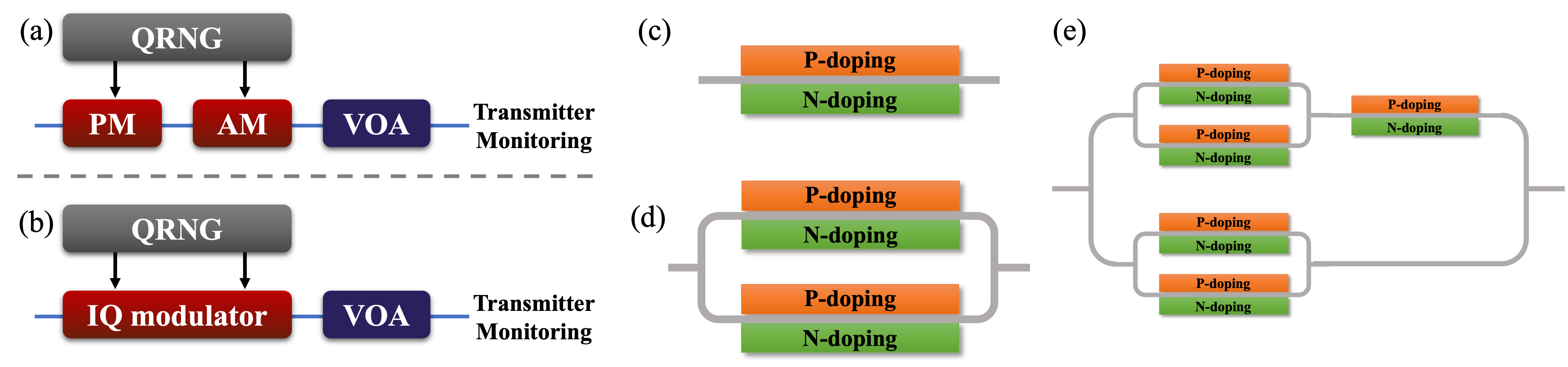}% Here is how to import EPS art
  \caption{\label{fig:Module_Modulation2} (a) The structure of modulation module for quantum signals. QRNG is used for providing modulation data, AM and PM is used to load the modulation information, and variable optical attenuator is used to adjust the intensity of quantum signal to quantum level. (b) The modulation module using IQ modulator. (c) A simple structure of the phase modulator, which consists of a PIN junction. (d) The structure of the amplitude modulator, which consists of two phase modulators. (e) The structure of the IQ modulator, with two amplitude modulators and one phase modulator.}
\end{figure*} 

The function of the transmitter of a CV-QKD system is to prepare the modulated quantum signals and the classical auxiliary signals such as the LO and pilot tones. It usually consists of the laser source, the modulation  module and the monitoring module. 

% \begin{figure}[t]
%   \includegraphics[width=0.25\textwidth]{Module_LD}% Here is how to import EPS art
%   \caption{\label{fig:Module_LD} The structure of a pulsed laser source in CV-QKD systems, usually consists of a laser diode and one or two amplitude modulators. AM: amplitude modulator.}
% \end{figure}

% \begin{table}[t]
%   \renewcommand{\arraystretch}{1.8}
%   \caption{\label{tab:LD}The laser source in CV-QKD systems.}
%   \begin{ruledtabular}
%   \begin{center}
%   \begin{tabular}{c c c c c c}
  
%   Source & Reference & Year & Rep. freq. & Pulse generation & LO 
%   \\
%   \hline
%   \multirow{4}*{Pulsed} & \cite{Jouguet_NatPhotonics_2013} & 2013 & 1 MHz & Not mentioned & In-line
%   \\
%   \cline{2-6}
%    ~ & \cite{Soh_PhysRevX_2015} & 2015 & 250 kHz & An AM &  Local
%   \\
%   \cline{2-6}
%   ~ & \cite{Wang_OptExpress_2018} & 2018 & 50 MHz & An AM & Local 
%   \\
%   \cline{2-6}
%   ~ & \cite{Zhang_PhysRevLett_2020} & 2020 & 5 MHz & Two AMs & In-line 
%   \\
%   \hline
%   \multirow{4}*{\tabincell{c}{Continuous-\\wave}} & \cite{Brunner_ICTON_2017} & 2017 & 10 Mbps & -- & Local 
%   \\
%   \cline{2-6}
%   ~ & \cite{Kleis_OptLett_2017} & 2022 & 40 MBaud & -- & Local 
%   \\
%   \cline{2-6}
%   ~ & \cite{SubGbps} & 2022 & 5 GBaud & -- & Local 
%   \\
%   \cline{2-6}
%   ~ & \cite{jain2022practical} & 2022 & 100 MHz & -- & Local 
%   \\
%   \end{tabular}
%   \end{center}
%   \end{ruledtabular}
%   \end{table}

\subsubsection{Source}
  Laser source usually requires the narrow linewidth and low relative intensity noise to achieve high-performance and stability.
  Among the wide range of lasers, fiber lasers are the most commonly used, which is advanced in monochromaticity, directionality, and stability. \hl{The distributed feedback lasers and external cavity lasers are also adopted in some systems} \cite{Kumar_NewJPhys_2015, Eriksson_CommunPhys_2019, Huang_OptLett_2015, aldama2023small}.

  The laser source can be pulsed or continuous-wave form. The pulsed laser source usually consists of a laser diode with amplitude modulators, while the continuous-wave laser source only requires a narrow-linewidth laser diode. 
  In the early stage of CV-QKD system, pulsed light is naturally used since the separation in time domain provides a clear understanding of different quantum states. Moreover, it is suitable for time division multiplexing to suppress the crosstalk between quantum and classical auxiliary signals. 
  Since the necessary classical auxiliary signal is around $10^4$ to $10^8$ photons per pulse, to eliminate the leakage from the classical pulse to quantum signal pulse, the required extinction ratio is normally about 80~dB or more. Usually a conventional amplitude modulators has an extinction ratio less than $40$~dB, therefore, two or more cascaded amplitude modulators are required for pulse generation \cite{Wang_JQuantumElectron_2015}.

  As the system repetition frequency increases for higher performance, it is harder to achieve pulses with high extinction ratio and high repetition frequency simultaneously, resulting in the demand of directly using continous-wave laser source.
  Meanwhile, the local LO system reduces the crosstalk between quantum and classical signals, which relaxes the requirement of the extremely high isolation.
  Therefore, the continous-wave laser source without the complex cascaded amplitude modulators is widely used in the recent local LO systems \cite{Brunner_ICTON_2017,Kleis_OptLett_2017, Wang_OptExpress_2020, SubGbps, Pan2022DM, jain2022practical}. 
  However, for an accurate interference with signals from two different lasers, the laser diodes in an local LO system require narrower linewidth.
  In addition, a laser locking module can be optionally used to reduce the phase noise.
  % As shown in Table \ref{tab:LD}, most of the in-line LO system, and a part of the early local LO systems adopt the pulsed laser source, using one or two cascaded AMs for pulse generation. However, when heading towards the high-speed system, continuous-wave source is widely adopted.

\subsubsection{Modulation}

  The light output from the laser source is then modulated to encode the quantum information,  generate the classical signals such as frame signals, and the LO or reference signal, which are combined at the output of the transmitter through various multiplexing techniques.
  In this part, we will detail the modulation process in three steps, encoding information, controlling the average power of quantum signal and adding classical auxiliary signals.

\begin{figure}[b]
    \includegraphics[width=0.45\textwidth]{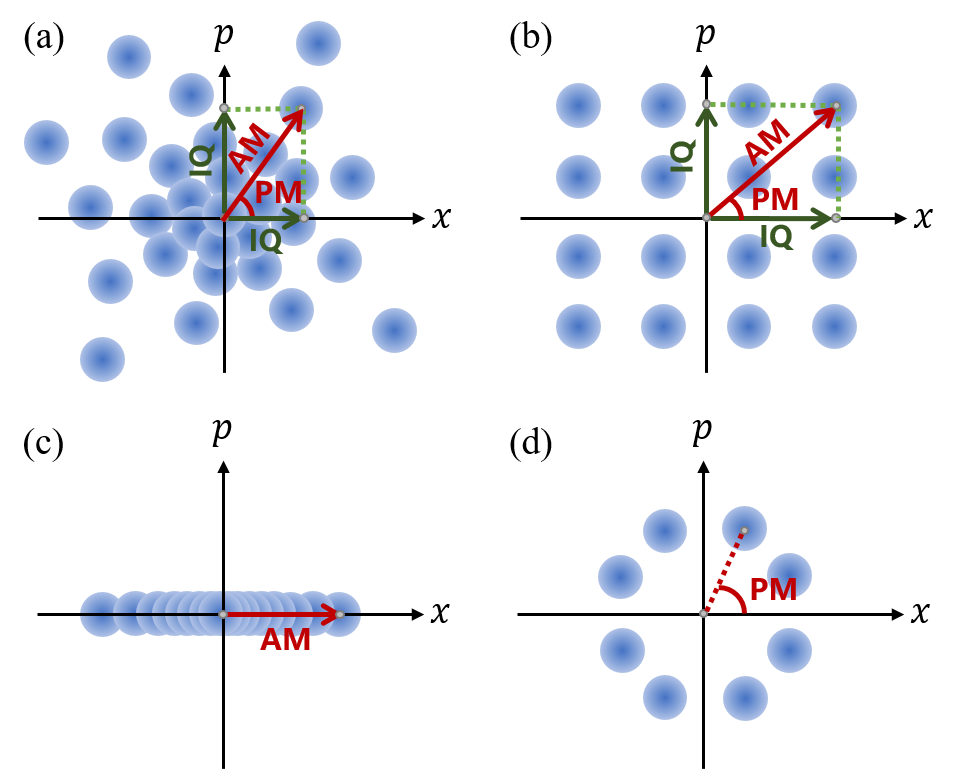}% Here is how to import EPS art
    \caption{\label{fig:Module_Modulation} (a) The Gaussian modulation format with phase and amplitude modulation or IQ modulation. The coherent states on phase space obeys a two-dimensional Gaussian distribution, where the $x$ and $p$ quadrature obeys independent identical Gaussian distribution. (b) The discrete-modulation format with phase and amplitude modulation or IQ modulation. Here a 16 QAM is specified. (c)  The unidimensional modulation on $x$ quadrature with only amplitude modulation. (d) The PSK modulation with only phase modulation, while the amplitude of each coherent state is equal.
    AM: amplitude modulator, PM: phase modulator, IQ: in-phase quadrature modulator.}
  \end{figure}

  For encoding information, two mainstream methods are adopted as shown in Fig. \ref{fig:Module_Modulation2} (a) and (b). One is the combination of a phase modulator and an amplitude modulator, and the other is the use of an in-phase quadrature (IQ) modulator.
  The structure of the modulators used here is detailed in Fig. \ref{fig:Module_Modulation2} (c), (d) and (e).
  The modulation format of a CV-QKD system includes Gaussian modulation and discrete modulation, as shown in Fig. \ref{fig:Module_Modulation} (a) and (b). Both of them require displacements of coherent states on two quadratures of the phase space. An exception is the unidimensional modulation shown in Fig.\ref{fig:Module_Modulation} (c), which only requires the modulation of one quadrature and can be realized with a single amplitude modulator.
  In addition, the Phase Shift Keying (PSK) discrete-modulation only requires a phase modulator for encoding, as shown in Fig. \ref{fig:Module_Modulation} (d).

  In fact, using the phase and amplitude modulators or using an IQ modulator, are the different perspectives of a two-dimensional distribution, as shown in Fig. \ref{fig:Module_Modulation} (a) and (b). 
  Take the Gaussian modulation as an example, for an IQ modulator, if we denote $I$ and $Q$ as the data loaded to the in-phase and quadrature path of the IQ modulator, which corresponds to the $x$ and $p$ quadrature on phase space, they should obey normal distribution as 
  \begin{equation}
    I\sim N(0,\sigma^2), Q\sim N(0,\sigma^2),
  \end{equation}
  where $\sigma^2$ is the variance, $N$ represents the normal distribution.
  For phase and amplitude modulators, the information is loaded with a polar coordinate system, therefore requires the Rayleigh distributed amplitude,
  \begin{equation}
    f(x,\sigma) = \frac{x}{\sigma^2}e^{-x^2/{2\sigma^2}},
  \end{equation}
  and the uniform distributed phase information.

  We remark that, sometimes the $x$ or $p$ quadrature is also called basis, but it is different to the basis in single-photon protocol since the security of CV-QKD is not determined by quadrature selection. Actually, the security of a coherent state CV-QKD system is ensured by the indistinguishability of coherent states on phase space. There also exists a noval CV-QKD protocol based on the basis encoding, where the secret information is encoded on the random choices of two measurement basis, and the security against individual attack has been proved \cite{huang2018quantum}. The scheme exhibits the potential to tolerate high excess noise.

  For practical implementations, we remark that these two methods require different acquisitions \cite{jouguet2012analysis}. Besides that, the working parameters of the modulators should be accurately adjusted, since the imperfect modulation may increase the excess noise and open security loopholes \cite{Liu_PhysRevA_2017}.
  
  After the encoding, one should adjust the average power of the prepared quantum states to a proper level.
  Although in principle, CV-QKD protocols can use the quantum states with high average power, the practical implementations always require the extremely low average power in a few photon number level, since the practical devices with limited resolution cannot measure the quantum characteristics of the high-power states.
  Using the variable optical attenuator with manual adjustment or automatic control is a general method to adjust the level of attenuation, as shown in Fig. \ref{fig:Module_Modulation2}. 
  Further, Y. Zhang et al. used an amplitude modulator to realize the real time attenuation control. It allows one to flexibly raise the launching power of the frame signals for better SNR, and reduce the power of the quantum signals \cite{Zhang_PhysRevLett_2020}. 

  In addition, the passive-state-preparation CV-QKD without modulation modules has also been proposed, where the transmitter consists of a thermal source, beamsplitters, optical attenuators, and homodyne detectors \cite{qi2018passive, qi2020experimental, huang2021experimental}. 
  
  % In an in-line system, the LOs are normally co-transmitted with the quantum signals.
  In the earliest experimental demonstrations, the quantum signals and the LOs are transmitted separately~\cite{Grosshans_Nature_2003}. Later, in order to avoid any polarization and phase drifts that may occur between the signal and LO over long-distance fiber transmissions, time-division multiplexing scheme was proposed to co-transmit the quantum signals and the LOs in the same fiber~\cite{Lodewyck_PhysRevA_2007}. In another literature, B. Qi et al. also co-transmitted the quantum signals with the LOs, but adopted a scheme combining polarization and frequency-division multiplexing~\cite{Qi_PhysRevA_2007}. 
  In later experimental demonstration, the co-transmission schemes basically adopt the combination scheme of time-division multiplexing and polarization multiplexing~\cite{Jouguet_NatPhotonics_2013,Huang_SciRep_2016,li2017continuous,Zhang_QuantumSciTechnol_2019,Zhang_PhysRevLett_2020}. In this way, the leakage from the LOs to the quantum signals can be maximally suppressed, while maintaining a relatively simple implementation structure.
  
  For local LO systems, though LO is generated by the receiver, a classical reference signal is still necessary for phase recovery. Early local LO systems using pulsed laser source is suitable with time division multiplexing, where the quantum signal and the classical pilot tone are alternately transmitted \cite{Qi_PhysRevX_2015,Soh_PhysRevX_2015}.  It can be combined with the polarization multiplexing to suppress the leakage of the pilot tone\cite{Wang_OptExpress_2018}.
  Towards high-speed and continuous-wave system, frequency division multiplexing of quantum signal and pilot tone is widely used\cite{Brunner_ICTON_2017,Kleis_OptLett_2017,jain2022practical}. Since the power of pilot tone is far lower than that of the LO, the requirement of isolation is relaxed. It can also be combined with polarization multiplexing for higher isolation \cite{Wang_OptExpress_2020,SubGbps}.  

  Besides the LO or reference signal, some classical frame signals are inserted in the quantum signal sequence to provide the essential information for synchronization, phase compensation, et al.. These frame signals are usually generated by the same modulation module of quantum signals, and time-division multiplexed.
  
  \subsubsection{Transmitter Monitoring}

  \begin{figure}[t]
    \includegraphics[width=0.4\textwidth]{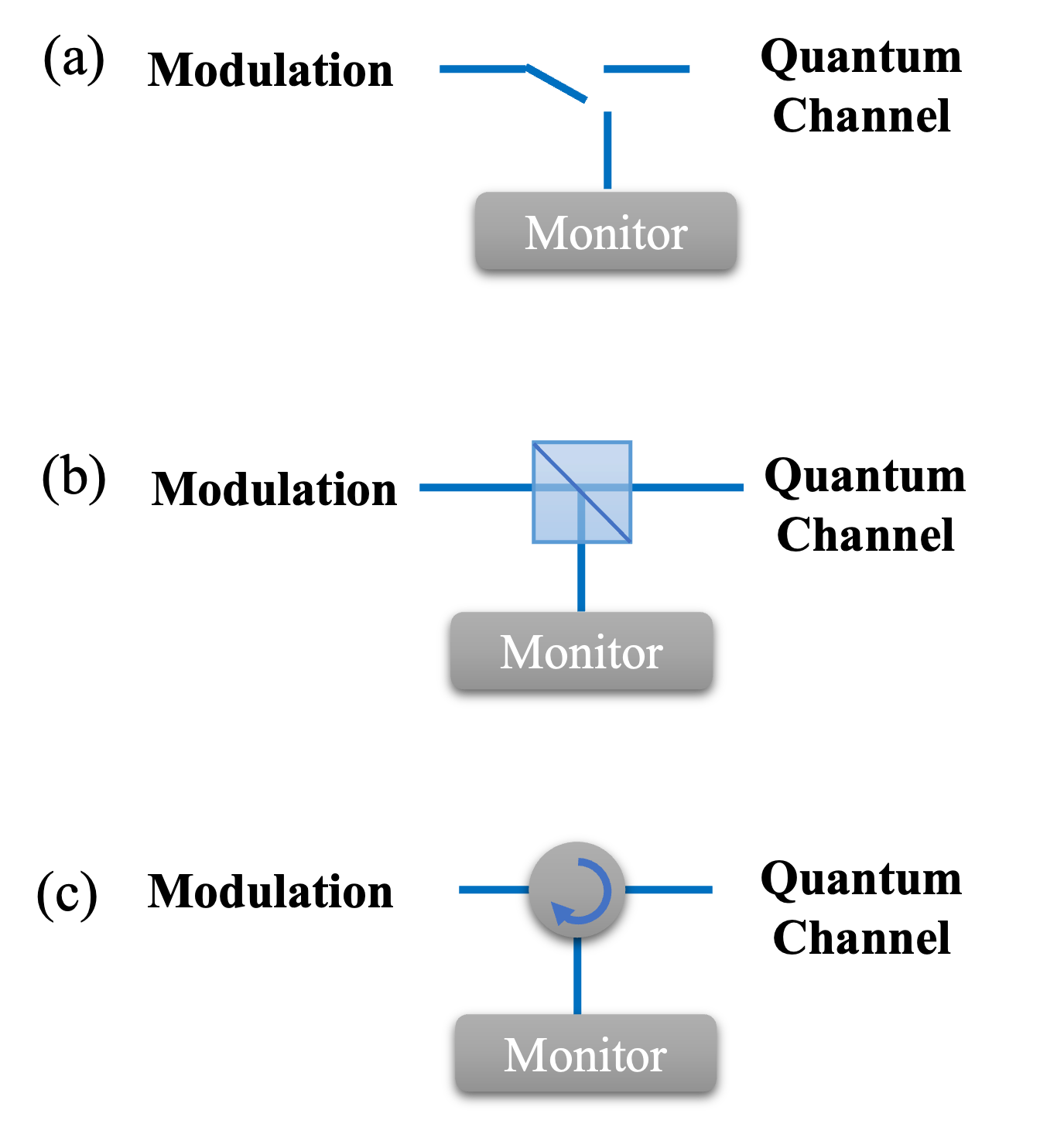}% Here is how to import EPS art
    \caption{\label{fig:Module_Monitor} (a) The active transmitter monitor. (b) The passive transmitter monitor. (c) The transmitter monitor aiming at monitoring the injected light.}
  \end{figure} 

  A source monitor is deployed after the modulation module to evaluate Alice's real output state by detecting a fraction of the signal before entering into the quantum channel. 
  It can be categorized into the active source monitor and the passive source monitor.
  As shown in Fig. \ref{fig:Module_Monitor} (a), the active source monitor adopts the optical switch for getting a part of the signal for monitor, while the passive source monitor uses a beamsplitter to divide the quantum signal for monitor, shown in Fig. \ref{fig:Module_Monitor} (b).
  
  The monitor module can be a homodyne or heterodyne detector, which provides the statistic of the modulated coherent states, including the modulation variance, which is related to the power of the modulated signals. 
  Besides, for Gaussian modulation, since the modulation variance corresponds to the average photon number, which is related to the optical power, an optical power meter can replace the homodyne or heterodyne detector for source monitor.

  Specifically, for Gaussian modulation, we can get
  \begin{equation}
    \langle \hat{n} \rangle = \frac{1}{4}(\langle x^2\rangle+\langle p^2\rangle)-\frac{1}{2}=\frac{1}{2}(V-1)=\frac{V_M}{2},
  \end{equation}
  where $V=V_M+1$ is the variance of the mode, and $V_M$ is defined as the modulation variance. 
  The average photon number of the output mode, $\hat{n}$, can be achieved by detecting the power of the output signal, denoted as $P_{out}$. Each photonic has the energy of  $E = h\nu$, where $h$ is the Plank constant  and $\nu$ is the frequency.
  Assuming $N$ pulses are detected, where $N$ is a sufficient large number, and the repetition frequency of the system is $f_{rep}$. The average  photon number can be achieved \hl{by}  
  \begin{equation}
    \langle \hat{n} \rangle = \frac{P_{out} }{h\nu f_{rep}} .
  \end{equation}
  Therefore, the modulation variance is $V_M = 2P_{out} / (h\nu f_{rep} )$.
  If we denote the modulation variance of the electrical data as $V_M^{e}$ and assume the modulator works linearly, the scale coefficient can be achieved as $c = \sqrt{V_M / V_M^{e}}$. Therefore, the output of the sender can be denoted by $(D_{x_{B_0}}, D_{p_{B_0}})=c~(D'_{x_{B_0}}, D'_{p_{B_0}})$. Here $(D'_{x_{B_0}}, D'_{p_{B_0}})$ is the electrical modulation data and $(D_{x_{B_0}}, D_{p_{B_0}})$ is the calibrated modulation data on phase space.

  In this way, one can define the data of state preparation on phase space corresponding to the electrical modulation data entered into the modulation module, and the estimated average power of the output quantum states, which is a crucial step for security analysis of a practical system.
  
  The source monitor also contributes to a tight estimation of channel parameters.
  In a practical system, the laser fluctuation~\cite{Laudenbach_AdvQuanTech_2018}, imperfect modulation~\cite{Liu_PhysRevA_2017} and other factors introduce source noise into state preparation stage~\cite{Chu_QuantumSciTech_2021}. It has been shown that the secret key rate may be undermined by the source noise~\cite{Filip_PhysRevA_2008,Weedbrook_PhysRevLett_2010,Weedbrook_PhysRevA_2012}. 
  Traditionally, source noise is ascribed into the channel noise to calculate the secret key rate, while in practice it is controlled neither by Eve, nor by legitimate users. So, this untrusted source noise model just overestimates Eve's power and leads to an untight security bound. 
  By real-time monitoring the modulated quantum signals, the source noise can be calibrated and removed from the channel noise, which helps to enhance the system performance.   

  In addition, for practical security considerations, the monitoring module of the source can help to resist the potential attacks on the system by injecting a strong light, such as the Trojan-horse attack \cite{Stiller_CLEO_2015}, which can open a side channel for Eve. 
  The monitoring structure is shown in Fig. \ref{fig:Module_Monitor} (c), where a circulator isolates the injected light from the components inside the transmitter, and directs it to the monitoring module, which can be a homodyne detector, or an optical power meter.
  The specific attack methods and countermeasures are detailed in the section of practical security.

\subsection{Receiver}
The receiver of a CV-QKD system should accurately measure the quadrature of the received state on phase space, which normally contains the monitoring, de-modulation and detection sub-modules, detailed as below.

\subsubsection{Receiver Monitoring}
The receiver monitoring module is mainly used to monitor the classical light from the channel, including the wavelength, frequency, power and amplitude of the LO in an in-line LO system, as well as the pilot tone in a local LO system \cite{Liu_OptExpress_2017}. 

Receiver monitoring is essential for an in-line LO system, since the LO is transmitted in the quantum channel, which may be manipulated by Eve, and directly participates in the homodyne or heterodyne detection. 
For an in-line LO system with polarization multiplexed quantum signal and LO, after the de-multiplexing of the received signal, the LO can be divided by an unbalanced beamsplitter. The stronger output is sent to the detector and the weaker one is detected by a photodiode for monitoring. 

Besides that, the monitor also need to ensure that the receiver is not affected by the potential injection of a strong light. Several efficient attacking strategies for the CV-QKD system are aiming at the receiver by injecting a strong light to disturb the detection or SNU calibration process. The attacks and countermeasures are detailed in the section of practical security.

\begin{figure*}[t]
  \includegraphics[width=0.6\textwidth]{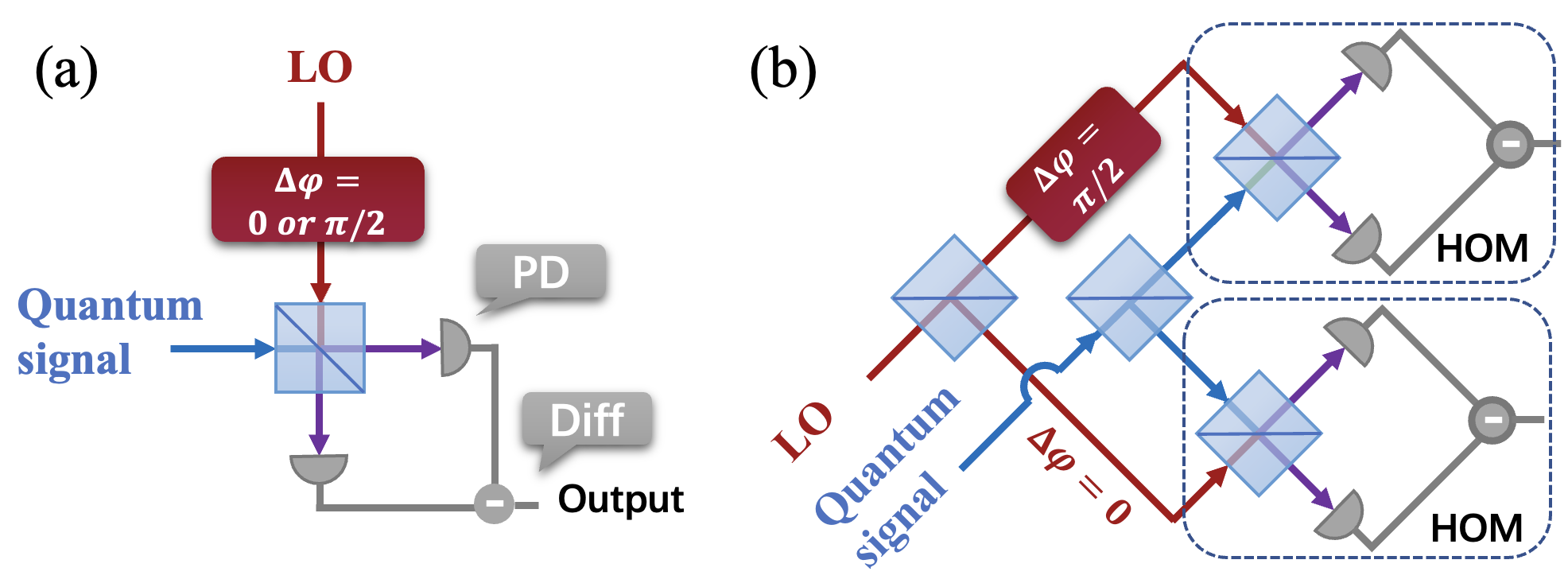}% Here is how to import EPS art
  \caption{\label{fig:Module_BHD} The two mainstream detection strategies in CV-QKD systems. (a) The homodyne detector, where the quantum signal and local oscillator are interfered by a 50:50 coupler, then the two output signals are detected by two photodiodes, finally the output of the two photodiodes, the differential current, is output. A phase shifter is necessary to switch the detection basis between $x$ and $p$ for ensuring the system security.
  (b) The heterodyne detector, where the quantum and local oscillator are divided respectively and detected by two homodyne detectors. A phase shift of $\pi/2$ between the two local oscillators is introduced to realize the detection of $x$ and $p$ quadratures of the quantum signals.
  LO: local oscillator, PD: photodiode, Diff: differentiator and HOM: homodyne detector.}
\end{figure*}

\subsubsection{De-modulation}
The de-modulation module responses for the compensation and de-multiplexing in optical path to well separate the quantum and classical signals, \hl{mainly including the polarization} \cite{wang_OptExp_2019, liu2020continuous} and phase compensation \cite{li_OptExp_2019,wang_PhysRevA_2019,xing_Photonics_2022,chin_OptFibCommCon_2023}, as well as the de-multiplexing of polarization and time. 
It also includes the synchronization in hardware layer \cite{lin_OptExp_2015,liu_InterConfer_2017,chen_Entropy_2019,dong_PhysRevA_2022,li_InterConfer_2022}.

% The compensation aims at suppressing the noise caused by the influence of external environment, such as fiber dither, external mechanical vibration, humidity and temperature, etc, which result in polarization mismatch and phase noise. 

Since the quantum signal is transmitted in single-mode fiber while the receiver is a polarization-maintaining system, the compensation of polarization is crucial for reducing the loss of quantum signals. More importantly, it can suppress the excess noise caused by the crosstalk of quantum and classical auxiliary signals in a polarization-multiplexing system.
For proof-of-principle experiments, the polarization compensation is realized by a manual polarization controller, while dynamic polarization controller with feedback systems is adopted in long-term or field tests \cite{Zhang_QuantumSciTechnol_2019}. 
For simpler deployments, the digital polarization compensation is a crucial technique, where the recently proposed polarization tracking algorithm specially designed for CV-QKD system has shown the possibility of realizing a simple and effective digital polarization compensation for extremely weak quantum signals \cite{pan2023simple}.

Phase noise is inevitably introduced in a CV-QKD system, caused by the phase mismatch between quantum signal and LO. Though they are co-transmitted in an in-line system, the differences in transmission paths before and after multiplexing, as well as the disturbance of fiber link, result in different phases. The local LO system is worse, since quantum signal and LO are generated separately. 
Generally, the phase mismatch consists of the slow-fading and fast-fading ones. The slow-fading phase noise is normally caused by the optical path difference, while the fast-fading phase noise is caused by the channel disturbance and the frequency drift between the two lasers in local LO systems. The slow-fading ones can be compensated with the inserted frame signals, such as the four-state modulated signals. While more reference signals are required for fast-fading phase compensation, such as pilot tones. 
The phase compensation in optical path is normally realized with a phase modulator. Specifically, for the system with GG02 protocol, phase compensation can be integrated with the phase modulator for detection basis switching.

% Digital signal processing can directly realize the polarization compensation of a dual-polarization system, where both direction of polarization have signals for estimating the polarization rotation.
% The algorithms can then rotate the signals in digital domain, and suppressing the crosstalk between different polarization directions.    

In a pulsed CV-QKD system, a de-multiplexing operation is usually performed before the homodyne or heterodyne detection to seperate the quantum signals and LO (or pilot tone). 
After the polarization compensation at the beginning of the receiver, a polarization beamsplitter is used to separate the quantum and classical signals in orthogonal polarization directions. Then, for the time-division multiplexed signals, such as the quantum signal and LO, an optical delay line is adopted to align the quantum signal pulse and LO pulse for homodyne detection. In this structure, a phase modulator in the LO path realizes the basis switching and phase compensation as we mentioned before. Finally, the detection results can be used as the raw data.

The continous-wave system is different, where most of the de-multiplexing and compensation can be realized in digital domain, using the data after the detection. 
The architecture of the receiver is similar to the integrated coherent receiver in classical coherent optical communications, and lots of algorithms participate in the compensation and de-multiplexing process. We will detail this process in the part after detection.

Clock synchronization responses for the alignment of the sampling points at the transmitter and the receiver site.
If the sampling points of the receiver are mismatched to those of the transmitter, the correlation between the modulation data and the detection data will be reduced.
As for clock synchronization, there are many hardware layer \hl{solutions}, such as transmitting in a cable~\cite{Mo_OptLett_2005}, in a paired fiber~\cite{Silva_JMOe_2009}, multiplexing with quantum signal in one fiber by wave-length division multiplex (WDM)~\cite{Beveratos_PhysRevLett_2002}, and distilling the synchronization signal from LO. 
Each of the above options has its advantages and disadvantages.
The electric cable is not only expensive and also instable for distant transmission. Transmitting clock signal and quantum signal in the same fiber by WDM can compensate the difference of signal arrival time but may introduce excess noise generated by WDM crosstalk. Distilling the synchronization signal from local oscillator need divide the local oscillator beam. The division of LO power will decrease the maximal transmission distance or require higher gain of homodyne detection to compensate. Transmitting clock signal in another independent fiber will generate the fluctuation between quantum and clock signal.

\hl{The recent system normally adopts the clock synchronization in digital domain} \cite{wang2022continuous}, using algorithms to realize the above requirements, which will be detailed in DSP part.

% In proof-of-principle systems, the polarization compensation is normally performed by a manual polarization controller. Further, the automatic polarization control is used in long-term experiments, such as the dynamic polarization maintainer~\cite{Zhang_QuantumSciTechnol_2019}. 
% In homodyne detection systems, to make sure that the detection basis is stable, phase compensation is normally performed before the detection, realized by adjusting the phase shifter before the homodyne detector in real-time.

\subsubsection{Detection}

  Homodyne detection is the basis of measuring the quantum signals in a CV-QKD system \cite{yuen_IEEETransInforThe_1980, smithey_PhysRevLett_1993, ou_PhysRevA_1995, Hansen_OptLett_2001, Haderka_ApplOpt_2009,chi_NewJourPhys_2011, Kumar_OptComm_2012, Kumar_OptComm_2012, Wang2012UltrastableFT, Huang_ChinPhysLett_2013, Cooper_JModOpt_2013, Zhang_IEEEPhotonJ_2018, liu_OptExp_2022}. 
  As shown in Fig. \ref{fig:Module_BHD} (a), the quantum and LO signals are interfered by a 50:50 coupler, then the two output signals are detected by two photodiodes. The two branches of the photocurrent generated by the photodiodes are differentiated and output. The differential current contains the information on the quadrature of the quantum signal. 
  A simple derivation of homodyne detection is shown as below.

  We define the modes of quantum signal and LO as $\hat{a}_S$ and $\hat{a}_{L}$. Therefore, the output modes of after the coupler are
  \begin{equation}
    \begin{aligned}
      \hat{a}_{1} &= \frac{1}{\sqrt{2}}(\hat{a}_{S}+\hat{a}_L), \\ 
      \hat{a}_{2} &= \frac{1}{\sqrt{2}}(\hat{a}_{S}-\hat{a}_L).
    \end{aligned}
  \end{equation}
  Then, the photon number of the two output modes can be achieved
  \begin{equation}
    \begin{aligned}
      \hat{n}_{1} = \hat{a}_1^\dagger\hat{a}_1 = \frac{1}{2}(\hat{a}_S^\dagger+\hat{a}^\dagger_{L})(\hat{a}_S+\hat{a}_L), \\
      \hat{n}_{2} = \hat{a}^\dagger_2\hat{a}_2 = \frac{1}{2}(\hat{a}_S^\dagger-\hat{a}^\dagger_{L})(\hat{a}_S-\hat{a}_L).
    \end{aligned}
  \end{equation}
  The differential current can be written as 
  \begin{equation}
    \delta I = I_1-I_2 \propto (\hat{n}_{1}-\hat{n}_{2})=(\hat{a}_S^\dagger\hat{a}_L+\hat{a}_L^\dagger\hat{a}_S).
  \end{equation}
  The strong LO can be seen as a classical light, written as 
  \begin{equation}
  \hat{a}_L = |a_L|e^{i \theta}.
  \end{equation}
  Therefore, one can get
  \begin{equation}
    \delta I \propto |a_L| (\hat{a}_S^\dagger e^{i \theta} + \hat{a}_S e^{-i \theta}).
  \end{equation}
  We get the measurement result of $x$ quadrature when $\theta=0$, where
  \begin{equation}\label{Eq.HomX}
    \delta I (\theta=0) \propto|a_L|(\hat{a}_S^\dagger  + \hat{a}_S ) \propto \hat{x},
  \end{equation}
  and $p$ quadrature when $\theta = \pi/2$,
  \begin{equation}\label{Eq.HomP}
    \delta I (\theta=\pi/2) \propto i |a_L|(\hat{a}_S^\dagger  - \hat{a}_S ) \propto \hat{p}.
  \end{equation}

  A homodyne detector can only measure one quadrature at a time but the security analysis requires detection of both quadratures. Therefore a phase modulator is deployed on the path of the LO, for switching the phase difference between the LO and quantum signal between 0 and $\pi/2$ randomly. This can realize the function of switching the detection basis, satisfying the requirement of getting the statistic data of both quadratures.
  Further, by combining two homodyne detector together, a heterodyne detector which enables the detection of both quadratures of the quantum signal is  realized, as shown in Fig. \ref{fig:Module_BHD} (b). It can simultaneously detect the $x$ and $p$ quadrature of a quantum signal since a phase shift of $\pi/2$ is introduced to one path of the LO. 
  \hl{To avoid the low frequency noise, the heterodyne detection can also be realized by moving the quantum signal to the intermediate band in frequency domain} \cite{Brunner_ICTON_2017,Kleis_OptLett_2017,Wang_OptExpress_2020,SubGbps,jain2022practical,goncharov2022rationale}.
  Both quadratures can be simultaneously distilled with down conversion in analog domain or digitally, while a full architecture is still required to avoid the image band issue.
  Here we remark that the heterodyne detector in CV-QKD represents the dual-homodyne structure, which has different meanings of that in coherent optical communications.

  \begin{table}[t]
    \renewcommand{\arraystretch}{1.8}
    
    \caption{\label{tab:BHD} A comparison between BHDs. Here, BW means the 3 dB bandwidth, QCNR means quantum to classical noise ratio, the and CMRR means the common mode rejection ratio.}
    \begin{ruledtabular}
    \begin{center}
    
        \begin{tabular}{lllll}
                & Year & BW       & QCNR & CMRR    \\ \hline
        Bulk    & 2011 & 104 MHz  & 13 dB     & 46 dB   \cite{chi_NewJourPhys_2011}     \\
                & 2011 & 100 MHz  & 13 dB     & 52.4 dB \cite{Kumar_OptComm_2012}       \\
                & 2013 & 300 MHz  & 14 dB     & 54 dB   \cite{Huang_ChinPhysLett_2013}  \\
                & 2015 & 5 MHz    & 37 dB     & 75.2 dB \cite{jin_OptExp_2015}          \\
                & 2018 & 40 MHz   & 14.5 dB   &   //    \cite{du_JOSAB_2018}            \\
                & 2018 & 1.2 GHz  & 18.5 dB   & 57.9 dB \cite{Zhang_IEEEPhotonJ_2018}   \\ \hline
        On chip & 2019 & 1-10 MHz & 5 dB      & //      \cite{Zhang_NatPhotonics_2019}  \\
                & 2021 & 750 MHz  & 26.82 dB  & 40 dB   \cite{milovanvcev2021ultra}     \\
                & 2021 & 2.6 GHz  & 21.1dB    & 50 dB   \cite{honz2021broadband}        \\
                & 2021 & 1.5 GHz  & 28 dB     & 80 dB   \cite{bruynsteen2021integrated} \\
                & 2021 & 1.7 GHz  & 14 dB     & 52 dB   \cite{tasker2021silicon}        \\
                & 2023 &   //     & 19.42 dB  & 86.9 dB \cite{jia2023silicon}
        \end{tabular}
    
    \end{center}
    \end{ruledtabular}
    
    \end{table}

  \hl{The homodyne detector can be divided into two types, with direct output and the integral output} \cite{du_JOSAB_2018, jia2023silicon, wang_JourLigTech_2023}.
  The integral-output homodyne detector is widely used in the early CV-QKD system since the signal and LO are both pulsed light.
  Therefore, it requires the integration of each laser pulse and output a field quadrature signal. At this stage, the homodyne detector is heading towards the high bandwidth, balance, and common mode rejection ratio \cite{Wang2012UltrastableFT,jin_OptExp_2015, jian_ActaPhysSin_2016, tang_OptComm_2020, pereira_EPJQuanTech_2021}.
  Later, since the CV-QKD system adopts the continuous-wave light, the direct-output homodyne detector is enough, which has less requirement on the balance, and it is easier to realize a high-speed homodyne detection.  

  The bandwidth, detection efficiency, electronic noise and quantum to classical noise ratio (QCNR) are the key parameters of homodyne detectors.  
  The bandwidth of the receiver directly affects the settings of the system in frequency domain, including the multiplexing and processing of the quantum and phase reference signals.  
  The detection efficiency is another crucial parameter of the practical homodyne detector, corresponding to the detector parameter $\eta$ in Eq. (\ref{eq:TruDet}). It is limited by the photodiodes, and the balance of the two arms of the homodyne detectors. 
  The electronic noise of the practical homodyne detector is vital in the CV-QKD systems, which corresponds to $\nu_{ele}$ in Eq. (\ref{eq:TruDet}). Since the power of the quantum signal is extremely small, a high electronic noise can significantly reduce the SNR, therefore resulting in a low-quality detection.
  The last parameter that are normally used to describe the homodyne detector is the \hl{QCNR}. The QCNR demonstrate how good the weak quantum signals can be amplified while suppressing the electronic noise. The general requirement for the QCNR is that it should be at least 10 dB.
  We also list some reported homodyne detectors with their parameters in Table.~\ref{tab:BHD}, common-mode rejection ratio is introduced to reflect the balance of the detector.
  
  \begin{figure}[t]
    \includegraphics[width=0.45\textwidth]{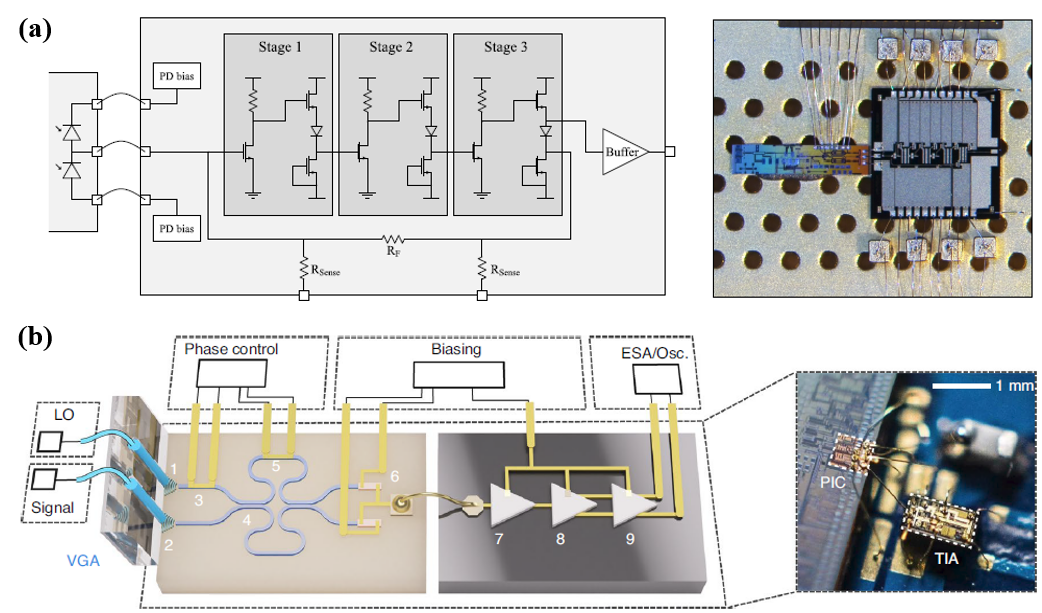}% Here is how to import EPS art
    \caption{\label{fig:Module_ChipDetector} (a) The chip-based homodyne detector for beyond 20 GHz shot-noise-limited measurements. From C. Bruynsteen et al. \cite{bruynsteen2021integrated}. (b) The chip-based homodyne detector for 9 GHz measurement of squeezed light. From J. Tasker et al. \cite{tasker2021silicon}.
    (a) Reproduced with permission from Optica 8, 1146 (2021). Copyright 2021 Optica Publishing Group. 
    (b) Reproduced with permission from Nat. Photonics 15, 11 (2021). Copyright 2021 Springer Nature. 
    }
  \end{figure} 

  The homodyne detector can be integrated on chip driven by the photonics integrated circuit techniques, as shown in Fig. \ref{fig:Module_ChipDetector}.
  The highly balanced photonics circuit, photodiode with high detection efficiency, and the low noise transimpedance amplifier are the core issues.
  The 3 dB bandwidth of the chip-based homodyne detector can break 1.5 GHz, the clearance between the shot noise and the electronic noise can reach 28 dB, and the common-mode rejection ratio can reach 80 dB \cite{bruynsteen_Optica_2021}. These meaningful parameters show that the homodyne detectors on chip have achieved high-speed, low electronic noise and excellent balance.

  % In the process of QCNR estimation, we prefer CW LO as an optical source. Since the electronic noise and quantum noise follow Gaussian distribution and the BHD is AC-coupled, the mean of the noise is zero and the variance of noise is equal to the noise power. We are required to verify the behavior of QCNR between frequency and time domain. Note that, in the following statements, the total noise is contributed by electronic noise and quantum noise. The electronic noise, that includes background noise of instrument and electronic noise of BHD is dominated by electronic noise of BHD. The quantum noise is calculated by subtracting the electronic noise from total noise.

  % Early systems where the quantum signal and LO have the same frequency usually have relatively high requirements for the balance of a homodyne detector. 
  % For practical systems, the imbalance is introduced by the beamsplitter, where the splitting ratio is not a proper 0.5, and the photodiodes, where the detection efficiencies of each photodiodes are not same. The imbalance results in extremely high noise with low frequency, which significantly affects the system performance. The countermeasure is to compensate the imbalance from different optical paths with a VOA placed at one optical path, which can slightly adjust the optical power.
  % Further, the auto-balance detection scheme helps to balance the optical powers of the two optical paths automatically.

  \subsection{Shot noise unit calibration}
  
  \begin{figure}[t]
    \includegraphics[width=0.45\textwidth]{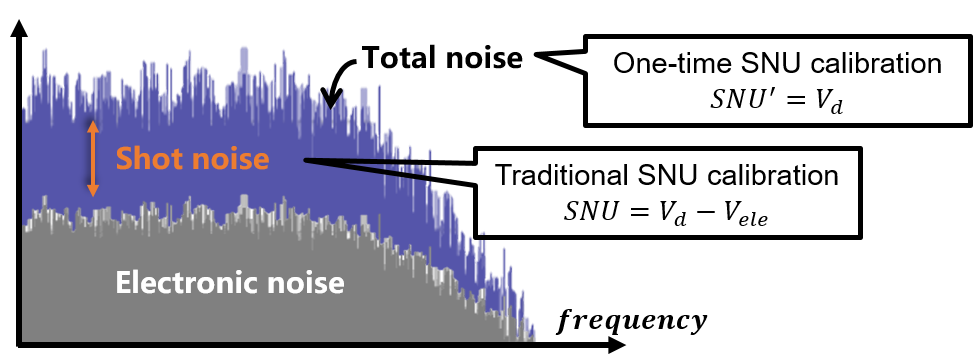}% Here is how to import EPS art
    \caption{\label{fig:Module_SNU} The power density of measurement. The electronic noise and the superposition of electronic and shot noise (vacuum noise) can be achieved in practical measurements. $V_d$: the variance of total noise, $V_{ele}$: the variance of the electronic noise. SNU: shot noise unit, which is the variance of the shot noise.}
  \end{figure}

  \begin{figure}[b]
    \includegraphics[width=0.45\textwidth]{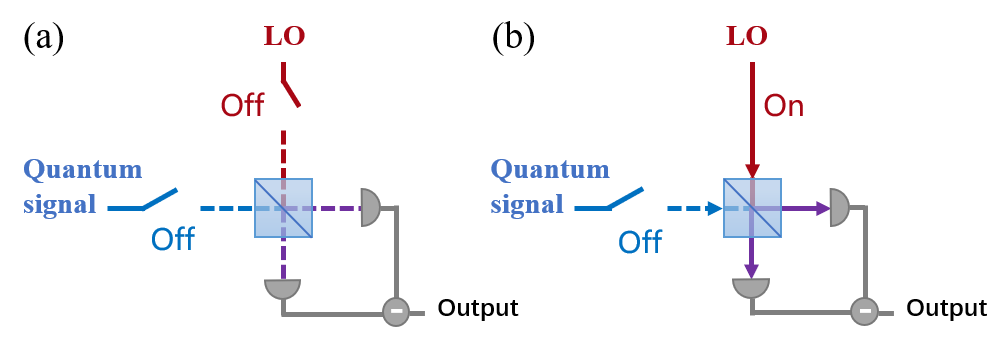}% Here is how to import EPS art
    \caption{\label{fig:Module_BHD2} The calibration of electronic noise and total noise with a practical balanced homodyne detector (BHD). (a) The calibration of electronic noise, where the signal and LO path are cut off. (b) The calibration of the total noise with LO inputted.}
  \end{figure}

  As we mentioned before, the electrical modulation data is transformed to the phase space based on the source monitoring. Correspondingly, the electrical detection data also requires the transformation, which is realized by SNU calibration.
  
  SNU is defined as the variance of the shot noise. 
  In security analysis, the variance of the quantum fluctuation on phase space is defined as unity, while in a practical system, this fluctuation can be measured and recorded as electrical signals, simply by measuring the vacuum.
  With enough electrical detection data of vacuum state, the variance of the shot noise measurement result can be estimated in a practical system and used for the normalization of the detection data of coherent states. In this way, the electrical detection output can be transformed to the data for security analysis. \hl{Therefore, accurate SNU calibration is crucial for the security of the CV-QKD system} \cite{wang_JourLigTech_2023}.
  % Further, by measuring a fraction of the modulation data and normalize it with SNU, the variance of the modulated coherent states on  phase space can be achieved, which is the coefficient for converting the electrical modulation data to the data on phase space. [This part can be expanded later.]
  
  However, SNU calibration of a practical system needs to consider the additive electronic noise, which is introduced inevitably by a practical measurement, shown in Fig. \ref{fig:Module_SNU}.
  Usually, the variance of electronic noise, $V_{ele}$, can be calibrated independently. Therefore, SNU can be achieved in a practical CV-QKD system with twice measurements. 
  As shown in Fig. \ref{fig:Module_BHD2} (a), the electronic noise in the CV-QKD systems can be directly calibrated as follows: fisrt, turn on the electric power of the homodyne detector, and cut off the two optical input ports, the measured variance is the raw electronic noise. 
  Then, total noise can be calibrated with LO on, shown in Fig. \ref{fig:Module_BHD2} (b).
  The difference of the two variables is the SNU, denoted as $u$.

\begin{figure}[b]
    \includegraphics[width=0.4\textwidth]{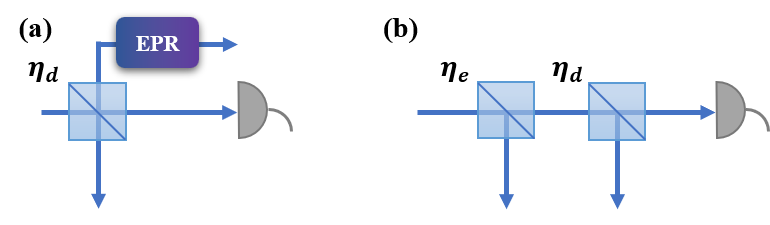}% Here is how to import EPS art
    \caption{\label{fig:Module_SNU2} The traditional trusted detector module (a) and the trusted detector module of one-time SNU calibration (b).}
  \end{figure}

  Specifically, considering the limited detection efficiency $\eta_d$, the detection results of the quantum signal before normalization can be written as
  \begin{equation}
      X_{out}=AX_{LO}(\sqrt{\eta_d}\hat{x}_{s}+\sqrt{1-\eta_d}\hat{x}_{v})+X_{ele}.
  \end{equation}
  Here, $\hat{x}_{s}$  $\hat{x}_{v}$ and $X_{ele}$ are the quadrature information of quantum signal, vacuum state (shot noise), and electronic noise. $X_{LO}$ represents the effect of LO, and $A$ is the amplification coefficient of circuits.
  Therefore, the SNU calibrated with the method we mentioned before is $(AX_{LO})^2$.
  In this way, after normalization we can get 
  \begin{equation}
      x_{out}=\frac{X_{out}}{\sqrt{u}}=(\sqrt{\eta_d}\hat{x}_{s}+\sqrt{1-\eta_d}\hat{x}_{v})+\frac{X_{ele}}{AX_{LO}}.
  \end{equation}
  Based on this normalized output, we can establish the trusted detector module detailed in Section \uppercase\expandafter{\romannumeral2}. The variance of the electronic noise after normalization with SNU, $\nu_{ele}$, is the variance of $X_{ele}/(AX_{LO})$. The scheme of the trusted detector is shown in Fig. \ref{fig:Module_SNU2} (a).

  \begin{table}[t]
    \renewcommand{\arraystretch}{1.8}
    \caption{\label{tab:SNU}The SNU calibration schemes in CV-QKD systems.}
    \begin{ruledtabular}
    \begin{center}
    \begin{tabular}{c c c}
    
    Calibration & Strategy & Procedures
    \\
    \hline
     \tabincell{c}{Traditional\\ two-time \cite{Fossier_NewJPhys_2009}} & \tabincell{c}{Pre-\\calibration} & \tabincell{c}{1) Calibration of electronic noise \\ 2) Calibration of total noise  \\ (assuming a stable SNU)} 
    \\
    \hline
     \tabincell{c}{Traditional \\ two-time \cite{Jouguet_OptExpress_2012,Jouguet_NatPhotonics_2013}}  & Real-time & \tabincell{c}{1) Calibration of electronic noise \\ 2) Calibration of total noise \\ with different LO power \\ (monitoring the real-time LO power) } 
    \\
    \hline
    One-time \cite{Zhang_PhysRevApplied_2020,Zhang_QuantumSciTechnol_2019} & \hl{Real-time} & \tabincell{c}{1) Calibration of total noise with different \\ LO power before system operation \\ (monitoring the real-time LO power) } 
    \\
    \end{tabular}
    \end{center}
    \end{ruledtabular}
  
  \end{table}  

  \begin{figure*}[t]
    \includegraphics[width=0.8\textwidth]{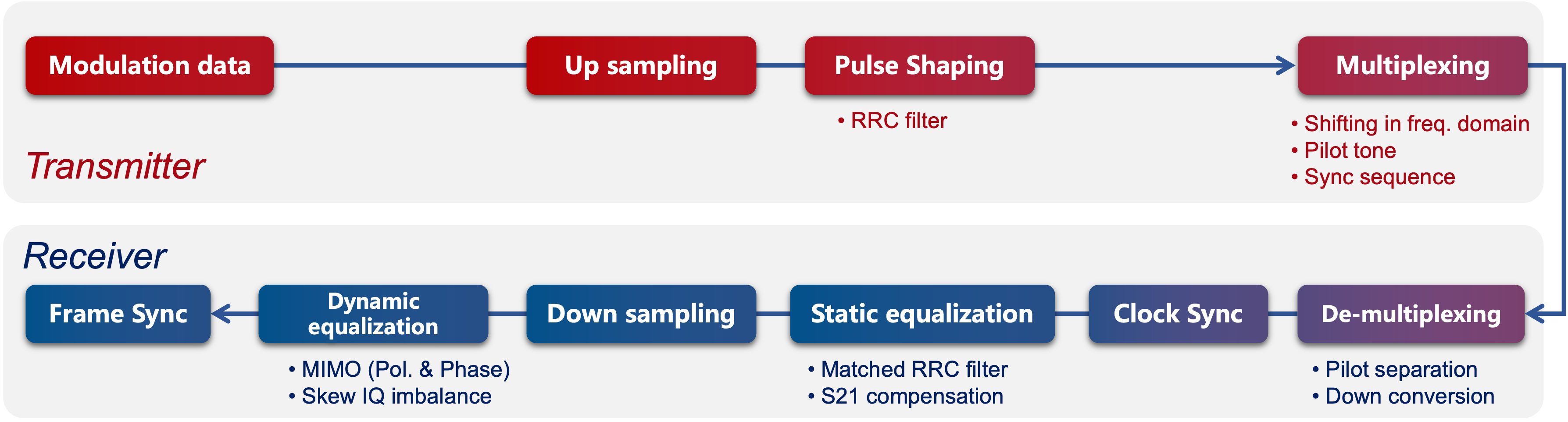}% Here is how to import EPS art
    \caption{\label{fig:Module_DSP} The DSP routine of a CV-QKD system, which includes upsampling, pulse shaping and multiplexing in transmitter side, as well as de-multiplexing, clock synchronization (sync), static equalization, dwon sampling, dynamic equalization and frame sync in receiver side. RRC means Root-Rasied-Cosine, freq. means frequency, and Pol. means polarization.}
  \end{figure*}
  
  In fact, this SNU calibration method has been widely applied in the early experimental demonstrations~\cite{Fossier_NewJPhys_2009,Jouguet_OptExpress_2012} through different implementation schemes.
  However, the LO power and electronic noise are not constant in the practical CV-QKD operations, it tends to drift with the time or temperature changes. Thus, the SNU calibration should be performed in real-time.
  For example, in the Ref.~\cite{Fossier_NewJPhys_2009}, the SNU calibration is performed before the CV-QKD operation, which is called a pre-calibration scheme. The SNU would be calibrated through the above method, the raw data which is obtained in the later running of the CV-QKD can be normalized by using the calibrated SNU. This implementation scheme is surely fine in the proof-of-principle demonstrations, but in reality, the fading of SNU may  may cause severe security issues. Therefore, monitor scheme that can trace the SNU change should be applied.

  \begin{figure*}[t]
    \includegraphics[width=0.85\textwidth]{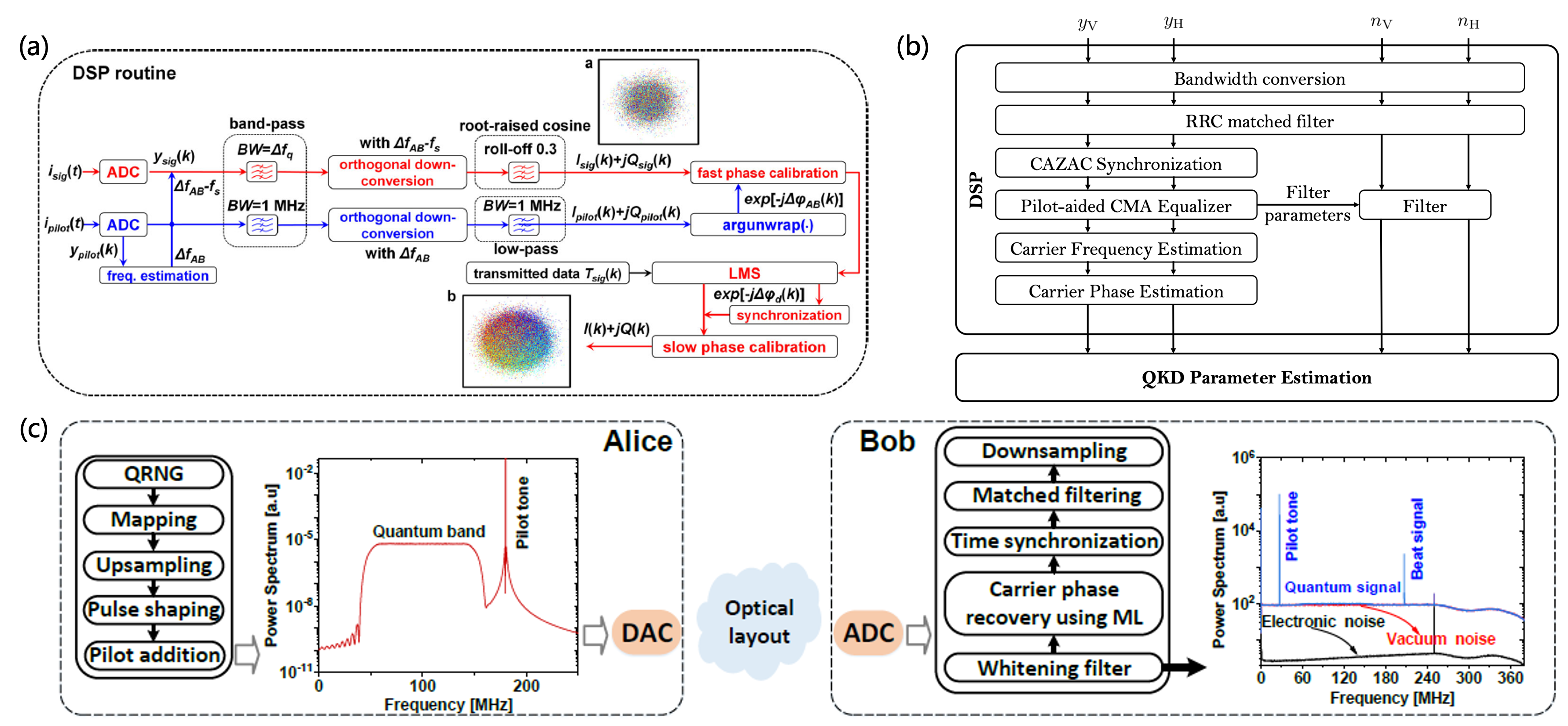}% Here is how to import EPS art
    \caption{\label{fig:System_LLO_DSP} The state-of-the-art digital signal processing routines of different local LO CV-QKD systems. (a) The routine of a discrete modulated system with polarization and frequency division multiplexing of quantum and pilot signals \cite{SubGbps}. (b) The routine of a discrete modulated dual-polarization system where the polarization compensation is finished digitally \cite{roumestan2022experimental}. (c) The routine of a Gaussian modulated system with frequency division multiplexing of quantum and pilot signals. Machine learning (ML) is used for high-quality phase compensation with a wide range of pilot signal-to-noise ratios \cite{Chin2021Machine, hajomer2023longdistance}.
    (a) H. Wang et al., Commun. Phys., 5, 162, 2022; licensed under a Creative Commons Attribution (CC BY) license.
    (b) F. Roumestan et al., arXiv, 2207.11702, 2022; licensed under a Creative Commons Attribution (CC BY) license.
    (c) H. Chin et al., npj Quantum Inf., 7, 20, 2021; licensed under a Creative Commons Attribution (CC BY) license.
    }
  \end{figure*}
  
  The implementation scheme based on the combination of the pre-calibration scheme as well as the LO monitor scheme is adopted~\cite{Jouguet_OptExpress_2012,Jouguet_NatPhotonics_2013}. In this scheme, the pre-calibration still runs before the CV-QKD operation. Also, before the system running, Bob will measure a series of SNU according to different powers of the LO transmitted from Alice. These data would later form a linear relation between the optical power and the SNU. During the CV-QKD system operating, a small portion of the LO is separated and constantly measured. Then, based on the obtained optical power, Bob adjusts the SNU according to the linear relation that has been formed previously. The raw data obtained from the CV-QKD operation can be then normalized by the modified SNU.

  Although the conventional SNU calibration method has been applied in many experimental demonstrations, there are still several issues that are unsolved. First, the SNU calibration process is rather complicated, two steps are required at the optical paths. What's more, with the deeper studies on the practical security of CV-QKD, it makes us realize that the existing SNU calibration method can have security loopholes.
  
  Recently, the SNU calibration method has been improved to meet the needs of the practical implementations of CV-QKD systems, and to reduce the complexity of SNU calibration. An improved security analysis with one-time SNU calibration method has been proposed by redefining SNU as 
  \begin{equation}
    u' = V_{d} = u+V_{ele}.
  \end{equation}
  In this way, SNU can be achieved by just measuring the total noise, which significantly simplifies the procedure \cite{Zhang_PhysRevApplied_2020,Huang_JPhysB_2020,Huang2021ContinuousVariableMQ,Chu2021PracticalSM}.
  Therefore, we get 
  \begin{equation}
    u' = (AX_{LO})^2+V_{ele}.
  \end{equation}
  Here, $V_{ele}$ is the variance of the electronic noise before normalization.
  Further, we can get the output after the normalization with $SNU'$ as $x_{out}'=X_{out}/\sqrt{u'}.$
  If we use another vacuum state to represent the electronic noise as $X_{ele} = \sqrt{V_{ele}}\hat{x}_{v}'$, then we can get 
  \begin{equation}
      \begin{split}
          x_{out}'&=\frac{AX_{LO}}{\sqrt{(AX_{LO})^2+V_{ele}}}(\sqrt{\eta_d}\hat{x}_{s}+\sqrt{1-\eta_d}\hat{x}_{v})\\ 
          &+\sqrt{\frac{V_{ele}}{{(AX_{LO})^2+V_{ele}}}}\hat{x}_{v}'\\ 
          & = \sqrt{\eta_e}(\sqrt{\eta_d}\hat{x}_{s}+\sqrt{1-\eta_d}\hat{x}_{v})+\sqrt{1-\eta_e}\hat{x}_{v}',
      \end{split}
  \end{equation}
  where 
  \begin{equation}
      \eta_e=\frac{(AX_{LO})^2}{{(AX_{LO})^2+V_{ele}}}=\frac{1}{{1+\nu_{ele}}}.
  \end{equation}
  This means the imperfect detector scheme in security analysis can be built with two beamsplitters, one represents the detection efficiency, the other one represents the electronic noise, as shown in Fig. \ref{fig:Module_SNU2} (b). 
  The beamsplitters represent the electronic noise and the detection efficiency can be exchanged in order.
  Note that, since electronic noise is not estimated in one-time SNU calibration, the output mode of the beamsplitter representing the electronic noise cannot be seen as trusted. This means the loss introduced by electronic noise is untrusted, which is the reason that the performance of the protocol with one-time SNU calibration is slightly lower than that using the traditional calibration method \cite{Zhang_PhysRevApplied_2020}.

  With this novel SNU calibration method, it is possible to achieve a real-time SNU calibration, since we only need to monitor the power of the split LO, without the demand of constantly measuring the electronic noise to finish the calibration procedure. Several experimental demonstrations~\cite{Zhang_PhysRevApplied_2020,Zhang_QuantumSciTechnol_2019} have successfully adopted this SNU calibration methods. A conclusion of the SNU calibration methods in CV-QKD systems is shown in Table \ref{tab:SNU}.

\subsection{Digital signal processing}  

For the early systems, DSP is mainly used for the compensation in optical layer, such as providing the feedback information for phase and polarization control.
Later in the high-speed local LO systems, the DSP is gradually becoming widely used, which is applied to the transmitter and receiver, including the clock synchronization, the static equalization, the dynamic equalization and the frame synchronization.
The ultimate purpose of DSP in a CV-QKD system is to maximize the data correlation between the transmitter and receiver, which is consistent with the traditional optical communication algorithm to improve SNR. Therefore, CV-QKD system can widely use the algorithms in classical optical communications, and finally form the current routine, as shown in Fig. \ref{fig:Module_DSP}.

At the transmitter site, for raising the accuracy of the modulation, upsampling is performed before the Root-Raised-Cosine (RRC) pulse shaping \cite{Cubukcu2012RootRC}.  
Then, the pulse shaping of the quantum signal is processed to reduce the correlation between each quantum signal and satisfying the definition of a quantum pulse in a CV-QKD protocol.
Practically, the signal pulse should be bounded in frequency domain so that the modulation and detection with limited bandwidth will not affect the shape of the signal in time domain. But, this will result in the infinite expansion in time domain. 
The mostly used solution is the RRC filter, where the integration of each two quantum signal pulse is 0 in time domain while the frequency band is limited, which both satisfies the requirements of the temporal mode of a CV-QKD protocol and the practical implementation. 
After that, the quantum signal is digitally shifted in frequency domain at the transmitter site, and the pilot tone for phase reference is added. 
Also, for frame synchronization, a Const Amplitude Zero Auto-Corelation (CAZAC) sequence is added at the beginning of each frame. The widely used CAZAC sequence is the Zadoff-Chu sequence.

For the receiver, the \hl{frequency-division} multiplexed quantum and pilot signal are firstly separated from the output of the detection data, then the quantum signals are down converted to the baseband corresponding to the frequency shift at the transmitter. Secondly, clock synchronization is processed, which is a digital alternative to the hardware solution mentioned before. Subsequently, the static equalization algorithms are performed to compensate the impairments, including the IQ imbalance compensation and S21 compensation, and to recover the signal pulse with matched RRC filter. Down-sampling at the best sampling point is then performed to achieve the information with the best SNR.
Then, the dynamic equalization is processed to compensate the phase and polarization mismatch, normally using Multiple-Input Multiple-Output (MIMO) algorithms \cite{Marie_PhysRevA_2017,Corvaja_PhysRevA_2017,Qi_PhysRevApplied_2018,Zou_JApplPhys_2019, wang_PhysRevA_2019, li_OptExp_2019, xing_Photonics_2022}. Finally, frame synchronization is performed to align the beginning of the modulation and detection data. 
% The fast-fading and the slow-fading phase noise are then compensated digitally . 
% The polarization can also be compensated in digital domain\cite{wang_OptExp_2019, chin_OptFibCommCon_2023}.
The final output of the DSP is treated as the raw data, which is then sent to the postprocessing process.

The order of the DSP steps can be exchanged for the linearity of the algorithms.
The coefficients in DSP is normally adjusted dynamically with the system situation for better performance, especially in long-distance or high-noise scene. For this purpose, machine learning is recently introduced to achieve consistently excellent phase estimation under a wide range of pilot SNR \cite{Chin2021Machine, hajomer2023longdistance}.
The compatibility with classical coherent optical algorithms significantly promotes the development of CV-QKD, which provides a large number of tools and experience when developing towards high speed and long distance.

  \begin{figure*}[t]
    \includegraphics[width=0.85\textwidth]{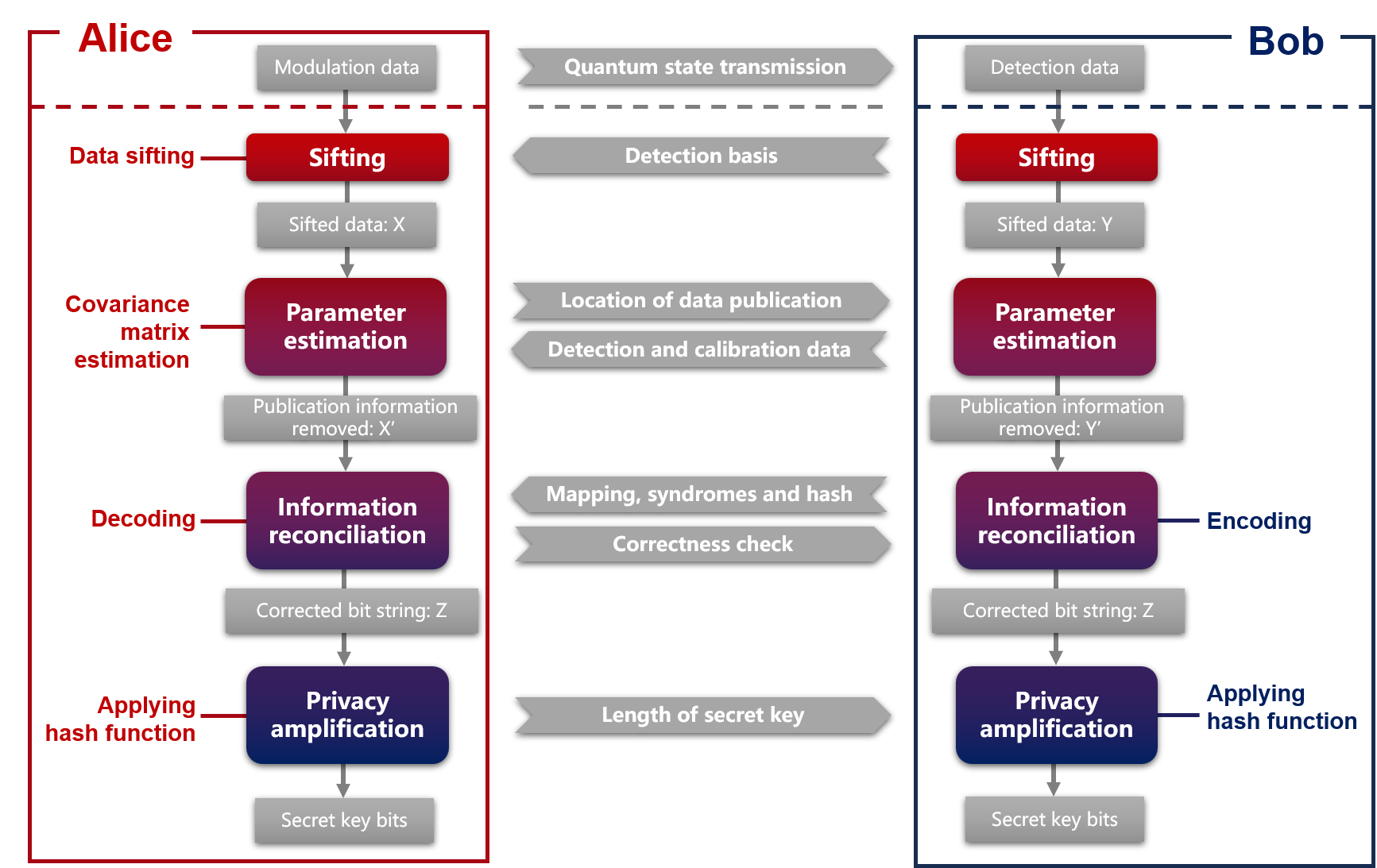}% Here is how to import EPS art
    \caption{\label{fig:Module_PP} 
    The postprocessing steps in CV-QKD systems, where reverse reconciliation is adopted.
    Before postprocessing, Alice and Bob establish correlation by means of modulation, transmission and detection of quantum states. 
    They then process sifting, parameter estimation, information reconciliation and privacy amplification to obtain the final secret key bits.
    }
  \end{figure*}

  \subsection{Postprocessing}
  The quantum stage is followed by classical data processing steps (normally called postprocessing) as is illustrated in Fig. \ref{fig:Module_PP}, which includes four steps: base sifting, information reconciliation, parameter estimation, and \hl{privacy amplification} \cite{van_IEEETransInforTheory_2004,mountogiannakis_PhysRevA_2022, yang2023information}. For this purpose, Alice and Bob use an authenticated channel on which Eve cannot modify the communicated messages but can learn their content. 
  Also, after the steps in postprocessing process, Alice and Bob will perform verification to ensure that the step is successful processed \cite{ma2011universally}. 

  % Base sifting refers to Bob sending a randomly selected measurement base to Alice. Then Alice keeps the correlated raw data according to the measurement bases. During information reconciliation, Alice and Bob can extract available common sequences from their correlated raw data. The purposes of the parameter estimation procedure are to determine quantum-channel parameters and estimate the secret key rate. After the above steps, Eve may have collected sufficient information during her observations of the quantum and classical channels. Hence, privacy amplification is an indispensable step, which is used to distill the final secret keys from the common sequence between Alice and Bob.

\subsubsection{Sifting}

Base sifting is required for systems with homodyne detection or squeezed states, where only one quadrature can used for each measurement. Bob announces the detection basis after a round, and Alice save the corresponding quadrature data. Further developments of CV-QKD systems such as the heterodyne detection scheme can save the data of both quadratures, resulting in a system without sifting. The procedure is significantly simplified since the switching and sifting of detection basis is removed.

\subsubsection{Parameter estimation}

The parameter estimation requires Alice and Bob to estimate the security parameters of the system for getting the secret key rate, based on the modulation and detection data.
Take the Gaussian modulated CV-QKD system as an example, the key point is to obtain the covariance matrix we detailed in Eq. (\ref{eq:MatSecurity}). 
This requires firstly converting the electrical data to the information on phase space in a PM scheme (detailed in transmitter monitoring and SNU calibration), and then the conversion between a PM scheme and EB scheme.
Specifically, we denote the modulation data as $(D_{x_{B_0}}, D_{p_{B_0}})$ and the normalized detection data as $(D_{x_B}, D_{p_B})$. 
For EB scheme, the sender performs heterodyne detection on one mode of the EPR state. If the detection results is $(x_{A_x},  p_{A_p})$, the other mode $B_0$ is projected onto a Gaussian state with 
\begin{equation}
  (x_{B_0},  p_{B_0}) = \sqrt{2\frac{V-1}{V+1}}(x_{A_x},p_{A_p}).
\end{equation}
This means  $(x_{B_0},  p_{B_0})$ and $(x_{A_x},p_{A_p})$ has a linear transformation relationship.
Note that,  here the $x$ and $p$ quadrature corresponds to the modes after a 50:50 beamsplitter, where
\begin{equation}
  \begin{aligned}
    {A}_x &= \frac{1}{\sqrt{2}}{A}+\frac{1}{\sqrt{2}}{N},\\
    {A}_p &= \frac{1}{\sqrt{2}}{A}-\frac{1}{\sqrt{2}}{N}.
  \end{aligned}
\end{equation}
Here, ${A}$ is one mode of the EPR state, ${N}$ is the mode of vacuum state. ${A}_x$ and ${A}_p$ are the two modes after the beamsplitter. $(x_{A_x},  p_{A_p})$ are the $x$ and $p$ quadrature of ${A}_x$ and ${A}_p$ respectively. Therefore the calibrated data $(D_{x_{B_0}}, D_{p_{B_0}})$ can be converted to $(D_{x_{A_x}}, D_{p_{A_p}})$.
With $(D_{x_{A_x}}, D_{p_{A_p}})$ and $(D_{x_B}, D_{p_B})$, the covariance and variance of $x_{A_x}$, $p_{A_p}$, ${x_B}$ and ${p_B}$ can be estimated. Since mode $A_x$ and $A_p$ are symmetric, the covariance and variance of $x_{A}$, $p_{A}$, ${x_B}$ and ${p_B}$ can be further estimated, which results in the covariance matrix $\gamma_{AB}$.

\begin{figure}[t]
  \includegraphics[width=0.45\textwidth]{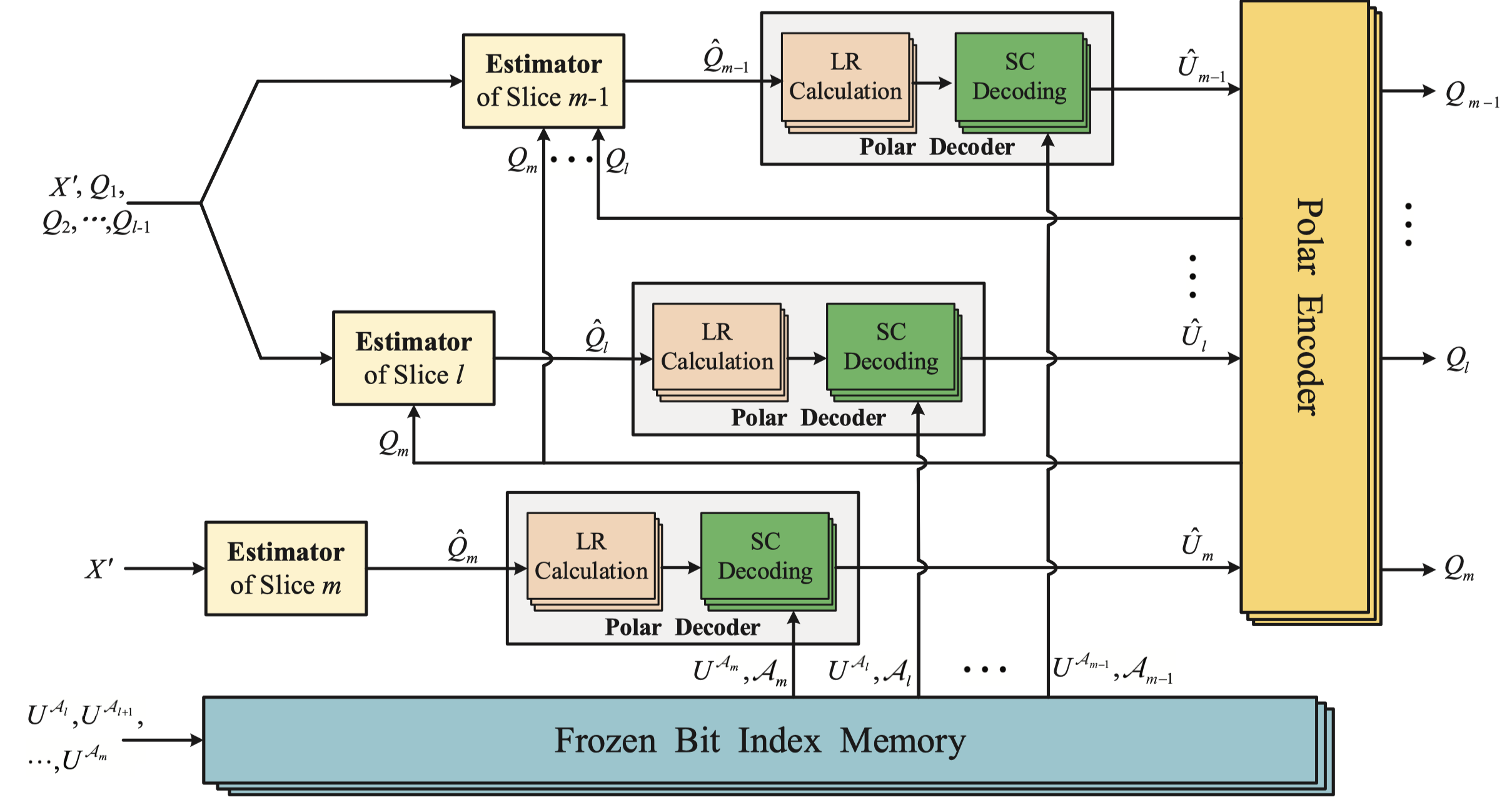}% Here is how to import EPS art
  \caption{\label{fig:Slice} The slice reconciliation in CV-QKD, where the information are sliced. From X. Wen et al. \cite{wen2021improved}. X. Wen et al., Entropy, 23, 1317, 2021; licensed under a Creative Commons Attribution (CC BY) license.}
\end{figure}

Further, for a physical existed convariance matrix, the variance of mode $B$ can be expressed as below
\begin{equation}
  V_B = T V_M +\xi +1= T(V_M+\varepsilon)+1,
  \label{eq_V_B}
\end{equation}
where $T$ and $\varepsilon$ are the two parameters ralated to the system security, estimated from the modulation and detection data. 

% Specificially, we denote the modulation data as $X$ and the detection data as $Y$. The transmission of the quantum signal in a physical-exist quantum channel can be represented by $Y = t X + z$. Here $z$ is the noise with a Gaussian distribution, and $t$ represents the loss of the channel, where $t = \sqrt{T}$.
% If we denote the covariance of $X$ and $Y$ as $\langle X,Y\rangle$, we can get 
% \begin{equation}
%   \langle X,Y\rangle = \langle X,t X + z\rangle=t\langle X^2\rangle.
% \end{equation}
% Therefore, we can get 
% \begin{equation}
%   T = t^2=(\langle X,Y \rangle / \langle X^{2}\rangle)^2.
% \end{equation}
% The excess noise $\varepsilon$ can be estimated with
% \begin{equation}
%   \varepsilon = (\langle Y^2\rangle-1)/T-\langle X^{2}\rangle.
% \end{equation}
% For the No-Switching protocol, the transmittance of the channel includes a trusted beamsplitter at the receiver site.
% Moreover, for the practical case with trusted detection efficiency and electronic noise, the estimation should further consider the trusted detector module.
Based on the security analysis with analytical solution we detailed in Section \uppercase\expandafter{\romannumeral2}, the secret key rate can be calculated.
For a fiber channel, the loss and noise corresponds to the physical meanings of $T$ and $\varepsilon$, which can be used for simulating the system performance.

The estimation above is based on the modulation and detection data with infinite size. However, in the practical CV-QKD systems, the length of the data should be limited. The finite-size data length leads to a fluctuation of the estimated parameter.
The secret key rate under the finite-size analysis is 
\begin{equation}
  K=\frac{n}{N}[\beta I_{AB}-\chi^{\epsilon_{PE}}_{BE}-\Delta(n)].
\end{equation}
Here, $n/N$ represents the proportion of the preserved data, since part of the data, $m=N-n$, is publicized for parameter estimation.
$\chi^{\epsilon_{PE}}_{BE}$ is the Holevo bound considering the effect of finite size, and the probability that the expression is wrong is $\epsilon_{PE}$, meaning that the true parameters lie within a certain confidence interval around the estimated channel parameters.

The estimation of $\chi^{\epsilon_{PE}}_{BE}$ is based on the parameters, $t_{min}$ and $\sigma^2_{max}$, which lead to the worst-case secret key rate. Here, $t_{min}$ means the minimum square root of the channel transmittance, and $\sigma^2_{max}$ means the maximum of the variance of noise. Specifically,
\begin{equation}
  \begin{aligned}
    t_{min} \approx         & \sqrt{T}-z_{\epsilon_{PE}/2}\sqrt{\frac{1+T\varepsilon}{mV_M}}, \\
    \sigma^2_{max} \approx  & 1 +T\varepsilon+z_{\epsilon_{PE}/2}\frac{(1+T\varepsilon)\sqrt{2}}{\sqrt{m}}.
  \end{aligned}
\end{equation}
Here, $z_{\epsilon_{PE}/2}$ should satisfy 
\begin{equation}
  1-erf(z_{\epsilon_{PE}/2}/\sqrt{2})/2=\epsilon_{PE}/2, 
\end{equation}
where $erf$ is the error function.

$\Delta (n)$ is related to the security of privacy amplification, which can be calculated as
\begin{equation}
  \Delta(n)=(2dim H_X+3)\sqrt{\frac{log_2(2/\overline{\epsilon})}{n}}+\frac{2}{n}log_2(1/\epsilon_{PA}).
\end{equation}
Here, $H_X$ corresponds to the Hilbert space of the raw key, $\overline{\epsilon}$ is a smoothing parameter, $\epsilon_{PA}$ is the failure probability \cite{Leverrier_PRA_Finite}.

\subsubsection{Information reconciliation}

Information reconciliation is an essential part of CV-QKD postprocessing, which makes sure that the transmitter and receiver share a same bit string. It consists of two parts, mapping the quadrature results to several bits, and error correction. 
The reconciliation in practical systems requires the public transmission of part of the information, therefore causes the loss of the secret information, leading to the efficiency lower than 100 \%, which can be written as $\beta$,
\begin{equation}
  \beta = \frac{H(X)-leak}{I_{AB}}.
\end{equation}
Here, $H(X)$ is the Shannon information of the target bit string, $leak$ represents the information leakage during the public transmission (usually the syndrome), and $I_{AB}$ is the mutual information between Alice and Bob.
The efficiency of information reconciliation plays a crucial role in the final system's secret key rate and the maximal transmission distance, which are seriously affected by the reconciliation strategy.
We remark that, the purpose of the reconciliation in a practical system is to optimize the whole system, rather than simply chasing the highest efficiency. It is necessary to balance all resources of the system under a given hardware condition to maximize the transmission distance or secret key rate. The comprehensive optimization method proposed by L. Ma et al. which considers the simultaneous optimization of the modulation variance and error correction matrix achieves a significant improvement of the system performance by at least 24 \% compared with the previously used optimization methods \cite{ma_SciChinaInforSci_2023}.

The early reconciliation strategy corrects the detection data to make it consistent with the modulation data, so called the direct reconciliation\cite{Grosshans_PhysRevLett_2002}. However, the system cannot go beyond 3 dB loss since the potentially leaked information is larger than the information can be utilized be the receiver, which corresponds to the 15 km fiber transmission distance.
Aiming at this problem, reverse reconciliation is proposed, where the modulation data is corrected to match the detection data. This enables the system to break the 3 dB limit, making reverse reconciliation the most common information reconciliation strategy in CV-QKD system.
The main parameters of the information reconciliation in a CV-QKD system includes the reconciliation efficiency, the frame error rate, the throughput and the SNR range. 
Different approaches have been explored to increase the reconciliation efficiency for a Gaussian modulation, especially in the regime of a low SNR, detailed in Table \ref{tab:Reconciliation}. 

Usually the detection data is converted to discrete format first, then the error correction code for discrete data is used.   
A first approach is the slice reconciliation using multilevel coding and multistage decoding, as shown in Fig. \ref{fig:Slice}. It is suitable for the detection signal with large SNR, normally more than 0 dB.
In principle, this method can extract more than 1 bit of information per pulse.
When slice reconciliation is used, the SNR of each layer of data is different, so different error correction codes need to be used for subsequent error correction steps, such as the LDPC code or the Polar code. 
% For a practical system with the modulation variance of 4 SNU, the detection efficiency of 0.6, the electronic noise of 0.15 SNU, and the excess noise of 0.1 SNU, the channel loss is \hl{3.54 dB} corresponding to the SNR more than 0 dB, which means the slice reconciliation can only well support the transmission distance within 17 km, which makes the application of this method somewhat limited.
\begin{figure}[t]
  \includegraphics[width=0.25\textwidth]{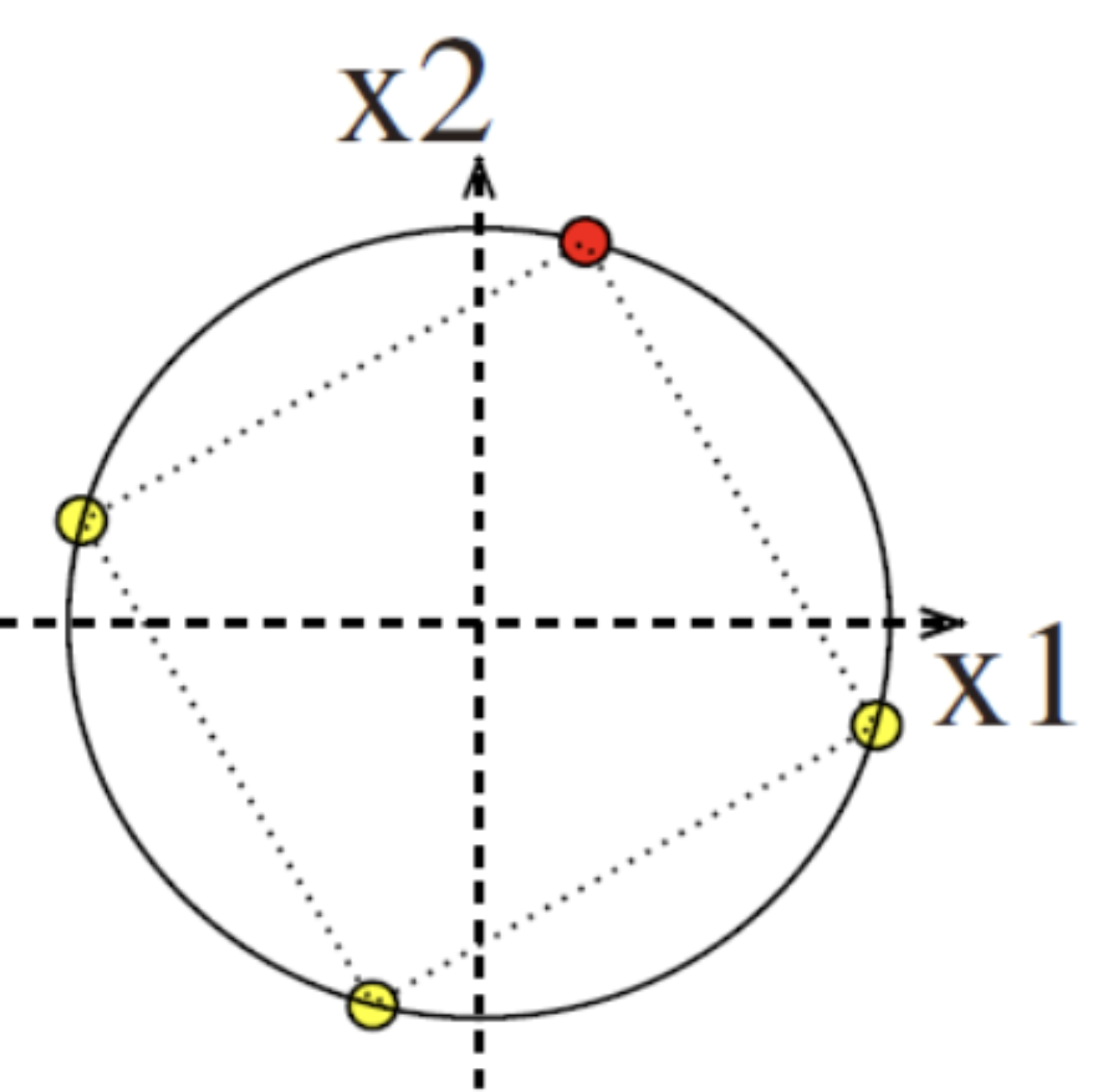}% Here is how to import EPS art
  \caption{\label{fig:MDReconciliation} The multidimensional reconciliation in CV-QKD, where the information are mapped to the sphere, which can be well separated and the symmetry is preserved. From A. Leverrier et al. \cite{leverrier_PhysRevA_2008}. Reproduced with permission from Phys. Rev. A 77, 042325 (2008). Copyright 2008 American Physical Society.}
\end{figure}

\begin{table*}[t]
  \renewcommand{\arraystretch}{1.8}
  \caption{\label{tab:Reconciliation}The reconciliation methods and performance in CV-QKD systems.}
  \begin{ruledtabular}
  \begin{center}

  \begin{tabular}{llllll}
  Reference                                               &Year & Method         & SNR(dB)                                                                                     & Reconciliation   efficiency                                                                                         & FER                                  \\
  \hline
  V. Assche et al. \cite{van_IEEETransInforTheory_2004}   & 2004                & Slice,   Turbo & -                                                                                           & -                                                                                                                  & -                                    \\
  \hline
  J. Lodewyck et al. \cite{Lodewyck_PhysRevA_2007}        & 2007         & Slice,   LDPC  & -                                                                                           & 88.7\%                                                                                                             & -                                   \\
  \hline
  P. Jouguet et al. \cite{jouguet_PhysRevA_2011}          & 2011      & MD, LDPC       & \begin{tabular}[c]{@{}l@{}}0.4,   -7.93, -11.25, \\      -15.38, -18.39, -21.4\end{tabular} & \begin{tabular}[c]{@{}l@{}}93.6\%,   93.1\%, 95.8\%, \\      96.9\%, 96.6\%, 95.9\%\end{tabular}                   & -                                  \\
  \hline
  P. Jouguet et al. \cite{jouguet_PhysRevA_2014}          & 2014      & Slice, LDPC    & \begin{tabular}[c]{@{}l@{}}0,   4.77, 7.09, \\      11.63, 18.2\end{tabular}                & \begin{tabular}[c]{@{}l@{}}94.2\%,   94.1\%, 94.40\%, \\      95.80\%, 94.8\%\end{tabular}                         & -                                \\
  \hline
  Z. Bai et al. \cite{Bai2017HighefficiencyRF}            & 2017         & Slice, LDPC    & 0, 4.77                                                                                        & \hl{95.02\%, 95.26\%}                                                                                                             & -                                  \\
  \hline
  X. Wang et al. \cite{wang_SciRep_2018}                  & 2018      & MD, LDPC       & -15.24                                                                                      & 96.46\%                                                                                                            & -                                 \\
  \hline
  M. Milicevic et   al. \cite{milicevic_npjQuanInfor_2018} & 2018             & MD, LDPC       & -15.47, -7.93                                                                               & 99.0\%, 93.0\%                                                                                                     & 0.883, 0.04                          \\
  \hline
  S. Zhao et al.  \cite{zhao_SciChinaPhys_2018}           & 2018       & MD, Polar      & -0.46, -0.97, -1.55                                                                         & 81\%, 88.4\%, 97.9\%                                                                                               & 0.013, 0.019, 0.04                  \\
  \hline
  C. Zhou et al.   \cite{zhou_PhysRevAppl_2019}           & 2019        & MD, Raptor     & \begin{tabular}[c]{@{}l@{}}0,   -4, -8, \\      -12, -16, -20\end{tabular}                  & \begin{tabular}[c]{@{}l@{}}95.0\%,   95.0\%, $\sim$96.0\%, \\      $\sim$96.0\%, $\sim$97.0\%, 98.0\%\end{tabular} & -                                     \\
  \hline
  Y. Li et al. \cite{li_SciRep_2020}                      & 2020        & MD, LDPC       & -7.93, -11.19, -15.23                                                                       & 92.9\%   , 94.6\% , 93.8\%                                                                                         & 0.17, 0.25, 0.32         \\
  \hline
  S. Yang et al.   \cite{yang_jourLightTech_2020}         & 2020          & Slice, LDPC    & 0, 4.77                                                                                     & 93.0\%, 93.06\%                                                                                                    & 0.14,  0.09                         \\
  \hline
  \multirow{3}{*}{Y. Zhang   et al. \cite{Zhang_PhysRevLett_2020}} & \multirow{3}{*}{2020} & MD, Raptor     & -26.38                                                                                      & 98\%                                                                                                               & 0.9                 \\
                                         &  & MD, LDPC       & -15.11, -7.43, -3.35                                                                        & 96.0\%, 96.0\%,   96.0\%                                                                                           & 0.1,   0.1, 0.1                           \\
                                         &  & Slice, Polar   & 0.3, 4.48                                                                                   & 95\%,  95\%                                                                                                        & 0.5,   0.5                               \\
  \hline
  H. Mani et al.   \cite{mani_PhysRevA_2021}              & 2021  & MD, LDPC       & \begin{tabular}[c]{@{}l@{}}-8.16,   -11.34, -15.46, \\      -18.45\end{tabular}             & \begin{tabular}[c]{@{}l@{}}97.5\%,   97.8\%, 98.8\%, \\      97.7\%\end{tabular}                                   & -                                   \\
  \hline
  S. Jeong et al.   \cite{jeong_npjQuanInfor_2022}        & 2022        & MD, LDPC       & -15.25, -14.25, -14.2                                                                       & -                                                                                                                  & -                                   \\
  \end{tabular}

  \end{center}
  \end{ruledtabular}

\end{table*}

The other method called multidimensional reconciliation was proposed to be employed for low SNRs, i.e., below 0 dB \cite{leverrier_PhysRevA_2008,li_QuanInforProc_2019}, which reduces the Gaussian variables reconciliation problem to the discrete variable channel coding problem, shown in Fig. \ref{fig:MDReconciliation}.
It is suitable for the SNR lower than 0, even to -26 dB, which is the crucial technique for a long-distance CV-QKD system.
The multidimensional reconciliation can be combined with the LDPC code or the other codes such as the Raptor codes \cite{Zhou2019ContinuousVariableQK}.
The final reconciliation efficiency one obtains with such a scheme depends on two things: The intrinsic efficiency of the error correcting code used on the virtual channel on the Binary Input Additive White-Gaussian-Noise Channel (BIAWGNC). The quality of the approximation between the virtual channel and the BIAWGNC. One can therefore improve the reconciliation efficiency based on these two points.

\begin{figure}[t]
  \includegraphics[width=0.45\textwidth]{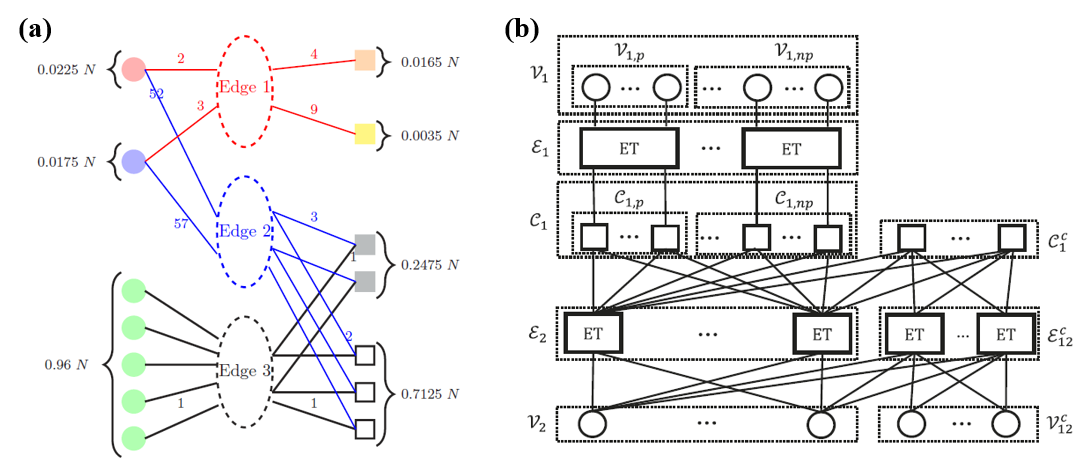}% Here is how to import EPS art
  \caption{\label{fig:Module_METLDPC} The graphical representation of typical MET-LDPC. (a) from H. Mani et al. \cite{mani_PhysRevA_2021}. (b) from S. Jeong et al. \cite{jeong_npjQuanInfor_2022}.
  (a) Reproduced with permission from Phys. Rev. A  103, 062419 (2021). Copyright 2021 American Physical Society.
  (b) S. Jeong, npj Quantum Inf, 8, 6, 2022; licensed under a Creative Commons Attribution (CC BY) license.
  }
\end{figure}

In a long distance system where the SNR is low, LDPC code is usually used \cite{jiang_IEEEPhotJour_2018,zhang_Entropy_2020,guo_QuanInforProc_2020,yang_IEEEAcc_2021,xie_Photonics_2022,sun_OptEng_2023,zhou2023high}. The graphical representation of a typical multi-edge-type LDPC code is shown in Fig. \ref{fig:Module_METLDPC}. 
\hl{The state-of-the-art decoding throughput can reach 1.44 Gbps and 0.78 Gbps for the code rates 0.2 and 0.1, which can support the real-time secret key generation with the speed of 71.89 Mbps and 9.97 Mbp in 25 km and 50 km} \cite{zhou2022high,zhou2023high}.
Towards the practical application, the rate adaptive error correction is proposed to support \hl{the system with fading SNR} \cite{jiang_PhysRevA_2017,zhou_SciChinaPhys_2021,cao_PhysRevAppl_2023,yang2024high}.
In addition, to raise the utilization rate of the raw data, exchanging the order of parameter estimation and reconciliation is proposed. One can process the parameter estimation after the error correction. In this way, the abandoned data for parameter estimation can be reduced.

\begin{figure*}[t]
  \includegraphics[width=0.75\textwidth]{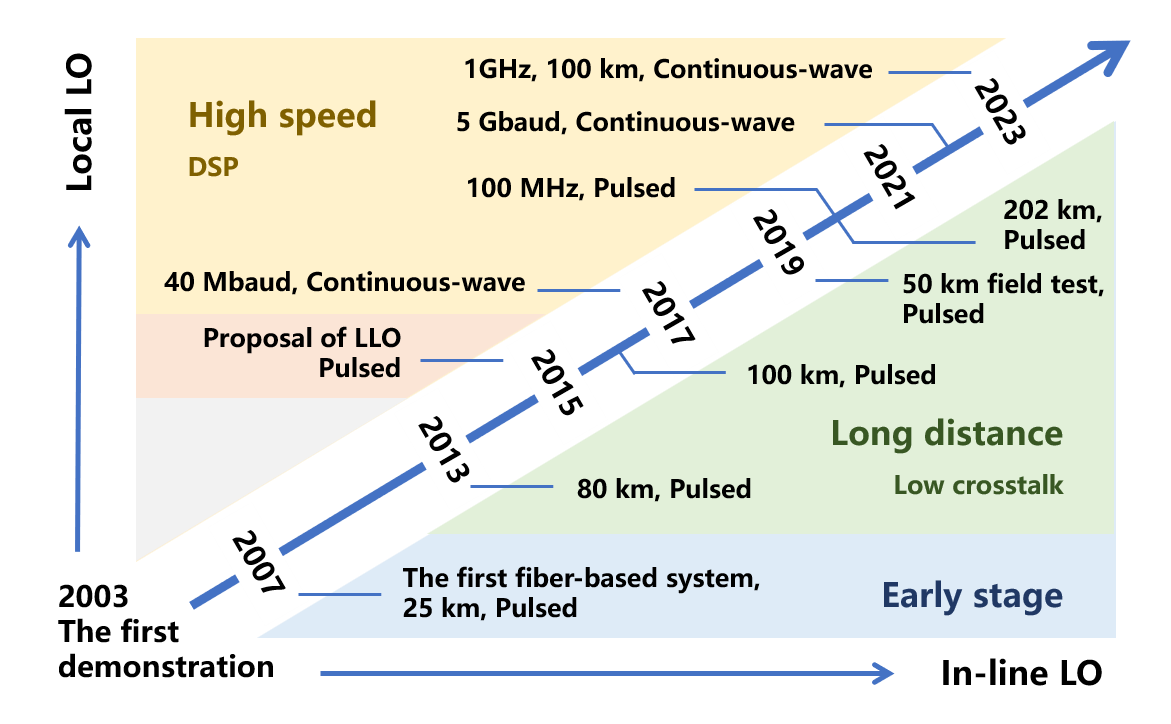}% Here is how to import EPS art
  \caption{\label{fig:System_Dev} The overview of the  CV-QKD system development. Since the first demonstration in 2003, the in-line LO system has been subsequently developed. After the early stage development, the in-line LO scheme has effectively supported the long-distance system, by reducing the crosstalk from LO to quantum signals.
  In 2015, the local LO scheme was proposed, which then heads towards high speed system with advance DSP technique.}
\end{figure*}

\subsubsection{Privacy amplification}
The output of the error correction is a same bit string between Alice and Bob, which can be denoted as $K_n$ with $n$ bits.
Privacy amplification is finally processed to realize the distillation of secret key, of which the requirements are consistent between CV-QKD and DV-QKD. The initial proposed method is aiming at the case of asymptotic limit \cite{Bennett1994GeneralizedPA}, after that, the leftover hashing is used which extends the it to the finite-size case \cite{Tomamichel2010LeftoverHA}.
We denote this operation as $G$, which is a universal hash function mapping the bit string with length $N$ to $L$. 
If the mutual information between eavesdropper and $K_n$ is known, a proper privacy amplification can make the eavesdropper's knowledge on the final secret key, $K_l=G(K_n)$, close to 0. 
For practical implementation, the speed of privacy amplification is crucial \cite{wang_IEEEPhotJour_2018,yan_JourLightTech_2022}.

The most common method is using Toeplitz matrix, which can be written as 
\begin{align}
  \gamma_{Toepliz}=\left(\begin{matrix}\begin{matrix}t_0&t_n\\t_1&t_0\\\end{matrix}&\begin{matrix}t_{n+1}&\begin{matrix}\ldots&t_{2n-2}\\\end{matrix}\\t_n&\begin{matrix}\ldots&\ldots\\\end{matrix}\\\end{matrix}\\\begin{matrix}t_2&t_1\\\begin{matrix}\ldots\\t_{n-1}\\\end{matrix}&\begin{matrix}\ldots\\\ldots\\\end{matrix}\\\end{matrix}&\begin{matrix}\ldots&\begin{matrix}t_n&t_{n+1}\\\end{matrix}\\\begin{matrix}t_1\\t_2\\\end{matrix}&\begin{matrix}\begin{matrix}t_0\\t_1\\\end{matrix}&\begin{matrix}t_n\\t_0\\\end{matrix}\\\end{matrix}\\\end{matrix}\\\end{matrix}\right)
\end{align}
The elements on the main diagonal of the Toeplitz matrix are equal, and the elements on the lines parallel to the main diagonal are also equal. In this way, a Toeplitz matrix can be easily constructed.
With the calculated secret key rate after parameter estimation, the length of the secret key bits can be distilled is known. Assuming the length of K'is L' and the length of K is L, G can be a L $\times$ L' Toeplitz matrix.

\section{MAINSTREAM IMPLEMENTATIONS}\label{sec:4}

The main purpose of the early CV-QKD systems is to demonstrate the feasibility of the Gaussian modulation and shot-noise limited homodyne detection \cite{Grosshans_Nature_2003}. 
After that, the enhancement of system performance begins. 
The efficient error correction successfully supports the long distance systems, from 80 km and up to 202 km. Meanwhile, different system strctures are proposed and implemented to suppress the excess noise, where the most representative one is the local LO scheme. The overview of the CV-QKD system development is shown in Fig. \ref{fig:System_Dev}.

% The main direction for improving the system performance is raising the maximum transmission distance and secret key rate. Two key techniques, efficient error correction and excess noise suppression, determines the system performance. The enhancement of reconciliation plays an important role since the efficiency of error correction limits the maximum transmission distance. The proposal of reverse reconciliation enables the CV-QKD system to overcome the 3 dB limits, and the later proposed multidimensional reconciliation successfully supports the long distance systems, over 80 km and up to 202 km. 
% The error correction module is relatively independent, where the signal-to-noise ratio of the received data is the main concern. 
% Compared with the error correction module, the suppression of the excess noise of the system is more complicated, which is determined by the modulation, detection and the compensation module.  
% The optimization between different modules may have constraints, for example, raising the repetition frequency can  improve the effect of phase recovery, but it also raise the requirement of the system hardware, which may introduce more excess noise in turn.
% Therefore, the effect of the excess noise suppression is highly related to the system structure. 

The CV-QKD system is heading towards high secret key rate at long distance.
For the early CV-QKD systems, the quantum signal and LO are generated by the same laser in the transmitter and co-propagate through the quantum channel. 
However, the longer transmission distance raises the requirement of the power of LO, which leads to more excess noise caused by the LO leakage. Meanwhile, the transmission of LO in an unsecure quantum channel makes it easy for the manipulation of LO by an eavesdropper, leading to a security loophole.
To solve the issues above, the local LO system with the LO generated at the receiver side is proposed. Here the LO no longer needs to be transmitted through the quantum channel, therefore it never affects the quality of quantum signals and  the practical security of the system. However, the system is seriously affected by the  phase mismatch between the LO and quantum signal since they are not generated from the same laser source, thus a classical pilot tone generated by the transmitter is necessary in most of the local LO systems for phase recovery. 

Though the local LO system still requires the co-propagation of quantum and classical signals, the leakage of the pilot tone will not seriously affect the excess noise, since the requirement of the power of the pilot tone is much lower than the LO. Therefore, the multiplexing of the quantum signal and pilot tone is more flexible. Various multiplexing schemes are proposed to adapt to different hardware configurations and system requirements, such as using the frequency division multiplexing, or further combined with polarization multiplexing in high isolation scenarios.
Naturally, raising the repetition frequency contributes to the accuracy of phase tracking and phase noise suppression as we mentioned before. To compensate the influence from the limited hardware resources, DSP is introduced. A highly digitized system can compensate both of the polarization and phase change in digital domain, which significantly simplifies the system structure and leads to a more practical system.
In this section, we review the development of the in-line LO and local LO systems, as well as the system co-existed with classical network and the other system schemes such as the free space system and the entanglement-based systems.

% Recently, the local LO CV-QKD is heading towards a digital system. Since the power of pilot tone is significantly lower than the LO signal, the  transmitter can  generate the quantum and pilot tone simultaneously with the same modulation module using frequency division multiplexing. Meanwhile, the receiver can realize the digital heterodyne detection of two quadratures with only one homodyne detector by introducing frequency difference between quantum signal and LO. This further simplifies the complexity of the system and makes it more compatible with classical communication systems, which seems open the way for large scale deployments.

\subsection{In-line LO systems}

\begin{table*}[t]
  \renewcommand{\arraystretch}{1.8}
  \caption{\label{tab:inline}A Comparison Between Different in-line LO CV-QKD Systems. $\beta$ is the reconciliation efficiency, $L_{max}$ is the maximum transmission distance or loss, and SKR is the secret key rate corresponds to $L_{max}$. MD means multidimensional reconciliation.}
  \begin{ruledtabular}
  \begin{center}
  
    \begin{tabular}{lcccccc}
                                       & \multirow{2}{*}{Years} & \multicolumn{2}{c}{Key   modules}               & \multicolumn{3}{c}{Key   indicators}                                                             \\ \cline{3-7} 
                                       &                        & Multiplexing                 & Reconciliation   & $\beta$       & $L_{max}$   & SKR                                                                \\ \hline
    \multirow{6}{*}{Lab system}        & 2003                   & Transmitting   separately    & Slice            & 85   \%       & 3.1   dB    & 75   kbps \cite{Grosshans_Nature_2003}          \\
                                       & 2007                   & Time                         & Slice            & 89.8   \%     & 25   km     & 2   kbps \cite{Lodewyck_PhysRevA_2007}          \\
                                       & 2007                   & Polarization   and frequency & Slice            & 89.8   \%     & 5   km      & 0.3   bit/pulse \cite{Qi_PhysRevA_2007}         \\
                                       & 2013                   & Time   and polarization      & MD & 95   \%       & 80   km     & 0.2   kbps \cite{Jouguet_NatPhotonics_2013}     \\
                                       & 2016                   & Time   and polarization      & MD & 95.6   \%     & 100   km    & 0.5   kbps \cite{Huang_SciRep_2016}             \\
                                       & 2020                   & Time   and polarization      & MD & 98   \%       & 202.81   km & 6.214   bps \cite{Zhang_PhysRevLett_2020}       \\ \hline
    \multirow{3}{*}{Field test system} & 2012                   & Time   and polarization      & Slice            & $\sim$90   \% & 17.7   km   & 0.3   kbps \cite{Jouguet_OptExpress_2012}       \\
                                       & 2016                   & Time   and polarization      & MD & $\sim$  95 \% & 17.52   km  & 0.2   kbps \cite{Huang_OptLett_2016}            \\
                                       & 2019                   & Time   and polarization      & MD & 95.1   \%     & 49.85   km  & 5.77   kbps \cite{Zhang_QuantumSciTechnol_2019} \\ \hline
    Chip-based system                  & 2019                   & Time   and polarization      & MD & 97.99   \%    & 2 m         & 0.25   Mbps \cite{Zhang_NatPhotonics_2019}     
    \end{tabular}
  
  \end{center}
  \end{ruledtabular}
    
\end{table*}

The main feature of the in-line LO CV-QKD system is that the quantum signal and LO are generated by the same laser, therefore the interference at the receiver is not seriously affected by the signal mismatch. However, the large power difference between the quantum signal and LO makes the leakage of LO significantly affect the quantum signal, which is the main noise source of the system. For instance, the average photon number of a quantum signal is no more than 20, but the average photon number of LO is usually higher than $10^7$ at the receiver site to provide sufficient power for a high-quality shot-noise limited homodyne detection.
The leakage of LO leads to the increase of excess noise. Moreover, longer transmission distance requires higher LO launch power. Therefore, higher isolation between quantum signal and LO is required by the long-distance in-line LO system. 

The key to reduce the LO leakage is using multiplexing including time-division multiplexing, the polarization multiplexing and the frequency-division multiplexing. The quantum signal and LO are encoded on different dimensions for co-transmission in fiber (that's why we call the in-line LO), \hl{and de-multiplexed in physical layer at the receiver} \cite{Grosshans_Nature_2003,Lodewyck_PhysRevA_2007,Qi_PhysRevA_2007,Jouguet_OptExpress_2012,Jouguet_NatPhotonics_2013,Huang_OptExpress_2015,Huang_SciRep_2016,Huang_OptLett_2016,Huang_PhysRevA_2016,Li_ChinPhysB_2017,Wang_OptExpress_2019,Zhang_QuantumSciTechnol_2019,Zhang_NatPhotonics_2019,Zhang_PhysRevLett_2020}. 
Now we review the development of in-line LO CV-QKD systems, including the early systems, the long-distance achievements, the chip-based systems, and the field tests. The details of the representative in-line LO system experiments are concluded in Table. \ref{tab:inline}.

\subsubsection{Early systems}

\begin{figure*}[t]
  \includegraphics[width=0.75\textwidth]{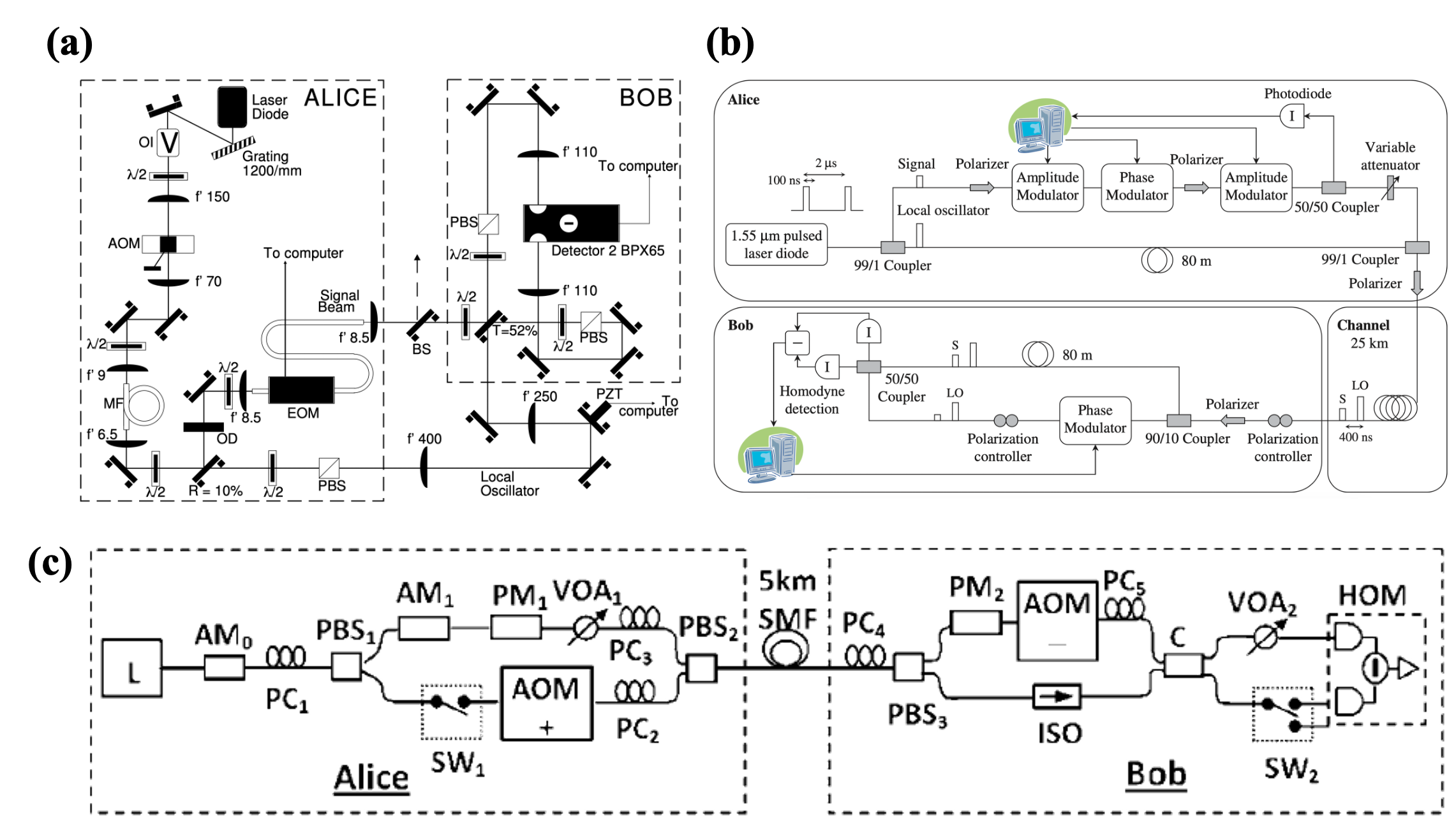}% Here is how to import EPS art
  \caption{\label{fig:iLOearly} The early in-line LO CV-QKD systems. (a) The first CV-QKD system, from F. Grosshans et al. \cite{Grosshans_Nature_2003}. (b) The first all-fiber CV-QKD system with time division multiplexing, from J. Lodewyck et al. \cite{Lodewyck_PhysRevA_2007}. (c) The first all-fiber CV-QKD system with polarization and frequency-division multiplexing, from B. Qi et al. \cite{Qi_PhysRevA_2007}. 
  (a) Reproduced with permission from Nature 421, 238-241 (2003). Copyright 2003 Nature Publishing Group.
  (b) Reproduced with permission from Phys. Rev. A 76, 042305 (2007). Copyright 2007 American Physical Society.
  (c) Reproduced with permission from Phys. Rev. A 76, 052323 (2007). Copyright 2007 American Physical Society.
  }
\end{figure*}

The first CV-QKD system implementation is realized in 2003, as shown in Fig. \ref{fig:iLOearly} (a), where the secret key rate can reach 1.7 Mbps in a loss free channel and 75 kbps in a channel with 3.1 dB loss \cite{Grosshans_Nature_2003}. 
The system is based on the free space optical devices, working at the wavelength of 780 nm.
Reverse reconciliation is firsly implemented to support a secret key distillation beyond 3-dB limit. 
The quantum signal and LO are generated from the same laser, where the light outputs from Alice's laser is firstly divided, part of the light is Gaussian modulated as the quantum signal of coherent states, and the other part of the light is used to provide the LO signal. 
The Gaussian modulation is realized based on the amplitude and phase modulation method with pulsed light. 
To compensate the phase difference between quantum signal and LO since they go through different path, training sequence is introduced. 
These features are reserved and improved in the later fiber based CV-QKD systems.

The first all-fiber CV-QKD system is realized in 2007, where 2 kbps secret key rate is achieved with a 25 km (5.2 dB) optical fiber channel \cite{Lodewyck_PhysRevA_2007}. 
The most important enhancement is that the quantum signal and the LO are co-transmitted in the same fiber to reduce the accumulation of phase noise caused by the seperate transmission. 
As shown in Fig. \ref{fig:iLOearly} (b), the multiplexing strategy is the time division multiplexing using an unbalanced Mach Zender interferometer (MZI) structure. 
Specifically, the light from the pulsed laser is divided by a 1:99 beamsplitter, where the 1 \% weak light path is used to generate the modulated coherent states with amplitude and phase modulators, and the strong light is used as the LO. With a variable optical attenuator, the modulated coherent states are attenuated to the quantum level, which is then time-division multiplexed with the LO by combining with a 99:1 coupler, and then transmitted to the receiver side. The optical delay line deployed in the LO path is used to adjust the gap between quantum signal and LO. 
This time division multiplexing strategy is reserved in the long distance system to enhance the isolation between quantum signal and LO.
The de-multiplexing is realized with an unbalanced beamsplitter (10:90) to reduce the loss of quantum signal, where the quantum signal with high power is interfered with the low power LO after the beamsplitter. A delay line is deployed at the quantum signal path to align the signal pulses. 
Besides, the automatic system control is introduced into the system, where the average power of the quantum signal is monitored and real-time adjusted to adapt the SNR requirement of error correction.  
The training sequence is used for synchronizing Alice and Bob, and determining the relative phase between the signal and the LO. An automatic adjustment of the bias voltages that need to be applied to the amplitude modulators in Alice's site is performed in every 10 s. 
The repetition rate of the system is 350 kHz, the detection efficiency of the detector is 0.606. The modulation variance of the system is 18.5, and the excess noise is 0.005. 

B. Qi et al. realized a CV-QKD system over 5 km fibers using polarization and frequency division multiplexing to co-transmit the quantum signals and LO \cite{Qi_PhysRevA_2007}. 
The final secret key rate at 5 km is 0.3 bit/pulse. 
As shown in Fig. \ref{fig:iLOearly} (c), the quantum signals and LO in different polarizations are combined with a polarization coupler, and de-multiplexed at the receiver by another polarization beamsplitter. 
The frequency division multiplexing is realized with an acousto-optic modulator.
By applying both polarization and frequency division multiplexing, the isolation between quantum signal and LO comes up to 70 dB, well preventing the LO leakage.
An isolator is placed in the signal arm of Bob's MZI to reduce the noise due to multiple reflections of LO.
The phase compensation on hardware layer is replaced by a digital compensation at Alice's site. 
The noise from the receiver is assumed to be independent from Eve's control, which contributes to the performance enhancement.

\subsubsection{Long distance achievements}

\begin{figure}[t]
  \includegraphics[width=0.45\textwidth]{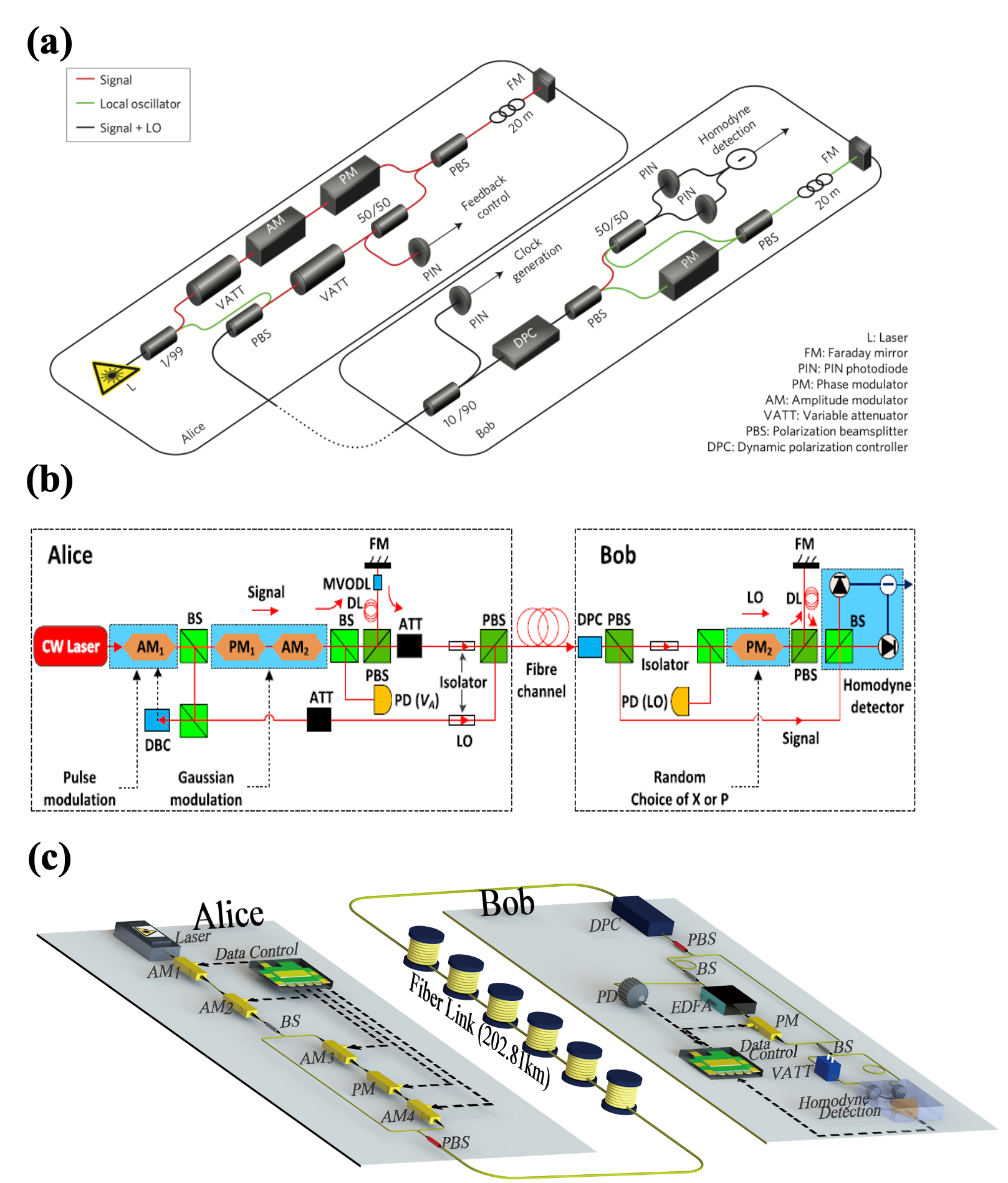}% Here is how to import EPS art
  \caption{\label{fig:iLOlong} The in-line LO CV-QKD systems with long distance. (a) 80 km CV-QKD system from P. Jouguet et, al. \cite{Jouguet_NatPhotonics_2013}. (b) 100 km CV-QKD system from D. Huang et al. \cite{Huang_SciRep_2016}. (c) 202.81 km CV-QKD system from Y. Zhang et al. \cite{Zhang_PhysRevLett_2020}.
  (a) Reproduced with permission from Nat. Photonics 7, 378-381 (2013). Copyright 2013 Springer Nature.
  (b) D. Huang, Sci. Rep., 6, 1-9, 2016; licensed under a Creative Commons Attribution (CC BY) license.
  (c) Reproduced with permission from Phys. Rev. Lett. 125, 010502 (2020). Copyright 2020 American Physical Society.
  }
\end{figure}

In 2013, the first long-distance CV-QKD system shown in Fig. \ref{fig:iLOlong} (a) is realized by P. Jouguet et al., which achieves the secret key rate of over 100 bps at 80 km \cite{Jouguet_NatPhotonics_2013}. Such an improvement is supported by the optimized reconciliation strategy and system architecture.
The multi-dimensional reconciliation firsly introduced into the practical experiment significantly enhances the reconciliation efficiency of a CV-QKD system, from no more than 90 \% to 95 \%. This promotes the developments of long-distance CV-QKD system and supports all of the long-distance CV-QKD system until now. 
The main features of this experimental system include the polarization and time division multiplexing, the trusted detection noise, the polarization control using a dynamic polarization controller and the clock synchronization with part of the LO.
The multiplexing scheme used in this system is widely used for achieving high isolation between quantum signal and LO in most of the long-distance systems.

As shown in Fig. \ref{fig:iLOlong} (b), D. Huang et al. later realized a CV-QKD system with transmission distance over 100 km and more than 300 bps secret key rate by controlling the excess noise to low level \cite{Huang_SciRep_2016}. An efficient scheme is proposed to perform high-precision phase compensation under low SNR conditions which contributes to the excess noise of 0.015. 
For system hardware, they developed a low-noise detector, which reduces the requirement of the high LO power. 

As shown in Fig. \ref{fig:iLOlong} (c), the CV-QKD system with the longest transmission distance is realized by Y. Zhang et al. at 2020, where the transmission distance can reach 202.81 km, which doubles the previous transmission distance record \cite{Zhang_PhysRevLett_2020}. 
In this work, two amplitude modulators are used for generating pulsed light, then an amplitude and a phase modulator are used for Gaussian modulation, and an amplitude modulator is used to attenuate the modulated light signal to quantum level, as well as enhancing the SNR of frame sequence. Time-division multiplexing and polarization multiplexing are used to ensure sufficient isolation between the co-transmit of quantum signals and LO.
Moreover, LO is amplified at the receiver site therefore the requirement of the launch power of LO at the transmitter site is reduced, which is beneficial for reducing the cross talk. 
Automatic feedback systems are used to overcome the channel perturbations, and high-precision phase compensation is adopted to supress the excess noise and highly efficient postprocessing is realized to achieve long transmission distances at sufficiently high secret key rates. Besides the transmission distance of 202.81 km, the system was also tested with the link distance of 27.27, 49.30, 69.53, 99.31 and 140.52 km, where the secret key rate reached 278, 62, 4.28, 1.18 and 0.318 kbps. 

The key modules of an in-line LO CV-QKD system are the multiplexing module and the reconciliation module. 
It can be seen that the development of the reconciliation directly supports the long-distance transmission, where the long transmission distance always combines with high reconciliation efficiency.
The multiplexing scheme finally evolves into the time and polarization multiplexing, since this is the simplest method to achieve sufficient isolation between quantum signal and LO.

\begin{figure}[t]
  \includegraphics[width=0.45\textwidth]{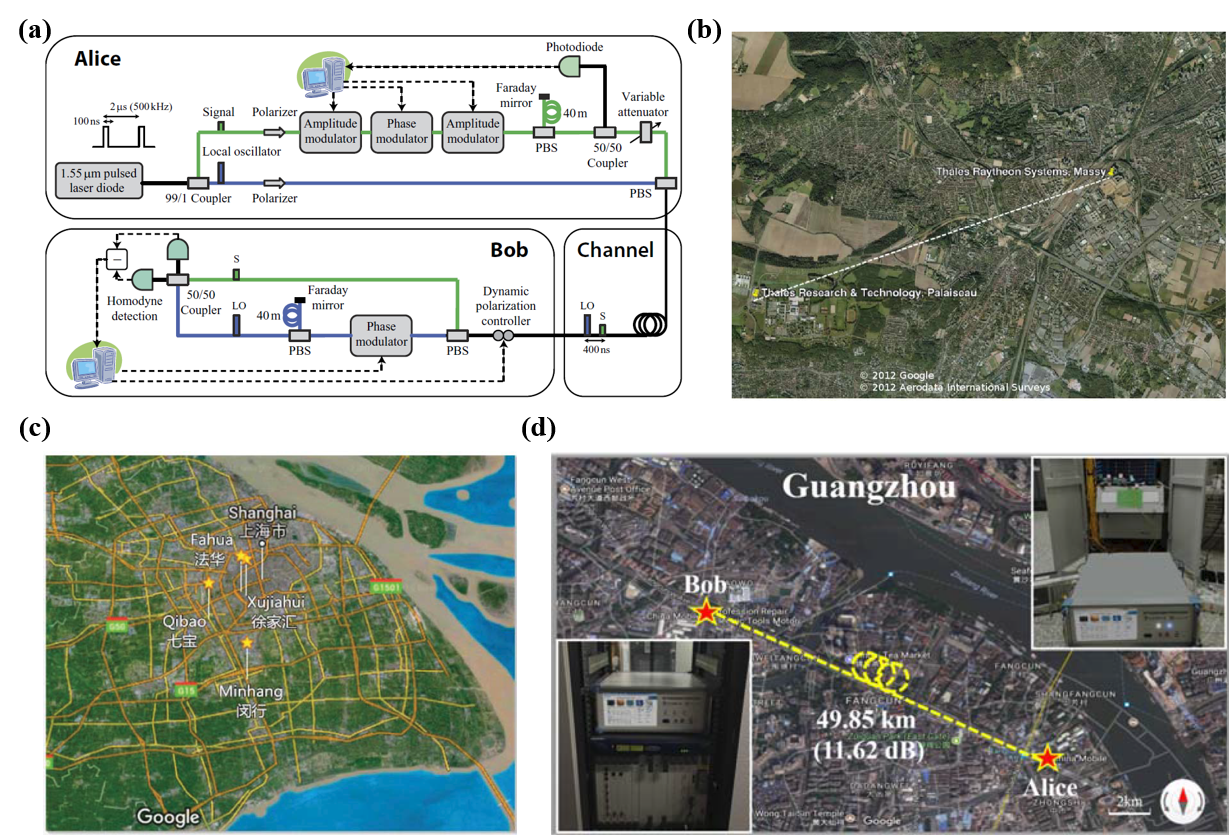}% Here is how to import EPS art
  \caption{\label{fig:iLOField} The field tests of in-line LO CV-QKD systems. (a) The first field test of CV-QKD system, from S. Fossier et al. \cite{Fossier_NewJPhys_2009}. (b) The field test of using CV-QKD for classical symmetric encryption, from P. Jouguet et al. \cite{Jouguet_OptExpress_2012} ©2012 Google. (c) Field demonstration of a CV-QKD network, from D. Huang et al. \cite{Huang_OptLett_2016}. (d) The longest field test of CV-QKD with 50 km commercial fiber, from Y. Zhang et al. \cite{Zhang_QuantumSciTechnol_2019}. 
  (a) Reproduced with permission from New J. Phys. 11, 045023 (2009). Copyright 2009 IOP Publishing Ltd.
  (b) Copyright 2012 Google.
  (c) Copyright 2016 Google.
  (d) Reproduced with permission from Quantum Sci. Technol. 4, 035006 (2019). Copyright 2019 IOP Publishing Ltd. 
  }
\end{figure}

\begin{figure*}[htbp]
  \includegraphics[width=0.9\textwidth]{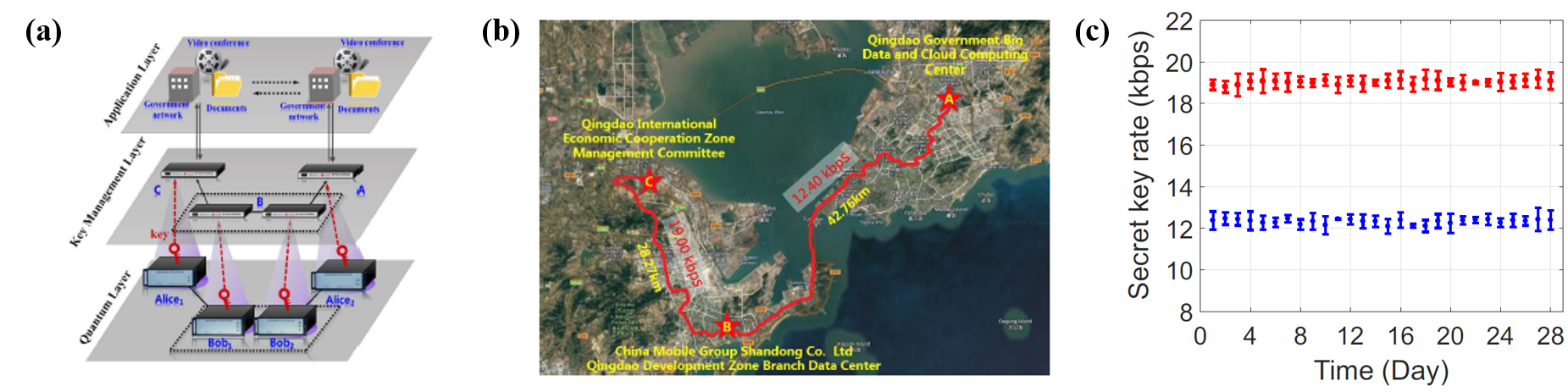}% Here is how to import EPS art
  \caption{\label{fig:iLOField2} The long-term field test in Qingdao, China, from Y. Zhang et al.\cite{zhang2020continuous}. (a) The layer structure of the 3-node network. (b) A Bird's-eye view of the field test environment. (c) 28-day continuous test results.}
\end{figure*}

Besides the reconciliation and multiplexing, the modulation, monitoring, sampling and compensation techniques are also well developed. 
The generation of light pulses with high extinction ratio is demonstrated in 2015 \cite{Wang_JQuantumElectron_2015}, where the extinction ratio overpasses 80 dB  by using a \hl{double} cascaded MZI modulation.
Later, the imperfect quantum state preparation is theoretically and experimentally investigated in 2017, which demonstrates that the incorrect calibration of the working parameters for the amplitude modulator and phase modulator can lead to a significant increase of the excess noise and misestimate of the channel loss. Schemes for calibrating the working parameters of the modulators are proposed and demonstrated to solve the imperfect state preparation issue.
An LO monitoring scheme is proposed to enhance the practical security of a CV-QKD system \cite{Liu_OptExpress_2017}. The LO is monitored by the balanced photodiode, and the SNU is real-time calibrated.
The excess noise in the system and its impact on the performance is also investigated \cite{Wang_OptExpress_2019}.
Finite sampling bandwidth of the analog-to-digital (AD) converter may lead to inaccurate results of pulse peak sampling, which is solved by a dynamic delay adjusting module and a statistical power feedback-control algorithm \cite{Li_OptExpress_2015}.

As for polarization compensation, a feedback algorithm is proposed to stable the system, where a polarization feedback signal is produced by an amplified Root Mean Square to Direct Current conversion by picking out a 10 \% portion of the LO light in real time \cite{Wang_SciRep_2015}. With the output data, one can estimate the mean value and the standard deviation of the polarization drift. Then a dynamic polarization controller is deployed at the receiver's site to stable the polarization automatically. 
For phase noise compensation, a widely used scheme is to insert one frame of training sequence into the quantum signal path, to calculate the expectations and evaluate the phase difference \cite{Wang_SciRep_2015}. Long term stable phase locking is employed to seperately compensate the fast-fading and the slow-fading phase mismatch by adjusting the phase modulator and fiber length \cite{Shen_SciRep_2015}. Later, a novel phase compensation scheme based on an optimal iteration algorithm is proposed to realize the fast-fading phase compensation accurately \cite{Li_OptExpress_2019}. 

\subsubsection{Field tests}

Besides the system in laboratory, field tests of the in-line CV-QKD system are also widely performed.
The first field test CV-QKD system is realized by S. Fossier in 2009 \cite{Fossier_NewJPhys_2009}, the system structure is shown in Fig. \ref{fig:iLOField} (a). It was automatically operated over 57 h, and achieved a secret key rate of 8 kbps over a 3 dB loss optical fiber. The system is part of the SECOQC network \cite{Peev_NewJPhys_2009}, where its practical fiber length is 9 km.
Later, the system was used for the encryption of point-to-point communications, which demonstrated the reliability of a CV-QKD system over a long period of time in a server room environment \cite{Jouguet_OptExpress_2012}. The map of the CV-QKD link is shown in Fig. \ref{fig:iLOField} (b).
The stability of this field test system is studied in detail, and the results show that, the birefringence, and consequently the polarization, in the installed fibre typically varied ten times slower than a laboratory fibre spool of equivalent length, while the phase drift linked to temperature changes in the devices is typically of $2\pi$ every 30 s and these vibrations have a typical frequency of 50-1000 Hz.

A full-mesh 4-node CV-QKD field test is realized in 2016 by D. Huang et al. \cite{Huang_OptLett_2016}, where 6 point-to-point CV-QKD links with distances of 19.92, 35.35, 37.44, 15.34, 17.52 and 2.08 km connect 4 nodes together, as shown in Fig. \ref{fig:iLOField} (c). 
This work adopts wavelength division multiplexing to co-transmit the quantum signal and the essential classical signals for clock synchronization and the forward and backward classical data communication. The reflection caused by the connectors of the field fiber links is well studied, and the results show that the connectors feature a nominal reflectance of -40 dB, and more than 20 reflective events are measured in the experiments.

The longest field test of CV-QKD system is realized in 2019, by Y. Zhang \cite{Zhang_QuantumSciTechnol_2019}, through 49.85 km commercial fiber, as shown in Fig. \ref{fig:iLOField} (d). 
By applying an efficient calibration model with one-time evaluation, a rate-adaptive reconciliation method which maintains high reconciliation efficiency with high success probability in fluctuated environments, and a fully automatic control system which stabilizes system noise, a secret key rate which is two orders-of-magnitude higher than the previous field test demonstrations is achieved. 

Later, the first network application demonstration with clear application scenarios over a long period of time through existing commercial optical fiber links is tested in Qingdao, China, as shown in Fig. \ref{fig:iLOField2} (a) and (b) \cite{zhang2020continuous}.
The performance of the 3-node network is tested for a month, where the the total length of the application demonstration link is 71.03 km, with a trusted relay in the middle.  As shown in Fig. \ref{fig:iLOField2} (c), the average secret key rate achieves higher than 12.00 kbps over 71.03 km optical fiber line, which paves the way to deploy CV-QKD in metropolitan settings.

\subsubsection{Systems on chip}

\begin{figure}[b]
  \includegraphics[width=0.4\textwidth]{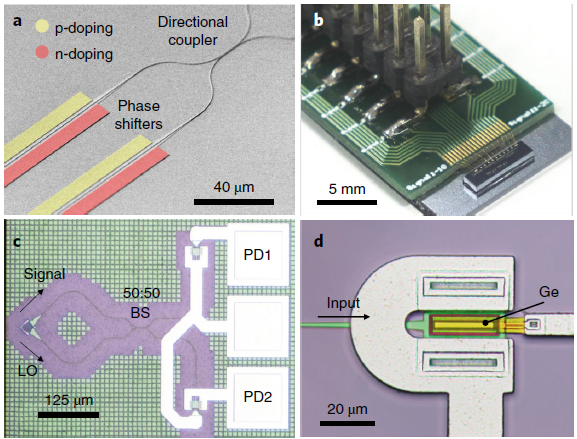}
  \caption{\label{fig:System_Chip1}  The first chip-based CV-QKD system. From G. Zhang et al.  \cite{Zhang_NatPhotonics_2019}. 
  Reproduced with permission from Nat. Photonics 13, 839 (2019). Copyright 2019 Springer Nature.
  }
\end{figure}

Photonic integrated circuit is a promising way to realize a large scale and cost effective system. After the design is finalized, the production cost of the chip will sharply decrease with the increase of production. Therefore, in addition to miniaturizing the system, chipization can also greatly promote the low-cost mass production for the cost sensitive CV-QKD system.
The earliest attempt to chip-based CV-QKD systems is in 2015, where a silicon photonic chip comprising all major CV-QKD components as well as complete subsystems \hl{are} designed and fabricated \cite{ziebell2015towards}. 

Later the  first chip-based CV-QKD platform is demonstrated in 2019 \cite{Zhang_NatPhotonics_2019}, where most of the active devices such as the phase modulator, amplitude modulator, optical variable attenuator and homodyne detector are integrated on Silicon-On-Insulator chip, as shown in Fig. \ref{fig:System_Chip1}. The phase and amplitude modulator have  a 90 \% switching time of 2.5 ns, corresponding to a 200-MHz modulation frequency. However, the homodyne detector limits the bandwidth of the system to 10 MHz, mainly affected by the two-stage transimpedance amplifier. The shot noise is 5 dB higher than the electronic noise, and the detection efficiency is 0.498. 
With these on-chip devices, an in-line LO system is demonstrated. 
A grating coupler introduces the light from an external laser source into the chip, then an 1:99 directional coupler splits it into two path, where the weak one is used for quantum signal modulation, while the stronger one is the LO. After that, the Gaussian modulated quantum signal and LO are multiplexed into two orthogonal polarization states with a 2D grating coupler, and output to the channel. The receiver uses a 2D grating coupler to seperate the quantum signal and LO, then the quantum signal is homodyne detected. The system is tested with a 2 m fiber for proof-of-principle demonstration, the secret  key rate can reach 0.25 Mbps. 

\begin{figure}[t]
  \includegraphics[width=0.45\textwidth]{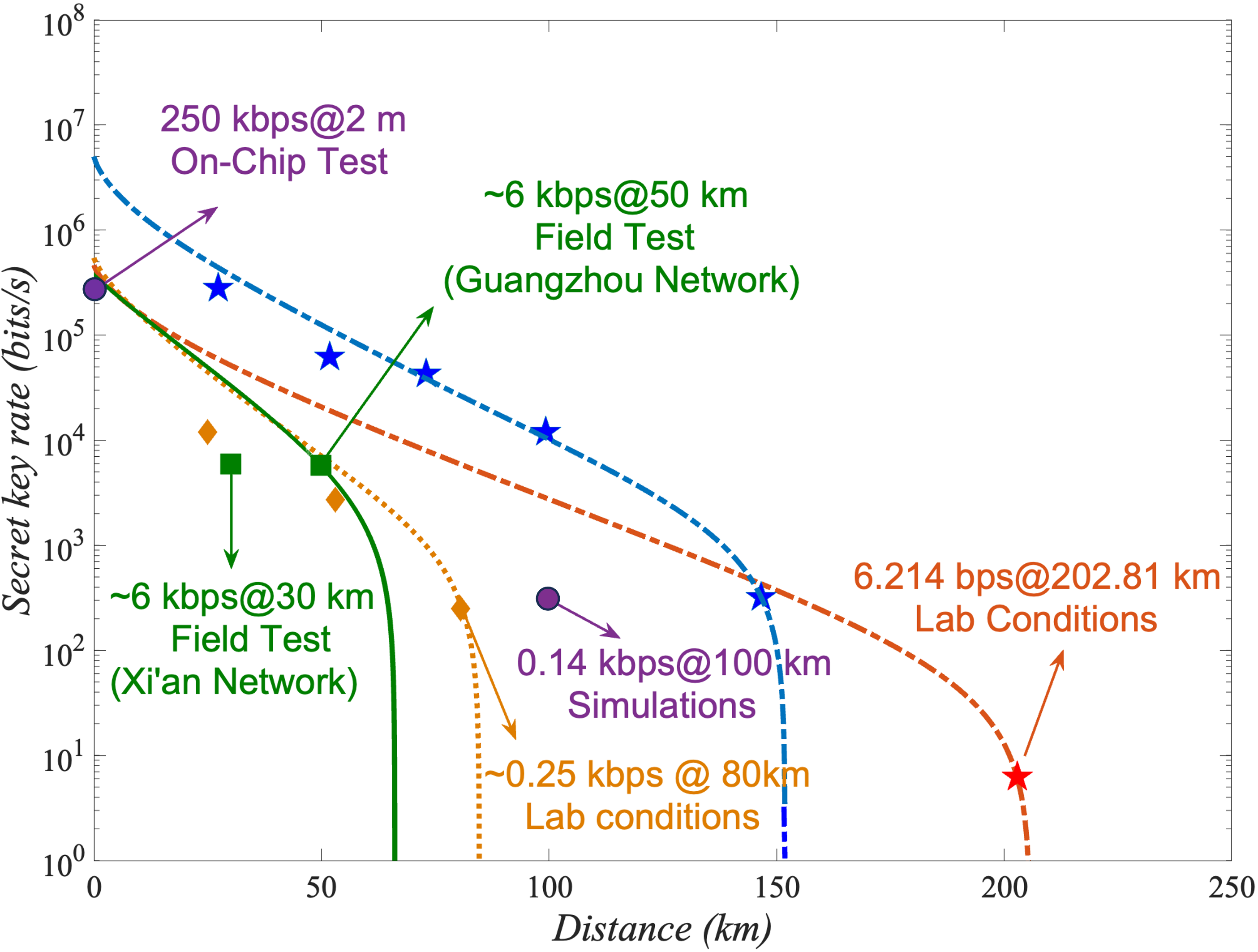}% Here is how to import EPS art
  \caption{\label{fig:iLOKeyRate} The overview of the key achievements of the secret key rate in the in-line CV-QKD systems. The longest transmission distance is 202.81 km in laboratory, and 50 km with field test. The back-to-back test of the on-chip system is demonstrated, and the estimated parameters can support the 100 km system.}
\end{figure}

This work \hl{demonstrates} that the Silicon-On-Insulator platform can basically satisfy the requirement of a CV-QKD system. 
A two-dimensional grating coupler integrated on chip can be directly used for the polarization multiplexing and de-multiplexing. The pulse generation, the Gaussian modulation, the variable optical attenuator and the detector are all  realized with chip-based components. The insertion loss of the grating coupler, the detection efficiency and the bandwidth of the chip-based detector are the concerns, where a part of these issues are solved in the later works, and we detailed them in the local LO system part.
A comparison of the secret key rate of different systems is shown in Fig. \ref{fig:iLOKeyRate}.

\subsubsection{Other in-line LO systems}
\begin{figure}[t]
  \includegraphics[width=0.45\textwidth]{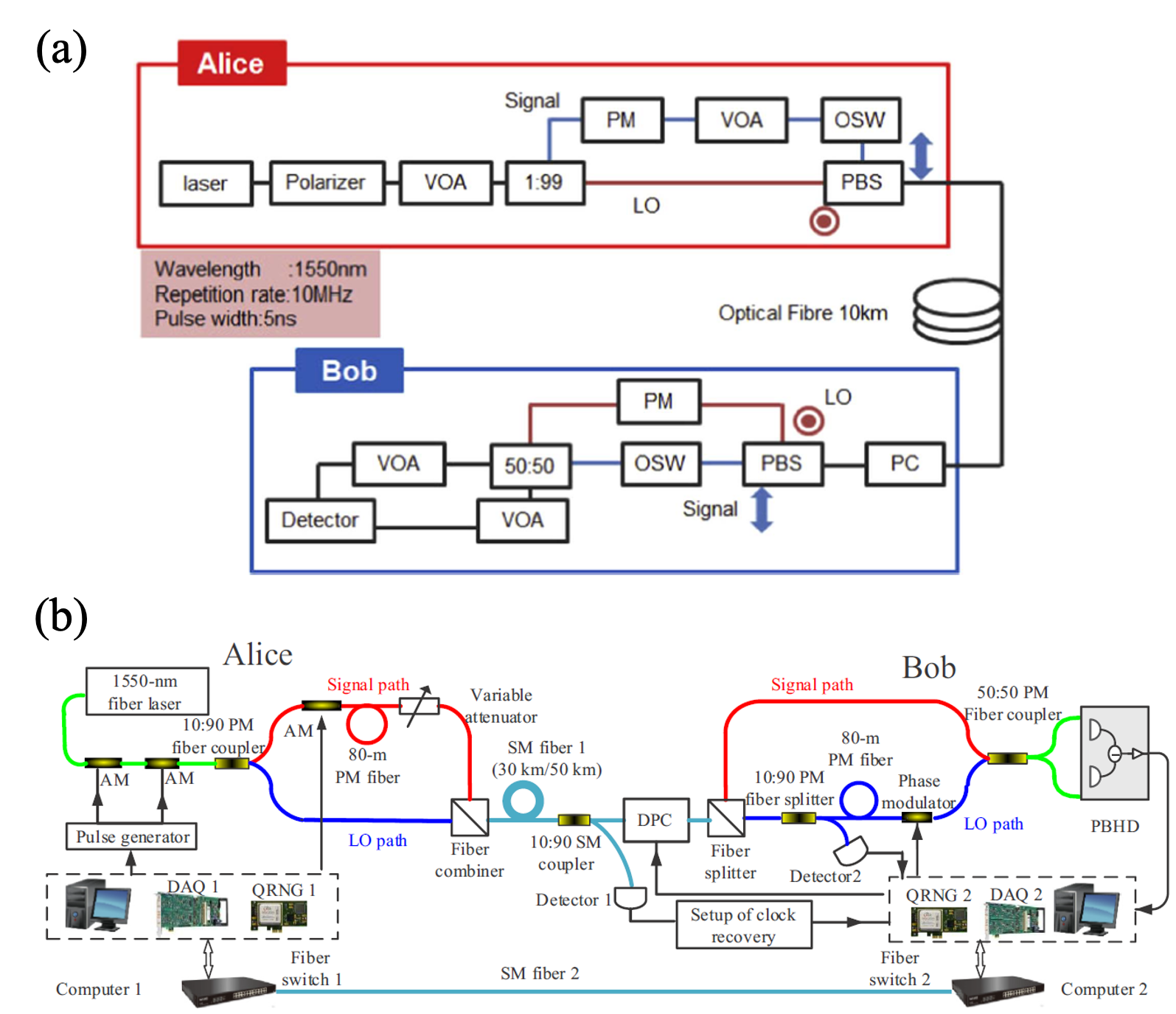}% Here is how to import EPS art
  \caption{\label{fig:iLO_DMandUD} The in-line LO CV-QKD systems with particular modulation format. (a) The 4-state modulated in-line LO CV-QKD system \cite{hirano2017implementation}, from T. Hirano et al. (b) The unidimensional modulated in-line LO CV-QKD system \cite{Wang_PhysRevA_2017}, from X. Wang et al.
  (a) T. Hirano, Quantum Sci. Technol., 2, 024010, 2017; licensed under a Creative Commons Attribution (CC BY) license.
  (b) Reproduced with permission from Phys. Rev. A  95, 062330 (2017). Copyright 2017 American Physical Society.
  }
\end{figure}

Besides Gaussian modulation, discrete modulation and unidimensional modulation is also realized with in-line LO system, as shown in Fig. \ref{fig:iLO_DMandUD}. The discrete modulation format has lower requirement for digital to analog conversion, and  the unidimensional modulation can be realized with a single amplitude modulator, which are suitable for cost-effective applications.  Polarization multiplexing is also used in these systems. 
The four-state modulation CV-QKD system from X. Wang et al. can reach 1 kbps secret key rate with the transmission distance of 30.2 km \cite{Wang_ChinesePhysLett_2013}, the four-state modulation CV-QKD system from \hl{T. Hirano et al.} can reach 50 kbps secret key rate with the transmission distance of 10 km \cite{hirano2017implementation}, and the unidimensional modulated system achieves 5.4 kbps and 0.7 kbps secret key rate at 30 and 50 km \cite{Wang_PhysRevA_2017}. 

% The in-line LO CV-QKD system has developed for a long time, and different multiplexing schemes are proposed to suppress the cross talk from LO to quantum signal. However, when we hope to achieve high key rate by raising the repetition frequency, the cross talk will increase, which weakens the system performance.

\begin{table*}[t]
  \renewcommand{\arraystretch}{1.8}
  
  \caption{\label{tab:LLO}A Comparison Between Different Local LO CV-QKD Systems. Here, $f$ means the repetition frequency, $L_{max}$ means the maximum transmission distance and SKR is the secret key rate corresponding to $L_{max}$. }
  \begin{ruledtabular}
  \begin{center}
  
  \begin{tabular}{cccccccc}
    & \multirow{2}{*}{Years} & \multicolumn{3}{c}{Key modules}                                & \multicolumn{3}{c}{Key indicators}                                                     \\ \cline{3-8} 
    &                        & Modulation format & Modulator    & Multiplexing                & $f$        & $L_{max}$ & SKR                                                           \\ \hline
  \multirow{8}{*}{Lab systems}        & 2017                   & 8-PSK             & DP-MZM       & Frequency                   & 40 MBaud   & 40 km     & 0.006 bit/symbol \cite{Kleis_OptLett_2017} \\
    & 2020                   & Gaussian          & AM+PM        & Frequency  and polarization & 100 MHz    & 25 km     & 7.04 Mbps  \cite{Wang_OptExpress_2020}     \\
    & 2022                   & QPSK              & IQ modulator & Frequency  and polarization & 5 GBaud    & 10 km     & 133.6 Mbps \cite{SubGbps}                     \\
    & 2022                   & 256 QAM           & IQ modulator & Frequency                   & 600 MBaud  & 25 km     & 24 Mbps \cite{roumestan2022experimental}     \\
    & 2022                   & 256 QAM           & IQ modulator & Frequency  and polarization & 1 GBaud    & 50 km     & 9.212 Mbps \cite{Pan2022DM}                  \\
    & 2022                   & Gaussian          & \hl{Sagnac fiber loop} & Time                   & 10 MHz     & 50 km     & 0.08 Mbps \cite{zhao2022simple}                  \\
    & 2023                   & Gaussian          & IQ modulator & Frequency                   & 100 MHz    & 20 km     & 0.0471 bit/symbol \cite{jain2022practical}   \\
    & 2023                   & 16 state          & IQ modulator & Frequency                   & 2.5 GBaud  & 80 km     & 2.11 Mbps \cite{Tian2023DM}                  \\
    & 2023                   & Gaussian          & IQ modulator & Frequency  and polarization & 1 GHz      & 100 km    & 0.51 Mbps \cite{Pi2023SubMbps}     \\ 
    & 2024                   & Gaussian          & IQ modulator & Frequency                   & 100 MBaud  & 100 km    & 0.0254 Mbps \cite{hajomer2024long}     \\ \cline{1-8} 
    \multirow{3}{*}{Field tests} & 2019          & //           & //        &   //            & //      & 3.9 km    & 0.07 Mbps \cite{aguado2019engineering}     \\ 
    & 2023                  & Gaussian           & //           & //                   & 12.5 MBaud      & 22.5 dB    & 0.01 kbps \cite{Brunner2023DemonstrationOA}     \\ 
    & 2023                   & Gaussian          & Sagnac fiber loop & Time                   & 50 kHz      & 10.4 km    & ~1.6 kbps \cite{williams2023continuousvariable}     \\ \cline{1-8}
    \multirow{3}{*}{Chip-based systems} & 2023                   & Gaussian          & IQ modulator & Frequency                   & 100 MHz    & 6.9 km    & 0.28 Mbps  \cite{pietri2023cv}                                                    \\
    & 2023                   & Gaussian          & IQ modulator & Time                        & 8 MHz      & 11 km     & 0.4 Mbps  \cite{aldama2023inp}                                                    \\
    & 2023                   & Gaussian          & IQ modulator & Time                        & 0.25 GBaud & 50 km     & 0.75 Mbps \cite{li2023continuous}                                                   
  \end{tabular}  
  \end{center}
  \end{ruledtabular}
\end{table*}

\subsection{Local LO systems}
Though the in-line LO system has developed for a long time, some problems are still inevitable. The most critical issue is the security loophole caused by the LO accessible to a potential eavesdropper \cite{ma2013local, ma2013wavelength, huang2013quantum, jouguet2013preventing}. The manipulation of LO can make the sender and receiver perform parameter estimation mistakenly, leading to an overestimate of secret key rate. Monitoring the LO can defend part of the attacks, but the loophole still exists and more attacking strategies aiming at it can be continuously developed. 
Besides, from the system development, when we hope to achieve high key rate by raising the repetition frequency or achieve long distance overcoming the large channel loss, the crosstalk from LO to quantum signal will increase, which weakens the system performance. Therefore, after more than a decade of development, the local LO CV-QKD system without the transmission of LO is proposed.

% Secondly, the co-transmission of quantum and LO signal results in the insufficient LO intensity at the receiver side. Since a high power LO contributes to more cross-talk and noise, the power of LO cannot continuously increase. Optical amplifiers can raise the power to a sufficient level, but it introduces extra noise and makes the system complex.
% Since the LO is not required to be transmitted through the quantum channel, sufficient LO power can be provided without limited by the transmission distance, which contributes to the high performance system.
The local LO system is firstly proposed in 2015, intends to solve the problems above by generating LO inside the receiver, which is a once and for all solution. Since the quantum signal and the LO are generated by different lasers, a fast-fading phase noise is introduced, leading to the increase of excess noise.
The key of a local LO system is to establish a reliable phase reference between the sender and receiver, usually realized by a classical reference signal, namely pilot tone. 
When heading towards high-speed system with high repetition rate, the pulsed system with time division multiplexing is no more effective. Considering that the power of pilot tone is low enough so that its leakage is not as much as the in-line transmitted LO, the frequency division multiplexed quantum and pilot tone signals with continuous-wave light becomes the mainstream scheme of the local LO system. In addition, the DSP is introduced to the local LO system for more accurate phase recovery, high speed modulation with continuous-wave light and simpler detection. A review of the evolution of local LO CV-QKD system is shown in Table. \ref{tab:LLO}, \hl{including the typical systems with different settings, and there are also some noteworthy local LO systems.} \cite{grande2021adaptable,sarmiento2022continuous,aldama2022integrated}

% \begin{figure}[t]
%   \includegraphics[width=0.45\textwidth]{System_LLOProposed.png}% Here is how to import EPS art
%   \caption{\label{fig:LLOProposed} The first two local LO CV-QKD systems. (a) The local LO CV-QKD system with heterodyne detection. From Qi et al. \cite{Qi_PhysRevX_2015}. (b) The local LO CV-QKD system with one laser to support quantum signal and LO, and homodyne detection is used. From Soh et al. \cite{Soh_PhysRevX_2015}.}
% \end{figure}

\subsubsection{Early systems}
The early experiments adopt a pulsed laser source with LiNbO3 amplitude and phase modulators for generating quantum and pilot tone signals \cite{Qi_PhysRevX_2015,Soh_PhysRevX_2015}. In the work finished by B. Qi et al.\cite{Qi_PhysRevX_2015}, the quantum signals and pilot tones are time-division multiplexed and then detected by receivers using heterodyne detection. The sender's and receiver's LO laser sources are free running without any connections, and the heterodyne detection results of the pilot tone provide the phase reference for data rotation. 
In D. Soh's work \cite{Soh_PhysRevX_2015}, the signal and LO are generated from one laser for proof-of-principle demonstration. Homodyne detection is used for detecting the quantum signal and pilot tone. Each pilot tone is sent twice in a pair for the receiver to get both quadratures with homodyne detection.

In the above works, continuous-wave LO signals are adopted, while in Huang's work \cite{Huang_OptLett_2015}, an amplitude modulator is deployed inside the receiver side for generating pulsed LO. The quantum and pilot tone signals are also generated by the same modulation module with time division multiplexing. 
The commonalities of these early systems are using pulsed light, and time-division multiplexed quantum and pilot tone signals are generated by the same modulation module. The key role of these early systems is verifying the possibility of compensating the phase mismatch between the quantum and LO signal with high precision required in a CV-QKD system using a pilot tone.

\begin{figure}[t]
  \includegraphics[width=0.45\textwidth]{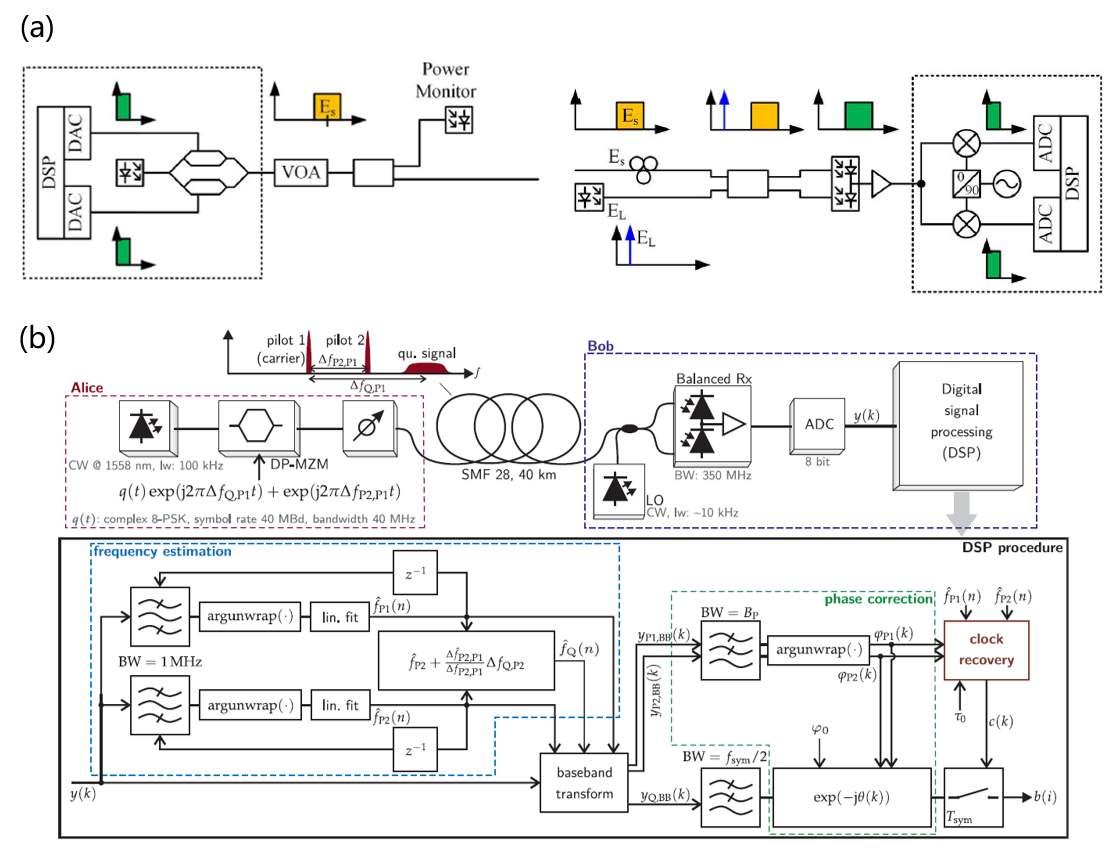}% Here is how to import EPS art
  \caption{\label{fig:System_LLODigital} The continuous-wave local LO CV-QKD system. (a) The software defined CV-QKD transmitter  and receiver.  From H. H. Brunner et al. \cite{Brunner_ICTON_2017}. (b) The local LO CV-QKD system with DSP where the quantum signal and two pilot tones are frequency division multiplexed. From S. Kleis et al. \cite{Kleis_OptLett_2017}.
  (a) Reproduced with permission from ICTON 1-4 (2017). Copyright 2017 IEEE.
  (b) Reproduced with permission from Opt. Lett. 42, 1588 (2017). Copyright 2017 Optical Society of America.
  }
\end{figure}

% \begin{figure}[b]
%   \includegraphics[width=0.35\textwidth]{System_Quanti.png}% Here is how to import EPS art
%   \caption{\label{fig:System_Quanti} The quantification of quantum and pilot signal. When modulating or detecting quantum and pilot signal simultaneously with different center frequency, the quantification interval should cover both quantum and the much stronger pilot tone. The quantum signal can only use part of the quantification interval.}
% \end{figure}

\subsubsection{Systems with continuous-wave light}
As we mentioned above, the most crucial issue of a local LO CV-QKD system is the phase recovery with pilot tone. To improve the accuracy of the phase recovery, raising the repetition frequency of the quantum and pilot tone signals for a more accurate track of the phase shift is a direct and effective way.
However, the pulsed system with time division multiplexing significantly limits the system repetition frequency. Therefore, it is a general trend to use continuous-wave light instead of pulsed light to support a high speed system \cite{Brunner_ICTON_2017,Kleis_OptLett_2017}. Without time division multiplexing, the repetition rate of the system is significantly enhanced, resulting in a high-rate system with 1 GHz repetition rate \cite{Wang_OptExpress_2020, SubGbps, Pan2022DM}. 

Besides, for the multiplexing of quantum signal and pilot tone, since the high power pilot tone is not necessary required, the quantum signal and pilot tone can be modulated simultaneously with the same modulation module in a frequency division multiplexed scheme \cite{Brunner_ICTON_2017,Kleis_OptLett_2017}. Naturally, by introducing a frequency difference between the quantum signal and LO, a digital heterodyne detection can be realized, where we can use the detection results from one homodyne detector to recover both information of quadratures \cite{Brunner_ICTON_2017,Kleis_OptLett_2017,Wang_OptExpress_2020,SubGbps, Pan2022DM,Pi2023SubMbps}. In this way, the experimental demonstration of the system is significantly simplified. 

H. H. Brunner et al. demonstrated a local LO CV-QKD system as shown in Fig. \ref{fig:System_LLODigital} (a) \cite{Brunner_ICTON_2017}. The 4-state discrete modulation of coherent states is realized with quadrature phase-shift keying modulation used in classical communications, and a variable optical attenuator is used to adjust the power to the quantum level. No amplitude modulator for pulse generation is used, instead, a RRC filter in the digital domain completes the pulse shaping. Moreover, combined with an analog electronic low-pass filter at the output of the digital-to-analog (DA) convertor, the quantum signal is concentrated in the 10 MHz bandwidth. Digitally, the quantum signal is up-converted and combined with a pilot tone. To reduce the quantization noise since the weak quantum signal is produced and detected together with a much stronger pilot tone, the DA and AD convertor bit width are 16 bits and 14 bits respectively. The heterodyne detection is also performed digitally, the LO is set to have a frequency difference with the quantum signal, and a down conversion in digital domain is then performed to recover the  information on both quadratures.

As shown in Fig. \ref{fig:System_LLODigital} (b), S. Kleis et al. realized a complete system with the simultaneous modulation of 8 phase-shift keying discrete modulated quantum signal and two pilot tones using a dual-parallel Mach-Zehnder modulator \cite{Kleis_OptLett_2017}. Digital heterodyne detection is used to get both quadratures with a single homodyne detector. This work achieved the secret key rate of $6 \times 10^{-3}$ bit/symbol at 40 km, which shows the way of low-complexity QKD system demonstration within metropolitan distances. 
The core ideas of the above works is to realize CV-QKD system with continuous-wave light and a structure similar to the classical coherent communications, DSP is widely used in the system, which opens the new way of CV-QKD systems.

% Despite of the simplified system structure, the simultaneous modulation and detection of weak quantum signal and stronger pilot tone raise the requirements of digital-to-analog and analog-to-digital conversion. As shown in Fig. \ref{fig:System_Quanti}, when simultaneously modulate or detect the frequency division multiplexed quantum signal and pilot tone with one modulator, most of the quantification bits are used by the much stronger pilot tone. Only a small part of  is left for quantum signal, which reduces the precision and introduces quantification noise. If the quantification bit cannot be higher, using discrete-modulation format rather than the Gaussian modulation format is more suitable since the quantification requirement is reduced. 

% As we can see, the early continuous-wave systems both adopted the discrete modulation for a simpler modulation module and a structure similar to the classical coherent communication system. 
Subsequently, the discrete local LO CV-QKD system is extended to high-order modulation formats for better performance. 
As theoretically analyzed, the increasing of modulation order results in higher secret key rate with most of the security framework of discrete-modulation CV-QKD. Moreover, when the constellation is similar to the Gaussian distribution, the secret key rate of the system can be better than the constellation with uniform distribution. 
In this guidance, a high-order discrete modulated CV-QKD system is realized with 16-order two-ring phase shift key modulated coherent states \cite{Tian2023DM}. Compared with the high-order quadrature amplitude modulation, 16-order two-ring phase shift key modulation can be realized with less DA requirements, contributing to suppressing the modulation noise. The achieved secret key rates are 49.02 Mbps, 11.86 Mbps and 2.11 Mbps over 25-km, 50-km, and 80-km optical fiber. 67.4 \%, 70.0  \% and 66.5 \%  of the performance of a Gaussian modulated protocol can be achieved with this simpler scheme.

\subsubsection{Systems with polarization multiplexing}
Besides modulating the quantum signals together with pilot tone, the quantum signals can also be modulated individually and combined with the pilot tone in different polarization directions \cite{Wang_OptExpress_2018, Wang_PhysRevA_2018, laudenbach2019pilot, Wang_OptExpress_2020, Pi2023SubMbps}. 

% As shown in Fig. \ref{fig:System_LLOMul} (a), a local LO CV-QKD system with time division multiplexed and polarization multiplexed quantum and pilot tone signals is realized. The light from the continuous-wave laser is pulsed with an amplitued modulator, then splitted. One output is used for quantum signal modulation, while the other is used as the pilot tone. These two path of lights are then combined with a polarization beam combiner and output into the channel. 
% The receiver uses a polarization controller to adjust the polarization into a proper angle, then a polarization beamsplitter seperates the quantum and pilot signal. The quantum signal is homodyne detected, while the  pilot tone is seperately heterodyne detected for getting phase information. This work achieved 3.14 Mbps secret key generation rate at 25 km, with a repetition rate of 50 MHz. A continuing work is subsequently finished with heterodyne detection for  both quantum and pilot signal detection, as shown in Fig. \ref{fig:System_LLOMul}  (b).

\begin{figure}[t]
  \includegraphics[width=0.45\textwidth]{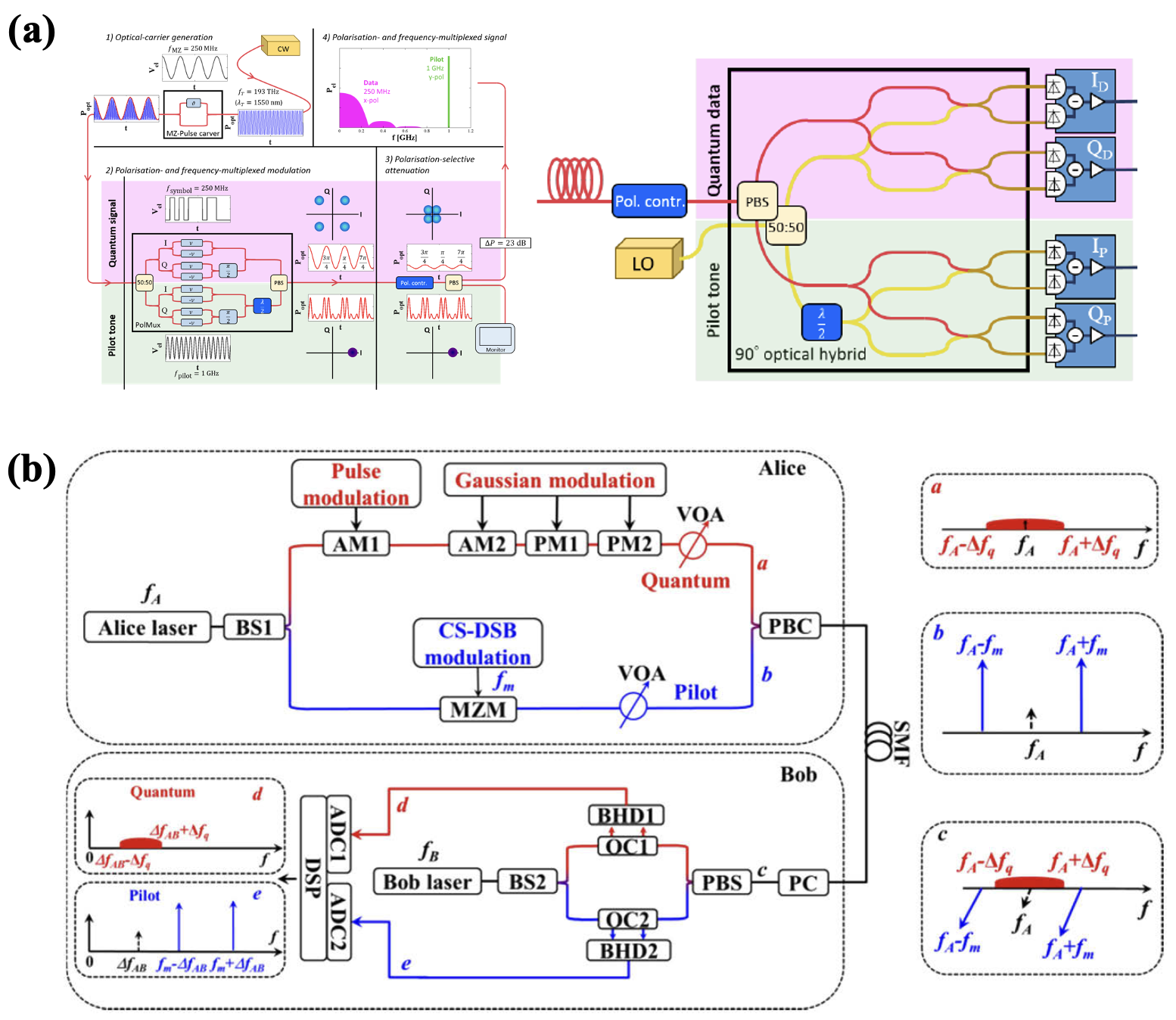}
  \caption{\label{fig:System_LLOMul} Local LO CV-QKD system with polarization multiplexed quantum signal and pilot tone.  (a) From F. Laudenbach et al. \cite{laudenbach2019pilot}.  (b) From H. Wang et al. \cite{Wang_OptExpress_2020}.
  (a) F. Laudenbach, Quantum, 3, 193, 2019; licensed under a Creative Commons Attribution (CC BY) license.
  (b) Reproduced with permission from Opt. Express 28, 32882 (2020). Copyright 2020 Optical Society of America.
  }
\end{figure}

As shown in Fig. \ref{fig:System_LLOMul} (a), the optical carrier is firstly amplitude modulated to generate 250 MHz signal pulses, then sent into the dual-polarization IQ modulator for the modulation of the discrete coherent states and the pilot tone. The quantum signal and pilot tone are multiplexed in different frequency and polarization directions.
The receiver performs heterodyne detection of both polarizations after the compensation of a polarization controller.
This scheme can help to reduce the excess noise since the isolation between the quantum signal and pilot tone is more sufficient than the frequency division multiplexed method.

Further, a Gaussian modulated system with frequency division multiplexing and polarization multiplexing is realized by H. Wang et al. \cite{Wang_OptExpress_2020}, shown in Fig. \ref{fig:System_LLOMul} (b). Digital heterodyne detection is used for detecting quantum and pilot tone signals respectively. It can achieve 7.04 Mbps asymptotic-limit secret key rate at 25 km  and 1.85  Mbps with finite-size.

% \begin{figure*}[t]
%   \includegraphics[width=0.85\textwidth]{System_LLO.png}% Here is how to import EPS art
%   \caption{\label{fig:System_LLO} The evolution of local LO CV-QKD system. TDM: time division multiplexing, FDM: frequency division multiplexing, Het: heterodyne detection, Hom: homodyne detection, Pol.: polarization multiplexing, and CW: continuous wave.}
% \end{figure*}

\subsubsection{Recent \hl{progresses}}
Based on the digital modulation and detection scheme, local LO CV-QKD system is heading towards high secret key rate \cite{SubGbps}, composable security\cite{jain2022practical} and long distance \cite{hajomer2023longdistance}. 

\begin{figure}[t]
  \includegraphics[width=0.45\textwidth]{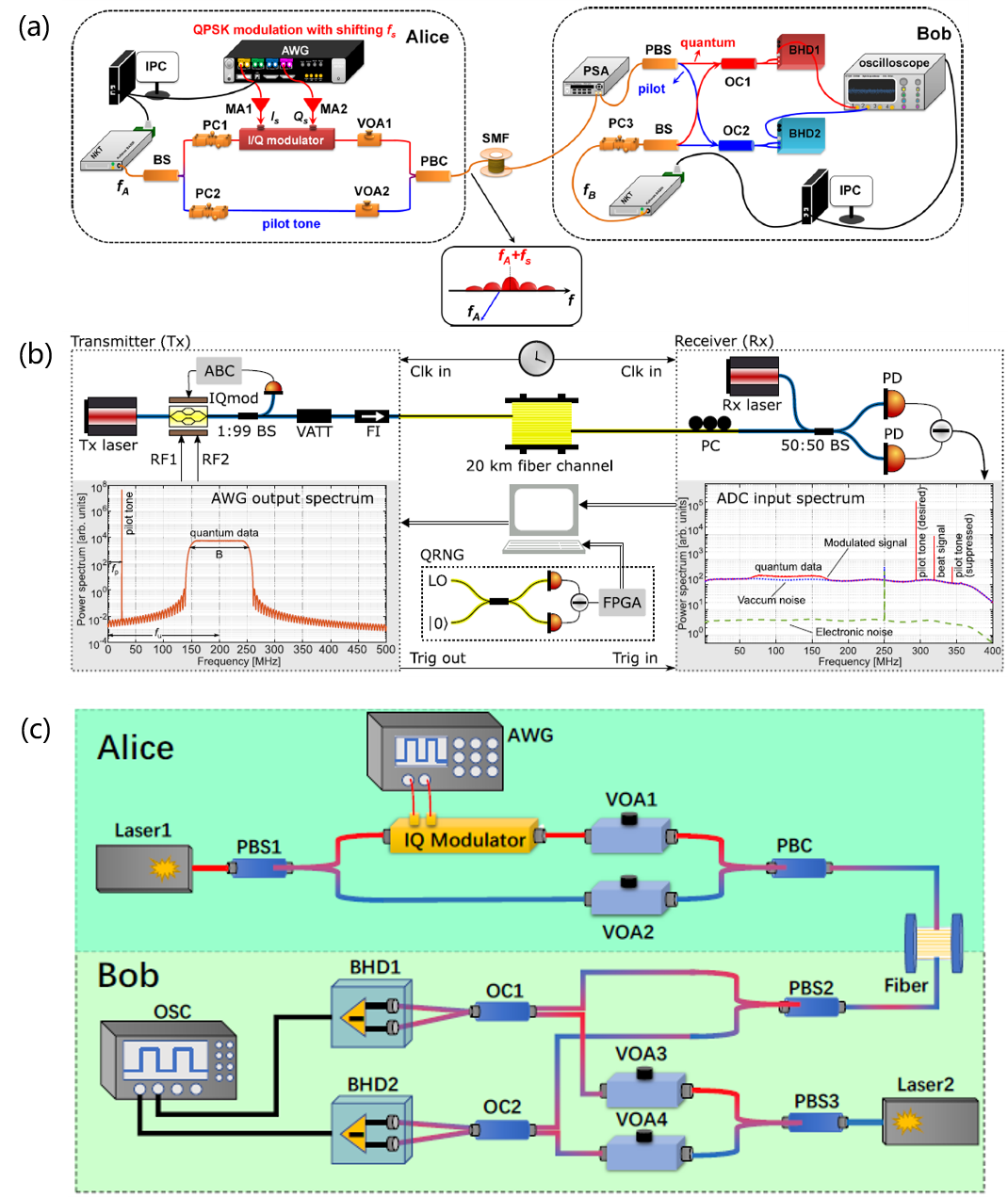}% Here is how to import EPS art
  \caption{\label{fig:System_LLO_New} The high-speed, composable-security and long-distance local LO CV-QKD systems. (a) The high-speed system with 4-state modulation format.  From H. Wang et al. \cite{SubGbps}. (b) The system which generates composable secret keys. From N. Jain et al.  \cite{jain2022practical}.(c) The system supporting 100 km transmission distance considering finite-size effect. From Y. Pi et al. \cite{Pi2023SubMbps}.
  (a) H. Wang et al., Commun. Phys., 5, 162, 2022; licensed under a Creative Commons Attribution (CC BY) license.
  (b) N. Jain et al., Nat. Commun., 13, 4740, 2022; licensed under a Creative Commons Attribution (CC BY) license.
  (c) Reproduced with permission from Opt. Lett. 48, 1766 (2023). Copyright 2023 Optica Publishing Group.
  }
\end{figure}

As shown in Fig. \ref{fig:System_LLO_New} (a), in H. Wang's work \cite{SubGbps}, using frequency division multiplexing and polarization multiplexing to transmit the quantum and pilot signal, as well as two digital heterodyne detector for measuring, a system with 4-state discrete modulated coherent states achieved 233.87 Mbps, 133.6 Mbps and 21.53 Mbps secret key rate at 5 km, 10 km and 25 km. This increases the asymptotic secret key rate to sub-Gbps level, which can satisfy the one-time pad cryptographic task. The further investigation on optimizing the practical system parameters shows an effective way to the high-performance system \cite{ma_SciChinaInforSci_2023}.

N. Jain's system with composable security realized 0.0471 bits/symbol secret key rate at 20.3 km with extremely simple devices shown in Fig. \ref{fig:System_LLO_New} (b), an IQ modulator produces frequency division multiplexed quantum and pilot signals which are detected by a digital heterodyne detector \cite{jain2022practical}. The system is able to generate composable secret keys with $2 \times 10^8$ coherent states,  which is far less than the previous requirements due to improvements to the security proof. 

A recent work with the structure shown in Fig. \ref{fig:System_LLO_New} (c) can realize a transmission distance over 100  km, and the asymptotic secret key rate can reach 7.55 Mbps, 1.87 Mbps and 0.51 Mbps over transmission distance of 50 km, 75 km and 100 km \cite{Pi2023SubMbps}.  This work significantly increases the transmission distances of a local LO CV-QKD system, and shows a promising way to realize  long-distance and high-speed QKD using telecom devices.

\subsection{CV-QKD systems co-existed with classical communication environment}

\begin{figure}[t]
  \includegraphics[width=0.45\textwidth]{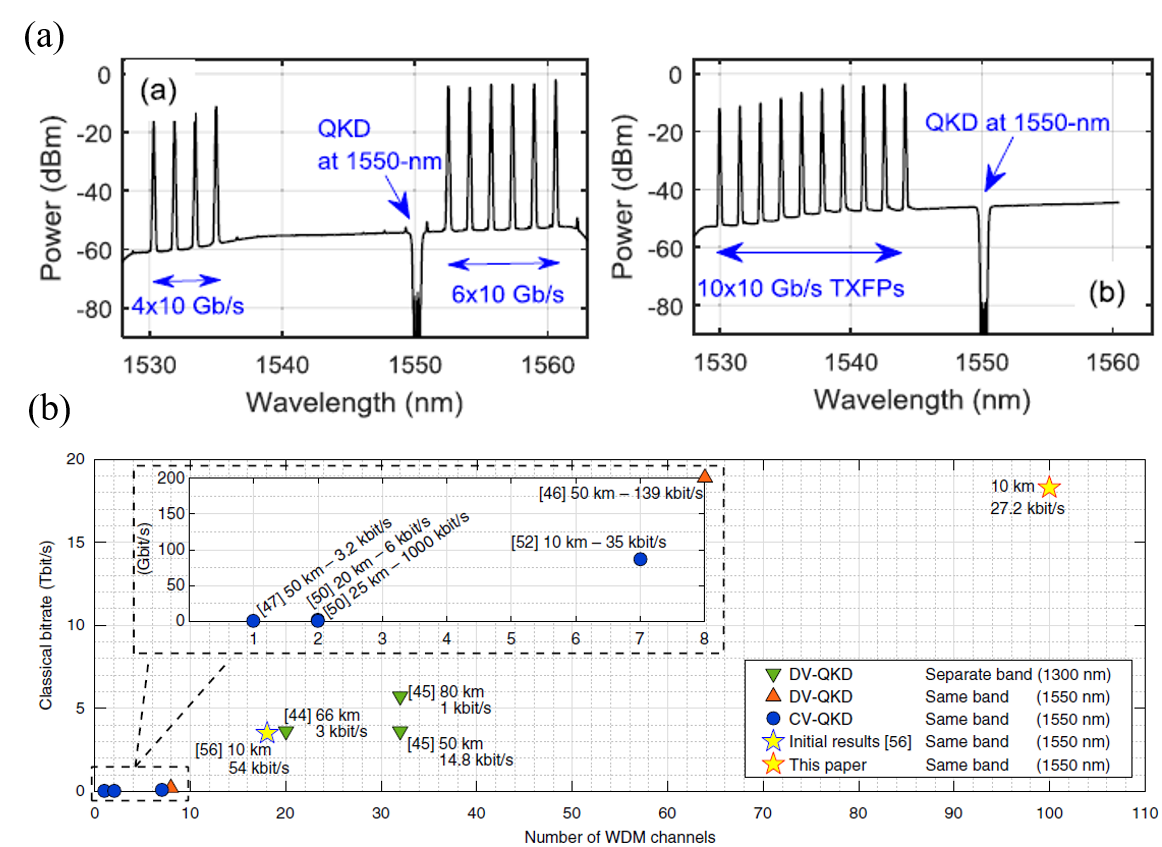}% Here is how to import EPS art
  \caption{\label{fig:System_coExist3} The CV-QKD systems coexisting with classical signals. (a) The frequency spectrum of the CV-QKD system co-existing with classical communication systems, from F. Karinou et al. \cite{Karinou_IEEE_2018}. (b) The performance of the CV-QKD system co-existing with 18.3 Tbps data channels, from T. A. Eriksson et al. \cite{Eriksson_CommunPhys_2019}. 
  (a) Reproduced with permission from IEEE Photon. Technol. Lett. (2018). Copyright 2018 IEEE.
  (b) T. Eriksson, Commun. Phys., 2, 1-8, 2019; licensed under a Creative Commons Attribution (CC BY) license.
  }
\end{figure}

The homodyne and heterodyne detection in CV-QKD system act as a matched filter since the LO naturally \hl{introduces} a frequency selection on the received signal, therefore the filter in time and frequency domain is unnecessary.
Moreover, the devices in a CV-QKD system is compatible with the classical coherent optical communications.
Therefore, CV-QKD is suitable for coexisting with classical signals, which is easy for deployments. 

\begin{figure*}[t]
  \includegraphics[width=0.7\textwidth]{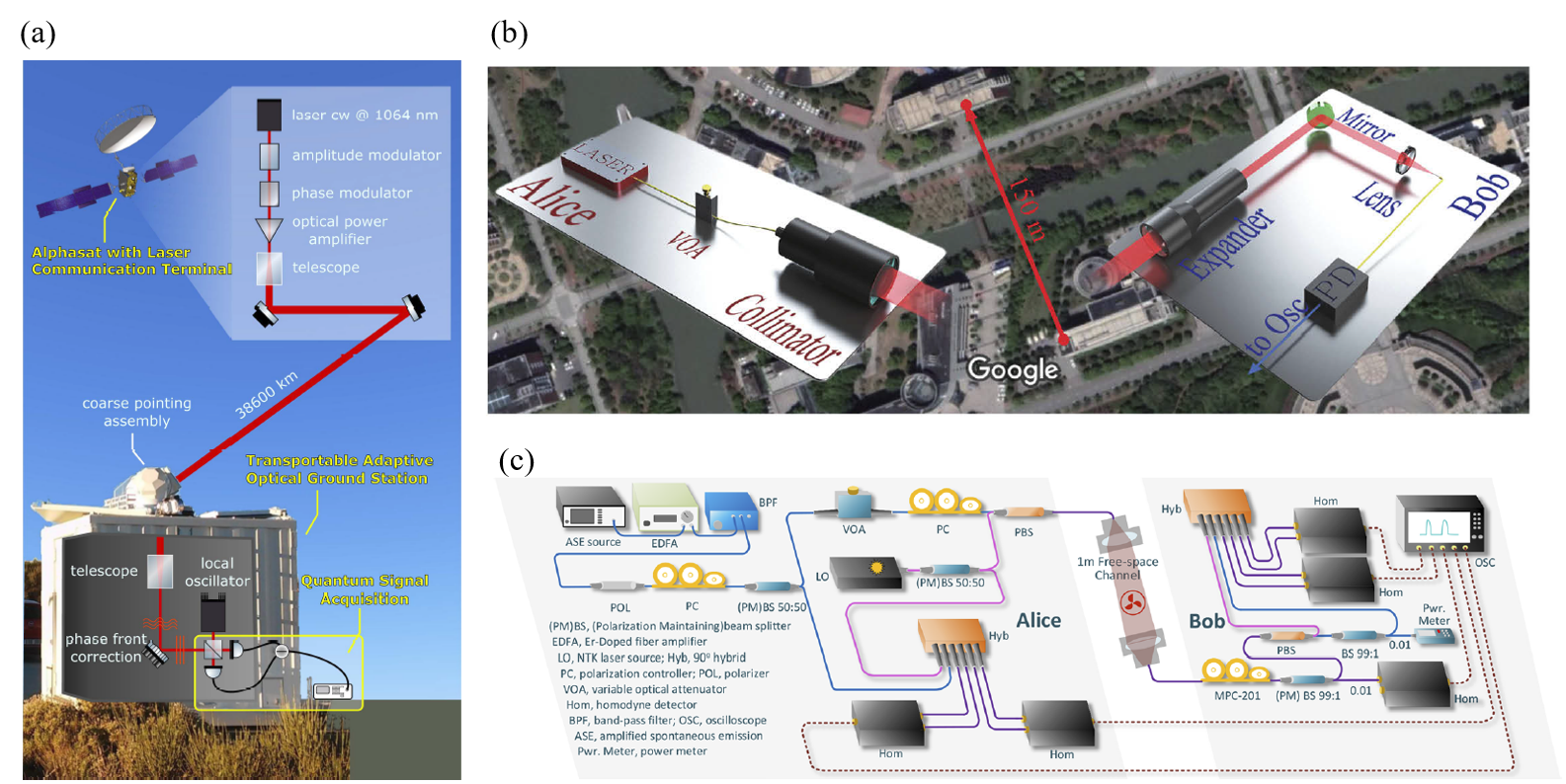}% Here is how to import EPS art
  \caption{\label{fig:System_FreeSpace} The free space CV-QKD key techniques and systems. (a) The demonstration of the satellite to ground quantum limited experiment \cite{gunthner2017quantum}. (b) Phase compensation for a free space CV-QKD system \cite{wang2020phase}. (c) Passive-state-preparation CV-QKD system with free space channel \cite{zhang2023experimental}.
  (a) Reproduced with permission from Optica. 4, 611 (2017). Copyright 2017 Optical Society of America.
  (b) Reproduced with permission from Opt. Express 28, 10737 (2020). Copyright 2020 Optical Society of America.
  (c) Reproduced with permission from Opt. Lett. 48, 1184 (2023). Copyright 2023 Optica Publishing Group.
  }
\end{figure*}

The test of the system co-existing with classical channels was firstly demonstrated in 2010 \cite{Qi_NewJPhys_2010}, and \hl{has been further developed in recent years} \cite{li2014influence,Kumar_NewJPhys_2015,Karinou_IEEE_2018,Eriksson_JLightwaveTechnol_2020,Du_PhysRevApplied_2020,Milovančev_2021_High}.
The further test results show that, over a 25 km fiber, a CV-QKD operated over the 1530.12 nm channel can tolerate the noise arising from up to 11.5 dBm classical channel at 1550.12 nm in the forward direction and 9.7 dBm in backward \cite{Kumar_NewJPhys_2015}. The system with 75 km transmission distance can work in the channel with -3 dBm forward classical signals or -9 dBm backward classical signals. These results demonstrate the outstanding capacity of CV-QKD to coexist with classical signals of realistic intensity in optical networks.
In 2018, the spontaneous Raman scattering noise of a CV-QKD system co-exisiting with classical channels is investigated, which is the most dominant impairment in a wavelength division multiplexed co-existence environment for CV-QKD \cite{Karinou_IEEE_2018}. 
The setting of the quantum signal and the classical signals in frequency domain is shown in Fig. \ref{fig:System_coExist3}.
The influence of the spontaneous Raman scattering noise on a CV-QKD system under different transmission situation is investigated, resulting in a scheme which can support a secret key rate of 90 kbps over 20 km, for an ideal QKD system multiplexed with 2 mW optical power.

A CV-QKD system co-propagates with large-scale C-band DWDM channels is investigated in 2019 \cite{Kleis_OE_2019}. By operating the quantum signals in S- or L-band, the number of co-propagating channels is doubled. 56 density WDM channels with a total launch power of 14.5 dBm are co-propagated with the quantum signal at the distance of 25 km.
Meanwhile, CV-QKD system co-existing with 18.3 Tbps data channels is tested, which contains 100 WDM channels, more than 90 times higher classical bit rate than the previous results \cite{Eriksson_CommunPhys_2019}, shown in Fig. \ref{fig:System_coExist3} (b).
In 2020, B. Chu et al. investigated the LO quality of an in-line LO system in a co-existing environment, which is normally ignored in former studies. 
It is shown that, four-wave mixing in excess noise analysis is a visible factor causing LO fluctuation characterized by the statistical properties
of the power evolution of LO \cite{Chu_2020}.

\subsection{Others}
In this part, we introduce the free space and the entanglement based CV-QKD systems. 
The free space CV-QKD is crucial for the wide range CV-QKD, which is a promising way to connect two distant parties by using the satellite as the relay.
On the other hand, for accessing multiple users in complex environment, free space links can be used to construct an access network with simple system structure.
For the value in the long-distance and access network applications, the free space CV-QKD is studied, and some results have been achieved.
The entanglement-based CV-QKD systems are suitable to distill secret key bits against the high excess noise, which can be used in the scenario with worse channel situation.

\subsubsection{Free space systems}

Compared to the fiber link, the free space link suffers from the disturbance of atmosphere, leading to the fading channel and the beam wandering. The fluctuation of transmission efficiency and induced extra excess noise will seriously deteriorate the secure key generation. The atmospheric effects on CV-QKD are studied \cite{wang2018atmospheric,papanastasiou2018continuous,pirandola2021satellite,pirandola2021limits,ghalaii2023continuous}, where beam wandering, broadening, deformation, and scintillation are found to be the primary effects to lead to transmittance fluctuation of horizontal link within the boundary layer, and effect of arrival time fluctuations will induce phase excess noise. Except for the fast fading of the channel parameters such as the channel loss and excess noise, the fading of phase also increases the difficulty of phase recovery. Therefore, new coding strategy and phase compensation methods should be developed.

In 2014, a free space continuous-variable quantum communication is demonstrated with a point-to-point free space link of 1.6 km in urban conditions \cite{heim2014atmospheric}. 
% The information is polarization encoded and sent to the receiver for homodyne detection. This experiment preliminarily verified the feasibility.
Later, a satellite to ground experiment on the quantum limited measurement of the quadrature information is demonstrated \cite{gunthner2017quantum}, as shown in Fig. \ref{fig:System_FreeSpace} (a).
A series of technologies are developed to support the measurement of the coherent state generated by the laser communication terminal on the geostationary Earth orbit, including the pointing, acquisition, and tracking system of the Transportable Adaptive Optical Ground Station, the adaptive optics system to process the phase front distortions for launching the beam into a single mode fiber, and an optical phase lock loop to lock the phase between the signal and the LO, where the LO is generated by the laser source inside the ground station.  The feasibility of secret key establishment in a satellite-to-ground downlink configuration based on CV-QKD is further examined theoretically, where positive secret key rate can be achieved for low-Earth-orbit scenario. While for higher orbits, no secure keys will be generated when considering the finite-size effects \cite{dequal2021feasibility}.

In 2019, the effective resistance against background noise of CV-QKD is theoretically and experimentally demonstrated \cite{wang2019feasibility}. 
In 2020, as shown in Fig. \ref{fig:System_FreeSpace} (b), a phase compensation strategy is developed and experimental demonstrated \cite{wang2020phase}.
This work checks the correlation between data held by Alice and Bob in a free space channel, and proves that the fluctuation of transmittance disappears in the correlation, thus enabling phase compensation for signals over fluctuant channels. 
Later, a series of methods for free space polarization compensation \cite{wang2020dynamic}, data and transmittance synchronization \cite{wang2021robust}, data acquisition \cite{wei2023high} are proposed, and the feasibility of secure key distribution with free space CV-QKD through fog is experimentally demonstrated \cite{wang2021feasibility}.
Recently, a passive-state-preparation CV-QKD system shown in Fig. \ref{fig:System_FreeSpace} (c) with free space channel is experimentally demonstrated \cite{zhang2023experimental}. Thermal-state polarization multiplexing transmitted LO, synchronized channel transmittance monitoring and fine-grained phase compensation techniques are proposed to support a secret key rate of 1.015 Mbps with a free space channel of -15 dB simulated transmittance.

\subsubsection{Entanglement-based systems}
\hl{The entanglement-based systems have also been developed in recent years as an alternative approach of the CV-QKD system implementation} \cite{su_EurophLett_2009, madsen_NatComm_2012, Gehring_NatCommun_2015, wang2018long, Du_PhysRevApplied_2020, ren_SciChinaInforSci_2022, feng_OptLett_2017,du2023continuous}.
The first entanglement-based CV-QKD system is realized in 2009, a pair of bright EPR entangled beams produced from a non-degenerate optical parametric amplifier is used as the source. The secret key rates of 84 kbps and 3 kbps are achieved agianst collective attack with the channel the transmission efficiency of 80 \% and 40 \% \cite{su_EurophLett_2009}.
Later in 2012, a system using modulated fragile entangled states is realized.
The system can generate secret key bits against the excess noise of 0.45, which is unable for any coherent state protocols to distill secret key bits \cite{madsen_NatComm_2012}.
In 2018, the performance of the entanglement-based CV-QKD system is further enhanced, which can achieve an asymptotic secret key rate of 0.03 (0.01) bit per sample at a channel excess noise level of 0.01 (0.1) \cite{wang2018long}.

% \begin{figure}[t]
%   \includegraphics[width=0.45\textwidth]{System_EBExp.png}% Here is how to import EPS art
%   \caption{\label{fig:System_EBExp} The CV-QKD system with entanglement states \cite{wang2018long}.}
% \end{figure}

\section{THE ADVANCED CV-QKD SYSTEM PROGRESS FOR FUTURE APPLICATIONS}

The future CV-QKD system will head towards the high-speed and compact integration, rely on full scale DSP and integration with photonic chip. Besides that, the point-to-multipoint CV-QKD system will be widely applied to support a high-speed quantum access network for end-user access.

\subsection{Digital CV-QKD systems}
Impairment compensation on digital domain can significantly simplify the system structure, contributing to a simple and stable system. The study on classical coherent communications promotes the DSP algorithms in a CV-QKD system. 
For instance, at the transmitter side, the pulse shaping algorithm is used to raise the availability of the frequency band. For the receiver, the frequency shift is recognized and the down conversion is completed in digital domain, time recovery algorithm is used to obtain the optimal sampling points, digital filter is used to reduce the spectrum mismatch between the transmitter and receiver site, and various of algorithms are developed to distill the parameters of polarization compensation and phase compensation.
Thanks to the wide application of DSP in CV-QKD systems, the speed of the system is becoming higher and higher, the latest achievement \hl{reaches} the baud rate of 10 GBaud \cite{hajomer2023continuous}.

With more DSP algorithms being introduced to the CV-QKD system, the security of the DSP is getting more concerns. 
The  security of the linear DSP algorithms in CV-QKD application is proved in 2023, based on the continous-mode quantum optics theroy \cite{chen2023continuous}. 

For the transmitter, the modulation based on the continuous-wave light and the pulse shaping can be seen as a sequence of coherent states with different temporal modes, which raises the requirement that the pulse shaping function in different period should be integrated orthogonality. The commonly used RRC pulse shaping function can satisfy this defination.

\begin{figure}[t]
  \includegraphics[width=0.45\textwidth]{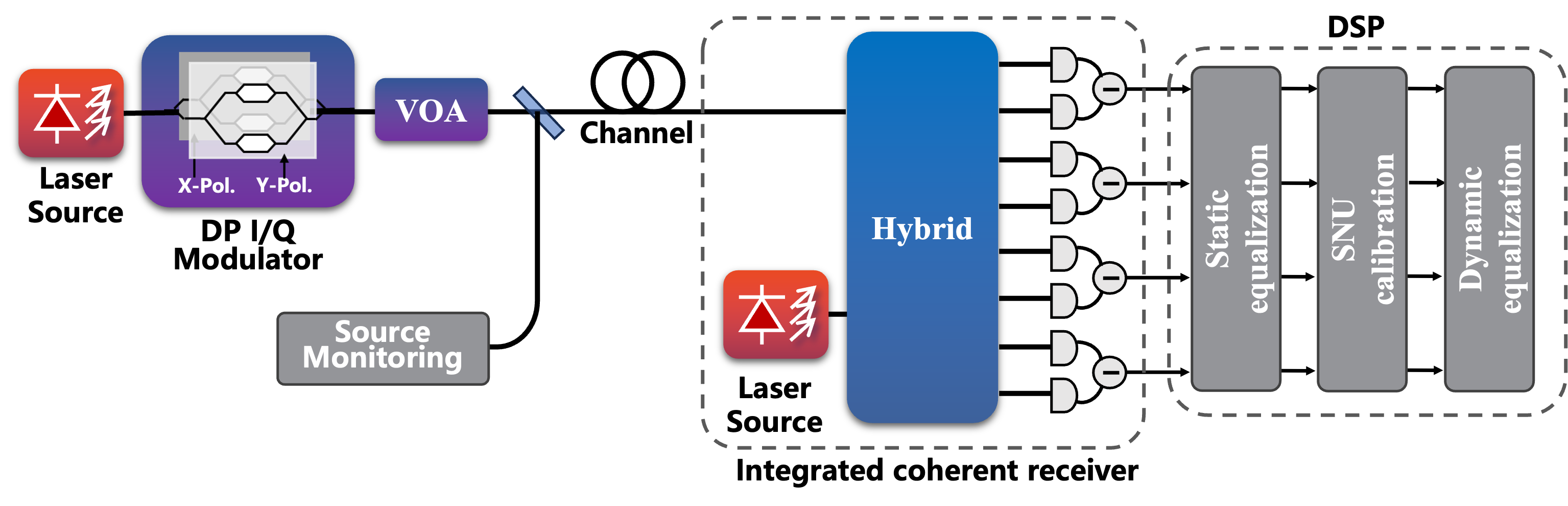}% Here is how to import EPS art
  \caption{\label{fig:System_DP} The advanced point-to-point dual-polarization CV-QKD systems with digital technologies. Here, both of the polarization direction is modulated with quantum signals, the DSP is used to compensate the phase mismatch and polarization rotation. The polarization controller in the optical path is not required. DP IQ modulator: Dual polarization IQ modulator. }
\end{figure}

For the receiver, the pilot tone is individually processed for distilling the parameter for impairment compensation of quantum signals. 
Various algorithms can be used to raise the accuracy of distilling compensation parameters, since the process of pilot tone does not affect the quantum signal, only how to use the parameters for compensation matters. 
While the processing of quantum signals has two main steps, including the static and dynamic equalization. 
% This process manipulates the sampled measurement results of quantum states, which is the most concerning part in security analysis. 
The static equalization aims at compensating the imperfections of the measurement process to provide the right quadrature measurement results, in which a proper SNU normalization is crucial. After that, the dynamic equalization is performed to compensate the mismatch of the polarization and phase during the transmission in quantum channel. This can be easily mapped to linear quantum optics, for instance, the phase or polarization rotation and beamsplitters (attenuation). Therefore, the key of the security is the static equalization, to obtain a reasonable quadrature measurement result.

% The algorithm applied in this part should ensure that the final output is a legal measurement result of the quadrature information, which corresponds to a physical existed result.
% The third part is the compensation of polarization and phase based on the measurement results of the quadrature information.
% The commonly used algorithms are usually unitary, which can be equivalent to the polarization rotation and phase rotation before the  homodyne detection. These rotation operations are secure as long as the output results of the second part is secure. Therefore, the key point of the security of DSP is based on the equalization.

The security of static equalization algorithms depends on proper calibration of SNU. The measurement result of $t_j$ period, $ \hat{D}_{t_j}^{N}$, is defined by a linear function of multiple sampled data $\{ \hat{D}_{t_{j-k+i}} \}$, 
\begin{equation}
  \hat{D}_{t_j}^{N}=f_{dsp}(\hat{D}_{t_{j-k+1}},...,\hat{D}_{t_{j-k+N}})=\Sigma_{i=1}^{N}f_{dsp}^{i}\hat{D}_{t_{j-k+i}}.
\end{equation}
If SNU is calibrated from vacuum input following the same equalization procedure, then after SNU normalization, the measurement result forms a single-mode quadrature measurement result (with relative phase $\theta$ to LO) with certain temporal mode, $\Xi_{DSP}^{t_j}$, defined by the hardware features and equalization algorithms.  
\begin{equation}
  \hat{D} ^{SNU} _{t_j} =\hat{A}^{\dagger}_{\Xi_{DSP}^{t_j}} exp(i\theta) + \hat{A}_{\Xi_{DSP}^{t_j}} exp(-i\theta) = \hat{X}^\theta_{\Xi_{DSP}^{t_j}}.
\end{equation}

\begin{figure}[b]
  \includegraphics[width=0.45\textwidth]{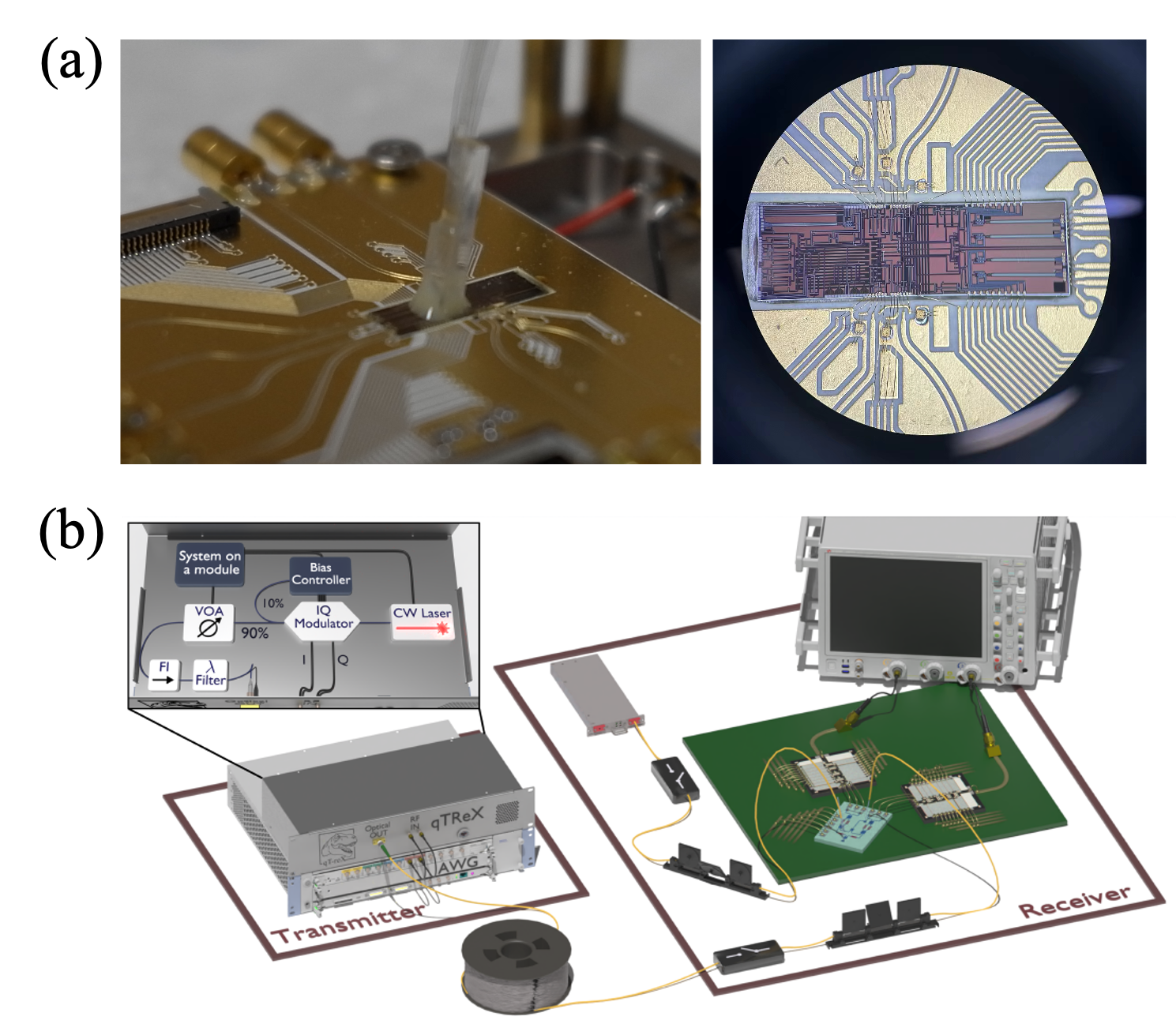}
  \caption{\label{fig:System_Chip2} The local LO CV-QKD systems with chip-based devices. (a) A silicon based high-speed CV-QKD receiver. From Y. Bian et al. \cite{bian2023chip}. (b) The 10 GBaud high-speed CV-QKD system with chip-based receiver. From A. Hajomer et al. \cite{hajomer2023continuous}.
  (b) Reproduced with permission from arXiv 2305.19642 (2023). Copyright 2023 The Authors.
  }
\end{figure}

With digital polarization compensation, the polarization-diversity integrated coherent receiver (ICR) can be used to simplfy the system, in which optical polarization controller is no longer required, as shown in Fig. \ref{fig:System_DP}. 
Recently, a dual-polarization local LO CV-QKD system is experimentally demonstrated \cite{roumestan2022experimental}. 
It performs the probability-shaping discrete-modulated 64 and 256 QAM with dual-polarization IQ modulator, and the polarization-diversity ICR is used at receiver side. With all compensation finished in digital domain, it can achieve 91.8  Mbps of secret key rate at 9.5 km and 24  Mbps at 25 km, which can support the high-speed connection within metropolitan distances. 

\begin{figure*}[t]
  \includegraphics[width=0.8\textwidth]{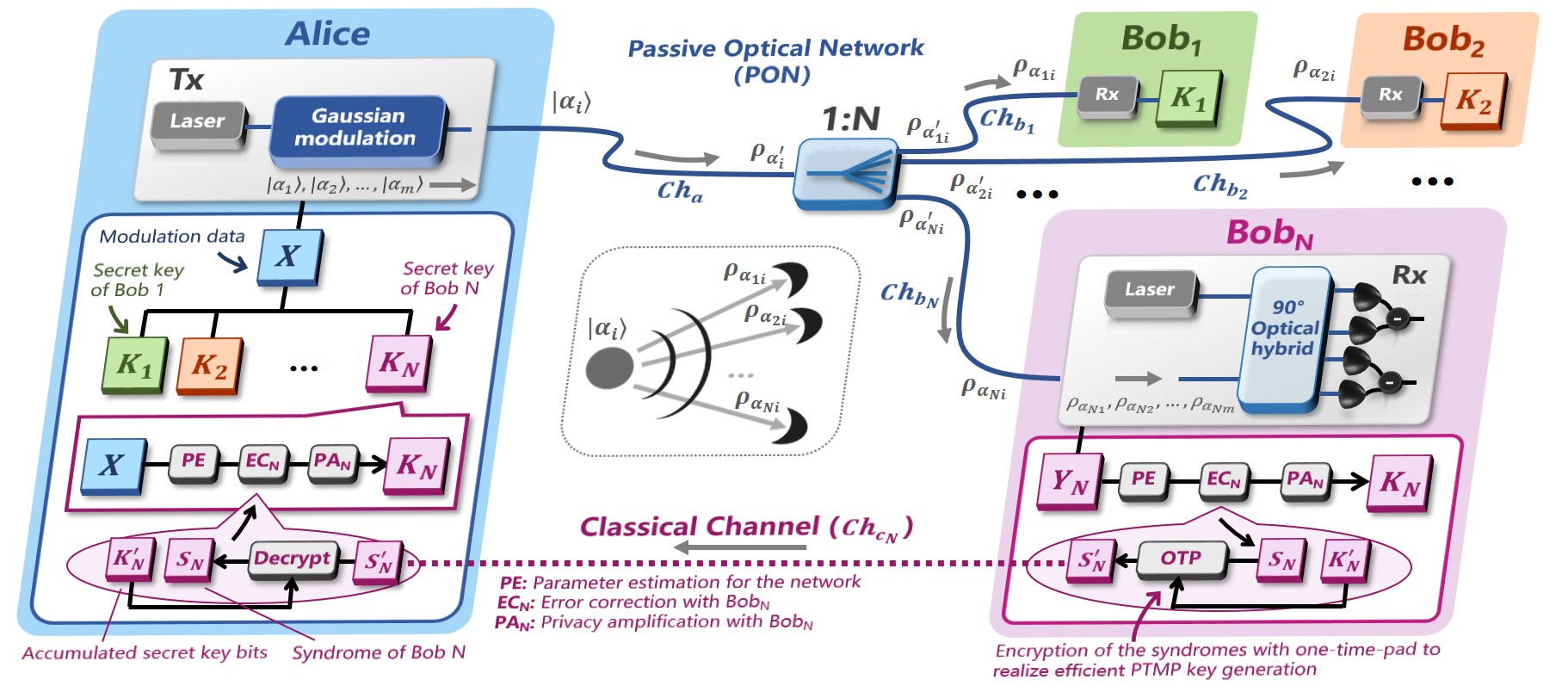}% Here is how to import EPS art
  \caption{\label{fig:PTMPProtocol} The prepare-and-measure scheme of the point-to-multipoint CV-QKD protocol. From Y. Bian et al. \cite{bian2023high}. Each coherent state prepared by Alice can be responded by all Bobs, for security analysis and secret key distillation. The syndrome is secretly transmitted to avoid the cross information leakage between different Bobs. Reproduced with permission from arXiv 2302.02391 (2023). Copyright 2023 arXiv.}
\end{figure*}

\subsection{Chip-based local LO systems}

The advanced local LO CV-QKD system is developing towards compact module with photonics integration, benifiting from stability and scalability, which enables cost-effective deployments in large-scale.
% The key challenges including the high-precision Gaussian modulation and shot-noise limited homodyne detection has been demonstrated in early in-line LO chip-based system \cite{Zhang_NatPhotonics_2019}. 
\hl{Recently, a seires of investigations have been carried out on high-performance chip-based transmitter and receiver for local LO CV-QKD system} \cite{wang2020integrated,li2021practical,wang2022silicon,aldama2023inp,pietri2023cv,hajomer2023continuous,li2023continuous,jia2023silicon,luo2023recent}. There are two mainstream fabrication platforms, Silicon-On-Insulator and \uppercase\expandafter{\romannumeral3}-\uppercase\expandafter{\romannumeral5}. 
% have demonstrated the possibility of the high-performance transmitter and receiver. Foreseeable, a fully integrated local LO system is not far away.

% recently, a Silicon-On-Insulator CV-QKD receiver shown in Fig.\ref{fig:System_Chip1} (a) for a digital system is tested. The bandwidth is significantly enhanced to 250 MHz which makes detect the frequency division multiplexed quantum and pilot signal possible \cite{pietri2023cv}. The maximal detection efficiency of the overall receiver is 0.26, with a shot-noise-to-electronic-noise ratio of 20 dB at low frequencies and more than 7 dB for 250 MHz. Under the untrusted loss of 1.38 dB, a secret key rate of 280 kbps is achieved with excess noise of 0.1102 at Alice's side.

The Silicon-On-Insulator platform has the advantages of low cost and good ductility, which can utilize mature silicon CMOS processes to manufacture optical devices. The refractive index of the silicon waveguide is 3.42, which can form a significant refractive index difference with silicon dioxide, ensuring that the silicon waveguide can have a smaller waveguide bending radius, which is beneficial for high-density device integration. 
In 2023, a Silicon-On-Insulator CV-QKD receiver for a digital system is tested. The bandwidth is significantly enhanced to 250 MHz which makes detect the frequency division multiplexed quantum and pilot signal possible \cite{pietri2023cv}. The maximal detection efficiency of the overall receiver is 0.26, with a shot noise-to-electronic-noise ratio of 20 dB at low frequencies and more than 7 dB for 250 MHz. Under the untrusted loss of 1.38 dB, a secret key rate of 280 kbps is achieved with excess noise of 0.1102 at Alice's side. 
The photonics chip shown in Fig. \ref{fig:System_Chip2} (a) is the CV-QKD receiver designed by Y. Bian et al., which achieves the bandwidth of over 2 GHz and the quantum to classical noise ratio of 5 dB at 2 GHz \cite{bian2023chip}.
Using the silicon based on-chip receiver with high-efficiency Ge photodiodes, a high-speed system at 10 GBaud is realized \cite{hajomer2023continuous}, as shown in Fig. \ref{fig:System_Chip2} (b). The system is able to generate high secret key rates exceeding 0.7 Gbps over a distance of \hl{5 km} and 0.3 Gbps over a distance of 10km, paving the way of the high-performance and compact CV-QKD system.

However, the Silicon-On-Insulator platform also suffers some issues, such as the low coupling efficiency for the optical signal input, which limits the detection efficiency, and the lack of an integrated high performance light source. The first issue can be solved through device optimization, such as low loss grating couplers and edge couplers. The latter issue is quite challenging, since silicon is not suitable for producing high performance integrated lasers.

One promising solution is using \uppercase\expandafter{\romannumeral3}-\uppercase\expandafter{\romannumeral5} platform such as InP, the other way is making heterogeneous integration, where the \uppercase\expandafter{\romannumeral3}-\uppercase\expandafter{\romannumeral5}  chip-based laser is combined with the silicon chip via bonding or growing. 
Some recent researches have shown the feasibility of implementing the above two technical routines. J. Aldama et al. has demonstrated a InP CV-QKD transmitter \cite{aldama2023inp} including an electro-absorption modulator, an IQ modulator and a variable optical attenuator. These three devices are cascaded and integrated on one chip, supporting more than 1 GHz bandwidth. 0.4 and 2.3 Mbps secret key rate is achieved via a test with 11 km fiber and back-to-back connection, which verfies the possibility of using InP platform  to produce a transmitter satisfying the requirement of CV-QKD. Once the laser diode is integrated to the InP chip, a fully integrated CV-QKD chip can be realized.

\hl{For chip-based laser diode, L. Li et al. have demonstrated two high-performance on-chip external cavity lasers based on Si3N4 platform for local LO CV-QKD system} \cite{li2023continuous}. The secret key rate can reach 0.75 Mbps within 50 km fiber and the excess noise is controlled at 0.0579. In conclusion, the last obstacle to the fully chip-based CV-QKD system, which is the integrated light source, is expected to be solved through \uppercase\expandafter{\romannumeral3}-\uppercase\expandafter{\romannumeral5} platform with an overall  or heterogeneous integration.
The chip-based CV-QKD system is a promising way to realize large-scale, small-size and cost-effective applications. 

\subsection{Point-to-multipoint systems}\label{sec:6}

Quantum access network is an efficient way to realize the point-to-multipoint connection between a  network  node and massive end users \cite{QNetNature2013}. There are two mainstream routines, using optical switch with active time-multiplexing control between different end users \cite{Brunner2023DemonstrationOA}, or passive optical nework (PON) with simpler network facilities.
However, the $1\times N$ beamsplitter in a PON significantly increases the equivalent channel loss for individual end users. Moreover, multiplexing techniques are required to seperate different users for upstream configuration \cite{huang2020experimental, xu2023round}. 
Recently, a downstream point-to-multipoint CV-QKD scheme based on PON is proposed to solve the problems above \cite{bian2023high}. The performance is significantly improved with a multiuser protocol, shown in Fig. \ref{fig:PTMPProtocol}.

\begin{figure}[t]
  \includegraphics[width=0.45\textwidth]{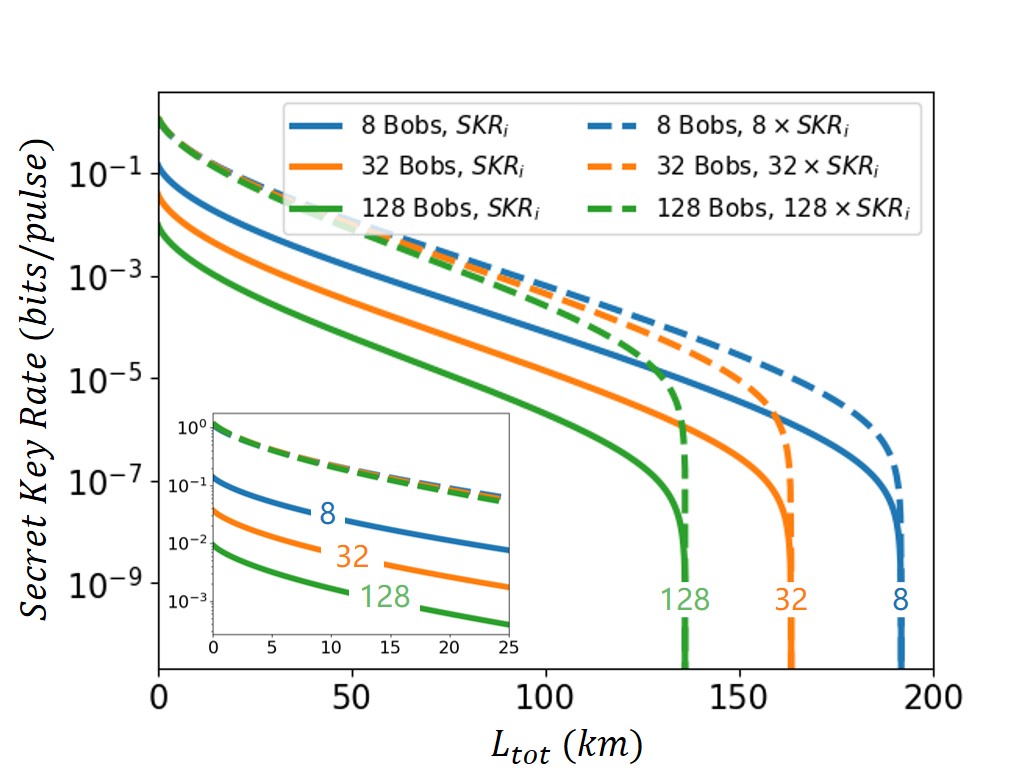}% Here is how to import EPS art
  \caption{\label{fig:PTMPSimulation} Secret key rate versus transmission distance when the number of the Bobs is 8 (blue), 32 (orange) and 128 (green), including the secret key rate of a single Bob (solid line) and the secret key rate of the overall protocol (dashed line).
  The simulation parameters are as below, number of Bobs $N=8$, modulation variance $V_{M}=4$, reconciliation efficiency $\beta=95.6 \%$, $\varepsilon_{tot}=0.0383$, the detection efficiency $\eta_d=60 \%$, the detection noise $v_{el}=0.15$.}
\end{figure}

\begin{figure}[b]
  \includegraphics[width=0.45\textwidth]{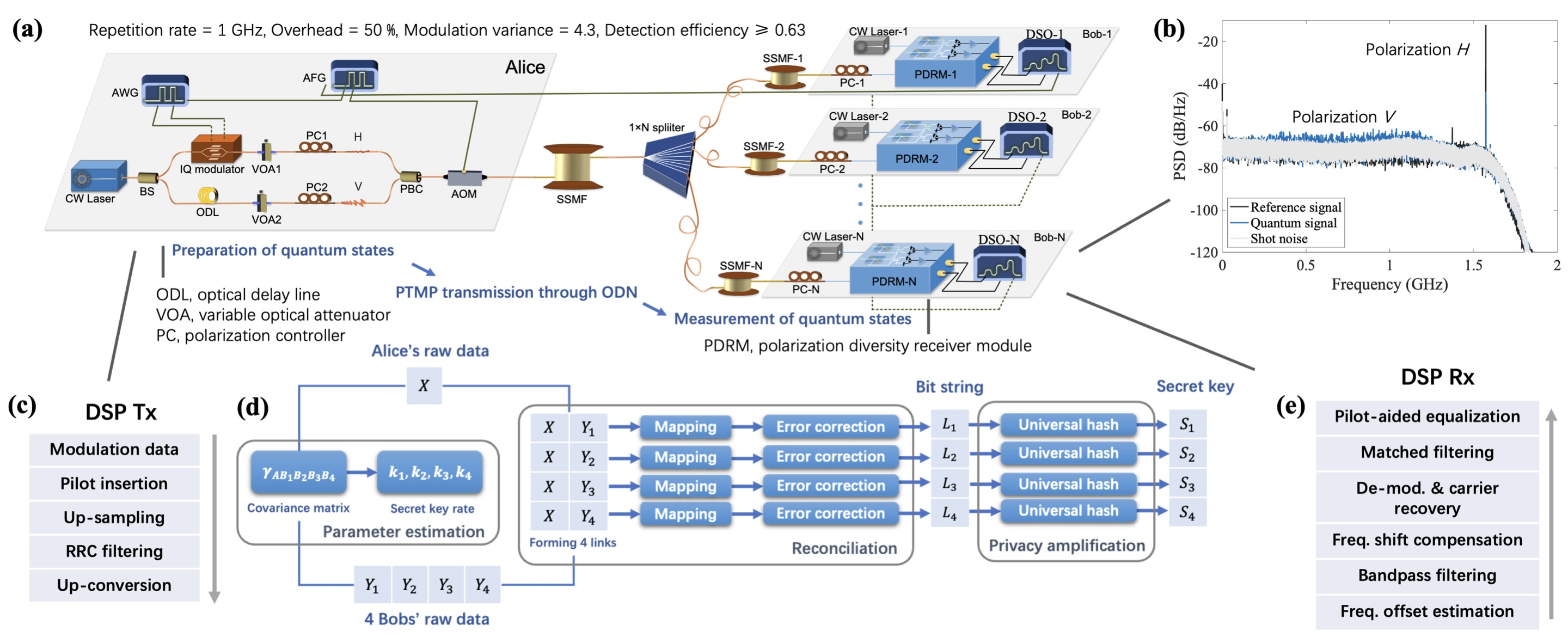}% Here is how to import EPS art
  \caption{\label{fig:PTMPExp} The CV-QKD quantum access network using multiuser protocol. From Y. Bian et al. \cite{bian2023first}. The network is based on the local LO scheme, where the transmitter generates quantum signal and pilot tone, which are multiplexed in different polarization directions and frequency. After the transmission in a passive optical network with an optical power splitter, different receivers at different site detects the signal independently, and distill secret key bits.
  Reproduced with permission from the authors. Copyright 2023 The Authors.}
\end{figure}

\begin{figure}[t]
  \includegraphics[width=0.45\textwidth]{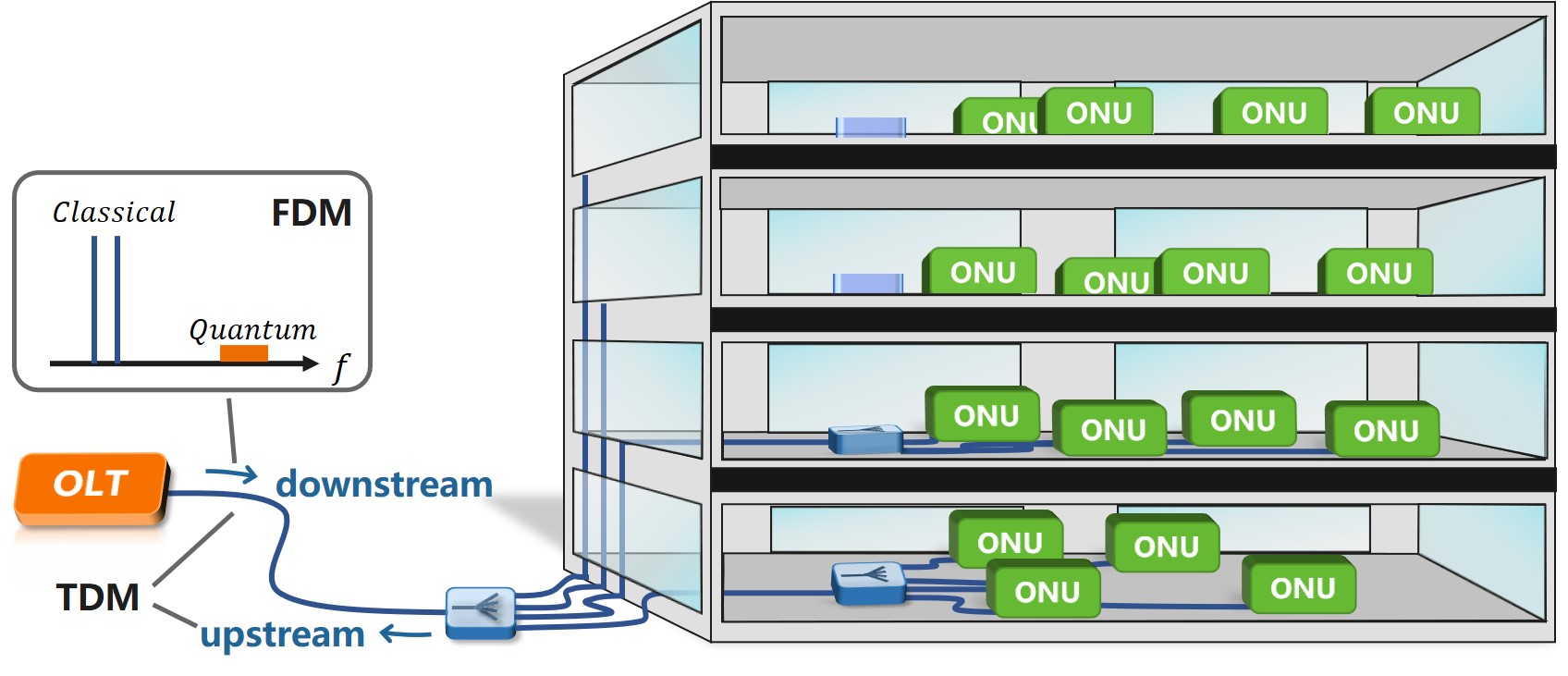}% Here is how to import EPS art
  \caption{\label{fig:PTMPFuture} The future quantum access network with point-to-multipoint protocol.  The OLT means the optical line terminal, and the ONU means the optical network unit. The network can be an optical distribution network, which is widely deployed by the fiber-based classical optical communication network, using as the access network. The quantum signal and classical signals can be co-transmitted with frequency division multiplexing (FDM), while the downstream and upstream signals can be transmitted with time division multiplexing. }
\end{figure}

The state preparation and measurement are similar to the Gaussian modulated protocols. One of the key points of this protocol is the multi-user parameter estimation and security analysis, in which all Bobs work together with Alice for a tighter estimation of the potential channel eavesdropping behavior. Specifically, Alice discloses part of the modulation data, denoted as $X^{est}$, and all Bobs disclose their corresponding detection data, $Y^{est}_1, Y^{est}_2, ..., Y^{est}_N$. With these data, the correlations between Alice and Bobs can be estimated, which forms a covariance matrix $\gamma_{AB_1B_2...B_N}$ consists of all trusted modes. 
Further, the secret key rate between Alice and each Bob can be calculated with $\gamma_{AB_1B_2...B_N}$. 

  \begin{table*}[t]
    \renewcommand{\arraystretch}{1.8}
    \caption{\label{tab:Attack}The attacks and the corresponding countermeasures of CV-QKD systems.}
    \begin{ruledtabular}
    \begin{center}
  
        \begin{tabular}{lllll}
        \textbf{Attack}                     & \textbf{Year} & \textbf{Transmitter / Receiver}  & \textbf{Target component} & \textbf{Countermeasures} \\ \hline
        Imperfect coherent source attack \cite{huang2013bound}    & 2013                           & Transmitter                                        & Coherent source                            & Monitoring Gaussian modulation            \\
        LO fluctuation attack \cite{ma2013local} & 2013                           & Receiver                                     & LO                           & Monitoring LO intensity     \\
        Calibration attack \cite{jouguet2013preventing}                  & 2013                           & Receiver                                     & LO                           & Monitoring LO intensity     \\
        Wavelength attack \cite{huang2013quantum}                  & 2013                           & Receiver                                     & Detector                                   & Monitoring LO wavelength    \\
        Trojan-horse attack \cite{Stiller:15}                & 2015                           & Transmitter                                        & Back reflection light                      & Adding isolators                          \\
        Saturation attack \cite{Qin2016QuantumHacking}                   & 2016                           & Receiver                                     & Detector                                   & Monitoring detector status                \\
        Polarization attack \cite{zhao2018polarization}                & 2018                           & Receiver                                     & LO                           & Monitoring SNU calibration         \\
        Homodyne-detector-blinding attack \cite{Qin2018Homodyne}  & 2018                           & Receiver                                     & Detector                                   & Monitoring detector status                                \\
        Laser seeding attack \cite{zheng2019security}               & 2019                           & Transmitter                                        & Laser diode                                & Monitoring output signal intensity        \\
        Reduced optical attenuation attack \cite{Zheng_PhysRevA_2019} & 2019                           & Transmitter                                        & Variable optical attenuator               & Monitoring signal attenuation level       \\
        Reference pulse attack \cite{ren2019reference,Shao2022Phase,Huang_OptExpress_2019}              & 2019                           & Receiver                                     & Reference pulse                            & Monitoring SNU calibration \\
        Modified LO fluctuation attack \cite{Fan2023Quantum} & 2023                           & Receiver                                     & LO / pilot tone                          & Real-time LO intensity monitoring    \\     
        \end{tabular}
  
    \end{center}
    \end{ruledtabular}
  \end{table*}

The other key point is the parallel key distillation for all end users, which enhances the secret key rate of the overall network. 
% In single-photon scheme, the QKD network with optical power splitter adopts time division multiplexing to make the quantum state preparation and measurement of different QKD links staggered. 
In this protocol, each prepared quantum state can be measured by all users. Several techniques are developed to suppress the negative influence of the correlation between different end users on system performance. Therefore, all users can generate independent secret keys with the sender at the same time. 
% In protocol design, all users are asked to measure the received states independently, they are forbidden to communicate with each other which may amplify their correlation.
% Since receivers' are legitimate parties, it is reasonable to assume that they would perform the required operation in protocol design, which results in classical correlations between each receiver. 
% The results in Y. Bian et al. show that the mutual information between receivers is low enough to support the parallel secret key generation. 
% To achieve this, the protocol is designed with improved error correction to protect the classical correlation between different receivers from using by the potential eavesdropper. 
The simulation results in Y. Bian et al. show the ability of supporting 128 end users with more than 100 km distance (see Fig. \ref{fig:PTMPSimulation}), which can well satisfy the multiuser interconnection requirements within metropolitan distances \cite{bian2023high}.

A high-rate CV-QKD access network is realized, as shown in Fig. \ref{fig:PTMPExp}. 
For the transmitter, quantum signal is generated with an IQ modulator, which is multiplexed with pilot tone signal with frequency division multiplexing and polarization multiplexing. 
For the receiver, the de-multiplexing is realized by a polarization controller and a polarization beamsplitter. Then, quantum signal and pilot tone are seperately detected by heterodyne detection.
The average secret key rate of each user can reach 4.1 Mbps at 15 km when the network capacity is 4, with the repetition frequency of 500 MHz \cite{PTMPAccessNetwork}. 
When the capacity of the network is extended to 8, the secret key rate for each user can reach 7.44 Mbps at 6 km \cite{bian2023first}, \hl{and 3.20 Mbps at 15 km} \cite{PanHigh2024}.
This result can well support the quantum access network, such as linking the end users within a campus (see Fig. \ref{fig:PTMPFuture}).

\section{PRACTICAL SECURITY}
\hl{The information-theoretical security of the CV-QKD protocols using ideal devices has been strictly proved.} However, in practical implementations, the unperfect devices may introduce security loopholes, which can be used by Eve to attack the systems. Correspondingly, countermeasures are proposed to defense the attacks and close the security loopholes, which ensure the practical security. 

\subsection{Attacks and countermeasures}
The theoretical security analysis of CV-QKD is based on the assumption that Alice and Bob are both trusted, where the attack by the eavesdropper can only be performed in channel, without affecting the devices of the legitimate parties.
However, the source and the detector of a practical CV-QKD system cannot satisfy the requirement that the devices can be perfect and fully trusted, since the actual optical and electrical devices inevitably introduce the imperfections. 
These imperfections can be used by the eavesdropper to increase the knowledge to the legitimate parties, which weakens the security of the system. 
According to the different attack targets, the hacking schemes against the CV-QKD system can be classified into hacking schemes against the LO, the source, and the measurement devices. The related attack and defense schemes are summarized in Table \ref{tab:Attack}. At present, all the hacking schemes can be defended, and the research of hacking is mainly to better improve the practical security of the system. 

% \subsubsection{Attacks at the local oscillator}
By manipulating the LO, the representative attack strateiges against the CV-QKD systems, such as the local oscillator fluctuation attack \cite{ma2013local, Fan2023Quantum}, calibration attack \cite{jouguet2013preventing}, polarization attack \cite{zhao2018polarization} and so on, can be performed. It is assumed in the theoretical security analysis of CV-QKD that the LO is trustworthy and will not be controlled by the eavesdropper, but in the actual in-line LO system, since the LO needs to be transmitted over the channel, it is very likely to be controlled by an eavesdropper. Variations in the intensity of the LO can cause the SNU calibrated with the measurement results to be different from the system's true value, which leads to a biased estimation of the system's excess noise, leaving a security loophole. These attack schemes can be defended by monitoring the mean and variance of the intensity of the LO.

The local LO system is proposed to defend the attack at the LO by generating LO locally. However, \hl{since the classical pilot tone is transmitted in the unsecured channel} \cite{shao2021phase}, some attacking schemes at the pilot tone are proposed such as the reference pulse attack and so on \cite{ren2019reference,Shao2022Phase,Huang_OptExpress_2019}.

% \subsubsection{Attacks at the source}
Aiming at the source, the imperfect coherent source attack \cite{huang2013bound}, Trojan-horse attack \cite{Stiller:15}, laser seeding attack \cite{zheng2019security} and reduced optical attenuation attack \cite{Zheng_PhysRevA_2019} are developed.
In a Trojan-horse attack, Eve detects modulation information in the CV-QKD system by injecting bright light pulses and analyzing the reflected pulses. These reflected pulses originate from refractive index changes, such as density fluctuations at the junction between two optical devices or inside the optical devices. The reflected pulse signals contain the markings of the optical modulator (used to encode the quantum state), which allows Eve, the eavesdropper, to extract part of the original key.
The counter measure of the source attack is using an optical isolator to prevent the injected light and a source monitoring detector to achieve the output of the system.

% \subsubsection{Attacks at the detection}
Exploiting the unlinear behavior of the homodyne or heterodyne detectors, the detector saturation attack \cite{Qin2016QuantumHacking} and the homodyne-detector-blinding attack \cite{Qin2018Homodyne} are developed. 
A detector saturation attack involves saturating the detector by increasing the light intensity so as to reduce the output signal variance, which reduces the increase in excess noise caused by interception-retransmission operations, making it impossible for the legitimate parties to detect the eavesdropping behavior. 
A strategy to defend against the detector saturation attack scheme can be realized by monitoring the operating state of the detector by detecting the relationship between the output signal and the channel transmittance.

\begin{table}[t]
  \renewcommand{\arraystretch}{1.8}
  
  \caption{\label{tab:Counter}The grading levels and methodologies of countermeasures.}
  \begin{ruledtabular}
  \begin{center}
    \begin{tabular}{m{1cm} m{2cm}<{\centering} m{5cm}<{\centering}}  
      \textbf{Level} & \textbf{Evaluation} & \textbf{Description} \\ \hline
      $C_3$ & Solution secure & The imperfections of devices in the CV-QKD system has been included in the security analysis, or there is no security risk.  \\ 
      $C_2$ & Solution robust & It can effectively defense the specific attack in experiments, but it has not yet been included in the security analysis.  \\  
      $C_1$ & Solution partially effective & This solution is effective only for certain attack strategies, but is ineffective for others or modified versions of original attack.  \\   
      $C_0$ & Insecure & Security vulnerabilities have been proven to exist, but there is currently no effective countermeasures.  \\   
      $C_X$ & No test & The imperfections of the practical devices are suspected to exist, but testing and verification have not yet been conducted.  \\    
    \end{tabular}   
 
  \end{center}
  \end{ruledtabular}
\end{table}

Taking advantage of the vulnerability that the beam splitting ratio of the beamsplitter at both output ends depends on the wavelength, a wavelength attack scheme for real CV-QKD systems has been proposed. The strategy to defend against this attack scheme can be realized by adding filters in front of the detector and monitoring the intensity of the LO in real time.

% \subsection{The evaluation of countermeasures}
Researchers have also proposed a series of corresponding countermeasures to address the potential security loopholes in the practical CV-QKD system, \hl{including recognizing whether the system is under the attack} \cite{liao2022detecting}. In order to better evaluate the degree of harm caused by different attack strategies and the effectiveness of corresponding countermeasures, it is necessary to conduct grading evaluation for different countermeasures, and quantify their defense performance \cite{sajeed2021approach}. The effectiveness of countermeasures is divided into four levels, with specific levels and grading methodologies as shown in Table. \ref{tab:Counter}.

The above grading scheme for countermeasures mainly focuses on their effectiveness, without paying attention to the ``cost'' of countermeasures. We focuses on the attacker's point of view, and is more concerned with the effect of countermeasures on the risk of attack. The ``cost'' is mainly related to needs of users whose devices are being attacked. For example, when the security demand for users is high, that is, they need to protect to national-level secrets, the defender will not consider the ``cost'' of defense, but rather the the effectiveness of countermeasures as the main consideration. When the user demand is low, that is, only personal information needs to be protected, defenders need to primarily consider the ``cost'' of countermeasures. Here, a certain amount of cost is discussed concerning practical CV-QKD devices. The CV-QKD equipment is a set of communication equipment that integrates electronics and optical devices, and the ``cost'' of countermeasures can be discussed and analyzed in the following aspects: the price of the equipment, the difficulty of integrating the defense scheme, the difficulty of the technical implementation, the computational power and the number of original keys sacrificed in the postprocessing process required for defense. 

In summary, the relationship between defense ``cost'' and defense effectiveness is a trade-off, requiring experts from multiple fields to combine different practical application scenarios to form a more comprehensive evaluation system.

\subsection{Measurement-device-independent systems}

In order to completely eliminate the security loopholes due to the imperfection of devices, completely device-independent QKD protocols have been proposed. This type of protocol does not rely on any security assumptions for QKD devices, and thus is one of the most secure protocols that can guarantee its security even when the device is completely controlled by an eavesdropper. However, completely device-independent protocol face significant experimental challenges since they require key acquisition using the loophole-free Bell test. The key rate of such protocols is too low under the existing experimental conditions, which makes it difficult to meet the usage requirements. 
As a result, the concept of semi-device-independent CV-QKD protocols has been proposed, which is a good compromise between security and performance, and can have high security and high key generation rate at the same time. The main idea of the semi-device-independent CV-QKD protocol is that the security of some devices in the system is not required, while other devices are considered to be trusted. Typical  semi-device-independent protocols are the CV-SDI QKD protocol and the CV-MDI QKD protocol.
In this part, we mainly introduce the process of the CV-MDI QKD system.

\begin{figure}[t]
  \includegraphics[width=0.45\textwidth]{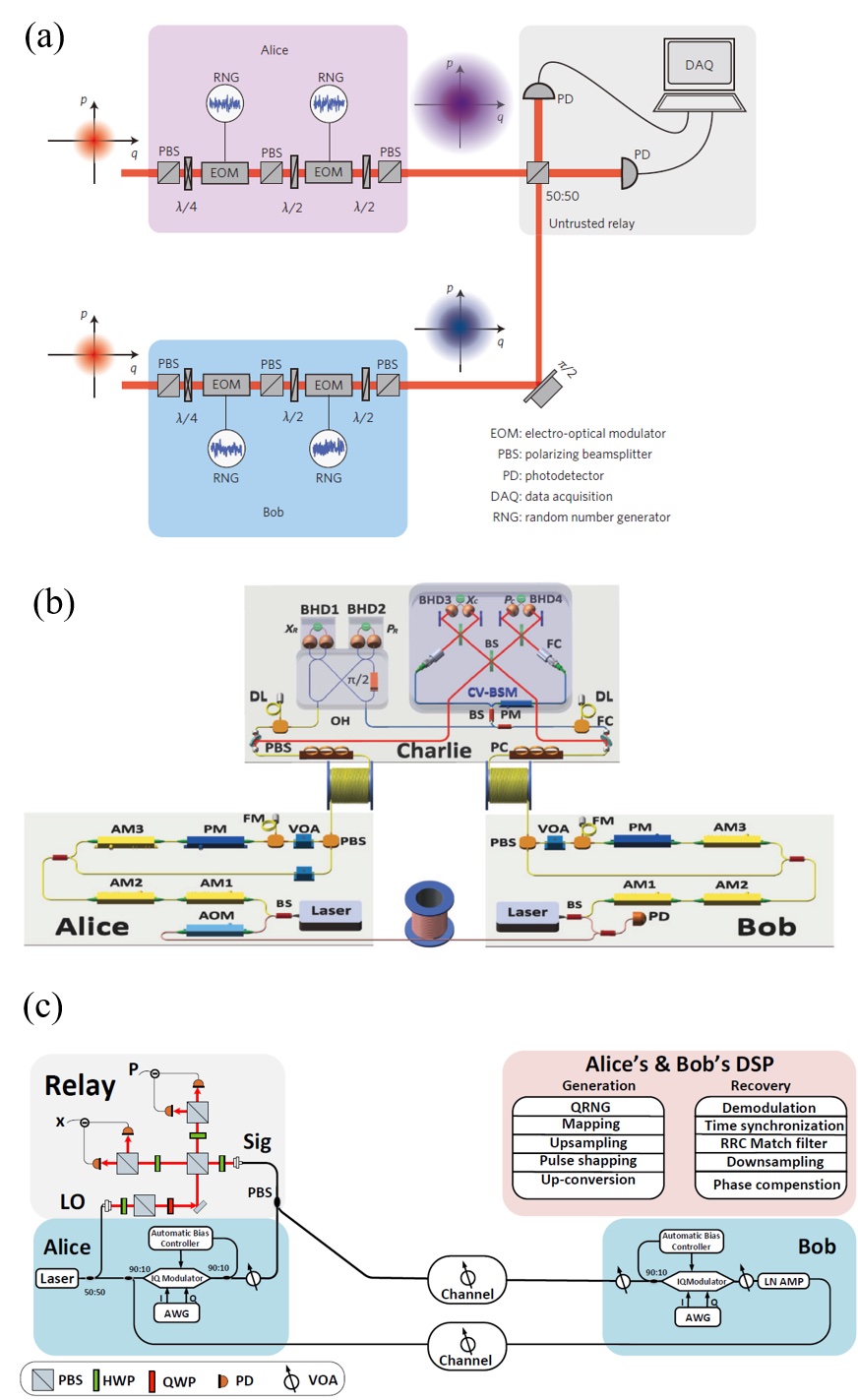}% Here is how to import EPS art
  \caption{\label{fig:MDI_Exp} The experimental schemes of the CV-MDI QKD. (a) The first experimental demonstration. From S. Pirandola et al. \cite{Pirandola_NatPhoton_2015}. (b) The first CV-MDI QKD with optical fiber links. From Y. Tian et al. \cite{tian2022experimental}. (c) The first experimental demonstration without locking systems. From A. Hajomer et al. \cite{hajomer2023high}.
  (a) Reproduced with permission from Nat. Photonics 9, 397 (2015). Copyright 2015 Springer Nature.
  (b) Reproduced with permission from Optica 9, 492 (2022). Copyright 2022 Optica Publishing Group.
  (c) Reproduced with permission from OFC. M2I.2 (2023). Copyright 2023 The Authors.
  }
\end{figure}

In an  CV-MDI QKD, both two legitimate parties, namely Alice and Bob, are the transmitter, while the detection is performed by the untrusted third party. Therefore, the protocol is naturally immune to any attacks at the receiver's side.
For the coherent state CV-MDI QKD, normally the legitimate parties adopt Gaussian modulation, they send the Gaussian modulated coherent states to the third party Charlie, who performs the continuous-variable Bell-state detection. Specifically, Charlie uses a beamsplitter to interfere the two received coherent state, then uses two homodyne detectors to measure the output results. After that, Charlie announces the detection results, and the legitimate parties correct their data with the results announced by Charlie. Finally, Alice and Bob use their corrected data to perform the postprocessing and get final secret keys.
During the above process, the legitimate parties can both correct their data, or correct the data by only one side. 
The theroetical analysis shows that the CV-MDI QKD is limited by the symmetry of the protocol.
When Charlie is deployed at the center, where the loss of the Alice-Charlie link is equal to that of the Bob-Charlie link, the transmisson distance of the protocol is the worst.
When Charlie is near one of the legitimate parties, the performance of the protocol is enhanced.

% \begin{figure}[t]
%   \includegraphics[width=0.45\textwidth]{MDI_EB.png}% Here is how to import EPS art
%   \caption{\label{fig:MDI_EB} The EB scheme of the MDI CV-QKD. From Li te al. \cite{Li_PhysRevA_2014}.}
% \end{figure}

The first proof-of-principle of CV-MDI QKD system is demonstrated in 2015, shown in Fig. \ref{fig:MDI_Exp} (a). With free space optical devices working in 1064 nm band, it demonstrates the high-rate characteristic of CV-MDI QKD \cite{Pirandola_NatPhoton_2015}.
Compared with the qubit-based protocols, the secret key rate increases by 3 orders of magnitude within metropolitan distances, providing a promising way of building the pratical secure metropolitan CV-QKD network.
The difficulties of the experimental CV-MDI QKD is the implementation of the CV Bell detection, which requires highly efficient photodetectors, locking systems, and the free space devices. 

Later, the recent progresses of CV-MDI QKD system solve the issues left above.
As shown in Fig. \ref{fig:MDI_Exp} (b), the CV-MDI QKD with optical fiber links is realized in 2022, where the optical phase locking, phase estimation, real-time phase feedback and quadrature remapping are developed to realize an accurate CV Bell-state measurement. 
In this work, the optical phase-locked loop is a key technique to make the two lasers used for Alice's and Bob's QKD laser source have the identical center frequency. Part of the Alice's laser beam is frequency-up-shifted by 80 MHz  and sent to Bob's station for frequency-locking. Moreover, a free space time-domain homodyne detector is developed, which can reach the detection efficiency of $99 \%$.
The secret key rate can break 0.19 bit/pulse over a 10 km optical fiber, which paves the way of a high-rate metropolitan MDI-QKD network \cite{tian2022experimental}.

\hl{In 2023, a simple and practical CV-MDI QKD system that can coexist with classical telecommunications channels is realized with a real-time frequency and phase locking system} \cite{hajomer2023real}. Later, the frequency and optical phase locking in a CV-MDI QKD system is removed by a new relay structure based on a polarization-based 90-degree optical hybrid and a well-designed DSP pipeline \cite{hajomer2023high}, shown in Fig. \ref{fig:MDI_Exp} (c).
Moreover, this system uses continuous-wave laser with digital pulse shaping and digital time synchronization, therefore the additional amplitude modulation for pulse generation and a delay line for time synchronization is unneccessary.
A secret key rate of 600 kbps over 2 dB loss channel is achieved.

\section{SUMMARY AND OUTLOOK}\label{sec:7}

\begin{figure*}[t]
  \includegraphics[width=0.8\textwidth]{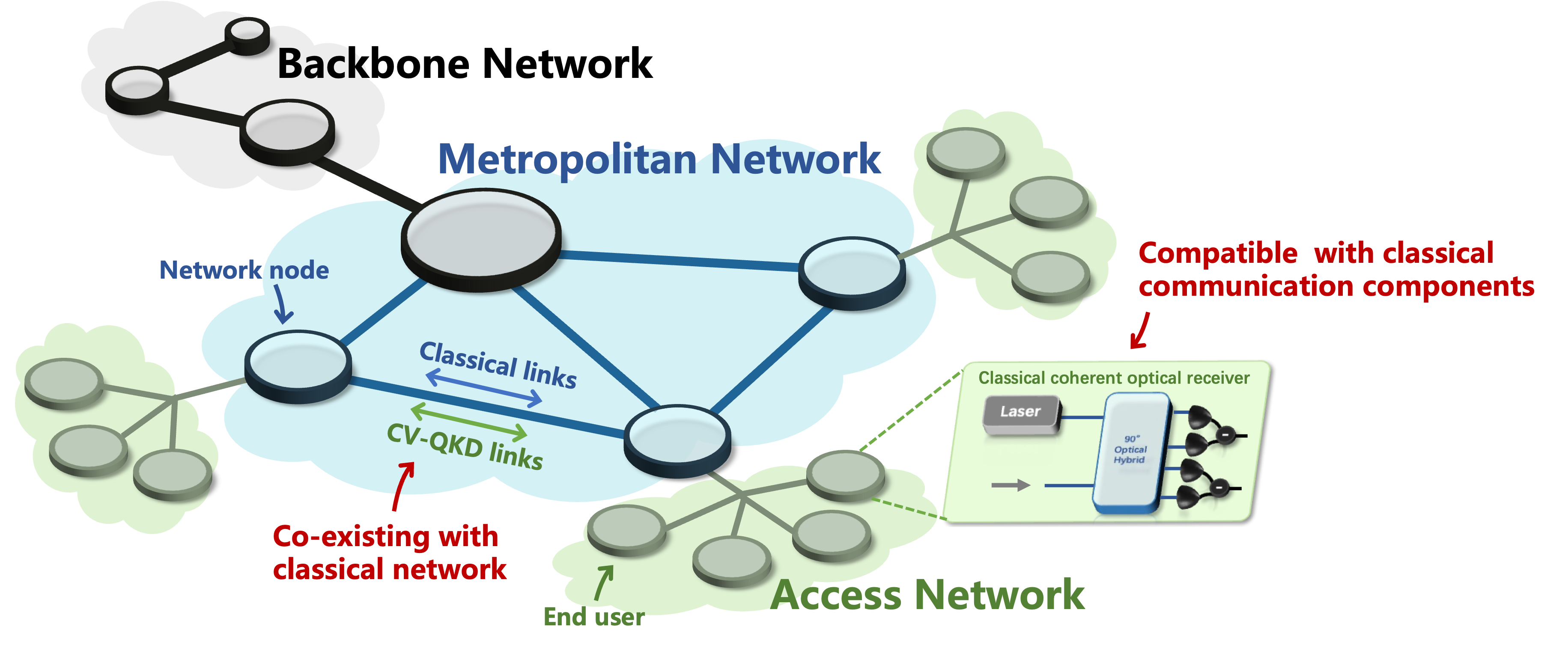}% Here is how to import EPS art
  \caption{\label{fig:Furture} Structure of the future quantum secure network, which consists of the backbone network, the metropolitan network and the access network. Here, the backbone network connects the main network node of different cities, the metropolitan network connects the nodes in a city, and the access network connects the network node with end users.}
\end{figure*}

In this review, we have presented the development of the continuous-variable quantum key distribution system, including basic protocols and security analysis, system structures and demonstrations, advanced system development directions, as well as practical security. The key challenges facing by different system schemes and solutions are summarized, showing the methodology towards a high-performance system. 

There is still the need to develop and design high-performance protocol, including the deep exploration on the insight between shot-noise-unit and security, and developing novel protocols with the secret key rate exceeding Gaussian protocols under same launch energy, which further approaches the fundamental rate-distance limit of point-to-point quantum key distribution while still keeps simple system structure. In practice, different secret key rates might be considered, for instance, with respect to individual, collective, or fully coherent attacks. The choice of these rates may also be associated with a specific sublevel of security to be reached.

Furthermore, the system implementations should move toward higher integration, higher speed and longer transmission distances. Firstly, miniaturization of the system by photonic integrated circuit is crucial, which significantly reduces the system size and broadens the application scenarios. Thanks to the well compatibility, the on-chip integration of CV-QKD systems can draw on classical chip-based optical communication systems. But for the receivers, the detection of weak quantum signals significantly raises the requirements on coherent detectors for high gain and low noise, where the performance of trans-impedance amplifier is the main bottleneck, which requires more attentions. Secondly, high speed system requires high repetition frequency quantum signal modulation, wide bandwidth detection and high throughout data processing. The high repetition frequency introduces new sources of excess noise, such as the dispersion, therefore compensations in physical layer or digital layer are further required for reduce the excess noise. It's crucial to design the better digital algorithms and frame structures to compensate the impairments caused by the low effective-number-of-bits AD converter, which works at high speed. In the aspect of data processing, the pressure on computational resources comes mainly from error correction, wherefore we will need to develop the high throughout and high-efficient error correction strategies in the suitable hardware platform. Thirdly, to achieve a long-distance system, we need to design the high-performance error correction codes that can operate at extremely lower SNR (such as -30 dB), and develop novel system structure to decrease the crosstalk and excess noise.

Finally, quantum hacking and countermeasures are an important and growing area. Closing the side channel at the transmitter site and the accurate SNU calibration are two crucial points for the practical security of the system.

In general, up to now, the potential of using continuous-variable quantum key distribution technique to support high-performance quantum networks within metropolitan and access distances has been demonstrated by the high-speed system implementations, chip-based integrations and field networking tests, and as shown in Fig. \ref{fig:Furture}, it is expected to be highly compatible with the existing optical network infrastructures, enabling the large-scale and cost-effective deployments, and will bring this technology a step closer to a wide range of applications within future quantum networks \cite{guo_FunRes_2021}.

\begin{acknowledgments}
  We are thankful for the enlightening discussions with and helpful comments from reviewers and numerous colleagues, including R. Goncharov, A. Hajomer, P. Huang, L. Huang, Y. Li, Q. Liao, T. Matsuura, S. Pirandola,  S. Sarmiento, A. Vidiella-Barranco, T. Wang, X. Wang and B. Xu. This work was supported by the National Natural Science Foundation of China (62001044, 62201013), the Basic Research Program of China (JCKY2021210B059), the Equipment Advance Research Field Foundation (315067206), and the Fund of State Key Laboratory of Information Photonics and Optical Communications (IPOC2021ZT02). 
\end{acknowledgments}

\nocite{*}
\bibliography{aipsamp20240111}% Produces the bibliography via BibTeX.

\providecommand{\noopsort}[1]{}\providecommand{\singleletter}[1]{#1}%
\begin{thebibliography}{476}%
\makeatletter
\providecommand \@ifxundefined [1]{%
 \@ifx{#1\undefined}
}%
\providecommand \@ifnum [1]{%
 \ifnum #1\expandafter \@firstoftwo
 \else \expandafter \@secondoftwo
 \fi
}%
\providecommand \@ifx [1]{%
 \ifx #1\expandafter \@firstoftwo
 \else \expandafter \@secondoftwo
 \fi
}%
\providecommand \natexlab [1]{#1}%
\providecommand \enquote  [1]{``#1''}%
\providecommand \bibnamefont  [1]{#1}%
\providecommand \bibfnamefont [1]{#1}%
\providecommand \citenamefont [1]{#1}%
\providecommand \href@noop [0]{\@secondoftwo}%
\providecommand \href [0]{\begingroup \@sanitize@url \@href}%
\providecommand \@href[1]{\@@startlink{#1}\@@href}%
\providecommand \@@href[1]{\endgroup#1\@@endlink}%
\providecommand \@sanitize@url [0]{\catcode `\\12\catcode `\$12\catcode `\&12\catcode `\#12\catcode `\^12\catcode `\_12\catcode `\%12\relax}%
\providecommand \@@startlink[1]{}%
\providecommand \@@endlink[0]{}%
\providecommand \url  [0]{\begingroup\@sanitize@url \@url }%
\providecommand \@url [1]{\endgroup\@href {#1}{\urlprefix }}%
\providecommand \urlprefix  [0]{URL }%
\providecommand \Eprint [0]{\href }%
\providecommand \doibase [0]{http://dx.doi.org/}%
\providecommand \selectlanguage [0]{\@gobble}%
\providecommand \bibinfo  [0]{\@secondoftwo}%
\providecommand \bibfield  [0]{\@secondoftwo}%
\providecommand \translation [1]{[#1]}%
\providecommand \BibitemOpen [0]{}%
\providecommand \bibitemStop [0]{}%
\providecommand \bibitemNoStop [0]{.\EOS\space}%
\providecommand \EOS [0]{\spacefactor3000\relax}%
\providecommand \BibitemShut  [1]{\csname bibitem#1\endcsname}%
\let\auto@bib@innerbib\@empty
%</preamble>
\bibitem [{\citenamefont {Bennett}\ and\ \citenamefont {Brassard}(1984)}]{Bennett_BB84_1984}%
  \BibitemOpen
  \bibfield  {author} {\bibinfo {author} {\bibfnamefont {C.~H.}\ \bibnamefont {Bennett}}\ and\ \bibinfo {author} {\bibfnamefont {G.}~\bibnamefont {Brassard}},\ }\bibfield  {title} {\enquote {\bibinfo {title} {Quantum cryptography: Public key distribution and coin tossing},}\ }\href {https://arxiv.org/abs/1906.01645} {\bibfield  {journal} {\bibinfo  {journal} {in Proc. IEEE Int. Conf. on Computers, Systems and Signal Processing}\ ,\ \bibinfo {pages} {175--179}} (\bibinfo {year} {1984})}\BibitemShut {NoStop}%
\bibitem [{\citenamefont {Ekert}(1991)}]{Ekert_PhysRevLett_1991}%
  \BibitemOpen
  \bibfield  {author} {\bibinfo {author} {\bibfnamefont {A.~K.}\ \bibnamefont {Ekert}},\ }\bibfield  {title} {\enquote {\bibinfo {title} {Quantum cryptography based on bell's theorem},}\ }\href {\doibase 10.1103/PhysRevLett.67.661} {\bibfield  {journal} {\bibinfo  {journal} {Phys. Rev. Lett.}\ }\textbf {\bibinfo {volume} {67}},\ \bibinfo {pages} {661--663} (\bibinfo {year} {1991})}\BibitemShut {NoStop}%
\bibitem [{\citenamefont {Gisin}\ \emph {et~al.}(2002)\citenamefont {Gisin}, \citenamefont {Ribordy}, \citenamefont {Tittel},\ and\ \citenamefont {Zbinden}}]{Gisin_RevModPhys_2002}%
  \BibitemOpen
  \bibfield  {author} {\bibinfo {author} {\bibfnamefont {N.}~\bibnamefont {Gisin}}, \bibinfo {author} {\bibfnamefont {G.}~\bibnamefont {Ribordy}}, \bibinfo {author} {\bibfnamefont {W.}~\bibnamefont {Tittel}}, \ and\ \bibinfo {author} {\bibfnamefont {H.}~\bibnamefont {Zbinden}},\ }\bibfield  {title} {\enquote {\bibinfo {title} {Quantum cryptography},}\ }\href {\doibase 10.1103/RevModPhys.74.145} {\bibfield  {journal} {\bibinfo  {journal} {Rev. Mod. Phys.}\ }\textbf {\bibinfo {volume} {74}},\ \bibinfo {pages} {145--195} (\bibinfo {year} {2002})}\BibitemShut {NoStop}%
\bibitem [{\citenamefont {Scarani}\ \emph {et~al.}(2009)\citenamefont {Scarani}, \citenamefont {Bechmann-Pasquinucci}, \citenamefont {Cerf}, \citenamefont {Du\ifmmode~\check{s}\else \v{s}\fi{}ek}, \citenamefont {L\"utkenhaus} \emph {et~al.}}]{Scarani_RevModPhys_2009}%
  \BibitemOpen
  \bibfield  {author} {\bibinfo {author} {\bibfnamefont {V.}~\bibnamefont {Scarani}}, \bibinfo {author} {\bibfnamefont {H.}~\bibnamefont {Bechmann-Pasquinucci}}, \bibinfo {author} {\bibfnamefont {N.~J.}\ \bibnamefont {Cerf}}, \bibinfo {author} {\bibfnamefont {M.}~\bibnamefont {Du\ifmmode~\check{s}\else \v{s}\fi{}ek}}, \bibinfo {author} {\bibfnamefont {N.}~\bibnamefont {L\"utkenhaus}},  \emph {et~al.},\ }\bibfield  {title} {\enquote {\bibinfo {title} {The security of practical quantum key distribution},}\ }\href {\doibase 10.1103/RevModPhys.81.1301} {\bibfield  {journal} {\bibinfo  {journal} {Rev. Mod. Phys.}\ }\textbf {\bibinfo {volume} {81}},\ \bibinfo {pages} {1301--1350} (\bibinfo {year} {2009})}\BibitemShut {NoStop}%
\bibitem [{\citenamefont {Pirandola}\ \emph {et~al.}(2020)\citenamefont {Pirandola}, \citenamefont {Andersen}, \citenamefont {Banchi}, \citenamefont {Berta}, \citenamefont {Bunandar} \emph {et~al.}}]{Pirandola_Advances_2020}%
  \BibitemOpen
  \bibfield  {author} {\bibinfo {author} {\bibfnamefont {S.}~\bibnamefont {Pirandola}}, \bibinfo {author} {\bibfnamefont {U.~L.}\ \bibnamefont {Andersen}}, \bibinfo {author} {\bibfnamefont {L.}~\bibnamefont {Banchi}}, \bibinfo {author} {\bibfnamefont {M.}~\bibnamefont {Berta}}, \bibinfo {author} {\bibfnamefont {D.}~\bibnamefont {Bunandar}},  \emph {et~al.},\ }\bibfield  {title} {\enquote {\bibinfo {title} {Advances in quantum cryptography},}\ }\href {\doibase 10.1364/AOP.361502} {\bibfield  {journal} {\bibinfo  {journal} {Adv. Opt. Photon.}\ }\textbf {\bibinfo {volume} {12}},\ \bibinfo {pages} {1012--1236} (\bibinfo {year} {2020})}\BibitemShut {NoStop}%
\bibitem [{\citenamefont {Xu}\ \emph {et~al.}(2020)\citenamefont {Xu}, \citenamefont {Ma}, \citenamefont {Zhang}, \citenamefont {Lo} \emph {et~al.}}]{Xu_RevModPhys_2020}%
  \BibitemOpen
  \bibfield  {author} {\bibinfo {author} {\bibfnamefont {F.}~\bibnamefont {Xu}}, \bibinfo {author} {\bibfnamefont {X.}~\bibnamefont {Ma}}, \bibinfo {author} {\bibfnamefont {Q.}~\bibnamefont {Zhang}}, \bibinfo {author} {\bibfnamefont {H.-K.}\ \bibnamefont {Lo}},  \emph {et~al.},\ }\bibfield  {title} {\enquote {\bibinfo {title} {Secure quantum key distribution with realistic devices},}\ }\href {\doibase 10.1103/RevModPhys.92.025002} {\bibfield  {journal} {\bibinfo  {journal} {Rev. Mod. Phys.}\ }\textbf {\bibinfo {volume} {92}},\ \bibinfo {pages} {025002} (\bibinfo {year} {2020})}\BibitemShut {NoStop}%
\bibitem [{\citenamefont {Cao}\ \emph {et~al.}(2022)\citenamefont {Cao}, \citenamefont {Zhao}, \citenamefont {Wang}, \citenamefont {Zhang}, \citenamefont {Ng} \emph {et~al.}}]{cao_IEEECommunSurvTutor_2022}%
  \BibitemOpen
  \bibfield  {author} {\bibinfo {author} {\bibfnamefont {Y.}~\bibnamefont {Cao}}, \bibinfo {author} {\bibfnamefont {Y.}~\bibnamefont {Zhao}}, \bibinfo {author} {\bibfnamefont {Q.}~\bibnamefont {Wang}}, \bibinfo {author} {\bibfnamefont {J.}~\bibnamefont {Zhang}}, \bibinfo {author} {\bibfnamefont {S.~X.}\ \bibnamefont {Ng}},  \emph {et~al.},\ }\bibfield  {title} {\enquote {\bibinfo {title} {The evolution of quantum key distribution networks: On the road to the qinternet},}\ }\href@noop {} {\bibfield  {journal} {\bibinfo  {journal} {IEEE Commun. Surv. Tutor.}\ }\textbf {\bibinfo {volume} {24}},\ \bibinfo {pages} {839--894} (\bibinfo {year} {2022})}\BibitemShut {NoStop}%
\bibitem [{\citenamefont {Portmann}\ and\ \citenamefont {Renner}(2022)}]{portmann_RevModPhys_2022}%
  \BibitemOpen
  \bibfield  {author} {\bibinfo {author} {\bibfnamefont {C.}~\bibnamefont {Portmann}}\ and\ \bibinfo {author} {\bibfnamefont {R.}~\bibnamefont {Renner}},\ }\bibfield  {title} {\enquote {\bibinfo {title} {Security in quantum cryptography},}\ }\href@noop {} {\bibfield  {journal} {\bibinfo  {journal} {Rev. Mod. Phys.}\ }\textbf {\bibinfo {volume} {94}},\ \bibinfo {pages} {025008} (\bibinfo {year} {2022})}\BibitemShut {NoStop}%
\bibitem [{\citenamefont {Wootters}, \citenamefont {Wootters},\ and\ \citenamefont {Zurek}(1982)}]{Wootters1982ASQ}%
  \BibitemOpen
  \bibfield  {author} {\bibinfo {author} {\bibfnamefont {W.~K.}\ \bibnamefont {Wootters}}, \bibinfo {author} {\bibfnamefont {W.~K.}\ \bibnamefont {Wootters}}, \ and\ \bibinfo {author} {\bibfnamefont {W.~H.}\ \bibnamefont {Zurek}},\ }\bibfield  {title} {\enquote {\bibinfo {title} {A single quantum cannot be cloned},}\ }\href {https://api.semanticscholar.org/CorpusID:4339227} {\bibfield  {journal} {\bibinfo  {journal} {Nature}\ }\textbf {\bibinfo {volume} {299}},\ \bibinfo {pages} {802--803} (\bibinfo {year} {1982})}\BibitemShut {NoStop}%
\bibitem [{\citenamefont {Diamanti}\ \emph {et~al.}(2016)\citenamefont {Diamanti}, \citenamefont {Lo}, \citenamefont {Qi},\ and\ \citenamefont {Yuan}}]{diamanti_npjQuantumInf_2016}%
  \BibitemOpen
  \bibfield  {author} {\bibinfo {author} {\bibfnamefont {E.}~\bibnamefont {Diamanti}}, \bibinfo {author} {\bibfnamefont {H.-K.}\ \bibnamefont {Lo}}, \bibinfo {author} {\bibfnamefont {B.}~\bibnamefont {Qi}}, \ and\ \bibinfo {author} {\bibfnamefont {Z.}~\bibnamefont {Yuan}},\ }\bibfield  {title} {\enquote {\bibinfo {title} {Practical challenges in quantum key distribution},}\ }\href@noop {} {\bibfield  {journal} {\bibinfo  {journal} {npj Quantum Inf.}\ }\textbf {\bibinfo {volume} {2}},\ \bibinfo {pages} {1--12} (\bibinfo {year} {2016})}\BibitemShut {NoStop}%
\bibitem [{\citenamefont {L{\"u}tkenhaus}(2000)}]{lutkenhaus2000security}%
  \BibitemOpen
  \bibfield  {author} {\bibinfo {author} {\bibfnamefont {N.}~\bibnamefont {L{\"u}tkenhaus}},\ }\bibfield  {title} {\enquote {\bibinfo {title} {Security against individual attacks for realistic quantum key distribution},}\ }\href@noop {} {\bibfield  {journal} {\bibinfo  {journal} {Phys. Rev. A}\ }\textbf {\bibinfo {volume} {61}},\ \bibinfo {pages} {052304} (\bibinfo {year} {2000})}\BibitemShut {NoStop}%
\bibitem [{\citenamefont {Inoue}, \citenamefont {Waks},\ and\ \citenamefont {Yamamoto}(2002)}]{inoue2002differential}%
  \BibitemOpen
  \bibfield  {author} {\bibinfo {author} {\bibfnamefont {K.}~\bibnamefont {Inoue}}, \bibinfo {author} {\bibfnamefont {E.}~\bibnamefont {Waks}}, \ and\ \bibinfo {author} {\bibfnamefont {Y.}~\bibnamefont {Yamamoto}},\ }\bibfield  {title} {\enquote {\bibinfo {title} {Differential phase shift quantum key distribution},}\ }\href@noop {} {\bibfield  {journal} {\bibinfo  {journal} {Phys. Rev. Lett.}\ }\textbf {\bibinfo {volume} {89}},\ \bibinfo {pages} {037902} (\bibinfo {year} {2002})}\BibitemShut {NoStop}%
\bibitem [{\citenamefont {Hwang}(2003)}]{hwang2003quantum}%
  \BibitemOpen
  \bibfield  {author} {\bibinfo {author} {\bibfnamefont {W.-Y.}\ \bibnamefont {Hwang}},\ }\bibfield  {title} {\enquote {\bibinfo {title} {Quantum key distribution with high loss: toward global secure communication},}\ }\href@noop {} {\bibfield  {journal} {\bibinfo  {journal} {Phys. Rev. Lett.}\ }\textbf {\bibinfo {volume} {91}},\ \bibinfo {pages} {057901} (\bibinfo {year} {2003})}\BibitemShut {NoStop}%
\bibitem [{\citenamefont {Scarani}\ \emph {et~al.}(2004)\citenamefont {Scarani}, \citenamefont {Acin}, \citenamefont {Ribordy},\ and\ \citenamefont {Gisin}}]{scarani2004quantum}%
  \BibitemOpen
  \bibfield  {author} {\bibinfo {author} {\bibfnamefont {V.}~\bibnamefont {Scarani}}, \bibinfo {author} {\bibfnamefont {A.}~\bibnamefont {Acin}}, \bibinfo {author} {\bibfnamefont {G.}~\bibnamefont {Ribordy}}, \ and\ \bibinfo {author} {\bibfnamefont {N.}~\bibnamefont {Gisin}},\ }\bibfield  {title} {\enquote {\bibinfo {title} {Quantum cryptography protocols robust against photon number splitting attacks for weak laser pulse implementations},}\ }\href@noop {} {\bibfield  {journal} {\bibinfo  {journal} {Phys. Rev. Lett.}\ }\textbf {\bibinfo {volume} {92}},\ \bibinfo {pages} {057901} (\bibinfo {year} {2004})}\BibitemShut {NoStop}%
\bibitem [{\citenamefont {Barrett}, \citenamefont {Hardy},\ and\ \citenamefont {Kent}(2005)}]{barrett2005no}%
  \BibitemOpen
  \bibfield  {author} {\bibinfo {author} {\bibfnamefont {J.}~\bibnamefont {Barrett}}, \bibinfo {author} {\bibfnamefont {L.}~\bibnamefont {Hardy}}, \ and\ \bibinfo {author} {\bibfnamefont {A.}~\bibnamefont {Kent}},\ }\bibfield  {title} {\enquote {\bibinfo {title} {No signaling and quantum key distribution},}\ }\href@noop {} {\bibfield  {journal} {\bibinfo  {journal} {Phys. Rev. Lett.}\ }\textbf {\bibinfo {volume} {95}},\ \bibinfo {pages} {010503} (\bibinfo {year} {2005})}\BibitemShut {NoStop}%
\bibitem [{\citenamefont {Stucki}\ \emph {et~al.}(2005)\citenamefont {Stucki}, \citenamefont {Brunner}, \citenamefont {Gisin}, \citenamefont {Scarani},\ and\ \citenamefont {Zbinden}}]{stucki2005fast}%
  \BibitemOpen
  \bibfield  {author} {\bibinfo {author} {\bibfnamefont {D.}~\bibnamefont {Stucki}}, \bibinfo {author} {\bibfnamefont {N.}~\bibnamefont {Brunner}}, \bibinfo {author} {\bibfnamefont {N.}~\bibnamefont {Gisin}}, \bibinfo {author} {\bibfnamefont {V.}~\bibnamefont {Scarani}}, \ and\ \bibinfo {author} {\bibfnamefont {H.}~\bibnamefont {Zbinden}},\ }\bibfield  {title} {\enquote {\bibinfo {title} {Fast and simple one-way quantum key distribution},}\ }\href@noop {} {\bibfield  {journal} {\bibinfo  {journal} {Appl. Phys. Lett.}\ }\textbf {\bibinfo {volume} {87}} (\bibinfo {year} {2005})}\BibitemShut {NoStop}%
\bibitem [{\citenamefont {Wang}(2005)}]{wang2005beating}%
  \BibitemOpen
  \bibfield  {author} {\bibinfo {author} {\bibfnamefont {X.-B.}\ \bibnamefont {Wang}},\ }\bibfield  {title} {\enquote {\bibinfo {title} {Beating the photon-number-splitting attack in practical quantum cryptography},}\ }\href@noop {} {\bibfield  {journal} {\bibinfo  {journal} {Phys. Rev. Lett.}\ }\textbf {\bibinfo {volume} {94}},\ \bibinfo {pages} {230503} (\bibinfo {year} {2005})}\BibitemShut {NoStop}%
\bibitem [{\citenamefont {Lo}, \citenamefont {Ma},\ and\ \citenamefont {Chen}(2005)}]{lo2005decoy}%
  \BibitemOpen
  \bibfield  {author} {\bibinfo {author} {\bibfnamefont {H.-K.}\ \bibnamefont {Lo}}, \bibinfo {author} {\bibfnamefont {X.}~\bibnamefont {Ma}}, \ and\ \bibinfo {author} {\bibfnamefont {K.}~\bibnamefont {Chen}},\ }\bibfield  {title} {\enquote {\bibinfo {title} {Decoy state quantum key distribution},}\ }\href@noop {} {\bibfield  {journal} {\bibinfo  {journal} {Phys. Rev. Lett.}\ }\textbf {\bibinfo {volume} {94}},\ \bibinfo {pages} {230504} (\bibinfo {year} {2005})}\BibitemShut {NoStop}%
\bibitem [{\citenamefont {Ac{\'\i}n}\ \emph {et~al.}(2007)\citenamefont {Ac{\'\i}n}, \citenamefont {Brunner}, \citenamefont {Gisin}, \citenamefont {Massar}, \citenamefont {Pironio} \emph {et~al.}}]{acin2007device}%
  \BibitemOpen
  \bibfield  {author} {\bibinfo {author} {\bibfnamefont {A.}~\bibnamefont {Ac{\'\i}n}}, \bibinfo {author} {\bibfnamefont {N.}~\bibnamefont {Brunner}}, \bibinfo {author} {\bibfnamefont {N.}~\bibnamefont {Gisin}}, \bibinfo {author} {\bibfnamefont {S.}~\bibnamefont {Massar}}, \bibinfo {author} {\bibfnamefont {S.}~\bibnamefont {Pironio}},  \emph {et~al.},\ }\bibfield  {title} {\enquote {\bibinfo {title} {Device-independent security of quantum cryptography against collective attacks},}\ }\href@noop {} {\bibfield  {journal} {\bibinfo  {journal} {Phys. Rev. Lett.}\ }\textbf {\bibinfo {volume} {98}},\ \bibinfo {pages} {230501} (\bibinfo {year} {2007})}\BibitemShut {NoStop}%
\bibitem [{\citenamefont {Inamori}, \citenamefont {L{\"u}tkenhaus},\ and\ \citenamefont {Mayers}(2007)}]{inamori2007unconditional}%
  \BibitemOpen
  \bibfield  {author} {\bibinfo {author} {\bibfnamefont {H.}~\bibnamefont {Inamori}}, \bibinfo {author} {\bibfnamefont {N.}~\bibnamefont {L{\"u}tkenhaus}}, \ and\ \bibinfo {author} {\bibfnamefont {D.}~\bibnamefont {Mayers}},\ }\bibfield  {title} {\enquote {\bibinfo {title} {Unconditional security of practical quantum key distribution},}\ }\href@noop {} {\bibfield  {journal} {\bibinfo  {journal} {Eur. Phys. J. D}\ }\textbf {\bibinfo {volume} {41}},\ \bibinfo {pages} {599--627} (\bibinfo {year} {2007})}\BibitemShut {NoStop}%
\bibitem [{\citenamefont {Pironio}\ \emph {et~al.}(2009)\citenamefont {Pironio}, \citenamefont {Ac{\'\i}n}, \citenamefont {Brunner}, \citenamefont {Gisin}, \citenamefont {Massar} \emph {et~al.}}]{pironio2009device}%
  \BibitemOpen
  \bibfield  {author} {\bibinfo {author} {\bibfnamefont {S.}~\bibnamefont {Pironio}}, \bibinfo {author} {\bibfnamefont {A.}~\bibnamefont {Ac{\'\i}n}}, \bibinfo {author} {\bibfnamefont {N.}~\bibnamefont {Brunner}}, \bibinfo {author} {\bibfnamefont {N.}~\bibnamefont {Gisin}}, \bibinfo {author} {\bibfnamefont {S.}~\bibnamefont {Massar}},  \emph {et~al.},\ }\bibfield  {title} {\enquote {\bibinfo {title} {Device-independent quantum key distribution secure against collective attacks},}\ }\href@noop {} {\bibfield  {journal} {\bibinfo  {journal} {New J. Phys.}\ }\textbf {\bibinfo {volume} {11}},\ \bibinfo {pages} {045021} (\bibinfo {year} {2009})}\BibitemShut {NoStop}%
\bibitem [{\citenamefont {Masanes}, \citenamefont {Pironio},\ and\ \citenamefont {Ac{\'\i}n}(2011)}]{masanes2011secure}%
  \BibitemOpen
  \bibfield  {author} {\bibinfo {author} {\bibfnamefont {L.}~\bibnamefont {Masanes}}, \bibinfo {author} {\bibfnamefont {S.}~\bibnamefont {Pironio}}, \ and\ \bibinfo {author} {\bibfnamefont {A.}~\bibnamefont {Ac{\'\i}n}},\ }\bibfield  {title} {\enquote {\bibinfo {title} {Secure device-independent quantum key distribution with causally independent measurement devices},}\ }\href@noop {} {\bibfield  {journal} {\bibinfo  {journal} {Nat. Commun.}\ }\textbf {\bibinfo {volume} {2}},\ \bibinfo {pages} {238} (\bibinfo {year} {2011})}\BibitemShut {NoStop}%
\bibitem [{\citenamefont {Braunstein}\ and\ \citenamefont {Pirandola}(2012)}]{braunstein2012side}%
  \BibitemOpen
  \bibfield  {author} {\bibinfo {author} {\bibfnamefont {S.~L.}\ \bibnamefont {Braunstein}}\ and\ \bibinfo {author} {\bibfnamefont {S.}~\bibnamefont {Pirandola}},\ }\bibfield  {title} {\enquote {\bibinfo {title} {Side-channel-free quantum key distribution},}\ }\href@noop {} {\bibfield  {journal} {\bibinfo  {journal} {Phys. Rev. Lett.}\ }\textbf {\bibinfo {volume} {108}},\ \bibinfo {pages} {130502} (\bibinfo {year} {2012})}\BibitemShut {NoStop}%
\bibitem [{\citenamefont {Lo}, \citenamefont {Curty},\ and\ \citenamefont {Qi}(2012)}]{lo2012measurement}%
  \BibitemOpen
  \bibfield  {author} {\bibinfo {author} {\bibfnamefont {H.-K.}\ \bibnamefont {Lo}}, \bibinfo {author} {\bibfnamefont {M.}~\bibnamefont {Curty}}, \ and\ \bibinfo {author} {\bibfnamefont {B.}~\bibnamefont {Qi}},\ }\bibfield  {title} {\enquote {\bibinfo {title} {Measurement-device-independent quantum key distribution},}\ }\href@noop {} {\bibfield  {journal} {\bibinfo  {journal} {Phys. Rev. Lett.}\ }\textbf {\bibinfo {volume} {108}},\ \bibinfo {pages} {130503} (\bibinfo {year} {2012})}\BibitemShut {NoStop}%
\bibitem [{\citenamefont {Sasaki}, \citenamefont {Yamamoto},\ and\ \citenamefont {Koashi}(2014)}]{sasaki2014practical}%
  \BibitemOpen
  \bibfield  {author} {\bibinfo {author} {\bibfnamefont {T.}~\bibnamefont {Sasaki}}, \bibinfo {author} {\bibfnamefont {Y.}~\bibnamefont {Yamamoto}}, \ and\ \bibinfo {author} {\bibfnamefont {M.}~\bibnamefont {Koashi}},\ }\bibfield  {title} {\enquote {\bibinfo {title} {Practical quantum key distribution protocol without monitoring signal disturbance},}\ }\href@noop {} {\bibfield  {journal} {\bibinfo  {journal} {Nature}\ }\textbf {\bibinfo {volume} {509}},\ \bibinfo {pages} {475--478} (\bibinfo {year} {2014})}\BibitemShut {NoStop}%
\bibitem [{\citenamefont {Lucamarini}\ \emph {et~al.}(2018)\citenamefont {Lucamarini}, \citenamefont {Yuan}, \citenamefont {Dynes},\ and\ \citenamefont {Shields}}]{lucamarini2018overcoming}%
  \BibitemOpen
  \bibfield  {author} {\bibinfo {author} {\bibfnamefont {M.}~\bibnamefont {Lucamarini}}, \bibinfo {author} {\bibfnamefont {Z.~L.}\ \bibnamefont {Yuan}}, \bibinfo {author} {\bibfnamefont {J.~F.}\ \bibnamefont {Dynes}}, \ and\ \bibinfo {author} {\bibfnamefont {A.~J.}\ \bibnamefont {Shields}},\ }\bibfield  {title} {\enquote {\bibinfo {title} {Overcoming the rate--distance limit of quantum key distribution without quantum repeaters},}\ }\href@noop {} {\bibfield  {journal} {\bibinfo  {journal} {Nature}\ }\textbf {\bibinfo {volume} {557}},\ \bibinfo {pages} {400--403} (\bibinfo {year} {2018})}\BibitemShut {NoStop}%
\bibitem [{\citenamefont {Arnon-Friedman}\ \emph {et~al.}(2018)\citenamefont {Arnon-Friedman}, \citenamefont {Dupuis}, \citenamefont {Fawzi}, \citenamefont {Renner},\ and\ \citenamefont {Vidick}}]{arnon2018practical}%
  \BibitemOpen
  \bibfield  {author} {\bibinfo {author} {\bibfnamefont {R.}~\bibnamefont {Arnon-Friedman}}, \bibinfo {author} {\bibfnamefont {F.}~\bibnamefont {Dupuis}}, \bibinfo {author} {\bibfnamefont {O.}~\bibnamefont {Fawzi}}, \bibinfo {author} {\bibfnamefont {R.}~\bibnamefont {Renner}}, \ and\ \bibinfo {author} {\bibfnamefont {T.}~\bibnamefont {Vidick}},\ }\bibfield  {title} {\enquote {\bibinfo {title} {Practical device-independent quantum cryptography via entropy accumulation},}\ }\href@noop {} {\bibfield  {journal} {\bibinfo  {journal} {Nat. Commun.}\ }\textbf {\bibinfo {volume} {9}},\ \bibinfo {pages} {459} (\bibinfo {year} {2018})}\BibitemShut {NoStop}%
\bibitem [{\citenamefont {Vazirani}\ and\ \citenamefont {Vidick}(2019)}]{vazirani2019fully}%
  \BibitemOpen
  \bibfield  {author} {\bibinfo {author} {\bibfnamefont {U.}~\bibnamefont {Vazirani}}\ and\ \bibinfo {author} {\bibfnamefont {T.}~\bibnamefont {Vidick}},\ }\bibfield  {title} {\enquote {\bibinfo {title} {Fully device independent quantum key distribution},}\ }\href@noop {} {\bibfield  {journal} {\bibinfo  {journal} {Commun. ACM}\ }\textbf {\bibinfo {volume} {62}},\ \bibinfo {pages} {133--133} (\bibinfo {year} {2019})}\BibitemShut {NoStop}%
\bibitem [{\citenamefont {Zhao}\ \emph {et~al.}(2006)\citenamefont {Zhao}, \citenamefont {Qi}, \citenamefont {Ma}, \citenamefont {Lo},\ and\ \citenamefont {Qian}}]{zhao2006experimental}%
  \BibitemOpen
  \bibfield  {author} {\bibinfo {author} {\bibfnamefont {Y.}~\bibnamefont {Zhao}}, \bibinfo {author} {\bibfnamefont {B.}~\bibnamefont {Qi}}, \bibinfo {author} {\bibfnamefont {X.}~\bibnamefont {Ma}}, \bibinfo {author} {\bibfnamefont {H.-K.}\ \bibnamefont {Lo}}, \ and\ \bibinfo {author} {\bibfnamefont {L.}~\bibnamefont {Qian}},\ }\bibfield  {title} {\enquote {\bibinfo {title} {Experimental quantum key distribution with decoy states},}\ }\href@noop {} {\bibfield  {journal} {\bibinfo  {journal} {Phys. Rev. Lett.}\ }\textbf {\bibinfo {volume} {96}},\ \bibinfo {pages} {070502} (\bibinfo {year} {2006})}\BibitemShut {NoStop}%
\bibitem [{\citenamefont {Peng}\ \emph {et~al.}(2007)\citenamefont {Peng}, \citenamefont {Zhang}, \citenamefont {Yang}, \citenamefont {Gao}, \citenamefont {Ma} \emph {et~al.}}]{peng2007experimental}%
  \BibitemOpen
  \bibfield  {author} {\bibinfo {author} {\bibfnamefont {C.-Z.}\ \bibnamefont {Peng}}, \bibinfo {author} {\bibfnamefont {J.}~\bibnamefont {Zhang}}, \bibinfo {author} {\bibfnamefont {D.}~\bibnamefont {Yang}}, \bibinfo {author} {\bibfnamefont {W.-B.}\ \bibnamefont {Gao}}, \bibinfo {author} {\bibfnamefont {H.-X.}\ \bibnamefont {Ma}},  \emph {et~al.},\ }\bibfield  {title} {\enquote {\bibinfo {title} {Experimental long-distance decoy-state quantum key distribution based on polarization encoding},}\ }\href@noop {} {\bibfield  {journal} {\bibinfo  {journal} {Phys. Rev. Lett.}\ }\textbf {\bibinfo {volume} {98}},\ \bibinfo {pages} {010505} (\bibinfo {year} {2007})}\BibitemShut {NoStop}%
\bibitem [{\citenamefont {Rosenberg}\ \emph {et~al.}(2007)\citenamefont {Rosenberg}, \citenamefont {Harrington}, \citenamefont {Rice}, \citenamefont {Hiskett}, \citenamefont {Peterson} \emph {et~al.}}]{rosenberg2007long}%
  \BibitemOpen
  \bibfield  {author} {\bibinfo {author} {\bibfnamefont {D.}~\bibnamefont {Rosenberg}}, \bibinfo {author} {\bibfnamefont {J.~W.}\ \bibnamefont {Harrington}}, \bibinfo {author} {\bibfnamefont {P.~R.}\ \bibnamefont {Rice}}, \bibinfo {author} {\bibfnamefont {P.~A.}\ \bibnamefont {Hiskett}}, \bibinfo {author} {\bibfnamefont {C.~G.}\ \bibnamefont {Peterson}},  \emph {et~al.},\ }\bibfield  {title} {\enquote {\bibinfo {title} {Long-distance decoy-state quantum key distribution in optical fiber},}\ }\href@noop {} {\bibfield  {journal} {\bibinfo  {journal} {Phys. Rev. Lett.}\ }\textbf {\bibinfo {volume} {98}},\ \bibinfo {pages} {010503} (\bibinfo {year} {2007})}\BibitemShut {NoStop}%
\bibitem [{\citenamefont {Schmitt-Manderbach}\ \emph {et~al.}(2007)\citenamefont {Schmitt-Manderbach}, \citenamefont {Weier}, \citenamefont {F{\"u}rst}, \citenamefont {Ursin}, \citenamefont {Tiefenbacher} \emph {et~al.}}]{schmitt2007experimental}%
  \BibitemOpen
  \bibfield  {author} {\bibinfo {author} {\bibfnamefont {T.}~\bibnamefont {Schmitt-Manderbach}}, \bibinfo {author} {\bibfnamefont {H.}~\bibnamefont {Weier}}, \bibinfo {author} {\bibfnamefont {M.}~\bibnamefont {F{\"u}rst}}, \bibinfo {author} {\bibfnamefont {R.}~\bibnamefont {Ursin}}, \bibinfo {author} {\bibfnamefont {F.}~\bibnamefont {Tiefenbacher}},  \emph {et~al.},\ }\bibfield  {title} {\enquote {\bibinfo {title} {Experimental demonstration of free-space decoy-state quantum key distribution over 144 km},}\ }\href@noop {} {\bibfield  {journal} {\bibinfo  {journal} {Phys. Rev. Lett.}\ }\textbf {\bibinfo {volume} {98}},\ \bibinfo {pages} {010504} (\bibinfo {year} {2007})}\BibitemShut {NoStop}%
\bibitem [{\citenamefont {Yuan}, \citenamefont {Sharpe},\ and\ \citenamefont {Shields}(2007)}]{yuan2007unconditionally}%
  \BibitemOpen
  \bibfield  {author} {\bibinfo {author} {\bibfnamefont {Z.}~\bibnamefont {Yuan}}, \bibinfo {author} {\bibfnamefont {A.}~\bibnamefont {Sharpe}}, \ and\ \bibinfo {author} {\bibfnamefont {A.}~\bibnamefont {Shields}},\ }\bibfield  {title} {\enquote {\bibinfo {title} {Unconditionally secure one-way quantum key distribution using decoy pulses},}\ }\href@noop {} {\bibfield  {journal} {\bibinfo  {journal} {Appl. Phys. Lett.}\ }\textbf {\bibinfo {volume} {90}} (\bibinfo {year} {2007})}\BibitemShut {NoStop}%
\bibitem [{\citenamefont {Yin}\ \emph {et~al.}(2008)\citenamefont {Yin}, \citenamefont {Han}, \citenamefont {Chen}, \citenamefont {Xu}, \citenamefont {Wu} \emph {et~al.}}]{yin2007experimental}%
  \BibitemOpen
  \bibfield  {author} {\bibinfo {author} {\bibfnamefont {Z.-Q.}\ \bibnamefont {Yin}}, \bibinfo {author} {\bibfnamefont {Z.-F.}\ \bibnamefont {Han}}, \bibinfo {author} {\bibfnamefont {W.}~\bibnamefont {Chen}}, \bibinfo {author} {\bibfnamefont {F.-X.}\ \bibnamefont {Xu}}, \bibinfo {author} {\bibfnamefont {Q.-L.}\ \bibnamefont {Wu}},  \emph {et~al.},\ }\bibfield  {title} {\enquote {\bibinfo {title} {Experimental decoy state quantum key distribution over 120 km fibre},}\ }\href@noop {} {\bibfield  {journal} {\bibinfo  {journal} {Chinese Phys. Lett.}\ }\textbf {\bibinfo {volume} {25}},\ \bibinfo {pages} {3547} (\bibinfo {year} {2008})}\BibitemShut {NoStop}%
\bibitem [{\citenamefont {Wang}\ \emph {et~al.}(2008)\citenamefont {Wang}, \citenamefont {Chen}, \citenamefont {Xavier}, \citenamefont {Swillo}, \citenamefont {Zhang} \emph {et~al.}}]{wang2008experimental}%
  \BibitemOpen
  \bibfield  {author} {\bibinfo {author} {\bibfnamefont {Q.}~\bibnamefont {Wang}}, \bibinfo {author} {\bibfnamefont {W.}~\bibnamefont {Chen}}, \bibinfo {author} {\bibfnamefont {G.}~\bibnamefont {Xavier}}, \bibinfo {author} {\bibfnamefont {M.}~\bibnamefont {Swillo}}, \bibinfo {author} {\bibfnamefont {T.}~\bibnamefont {Zhang}},  \emph {et~al.},\ }\bibfield  {title} {\enquote {\bibinfo {title} {Experimental decoy-state quantum key distribution with a sub-poissionian heralded single-photon source},}\ }\href@noop {} {\bibfield  {journal} {\bibinfo  {journal} {Phys. Rev. Lett.}\ }\textbf {\bibinfo {volume} {100}},\ \bibinfo {pages} {090501} (\bibinfo {year} {2008})}\BibitemShut {NoStop}%
\bibitem [{\citenamefont {Dixon}\ \emph {et~al.}(2008)\citenamefont {Dixon}, \citenamefont {Yuan}, \citenamefont {Dynes}, \citenamefont {Sharpe},\ and\ \citenamefont {Shields}}]{dixon2008gigahertz}%
  \BibitemOpen
  \bibfield  {author} {\bibinfo {author} {\bibfnamefont {A.}~\bibnamefont {Dixon}}, \bibinfo {author} {\bibfnamefont {Z.}~\bibnamefont {Yuan}}, \bibinfo {author} {\bibfnamefont {J.}~\bibnamefont {Dynes}}, \bibinfo {author} {\bibfnamefont {A.}~\bibnamefont {Sharpe}}, \ and\ \bibinfo {author} {\bibfnamefont {A.}~\bibnamefont {Shields}},\ }\bibfield  {title} {\enquote {\bibinfo {title} {Gigahertz decoy quantum key distribution with 1 mbit/s secure key rate},}\ }\href@noop {} {\bibfield  {journal} {\bibinfo  {journal} {Opt. Express}\ }\textbf {\bibinfo {volume} {16}},\ \bibinfo {pages} {18790--18797} (\bibinfo {year} {2008})}\BibitemShut {NoStop}%
\bibitem [{\citenamefont {Rosenberg}\ \emph {et~al.}(2009)\citenamefont {Rosenberg}, \citenamefont {Peterson}, \citenamefont {Harrington}, \citenamefont {Rice}, \citenamefont {Dallmann} \emph {et~al.}}]{rosenberg2009practical}%
  \BibitemOpen
  \bibfield  {author} {\bibinfo {author} {\bibfnamefont {D.}~\bibnamefont {Rosenberg}}, \bibinfo {author} {\bibfnamefont {C.~G.}\ \bibnamefont {Peterson}}, \bibinfo {author} {\bibfnamefont {J.}~\bibnamefont {Harrington}}, \bibinfo {author} {\bibfnamefont {P.~R.}\ \bibnamefont {Rice}}, \bibinfo {author} {\bibfnamefont {N.}~\bibnamefont {Dallmann}},  \emph {et~al.},\ }\bibfield  {title} {\enquote {\bibinfo {title} {Practical long-distance quantum key distribution system using decoy levels},}\ }\href@noop {} {\bibfield  {journal} {\bibinfo  {journal} {New J. Phys.}\ }\textbf {\bibinfo {volume} {11}},\ \bibinfo {pages} {045009} (\bibinfo {year} {2009})}\BibitemShut {NoStop}%
\bibitem [{\citenamefont {Yuan}\ \emph {et~al.}(2009)\citenamefont {Yuan}, \citenamefont {Dixon}, \citenamefont {Dynes}, \citenamefont {Sharpe},\ and\ \citenamefont {Shields}}]{yuan2009practical}%
  \BibitemOpen
  \bibfield  {author} {\bibinfo {author} {\bibfnamefont {Z.}~\bibnamefont {Yuan}}, \bibinfo {author} {\bibfnamefont {A.}~\bibnamefont {Dixon}}, \bibinfo {author} {\bibfnamefont {J.}~\bibnamefont {Dynes}}, \bibinfo {author} {\bibfnamefont {A.}~\bibnamefont {Sharpe}}, \ and\ \bibinfo {author} {\bibfnamefont {A.}~\bibnamefont {Shields}},\ }\bibfield  {title} {\enquote {\bibinfo {title} {Practical gigahertz quantum key distribution based on avalanche photodiodes},}\ }\href@noop {} {\bibfield  {journal} {\bibinfo  {journal} {New J. Phys.}\ }\textbf {\bibinfo {volume} {11}},\ \bibinfo {pages} {045019} (\bibinfo {year} {2009})}\BibitemShut {NoStop}%
\bibitem [{\citenamefont {Liu}\ \emph {et~al.}(2010)\citenamefont {Liu}, \citenamefont {Chen}, \citenamefont {Wang}, \citenamefont {Cai}, \citenamefont {Wan} \emph {et~al.}}]{liu2010decoy}%
  \BibitemOpen
  \bibfield  {author} {\bibinfo {author} {\bibfnamefont {Y.}~\bibnamefont {Liu}}, \bibinfo {author} {\bibfnamefont {T.-Y.}\ \bibnamefont {Chen}}, \bibinfo {author} {\bibfnamefont {J.}~\bibnamefont {Wang}}, \bibinfo {author} {\bibfnamefont {W.-Q.}\ \bibnamefont {Cai}}, \bibinfo {author} {\bibfnamefont {X.}~\bibnamefont {Wan}},  \emph {et~al.},\ }\bibfield  {title} {\enquote {\bibinfo {title} {Decoy-state quantum key distribution with polarized photons over 200 km},}\ }\href@noop {} {\bibfield  {journal} {\bibinfo  {journal} {Opt. Express}\ }\textbf {\bibinfo {volume} {18}},\ \bibinfo {pages} {8587--8594} (\bibinfo {year} {2010})}\BibitemShut {NoStop}%
\bibitem [{\citenamefont {Chen}\ \emph {et~al.}(2010)\citenamefont {Chen}, \citenamefont {Wang}, \citenamefont {Liang}, \citenamefont {Liu}, \citenamefont {Liu} \emph {et~al.}}]{chen2010metropolitan}%
  \BibitemOpen
  \bibfield  {author} {\bibinfo {author} {\bibfnamefont {T.-Y.}\ \bibnamefont {Chen}}, \bibinfo {author} {\bibfnamefont {J.}~\bibnamefont {Wang}}, \bibinfo {author} {\bibfnamefont {H.}~\bibnamefont {Liang}}, \bibinfo {author} {\bibfnamefont {W.-Y.}\ \bibnamefont {Liu}}, \bibinfo {author} {\bibfnamefont {Y.}~\bibnamefont {Liu}},  \emph {et~al.},\ }\bibfield  {title} {\enquote {\bibinfo {title} {Metropolitan all-pass and inter-city quantum communication network},}\ }\href@noop {} {\bibfield  {journal} {\bibinfo  {journal} {Opt. Express}\ }\textbf {\bibinfo {volume} {18}},\ \bibinfo {pages} {27217--27225} (\bibinfo {year} {2010})}\BibitemShut {NoStop}%
\bibitem [{\citenamefont {Wang}\ \emph {et~al.}(2013{\natexlab{a}})\citenamefont {Wang}, \citenamefont {Yang}, \citenamefont {Liao}, \citenamefont {Zhang}, \citenamefont {Shen} \emph {et~al.}}]{wang2013direct}%
  \BibitemOpen
  \bibfield  {author} {\bibinfo {author} {\bibfnamefont {J.-Y.}\ \bibnamefont {Wang}}, \bibinfo {author} {\bibfnamefont {B.}~\bibnamefont {Yang}}, \bibinfo {author} {\bibfnamefont {S.-K.}\ \bibnamefont {Liao}}, \bibinfo {author} {\bibfnamefont {L.}~\bibnamefont {Zhang}}, \bibinfo {author} {\bibfnamefont {Q.}~\bibnamefont {Shen}},  \emph {et~al.},\ }\bibfield  {title} {\enquote {\bibinfo {title} {Direct and full-scale experimental verifications towards ground--satellite quantum key distribution},}\ }\href@noop {} {\bibfield  {journal} {\bibinfo  {journal} {Nat. Photonics}\ }\textbf {\bibinfo {volume} {7}},\ \bibinfo {pages} {387--393} (\bibinfo {year} {2013}{\natexlab{a}})}\BibitemShut {NoStop}%
\bibitem [{\citenamefont {Boaron}\ \emph {et~al.}(2018)\citenamefont {Boaron}, \citenamefont {Boso}, \citenamefont {Rusca}, \citenamefont {Vulliez}, \citenamefont {Autebert} \emph {et~al.}}]{boaron2018secure}%
  \BibitemOpen
  \bibfield  {author} {\bibinfo {author} {\bibfnamefont {A.}~\bibnamefont {Boaron}}, \bibinfo {author} {\bibfnamefont {G.}~\bibnamefont {Boso}}, \bibinfo {author} {\bibfnamefont {D.}~\bibnamefont {Rusca}}, \bibinfo {author} {\bibfnamefont {C.}~\bibnamefont {Vulliez}}, \bibinfo {author} {\bibfnamefont {C.}~\bibnamefont {Autebert}},  \emph {et~al.},\ }\bibfield  {title} {\enquote {\bibinfo {title} {Secure quantum key distribution over 421 km of optical fiber},}\ }\href@noop {} {\bibfield  {journal} {\bibinfo  {journal} {Phys. Rev. Lett.}\ }\textbf {\bibinfo {volume} {121}},\ \bibinfo {pages} {190502} (\bibinfo {year} {2018})}\BibitemShut {NoStop}%
\bibitem [{\citenamefont {Rubenok}\ \emph {et~al.}(2013)\citenamefont {Rubenok}, \citenamefont {Slater}, \citenamefont {Chan}, \citenamefont {Lucio-Martinez},\ and\ \citenamefont {Tittel}}]{rubenok2013real}%
  \BibitemOpen
  \bibfield  {author} {\bibinfo {author} {\bibfnamefont {A.}~\bibnamefont {Rubenok}}, \bibinfo {author} {\bibfnamefont {J.~A.}\ \bibnamefont {Slater}}, \bibinfo {author} {\bibfnamefont {P.}~\bibnamefont {Chan}}, \bibinfo {author} {\bibfnamefont {I.}~\bibnamefont {Lucio-Martinez}}, \ and\ \bibinfo {author} {\bibfnamefont {W.}~\bibnamefont {Tittel}},\ }\bibfield  {title} {\enquote {\bibinfo {title} {Real-world two-photon interference and proof-of-principle quantum key distribution immune to detector attacks},}\ }\href@noop {} {\bibfield  {journal} {\bibinfo  {journal} {Phys. Rev. Lett.}\ }\textbf {\bibinfo {volume} {111}},\ \bibinfo {pages} {130501} (\bibinfo {year} {2013})}\BibitemShut {NoStop}%
\bibitem [{\citenamefont {Liu}\ \emph {et~al.}(2013)\citenamefont {Liu}, \citenamefont {Chen}, \citenamefont {Wang}, \citenamefont {Liang}, \citenamefont {Shentu} \emph {et~al.}}]{liu2013experimental}%
  \BibitemOpen
  \bibfield  {author} {\bibinfo {author} {\bibfnamefont {Y.}~\bibnamefont {Liu}}, \bibinfo {author} {\bibfnamefont {T.-Y.}\ \bibnamefont {Chen}}, \bibinfo {author} {\bibfnamefont {L.-J.}\ \bibnamefont {Wang}}, \bibinfo {author} {\bibfnamefont {H.}~\bibnamefont {Liang}}, \bibinfo {author} {\bibfnamefont {G.-L.}\ \bibnamefont {Shentu}},  \emph {et~al.},\ }\bibfield  {title} {\enquote {\bibinfo {title} {Experimental measurement-device-independent quantum key distribution},}\ }\href@noop {} {\bibfield  {journal} {\bibinfo  {journal} {Phys. Rev. Lett.}\ }\textbf {\bibinfo {volume} {111}},\ \bibinfo {pages} {130502} (\bibinfo {year} {2013})}\BibitemShut {NoStop}%
\bibitem [{\citenamefont {Da~Silva}\ \emph {et~al.}(2013)\citenamefont {Da~Silva}, \citenamefont {Vitoreti}, \citenamefont {Xavier}, \citenamefont {Do~Amaral}, \citenamefont {Tempor{\~a}o} \emph {et~al.}}]{da2013proof}%
  \BibitemOpen
  \bibfield  {author} {\bibinfo {author} {\bibfnamefont {T.~F.}\ \bibnamefont {Da~Silva}}, \bibinfo {author} {\bibfnamefont {D.}~\bibnamefont {Vitoreti}}, \bibinfo {author} {\bibfnamefont {G.}~\bibnamefont {Xavier}}, \bibinfo {author} {\bibfnamefont {G.}~\bibnamefont {Do~Amaral}}, \bibinfo {author} {\bibfnamefont {G.}~\bibnamefont {Tempor{\~a}o}},  \emph {et~al.},\ }\bibfield  {title} {\enquote {\bibinfo {title} {Proof-of-principle demonstration of measurement-device-independent quantum key distribution using polarization qubits},}\ }\href@noop {} {\bibfield  {journal} {\bibinfo  {journal} {Phys. Rev. A}\ }\textbf {\bibinfo {volume} {88}},\ \bibinfo {pages} {052303} (\bibinfo {year} {2013})}\BibitemShut {NoStop}%
\bibitem [{\citenamefont {Tang}\ \emph {et~al.}(2014{\natexlab{a}})\citenamefont {Tang}, \citenamefont {Liao}, \citenamefont {Xu}, \citenamefont {Qi}, \citenamefont {Qian} \emph {et~al.}}]{tang2014experimental}%
  \BibitemOpen
  \bibfield  {author} {\bibinfo {author} {\bibfnamefont {Z.}~\bibnamefont {Tang}}, \bibinfo {author} {\bibfnamefont {Z.}~\bibnamefont {Liao}}, \bibinfo {author} {\bibfnamefont {F.}~\bibnamefont {Xu}}, \bibinfo {author} {\bibfnamefont {B.}~\bibnamefont {Qi}}, \bibinfo {author} {\bibfnamefont {L.}~\bibnamefont {Qian}},  \emph {et~al.},\ }\bibfield  {title} {\enquote {\bibinfo {title} {Experimental demonstration of polarization encoding measurement-device-independent quantum key distribution},}\ }\href@noop {} {\bibfield  {journal} {\bibinfo  {journal} {Phys. Rev. Lett.}\ }\textbf {\bibinfo {volume} {112}},\ \bibinfo {pages} {190503} (\bibinfo {year} {2014}{\natexlab{a}})}\BibitemShut {NoStop}%
\bibitem [{\citenamefont {Tang}\ \emph {et~al.}(2014{\natexlab{b}})\citenamefont {Tang}, \citenamefont {Yin}, \citenamefont {Chen}, \citenamefont {Liu}, \citenamefont {Zhang} \emph {et~al.}}]{tang2014measurement}%
  \BibitemOpen
  \bibfield  {author} {\bibinfo {author} {\bibfnamefont {Y.-L.}\ \bibnamefont {Tang}}, \bibinfo {author} {\bibfnamefont {H.-L.}\ \bibnamefont {Yin}}, \bibinfo {author} {\bibfnamefont {S.-J.}\ \bibnamefont {Chen}}, \bibinfo {author} {\bibfnamefont {Y.}~\bibnamefont {Liu}}, \bibinfo {author} {\bibfnamefont {W.-J.}\ \bibnamefont {Zhang}},  \emph {et~al.},\ }\bibfield  {title} {\enquote {\bibinfo {title} {Measurement-device-independent quantum key distribution over 200 km},}\ }\href@noop {} {\bibfield  {journal} {\bibinfo  {journal} {Phys. Rev. Lett.}\ }\textbf {\bibinfo {volume} {113}},\ \bibinfo {pages} {190501} (\bibinfo {year} {2014}{\natexlab{b}})}\BibitemShut {NoStop}%
\bibitem [{\citenamefont {Tang}\ \emph {et~al.}(2014{\natexlab{c}})\citenamefont {Tang}, \citenamefont {Yin}, \citenamefont {Chen}, \citenamefont {Liu}, \citenamefont {Zhang} \emph {et~al.}}]{tang2014field}%
  \BibitemOpen
  \bibfield  {author} {\bibinfo {author} {\bibfnamefont {Y.-L.}\ \bibnamefont {Tang}}, \bibinfo {author} {\bibfnamefont {H.-L.}\ \bibnamefont {Yin}}, \bibinfo {author} {\bibfnamefont {S.-J.}\ \bibnamefont {Chen}}, \bibinfo {author} {\bibfnamefont {Y.}~\bibnamefont {Liu}}, \bibinfo {author} {\bibfnamefont {W.-J.}\ \bibnamefont {Zhang}},  \emph {et~al.},\ }\bibfield  {title} {\enquote {\bibinfo {title} {Field test of measurement-device-independent quantum key distribution},}\ }\href@noop {} {\bibfield  {journal} {\bibinfo  {journal} {IEEE J. Quantum Electron.}\ }\textbf {\bibinfo {volume} {21}},\ \bibinfo {pages} {116--122} (\bibinfo {year} {2014}{\natexlab{c}})}\BibitemShut {NoStop}%
\bibitem [{\citenamefont {Wang}\ \emph {et~al.}(2015{\natexlab{a}})\citenamefont {Wang}, \citenamefont {Song}, \citenamefont {Yin}, \citenamefont {Wang}, \citenamefont {Chen} \emph {et~al.}}]{wang2015phase}%
  \BibitemOpen
  \bibfield  {author} {\bibinfo {author} {\bibfnamefont {C.}~\bibnamefont {Wang}}, \bibinfo {author} {\bibfnamefont {X.-T.}\ \bibnamefont {Song}}, \bibinfo {author} {\bibfnamefont {Z.-Q.}\ \bibnamefont {Yin}}, \bibinfo {author} {\bibfnamefont {S.}~\bibnamefont {Wang}}, \bibinfo {author} {\bibfnamefont {W.}~\bibnamefont {Chen}},  \emph {et~al.},\ }\bibfield  {title} {\enquote {\bibinfo {title} {Phase-reference-free experiment of measurement-device-independent quantum key distribution},}\ }\href@noop {} {\bibfield  {journal} {\bibinfo  {journal} {Phys. Rev. Lett.}\ }\textbf {\bibinfo {volume} {115}},\ \bibinfo {pages} {160502} (\bibinfo {year} {2015}{\natexlab{a}})}\BibitemShut {NoStop}%
\bibitem [{\citenamefont {Valivarthi}\ \emph {et~al.}(2015)\citenamefont {Valivarthi}, \citenamefont {Lucio-Martinez}, \citenamefont {Chan}, \citenamefont {Rubenok}, \citenamefont {John} \emph {et~al.}}]{valivarthi2015measurement}%
  \BibitemOpen
  \bibfield  {author} {\bibinfo {author} {\bibfnamefont {R.}~\bibnamefont {Valivarthi}}, \bibinfo {author} {\bibfnamefont {I.}~\bibnamefont {Lucio-Martinez}}, \bibinfo {author} {\bibfnamefont {P.}~\bibnamefont {Chan}}, \bibinfo {author} {\bibfnamefont {A.}~\bibnamefont {Rubenok}}, \bibinfo {author} {\bibfnamefont {C.}~\bibnamefont {John}},  \emph {et~al.},\ }\bibfield  {title} {\enquote {\bibinfo {title} {Measurement-device-independent quantum key distribution: from idea towards application},}\ }\href@noop {} {\bibfield  {journal} {\bibinfo  {journal} {J. Mod. Opt.}\ }\textbf {\bibinfo {volume} {62}},\ \bibinfo {pages} {1141--1150} (\bibinfo {year} {2015})}\BibitemShut {NoStop}%
\bibitem [{\citenamefont {Yin}\ \emph {et~al.}(2016)\citenamefont {Yin}, \citenamefont {Chen}, \citenamefont {Yu}, \citenamefont {Liu}, \citenamefont {You} \emph {et~al.}}]{yin2016measurement}%
  \BibitemOpen
  \bibfield  {author} {\bibinfo {author} {\bibfnamefont {H.-L.}\ \bibnamefont {Yin}}, \bibinfo {author} {\bibfnamefont {T.-Y.}\ \bibnamefont {Chen}}, \bibinfo {author} {\bibfnamefont {Z.-W.}\ \bibnamefont {Yu}}, \bibinfo {author} {\bibfnamefont {H.}~\bibnamefont {Liu}}, \bibinfo {author} {\bibfnamefont {L.-X.}\ \bibnamefont {You}},  \emph {et~al.},\ }\bibfield  {title} {\enquote {\bibinfo {title} {Measurement-device-independent quantum key distribution over a 404 km optical fiber},}\ }\href@noop {} {\bibfield  {journal} {\bibinfo  {journal} {Phys. Rev. Lett.}\ }\textbf {\bibinfo {volume} {117}},\ \bibinfo {pages} {190501} (\bibinfo {year} {2016})}\BibitemShut {NoStop}%
\bibitem [{\citenamefont {Tang}\ \emph {et~al.}(2016{\natexlab{a}})\citenamefont {Tang}, \citenamefont {Yin}, \citenamefont {Zhao}, \citenamefont {Liu}, \citenamefont {Sun} \emph {et~al.}}]{tang2016measurement}%
  \BibitemOpen
  \bibfield  {author} {\bibinfo {author} {\bibfnamefont {Y.-L.}\ \bibnamefont {Tang}}, \bibinfo {author} {\bibfnamefont {H.-L.}\ \bibnamefont {Yin}}, \bibinfo {author} {\bibfnamefont {Q.}~\bibnamefont {Zhao}}, \bibinfo {author} {\bibfnamefont {H.}~\bibnamefont {Liu}}, \bibinfo {author} {\bibfnamefont {X.-X.}\ \bibnamefont {Sun}},  \emph {et~al.},\ }\bibfield  {title} {\enquote {\bibinfo {title} {Measurement-device-independent quantum key distribution over untrustful metropolitan network},}\ }\href@noop {} {\bibfield  {journal} {\bibinfo  {journal} {Phys. Rev. X}\ }\textbf {\bibinfo {volume} {6}},\ \bibinfo {pages} {011024} (\bibinfo {year} {2016}{\natexlab{a}})}\BibitemShut {NoStop}%
\bibitem [{\citenamefont {Tang}\ \emph {et~al.}(2016{\natexlab{b}})\citenamefont {Tang}, \citenamefont {Sun}, \citenamefont {Xu}, \citenamefont {Chen}, \citenamefont {Li} \emph {et~al.}}]{tang2016experimental}%
  \BibitemOpen
  \bibfield  {author} {\bibinfo {author} {\bibfnamefont {G.-Z.}\ \bibnamefont {Tang}}, \bibinfo {author} {\bibfnamefont {S.-H.}\ \bibnamefont {Sun}}, \bibinfo {author} {\bibfnamefont {F.}~\bibnamefont {Xu}}, \bibinfo {author} {\bibfnamefont {H.}~\bibnamefont {Chen}}, \bibinfo {author} {\bibfnamefont {C.-Y.}\ \bibnamefont {Li}},  \emph {et~al.},\ }\bibfield  {title} {\enquote {\bibinfo {title} {Experimental asymmetric plug-and-play measurement-device-independent quantum key distribution},}\ }\href@noop {} {\bibfield  {journal} {\bibinfo  {journal} {Phys. Rev. A}\ }\textbf {\bibinfo {volume} {94}},\ \bibinfo {pages} {032326} (\bibinfo {year} {2016}{\natexlab{b}})}\BibitemShut {NoStop}%
\bibitem [{\citenamefont {Comandar}\ \emph {et~al.}(2016)\citenamefont {Comandar}, \citenamefont {Lucamarini}, \citenamefont {Fr{\"o}hlich}, \citenamefont {Dynes}, \citenamefont {Sharpe} \emph {et~al.}}]{comandar2016quantum}%
  \BibitemOpen
  \bibfield  {author} {\bibinfo {author} {\bibfnamefont {L.}~\bibnamefont {Comandar}}, \bibinfo {author} {\bibfnamefont {M.}~\bibnamefont {Lucamarini}}, \bibinfo {author} {\bibfnamefont {B.}~\bibnamefont {Fr{\"o}hlich}}, \bibinfo {author} {\bibfnamefont {J.}~\bibnamefont {Dynes}}, \bibinfo {author} {\bibfnamefont {A.}~\bibnamefont {Sharpe}},  \emph {et~al.},\ }\bibfield  {title} {\enquote {\bibinfo {title} {Quantum key distribution without detector vulnerabilities using optically seeded lasers},}\ }\href@noop {} {\bibfield  {journal} {\bibinfo  {journal} {Nat. Photonics}\ }\textbf {\bibinfo {volume} {10}},\ \bibinfo {pages} {312--315} (\bibinfo {year} {2016})}\BibitemShut {NoStop}%
\bibitem [{\citenamefont {Kaneda}\ \emph {et~al.}(2017)\citenamefont {Kaneda}, \citenamefont {Xu}, \citenamefont {Chapman},\ and\ \citenamefont {Kwiat}}]{kaneda2017quantum}%
  \BibitemOpen
  \bibfield  {author} {\bibinfo {author} {\bibfnamefont {F.}~\bibnamefont {Kaneda}}, \bibinfo {author} {\bibfnamefont {F.}~\bibnamefont {Xu}}, \bibinfo {author} {\bibfnamefont {J.}~\bibnamefont {Chapman}}, \ and\ \bibinfo {author} {\bibfnamefont {P.~G.}\ \bibnamefont {Kwiat}},\ }\bibfield  {title} {\enquote {\bibinfo {title} {Quantum-memory-assisted multi-photon generation for efficient quantum information processing},}\ }\href@noop {} {\bibfield  {journal} {\bibinfo  {journal} {Optica}\ }\textbf {\bibinfo {volume} {4}},\ \bibinfo {pages} {1034--1037} (\bibinfo {year} {2017})}\BibitemShut {NoStop}%
\bibitem [{\citenamefont {Wang}\ \emph {et~al.}(2017{\natexlab{a}})\citenamefont {Wang}, \citenamefont {Yin}, \citenamefont {Wang}, \citenamefont {Chen}, \citenamefont {Guo} \emph {et~al.}}]{wang2017measurement}%
  \BibitemOpen
  \bibfield  {author} {\bibinfo {author} {\bibfnamefont {C.}~\bibnamefont {Wang}}, \bibinfo {author} {\bibfnamefont {Z.-Q.}\ \bibnamefont {Yin}}, \bibinfo {author} {\bibfnamefont {S.}~\bibnamefont {Wang}}, \bibinfo {author} {\bibfnamefont {W.}~\bibnamefont {Chen}}, \bibinfo {author} {\bibfnamefont {G.-C.}\ \bibnamefont {Guo}},  \emph {et~al.},\ }\bibfield  {title} {\enquote {\bibinfo {title} {Measurement-device-independent quantum key distribution robust against environmental disturbances},}\ }\href@noop {} {\bibfield  {journal} {\bibinfo  {journal} {Optica}\ }\textbf {\bibinfo {volume} {4}},\ \bibinfo {pages} {1016--1023} (\bibinfo {year} {2017}{\natexlab{a}})}\BibitemShut {NoStop}%
\bibitem [{\citenamefont {Valivarthi}\ \emph {et~al.}(2017)\citenamefont {Valivarthi}, \citenamefont {Zhou}, \citenamefont {John}, \citenamefont {Marsili}, \citenamefont {Verma} \emph {et~al.}}]{valivarthi2017cost}%
  \BibitemOpen
  \bibfield  {author} {\bibinfo {author} {\bibfnamefont {R.}~\bibnamefont {Valivarthi}}, \bibinfo {author} {\bibfnamefont {Q.}~\bibnamefont {Zhou}}, \bibinfo {author} {\bibfnamefont {C.}~\bibnamefont {John}}, \bibinfo {author} {\bibfnamefont {F.}~\bibnamefont {Marsili}}, \bibinfo {author} {\bibfnamefont {V.~B.}\ \bibnamefont {Verma}},  \emph {et~al.},\ }\bibfield  {title} {\enquote {\bibinfo {title} {A cost-effective measurement-device-independent quantum key distribution system for quantum networks},}\ }\href@noop {} {\bibfield  {journal} {\bibinfo  {journal} {Quantum Sci. Technol.}\ }\textbf {\bibinfo {volume} {2}},\ \bibinfo {pages} {04LT01} (\bibinfo {year} {2017})}\BibitemShut {NoStop}%
\bibitem [{\citenamefont {Liu}\ \emph {et~al.}(2018{\natexlab{a}})\citenamefont {Liu}, \citenamefont {Wang}, \citenamefont {Ma},\ and\ \citenamefont {Sun}}]{liu2018polarization}%
  \BibitemOpen
  \bibfield  {author} {\bibinfo {author} {\bibfnamefont {H.}~\bibnamefont {Liu}}, \bibinfo {author} {\bibfnamefont {J.}~\bibnamefont {Wang}}, \bibinfo {author} {\bibfnamefont {H.}~\bibnamefont {Ma}}, \ and\ \bibinfo {author} {\bibfnamefont {S.}~\bibnamefont {Sun}},\ }\bibfield  {title} {\enquote {\bibinfo {title} {Polarization-multiplexing-based measurement-device-independent quantum key distribution without phase reference calibration},}\ }\href@noop {} {\bibfield  {journal} {\bibinfo  {journal} {Optica}\ }\textbf {\bibinfo {volume} {5}},\ \bibinfo {pages} {902--909} (\bibinfo {year} {2018}{\natexlab{a}})}\BibitemShut {NoStop}%
\bibitem [{\citenamefont {Wei}\ \emph {et~al.}(2020)\citenamefont {Wei}, \citenamefont {Li}, \citenamefont {Tan}, \citenamefont {Li}, \citenamefont {Min} \emph {et~al.}}]{wei2020high}%
  \BibitemOpen
  \bibfield  {author} {\bibinfo {author} {\bibfnamefont {K.}~\bibnamefont {Wei}}, \bibinfo {author} {\bibfnamefont {W.}~\bibnamefont {Li}}, \bibinfo {author} {\bibfnamefont {H.}~\bibnamefont {Tan}}, \bibinfo {author} {\bibfnamefont {Y.}~\bibnamefont {Li}}, \bibinfo {author} {\bibfnamefont {H.}~\bibnamefont {Min}},  \emph {et~al.},\ }\bibfield  {title} {\enquote {\bibinfo {title} {High-speed measurement-device-independent quantum key distribution with integrated silicon photonics},}\ }\href@noop {} {\bibfield  {journal} {\bibinfo  {journal} {Phys. Rev. X}\ }\textbf {\bibinfo {volume} {10}},\ \bibinfo {pages} {031030} (\bibinfo {year} {2020})}\BibitemShut {NoStop}%
\bibitem [{\citenamefont {Minder}\ \emph {et~al.}(2019)\citenamefont {Minder}, \citenamefont {Pittaluga}, \citenamefont {Roberts}, \citenamefont {Lucamarini}, \citenamefont {Dynes} \emph {et~al.}}]{minder2019experimental}%
  \BibitemOpen
  \bibfield  {author} {\bibinfo {author} {\bibfnamefont {M.}~\bibnamefont {Minder}}, \bibinfo {author} {\bibfnamefont {M.}~\bibnamefont {Pittaluga}}, \bibinfo {author} {\bibfnamefont {G.~L.}\ \bibnamefont {Roberts}}, \bibinfo {author} {\bibfnamefont {M.}~\bibnamefont {Lucamarini}}, \bibinfo {author} {\bibfnamefont {J.~F.}\ \bibnamefont {Dynes}},  \emph {et~al.},\ }\bibfield  {title} {\enquote {\bibinfo {title} {Experimental quantum key distribution beyond the repeaterless secret key capacity},}\ }\href@noop {} {\bibfield  {journal} {\bibinfo  {journal} {Nat. Photonics}\ }\textbf {\bibinfo {volume} {13}},\ \bibinfo {pages} {334--338} (\bibinfo {year} {2019})}\BibitemShut {NoStop}%
\bibitem [{\citenamefont {Wang}\ \emph {et~al.}(2019{\natexlab{a}})\citenamefont {Wang}, \citenamefont {He}, \citenamefont {Yin}, \citenamefont {Lu}, \citenamefont {Cui} \emph {et~al.}}]{wang2019beating}%
  \BibitemOpen
  \bibfield  {author} {\bibinfo {author} {\bibfnamefont {S.}~\bibnamefont {Wang}}, \bibinfo {author} {\bibfnamefont {D.-Y.}\ \bibnamefont {He}}, \bibinfo {author} {\bibfnamefont {Z.-Q.}\ \bibnamefont {Yin}}, \bibinfo {author} {\bibfnamefont {F.-Y.}\ \bibnamefont {Lu}}, \bibinfo {author} {\bibfnamefont {C.-H.}\ \bibnamefont {Cui}},  \emph {et~al.},\ }\bibfield  {title} {\enquote {\bibinfo {title} {Beating the fundamental rate-distance limit in a proof-of-principle quantum key distribution system},}\ }\href@noop {} {\bibfield  {journal} {\bibinfo  {journal} {Phys. Rev. X}\ }\textbf {\bibinfo {volume} {9}},\ \bibinfo {pages} {021046} (\bibinfo {year} {2019}{\natexlab{a}})}\BibitemShut {NoStop}%
\bibitem [{\citenamefont {Liu}\ \emph {et~al.}(2019)\citenamefont {Liu}, \citenamefont {Yu}, \citenamefont {Zhang}, \citenamefont {Guan}, \citenamefont {Chen} \emph {et~al.}}]{liu2019experimental}%
  \BibitemOpen
  \bibfield  {author} {\bibinfo {author} {\bibfnamefont {Y.}~\bibnamefont {Liu}}, \bibinfo {author} {\bibfnamefont {Z.-W.}\ \bibnamefont {Yu}}, \bibinfo {author} {\bibfnamefont {W.}~\bibnamefont {Zhang}}, \bibinfo {author} {\bibfnamefont {J.-Y.}\ \bibnamefont {Guan}}, \bibinfo {author} {\bibfnamefont {J.-P.}\ \bibnamefont {Chen}},  \emph {et~al.},\ }\bibfield  {title} {\enquote {\bibinfo {title} {Experimental twin-field quantum key distribution through sending or not sending},}\ }\href@noop {} {\bibfield  {journal} {\bibinfo  {journal} {Phys. Rev. Lett.}\ }\textbf {\bibinfo {volume} {123}},\ \bibinfo {pages} {100505} (\bibinfo {year} {2019})}\BibitemShut {NoStop}%
\bibitem [{\citenamefont {Zhong}\ \emph {et~al.}(2019)\citenamefont {Zhong}, \citenamefont {Hu}, \citenamefont {Curty}, \citenamefont {Qian},\ and\ \citenamefont {Lo}}]{zhong2019proof}%
  \BibitemOpen
  \bibfield  {author} {\bibinfo {author} {\bibfnamefont {X.}~\bibnamefont {Zhong}}, \bibinfo {author} {\bibfnamefont {J.}~\bibnamefont {Hu}}, \bibinfo {author} {\bibfnamefont {M.}~\bibnamefont {Curty}}, \bibinfo {author} {\bibfnamefont {L.}~\bibnamefont {Qian}}, \ and\ \bibinfo {author} {\bibfnamefont {H.-K.}\ \bibnamefont {Lo}},\ }\bibfield  {title} {\enquote {\bibinfo {title} {Proof-of-principle experimental demonstration of twin-field type quantum key distribution},}\ }\href@noop {} {\bibfield  {journal} {\bibinfo  {journal} {Phys. Rev. Lett.}\ }\textbf {\bibinfo {volume} {123}},\ \bibinfo {pages} {100506} (\bibinfo {year} {2019})}\BibitemShut {NoStop}%
\bibitem [{\citenamefont {Fang}\ \emph {et~al.}(2020)\citenamefont {Fang}, \citenamefont {Zeng}, \citenamefont {Liu}, \citenamefont {Zou}, \citenamefont {Wu} \emph {et~al.}}]{fang2020implementation}%
  \BibitemOpen
  \bibfield  {author} {\bibinfo {author} {\bibfnamefont {X.-T.}\ \bibnamefont {Fang}}, \bibinfo {author} {\bibfnamefont {P.}~\bibnamefont {Zeng}}, \bibinfo {author} {\bibfnamefont {H.}~\bibnamefont {Liu}}, \bibinfo {author} {\bibfnamefont {M.}~\bibnamefont {Zou}}, \bibinfo {author} {\bibfnamefont {W.}~\bibnamefont {Wu}},  \emph {et~al.},\ }\bibfield  {title} {\enquote {\bibinfo {title} {Implementation of quantum key distribution surpassing the linear rate-transmittance bound},}\ }\href@noop {} {\bibfield  {journal} {\bibinfo  {journal} {Nat. Photonics}\ }\textbf {\bibinfo {volume} {14}},\ \bibinfo {pages} {422--425} (\bibinfo {year} {2020})}\BibitemShut {NoStop}%
\bibitem [{\citenamefont {Chen}\ \emph {et~al.}(2020)\citenamefont {Chen}, \citenamefont {Zhang}, \citenamefont {Liu}, \citenamefont {Jiang}, \citenamefont {Zhang} \emph {et~al.}}]{chen2020sending}%
  \BibitemOpen
  \bibfield  {author} {\bibinfo {author} {\bibfnamefont {J.-P.}\ \bibnamefont {Chen}}, \bibinfo {author} {\bibfnamefont {C.}~\bibnamefont {Zhang}}, \bibinfo {author} {\bibfnamefont {Y.}~\bibnamefont {Liu}}, \bibinfo {author} {\bibfnamefont {C.}~\bibnamefont {Jiang}}, \bibinfo {author} {\bibfnamefont {W.}~\bibnamefont {Zhang}},  \emph {et~al.},\ }\bibfield  {title} {\enquote {\bibinfo {title} {Sending-or-not-sending with independent lasers: Secure twin-field quantum key distribution over 509 km},}\ }\href@noop {} {\bibfield  {journal} {\bibinfo  {journal} {Phys. Rev. Lett.}\ }\textbf {\bibinfo {volume} {124}},\ \bibinfo {pages} {070501} (\bibinfo {year} {2020})}\BibitemShut {NoStop}%
\bibitem [{\citenamefont {Pittaluga}\ \emph {et~al.}(2020)\citenamefont {Pittaluga}, \citenamefont {Minder}, \citenamefont {Lucamarini}, \citenamefont {Sanzaro}, \citenamefont {Woodward} \emph {et~al.}}]{Pittaluga2020600kmRQ}%
  \BibitemOpen
  \bibfield  {author} {\bibinfo {author} {\bibfnamefont {M.}~\bibnamefont {Pittaluga}}, \bibinfo {author} {\bibfnamefont {M.}~\bibnamefont {Minder}}, \bibinfo {author} {\bibfnamefont {M.}~\bibnamefont {Lucamarini}}, \bibinfo {author} {\bibfnamefont {M.}~\bibnamefont {Sanzaro}}, \bibinfo {author} {\bibfnamefont {R.~I.}\ \bibnamefont {Woodward}},  \emph {et~al.},\ }\bibfield  {title} {\enquote {\bibinfo {title} {600-km repeater-like quantum communications with dual-band stabilization},}\ }\href {https://api.semanticscholar.org/CorpusID:229923162} {\bibfield  {journal} {\bibinfo  {journal} {Nat. Photonics}\ }\textbf {\bibinfo {volume} {15}},\ \bibinfo {pages} {530 -- 535} (\bibinfo {year} {2020})}\BibitemShut {NoStop}%
\bibitem [{\citenamefont {Chen}\ \emph {et~al.}(2021{\natexlab{a}})\citenamefont {Chen}, \citenamefont {Zhang}, \citenamefont {Liu}, \citenamefont {Jiang}, \citenamefont {Zhang} \emph {et~al.}}]{chen2021twin}%
  \BibitemOpen
  \bibfield  {author} {\bibinfo {author} {\bibfnamefont {J.-P.}\ \bibnamefont {Chen}}, \bibinfo {author} {\bibfnamefont {C.}~\bibnamefont {Zhang}}, \bibinfo {author} {\bibfnamefont {Y.}~\bibnamefont {Liu}}, \bibinfo {author} {\bibfnamefont {C.}~\bibnamefont {Jiang}}, \bibinfo {author} {\bibfnamefont {W.-J.}\ \bibnamefont {Zhang}},  \emph {et~al.},\ }\bibfield  {title} {\enquote {\bibinfo {title} {Twin-field quantum key distribution over a 511 km optical fibre linking two distant metropolitan areas},}\ }\href@noop {} {\bibfield  {journal} {\bibinfo  {journal} {Nat. Photonics}\ }\textbf {\bibinfo {volume} {15}},\ \bibinfo {pages} {570--575} (\bibinfo {year} {2021}{\natexlab{a}})}\BibitemShut {NoStop}%
\bibitem [{\citenamefont {Wang}\ \emph {et~al.}(2022{\natexlab{a}})\citenamefont {Wang}, \citenamefont {Yin}, \citenamefont {He}, \citenamefont {Chen}, \citenamefont {Wang} \emph {et~al.}}]{wang2022twin}%
  \BibitemOpen
  \bibfield  {author} {\bibinfo {author} {\bibfnamefont {S.}~\bibnamefont {Wang}}, \bibinfo {author} {\bibfnamefont {Z.-Q.}\ \bibnamefont {Yin}}, \bibinfo {author} {\bibfnamefont {D.-Y.}\ \bibnamefont {He}}, \bibinfo {author} {\bibfnamefont {W.}~\bibnamefont {Chen}}, \bibinfo {author} {\bibfnamefont {R.-Q.}\ \bibnamefont {Wang}},  \emph {et~al.},\ }\bibfield  {title} {\enquote {\bibinfo {title} {Twin-field quantum key distribution over 830-km fibre},}\ }\href@noop {} {\bibfield  {journal} {\bibinfo  {journal} {Nat. Photonics}\ }\textbf {\bibinfo {volume} {16}},\ \bibinfo {pages} {154--161} (\bibinfo {year} {2022}{\natexlab{a}})}\BibitemShut {NoStop}%
\bibitem [{\citenamefont {Liu}\ \emph {et~al.}(2023)\citenamefont {Liu}, \citenamefont {Zhang}, \citenamefont {Jiang}, \citenamefont {Chen}, \citenamefont {Zhang} \emph {et~al.}}]{liu2023experimental}%
  \BibitemOpen
  \bibfield  {author} {\bibinfo {author} {\bibfnamefont {Y.}~\bibnamefont {Liu}}, \bibinfo {author} {\bibfnamefont {W.-J.}\ \bibnamefont {Zhang}}, \bibinfo {author} {\bibfnamefont {C.}~\bibnamefont {Jiang}}, \bibinfo {author} {\bibfnamefont {J.-P.}\ \bibnamefont {Chen}}, \bibinfo {author} {\bibfnamefont {C.}~\bibnamefont {Zhang}},  \emph {et~al.},\ }\bibfield  {title} {\enquote {\bibinfo {title} {Experimental twin-field quantum key distribution over 1000 km fiber distance},}\ }\href@noop {} {\bibfield  {journal} {\bibinfo  {journal} {Phys. Rev. Lett.}\ }\textbf {\bibinfo {volume} {130}},\ \bibinfo {pages} {210801} (\bibinfo {year} {2023})}\BibitemShut {NoStop}%
\bibitem [{\citenamefont {Ma}\ \emph {et~al.}(2016{\natexlab{a}})\citenamefont {Ma}, \citenamefont {Sacher}, \citenamefont {Tang}, \citenamefont {Mikkelsen}, \citenamefont {Yang} \emph {et~al.}}]{ma2016silicon}%
  \BibitemOpen
  \bibfield  {author} {\bibinfo {author} {\bibfnamefont {C.}~\bibnamefont {Ma}}, \bibinfo {author} {\bibfnamefont {W.~D.}\ \bibnamefont {Sacher}}, \bibinfo {author} {\bibfnamefont {Z.}~\bibnamefont {Tang}}, \bibinfo {author} {\bibfnamefont {J.~C.}\ \bibnamefont {Mikkelsen}}, \bibinfo {author} {\bibfnamefont {Y.}~\bibnamefont {Yang}},  \emph {et~al.},\ }\bibfield  {title} {\enquote {\bibinfo {title} {Silicon photonic transmitter for polarization-encoded quantum key distribution},}\ }\href@noop {} {\bibfield  {journal} {\bibinfo  {journal} {Optica}\ }\textbf {\bibinfo {volume} {3}},\ \bibinfo {pages} {1274--1278} (\bibinfo {year} {2016}{\natexlab{a}})}\BibitemShut {NoStop}%
\bibitem [{\citenamefont {Sibson}\ \emph {et~al.}(2017{\natexlab{a}})\citenamefont {Sibson}, \citenamefont {Erven}, \citenamefont {Godfrey}, \citenamefont {Miki}, \citenamefont {Yamashita} \emph {et~al.}}]{sibson2017chip}%
  \BibitemOpen
  \bibfield  {author} {\bibinfo {author} {\bibfnamefont {P.}~\bibnamefont {Sibson}}, \bibinfo {author} {\bibfnamefont {C.}~\bibnamefont {Erven}}, \bibinfo {author} {\bibfnamefont {M.}~\bibnamefont {Godfrey}}, \bibinfo {author} {\bibfnamefont {S.}~\bibnamefont {Miki}}, \bibinfo {author} {\bibfnamefont {T.}~\bibnamefont {Yamashita}},  \emph {et~al.},\ }\bibfield  {title} {\enquote {\bibinfo {title} {Chip-based quantum key distribution},}\ }\href@noop {} {\bibfield  {journal} {\bibinfo  {journal} {Nat. Commun.}\ }\textbf {\bibinfo {volume} {8}},\ \bibinfo {pages} {13984} (\bibinfo {year} {2017}{\natexlab{a}})}\BibitemShut {NoStop}%
\bibitem [{\citenamefont {Sibson}\ \emph {et~al.}(2017{\natexlab{b}})\citenamefont {Sibson}, \citenamefont {Kennard}, \citenamefont {Stanisic}, \citenamefont {Erven}, \citenamefont {O'Brien} \emph {et~al.}}]{sibson2017integrated}%
  \BibitemOpen
  \bibfield  {author} {\bibinfo {author} {\bibfnamefont {P.}~\bibnamefont {Sibson}}, \bibinfo {author} {\bibfnamefont {J.~E.}\ \bibnamefont {Kennard}}, \bibinfo {author} {\bibfnamefont {S.}~\bibnamefont {Stanisic}}, \bibinfo {author} {\bibfnamefont {C.}~\bibnamefont {Erven}}, \bibinfo {author} {\bibfnamefont {J.~L.}\ \bibnamefont {O'Brien}},  \emph {et~al.},\ }\bibfield  {title} {\enquote {\bibinfo {title} {Integrated silicon photonics for high-speed quantum key distribution},}\ }\href@noop {} {\bibfield  {journal} {\bibinfo  {journal} {Optica}\ }\textbf {\bibinfo {volume} {4}},\ \bibinfo {pages} {172--177} (\bibinfo {year} {2017}{\natexlab{b}})}\BibitemShut {NoStop}%
\bibitem [{\citenamefont {Bunandar}\ \emph {et~al.}(2018)\citenamefont {Bunandar}, \citenamefont {Lentine}, \citenamefont {Lee}, \citenamefont {Cai}, \citenamefont {Long} \emph {et~al.}}]{bunandar2018metropolitan}%
  \BibitemOpen
  \bibfield  {author} {\bibinfo {author} {\bibfnamefont {D.}~\bibnamefont {Bunandar}}, \bibinfo {author} {\bibfnamefont {A.}~\bibnamefont {Lentine}}, \bibinfo {author} {\bibfnamefont {C.}~\bibnamefont {Lee}}, \bibinfo {author} {\bibfnamefont {H.}~\bibnamefont {Cai}}, \bibinfo {author} {\bibfnamefont {C.~M.}\ \bibnamefont {Long}},  \emph {et~al.},\ }\bibfield  {title} {\enquote {\bibinfo {title} {Metropolitan quantum key distribution with silicon photonics},}\ }\href@noop {} {\bibfield  {journal} {\bibinfo  {journal} {Phys. Rev. X}\ }\textbf {\bibinfo {volume} {8}},\ \bibinfo {pages} {021009} (\bibinfo {year} {2018})}\BibitemShut {NoStop}%
\bibitem [{\citenamefont {Ding}\ \emph {et~al.}(2017)\citenamefont {Ding}, \citenamefont {Bacco}, \citenamefont {Dalgaard}, \citenamefont {Cai}, \citenamefont {Zhou} \emph {et~al.}}]{ding2017high}%
  \BibitemOpen
  \bibfield  {author} {\bibinfo {author} {\bibfnamefont {Y.}~\bibnamefont {Ding}}, \bibinfo {author} {\bibfnamefont {D.}~\bibnamefont {Bacco}}, \bibinfo {author} {\bibfnamefont {K.}~\bibnamefont {Dalgaard}}, \bibinfo {author} {\bibfnamefont {X.}~\bibnamefont {Cai}}, \bibinfo {author} {\bibfnamefont {X.}~\bibnamefont {Zhou}},  \emph {et~al.},\ }\bibfield  {title} {\enquote {\bibinfo {title} {High-dimensional quantum key distribution based on multicore fiber using silicon photonic integrated circuits},}\ }\href@noop {} {\bibfield  {journal} {\bibinfo  {journal} {npj Quantum Inf.}\ }\textbf {\bibinfo {volume} {3}},\ \bibinfo {pages} {25} (\bibinfo {year} {2017})}\BibitemShut {NoStop}%
\bibitem [{\citenamefont {Para{\"\i}so}\ \emph {et~al.}(2019)\citenamefont {Para{\"\i}so}, \citenamefont {De~Marco}, \citenamefont {Roger}, \citenamefont {Marangon}, \citenamefont {Dynes} \emph {et~al.}}]{paraiso2019modulator}%
  \BibitemOpen
  \bibfield  {author} {\bibinfo {author} {\bibfnamefont {T.~K.}\ \bibnamefont {Para{\"\i}so}}, \bibinfo {author} {\bibfnamefont {I.}~\bibnamefont {De~Marco}}, \bibinfo {author} {\bibfnamefont {T.}~\bibnamefont {Roger}}, \bibinfo {author} {\bibfnamefont {D.~G.}\ \bibnamefont {Marangon}}, \bibinfo {author} {\bibfnamefont {J.~F.}\ \bibnamefont {Dynes}},  \emph {et~al.},\ }\bibfield  {title} {\enquote {\bibinfo {title} {A modulator-free quantum key distribution transmitter chip},}\ }\href@noop {} {\bibfield  {journal} {\bibinfo  {journal} {npj Quantum Inf.}\ }\textbf {\bibinfo {volume} {5}},\ \bibinfo {pages} {42} (\bibinfo {year} {2019})}\BibitemShut {NoStop}%
\bibitem [{\citenamefont {Lee}\ \emph {et~al.}(2014)\citenamefont {Lee}, \citenamefont {Zhang}, \citenamefont {Steinbrecher}, \citenamefont {Zhou}, \citenamefont {Mower} \emph {et~al.}}]{lee2014entanglement}%
  \BibitemOpen
  \bibfield  {author} {\bibinfo {author} {\bibfnamefont {C.}~\bibnamefont {Lee}}, \bibinfo {author} {\bibfnamefont {Z.}~\bibnamefont {Zhang}}, \bibinfo {author} {\bibfnamefont {G.~R.}\ \bibnamefont {Steinbrecher}}, \bibinfo {author} {\bibfnamefont {H.}~\bibnamefont {Zhou}}, \bibinfo {author} {\bibfnamefont {J.}~\bibnamefont {Mower}},  \emph {et~al.},\ }\bibfield  {title} {\enquote {\bibinfo {title} {Entanglement-based quantum communication secured by nonlocal dispersion cancellation},}\ }\href@noop {} {\bibfield  {journal} {\bibinfo  {journal} {Phys. Rev. A}\ }\textbf {\bibinfo {volume} {90}},\ \bibinfo {pages} {062331} (\bibinfo {year} {2014})}\BibitemShut {NoStop}%
\bibitem [{\citenamefont {Guan}\ \emph {et~al.}(2015)\citenamefont {Guan}, \citenamefont {Cao}, \citenamefont {Liu}, \citenamefont {Shen-Tu}, \citenamefont {Pelc} \emph {et~al.}}]{guan2015experimental}%
  \BibitemOpen
  \bibfield  {author} {\bibinfo {author} {\bibfnamefont {J.-Y.}\ \bibnamefont {Guan}}, \bibinfo {author} {\bibfnamefont {Z.}~\bibnamefont {Cao}}, \bibinfo {author} {\bibfnamefont {Y.}~\bibnamefont {Liu}}, \bibinfo {author} {\bibfnamefont {G.-L.}\ \bibnamefont {Shen-Tu}}, \bibinfo {author} {\bibfnamefont {J.~S.}\ \bibnamefont {Pelc}},  \emph {et~al.},\ }\bibfield  {title} {\enquote {\bibinfo {title} {Experimental passive round-robin differential phase-shift quantum key distribution},}\ }\href@noop {} {\bibfield  {journal} {\bibinfo  {journal} {Phys. Rev. Lett.}\ }\textbf {\bibinfo {volume} {114}},\ \bibinfo {pages} {180502} (\bibinfo {year} {2015})}\BibitemShut {NoStop}%
\bibitem [{\citenamefont {Takesue}\ \emph {et~al.}(2015)\citenamefont {Takesue}, \citenamefont {Sasaki}, \citenamefont {Tamaki},\ and\ \citenamefont {Koashi}}]{takesue2015experimental}%
  \BibitemOpen
  \bibfield  {author} {\bibinfo {author} {\bibfnamefont {H.}~\bibnamefont {Takesue}}, \bibinfo {author} {\bibfnamefont {T.}~\bibnamefont {Sasaki}}, \bibinfo {author} {\bibfnamefont {K.}~\bibnamefont {Tamaki}}, \ and\ \bibinfo {author} {\bibfnamefont {M.}~\bibnamefont {Koashi}},\ }\bibfield  {title} {\enquote {\bibinfo {title} {Experimental quantum key distribution without monitoring signal disturbance},}\ }\href@noop {} {\bibfield  {journal} {\bibinfo  {journal} {Nat. Photonics}\ }\textbf {\bibinfo {volume} {9}},\ \bibinfo {pages} {827} (\bibinfo {year} {2015})}\BibitemShut {NoStop}%
\bibitem [{\citenamefont {Wang}\ \emph {et~al.}(2015{\natexlab{b}})\citenamefont {Wang}, \citenamefont {Yin}, \citenamefont {Chen}, \citenamefont {He}, \citenamefont {Song} \emph {et~al.}}]{wang2015experimental}%
  \BibitemOpen
  \bibfield  {author} {\bibinfo {author} {\bibfnamefont {S.}~\bibnamefont {Wang}}, \bibinfo {author} {\bibfnamefont {Z.-Q.}\ \bibnamefont {Yin}}, \bibinfo {author} {\bibfnamefont {W.}~\bibnamefont {Chen}}, \bibinfo {author} {\bibfnamefont {D.-Y.}\ \bibnamefont {He}}, \bibinfo {author} {\bibfnamefont {X.-T.}\ \bibnamefont {Song}},  \emph {et~al.},\ }\bibfield  {title} {\enquote {\bibinfo {title} {Experimental demonstration of a quantum key distribution without signal disturbance monitoring},}\ }\href@noop {} {\bibfield  {journal} {\bibinfo  {journal} {Nat. Photonics}\ }\textbf {\bibinfo {volume} {9}},\ \bibinfo {pages} {832--836} (\bibinfo {year} {2015}{\natexlab{b}})}\BibitemShut {NoStop}%
\bibitem [{\citenamefont {Zhong}\ \emph {et~al.}(2015)\citenamefont {Zhong}, \citenamefont {Zhou}, \citenamefont {Horansky}, \citenamefont {Lee}, \citenamefont {Verma} \emph {et~al.}}]{zhong2015photon}%
  \BibitemOpen
  \bibfield  {author} {\bibinfo {author} {\bibfnamefont {T.}~\bibnamefont {Zhong}}, \bibinfo {author} {\bibfnamefont {H.}~\bibnamefont {Zhou}}, \bibinfo {author} {\bibfnamefont {R.~D.}\ \bibnamefont {Horansky}}, \bibinfo {author} {\bibfnamefont {C.}~\bibnamefont {Lee}}, \bibinfo {author} {\bibfnamefont {V.~B.}\ \bibnamefont {Verma}},  \emph {et~al.},\ }\bibfield  {title} {\enquote {\bibinfo {title} {Photon-efficient quantum key distribution using time--energy entanglement with high-dimensional encoding},}\ }\href@noop {} {\bibfield  {journal} {\bibinfo  {journal} {New J. Phys.}\ }\textbf {\bibinfo {volume} {17}},\ \bibinfo {pages} {022002} (\bibinfo {year} {2015})}\BibitemShut {NoStop}%
\bibitem [{\citenamefont {Korzh}\ \emph {et~al.}(2015)\citenamefont {Korzh}, \citenamefont {Lim}, \citenamefont {Houlmann}, \citenamefont {Gisin}, \citenamefont {Li} \emph {et~al.}}]{korzh2015provably}%
  \BibitemOpen
  \bibfield  {author} {\bibinfo {author} {\bibfnamefont {B.}~\bibnamefont {Korzh}}, \bibinfo {author} {\bibfnamefont {C.~C.~W.}\ \bibnamefont {Lim}}, \bibinfo {author} {\bibfnamefont {R.}~\bibnamefont {Houlmann}}, \bibinfo {author} {\bibfnamefont {N.}~\bibnamefont {Gisin}}, \bibinfo {author} {\bibfnamefont {M.~J.}\ \bibnamefont {Li}},  \emph {et~al.},\ }\bibfield  {title} {\enquote {\bibinfo {title} {Provably secure and practical quantum key distribution over 307 km of optical fibre},}\ }\href@noop {} {\bibfield  {journal} {\bibinfo  {journal} {Nat. Photonics}\ }\textbf {\bibinfo {volume} {9}},\ \bibinfo {pages} {163--168} (\bibinfo {year} {2015})}\BibitemShut {NoStop}%
\bibitem [{\citenamefont {Mirhosseini}\ \emph {et~al.}(2015)\citenamefont {Mirhosseini}, \citenamefont {Maga{\~n}a-Loaiza}, \citenamefont {O'Sullivan}, \citenamefont {Rodenburg}, \citenamefont {Malik} \emph {et~al.}}]{mirhosseini2015high}%
  \BibitemOpen
  \bibfield  {author} {\bibinfo {author} {\bibfnamefont {M.}~\bibnamefont {Mirhosseini}}, \bibinfo {author} {\bibfnamefont {O.~S.}\ \bibnamefont {Maga{\~n}a-Loaiza}}, \bibinfo {author} {\bibfnamefont {M.~N.}\ \bibnamefont {O'Sullivan}}, \bibinfo {author} {\bibfnamefont {B.}~\bibnamefont {Rodenburg}}, \bibinfo {author} {\bibfnamefont {M.}~\bibnamefont {Malik}},  \emph {et~al.},\ }\bibfield  {title} {\enquote {\bibinfo {title} {High-dimensional quantum cryptography with twisted light},}\ }\href@noop {} {\bibfield  {journal} {\bibinfo  {journal} {New J. Phys.}\ }\textbf {\bibinfo {volume} {17}},\ \bibinfo {pages} {033033} (\bibinfo {year} {2015})}\BibitemShut {NoStop}%
\bibitem [{\citenamefont {Sit}\ \emph {et~al.}(2017)\citenamefont {Sit}, \citenamefont {Bouchard}, \citenamefont {Fickler}, \citenamefont {Gagnon-Bischoff}, \citenamefont {Larocque} \emph {et~al.}}]{sit2017high}%
  \BibitemOpen
  \bibfield  {author} {\bibinfo {author} {\bibfnamefont {A.}~\bibnamefont {Sit}}, \bibinfo {author} {\bibfnamefont {F.}~\bibnamefont {Bouchard}}, \bibinfo {author} {\bibfnamefont {R.}~\bibnamefont {Fickler}}, \bibinfo {author} {\bibfnamefont {J.}~\bibnamefont {Gagnon-Bischoff}}, \bibinfo {author} {\bibfnamefont {H.}~\bibnamefont {Larocque}},  \emph {et~al.},\ }\bibfield  {title} {\enquote {\bibinfo {title} {High-dimensional intracity quantum cryptography with structured photons},}\ }\href@noop {} {\bibfield  {journal} {\bibinfo  {journal} {Optica}\ }\textbf {\bibinfo {volume} {4}},\ \bibinfo {pages} {1006--1010} (\bibinfo {year} {2017})}\BibitemShut {NoStop}%
\bibitem [{\citenamefont {Li}\ \emph {et~al.}(2016{\natexlab{a}})\citenamefont {Li}, \citenamefont {Cao}, \citenamefont {Dai}, \citenamefont {Lin}, \citenamefont {Zhang} \emph {et~al.}}]{li2016experimental}%
  \BibitemOpen
  \bibfield  {author} {\bibinfo {author} {\bibfnamefont {Y.-H.}\ \bibnamefont {Li}}, \bibinfo {author} {\bibfnamefont {Y.}~\bibnamefont {Cao}}, \bibinfo {author} {\bibfnamefont {H.}~\bibnamefont {Dai}}, \bibinfo {author} {\bibfnamefont {J.}~\bibnamefont {Lin}}, \bibinfo {author} {\bibfnamefont {Z.}~\bibnamefont {Zhang}},  \emph {et~al.},\ }\bibfield  {title} {\enquote {\bibinfo {title} {Experimental round-robin differential phase-shift quantum key distribution},}\ }\href@noop {} {\bibfield  {journal} {\bibinfo  {journal} {Phys. Rev. A}\ }\textbf {\bibinfo {volume} {93}},\ \bibinfo {pages} {030302} (\bibinfo {year} {2016}{\natexlab{a}})}\BibitemShut {NoStop}%
\bibitem [{\citenamefont {Yuan}\ \emph {et~al.}(2016)\citenamefont {Yuan}, \citenamefont {Fr{\"o}hlich}, \citenamefont {Lucamarini}, \citenamefont {Roberts}, \citenamefont {Dynes},\ and\ \citenamefont {Shields}}]{yuan2016directly}%
  \BibitemOpen
  \bibfield  {author} {\bibinfo {author} {\bibfnamefont {Z.}~\bibnamefont {Yuan}}, \bibinfo {author} {\bibfnamefont {B.}~\bibnamefont {Fr{\"o}hlich}}, \bibinfo {author} {\bibfnamefont {M.}~\bibnamefont {Lucamarini}}, \bibinfo {author} {\bibfnamefont {G.}~\bibnamefont {Roberts}}, \bibinfo {author} {\bibfnamefont {J.}~\bibnamefont {Dynes}}, \ and\ \bibinfo {author} {\bibfnamefont {A.}~\bibnamefont {Shields}},\ }\bibfield  {title} {\enquote {\bibinfo {title} {Directly phase-modulated light source},}\ }\href@noop {} {\bibfield  {journal} {\bibinfo  {journal} {Phys. Rev. X}\ }\textbf {\bibinfo {volume} {6}},\ \bibinfo {pages} {031044} (\bibinfo {year} {2016})}\BibitemShut {NoStop}%
\bibitem [{\citenamefont {Islam}\ \emph {et~al.}(2017)\citenamefont {Islam}, \citenamefont {Lim}, \citenamefont {Cahall}, \citenamefont {Kim},\ and\ \citenamefont {Gauthier}}]{islam2017provably}%
  \BibitemOpen
  \bibfield  {author} {\bibinfo {author} {\bibfnamefont {N.~T.}\ \bibnamefont {Islam}}, \bibinfo {author} {\bibfnamefont {C.~C.~W.}\ \bibnamefont {Lim}}, \bibinfo {author} {\bibfnamefont {C.}~\bibnamefont {Cahall}}, \bibinfo {author} {\bibfnamefont {J.}~\bibnamefont {Kim}}, \ and\ \bibinfo {author} {\bibfnamefont {D.~J.}\ \bibnamefont {Gauthier}},\ }\bibfield  {title} {\enquote {\bibinfo {title} {Provably secure and high-rate quantum key distribution with time-bin qudits},}\ }\href@noop {} {\bibfield  {journal} {\bibinfo  {journal} {Sci. Adv.}\ }\textbf {\bibinfo {volume} {3}},\ \bibinfo {pages} {e1701491} (\bibinfo {year} {2017})}\BibitemShut {NoStop}%
\bibitem [{\citenamefont {Pirandola}\ \emph {et~al.}(2017)\citenamefont {Pirandola}, \citenamefont {Laurenza}, \citenamefont {Ottaviani},\ and\ \citenamefont {Banchi}}]{pirandola2017fundamental}%
  \BibitemOpen
  \bibfield  {author} {\bibinfo {author} {\bibfnamefont {S.}~\bibnamefont {Pirandola}}, \bibinfo {author} {\bibfnamefont {R.}~\bibnamefont {Laurenza}}, \bibinfo {author} {\bibfnamefont {C.}~\bibnamefont {Ottaviani}}, \ and\ \bibinfo {author} {\bibfnamefont {L.}~\bibnamefont {Banchi}},\ }\bibfield  {title} {\enquote {\bibinfo {title} {Fundamental limits of repeaterless quantum communications},}\ }\href@noop {} {\bibfield  {journal} {\bibinfo  {journal} {Nature communications}\ }\textbf {\bibinfo {volume} {8}},\ \bibinfo {pages} {15043} (\bibinfo {year} {2017})}\BibitemShut {NoStop}%
\bibitem [{\citenamefont {Tanaka}\ \emph {et~al.}(2012)\citenamefont {Tanaka}, \citenamefont {Fujiwara}, \citenamefont {Yoshino}, \citenamefont {Takahashi}, \citenamefont {Nambu} \emph {et~al.}}]{Tanaka2012High}%
  \BibitemOpen
  \bibfield  {author} {\bibinfo {author} {\bibfnamefont {A.}~\bibnamefont {Tanaka}}, \bibinfo {author} {\bibfnamefont {M.}~\bibnamefont {Fujiwara}}, \bibinfo {author} {\bibfnamefont {K.-i.}\ \bibnamefont {Yoshino}}, \bibinfo {author} {\bibfnamefont {S.}~\bibnamefont {Takahashi}}, \bibinfo {author} {\bibfnamefont {Y.}~\bibnamefont {Nambu}},  \emph {et~al.},\ }\bibfield  {title} {\enquote {\bibinfo {title} {High-speed quantum key distribution system for 1-mbps real-time key generation},}\ }\href {\doibase 10.1109/JQE.2012.2187327} {\bibfield  {journal} {\bibinfo  {journal} {IEEE J. Quantum Electron.}\ }\textbf {\bibinfo {volume} {48}},\ \bibinfo {pages} {542--550} (\bibinfo {year} {2012})}\BibitemShut {NoStop}%
\bibitem [{\citenamefont {Lucamarini}\ \emph {et~al.}(2013)\citenamefont {Lucamarini}, \citenamefont {Patel}, \citenamefont {Dynes}, \citenamefont {Fr\"{o}hlich}, \citenamefont {Sharpe} \emph {et~al.}}]{Lucamarini2013Efficient}%
  \BibitemOpen
  \bibfield  {author} {\bibinfo {author} {\bibfnamefont {M.}~\bibnamefont {Lucamarini}}, \bibinfo {author} {\bibfnamefont {K.~A.}\ \bibnamefont {Patel}}, \bibinfo {author} {\bibfnamefont {J.~F.}\ \bibnamefont {Dynes}}, \bibinfo {author} {\bibfnamefont {B.}~\bibnamefont {Fr\"{o}hlich}}, \bibinfo {author} {\bibfnamefont {A.~W.}\ \bibnamefont {Sharpe}},  \emph {et~al.},\ }\bibfield  {title} {\enquote {\bibinfo {title} {Efficient decoy-state quantum key distribution with quantified security},}\ }\href {\doibase 10.1364/OE.21.024550} {\bibfield  {journal} {\bibinfo  {journal} {Opt. Express}\ }\textbf {\bibinfo {volume} {21}},\ \bibinfo {pages} {24550--24565} (\bibinfo {year} {2013})}\BibitemShut {NoStop}%
\bibitem [{\citenamefont {Fr\"{o}hlich}\ \emph {et~al.}(2017)\citenamefont {Fr\"{o}hlich}, \citenamefont {Lucamarini}, \citenamefont {Dynes}, \citenamefont {Comandar}, \citenamefont {Tam} \emph {et~al.}}]{Frohlich2017Long}%
  \BibitemOpen
  \bibfield  {author} {\bibinfo {author} {\bibfnamefont {B.}~\bibnamefont {Fr\"{o}hlich}}, \bibinfo {author} {\bibfnamefont {M.}~\bibnamefont {Lucamarini}}, \bibinfo {author} {\bibfnamefont {J.~F.}\ \bibnamefont {Dynes}}, \bibinfo {author} {\bibfnamefont {L.~C.}\ \bibnamefont {Comandar}}, \bibinfo {author} {\bibfnamefont {W.~W.-S.}\ \bibnamefont {Tam}},  \emph {et~al.},\ }\bibfield  {title} {\enquote {\bibinfo {title} {Long-distance quantum key distribution secure against coherent attacks},}\ }\href {\doibase 10.1364/OPTICA.4.000163} {\bibfield  {journal} {\bibinfo  {journal} {Optica}\ }\textbf {\bibinfo {volume} {4}},\ \bibinfo {pages} {163--167} (\bibinfo {year} {2017})}\BibitemShut {NoStop}%
\bibitem [{\citenamefont {Yuan}\ \emph {et~al.}(2018)\citenamefont {Yuan}, \citenamefont {Plews}, \citenamefont {Takahashi}, \citenamefont {Doi}, \citenamefont {Tam} \emph {et~al.}}]{Yuan201810Mbps}%
  \BibitemOpen
  \bibfield  {author} {\bibinfo {author} {\bibfnamefont {Z.}~\bibnamefont {Yuan}}, \bibinfo {author} {\bibfnamefont {A.}~\bibnamefont {Plews}}, \bibinfo {author} {\bibfnamefont {R.}~\bibnamefont {Takahashi}}, \bibinfo {author} {\bibfnamefont {K.}~\bibnamefont {Doi}}, \bibinfo {author} {\bibfnamefont {W.}~\bibnamefont {Tam}},  \emph {et~al.},\ }\bibfield  {title} {\enquote {\bibinfo {title} {10-mb/s quantum key distribution},}\ }\href {\doibase 10.1109/JLT.2018.2843136} {\bibfield  {journal} {\bibinfo  {journal} {J. Light. Technol.}\ }\textbf {\bibinfo {volume} {36}},\ \bibinfo {pages} {3427--3433} (\bibinfo {year} {2018})}\BibitemShut {NoStop}%
\bibitem [{\citenamefont {Li}\ \emph {et~al.}(2023{\natexlab{a}})\citenamefont {Li}, \citenamefont {Zhang}, \citenamefont {Tan}, \citenamefont {Lu}, \citenamefont {Liao} \emph {et~al.}}]{li2023high}%
  \BibitemOpen
  \bibfield  {author} {\bibinfo {author} {\bibfnamefont {W.}~\bibnamefont {Li}}, \bibinfo {author} {\bibfnamefont {L.}~\bibnamefont {Zhang}}, \bibinfo {author} {\bibfnamefont {H.}~\bibnamefont {Tan}}, \bibinfo {author} {\bibfnamefont {Y.}~\bibnamefont {Lu}}, \bibinfo {author} {\bibfnamefont {S.-K.}\ \bibnamefont {Liao}},  \emph {et~al.},\ }\bibfield  {title} {\enquote {\bibinfo {title} {High-rate quantum key distribution exceeding 110 mb s--1},}\ }\href@noop {} {\bibfield  {journal} {\bibinfo  {journal} {Nat. Photonics}\ }\textbf {\bibinfo {volume} {17}},\ \bibinfo {pages} {416--421} (\bibinfo {year} {2023}{\natexlab{a}})}\BibitemShut {NoStop}%
\bibitem [{\citenamefont {Elliott}\ \emph {et~al.}(2005)\citenamefont {Elliott}, \citenamefont {Colvin}, \citenamefont {Pearson}, \citenamefont {Pikalo}, \citenamefont {Schlafer} \emph {et~al.}}]{elliott2005current}%
  \BibitemOpen
  \bibfield  {author} {\bibinfo {author} {\bibfnamefont {C.}~\bibnamefont {Elliott}}, \bibinfo {author} {\bibfnamefont {A.}~\bibnamefont {Colvin}}, \bibinfo {author} {\bibfnamefont {D.}~\bibnamefont {Pearson}}, \bibinfo {author} {\bibfnamefont {O.}~\bibnamefont {Pikalo}}, \bibinfo {author} {\bibfnamefont {J.}~\bibnamefont {Schlafer}},  \emph {et~al.},\ }\bibfield  {title} {\enquote {\bibinfo {title} {Current status of the darpa quantum network},}\ }in\ \href@noop {} {\emph {\bibinfo {booktitle} {Quantum Information and Computation III}}},\ Vol.\ \bibinfo {volume} {5815}\ (\bibinfo {organization} {SPIE},\ \bibinfo {year} {2005})\ pp.\ \bibinfo {pages} {138--149}\BibitemShut {NoStop}%
\bibitem [{\citenamefont {Peev}\ \emph {et~al.}(2009)\citenamefont {Peev}, \citenamefont {Pacher}, \citenamefont {All{\'e}aume}, \citenamefont {Barreiro}, \citenamefont {Bouda} \emph {et~al.}}]{Peev_NewJPhys_2009}%
  \BibitemOpen
  \bibfield  {author} {\bibinfo {author} {\bibfnamefont {M.}~\bibnamefont {Peev}}, \bibinfo {author} {\bibfnamefont {C.}~\bibnamefont {Pacher}}, \bibinfo {author} {\bibfnamefont {R.}~\bibnamefont {All{\'e}aume}}, \bibinfo {author} {\bibfnamefont {C.}~\bibnamefont {Barreiro}}, \bibinfo {author} {\bibfnamefont {J.}~\bibnamefont {Bouda}},  \emph {et~al.},\ }\bibfield  {title} {\enquote {\bibinfo {title} {The secoqc quantum key distribution network in vienna},}\ }\href@noop {} {\bibfield  {journal} {\bibinfo  {journal} {New J. Phys.}\ }\textbf {\bibinfo {volume} {11}},\ \bibinfo {pages} {075001} (\bibinfo {year} {2009})}\BibitemShut {NoStop}%
\bibitem [{\citenamefont {Chen}\ \emph {et~al.}(2009)\citenamefont {Chen}, \citenamefont {Liang}, \citenamefont {Liu}, \citenamefont {Cai}, \citenamefont {Ju} \emph {et~al.}}]{chen2009field}%
  \BibitemOpen
  \bibfield  {author} {\bibinfo {author} {\bibfnamefont {T.-Y.}\ \bibnamefont {Chen}}, \bibinfo {author} {\bibfnamefont {H.}~\bibnamefont {Liang}}, \bibinfo {author} {\bibfnamefont {Y.}~\bibnamefont {Liu}}, \bibinfo {author} {\bibfnamefont {W.-Q.}\ \bibnamefont {Cai}}, \bibinfo {author} {\bibfnamefont {L.}~\bibnamefont {Ju}},  \emph {et~al.},\ }\bibfield  {title} {\enquote {\bibinfo {title} {Field test of a practical secure communication network with decoy-state quantum cryptography},}\ }\href@noop {} {\bibfield  {journal} {\bibinfo  {journal} {Opt. Express}\ }\textbf {\bibinfo {volume} {17}},\ \bibinfo {pages} {6540--6549} (\bibinfo {year} {2009})}\BibitemShut {NoStop}%
\bibitem [{\citenamefont {Sasaki}\ \emph {et~al.}(2011)\citenamefont {Sasaki}, \citenamefont {Fujiwara}, \citenamefont {Ishizuka}, \citenamefont {Klaus}, \citenamefont {Wakui} \emph {et~al.}}]{sasaki2011field}%
  \BibitemOpen
  \bibfield  {author} {\bibinfo {author} {\bibfnamefont {M.}~\bibnamefont {Sasaki}}, \bibinfo {author} {\bibfnamefont {M.}~\bibnamefont {Fujiwara}}, \bibinfo {author} {\bibfnamefont {H.}~\bibnamefont {Ishizuka}}, \bibinfo {author} {\bibfnamefont {W.}~\bibnamefont {Klaus}}, \bibinfo {author} {\bibfnamefont {K.}~\bibnamefont {Wakui}},  \emph {et~al.},\ }\bibfield  {title} {\enquote {\bibinfo {title} {Field test of quantum key distribution in the tokyo qkd network},}\ }\href@noop {} {\bibfield  {journal} {\bibinfo  {journal} {Opt. Express}\ }\textbf {\bibinfo {volume} {19}},\ \bibinfo {pages} {10387--10409} (\bibinfo {year} {2011})}\BibitemShut {NoStop}%
\bibitem [{\citenamefont {Chen}\ \emph {et~al.}(2021{\natexlab{b}})\citenamefont {Chen}, \citenamefont {Jiang}, \citenamefont {Tang}, \citenamefont {Zhou}, \citenamefont {Yuan} \emph {et~al.}}]{chen2021implementation}%
  \BibitemOpen
  \bibfield  {author} {\bibinfo {author} {\bibfnamefont {T.-Y.}\ \bibnamefont {Chen}}, \bibinfo {author} {\bibfnamefont {X.}~\bibnamefont {Jiang}}, \bibinfo {author} {\bibfnamefont {S.-B.}\ \bibnamefont {Tang}}, \bibinfo {author} {\bibfnamefont {L.}~\bibnamefont {Zhou}}, \bibinfo {author} {\bibfnamefont {X.}~\bibnamefont {Yuan}},  \emph {et~al.},\ }\bibfield  {title} {\enquote {\bibinfo {title} {Implementation of a 46-node quantum metropolitan area network},}\ }\href@noop {} {\bibfield  {journal} {\bibinfo  {journal} {npj Quantum Inf.}\ }\textbf {\bibinfo {volume} {7}},\ \bibinfo {pages} {134} (\bibinfo {year} {2021}{\natexlab{b}})}\BibitemShut {NoStop}%
\bibitem [{\citenamefont {Yin}\ \emph {et~al.}(2017{\natexlab{a}})\citenamefont {Yin}, \citenamefont {Cao}, \citenamefont {Li}, \citenamefont {Liao}, \citenamefont {Zhang} \emph {et~al.}}]{Yin2017SatellitebasedED}%
  \BibitemOpen
  \bibfield  {author} {\bibinfo {author} {\bibfnamefont {J.}~\bibnamefont {Yin}}, \bibinfo {author} {\bibfnamefont {Y.}~\bibnamefont {Cao}}, \bibinfo {author} {\bibfnamefont {Y.}~\bibnamefont {Li}}, \bibinfo {author} {\bibfnamefont {S.}~\bibnamefont {Liao}}, \bibinfo {author} {\bibfnamefont {L.}~\bibnamefont {Zhang}},  \emph {et~al.},\ }\bibfield  {title} {\enquote {\bibinfo {title} {Satellite-based entanglement distribution over 1200 kilometers},}\ }\href {https://api.semanticscholar.org/CorpusID:5206894} {\bibfield  {journal} {\bibinfo  {journal} {Science}\ }\textbf {\bibinfo {volume} {356}},\ \bibinfo {pages} {1140 -- 1144} (\bibinfo {year} {2017}{\natexlab{a}})}\BibitemShut {NoStop}%
\bibitem [{\citenamefont {Liao}\ \emph {et~al.}(2017)\citenamefont {Liao}, \citenamefont {Cai}, \citenamefont {Liu}, \citenamefont {Zhang}, \citenamefont {Li} \emph {et~al.}}]{liao2017satellite}%
  \BibitemOpen
  \bibfield  {author} {\bibinfo {author} {\bibfnamefont {S.-K.}\ \bibnamefont {Liao}}, \bibinfo {author} {\bibfnamefont {W.-Q.}\ \bibnamefont {Cai}}, \bibinfo {author} {\bibfnamefont {W.-Y.}\ \bibnamefont {Liu}}, \bibinfo {author} {\bibfnamefont {L.}~\bibnamefont {Zhang}}, \bibinfo {author} {\bibfnamefont {Y.}~\bibnamefont {Li}},  \emph {et~al.},\ }\bibfield  {title} {\enquote {\bibinfo {title} {Satellite-to-ground quantum key distribution},}\ }\href@noop {} {\bibfield  {journal} {\bibinfo  {journal} {Nature}\ }\textbf {\bibinfo {volume} {549}},\ \bibinfo {pages} {43--47} (\bibinfo {year} {2017})}\BibitemShut {NoStop}%
\bibitem [{\citenamefont {Ren}\ \emph {et~al.}(2017)\citenamefont {Ren}, \citenamefont {Xu}, \citenamefont {Yong}, \citenamefont {Zhang}, \citenamefont {Liao} \emph {et~al.}}]{Ren2017GroundtosatelliteQT}%
  \BibitemOpen
  \bibfield  {author} {\bibinfo {author} {\bibfnamefont {J.-G.}\ \bibnamefont {Ren}}, \bibinfo {author} {\bibfnamefont {P.}~\bibnamefont {Xu}}, \bibinfo {author} {\bibfnamefont {H.-L.}\ \bibnamefont {Yong}}, \bibinfo {author} {\bibfnamefont {L.}~\bibnamefont {Zhang}}, \bibinfo {author} {\bibfnamefont {S.}~\bibnamefont {Liao}},  \emph {et~al.},\ }\bibfield  {title} {\enquote {\bibinfo {title} {Ground-to-satellite quantum teleportation},}\ }\href {https://api.semanticscholar.org/CorpusID:4468803} {\bibfield  {journal} {\bibinfo  {journal} {Nature}\ }\textbf {\bibinfo {volume} {549}},\ \bibinfo {pages} {70--73} (\bibinfo {year} {2017})}\BibitemShut {NoStop}%
\bibitem [{\citenamefont {Yin}\ \emph {et~al.}(2017{\natexlab{b}})\citenamefont {Yin}, \citenamefont {Cao}, \citenamefont {Li}, \citenamefont {Ren}, \citenamefont {Liao} \emph {et~al.}}]{Yin2017SatellitetoGroundEQ}%
  \BibitemOpen
  \bibfield  {author} {\bibinfo {author} {\bibfnamefont {J.}~\bibnamefont {Yin}}, \bibinfo {author} {\bibfnamefont {Y.}~\bibnamefont {Cao}}, \bibinfo {author} {\bibfnamefont {Y.}~\bibnamefont {Li}}, \bibinfo {author} {\bibfnamefont {J.-G.}\ \bibnamefont {Ren}}, \bibinfo {author} {\bibfnamefont {S.}~\bibnamefont {Liao}},  \emph {et~al.},\ }\bibfield  {title} {\enquote {\bibinfo {title} {Satellite-to-ground entanglement-based quantum key distribution},}\ }\href {https://api.semanticscholar.org/CorpusID:42244971} {\bibfield  {journal} {\bibinfo  {journal} {Phys. Rev. Lett.}\ }\textbf {\bibinfo {volume} {119 20}},\ \bibinfo {pages} {200501} (\bibinfo {year} {2017}{\natexlab{b}})}\BibitemShut {NoStop}%
\bibitem [{\citenamefont {Liao}\ \emph {et~al.}(2018{\natexlab{a}})\citenamefont {Liao}, \citenamefont {Cai}, \citenamefont {Handsteiner}, \citenamefont {Liu}, \citenamefont {Yin} \emph {et~al.}}]{Liao2018SatelliteRelayedIQ}%
  \BibitemOpen
  \bibfield  {author} {\bibinfo {author} {\bibfnamefont {S.}~\bibnamefont {Liao}}, \bibinfo {author} {\bibfnamefont {W.}~\bibnamefont {Cai}}, \bibinfo {author} {\bibfnamefont {J.}~\bibnamefont {Handsteiner}}, \bibinfo {author} {\bibfnamefont {B.}~\bibnamefont {Liu}}, \bibinfo {author} {\bibfnamefont {J.}~\bibnamefont {Yin}},  \emph {et~al.},\ }\bibfield  {title} {\enquote {\bibinfo {title} {Satellite-relayed intercontinental quantum network.}}\ }\href {https://api.semanticscholar.org/CorpusID:206306725} {\bibfield  {journal} {\bibinfo  {journal} {Phys. Rev. Lett.}\ }\textbf {\bibinfo {volume} {120 3}},\ \bibinfo {pages} {030501} (\bibinfo {year} {2018}{\natexlab{a}})}\BibitemShut {NoStop}%
\bibitem [{\citenamefont {Chen}\ \emph {et~al.}(2021{\natexlab{c}})\citenamefont {Chen}, \citenamefont {Zhang}, \citenamefont {Chen}, \citenamefont {Cai}, \citenamefont {Liao} \emph {et~al.}}]{chen2021integrated}%
  \BibitemOpen
  \bibfield  {author} {\bibinfo {author} {\bibfnamefont {Y.-A.}\ \bibnamefont {Chen}}, \bibinfo {author} {\bibfnamefont {Q.}~\bibnamefont {Zhang}}, \bibinfo {author} {\bibfnamefont {T.-Y.}\ \bibnamefont {Chen}}, \bibinfo {author} {\bibfnamefont {W.-Q.}\ \bibnamefont {Cai}}, \bibinfo {author} {\bibfnamefont {S.-K.}\ \bibnamefont {Liao}},  \emph {et~al.},\ }\bibfield  {title} {\enquote {\bibinfo {title} {An integrated space-to-ground quantum communication network over 4,600 kilometres},}\ }\href@noop {} {\bibfield  {journal} {\bibinfo  {journal} {Nature}\ }\textbf {\bibinfo {volume} {589}},\ \bibinfo {pages} {214--219} (\bibinfo {year} {2021}{\natexlab{c}})}\BibitemShut {NoStop}%
\bibitem [{\citenamefont {Ralph}(1999)}]{ralph1999continuous}%
  \BibitemOpen
  \bibfield  {author} {\bibinfo {author} {\bibfnamefont {T.~C.}\ \bibnamefont {Ralph}},\ }\bibfield  {title} {\enquote {\bibinfo {title} {Continuous variable quantum cryptography},}\ }\href@noop {} {\bibfield  {journal} {\bibinfo  {journal} {Phys. Rev. A}\ }\textbf {\bibinfo {volume} {61}},\ \bibinfo {pages} {010303} (\bibinfo {year} {1999})}\BibitemShut {NoStop}%
\bibitem [{\citenamefont {Cerf}, \citenamefont {L\'evy},\ and\ \citenamefont {Assche}(2001)}]{Cerf_PhysRevA_2001}%
  \BibitemOpen
  \bibfield  {author} {\bibinfo {author} {\bibfnamefont {N.~J.}\ \bibnamefont {Cerf}}, \bibinfo {author} {\bibfnamefont {M.}~\bibnamefont {L\'evy}}, \ and\ \bibinfo {author} {\bibfnamefont {G.~V.}\ \bibnamefont {Assche}},\ }\bibfield  {title} {\enquote {\bibinfo {title} {Quantum distribution of gaussian keys using squeezed states},}\ }\href {\doibase 10.1103/PhysRevA.63.052311} {\bibfield  {journal} {\bibinfo  {journal} {Phys. Rev. A}\ }\textbf {\bibinfo {volume} {63}},\ \bibinfo {pages} {052311} (\bibinfo {year} {2001})}\BibitemShut {NoStop}%
\bibitem [{\citenamefont {Grosshans}\ and\ \citenamefont {Grangier}(2002)}]{Grosshans_PhysRevLett_2002}%
  \BibitemOpen
  \bibfield  {author} {\bibinfo {author} {\bibfnamefont {F.}~\bibnamefont {Grosshans}}\ and\ \bibinfo {author} {\bibfnamefont {P.}~\bibnamefont {Grangier}},\ }\bibfield  {title} {\enquote {\bibinfo {title} {Continuous variable quantum cryptography using coherent states},}\ }\href {\doibase 10.1103/PhysRevLett.88.057902} {\bibfield  {journal} {\bibinfo  {journal} {Phys. Rev. Lett.}\ }\textbf {\bibinfo {volume} {88}},\ \bibinfo {pages} {057902} (\bibinfo {year} {2002})}\BibitemShut {NoStop}%
\bibitem [{\citenamefont {Weedbrook}\ \emph {et~al.}(2004)\citenamefont {Weedbrook}, \citenamefont {Lance}, \citenamefont {Bowen}, \citenamefont {Symul}, \citenamefont {Ralph} \emph {et~al.}}]{Weedbrook_PhysRevLett_2004}%
  \BibitemOpen
  \bibfield  {author} {\bibinfo {author} {\bibfnamefont {C.}~\bibnamefont {Weedbrook}}, \bibinfo {author} {\bibfnamefont {A.~M.}\ \bibnamefont {Lance}}, \bibinfo {author} {\bibfnamefont {W.~P.}\ \bibnamefont {Bowen}}, \bibinfo {author} {\bibfnamefont {T.}~\bibnamefont {Symul}}, \bibinfo {author} {\bibfnamefont {T.~C.}\ \bibnamefont {Ralph}},  \emph {et~al.},\ }\bibfield  {title} {\enquote {\bibinfo {title} {Quantum cryptography without switching},}\ }\href {\doibase 10.1103/PhysRevLett.93.170504} {\bibfield  {journal} {\bibinfo  {journal} {Phys. Rev. Lett.}\ }\textbf {\bibinfo {volume} {93}},\ \bibinfo {pages} {170504} (\bibinfo {year} {2004})}\BibitemShut {NoStop}%
\bibitem [{\citenamefont {Braunstein}\ and\ \citenamefont {van Loock}(2005)}]{Braunstein_RevModPhys_2005}%
  \BibitemOpen
  \bibfield  {author} {\bibinfo {author} {\bibfnamefont {S.~L.}\ \bibnamefont {Braunstein}}\ and\ \bibinfo {author} {\bibfnamefont {P.}~\bibnamefont {van Loock}},\ }\bibfield  {title} {\enquote {\bibinfo {title} {Quantum information with continuous variables},}\ }\href {\doibase 10.1103/RevModPhys.77.513} {\bibfield  {journal} {\bibinfo  {journal} {Rev. Mod. Phys.}\ }\textbf {\bibinfo {volume} {77}},\ \bibinfo {pages} {513--577} (\bibinfo {year} {2005})}\BibitemShut {NoStop}%
\bibitem [{\citenamefont {Wang}\ \emph {et~al.}(2007)\citenamefont {Wang}, \citenamefont {Hiroshima}, \citenamefont {Tomita},\ and\ \citenamefont {Hayashi}}]{Wang_PhysRep_2007}%
  \BibitemOpen
  \bibfield  {author} {\bibinfo {author} {\bibfnamefont {X.-B.}\ \bibnamefont {Wang}}, \bibinfo {author} {\bibfnamefont {T.}~\bibnamefont {Hiroshima}}, \bibinfo {author} {\bibfnamefont {A.}~\bibnamefont {Tomita}}, \ and\ \bibinfo {author} {\bibfnamefont {M.}~\bibnamefont {Hayashi}},\ }\bibfield  {title} {\enquote {\bibinfo {title} {Quantum information with gaussian states},}\ }\href@noop {} {\bibfield  {journal} {\bibinfo  {journal} {Phys. Rep.}\ }\textbf {\bibinfo {volume} {448}},\ \bibinfo {pages} {1--111} (\bibinfo {year} {2007})}\BibitemShut {NoStop}%
\bibitem [{\citenamefont {Weedbrook}\ \emph {et~al.}(2012)\citenamefont {Weedbrook}, \citenamefont {Pirandola}, \citenamefont {Garc\'{\i}a-Patr\'on}, \citenamefont {Cerf}, \citenamefont {Ralph} \emph {et~al.}}]{Weedbrook_RevModPhys_2012}%
  \BibitemOpen
  \bibfield  {author} {\bibinfo {author} {\bibfnamefont {C.}~\bibnamefont {Weedbrook}}, \bibinfo {author} {\bibfnamefont {S.}~\bibnamefont {Pirandola}}, \bibinfo {author} {\bibfnamefont {R.}~\bibnamefont {Garc\'{\i}a-Patr\'on}}, \bibinfo {author} {\bibfnamefont {N.~J.}\ \bibnamefont {Cerf}}, \bibinfo {author} {\bibfnamefont {T.~C.}\ \bibnamefont {Ralph}},  \emph {et~al.},\ }\bibfield  {title} {\enquote {\bibinfo {title} {Gaussian quantum information},}\ }\href {\doibase 10.1103/RevModPhys.84.621} {\bibfield  {journal} {\bibinfo  {journal} {Rev. Mod. Phys.}\ }\textbf {\bibinfo {volume} {84}},\ \bibinfo {pages} {621--669} (\bibinfo {year} {2012})}\BibitemShut {NoStop}%
\bibitem [{\citenamefont {Lam}\ and\ \citenamefont {Ralph}(2013)}]{Lam_NatPhotonics_2013}%
  \BibitemOpen
  \bibfield  {author} {\bibinfo {author} {\bibfnamefont {P.~K.}\ \bibnamefont {Lam}}\ and\ \bibinfo {author} {\bibfnamefont {T.~C.}\ \bibnamefont {Ralph}},\ }\bibfield  {title} {\enquote {\bibinfo {title} {Continuous improvement},}\ }\href {\doibase 10.1038/nphoton.2013.104} {\bibfield  {journal} {\bibinfo  {journal} {Nat. Photonics}\ }\textbf {\bibinfo {volume} {7}},\ \bibinfo {pages} {350} (\bibinfo {year} {2013})}\BibitemShut {NoStop}%
\bibitem [{\citenamefont {Diamanti}\ and\ \citenamefont {Leverrier}(2015)}]{Diamanti_Entropy_2015}%
  \BibitemOpen
  \bibfield  {author} {\bibinfo {author} {\bibfnamefont {E.}~\bibnamefont {Diamanti}}\ and\ \bibinfo {author} {\bibfnamefont {A.}~\bibnamefont {Leverrier}},\ }\bibfield  {title} {\enquote {\bibinfo {title} {Distributing secret keys with quantum continuous variables: principle, security and implementations},}\ }\href@noop {} {\bibfield  {journal} {\bibinfo  {journal} {Entropy}\ }\textbf {\bibinfo {volume} {17}},\ \bibinfo {pages} {6072--6092} (\bibinfo {year} {2015})}\BibitemShut {NoStop}%
\bibitem [{\citenamefont {Cerf}, \citenamefont {Leuchs},\ and\ \citenamefont {Polzik}(2007)}]{cerf2007quantum}%
  \BibitemOpen
  \bibfield  {author} {\bibinfo {author} {\bibfnamefont {N.~J.}\ \bibnamefont {Cerf}}, \bibinfo {author} {\bibfnamefont {G.}~\bibnamefont {Leuchs}}, \ and\ \bibinfo {author} {\bibfnamefont {E.~S.}\ \bibnamefont {Polzik}},\ }\href@noop {} {\emph {\bibinfo {title} {Quantum information with continuous variables of atoms and light}}}\ (\bibinfo  {publisher} {World Scientific},\ \bibinfo {year} {2007})\BibitemShut {NoStop}%
\bibitem [{\citenamefont {Furusawa}\ and\ \citenamefont {Takei}(2007)}]{furusawa2007quantum}%
  \BibitemOpen
  \bibfield  {author} {\bibinfo {author} {\bibfnamefont {A.}~\bibnamefont {Furusawa}}\ and\ \bibinfo {author} {\bibfnamefont {N.}~\bibnamefont {Takei}},\ }\bibfield  {title} {\enquote {\bibinfo {title} {Quantum teleportation for continuous variables and related quantum information processing},}\ }\href@noop {} {\bibfield  {journal} {\bibinfo  {journal} {Phys. Rep.}\ }\textbf {\bibinfo {volume} {443}},\ \bibinfo {pages} {97--119} (\bibinfo {year} {2007})}\BibitemShut {NoStop}%
\bibitem [{\citenamefont {Andersen}, \citenamefont {Leuchs},\ and\ \citenamefont {Silberhorn}(2010)}]{andersen2010continuous}%
  \BibitemOpen
  \bibfield  {author} {\bibinfo {author} {\bibfnamefont {U.~L.}\ \bibnamefont {Andersen}}, \bibinfo {author} {\bibfnamefont {G.}~\bibnamefont {Leuchs}}, \ and\ \bibinfo {author} {\bibfnamefont {C.}~\bibnamefont {Silberhorn}},\ }\bibfield  {title} {\enquote {\bibinfo {title} {Continuous-variable quantum information processing},}\ }\href@noop {} {\bibfield  {journal} {\bibinfo  {journal} {Laser Photonics Rev.}\ }\textbf {\bibinfo {volume} {4}},\ \bibinfo {pages} {337--354} (\bibinfo {year} {2010})}\BibitemShut {NoStop}%
\bibitem [{\citenamefont {Van~Meter}(2014)}]{van2014quantum}%
  \BibitemOpen
  \bibfield  {author} {\bibinfo {author} {\bibfnamefont {R.}~\bibnamefont {Van~Meter}},\ }\href@noop {} {\emph {\bibinfo {title} {Quantum networking}}}\ (\bibinfo  {publisher} {John Wiley \& Sons},\ \bibinfo {year} {2014})\BibitemShut {NoStop}%
\bibitem [{\citenamefont {Adesso}, \citenamefont {Ragy},\ and\ \citenamefont {Lee}(2014)}]{adesso2014continuous}%
  \BibitemOpen
  \bibfield  {author} {\bibinfo {author} {\bibfnamefont {G.}~\bibnamefont {Adesso}}, \bibinfo {author} {\bibfnamefont {S.}~\bibnamefont {Ragy}}, \ and\ \bibinfo {author} {\bibfnamefont {A.~R.}\ \bibnamefont {Lee}},\ }\bibfield  {title} {\enquote {\bibinfo {title} {Continuous variable quantum information: Gaussian states and beyond},}\ }\href@noop {} {\bibfield  {journal} {\bibinfo  {journal} {Open Syst. Inf. Dyn.}\ }\textbf {\bibinfo {volume} {21}},\ \bibinfo {pages} {1440001} (\bibinfo {year} {2014})}\BibitemShut {NoStop}%
\bibitem [{\citenamefont {Andersen}\ \emph {et~al.}(2015)\citenamefont {Andersen}, \citenamefont {Neergaard-Nielsen}, \citenamefont {Van~Loock},\ and\ \citenamefont {Furusawa}}]{andersen2015hybrid}%
  \BibitemOpen
  \bibfield  {author} {\bibinfo {author} {\bibfnamefont {U.~L.}\ \bibnamefont {Andersen}}, \bibinfo {author} {\bibfnamefont {J.~S.}\ \bibnamefont {Neergaard-Nielsen}}, \bibinfo {author} {\bibfnamefont {P.}~\bibnamefont {Van~Loock}}, \ and\ \bibinfo {author} {\bibfnamefont {A.}~\bibnamefont {Furusawa}},\ }\bibfield  {title} {\enquote {\bibinfo {title} {Hybrid discrete-and continuous-variable quantum information},}\ }\href@noop {} {\bibfield  {journal} {\bibinfo  {journal} {Nat. Phys.}\ }\textbf {\bibinfo {volume} {11}},\ \bibinfo {pages} {713--719} (\bibinfo {year} {2015})}\BibitemShut {NoStop}%
\bibitem [{\citenamefont {Kurizki}\ \emph {et~al.}(2015)\citenamefont {Kurizki}, \citenamefont {Bertet}, \citenamefont {Kubo}, \citenamefont {M{\o}lmer}, \citenamefont {Petrosyan} \emph {et~al.}}]{kurizki2015quantum}%
  \BibitemOpen
  \bibfield  {author} {\bibinfo {author} {\bibfnamefont {G.}~\bibnamefont {Kurizki}}, \bibinfo {author} {\bibfnamefont {P.}~\bibnamefont {Bertet}}, \bibinfo {author} {\bibfnamefont {Y.}~\bibnamefont {Kubo}}, \bibinfo {author} {\bibfnamefont {K.}~\bibnamefont {M{\o}lmer}}, \bibinfo {author} {\bibfnamefont {D.}~\bibnamefont {Petrosyan}},  \emph {et~al.},\ }\bibfield  {title} {\enquote {\bibinfo {title} {Quantum technologies with hybrid systems},}\ }\href@noop {} {\bibfield  {journal} {\bibinfo  {journal} {Proc. Natl. Acad. Sci.}\ }\textbf {\bibinfo {volume} {112}},\ \bibinfo {pages} {3866--3873} (\bibinfo {year} {2015})}\BibitemShut {NoStop}%
\bibitem [{\citenamefont {Serafini}(2017)}]{serafini2017quantum}%
  \BibitemOpen
  \bibfield  {author} {\bibinfo {author} {\bibfnamefont {A.}~\bibnamefont {Serafini}},\ }\href@noop {} {\emph {\bibinfo {title} {Quantum continuous variables: a primer of theoretical methods}}}\ (\bibinfo  {publisher} {CRC press},\ \bibinfo {year} {2017})\BibitemShut {NoStop}%
\bibitem [{\citenamefont {Garc\'{\i}a-Patr\'on}\ and\ \citenamefont {Cerf}(2006)}]{Garcia_PhysRevLett_2006}%
  \BibitemOpen
  \bibfield  {author} {\bibinfo {author} {\bibfnamefont {R.}~\bibnamefont {Garc\'{\i}a-Patr\'on}}\ and\ \bibinfo {author} {\bibfnamefont {N.~J.}\ \bibnamefont {Cerf}},\ }\bibfield  {title} {\enquote {\bibinfo {title} {Unconditional optimality of gaussian attacks against continuous-variable quantum key distribution},}\ }\href {\doibase 10.1103/PhysRevLett.97.190503} {\bibfield  {journal} {\bibinfo  {journal} {Phys. Rev. Lett.}\ }\textbf {\bibinfo {volume} {97}},\ \bibinfo {pages} {190503} (\bibinfo {year} {2006})}\BibitemShut {NoStop}%
\bibitem [{\citenamefont {Navascu\'es}, \citenamefont {Grosshans},\ and\ \citenamefont {Ac\'{\i}n}(2006)}]{Navascues_PhysRevLett_2006}%
  \BibitemOpen
  \bibfield  {author} {\bibinfo {author} {\bibfnamefont {M.}~\bibnamefont {Navascu\'es}}, \bibinfo {author} {\bibfnamefont {F.}~\bibnamefont {Grosshans}}, \ and\ \bibinfo {author} {\bibfnamefont {A.}~\bibnamefont {Ac\'{\i}n}},\ }\bibfield  {title} {\enquote {\bibinfo {title} {Optimality of gaussian attacks in continuous-variable quantum cryptography},}\ }\href {\doibase 10.1103/PhysRevLett.97.190502} {\bibfield  {journal} {\bibinfo  {journal} {Phys. Rev. Lett.}\ }\textbf {\bibinfo {volume} {97}},\ \bibinfo {pages} {190502} (\bibinfo {year} {2006})}\BibitemShut {NoStop}%
\bibitem [{\citenamefont {Leverrier}(2015)}]{Leverrier_PhysRevLett_2015}%
  \BibitemOpen
  \bibfield  {author} {\bibinfo {author} {\bibfnamefont {A.}~\bibnamefont {Leverrier}},\ }\bibfield  {title} {\enquote {\bibinfo {title} {Composable security proof for continuous-variable quantum key distribution with coherent states},}\ }\href {\doibase 10.1103/PhysRevLett.114.070501} {\bibfield  {journal} {\bibinfo  {journal} {Phys. Rev. Lett.}\ }\textbf {\bibinfo {volume} {114}},\ \bibinfo {pages} {070501} (\bibinfo {year} {2015})}\BibitemShut {NoStop}%
\bibitem [{\citenamefont {Leverrier}(2017)}]{Leverrier_PhysRevLett_2017}%
  \BibitemOpen
  \bibfield  {author} {\bibinfo {author} {\bibfnamefont {A.}~\bibnamefont {Leverrier}},\ }\bibfield  {title} {\enquote {\bibinfo {title} {Security of continuous-variable quantum key distribution via a gaussian de finetti reduction},}\ }\href {\doibase 10.1103/PhysRevLett.118.200501} {\bibfield  {journal} {\bibinfo  {journal} {Phys. Rev. Lett.}\ }\textbf {\bibinfo {volume} {118}},\ \bibinfo {pages} {200501} (\bibinfo {year} {2017})}\BibitemShut {NoStop}%
\bibitem [{\citenamefont {Li}, \citenamefont {Zhang},\ and\ \citenamefont {Guo}(2018)}]{Li_Arxiv_2018}%
  \BibitemOpen
  \bibfield  {author} {\bibinfo {author} {\bibfnamefont {Z.}~\bibnamefont {Li}}, \bibinfo {author} {\bibfnamefont {Y.-C.}\ \bibnamefont {Zhang}}, \ and\ \bibinfo {author} {\bibfnamefont {H.}~\bibnamefont {Guo}},\ }\href@noop {} {\enquote {\bibinfo {title} {User-defined quantum key distribution},}\ } (\bibinfo {year} {2018}),\ \Eprint {http://arxiv.org/abs/1805.04249} {arXiv:1805.04249 [quant-ph]} \BibitemShut {NoStop}%
\bibitem [{\citenamefont {Ghorai}\ \emph {et~al.}(2019)\citenamefont {Ghorai}, \citenamefont {Grangier}, \citenamefont {Diamanti},\ and\ \citenamefont {Leverrier}}]{Ghorai_PhysRevX_2019}%
  \BibitemOpen
  \bibfield  {author} {\bibinfo {author} {\bibfnamefont {S.}~\bibnamefont {Ghorai}}, \bibinfo {author} {\bibfnamefont {P.}~\bibnamefont {Grangier}}, \bibinfo {author} {\bibfnamefont {E.}~\bibnamefont {Diamanti}}, \ and\ \bibinfo {author} {\bibfnamefont {A.}~\bibnamefont {Leverrier}},\ }\bibfield  {title} {\enquote {\bibinfo {title} {Asymptotic security of continuous-variable quantum key distribution with a discrete modulation},}\ }\href {\doibase 10.1103/PhysRevX.9.021059} {\bibfield  {journal} {\bibinfo  {journal} {Phys. Rev. X}\ }\textbf {\bibinfo {volume} {9}},\ \bibinfo {pages} {021059} (\bibinfo {year} {2019})}\BibitemShut {NoStop}%
\bibitem [{\citenamefont {Lin}, \citenamefont {Upadhyaya},\ and\ \citenamefont {L\"utkenhaus}(2019)}]{Lin_PhysRevX_2019}%
  \BibitemOpen
  \bibfield  {author} {\bibinfo {author} {\bibfnamefont {J.}~\bibnamefont {Lin}}, \bibinfo {author} {\bibfnamefont {T.}~\bibnamefont {Upadhyaya}}, \ and\ \bibinfo {author} {\bibfnamefont {N.}~\bibnamefont {L\"utkenhaus}},\ }\bibfield  {title} {\enquote {\bibinfo {title} {Asymptotic security analysis of discrete-modulated continuous-variable quantum key distribution},}\ }\href {\doibase 10.1103/PhysRevX.9.041064} {\bibfield  {journal} {\bibinfo  {journal} {Phys. Rev. X}\ }\textbf {\bibinfo {volume} {9}},\ \bibinfo {pages} {041064} (\bibinfo {year} {2019})}\BibitemShut {NoStop}%
\bibitem [{\citenamefont {Denys}, \citenamefont {Brown},\ and\ \citenamefont {Leverrier}(2021)}]{Denys2021explicitasymptotic}%
  \BibitemOpen
  \bibfield  {author} {\bibinfo {author} {\bibfnamefont {A.}~\bibnamefont {Denys}}, \bibinfo {author} {\bibfnamefont {P.}~\bibnamefont {Brown}}, \ and\ \bibinfo {author} {\bibfnamefont {A.}~\bibnamefont {Leverrier}},\ }\bibfield  {title} {\enquote {\bibinfo {title} {Explicit asymptotic secret key rate of continuous-variable quantum key distribution with an arbitrary modulation},}\ }\href {\doibase 10.22331/q-2021-09-13-540} {\bibfield  {journal} {\bibinfo  {journal} {{Quantum}}\ }\textbf {\bibinfo {volume} {5}},\ \bibinfo {pages} {540} (\bibinfo {year} {2021})}\BibitemShut {NoStop}%
\bibitem [{\citenamefont {Matsuura}\ \emph {et~al.}(2021)\citenamefont {Matsuura}, \citenamefont {Maeda}, \citenamefont {Sasaki},\ and\ \citenamefont {Koashi}}]{matsuura2021finite}%
  \BibitemOpen
  \bibfield  {author} {\bibinfo {author} {\bibfnamefont {T.}~\bibnamefont {Matsuura}}, \bibinfo {author} {\bibfnamefont {K.}~\bibnamefont {Maeda}}, \bibinfo {author} {\bibfnamefont {T.}~\bibnamefont {Sasaki}}, \ and\ \bibinfo {author} {\bibfnamefont {M.}~\bibnamefont {Koashi}},\ }\bibfield  {title} {\enquote {\bibinfo {title} {Finite-size security of continuous-variable quantum key distribution with digital signal processing},}\ }\href@noop {} {\bibfield  {journal} {\bibinfo  {journal} {Nat. Commun.}\ }\textbf {\bibinfo {volume} {12}},\ \bibinfo {pages} {252} (\bibinfo {year} {2021})}\BibitemShut {NoStop}%
\bibitem [{\citenamefont {Grosshans}\ \emph {et~al.}(2003{\natexlab{a}})\citenamefont {Grosshans}, \citenamefont {Van~Assche}, \citenamefont {Wenger}, \citenamefont {Brouri}, \citenamefont {Cerf} \emph {et~al.}}]{Grosshans_Nature_2003}%
  \BibitemOpen
  \bibfield  {author} {\bibinfo {author} {\bibfnamefont {F.}~\bibnamefont {Grosshans}}, \bibinfo {author} {\bibfnamefont {G.}~\bibnamefont {Van~Assche}}, \bibinfo {author} {\bibfnamefont {J.}~\bibnamefont {Wenger}}, \bibinfo {author} {\bibfnamefont {R.}~\bibnamefont {Brouri}}, \bibinfo {author} {\bibfnamefont {N.~J.}\ \bibnamefont {Cerf}},  \emph {et~al.},\ }\bibfield  {title} {\enquote {\bibinfo {title} {Quantum key distribution using gaussian-modulated coherent states},}\ }\href@noop {} {\bibfield  {journal} {\bibinfo  {journal} {Nature}\ }\textbf {\bibinfo {volume} {421}},\ \bibinfo {pages} {238--241} (\bibinfo {year} {2003}{\natexlab{a}})}\BibitemShut {NoStop}%
\bibitem [{\citenamefont {Leverrier}\ \emph {et~al.}(2008)\citenamefont {Leverrier}, \citenamefont {All{\'e}aume}, \citenamefont {Boutros}, \citenamefont {Z{\'e}mor},\ and\ \citenamefont {Grangier}}]{leverrier_PhysRevA_2008}%
  \BibitemOpen
  \bibfield  {author} {\bibinfo {author} {\bibfnamefont {A.}~\bibnamefont {Leverrier}}, \bibinfo {author} {\bibfnamefont {R.}~\bibnamefont {All{\'e}aume}}, \bibinfo {author} {\bibfnamefont {J.}~\bibnamefont {Boutros}}, \bibinfo {author} {\bibfnamefont {G.}~\bibnamefont {Z{\'e}mor}}, \ and\ \bibinfo {author} {\bibfnamefont {P.}~\bibnamefont {Grangier}},\ }\bibfield  {title} {\enquote {\bibinfo {title} {Multidimensional reconciliation for a continuous-variable quantum key distribution},}\ }\href@noop {} {\bibfield  {journal} {\bibinfo  {journal} {Phys. Rev. A}\ }\textbf {\bibinfo {volume} {77}},\ \bibinfo {pages} {042325} (\bibinfo {year} {2008})}\BibitemShut {NoStop}%
\bibitem [{\citenamefont {Lodewyck}\ \emph {et~al.}(2007)\citenamefont {Lodewyck}, \citenamefont {Bloch}, \citenamefont {Garc{\'\i}a-Patr{\'o}n}, \citenamefont {Fossier}, \citenamefont {Karpov} \emph {et~al.}}]{Lodewyck_PhysRevA_2007}%
  \BibitemOpen
  \bibfield  {author} {\bibinfo {author} {\bibfnamefont {J.}~\bibnamefont {Lodewyck}}, \bibinfo {author} {\bibfnamefont {M.}~\bibnamefont {Bloch}}, \bibinfo {author} {\bibfnamefont {R.}~\bibnamefont {Garc{\'\i}a-Patr{\'o}n}}, \bibinfo {author} {\bibfnamefont {S.}~\bibnamefont {Fossier}}, \bibinfo {author} {\bibfnamefont {E.}~\bibnamefont {Karpov}},  \emph {et~al.},\ }\bibfield  {title} {\enquote {\bibinfo {title} {Quantum key distribution over 25 km with an all-fiber continuous-variable system},}\ }\href@noop {} {\bibfield  {journal} {\bibinfo  {journal} {Phys. Rev. A}\ }\textbf {\bibinfo {volume} {76}},\ \bibinfo {pages} {042305} (\bibinfo {year} {2007})}\BibitemShut {NoStop}%
\bibitem [{\citenamefont {Jouguet}\ \emph {et~al.}(2013)\citenamefont {Jouguet}, \citenamefont {Kunz-Jacques}, \citenamefont {Leverrier}, \citenamefont {Grangier},\ and\ \citenamefont {Diamanti}}]{Jouguet_NatPhotonics_2013}%
  \BibitemOpen
  \bibfield  {author} {\bibinfo {author} {\bibfnamefont {P.}~\bibnamefont {Jouguet}}, \bibinfo {author} {\bibfnamefont {S.}~\bibnamefont {Kunz-Jacques}}, \bibinfo {author} {\bibfnamefont {A.}~\bibnamefont {Leverrier}}, \bibinfo {author} {\bibfnamefont {P.}~\bibnamefont {Grangier}}, \ and\ \bibinfo {author} {\bibfnamefont {E.}~\bibnamefont {Diamanti}},\ }\bibfield  {title} {\enquote {\bibinfo {title} {Experimental demonstration of long-distance continuous-variable quantum key distribution},}\ }\href@noop {} {\bibfield  {journal} {\bibinfo  {journal} {Nat. Photonics}\ }\textbf {\bibinfo {volume} {7}},\ \bibinfo {pages} {378--381} (\bibinfo {year} {2013})}\BibitemShut {NoStop}%
\bibitem [{\citenamefont {Jouguet}\ \emph {et~al.}(2012{\natexlab{a}})\citenamefont {Jouguet}, \citenamefont {Kunz-Jacques}, \citenamefont {Debuisschert}, \citenamefont {Fossier}, \citenamefont {Diamanti}, \citenamefont {All{\'e}aume} \emph {et~al.}}]{Jouguet_OptExpress_2012}%
  \BibitemOpen
  \bibfield  {author} {\bibinfo {author} {\bibfnamefont {P.}~\bibnamefont {Jouguet}}, \bibinfo {author} {\bibfnamefont {S.}~\bibnamefont {Kunz-Jacques}}, \bibinfo {author} {\bibfnamefont {T.}~\bibnamefont {Debuisschert}}, \bibinfo {author} {\bibfnamefont {S.}~\bibnamefont {Fossier}}, \bibinfo {author} {\bibfnamefont {E.}~\bibnamefont {Diamanti}}, \bibinfo {author} {\bibfnamefont {R.}~\bibnamefont {All{\'e}aume}},  \emph {et~al.},\ }\bibfield  {title} {\enquote {\bibinfo {title} {Field test of classical symmetric encryption with continuous variables quantum key distribution},}\ }\href@noop {} {\bibfield  {journal} {\bibinfo  {journal} {Opt. Express}\ }\textbf {\bibinfo {volume} {20}},\ \bibinfo {pages} {14030--14041} (\bibinfo {year} {2012}{\natexlab{a}})}\BibitemShut {NoStop}%
\bibitem [{\citenamefont {Huang}\ \emph {et~al.}(2016{\natexlab{a}})\citenamefont {Huang}, \citenamefont {Huang}, \citenamefont {Li}, \citenamefont {Wang}, \citenamefont {Zhou} \emph {et~al.}}]{Huang_OptLett_2016}%
  \BibitemOpen
  \bibfield  {author} {\bibinfo {author} {\bibfnamefont {D.}~\bibnamefont {Huang}}, \bibinfo {author} {\bibfnamefont {P.}~\bibnamefont {Huang}}, \bibinfo {author} {\bibfnamefont {H.}~\bibnamefont {Li}}, \bibinfo {author} {\bibfnamefont {T.}~\bibnamefont {Wang}}, \bibinfo {author} {\bibfnamefont {Y.}~\bibnamefont {Zhou}},  \emph {et~al.},\ }\bibfield  {title} {\enquote {\bibinfo {title} {Field demonstration of a continuous-variable quantum key distribution network},}\ }\href@noop {} {\bibfield  {journal} {\bibinfo  {journal} {Opt. Lett.}\ }\textbf {\bibinfo {volume} {41}},\ \bibinfo {pages} {3511--3514} (\bibinfo {year} {2016}{\natexlab{a}})}\BibitemShut {NoStop}%
\bibitem [{\citenamefont {Huang}\ \emph {et~al.}(2016{\natexlab{b}})\citenamefont {Huang}, \citenamefont {Huang}, \citenamefont {Lin},\ and\ \citenamefont {Zeng}}]{Huang_SciRep_2016}%
  \BibitemOpen
  \bibfield  {author} {\bibinfo {author} {\bibfnamefont {D.}~\bibnamefont {Huang}}, \bibinfo {author} {\bibfnamefont {P.}~\bibnamefont {Huang}}, \bibinfo {author} {\bibfnamefont {D.}~\bibnamefont {Lin}}, \ and\ \bibinfo {author} {\bibfnamefont {G.}~\bibnamefont {Zeng}},\ }\bibfield  {title} {\enquote {\bibinfo {title} {Long-distance continuous-variable quantum key distribution by controlling excess noise},}\ }\href@noop {} {\bibfield  {journal} {\bibinfo  {journal} {Sci. Rep.}\ }\textbf {\bibinfo {volume} {6}},\ \bibinfo {pages} {1--9} (\bibinfo {year} {2016}{\natexlab{b}})}\BibitemShut {NoStop}%
\bibitem [{\citenamefont {Zhang}\ \emph {et~al.}(2019{\natexlab{a}})\citenamefont {Zhang}, \citenamefont {Li}, \citenamefont {Chen}, \citenamefont {Weedbrook}, \citenamefont {Zhao} \emph {et~al.}}]{Zhang_QuantumSciTechnol_2019}%
  \BibitemOpen
  \bibfield  {author} {\bibinfo {author} {\bibfnamefont {Y.}~\bibnamefont {Zhang}}, \bibinfo {author} {\bibfnamefont {Z.}~\bibnamefont {Li}}, \bibinfo {author} {\bibfnamefont {Z.}~\bibnamefont {Chen}}, \bibinfo {author} {\bibfnamefont {C.}~\bibnamefont {Weedbrook}}, \bibinfo {author} {\bibfnamefont {Y.}~\bibnamefont {Zhao}},  \emph {et~al.},\ }\bibfield  {title} {\enquote {\bibinfo {title} {Continuous-variable qkd over 50 km commercial fiber},}\ }\href@noop {} {\bibfield  {journal} {\bibinfo  {journal} {Quantum Sci. Technol.}\ }\textbf {\bibinfo {volume} {4}},\ \bibinfo {pages} {035006} (\bibinfo {year} {2019}{\natexlab{a}})}\BibitemShut {NoStop}%
\bibitem [{\citenamefont {Zhang}\ \emph {et~al.}(2019{\natexlab{b}})\citenamefont {Zhang}, \citenamefont {Haw}, \citenamefont {Cai}, \citenamefont {Xu}, \citenamefont {Assad} \emph {et~al.}}]{Zhang_NatPhotonics_2019}%
  \BibitemOpen
  \bibfield  {author} {\bibinfo {author} {\bibfnamefont {G.}~\bibnamefont {Zhang}}, \bibinfo {author} {\bibfnamefont {J.~Y.}\ \bibnamefont {Haw}}, \bibinfo {author} {\bibfnamefont {H.}~\bibnamefont {Cai}}, \bibinfo {author} {\bibfnamefont {F.}~\bibnamefont {Xu}}, \bibinfo {author} {\bibfnamefont {S.~M.}\ \bibnamefont {Assad}},  \emph {et~al.},\ }\bibfield  {title} {\enquote {\bibinfo {title} {An integrated silicon photonic chip platform for continuous-variable quantum key distribution},}\ }\href {\doibase 10.1038/s41566-019-0504-5} {\bibfield  {journal} {\bibinfo  {journal} {Nat. Photonics}\ }\textbf {\bibinfo {volume} {13}},\ \bibinfo {pages} {839--842} (\bibinfo {year} {2019}{\natexlab{b}})}\BibitemShut {NoStop}%
\bibitem [{\citenamefont {Zhang}\ \emph {et~al.}(2020{\natexlab{a}})\citenamefont {Zhang}, \citenamefont {Chen}, \citenamefont {Pirandola}, \citenamefont {Wang}, \citenamefont {Zhou} \emph {et~al.}}]{Zhang_PhysRevLett_2020}%
  \BibitemOpen
  \bibfield  {author} {\bibinfo {author} {\bibfnamefont {Y.}~\bibnamefont {Zhang}}, \bibinfo {author} {\bibfnamefont {Z.}~\bibnamefont {Chen}}, \bibinfo {author} {\bibfnamefont {S.}~\bibnamefont {Pirandola}}, \bibinfo {author} {\bibfnamefont {X.}~\bibnamefont {Wang}}, \bibinfo {author} {\bibfnamefont {C.}~\bibnamefont {Zhou}},  \emph {et~al.},\ }\bibfield  {title} {\enquote {\bibinfo {title} {Long-distance continuous-variable quantum key distribution over 202.81 km of fiber},}\ }\href {\doibase 10.1103/PhysRevLett.125.010502} {\bibfield  {journal} {\bibinfo  {journal} {Phys. Rev. Lett.}\ }\textbf {\bibinfo {volume} {125}},\ \bibinfo {pages} {010502} (\bibinfo {year} {2020}{\natexlab{a}})}\BibitemShut {NoStop}%
\bibitem [{\citenamefont {Zhang}\ \emph {et~al.}(2020{\natexlab{b}})\citenamefont {Zhang}, \citenamefont {Chen}, \citenamefont {Chu}, \citenamefont {Zhou}, \citenamefont {Wang} \emph {et~al.}}]{zhang2020continuous}%
  \BibitemOpen
  \bibfield  {author} {\bibinfo {author} {\bibfnamefont {Y.}~\bibnamefont {Zhang}}, \bibinfo {author} {\bibfnamefont {Z.}~\bibnamefont {Chen}}, \bibinfo {author} {\bibfnamefont {B.}~\bibnamefont {Chu}}, \bibinfo {author} {\bibfnamefont {C.}~\bibnamefont {Zhou}}, \bibinfo {author} {\bibfnamefont {X.}~\bibnamefont {Wang}},  \emph {et~al.},\ }\bibfield  {title} {\enquote {\bibinfo {title} {Continuous-variable qkd network in qingdao},}\ }\href@noop {} {\bibfield  {journal} {\bibinfo  {journal} {Bulletin of the American Physical Society}\ }\textbf {\bibinfo {volume} {65}} (\bibinfo {year} {2020}{\natexlab{b}})}\BibitemShut {NoStop}%
\bibitem [{\citenamefont {Qi}\ \emph {et~al.}(2015)\citenamefont {Qi}, \citenamefont {Lougovski}, \citenamefont {Pooser}, \citenamefont {Grice},\ and\ \citenamefont {Bobrek}}]{Qi_PhysRevX_2015}%
  \BibitemOpen
  \bibfield  {author} {\bibinfo {author} {\bibfnamefont {B.}~\bibnamefont {Qi}}, \bibinfo {author} {\bibfnamefont {P.}~\bibnamefont {Lougovski}}, \bibinfo {author} {\bibfnamefont {R.}~\bibnamefont {Pooser}}, \bibinfo {author} {\bibfnamefont {W.}~\bibnamefont {Grice}}, \ and\ \bibinfo {author} {\bibfnamefont {M.}~\bibnamefont {Bobrek}},\ }\bibfield  {title} {\enquote {\bibinfo {title} {Generating the local oscillator “locally” in continuous-variable quantum key distribution based on coherent detection},}\ }\href@noop {} {\bibfield  {journal} {\bibinfo  {journal} {Phys. Rev. X}\ }\textbf {\bibinfo {volume} {5}},\ \bibinfo {pages} {041009} (\bibinfo {year} {2015})}\BibitemShut {NoStop}%
\bibitem [{\citenamefont {Soh}\ \emph {et~al.}(2015)\citenamefont {Soh}, \citenamefont {Brif}, \citenamefont {Coles}, \citenamefont {L\"utkenhaus}, \citenamefont {Camacho} \emph {et~al.}}]{Soh_PhysRevX_2015}%
  \BibitemOpen
  \bibfield  {author} {\bibinfo {author} {\bibfnamefont {D.~B.~S.}\ \bibnamefont {Soh}}, \bibinfo {author} {\bibfnamefont {C.}~\bibnamefont {Brif}}, \bibinfo {author} {\bibfnamefont {P.~J.}\ \bibnamefont {Coles}}, \bibinfo {author} {\bibfnamefont {N.}~\bibnamefont {L\"utkenhaus}}, \bibinfo {author} {\bibfnamefont {R.~M.}\ \bibnamefont {Camacho}},  \emph {et~al.},\ }\bibfield  {title} {\enquote {\bibinfo {title} {Self-referenced continuous-variable quantum key distribution protocol},}\ }\href {\doibase 10.1103/PhysRevX.5.041010} {\bibfield  {journal} {\bibinfo  {journal} {Phys. Rev. X}\ }\textbf {\bibinfo {volume} {5}},\ \bibinfo {pages} {041010} (\bibinfo {year} {2015})}\BibitemShut {NoStop}%
\bibitem [{\citenamefont {Wang}\ \emph {et~al.}(2022{\natexlab{b}})\citenamefont {Wang}, \citenamefont {Li}, \citenamefont {Pi}, \citenamefont {Pan}, \citenamefont {Shao} \emph {et~al.}}]{SubGbps}%
  \BibitemOpen
  \bibfield  {author} {\bibinfo {author} {\bibfnamefont {H.}~\bibnamefont {Wang}}, \bibinfo {author} {\bibfnamefont {Y.}~\bibnamefont {Li}}, \bibinfo {author} {\bibfnamefont {Y.}~\bibnamefont {Pi}}, \bibinfo {author} {\bibfnamefont {Y.}~\bibnamefont {Pan}}, \bibinfo {author} {\bibfnamefont {Y.}~\bibnamefont {Shao}},  \emph {et~al.},\ }\bibfield  {title} {\enquote {\bibinfo {title} {Sub-gbps key rate four-state continuous-variable quantum key distribution within metropolitan area},}\ }\href@noop {} {\bibfield  {journal} {\bibinfo  {journal} {Commun. Phys.}\ }\textbf {\bibinfo {volume} {5}},\ \bibinfo {pages} {162} (\bibinfo {year} {2022}{\natexlab{b}})}\BibitemShut {NoStop}%
\bibitem [{\citenamefont {Aguado}\ \emph {et~al.}(2019)\citenamefont {Aguado}, \citenamefont {Lopez}, \citenamefont {Lopez}, \citenamefont {Peev}, \citenamefont {Poppe} \emph {et~al.}}]{aguado2019engineering}%
  \BibitemOpen
  \bibfield  {author} {\bibinfo {author} {\bibfnamefont {A.}~\bibnamefont {Aguado}}, \bibinfo {author} {\bibfnamefont {V.}~\bibnamefont {Lopez}}, \bibinfo {author} {\bibfnamefont {D.}~\bibnamefont {Lopez}}, \bibinfo {author} {\bibfnamefont {M.}~\bibnamefont {Peev}}, \bibinfo {author} {\bibfnamefont {A.}~\bibnamefont {Poppe}},  \emph {et~al.},\ }\bibfield  {title} {\enquote {\bibinfo {title} {The engineering of software-defined quantum key distribution networks},}\ }\href@noop {} {\bibfield  {journal} {\bibinfo  {journal} {IEEE Commun. Mag.}\ }\textbf {\bibinfo {volume} {57}},\ \bibinfo {pages} {20--26} (\bibinfo {year} {2019})}\BibitemShut {NoStop}%
\bibitem [{\citenamefont {Roumestan}\ \emph {et~al.}(2022)\citenamefont {Roumestan}, \citenamefont {Ghazisaeidi}, \citenamefont {Renaudier}, \citenamefont {Vidarte}, \citenamefont {Leverrier} \emph {et~al.}}]{roumestan2022experimental}%
  \BibitemOpen
  \bibfield  {author} {\bibinfo {author} {\bibfnamefont {F.}~\bibnamefont {Roumestan}}, \bibinfo {author} {\bibfnamefont {A.}~\bibnamefont {Ghazisaeidi}}, \bibinfo {author} {\bibfnamefont {J.}~\bibnamefont {Renaudier}}, \bibinfo {author} {\bibfnamefont {L.~T.}\ \bibnamefont {Vidarte}}, \bibinfo {author} {\bibfnamefont {A.}~\bibnamefont {Leverrier}},  \emph {et~al.},\ }\href@noop {} {\enquote {\bibinfo {title} {Experimental demonstration of discrete modulation formats for continuous variable quantum key distribution},}\ } (\bibinfo {year} {2022}),\ \Eprint {http://arxiv.org/abs/2207.11702} {arXiv:2207.11702 [quant-ph]} \BibitemShut {NoStop}%
\bibitem [{\citenamefont {Hajomer}\ \emph {et~al.}(2023{\natexlab{a}})\citenamefont {Hajomer}, \citenamefont {Bruynsteen}, \citenamefont {Derkach}, \citenamefont {Jain}, \citenamefont {Bomhals} \emph {et~al.}}]{hajomer2023continuous}%
  \BibitemOpen
  \bibfield  {author} {\bibinfo {author} {\bibfnamefont {A.~A.}\ \bibnamefont {Hajomer}}, \bibinfo {author} {\bibfnamefont {C.}~\bibnamefont {Bruynsteen}}, \bibinfo {author} {\bibfnamefont {I.}~\bibnamefont {Derkach}}, \bibinfo {author} {\bibfnamefont {N.}~\bibnamefont {Jain}}, \bibinfo {author} {\bibfnamefont {A.}~\bibnamefont {Bomhals}},  \emph {et~al.},\ }\bibfield  {title} {\enquote {\bibinfo {title} {Continuous-variable quantum key distribution at 10 gbaud using an integrated photonic-electronic receiver},}\ }\href@noop {} {\bibfield  {journal} {\bibinfo  {journal} {arXiv preprint arXiv:2305.19642}\ } (\bibinfo {year} {2023}{\natexlab{a}})}\BibitemShut {NoStop}%
\bibitem [{\citenamefont {Bian}\ \emph {et~al.}(2023{\natexlab{a}})\citenamefont {Bian}, \citenamefont {Zhang}, \citenamefont {Zhou}, \citenamefont {Yu}, \citenamefont {Li} \emph {et~al.}}]{bian2023high}%
  \BibitemOpen
  \bibfield  {author} {\bibinfo {author} {\bibfnamefont {Y.}~\bibnamefont {Bian}}, \bibinfo {author} {\bibfnamefont {Y.-C.}\ \bibnamefont {Zhang}}, \bibinfo {author} {\bibfnamefont {C.}~\bibnamefont {Zhou}}, \bibinfo {author} {\bibfnamefont {S.}~\bibnamefont {Yu}}, \bibinfo {author} {\bibfnamefont {Z.}~\bibnamefont {Li}},  \emph {et~al.},\ }\href@noop {} {\enquote {\bibinfo {title} {High-rate point-to-multipoint quantum key distribution using coherent states},}\ } (\bibinfo {year} {2023}{\natexlab{a}}),\ \Eprint {http://arxiv.org/abs/2302.02391} {arXiv:2302.02391 [quant-ph]} \BibitemShut {NoStop}%
\bibitem [{\citenamefont {Simon}, \citenamefont {Mukunda},\ and\ \citenamefont {Dutta}(1994)}]{Simon_1994_PRA}%
  \BibitemOpen
  \bibfield  {author} {\bibinfo {author} {\bibfnamefont {R.}~\bibnamefont {Simon}}, \bibinfo {author} {\bibfnamefont {N.}~\bibnamefont {Mukunda}}, \ and\ \bibinfo {author} {\bibfnamefont {B.}~\bibnamefont {Dutta}},\ }\bibfield  {title} {\enquote {\bibinfo {title} {Quantum-noise matrix for multimode systems: U(n) invariance, squeezing, and normal forms},}\ }\href {\doibase 10.1103/PhysRevA.49.1567} {\bibfield  {journal} {\bibinfo  {journal} {Phys. Rev. A}\ }\textbf {\bibinfo {volume} {49}},\ \bibinfo {pages} {1567--1583} (\bibinfo {year} {1994})}\BibitemShut {NoStop}%
\bibitem [{\citenamefont {Einstein}, \citenamefont {Podolsky},\ and\ \citenamefont {Rosen}(1935)}]{Einstein_PhysRev_Can}%
  \BibitemOpen
  \bibfield  {author} {\bibinfo {author} {\bibfnamefont {A.}~\bibnamefont {Einstein}}, \bibinfo {author} {\bibfnamefont {B.}~\bibnamefont {Podolsky}}, \ and\ \bibinfo {author} {\bibfnamefont {N.}~\bibnamefont {Rosen}},\ }\bibfield  {title} {\enquote {\bibinfo {title} {Can quantum-mechanical description of physical reality be considered complete?}}\ }\href {\doibase 10.1103/PhysRev.47.777} {\bibfield  {journal} {\bibinfo  {journal} {Phys. Rev.}\ }\textbf {\bibinfo {volume} {47}},\ \bibinfo {pages} {777--780} (\bibinfo {year} {1935})}\BibitemShut {NoStop}%
\bibitem [{\citenamefont {Grosshans}\ \emph {et~al.}(2003{\natexlab{b}})\citenamefont {Grosshans}, \citenamefont {Cerf}, \citenamefont {Wenger}, \citenamefont {Tualle-Brouri},\ and\ \citenamefont {Grangier}}]{Grosshans2003VirtualEA}%
  \BibitemOpen
  \bibfield  {author} {\bibinfo {author} {\bibfnamefont {F.}~\bibnamefont {Grosshans}}, \bibinfo {author} {\bibfnamefont {N.~J.}\ \bibnamefont {Cerf}}, \bibinfo {author} {\bibfnamefont {J.}~\bibnamefont {Wenger}}, \bibinfo {author} {\bibfnamefont {R.}~\bibnamefont {Tualle-Brouri}}, \ and\ \bibinfo {author} {\bibfnamefont {P.}~\bibnamefont {Grangier}},\ }\bibfield  {title} {\enquote {\bibinfo {title} {Virtual entanglement and reconciliation protocols for quantum cryptography with continuous variables},}\ }\href {https://api.semanticscholar.org/CorpusID:6870228} {\bibfield  {journal} {\bibinfo  {journal} {Quantum Inf. Comput.}\ }\textbf {\bibinfo {volume} {3}},\ \bibinfo {pages} {535--552} (\bibinfo {year} {2003}{\natexlab{b}})}\BibitemShut {NoStop}%
\bibitem [{\citenamefont {Renner}\ and\ \citenamefont {Cirac}(2009)}]{Renner_PhysRevLett_2009}%
  \BibitemOpen
  \bibfield  {author} {\bibinfo {author} {\bibfnamefont {R.}~\bibnamefont {Renner}}\ and\ \bibinfo {author} {\bibfnamefont {J.~I.}\ \bibnamefont {Cirac}},\ }\bibfield  {title} {\enquote {\bibinfo {title} {de finetti representation theorem for infinite-dimensional quantum systems and applications to quantum cryptography},}\ }\href {\doibase 10.1103/PhysRevLett.102.110504} {\bibfield  {journal} {\bibinfo  {journal} {Phys. Rev. Lett.}\ }\textbf {\bibinfo {volume} {102}},\ \bibinfo {pages} {110504} (\bibinfo {year} {2009})}\BibitemShut {NoStop}%
\bibitem [{\citenamefont {Pirandola}(2021{\natexlab{a}})}]{pirandola2021composable}%
  \BibitemOpen
  \bibfield  {author} {\bibinfo {author} {\bibfnamefont {S.}~\bibnamefont {Pirandola}},\ }\bibfield  {title} {\enquote {\bibinfo {title} {Composable security for continuous variable quantum key distribution: Trust levels and practical key rates in wired and wireless networks},}\ }\href@noop {} {\bibfield  {journal} {\bibinfo  {journal} {Phys. Rev. Research}\ }\textbf {\bibinfo {volume} {3}},\ \bibinfo {pages} {043014} (\bibinfo {year} {2021}{\natexlab{a}})}\BibitemShut {NoStop}%
\bibitem [{\citenamefont {Pirandola}(2021{\natexlab{b}})}]{pirandola2021limits}%
  \BibitemOpen
  \bibfield  {author} {\bibinfo {author} {\bibfnamefont {S.}~\bibnamefont {Pirandola}},\ }\bibfield  {title} {\enquote {\bibinfo {title} {Limits and security of free-space quantum communications},}\ }\href@noop {} {\bibfield  {journal} {\bibinfo  {journal} {Phys. Rev. Research}\ }\textbf {\bibinfo {volume} {3}},\ \bibinfo {pages} {013279} (\bibinfo {year} {2021}{\natexlab{b}})}\BibitemShut {NoStop}%
\bibitem [{\citenamefont {Pirandola}\ and\ \citenamefont {Papanastasiou}(2023)}]{pirandola2023improved}%
  \BibitemOpen
  \bibfield  {author} {\bibinfo {author} {\bibfnamefont {S.}~\bibnamefont {Pirandola}}\ and\ \bibinfo {author} {\bibfnamefont {P.}~\bibnamefont {Papanastasiou}},\ }\bibfield  {title} {\enquote {\bibinfo {title} {Improved composable key rates for cv-qkd},}\ }\href@noop {} {\bibfield  {journal} {\bibinfo  {journal} {arXiv preprint arXiv:2301.10270}\ } (\bibinfo {year} {2023})}\BibitemShut {NoStop}%
\bibitem [{\citenamefont {Leverrier}\ \emph {et~al.}(2013)\citenamefont {Leverrier}, \citenamefont {Garc\'{\i}a-Patr\'on}, \citenamefont {Renner},\ and\ \citenamefont {Cerf}}]{Leverrier_PhysRevLett_2013}%
  \BibitemOpen
  \bibfield  {author} {\bibinfo {author} {\bibfnamefont {A.}~\bibnamefont {Leverrier}}, \bibinfo {author} {\bibfnamefont {R.}~\bibnamefont {Garc\'{\i}a-Patr\'on}}, \bibinfo {author} {\bibfnamefont {R.}~\bibnamefont {Renner}}, \ and\ \bibinfo {author} {\bibfnamefont {N.~J.}\ \bibnamefont {Cerf}},\ }\bibfield  {title} {\enquote {\bibinfo {title} {Security of continuous-variable quantum key distribution against general attacks},}\ }\href {\doibase 10.1103/PhysRevLett.110.030502} {\bibfield  {journal} {\bibinfo  {journal} {Phys. Rev. Lett.}\ }\textbf {\bibinfo {volume} {110}},\ \bibinfo {pages} {030502} (\bibinfo {year} {2013})}\BibitemShut {NoStop}%
\bibitem [{\citenamefont {Furrer}(2014)}]{Furrer_PhysRevA_2014}%
  \BibitemOpen
  \bibfield  {author} {\bibinfo {author} {\bibfnamefont {F.}~\bibnamefont {Furrer}},\ }\bibfield  {title} {\enquote {\bibinfo {title} {Reverse-reconciliation continuous-variable quantum key distribution based on the uncertainty principle},}\ }\href {\doibase 10.1103/PhysRevA.90.042325} {\bibfield  {journal} {\bibinfo  {journal} {Phys. Rev. A}\ }\textbf {\bibinfo {volume} {90}},\ \bibinfo {pages} {042325} (\bibinfo {year} {2014})}\BibitemShut {NoStop}%
\bibitem [{\citenamefont {Furrer}\ \emph {et~al.}(2012)\citenamefont {Furrer}, \citenamefont {Franz}, \citenamefont {Berta}, \citenamefont {Leverrier}, \citenamefont {Scholz} \emph {et~al.}}]{Furrer_PhysRevLett_2012}%
  \BibitemOpen
  \bibfield  {author} {\bibinfo {author} {\bibfnamefont {F.}~\bibnamefont {Furrer}}, \bibinfo {author} {\bibfnamefont {T.}~\bibnamefont {Franz}}, \bibinfo {author} {\bibfnamefont {M.}~\bibnamefont {Berta}}, \bibinfo {author} {\bibfnamefont {A.}~\bibnamefont {Leverrier}}, \bibinfo {author} {\bibfnamefont {V.~B.}\ \bibnamefont {Scholz}},  \emph {et~al.},\ }\bibfield  {title} {\enquote {\bibinfo {title} {Continuous variable quantum key distribution: finite-key analysis of composable security against coherent attacks},}\ }\href {\doibase 10.1103/PhysRevLett.109.100502} {\bibfield  {journal} {\bibinfo  {journal} {Phys. Rev. Lett.}\ }\textbf {\bibinfo {volume} {109}},\ \bibinfo {pages} {100502} (\bibinfo {year} {2012})}\BibitemShut {NoStop}%
\bibitem [{\citenamefont {Leverrier}\ and\ \citenamefont {Grangier}(2009)}]{Leverrier_PhysRevLett_2009}%
  \BibitemOpen
  \bibfield  {author} {\bibinfo {author} {\bibfnamefont {A.}~\bibnamefont {Leverrier}}\ and\ \bibinfo {author} {\bibfnamefont {P.}~\bibnamefont {Grangier}},\ }\bibfield  {title} {\enquote {\bibinfo {title} {Unconditional security proof of long-distance continuous-variable quantum key distribution with discrete modulation},}\ }\href {\doibase 10.1103/PhysRevLett.102.180504} {\bibfield  {journal} {\bibinfo  {journal} {Phys. Rev. Lett.}\ }\textbf {\bibinfo {volume} {102}},\ \bibinfo {pages} {180504} (\bibinfo {year} {2009})}\BibitemShut {NoStop}%
\bibitem [{\citenamefont {Usenko}\ and\ \citenamefont {Grosshans}(2015)}]{Usenko_PhysRevA_2015}%
  \BibitemOpen
  \bibfield  {author} {\bibinfo {author} {\bibfnamefont {V.~C.}\ \bibnamefont {Usenko}}\ and\ \bibinfo {author} {\bibfnamefont {F.}~\bibnamefont {Grosshans}},\ }\bibfield  {title} {\enquote {\bibinfo {title} {Unidimensional continuous-variable quantum key distribution},}\ }\href {\doibase 10.1103/PhysRevA.92.062337} {\bibfield  {journal} {\bibinfo  {journal} {Phys. Rev. A}\ }\textbf {\bibinfo {volume} {92}},\ \bibinfo {pages} {062337} (\bibinfo {year} {2015})}\BibitemShut {NoStop}%
\bibitem [{\citenamefont {Liao}\ \emph {et~al.}(2018{\natexlab{b}})\citenamefont {Liao}, \citenamefont {Guo}, \citenamefont {Xie}, \citenamefont {Huang}, \citenamefont {Huang} \emph {et~al.}}]{Liao_QuantumInfProc_2018}%
  \BibitemOpen
  \bibfield  {author} {\bibinfo {author} {\bibfnamefont {Q.}~\bibnamefont {Liao}}, \bibinfo {author} {\bibfnamefont {Y.}~\bibnamefont {Guo}}, \bibinfo {author} {\bibfnamefont {C.}~\bibnamefont {Xie}}, \bibinfo {author} {\bibfnamefont {D.}~\bibnamefont {Huang}}, \bibinfo {author} {\bibfnamefont {P.}~\bibnamefont {Huang}},  \emph {et~al.},\ }\bibfield  {title} {\enquote {\bibinfo {title} {Composable security of unidimensional continuous-variable quantum key distribution},}\ }\href@noop {} {\bibfield  {journal} {\bibinfo  {journal} {Quantum Inf. Process.}\ }\textbf {\bibinfo {volume} {17}},\ \bibinfo {pages} {113} (\bibinfo {year} {2018}{\natexlab{b}})}\BibitemShut {NoStop}%
\bibitem [{\citenamefont {Garc\'{\i}a-Patr\'on}\ and\ \citenamefont {Cerf}(2009)}]{Garcia-Patron_PhysRevLett_2009}%
  \BibitemOpen
  \bibfield  {author} {\bibinfo {author} {\bibfnamefont {R.}~\bibnamefont {Garc\'{\i}a-Patr\'on}}\ and\ \bibinfo {author} {\bibfnamefont {N.~J.}\ \bibnamefont {Cerf}},\ }\bibfield  {title} {\enquote {\bibinfo {title} {Continuous-variable quantum key distribution protocols over noisy channels},}\ }\href {\doibase 10.1103/PhysRevLett.102.130501} {\bibfield  {journal} {\bibinfo  {journal} {Phys. Rev. Lett.}\ }\textbf {\bibinfo {volume} {102}},\ \bibinfo {pages} {130501} (\bibinfo {year} {2009})}\BibitemShut {NoStop}%
\bibitem [{\citenamefont {Vidiella-Barranco}\ and\ \citenamefont {Borelli}(2006)}]{vidiella2006continuous}%
  \BibitemOpen
  \bibfield  {author} {\bibinfo {author} {\bibfnamefont {A.}~\bibnamefont {Vidiella-Barranco}}\ and\ \bibinfo {author} {\bibfnamefont {L.}~\bibnamefont {Borelli}},\ }\bibfield  {title} {\enquote {\bibinfo {title} {Continuous variable quantum key distribution using polarized coherent states},}\ }\href@noop {} {\bibfield  {journal} {\bibinfo  {journal} {Int. J. Mod. Phys. B}\ }\textbf {\bibinfo {volume} {20}},\ \bibinfo {pages} {1287--1296} (\bibinfo {year} {2006})}\BibitemShut {NoStop}%
\bibitem [{\citenamefont {Filip}(2008)}]{Filip_PhysRevA_2008}%
  \BibitemOpen
  \bibfield  {author} {\bibinfo {author} {\bibfnamefont {R.}~\bibnamefont {Filip}},\ }\bibfield  {title} {\enquote {\bibinfo {title} {Continuous-variable quantum key distribution with noisy coherent states},}\ }\href {\doibase 10.1103/PhysRevA.77.022310} {\bibfield  {journal} {\bibinfo  {journal} {Phys. Rev. A}\ }\textbf {\bibinfo {volume} {77}},\ \bibinfo {pages} {022310} (\bibinfo {year} {2008})}\BibitemShut {NoStop}%
\bibitem [{\citenamefont {Papanastasiou}, \citenamefont {Ottaviani},\ and\ \citenamefont {Pirandola}(2018)}]{Papanastasiou_PhysRevA_2018}%
  \BibitemOpen
  \bibfield  {author} {\bibinfo {author} {\bibfnamefont {P.}~\bibnamefont {Papanastasiou}}, \bibinfo {author} {\bibfnamefont {C.}~\bibnamefont {Ottaviani}}, \ and\ \bibinfo {author} {\bibfnamefont {S.}~\bibnamefont {Pirandola}},\ }\bibfield  {title} {\enquote {\bibinfo {title} {Gaussian one-way thermal quantum cryptography with finite-size effects},}\ }\href {\doibase 10.1103/PhysRevA.98.032314} {\bibfield  {journal} {\bibinfo  {journal} {Phys. Rev. A}\ }\textbf {\bibinfo {volume} {98}},\ \bibinfo {pages} {032314} (\bibinfo {year} {2018})}\BibitemShut {NoStop}%
\bibitem [{\citenamefont {Fiur\'a\ifmmode~\check{s}\else \v{s}\fi{}ek}\ and\ \citenamefont {Cerf}(2012)}]{Fiurasek_PhysRevA_2012}%
  \BibitemOpen
  \bibfield  {author} {\bibinfo {author} {\bibfnamefont {J.}~\bibnamefont {Fiur\'a\ifmmode~\check{s}\else \v{s}\fi{}ek}}\ and\ \bibinfo {author} {\bibfnamefont {N.~J.}\ \bibnamefont {Cerf}},\ }\bibfield  {title} {\enquote {\bibinfo {title} {Gaussian postselection and virtual noiseless amplification in continuous-variable quantum key distribution},}\ }\href {\doibase 10.1103/PhysRevA.86.060302} {\bibfield  {journal} {\bibinfo  {journal} {Phys. Rev. A}\ }\textbf {\bibinfo {volume} {86}},\ \bibinfo {pages} {060302} (\bibinfo {year} {2012})}\BibitemShut {NoStop}%
\bibitem [{\citenamefont {Walk}\ \emph {et~al.}(2013)\citenamefont {Walk}, \citenamefont {Ralph}, \citenamefont {Symul},\ and\ \citenamefont {Lam}}]{Walk_PhysRevA_2013}%
  \BibitemOpen
  \bibfield  {author} {\bibinfo {author} {\bibfnamefont {N.}~\bibnamefont {Walk}}, \bibinfo {author} {\bibfnamefont {T.~C.}\ \bibnamefont {Ralph}}, \bibinfo {author} {\bibfnamefont {T.}~\bibnamefont {Symul}}, \ and\ \bibinfo {author} {\bibfnamefont {P.~K.}\ \bibnamefont {Lam}},\ }\bibfield  {title} {\enquote {\bibinfo {title} {Security of continuous-variable quantum cryptography with gaussian postselection},}\ }\href {\doibase 10.1103/PhysRevA.87.020303} {\bibfield  {journal} {\bibinfo  {journal} {Phys. Rev. A}\ }\textbf {\bibinfo {volume} {87}},\ \bibinfo {pages} {020303} (\bibinfo {year} {2013})}\BibitemShut {NoStop}%
\bibitem [{\citenamefont {Hosseinidehaj}\ \emph {et~al.}(2020)\citenamefont {Hosseinidehaj}, \citenamefont {Lance}, \citenamefont {Symul}, \citenamefont {Walk},\ and\ \citenamefont {Ralph}}]{Hosseinidehaj_PhysRevA_2020}%
  \BibitemOpen
  \bibfield  {author} {\bibinfo {author} {\bibfnamefont {N.}~\bibnamefont {Hosseinidehaj}}, \bibinfo {author} {\bibfnamefont {A.~M.}\ \bibnamefont {Lance}}, \bibinfo {author} {\bibfnamefont {T.}~\bibnamefont {Symul}}, \bibinfo {author} {\bibfnamefont {N.}~\bibnamefont {Walk}}, \ and\ \bibinfo {author} {\bibfnamefont {T.~C.}\ \bibnamefont {Ralph}},\ }\bibfield  {title} {\enquote {\bibinfo {title} {Finite-size effects in continuous-variable quantum key distribution with gaussian postselection},}\ }\href {\doibase 10.1103/PhysRevA.101.052335} {\bibfield  {journal} {\bibinfo  {journal} {Phys. Rev. A}\ }\textbf {\bibinfo {volume} {101}},\ \bibinfo {pages} {052335} (\bibinfo {year} {2020})}\BibitemShut {NoStop}%
\bibitem [{\citenamefont {Li}\ \emph {et~al.}(2016{\natexlab{b}})\citenamefont {Li}, \citenamefont {Zhang}, \citenamefont {Wang}, \citenamefont {Xu}, \citenamefont {Peng} \emph {et~al.}}]{Li_PhysRevA_2016}%
  \BibitemOpen
  \bibfield  {author} {\bibinfo {author} {\bibfnamefont {Z.}~\bibnamefont {Li}}, \bibinfo {author} {\bibfnamefont {Y.}~\bibnamefont {Zhang}}, \bibinfo {author} {\bibfnamefont {X.}~\bibnamefont {Wang}}, \bibinfo {author} {\bibfnamefont {B.}~\bibnamefont {Xu}}, \bibinfo {author} {\bibfnamefont {X.}~\bibnamefont {Peng}},  \emph {et~al.},\ }\bibfield  {title} {\enquote {\bibinfo {title} {Non-gaussian postselection and virtual photon subtraction in continuous-variable quantum key distribution},}\ }\href {\doibase 10.1103/PhysRevA.93.012310} {\bibfield  {journal} {\bibinfo  {journal} {Phys. Rev. A}\ }\textbf {\bibinfo {volume} {93}},\ \bibinfo {pages} {012310} (\bibinfo {year} {2016}{\natexlab{b}})}\BibitemShut {NoStop}%
\bibitem [{\citenamefont {Madsen}\ \emph {et~al.}(2012{\natexlab{a}})\citenamefont {Madsen}, \citenamefont {Usenko}, \citenamefont {Lassen}, \citenamefont {Filip},\ and\ \citenamefont {Andersen}}]{Madsen_NatCommun_2012}%
  \BibitemOpen
  \bibfield  {author} {\bibinfo {author} {\bibfnamefont {L.~S.}\ \bibnamefont {Madsen}}, \bibinfo {author} {\bibfnamefont {V.~C.}\ \bibnamefont {Usenko}}, \bibinfo {author} {\bibfnamefont {M.}~\bibnamefont {Lassen}}, \bibinfo {author} {\bibfnamefont {R.}~\bibnamefont {Filip}}, \ and\ \bibinfo {author} {\bibfnamefont {U.~L.}\ \bibnamefont {Andersen}},\ }\bibfield  {title} {\enquote {\bibinfo {title} {Continuous variable quantum key distribution with modulated entangled states},}\ }\href@noop {} {\bibfield  {journal} {\bibinfo  {journal} {Nat. Commun.}\ }\textbf {\bibinfo {volume} {3}},\ \bibinfo {pages} {1--6} (\bibinfo {year} {2012}{\natexlab{a}})}\BibitemShut {NoStop}%
\bibitem [{\citenamefont {Pirandola}\ \emph {et~al.}(2008)\citenamefont {Pirandola}, \citenamefont {Mancini}, \citenamefont {Lloyd},\ and\ \citenamefont {Braunstein}}]{Pirandola_NatPhys_2008}%
  \BibitemOpen
  \bibfield  {author} {\bibinfo {author} {\bibfnamefont {S.}~\bibnamefont {Pirandola}}, \bibinfo {author} {\bibfnamefont {S.}~\bibnamefont {Mancini}}, \bibinfo {author} {\bibfnamefont {S.}~\bibnamefont {Lloyd}}, \ and\ \bibinfo {author} {\bibfnamefont {S.~L.}\ \bibnamefont {Braunstein}},\ }\bibfield  {title} {\enquote {\bibinfo {title} {Continuous-variable quantum cryptography using two-way quantum communication},}\ }\href@noop {} {\bibfield  {journal} {\bibinfo  {journal} {Nat. Phys.}\ }\textbf {\bibinfo {volume} {4}},\ \bibinfo {pages} {726--730} (\bibinfo {year} {2008})}\BibitemShut {NoStop}%
\bibitem [{\citenamefont {Ghorai}, \citenamefont {Diamanti},\ and\ \citenamefont {Leverrier}(2019)}]{Ghorai_PhysRevA_2019}%
  \BibitemOpen
  \bibfield  {author} {\bibinfo {author} {\bibfnamefont {S.}~\bibnamefont {Ghorai}}, \bibinfo {author} {\bibfnamefont {E.}~\bibnamefont {Diamanti}}, \ and\ \bibinfo {author} {\bibfnamefont {A.}~\bibnamefont {Leverrier}},\ }\bibfield  {title} {\enquote {\bibinfo {title} {Composable security of two-way continuous-variable quantum key distribution without active symmetrization},}\ }\href {\doibase 10.1103/PhysRevA.99.012311} {\bibfield  {journal} {\bibinfo  {journal} {Phys. Rev. A}\ }\textbf {\bibinfo {volume} {99}},\ \bibinfo {pages} {012311} (\bibinfo {year} {2019})}\BibitemShut {NoStop}%
\bibitem [{\citenamefont {Sun}\ \emph {et~al.}(2012)\citenamefont {Sun}, \citenamefont {Peng}, \citenamefont {Shen},\ and\ \citenamefont {Guo}}]{Sun_IntJQuantumInform_2012}%
  \BibitemOpen
  \bibfield  {author} {\bibinfo {author} {\bibfnamefont {M.}~\bibnamefont {Sun}}, \bibinfo {author} {\bibfnamefont {X.}~\bibnamefont {Peng}}, \bibinfo {author} {\bibfnamefont {Y.}~\bibnamefont {Shen}}, \ and\ \bibinfo {author} {\bibfnamefont {H.}~\bibnamefont {Guo}},\ }\bibfield  {title} {\enquote {\bibinfo {title} {Security of a new two-way continuous-variable quantum key distribution protocol},}\ }\href@noop {} {\bibfield  {journal} {\bibinfo  {journal} {Int. J. Quantum Inf.}\ }\textbf {\bibinfo {volume} {10}},\ \bibinfo {pages} {1250059} (\bibinfo {year} {2012})}\BibitemShut {NoStop}%
\bibitem [{\citenamefont {Zhao}\ \emph {et~al.}(2017)\citenamefont {Zhao}, \citenamefont {Zhang}, \citenamefont {Li}, \citenamefont {Yu},\ and\ \citenamefont {Guo}}]{Zhao_QuantumInfProc_2017}%
  \BibitemOpen
  \bibfield  {author} {\bibinfo {author} {\bibfnamefont {Y.}~\bibnamefont {Zhao}}, \bibinfo {author} {\bibfnamefont {Y.}~\bibnamefont {Zhang}}, \bibinfo {author} {\bibfnamefont {Z.}~\bibnamefont {Li}}, \bibinfo {author} {\bibfnamefont {S.}~\bibnamefont {Yu}}, \ and\ \bibinfo {author} {\bibfnamefont {H.}~\bibnamefont {Guo}},\ }\bibfield  {title} {\enquote {\bibinfo {title} {Improvement of two-way continuous-variable quantum key distribution with virtual photon subtraction},}\ }\href@noop {} {\bibfield  {journal} {\bibinfo  {journal} {Quantum Inf. Process.}\ }\textbf {\bibinfo {volume} {16}},\ \bibinfo {pages} {184} (\bibinfo {year} {2017})}\BibitemShut {NoStop}%
\bibitem [{\citenamefont {Li}\ \emph {et~al.}(2016{\natexlab{c}})\citenamefont {Li}, \citenamefont {Miao}, \citenamefont {Gong}, \citenamefont {Guo},\ and\ \citenamefont {He}}]{li2016performance}%
  \BibitemOpen
  \bibfield  {author} {\bibinfo {author} {\bibfnamefont {C.}~\bibnamefont {Li}}, \bibinfo {author} {\bibfnamefont {R.}~\bibnamefont {Miao}}, \bibinfo {author} {\bibfnamefont {X.}~\bibnamefont {Gong}}, \bibinfo {author} {\bibfnamefont {Y.}~\bibnamefont {Guo}}, \ and\ \bibinfo {author} {\bibfnamefont {G.}~\bibnamefont {He}},\ }\bibfield  {title} {\enquote {\bibinfo {title} {Performance improvement of two-way quantum key distribution by using a heralded noiseless amplifier},}\ }\href@noop {} {\bibfield  {journal} {\bibinfo  {journal} {Int. J. Theor. Phys.}\ }\textbf {\bibinfo {volume} {55}},\ \bibinfo {pages} {2199--2211} (\bibinfo {year} {2016}{\natexlab{c}})}\BibitemShut {NoStop}%
\bibitem [{\citenamefont {Bian}, \citenamefont {Huang},\ and\ \citenamefont {Zhang}(2021)}]{bian2021unidimensional}%
  \BibitemOpen
  \bibfield  {author} {\bibinfo {author} {\bibfnamefont {Y.}~\bibnamefont {Bian}}, \bibinfo {author} {\bibfnamefont {L.}~\bibnamefont {Huang}}, \ and\ \bibinfo {author} {\bibfnamefont {Y.}~\bibnamefont {Zhang}},\ }\bibfield  {title} {\enquote {\bibinfo {title} {Unidimensional two-way continuous-variable quantum key distribution using coherent states},}\ }\href@noop {} {\bibfield  {journal} {\bibinfo  {journal} {Entropy}\ }\textbf {\bibinfo {volume} {23}},\ \bibinfo {pages} {294} (\bibinfo {year} {2021})}\BibitemShut {NoStop}%
\bibitem [{\citenamefont {Li}\ \emph {et~al.}(2014{\natexlab{a}})\citenamefont {Li}, \citenamefont {Zhang}, \citenamefont {Xu}, \citenamefont {Peng},\ and\ \citenamefont {Guo}}]{Li_PhysRevA_2014}%
  \BibitemOpen
  \bibfield  {author} {\bibinfo {author} {\bibfnamefont {Z.}~\bibnamefont {Li}}, \bibinfo {author} {\bibfnamefont {Y.-C.}\ \bibnamefont {Zhang}}, \bibinfo {author} {\bibfnamefont {F.}~\bibnamefont {Xu}}, \bibinfo {author} {\bibfnamefont {X.}~\bibnamefont {Peng}}, \ and\ \bibinfo {author} {\bibfnamefont {H.}~\bibnamefont {Guo}},\ }\bibfield  {title} {\enquote {\bibinfo {title} {Continuous-variable measurement-device-independent quantum key distribution},}\ }\href {\doibase 10.1103/PhysRevA.89.052301} {\bibfield  {journal} {\bibinfo  {journal} {Phys. Rev. A}\ }\textbf {\bibinfo {volume} {89}},\ \bibinfo {pages} {052301} (\bibinfo {year} {2014}{\natexlab{a}})}\BibitemShut {NoStop}%
\bibitem [{\citenamefont {Pirandola}\ \emph {et~al.}(2015)\citenamefont {Pirandola}, \citenamefont {Ottaviani}, \citenamefont {Spedalieri}, \citenamefont {Weedbrook}, \citenamefont {Braunstein} \emph {et~al.}}]{Pirandola_NatPhoton_2015}%
  \BibitemOpen
  \bibfield  {author} {\bibinfo {author} {\bibfnamefont {S.}~\bibnamefont {Pirandola}}, \bibinfo {author} {\bibfnamefont {C.}~\bibnamefont {Ottaviani}}, \bibinfo {author} {\bibfnamefont {G.}~\bibnamefont {Spedalieri}}, \bibinfo {author} {\bibfnamefont {C.}~\bibnamefont {Weedbrook}}, \bibinfo {author} {\bibfnamefont {S.~L.}\ \bibnamefont {Braunstein}},  \emph {et~al.},\ }\bibfield  {title} {\enquote {\bibinfo {title} {High-rate measurement-device-independent quantum cryptography},}\ }\href@noop {} {\bibfield  {journal} {\bibinfo  {journal} {Nat. Photonics}\ }\textbf {\bibinfo {volume} {9}},\ \bibinfo {pages} {397--402} (\bibinfo {year} {2015})}\BibitemShut {NoStop}%
\bibitem [{\citenamefont {Zhang}\ \emph {et~al.}(2017)\citenamefont {Zhang}, \citenamefont {Zhang}, \citenamefont {Zhao}, \citenamefont {Wang}, \citenamefont {Yu} \emph {et~al.}}]{Zhang_PhysRevA_2017}%
  \BibitemOpen
  \bibfield  {author} {\bibinfo {author} {\bibfnamefont {X.}~\bibnamefont {Zhang}}, \bibinfo {author} {\bibfnamefont {Y.}~\bibnamefont {Zhang}}, \bibinfo {author} {\bibfnamefont {Y.}~\bibnamefont {Zhao}}, \bibinfo {author} {\bibfnamefont {X.}~\bibnamefont {Wang}}, \bibinfo {author} {\bibfnamefont {S.}~\bibnamefont {Yu}},  \emph {et~al.},\ }\bibfield  {title} {\enquote {\bibinfo {title} {Finite-size analysis of continuous-variable measurement-device-independent quantum key distribution},}\ }\href {\doibase 10.1103/PhysRevA.96.042334} {\bibfield  {journal} {\bibinfo  {journal} {Phys. Rev. A}\ }\textbf {\bibinfo {volume} {96}},\ \bibinfo {pages} {042334} (\bibinfo {year} {2017})}\BibitemShut {NoStop}%
\bibitem [{\citenamefont {Lupo}\ \emph {et~al.}(2018)\citenamefont {Lupo}, \citenamefont {Ottaviani}, \citenamefont {Papanastasiou},\ and\ \citenamefont {Pirandola}}]{Lupo_PhysRevA_2018}%
  \BibitemOpen
  \bibfield  {author} {\bibinfo {author} {\bibfnamefont {C.}~\bibnamefont {Lupo}}, \bibinfo {author} {\bibfnamefont {C.}~\bibnamefont {Ottaviani}}, \bibinfo {author} {\bibfnamefont {P.}~\bibnamefont {Papanastasiou}}, \ and\ \bibinfo {author} {\bibfnamefont {S.}~\bibnamefont {Pirandola}},\ }\bibfield  {title} {\enquote {\bibinfo {title} {Continuous-variable measurement-device-independent quantum key distribution: Composable security against coherent attacks},}\ }\href {\doibase 10.1103/PhysRevA.97.052327} {\bibfield  {journal} {\bibinfo  {journal} {Phys. Rev. A}\ }\textbf {\bibinfo {volume} {97}},\ \bibinfo {pages} {052327} (\bibinfo {year} {2018})}\BibitemShut {NoStop}%
\bibitem [{\citenamefont {Zhang}\ \emph {et~al.}(2014)\citenamefont {Zhang}, \citenamefont {Li}, \citenamefont {Yu}, \citenamefont {Gu}, \citenamefont {Peng} \emph {et~al.}}]{Zhang_PhysRevA_2014}%
  \BibitemOpen
  \bibfield  {author} {\bibinfo {author} {\bibfnamefont {Y.-C.}\ \bibnamefont {Zhang}}, \bibinfo {author} {\bibfnamefont {Z.}~\bibnamefont {Li}}, \bibinfo {author} {\bibfnamefont {S.}~\bibnamefont {Yu}}, \bibinfo {author} {\bibfnamefont {W.}~\bibnamefont {Gu}}, \bibinfo {author} {\bibfnamefont {X.}~\bibnamefont {Peng}},  \emph {et~al.},\ }\bibfield  {title} {\enquote {\bibinfo {title} {Continuous-variable measurement-device-independent quantum key distribution using squeezed states},}\ }\href {\doibase 10.1103/PhysRevA.90.052325} {\bibfield  {journal} {\bibinfo  {journal} {Phys. Rev. A}\ }\textbf {\bibinfo {volume} {90}},\ \bibinfo {pages} {052325} (\bibinfo {year} {2014})}\BibitemShut {NoStop}%
\bibitem [{\citenamefont {Chen}\ \emph {et~al.}(2018)\citenamefont {Chen}, \citenamefont {Zhang}, \citenamefont {Wang}, \citenamefont {Li},\ and\ \citenamefont {Guo}}]{Chen_PhysRevA_2018}%
  \BibitemOpen
  \bibfield  {author} {\bibinfo {author} {\bibfnamefont {Z.}~\bibnamefont {Chen}}, \bibinfo {author} {\bibfnamefont {Y.}~\bibnamefont {Zhang}}, \bibinfo {author} {\bibfnamefont {G.}~\bibnamefont {Wang}}, \bibinfo {author} {\bibfnamefont {Z.}~\bibnamefont {Li}}, \ and\ \bibinfo {author} {\bibfnamefont {H.}~\bibnamefont {Guo}},\ }\bibfield  {title} {\enquote {\bibinfo {title} {Composable security analysis of continuous-variable measurement-device-independent quantum key distribution with squeezed states for coherent attacks},}\ }\href {\doibase 10.1103/PhysRevA.98.012314} {\bibfield  {journal} {\bibinfo  {journal} {Phys. Rev. A}\ }\textbf {\bibinfo {volume} {98}},\ \bibinfo {pages} {012314} (\bibinfo {year} {2018})}\BibitemShut {NoStop}%
\bibitem [{\citenamefont {Huang}\ \emph {et~al.}(2019)\citenamefont {Huang}, \citenamefont {Zhang}, \citenamefont {Chen},\ and\ \citenamefont {Yu}}]{Huang_Entropy_2019}%
  \BibitemOpen
  \bibfield  {author} {\bibinfo {author} {\bibfnamefont {L.}~\bibnamefont {Huang}}, \bibinfo {author} {\bibfnamefont {Y.}~\bibnamefont {Zhang}}, \bibinfo {author} {\bibfnamefont {Z.}~\bibnamefont {Chen}}, \ and\ \bibinfo {author} {\bibfnamefont {S.}~\bibnamefont {Yu}},\ }\bibfield  {title} {\enquote {\bibinfo {title} {Unidimensional continuous-variable quantum key distribution with untrusted detection under realistic conditions},}\ }\href@noop {} {\bibfield  {journal} {\bibinfo  {journal} {Entropy}\ }\textbf {\bibinfo {volume} {21}},\ \bibinfo {pages} {1100} (\bibinfo {year} {2019})}\BibitemShut {NoStop}%
\bibitem [{\citenamefont {Bai}\ \emph {et~al.}(2020)\citenamefont {Bai}, \citenamefont {Huang}, \citenamefont {Zhu}, \citenamefont {Ma}, \citenamefont {Xiao} \emph {et~al.}}]{Bai_QuantumInfProc_2020}%
  \BibitemOpen
  \bibfield  {author} {\bibinfo {author} {\bibfnamefont {D.}~\bibnamefont {Bai}}, \bibinfo {author} {\bibfnamefont {P.}~\bibnamefont {Huang}}, \bibinfo {author} {\bibfnamefont {Y.}~\bibnamefont {Zhu}}, \bibinfo {author} {\bibfnamefont {H.}~\bibnamefont {Ma}}, \bibinfo {author} {\bibfnamefont {T.}~\bibnamefont {Xiao}},  \emph {et~al.},\ }\bibfield  {title} {\enquote {\bibinfo {title} {Unidimensional continuous-variable measurement-device-independent quantum key distribution},}\ }\href@noop {} {\bibfield  {journal} {\bibinfo  {journal} {Quantum Inf. Process.}\ }\textbf {\bibinfo {volume} {19}},\ \bibinfo {pages} {53} (\bibinfo {year} {2020})}\BibitemShut {NoStop}%
\bibitem [{\citenamefont {Ma}\ \emph {et~al.}(2019)\citenamefont {Ma}, \citenamefont {Huang}, \citenamefont {Bai}, \citenamefont {Wang}, \citenamefont {Wang} \emph {et~al.}}]{Ma_PhysRevA_2019}%
  \BibitemOpen
  \bibfield  {author} {\bibinfo {author} {\bibfnamefont {H.-X.}\ \bibnamefont {Ma}}, \bibinfo {author} {\bibfnamefont {P.}~\bibnamefont {Huang}}, \bibinfo {author} {\bibfnamefont {D.-Y.}\ \bibnamefont {Bai}}, \bibinfo {author} {\bibfnamefont {T.}~\bibnamefont {Wang}}, \bibinfo {author} {\bibfnamefont {S.-Y.}\ \bibnamefont {Wang}},  \emph {et~al.},\ }\bibfield  {title} {\enquote {\bibinfo {title} {Long-distance continuous-variable measurement-device-independent quantum key distribution with discrete modulation},}\ }\href {\doibase 10.1103/PhysRevA.99.022322} {\bibfield  {journal} {\bibinfo  {journal} {Phys. Rev. A}\ }\textbf {\bibinfo {volume} {99}},\ \bibinfo {pages} {022322} (\bibinfo {year} {2019})}\BibitemShut {NoStop}%
\bibitem [{\citenamefont {Zhao}\ \emph {et~al.}(2018{\natexlab{a}})\citenamefont {Zhao}, \citenamefont {Zhang}, \citenamefont {Xu}, \citenamefont {Yu},\ and\ \citenamefont {Guo}}]{Zhao_PhysRevA_2018}%
  \BibitemOpen
  \bibfield  {author} {\bibinfo {author} {\bibfnamefont {Y.}~\bibnamefont {Zhao}}, \bibinfo {author} {\bibfnamefont {Y.}~\bibnamefont {Zhang}}, \bibinfo {author} {\bibfnamefont {B.}~\bibnamefont {Xu}}, \bibinfo {author} {\bibfnamefont {S.}~\bibnamefont {Yu}}, \ and\ \bibinfo {author} {\bibfnamefont {H.}~\bibnamefont {Guo}},\ }\bibfield  {title} {\enquote {\bibinfo {title} {Continuous-variable measurement-device-independent quantum key distribution with virtual photon subtraction},}\ }\href {\doibase 10.1103/PhysRevA.97.042328} {\bibfield  {journal} {\bibinfo  {journal} {Phys. Rev. A}\ }\textbf {\bibinfo {volume} {97}},\ \bibinfo {pages} {042328} (\bibinfo {year} {2018}{\natexlab{a}})}\BibitemShut {NoStop}%
\bibitem [{\citenamefont {Leverrier}\ and\ \citenamefont {Grangier}(2011)}]{Leverrier_PhysRevA_2011}%
  \BibitemOpen
  \bibfield  {author} {\bibinfo {author} {\bibfnamefont {A.}~\bibnamefont {Leverrier}}\ and\ \bibinfo {author} {\bibfnamefont {P.}~\bibnamefont {Grangier}},\ }\bibfield  {title} {\enquote {\bibinfo {title} {Continuous-variable quantum-key-distribution protocols with a non-gaussian modulation},}\ }\href {\doibase 10.1103/PhysRevA.83.042312} {\bibfield  {journal} {\bibinfo  {journal} {Phys. Rev. A}\ }\textbf {\bibinfo {volume} {83}},\ \bibinfo {pages} {042312} (\bibinfo {year} {2011})}\BibitemShut {NoStop}%
\bibitem [{\citenamefont {Su}\ \emph {et~al.}(2023)\citenamefont {Su}, \citenamefont {Wang}, \citenamefont {Cai}, \citenamefont {Guo}, \citenamefont {Wang},\ and\ \citenamefont {Li}}]{su2023experimental}%
  \BibitemOpen
  \bibfield  {author} {\bibinfo {author} {\bibfnamefont {Z.}~\bibnamefont {Su}}, \bibinfo {author} {\bibfnamefont {J.}~\bibnamefont {Wang}}, \bibinfo {author} {\bibfnamefont {D.}~\bibnamefont {Cai}}, \bibinfo {author} {\bibfnamefont {X.}~\bibnamefont {Guo}}, \bibinfo {author} {\bibfnamefont {D.}~\bibnamefont {Wang}}, \ and\ \bibinfo {author} {\bibfnamefont {Z.}~\bibnamefont {Li}},\ }\bibfield  {title} {\enquote {\bibinfo {title} {Experimental demonstration of phase-sensitive multimode continuous variable quantum key distribution with improved secure key rate},}\ }\href@noop {} {\bibfield  {journal} {\bibinfo  {journal} {Photonics Research}\ }\textbf {\bibinfo {volume} {11}},\ \bibinfo {pages} {1861--1869} (\bibinfo {year} {2023})}\BibitemShut {NoStop}%
\bibitem [{\citenamefont {Pirandola}(2021{\natexlab{c}})}]{pirandola2021satellite}%
  \BibitemOpen
  \bibfield  {author} {\bibinfo {author} {\bibfnamefont {S.}~\bibnamefont {Pirandola}},\ }\bibfield  {title} {\enquote {\bibinfo {title} {Satellite quantum communications: Fundamental bounds and practical security},}\ }\href@noop {} {\bibfield  {journal} {\bibinfo  {journal} {Phys. Rev. Research}\ }\textbf {\bibinfo {volume} {3}},\ \bibinfo {pages} {023130} (\bibinfo {year} {2021}{\natexlab{c}})}\BibitemShut {NoStop}%
\bibitem [{\citenamefont {Goncharov}\ \emph {et~al.}(2022)\citenamefont {Goncharov}, \citenamefont {Vorontsova}, \citenamefont {Kirichenko}, \citenamefont {Filipov}, \citenamefont {Adam} \emph {et~al.}}]{goncharov2022rationale}%
  \BibitemOpen
  \bibfield  {author} {\bibinfo {author} {\bibfnamefont {R.}~\bibnamefont {Goncharov}}, \bibinfo {author} {\bibfnamefont {I.}~\bibnamefont {Vorontsova}}, \bibinfo {author} {\bibfnamefont {D.}~\bibnamefont {Kirichenko}}, \bibinfo {author} {\bibfnamefont {I.}~\bibnamefont {Filipov}}, \bibinfo {author} {\bibfnamefont {I.}~\bibnamefont {Adam}},  \emph {et~al.},\ }\bibfield  {title} {\enquote {\bibinfo {title} {The rationale for the optimal continuous-variable quantum key distribution protocol},}\ }\href@noop {} {\bibfield  {journal} {\bibinfo  {journal} {Optics}\ }\textbf {\bibinfo {volume} {3}},\ \bibinfo {pages} {338--351} (\bibinfo {year} {2022})}\BibitemShut {NoStop}%
\bibitem [{\citenamefont {Bian}\ \emph {et~al.}(2020)\citenamefont {Bian}, \citenamefont {Huang}, \citenamefont {Zhang},\ and\ \citenamefont {Yu}}]{Bian_FiO_2020}%
  \BibitemOpen
  \bibfield  {author} {\bibinfo {author} {\bibfnamefont {Y.}~\bibnamefont {Bian}}, \bibinfo {author} {\bibfnamefont {L.}~\bibnamefont {Huang}}, \bibinfo {author} {\bibfnamefont {Y.}~\bibnamefont {Zhang}}, \ and\ \bibinfo {author} {\bibfnamefont {S.}~\bibnamefont {Yu}},\ }\bibfield  {title} {\enquote {\bibinfo {title} {Unidimensional two-way continuous-variable quantum key distribution},}\ }\href@noop {} {\bibfield  {journal} {\bibinfo  {journal} {OSA Frontiers in Optics + Laser Science APS/DLS}\ }\textbf {\bibinfo {volume} {FM7A.5}} (\bibinfo {year} {2020})}\BibitemShut {NoStop}%
\bibitem [{\citenamefont {Weedbrook}(2013)}]{Weedbrook_PhysRevA_2013}%
  \BibitemOpen
  \bibfield  {author} {\bibinfo {author} {\bibfnamefont {C.}~\bibnamefont {Weedbrook}},\ }\bibfield  {title} {\enquote {\bibinfo {title} {Continuous-variable quantum key distribution with entanglement in the middle},}\ }\href {\doibase 10.1103/PhysRevA.87.022308} {\bibfield  {journal} {\bibinfo  {journal} {Phys. Rev. A}\ }\textbf {\bibinfo {volume} {87}},\ \bibinfo {pages} {022308} (\bibinfo {year} {2013})}\BibitemShut {NoStop}%
\bibitem [{\citenamefont {Wang}, \citenamefont {Wang},\ and\ \citenamefont {Li}(2020)}]{wang2020continuous}%
  \BibitemOpen
  \bibfield  {author} {\bibinfo {author} {\bibfnamefont {P.}~\bibnamefont {Wang}}, \bibinfo {author} {\bibfnamefont {X.}~\bibnamefont {Wang}}, \ and\ \bibinfo {author} {\bibfnamefont {Y.}~\bibnamefont {Li}},\ }\bibfield  {title} {\enquote {\bibinfo {title} {Continuous-variable measurement-device-independent quantum key distribution with source-intensity errors},}\ }\href@noop {} {\bibfield  {journal} {\bibinfo  {journal} {Phys. Rev. A}\ }\textbf {\bibinfo {volume} {102}},\ \bibinfo {pages} {022609} (\bibinfo {year} {2020})}\BibitemShut {NoStop}%
\bibitem [{\citenamefont {Zhang}\ \emph {et~al.}(2020{\natexlab{c}})\citenamefont {Zhang}, \citenamefont {Chen}, \citenamefont {Weedbrook}, \citenamefont {Yu},\ and\ \citenamefont {Guo}}]{Zhang_SciRep_2020}%
  \BibitemOpen
  \bibfield  {author} {\bibinfo {author} {\bibfnamefont {Y.}~\bibnamefont {Zhang}}, \bibinfo {author} {\bibfnamefont {Z.}~\bibnamefont {Chen}}, \bibinfo {author} {\bibfnamefont {C.}~\bibnamefont {Weedbrook}}, \bibinfo {author} {\bibfnamefont {S.}~\bibnamefont {Yu}}, \ and\ \bibinfo {author} {\bibfnamefont {H.}~\bibnamefont {Guo}},\ }\bibfield  {title} {\enquote {\bibinfo {title} {Continuous-variable source-device-independent quantum key distribution against general attacks},}\ }\href@noop {} {\bibfield  {journal} {\bibinfo  {journal} {Sci. Rep.}\ }\textbf {\bibinfo {volume} {10}},\ \bibinfo {pages} {1--10} (\bibinfo {year} {2020}{\natexlab{c}})}\BibitemShut {NoStop}%
\bibitem [{\citenamefont {Qi}, \citenamefont {Evans},\ and\ \citenamefont {Grice}(2018)}]{qi2018passive}%
  \BibitemOpen
  \bibfield  {author} {\bibinfo {author} {\bibfnamefont {B.}~\bibnamefont {Qi}}, \bibinfo {author} {\bibfnamefont {P.~G.}\ \bibnamefont {Evans}}, \ and\ \bibinfo {author} {\bibfnamefont {W.~P.}\ \bibnamefont {Grice}},\ }\bibfield  {title} {\enquote {\bibinfo {title} {Passive state preparation in the gaussian-modulated coherent-states quantum key distribution},}\ }\href@noop {} {\bibfield  {journal} {\bibinfo  {journal} {Phys. Rev. A}\ }\textbf {\bibinfo {volume} {97}},\ \bibinfo {pages} {012317} (\bibinfo {year} {2018})}\BibitemShut {NoStop}%
\bibitem [{\citenamefont {Pirandola}\ \emph {et~al.}(2009)\citenamefont {Pirandola}, \citenamefont {Garc{\'\i}a-Patr{\'o}n}, \citenamefont {Braunstein},\ and\ \citenamefont {Lloyd}}]{pirandola2009direct}%
  \BibitemOpen
  \bibfield  {author} {\bibinfo {author} {\bibfnamefont {S.}~\bibnamefont {Pirandola}}, \bibinfo {author} {\bibfnamefont {R.}~\bibnamefont {Garc{\'\i}a-Patr{\'o}n}}, \bibinfo {author} {\bibfnamefont {S.~L.}\ \bibnamefont {Braunstein}}, \ and\ \bibinfo {author} {\bibfnamefont {S.}~\bibnamefont {Lloyd}},\ }\bibfield  {title} {\enquote {\bibinfo {title} {Direct and reverse secret-key capacities of a quantum channel},}\ }\href@noop {} {\bibfield  {journal} {\bibinfo  {journal} {Phys. Rev. Lett.}\ }\textbf {\bibinfo {volume} {102}},\ \bibinfo {pages} {050503} (\bibinfo {year} {2009})}\BibitemShut {NoStop}%
\bibitem [{\citenamefont {Devetak}\ and\ \citenamefont {Winter}(2003)}]{Devetak2003DistillationOS}%
  \BibitemOpen
  \bibfield  {author} {\bibinfo {author} {\bibfnamefont {I.}~\bibnamefont {Devetak}}\ and\ \bibinfo {author} {\bibfnamefont {A.~J.}\ \bibnamefont {Winter}},\ }\bibfield  {title} {\enquote {\bibinfo {title} {Distillation of secret key and entanglement from quantum states},}\ }\href {https://api.semanticscholar.org/CorpusID:15425899} {\bibfield  {journal} {\bibinfo  {journal} {Proceedings of the Royal Society A: Mathematical, Physical and Engineering Sciences}\ }\textbf {\bibinfo {volume} {461}},\ \bibinfo {pages} {207 -- 235} (\bibinfo {year} {2003})}\BibitemShut {NoStop}%
\bibitem [{\citenamefont {Pirandola}\ \emph {et~al.}(2018)\citenamefont {Pirandola}, \citenamefont {Braunstein}, \citenamefont {Laurenza}, \citenamefont {Ottaviani}, \citenamefont {Cope} \emph {et~al.}}]{pirandola2018theory}%
  \BibitemOpen
  \bibfield  {author} {\bibinfo {author} {\bibfnamefont {S.}~\bibnamefont {Pirandola}}, \bibinfo {author} {\bibfnamefont {S.~L.}\ \bibnamefont {Braunstein}}, \bibinfo {author} {\bibfnamefont {R.}~\bibnamefont {Laurenza}}, \bibinfo {author} {\bibfnamefont {C.}~\bibnamefont {Ottaviani}}, \bibinfo {author} {\bibfnamefont {T.~P.}\ \bibnamefont {Cope}},  \emph {et~al.},\ }\bibfield  {title} {\enquote {\bibinfo {title} {Theory of channel simulation and bounds for private communication},}\ }\href@noop {} {\bibfield  {journal} {\bibinfo  {journal} {Quantum Sci. Technol.}\ }\textbf {\bibinfo {volume} {3}},\ \bibinfo {pages} {035009} (\bibinfo {year} {2018})}\BibitemShut {NoStop}%
\bibitem [{\citenamefont {Holevo}(1973)}]{Holevo1973BoundsFT}%
  \BibitemOpen
  \bibfield  {author} {\bibinfo {author} {\bibfnamefont {A.~S.}\ \bibnamefont {Holevo}},\ }\bibfield  {title} {\enquote {\bibinfo {title} {Bounds for the quantity of information transmitted by a quantum communication channel},}\ }\href@noop {} {\bibfield  {journal} {\bibinfo  {journal} {Problemy Peredachi Informatsii}\ }\textbf {\bibinfo {volume} {9}},\ \bibinfo {pages} {3--11} (\bibinfo {year} {1973})}\BibitemShut {NoStop}%
\bibitem [{\citenamefont {Wolf}, \citenamefont {Giedke},\ and\ \citenamefont {Cirac}(2006)}]{Wolf_2006_PRL}%
  \BibitemOpen
  \bibfield  {author} {\bibinfo {author} {\bibfnamefont {M.~M.}\ \bibnamefont {Wolf}}, \bibinfo {author} {\bibfnamefont {G.}~\bibnamefont {Giedke}}, \ and\ \bibinfo {author} {\bibfnamefont {J.~I.}\ \bibnamefont {Cirac}},\ }\bibfield  {title} {\enquote {\bibinfo {title} {Extremality of gaussian quantum states},}\ }\href {\doibase 10.1103/PhysRevLett.96.080502} {\bibfield  {journal} {\bibinfo  {journal} {Phys. Rev. Lett.}\ }\textbf {\bibinfo {volume} {96}},\ \bibinfo {pages} {080502} (\bibinfo {year} {2006})}\BibitemShut {NoStop}%
\bibitem [{\citenamefont {Almeida}\ \emph {et~al.}(2021)\citenamefont {Almeida}, \citenamefont {Pereira}, \citenamefont {Muga}, \citenamefont {Fac{\~a}o}, \citenamefont {Pinto} \emph {et~al.}}]{almeida2021secret}%
  \BibitemOpen
  \bibfield  {author} {\bibinfo {author} {\bibfnamefont {M.}~\bibnamefont {Almeida}}, \bibinfo {author} {\bibfnamefont {D.}~\bibnamefont {Pereira}}, \bibinfo {author} {\bibfnamefont {N.~J.}\ \bibnamefont {Muga}}, \bibinfo {author} {\bibfnamefont {M.}~\bibnamefont {Fac{\~a}o}}, \bibinfo {author} {\bibfnamefont {A.~N.}\ \bibnamefont {Pinto}},  \emph {et~al.},\ }\bibfield  {title} {\enquote {\bibinfo {title} {Secret key rate of multi-ring m-apsk continuous variable quantum key distribution},}\ }\href@noop {} {\bibfield  {journal} {\bibinfo  {journal} {Opt. Express}\ }\textbf {\bibinfo {volume} {29}},\ \bibinfo {pages} {38669--38682} (\bibinfo {year} {2021})}\BibitemShut {NoStop}%
\bibitem [{\citenamefont {Lupo}\ and\ \citenamefont {Ouyang}(2022)}]{lupo2022quantum}%
  \BibitemOpen
  \bibfield  {author} {\bibinfo {author} {\bibfnamefont {C.}~\bibnamefont {Lupo}}\ and\ \bibinfo {author} {\bibfnamefont {Y.}~\bibnamefont {Ouyang}},\ }\bibfield  {title} {\enquote {\bibinfo {title} {Quantum key distribution with nonideal heterodyne detection: composable security of discrete-modulation continuous-variable protocols},}\ }\href@noop {} {\bibfield  {journal} {\bibinfo  {journal} {PRX Quantum}\ }\textbf {\bibinfo {volume} {3}},\ \bibinfo {pages} {010341} (\bibinfo {year} {2022})}\BibitemShut {NoStop}%
\bibitem [{\citenamefont {Lin}\ and\ \citenamefont {L{\"u}tkenhaus}(2020)}]{lin2020trusted}%
  \BibitemOpen
  \bibfield  {author} {\bibinfo {author} {\bibfnamefont {J.}~\bibnamefont {Lin}}\ and\ \bibinfo {author} {\bibfnamefont {N.}~\bibnamefont {L{\"u}tkenhaus}},\ }\bibfield  {title} {\enquote {\bibinfo {title} {Trusted detector noise analysis for discrete modulation schemes of continuous-variable quantum key distribution},}\ }\href@noop {} {\bibfield  {journal} {\bibinfo  {journal} {Phys. Rev. Appl.}\ }\textbf {\bibinfo {volume} {14}},\ \bibinfo {pages} {064030} (\bibinfo {year} {2020})}\BibitemShut {NoStop}%
\bibitem [{\citenamefont {Liu}\ \emph {et~al.}(2021{\natexlab{a}})\citenamefont {Liu}, \citenamefont {Li}, \citenamefont {Xie}, \citenamefont {Weng}, \citenamefont {Gu} \emph {et~al.}}]{liu2021homodyne}%
  \BibitemOpen
  \bibfield  {author} {\bibinfo {author} {\bibfnamefont {W.-B.}\ \bibnamefont {Liu}}, \bibinfo {author} {\bibfnamefont {C.-L.}\ \bibnamefont {Li}}, \bibinfo {author} {\bibfnamefont {Y.-M.}\ \bibnamefont {Xie}}, \bibinfo {author} {\bibfnamefont {C.-X.}\ \bibnamefont {Weng}}, \bibinfo {author} {\bibfnamefont {J.}~\bibnamefont {Gu}},  \emph {et~al.},\ }\bibfield  {title} {\enquote {\bibinfo {title} {Homodyne detection quadrature phase shift keying continuous-variable quantum key distribution with high excess noise tolerance},}\ }\href@noop {} {\bibfield  {journal} {\bibinfo  {journal} {PRX Quantum}\ }\textbf {\bibinfo {volume} {2}},\ \bibinfo {pages} {040334} (\bibinfo {year} {2021}{\natexlab{a}})}\BibitemShut {NoStop}%
\bibitem [{\citenamefont {Upadhyaya}\ \emph {et~al.}(2021)\citenamefont {Upadhyaya}, \citenamefont {van Himbeeck}, \citenamefont {Lin},\ and\ \citenamefont {L{\"u}tkenhaus}}]{upadhyaya2021dimension}%
  \BibitemOpen
  \bibfield  {author} {\bibinfo {author} {\bibfnamefont {T.}~\bibnamefont {Upadhyaya}}, \bibinfo {author} {\bibfnamefont {T.}~\bibnamefont {van Himbeeck}}, \bibinfo {author} {\bibfnamefont {J.}~\bibnamefont {Lin}}, \ and\ \bibinfo {author} {\bibfnamefont {N.}~\bibnamefont {L{\"u}tkenhaus}},\ }\bibfield  {title} {\enquote {\bibinfo {title} {Dimension reduction in quantum key distribution for continuous-and discrete-variable protocols},}\ }\href@noop {} {\bibfield  {journal} {\bibinfo  {journal} {PRX Quantum}\ }\textbf {\bibinfo {volume} {2}},\ \bibinfo {pages} {020325} (\bibinfo {year} {2021})}\BibitemShut {NoStop}%
\bibitem [{\citenamefont {Kanitschar}\ and\ \citenamefont {Pacher}(2022)}]{kanitschar2022optimizing}%
  \BibitemOpen
  \bibfield  {author} {\bibinfo {author} {\bibfnamefont {F.}~\bibnamefont {Kanitschar}}\ and\ \bibinfo {author} {\bibfnamefont {C.}~\bibnamefont {Pacher}},\ }\bibfield  {title} {\enquote {\bibinfo {title} {Optimizing continuous-variable quantum key distribution with phase-shift keying modulation and postselection},}\ }\href@noop {} {\bibfield  {journal} {\bibinfo  {journal} {Phys. Rev. Appl.}\ }\textbf {\bibinfo {volume} {18}},\ \bibinfo {pages} {034073} (\bibinfo {year} {2022})}\BibitemShut {NoStop}%
\bibitem [{\citenamefont {Wang}\ \emph {et~al.}(2023{\natexlab{a}})\citenamefont {Wang}, \citenamefont {Zhang}, \citenamefont {Lu}, \citenamefont {Wang},\ and\ \citenamefont {Li}}]{wang2023discrete}%
  \BibitemOpen
  \bibfield  {author} {\bibinfo {author} {\bibfnamefont {P.}~\bibnamefont {Wang}}, \bibinfo {author} {\bibfnamefont {Y.}~\bibnamefont {Zhang}}, \bibinfo {author} {\bibfnamefont {Z.}~\bibnamefont {Lu}}, \bibinfo {author} {\bibfnamefont {X.}~\bibnamefont {Wang}}, \ and\ \bibinfo {author} {\bibfnamefont {Y.}~\bibnamefont {Li}},\ }\bibfield  {title} {\enquote {\bibinfo {title} {Discrete-modulation continuous-variable quantum key distribution with a high key rate},}\ }\href@noop {} {\bibfield  {journal} {\bibinfo  {journal} {New J. Phys.}\ }\textbf {\bibinfo {volume} {25}},\ \bibinfo {pages} {023019} (\bibinfo {year} {2023}{\natexlab{a}})}\BibitemShut {NoStop}%
\bibitem [{\citenamefont {Kanitschar}\ \emph {et~al.}(2023)\citenamefont {Kanitschar}, \citenamefont {George}, \citenamefont {Lin}, \citenamefont {Upadhyaya},\ and\ \citenamefont {L{\"u}tkenhaus}}]{kanitschar2023finite}%
  \BibitemOpen
  \bibfield  {author} {\bibinfo {author} {\bibfnamefont {F.}~\bibnamefont {Kanitschar}}, \bibinfo {author} {\bibfnamefont {I.}~\bibnamefont {George}}, \bibinfo {author} {\bibfnamefont {J.}~\bibnamefont {Lin}}, \bibinfo {author} {\bibfnamefont {T.}~\bibnamefont {Upadhyaya}}, \ and\ \bibinfo {author} {\bibfnamefont {N.}~\bibnamefont {L{\"u}tkenhaus}},\ }\bibfield  {title} {\enquote {\bibinfo {title} {Finite-size security for discrete-modulated continuous-variable quantum key distribution protocols},}\ }\href@noop {} {\bibfield  {journal} {\bibinfo  {journal} {arXiv preprint arXiv:2301.08686}\ } (\bibinfo {year} {2023})}\BibitemShut {NoStop}%
\bibitem [{\citenamefont {Yamano}\ \emph {et~al.}(2022)\citenamefont {Yamano}, \citenamefont {Matsuura}, \citenamefont {Kuramochi}, \citenamefont {Sasaki},\ and\ \citenamefont {Koashi}}]{yamano2022finite}%
  \BibitemOpen
  \bibfield  {author} {\bibinfo {author} {\bibfnamefont {S.}~\bibnamefont {Yamano}}, \bibinfo {author} {\bibfnamefont {T.}~\bibnamefont {Matsuura}}, \bibinfo {author} {\bibfnamefont {Y.}~\bibnamefont {Kuramochi}}, \bibinfo {author} {\bibfnamefont {T.}~\bibnamefont {Sasaki}}, \ and\ \bibinfo {author} {\bibfnamefont {M.}~\bibnamefont {Koashi}},\ }\bibfield  {title} {\enquote {\bibinfo {title} {Finite-size security proof of binary-modulation continuous-variable quantum key distribution using only heterodyne measurement},}\ }\href@noop {} {\bibfield  {journal} {\bibinfo  {journal} {arXiv preprint arXiv:2208.11983}\ } (\bibinfo {year} {2022})}\BibitemShut {NoStop}%
\bibitem [{\citenamefont {B{\"a}uml}\ \emph {et~al.}(2023)\citenamefont {B{\"a}uml}, \citenamefont {Garc{\'\i}a}, \citenamefont {Wright}, \citenamefont {Fawzi},\ and\ \citenamefont {Ac{\'\i}n}}]{bauml2023security}%
  \BibitemOpen
  \bibfield  {author} {\bibinfo {author} {\bibfnamefont {S.}~\bibnamefont {B{\"a}uml}}, \bibinfo {author} {\bibfnamefont {C.~P.}\ \bibnamefont {Garc{\'\i}a}}, \bibinfo {author} {\bibfnamefont {V.}~\bibnamefont {Wright}}, \bibinfo {author} {\bibfnamefont {O.}~\bibnamefont {Fawzi}}, \ and\ \bibinfo {author} {\bibfnamefont {A.}~\bibnamefont {Ac{\'\i}n}},\ }\bibfield  {title} {\enquote {\bibinfo {title} {Security of discrete-modulated continuous-variable quantum key distribution},}\ }\href@noop {} {\bibfield  {journal} {\bibinfo  {journal} {arXiv preprint arXiv:2303.09255}\ } (\bibinfo {year} {2023})}\BibitemShut {NoStop}%
\bibitem [{\citenamefont {Papanastasiou}\ and\ \citenamefont {Pirandola}(2021)}]{papanastasiou2021continuous}%
  \BibitemOpen
  \bibfield  {author} {\bibinfo {author} {\bibfnamefont {P.}~\bibnamefont {Papanastasiou}}\ and\ \bibinfo {author} {\bibfnamefont {S.}~\bibnamefont {Pirandola}},\ }\bibfield  {title} {\enquote {\bibinfo {title} {Continuous-variable quantum cryptography with discrete alphabets: Composable security under collective gaussian attacks},}\ }\href@noop {} {\bibfield  {journal} {\bibinfo  {journal} {Phys. Rev. Research}\ }\textbf {\bibinfo {volume} {3}},\ \bibinfo {pages} {013047} (\bibinfo {year} {2021})}\BibitemShut {NoStop}%
\bibitem [{\citenamefont {Pirandola}(2014)}]{pirandola2014quantum}%
  \BibitemOpen
  \bibfield  {author} {\bibinfo {author} {\bibfnamefont {S.}~\bibnamefont {Pirandola}},\ }\bibfield  {title} {\enquote {\bibinfo {title} {Quantum discord as a resource for quantum cryptography},}\ }\href@noop {} {\bibfield  {journal} {\bibinfo  {journal} {Sci. Rep.}\ }\textbf {\bibinfo {volume} {4}},\ \bibinfo {pages} {6956} (\bibinfo {year} {2014})}\BibitemShut {NoStop}%
\bibitem [{\citenamefont {Usenko}\ and\ \citenamefont {Filip}(2016)}]{usenko_Entropy_2016}%
  \BibitemOpen
  \bibfield  {author} {\bibinfo {author} {\bibfnamefont {V.~C.}\ \bibnamefont {Usenko}}\ and\ \bibinfo {author} {\bibfnamefont {R.}~\bibnamefont {Filip}},\ }\bibfield  {title} {\enquote {\bibinfo {title} {Trusted noise in continuous-variable quantum key distribution: a threat and a defense},}\ }\href@noop {} {\bibfield  {journal} {\bibinfo  {journal} {Entropy}\ }\textbf {\bibinfo {volume} {18}},\ \bibinfo {pages} {20} (\bibinfo {year} {2016})}\BibitemShut {NoStop}%
\bibitem [{\citenamefont {Fossier}\ \emph {et~al.}(2009{\natexlab{a}})\citenamefont {Fossier}, \citenamefont {Diamanti}, \citenamefont {Debuisschert}, \citenamefont {Tualle-Brouri},\ and\ \citenamefont {Grangier}}]{Fossier_JPhysBAtMolOptPhys_2009}%
  \BibitemOpen
  \bibfield  {author} {\bibinfo {author} {\bibfnamefont {S.}~\bibnamefont {Fossier}}, \bibinfo {author} {\bibfnamefont {E.}~\bibnamefont {Diamanti}}, \bibinfo {author} {\bibfnamefont {T.}~\bibnamefont {Debuisschert}}, \bibinfo {author} {\bibfnamefont {R.}~\bibnamefont {Tualle-Brouri}}, \ and\ \bibinfo {author} {\bibfnamefont {P.}~\bibnamefont {Grangier}},\ }\bibfield  {title} {\enquote {\bibinfo {title} {Improvement of continuous-variable quantum key distribution systems by using optical preamplifiers},}\ }\href@noop {} {\bibfield  {journal} {\bibinfo  {journal} {J. Phys. B: At., Mol. Opt. Phys.}\ }\textbf {\bibinfo {volume} {42}},\ \bibinfo {pages} {114014} (\bibinfo {year} {2009}{\natexlab{a}})}\BibitemShut {NoStop}%
\bibitem [{\citenamefont {Zhang}\ \emph {et~al.}(2013)\citenamefont {Zhang}, \citenamefont {Li}, \citenamefont {Weedbrook}, \citenamefont {Yu}, \citenamefont {Gu} \emph {et~al.}}]{Zhang2013ImprovementOT}%
  \BibitemOpen
  \bibfield  {author} {\bibinfo {author} {\bibfnamefont {Y.}~\bibnamefont {Zhang}}, \bibinfo {author} {\bibfnamefont {Z.}~\bibnamefont {Li}}, \bibinfo {author} {\bibfnamefont {C.}~\bibnamefont {Weedbrook}}, \bibinfo {author} {\bibfnamefont {S.}~\bibnamefont {Yu}}, \bibinfo {author} {\bibfnamefont {W.}~\bibnamefont {Gu}},  \emph {et~al.},\ }\bibfield  {title} {\enquote {\bibinfo {title} {Improvement of two-way continuous-variable quantum key distribution using optical amplifiers},}\ }\href {https://api.semanticscholar.org/CorpusID:110829110} {\bibfield  {journal} {\bibinfo  {journal} {J. Phys. B: At., Mol. Opt. Phys.}\ }\textbf {\bibinfo {volume} {47}} (\bibinfo {year} {2013})}\BibitemShut {NoStop}%
\bibitem [{\citenamefont {Zhang}\ \emph {et~al.}(2015)\citenamefont {Zhang}, \citenamefont {Li}, \citenamefont {Weedbrook}, \citenamefont {Marshall}, \citenamefont {Pirandola} \emph {et~al.}}]{Zhang2015NoiselessLA}%
  \BibitemOpen
  \bibfield  {author} {\bibinfo {author} {\bibfnamefont {Y.}~\bibnamefont {Zhang}}, \bibinfo {author} {\bibfnamefont {Z.}~\bibnamefont {Li}}, \bibinfo {author} {\bibfnamefont {C.}~\bibnamefont {Weedbrook}}, \bibinfo {author} {\bibfnamefont {K.}~\bibnamefont {Marshall}}, \bibinfo {author} {\bibfnamefont {S.}~\bibnamefont {Pirandola}},  \emph {et~al.},\ }\bibfield  {title} {\enquote {\bibinfo {title} {Noiseless linear amplifiers in entanglement-based continuous-variable quantum key distribution},}\ }\href {https://api.semanticscholar.org/CorpusID:12340887} {\bibfield  {journal} {\bibinfo  {journal} {Entropy}\ }\textbf {\bibinfo {volume} {17}},\ \bibinfo {pages} {4547--4562} (\bibinfo {year} {2015})}\BibitemShut {NoStop}%
\bibitem [{\citenamefont {Zhang}, \citenamefont {Yu},\ and\ \citenamefont {Guo}(2015)}]{Zhang2015ApplicationOP}%
  \BibitemOpen
  \bibfield  {author} {\bibinfo {author} {\bibfnamefont {Y.}~\bibnamefont {Zhang}}, \bibinfo {author} {\bibfnamefont {S.}~\bibnamefont {Yu}}, \ and\ \bibinfo {author} {\bibfnamefont {H.}~\bibnamefont {Guo}},\ }\bibfield  {title} {\enquote {\bibinfo {title} {Application of practical noiseless linear amplifier in no-switching continuous-variable quantum cryptography},}\ }\href {https://api.semanticscholar.org/CorpusID:42874028} {\bibfield  {journal} {\bibinfo  {journal} {Quantum Information Process.}\ }\textbf {\bibinfo {volume} {14}},\ \bibinfo {pages} {4339--4349} (\bibinfo {year} {2015})}\BibitemShut {NoStop}%
\bibitem [{\citenamefont {Wu}\ \emph {et~al.}(2022)\citenamefont {Wu}, \citenamefont {Liu}, \citenamefont {Zhang}, \citenamefont {Ruan},\ and\ \citenamefont {Guo}}]{Wu2022PerformanceAO}%
  \BibitemOpen
  \bibfield  {author} {\bibinfo {author} {\bibfnamefont {H.}~\bibnamefont {Wu}}, \bibinfo {author} {\bibfnamefont {X.}~\bibnamefont {Liu}}, \bibinfo {author} {\bibfnamefont {H.}~\bibnamefont {Zhang}}, \bibinfo {author} {\bibfnamefont {X.}~\bibnamefont {Ruan}}, \ and\ \bibinfo {author} {\bibfnamefont {Y.}~\bibnamefont {Guo}},\ }\bibfield  {title} {\enquote {\bibinfo {title} {Performance analysis of continuous variable quantum teleportation with noiseless linear amplifier in seawater channel},}\ }\href {https://api.semanticscholar.org/CorpusID:248824698} {\bibfield  {journal} {\bibinfo  {journal} {Symmetry}\ }\textbf {\bibinfo {volume} {14}},\ \bibinfo {pages} {997} (\bibinfo {year} {2022})}\BibitemShut {NoStop}%
\bibitem [{\citenamefont {Bera}\ \emph {et~al.}(2017)\citenamefont {Bera}, \citenamefont {Ac{\'\i}n}, \citenamefont {Ku{\'s}}, \citenamefont {Mitchell},\ and\ \citenamefont {Lewenstein}}]{bera_RepProgPhys_2017}%
  \BibitemOpen
  \bibfield  {author} {\bibinfo {author} {\bibfnamefont {M.~N.}\ \bibnamefont {Bera}}, \bibinfo {author} {\bibfnamefont {A.}~\bibnamefont {Ac{\'\i}n}}, \bibinfo {author} {\bibfnamefont {M.}~\bibnamefont {Ku{\'s}}}, \bibinfo {author} {\bibfnamefont {M.~W.}\ \bibnamefont {Mitchell}}, \ and\ \bibinfo {author} {\bibfnamefont {M.}~\bibnamefont {Lewenstein}},\ }\bibfield  {title} {\enquote {\bibinfo {title} {Randomness in quantum mechanics: philosophy, physics and technology},}\ }\href@noop {} {\bibfield  {journal} {\bibinfo  {journal} {Rep. Prog. Phys.}\ }\textbf {\bibinfo {volume} {80}},\ \bibinfo {pages} {124001} (\bibinfo {year} {2017})}\BibitemShut {NoStop}%
\bibitem [{\citenamefont {Ma}\ \emph {et~al.}(2016{\natexlab{b}})\citenamefont {Ma}, \citenamefont {Yuan}, \citenamefont {Cao}, \citenamefont {Qi},\ and\ \citenamefont {Zhang}}]{ma_npjQuantumInf_2016}%
  \BibitemOpen
  \bibfield  {author} {\bibinfo {author} {\bibfnamefont {X.}~\bibnamefont {Ma}}, \bibinfo {author} {\bibfnamefont {X.}~\bibnamefont {Yuan}}, \bibinfo {author} {\bibfnamefont {Z.}~\bibnamefont {Cao}}, \bibinfo {author} {\bibfnamefont {B.}~\bibnamefont {Qi}}, \ and\ \bibinfo {author} {\bibfnamefont {Z.}~\bibnamefont {Zhang}},\ }\bibfield  {title} {\enquote {\bibinfo {title} {Quantum random number generation},}\ }\href@noop {} {\bibfield  {journal} {\bibinfo  {journal} {npj Quantum Inf.}\ }\textbf {\bibinfo {volume} {2}},\ \bibinfo {pages} {1--9} (\bibinfo {year} {2016}{\natexlab{b}})}\BibitemShut {NoStop}%
\bibitem [{\citenamefont {Herrero-Collantes}\ and\ \citenamefont {Garcia-Escartin}(2017)}]{herrero_RevModPhys_2017}%
  \BibitemOpen
  \bibfield  {author} {\bibinfo {author} {\bibfnamefont {M.}~\bibnamefont {Herrero-Collantes}}\ and\ \bibinfo {author} {\bibfnamefont {J.~C.}\ \bibnamefont {Garcia-Escartin}},\ }\bibfield  {title} {\enquote {\bibinfo {title} {Quantum random number generators},}\ }\href@noop {} {\bibfield  {journal} {\bibinfo  {journal} {Rev. Mod. Phys.}\ }\textbf {\bibinfo {volume} {89}},\ \bibinfo {pages} {015004} (\bibinfo {year} {2017})}\BibitemShut {NoStop}%
\bibitem [{\citenamefont {Symul}, \citenamefont {Assad},\ and\ \citenamefont {Lam}(2011)}]{symul_ApplPhysLett_2011}%
  \BibitemOpen
  \bibfield  {author} {\bibinfo {author} {\bibfnamefont {T.}~\bibnamefont {Symul}}, \bibinfo {author} {\bibfnamefont {S.~M.}\ \bibnamefont {Assad}}, \ and\ \bibinfo {author} {\bibfnamefont {P.~K.}\ \bibnamefont {Lam}},\ }\bibfield  {title} {\enquote {\bibinfo {title} {Real time demonstration of high bitrate quantum random number generation with coherent laser light},}\ }\href@noop {} {\bibfield  {journal} {\bibinfo  {journal} {Appl. Phys. Lett.}\ }\textbf {\bibinfo {volume} {98}} (\bibinfo {year} {2011})}\BibitemShut {NoStop}%
\bibitem [{\citenamefont {Zheng}\ \emph {et~al.}(2019{\natexlab{a}})\citenamefont {Zheng}, \citenamefont {Zhang}, \citenamefont {Huang}, \citenamefont {Yu},\ and\ \citenamefont {Guo}}]{zheng20196}%
  \BibitemOpen
  \bibfield  {author} {\bibinfo {author} {\bibfnamefont {Z.}~\bibnamefont {Zheng}}, \bibinfo {author} {\bibfnamefont {Y.}~\bibnamefont {Zhang}}, \bibinfo {author} {\bibfnamefont {W.}~\bibnamefont {Huang}}, \bibinfo {author} {\bibfnamefont {S.}~\bibnamefont {Yu}}, \ and\ \bibinfo {author} {\bibfnamefont {H.}~\bibnamefont {Guo}},\ }\bibfield  {title} {\enquote {\bibinfo {title} {6 gbps real-time optical quantum random number generator based on vacuum fluctuation},}\ }\href@noop {} {\bibfield  {journal} {\bibinfo  {journal} {Rev. Sci. Instrum.}\ }\textbf {\bibinfo {volume} {90}} (\bibinfo {year} {2019}{\natexlab{a}})}\BibitemShut {NoStop}%
\bibitem [{\citenamefont {Bai}\ \emph {et~al.}(2021)\citenamefont {Bai}, \citenamefont {Huang}, \citenamefont {Qiao}, \citenamefont {Nie}, \citenamefont {Tang} \emph {et~al.}}]{bai_ApplPhysLett_2021}%
  \BibitemOpen
  \bibfield  {author} {\bibinfo {author} {\bibfnamefont {B.}~\bibnamefont {Bai}}, \bibinfo {author} {\bibfnamefont {J.}~\bibnamefont {Huang}}, \bibinfo {author} {\bibfnamefont {G.-R.}\ \bibnamefont {Qiao}}, \bibinfo {author} {\bibfnamefont {Y.-Q.}\ \bibnamefont {Nie}}, \bibinfo {author} {\bibfnamefont {W.}~\bibnamefont {Tang}},  \emph {et~al.},\ }\bibfield  {title} {\enquote {\bibinfo {title} {18.8 gbps real-time quantum random number generator with a photonic integrated chip},}\ }\href@noop {} {\bibfield  {journal} {\bibinfo  {journal} {Appl. Phys. Lett.}\ }\textbf {\bibinfo {volume} {118}} (\bibinfo {year} {2021})}\BibitemShut {NoStop}%
\bibitem [{\citenamefont {Gehring}\ \emph {et~al.}(2021)\citenamefont {Gehring}, \citenamefont {Lupo}, \citenamefont {Kordts}, \citenamefont {Solar~Nikolic}, \citenamefont {Jain} \emph {et~al.}}]{gehring_NatComm_2021}%
  \BibitemOpen
  \bibfield  {author} {\bibinfo {author} {\bibfnamefont {T.}~\bibnamefont {Gehring}}, \bibinfo {author} {\bibfnamefont {C.}~\bibnamefont {Lupo}}, \bibinfo {author} {\bibfnamefont {A.}~\bibnamefont {Kordts}}, \bibinfo {author} {\bibfnamefont {D.}~\bibnamefont {Solar~Nikolic}}, \bibinfo {author} {\bibfnamefont {N.}~\bibnamefont {Jain}},  \emph {et~al.},\ }\bibfield  {title} {\enquote {\bibinfo {title} {Homodyne-based quantum random number generator at 2.9 gbps secure against quantum side-information},}\ }\href@noop {} {\bibfield  {journal} {\bibinfo  {journal} {Nat. Commun.}\ }\textbf {\bibinfo {volume} {12}},\ \bibinfo {pages} {605} (\bibinfo {year} {2021})}\BibitemShut {NoStop}%
\bibitem [{\citenamefont {Bruynsteen}\ \emph {et~al.}(2023)\citenamefont {Bruynsteen}, \citenamefont {Gehring}, \citenamefont {Lupo}, \citenamefont {Bauwelinck},\ and\ \citenamefont {Yin}}]{bruynsteen_PRXQuantum_2023}%
  \BibitemOpen
  \bibfield  {author} {\bibinfo {author} {\bibfnamefont {C.}~\bibnamefont {Bruynsteen}}, \bibinfo {author} {\bibfnamefont {T.}~\bibnamefont {Gehring}}, \bibinfo {author} {\bibfnamefont {C.}~\bibnamefont {Lupo}}, \bibinfo {author} {\bibfnamefont {J.}~\bibnamefont {Bauwelinck}}, \ and\ \bibinfo {author} {\bibfnamefont {X.}~\bibnamefont {Yin}},\ }\bibfield  {title} {\enquote {\bibinfo {title} {100-gbit/s integrated quantum random number generator based on vacuum fluctuations},}\ }\href@noop {} {\bibfield  {journal} {\bibinfo  {journal} {PRX Quantum}\ }\textbf {\bibinfo {volume} {4}},\ \bibinfo {pages} {010330} (\bibinfo {year} {2023})}\BibitemShut {NoStop}%
\bibitem [{\citenamefont {Qi}\ \emph {et~al.}(2010{\natexlab{a}})\citenamefont {Qi}, \citenamefont {Chi}, \citenamefont {Lo},\ and\ \citenamefont {Qian}}]{qi_OptLett_2010}%
  \BibitemOpen
  \bibfield  {author} {\bibinfo {author} {\bibfnamefont {B.}~\bibnamefont {Qi}}, \bibinfo {author} {\bibfnamefont {Y.-M.}\ \bibnamefont {Chi}}, \bibinfo {author} {\bibfnamefont {H.-K.}\ \bibnamefont {Lo}}, \ and\ \bibinfo {author} {\bibfnamefont {L.}~\bibnamefont {Qian}},\ }\bibfield  {title} {\enquote {\bibinfo {title} {High-speed quantum random number generation by measuring phase noise of a single-mode laser},}\ }\href@noop {} {\bibfield  {journal} {\bibinfo  {journal} {Opt. Lett.}\ }\textbf {\bibinfo {volume} {35}},\ \bibinfo {pages} {312--314} (\bibinfo {year} {2010}{\natexlab{a}})}\BibitemShut {NoStop}%
\bibitem [{\citenamefont {Yang}\ \emph {et~al.}(2016)\citenamefont {Yang}, \citenamefont {Liu}, \citenamefont {Su}, \citenamefont {Li}, \citenamefont {Fan} \emph {et~al.}}]{yang_OptExp_2016}%
  \BibitemOpen
  \bibfield  {author} {\bibinfo {author} {\bibfnamefont {J.}~\bibnamefont {Yang}}, \bibinfo {author} {\bibfnamefont {J.}~\bibnamefont {Liu}}, \bibinfo {author} {\bibfnamefont {Q.}~\bibnamefont {Su}}, \bibinfo {author} {\bibfnamefont {Z.}~\bibnamefont {Li}}, \bibinfo {author} {\bibfnamefont {F.}~\bibnamefont {Fan}},  \emph {et~al.},\ }\bibfield  {title} {\enquote {\bibinfo {title} {5.4 gbps real time quantum random number generator with simple implementation},}\ }\href@noop {} {\bibfield  {journal} {\bibinfo  {journal} {Opt. Express}\ }\textbf {\bibinfo {volume} {24}},\ \bibinfo {pages} {27475--27481} (\bibinfo {year} {2016})}\BibitemShut {NoStop}%
\bibitem [{\citenamefont {Yuan}\ \emph {et~al.}(2014)\citenamefont {Yuan}, \citenamefont {Lucamarini}, \citenamefont {Dynes}, \citenamefont {Fr{\"o}hlich}, \citenamefont {Plews} \emph {et~al.}}]{yuan_ApplPhysLett_2014}%
  \BibitemOpen
  \bibfield  {author} {\bibinfo {author} {\bibfnamefont {Z.}~\bibnamefont {Yuan}}, \bibinfo {author} {\bibfnamefont {M.}~\bibnamefont {Lucamarini}}, \bibinfo {author} {\bibfnamefont {J.}~\bibnamefont {Dynes}}, \bibinfo {author} {\bibfnamefont {B.}~\bibnamefont {Fr{\"o}hlich}}, \bibinfo {author} {\bibfnamefont {A.}~\bibnamefont {Plews}},  \emph {et~al.},\ }\bibfield  {title} {\enquote {\bibinfo {title} {Robust random number generation using steady-state emission of gain-switched laser diodes},}\ }\href@noop {} {\bibfield  {journal} {\bibinfo  {journal} {Appl. Phys. Lett.}\ }\textbf {\bibinfo {volume} {104}} (\bibinfo {year} {2014})}\BibitemShut {NoStop}%
\bibitem [{\citenamefont {Nie}\ \emph {et~al.}(2015)\citenamefont {Nie}, \citenamefont {Huang}, \citenamefont {Liu}, \citenamefont {Payne}, \citenamefont {Zhang} \emph {et~al.}}]{nie_RevSciInstr_2015}%
  \BibitemOpen
  \bibfield  {author} {\bibinfo {author} {\bibfnamefont {Y.-Q.}\ \bibnamefont {Nie}}, \bibinfo {author} {\bibfnamefont {L.}~\bibnamefont {Huang}}, \bibinfo {author} {\bibfnamefont {Y.}~\bibnamefont {Liu}}, \bibinfo {author} {\bibfnamefont {F.}~\bibnamefont {Payne}}, \bibinfo {author} {\bibfnamefont {J.}~\bibnamefont {Zhang}},  \emph {et~al.},\ }\bibfield  {title} {\enquote {\bibinfo {title} {The generation of 68 gbps quantum random number by measuring laser phase fluctuations},}\ }\href@noop {} {\bibfield  {journal} {\bibinfo  {journal} {Rev. Sci. Instrum.}\ }\textbf {\bibinfo {volume} {86}} (\bibinfo {year} {2015})}\BibitemShut {NoStop}%
\bibitem [{\citenamefont {Roger}\ \emph {et~al.}(2019)\citenamefont {Roger}, \citenamefont {Paraiso}, \citenamefont {De~Marco}, \citenamefont {Marangon}, \citenamefont {Yuan} \emph {et~al.}}]{roger_JOSAB_2019}%
  \BibitemOpen
  \bibfield  {author} {\bibinfo {author} {\bibfnamefont {T.}~\bibnamefont {Roger}}, \bibinfo {author} {\bibfnamefont {T.}~\bibnamefont {Paraiso}}, \bibinfo {author} {\bibfnamefont {I.}~\bibnamefont {De~Marco}}, \bibinfo {author} {\bibfnamefont {D.~G.}\ \bibnamefont {Marangon}}, \bibinfo {author} {\bibfnamefont {Z.}~\bibnamefont {Yuan}},  \emph {et~al.},\ }\bibfield  {title} {\enquote {\bibinfo {title} {Real-time interferometric quantum random number generation on chip},}\ }\href@noop {} {\bibfield  {journal} {\bibinfo  {journal} {J. Opt. Soc. Am. B}\ }\textbf {\bibinfo {volume} {36}},\ \bibinfo {pages} {B137--B142} (\bibinfo {year} {2019})}\BibitemShut {NoStop}%
\bibitem [{\citenamefont {Imran}\ \emph {et~al.}(2021)\citenamefont {Imran}, \citenamefont {Sorianello}, \citenamefont {Fresi}, \citenamefont {Jalil}, \citenamefont {Romagnoli} \emph {et~al.}}]{imran_OptComm_2021}%
  \BibitemOpen
  \bibfield  {author} {\bibinfo {author} {\bibfnamefont {M.}~\bibnamefont {Imran}}, \bibinfo {author} {\bibfnamefont {V.}~\bibnamefont {Sorianello}}, \bibinfo {author} {\bibfnamefont {F.}~\bibnamefont {Fresi}}, \bibinfo {author} {\bibfnamefont {B.}~\bibnamefont {Jalil}}, \bibinfo {author} {\bibfnamefont {M.}~\bibnamefont {Romagnoli}},  \emph {et~al.},\ }\bibfield  {title} {\enquote {\bibinfo {title} {On-chip tunable soi interferometer for quantum random number generation based on phase diffusion in lasers},}\ }\href@noop {} {\bibfield  {journal} {\bibinfo  {journal} {Opt. Commun.}\ }\textbf {\bibinfo {volume} {485}},\ \bibinfo {pages} {126736} (\bibinfo {year} {2021})}\BibitemShut {NoStop}%
\bibitem [{\citenamefont {Yang}\ \emph {et~al.}(2023{\natexlab{a}})\citenamefont {Yang}, \citenamefont {Wu}, \citenamefont {Zhang}, \citenamefont {Liu}, \citenamefont {Fan} \emph {et~al.}}]{yang2023ultra}%
  \BibitemOpen
  \bibfield  {author} {\bibinfo {author} {\bibfnamefont {J.}~\bibnamefont {Yang}}, \bibinfo {author} {\bibfnamefont {M.}~\bibnamefont {Wu}}, \bibinfo {author} {\bibfnamefont {Y.}~\bibnamefont {Zhang}}, \bibinfo {author} {\bibfnamefont {J.}~\bibnamefont {Liu}}, \bibinfo {author} {\bibfnamefont {F.}~\bibnamefont {Fan}},  \emph {et~al.},\ }\bibfield  {title} {\enquote {\bibinfo {title} {An ultra-fast quantum random number generation scheme based on laser phase noise},}\ }\href@noop {} {\bibfield  {journal} {\bibinfo  {journal} {arXiv preprint arXiv:2311.17380}\ } (\bibinfo {year} {2023}{\natexlab{a}})}\BibitemShut {NoStop}%
\bibitem [{\citenamefont {Williams}\ \emph {et~al.}(2010)\citenamefont {Williams}, \citenamefont {Salevan}, \citenamefont {Li}, \citenamefont {Roy},\ and\ \citenamefont {Murphy}}]{williams_OptExp_2010}%
  \BibitemOpen
  \bibfield  {author} {\bibinfo {author} {\bibfnamefont {C.~R.}\ \bibnamefont {Williams}}, \bibinfo {author} {\bibfnamefont {J.~C.}\ \bibnamefont {Salevan}}, \bibinfo {author} {\bibfnamefont {X.}~\bibnamefont {Li}}, \bibinfo {author} {\bibfnamefont {R.}~\bibnamefont {Roy}}, \ and\ \bibinfo {author} {\bibfnamefont {T.~E.}\ \bibnamefont {Murphy}},\ }\bibfield  {title} {\enquote {\bibinfo {title} {Fast physical random number generator using amplified spontaneous emission},}\ }\href@noop {} {\bibfield  {journal} {\bibinfo  {journal} {Opt. Express}\ }\textbf {\bibinfo {volume} {18}},\ \bibinfo {pages} {23584--23597} (\bibinfo {year} {2010})}\BibitemShut {NoStop}%
\bibitem [{\citenamefont {Li}\ \emph {et~al.}(2011)\citenamefont {Li}, \citenamefont {Cohen}, \citenamefont {Murphy},\ and\ \citenamefont {Roy}}]{li_OptLett_2011}%
  \BibitemOpen
  \bibfield  {author} {\bibinfo {author} {\bibfnamefont {X.}~\bibnamefont {Li}}, \bibinfo {author} {\bibfnamefont {A.~B.}\ \bibnamefont {Cohen}}, \bibinfo {author} {\bibfnamefont {T.~E.}\ \bibnamefont {Murphy}}, \ and\ \bibinfo {author} {\bibfnamefont {R.}~\bibnamefont {Roy}},\ }\bibfield  {title} {\enquote {\bibinfo {title} {Scalable parallel physical random number generator based on a superluminescent led},}\ }\href@noop {} {\bibfield  {journal} {\bibinfo  {journal} {Opt. Lett.}\ }\textbf {\bibinfo {volume} {36}},\ \bibinfo {pages} {1020--1022} (\bibinfo {year} {2011})}\BibitemShut {NoStop}%
\bibitem [{\citenamefont {Yang}\ \emph {et~al.}(2020)\citenamefont {Yang}, \citenamefont {Fan}, \citenamefont {Liu}, \citenamefont {Su}, \citenamefont {Li} \emph {et~al.}}]{yang_QuanSciTech_2020}%
  \BibitemOpen
  \bibfield  {author} {\bibinfo {author} {\bibfnamefont {J.}~\bibnamefont {Yang}}, \bibinfo {author} {\bibfnamefont {F.}~\bibnamefont {Fan}}, \bibinfo {author} {\bibfnamefont {J.}~\bibnamefont {Liu}}, \bibinfo {author} {\bibfnamefont {Q.}~\bibnamefont {Su}}, \bibinfo {author} {\bibfnamefont {Y.}~\bibnamefont {Li}},  \emph {et~al.},\ }\bibfield  {title} {\enquote {\bibinfo {title} {Randomness quantification for quantum random number generation based on detection of amplified spontaneous emission noise},}\ }\href@noop {} {\bibfield  {journal} {\bibinfo  {journal} {Quantum Sci. Technol.}\ }\textbf {\bibinfo {volume} {6}},\ \bibinfo {pages} {015002} (\bibinfo {year} {2020})}\BibitemShut {NoStop}%
\bibitem [{\citenamefont {Jennewein}\ \emph {et~al.}(2000)\citenamefont {Jennewein}, \citenamefont {Achleitner}, \citenamefont {Weihs}, \citenamefont {Weinfurter},\ and\ \citenamefont {Zeilinger}}]{jennewein_RevSciInstr_2000}%
  \BibitemOpen
  \bibfield  {author} {\bibinfo {author} {\bibfnamefont {T.}~\bibnamefont {Jennewein}}, \bibinfo {author} {\bibfnamefont {U.}~\bibnamefont {Achleitner}}, \bibinfo {author} {\bibfnamefont {G.}~\bibnamefont {Weihs}}, \bibinfo {author} {\bibfnamefont {H.}~\bibnamefont {Weinfurter}}, \ and\ \bibinfo {author} {\bibfnamefont {A.}~\bibnamefont {Zeilinger}},\ }\bibfield  {title} {\enquote {\bibinfo {title} {A fast and compact quantum random number generator},}\ }\href@noop {} {\bibfield  {journal} {\bibinfo  {journal} {Rev. Sci. Instrum.}\ }\textbf {\bibinfo {volume} {71}},\ \bibinfo {pages} {1675--1680} (\bibinfo {year} {2000})}\BibitemShut {NoStop}%
\bibitem [{\citenamefont {Stefanov}\ \emph {et~al.}(2000)\citenamefont {Stefanov}, \citenamefont {Gisin}, \citenamefont {Guinnard}, \citenamefont {Guinnard},\ and\ \citenamefont {Zbinden}}]{stefanov_JourModOpt_2000}%
  \BibitemOpen
  \bibfield  {author} {\bibinfo {author} {\bibfnamefont {A.}~\bibnamefont {Stefanov}}, \bibinfo {author} {\bibfnamefont {N.}~\bibnamefont {Gisin}}, \bibinfo {author} {\bibfnamefont {O.}~\bibnamefont {Guinnard}}, \bibinfo {author} {\bibfnamefont {L.}~\bibnamefont {Guinnard}}, \ and\ \bibinfo {author} {\bibfnamefont {H.}~\bibnamefont {Zbinden}},\ }\bibfield  {title} {\enquote {\bibinfo {title} {Optical quantum random number generator},}\ }\href@noop {} {\bibfield  {journal} {\bibinfo  {journal} {J. Mod. Opt.}\ }\textbf {\bibinfo {volume} {47}},\ \bibinfo {pages} {595--598} (\bibinfo {year} {2000})}\BibitemShut {NoStop}%
\bibitem [{\citenamefont {Dynes}\ \emph {et~al.}(2008)\citenamefont {Dynes}, \citenamefont {Yuan}, \citenamefont {Sharpe},\ and\ \citenamefont {Shields}}]{dynes_ApplPhysLett_2008}%
  \BibitemOpen
  \bibfield  {author} {\bibinfo {author} {\bibfnamefont {J.~F.}\ \bibnamefont {Dynes}}, \bibinfo {author} {\bibfnamefont {Z.~L.}\ \bibnamefont {Yuan}}, \bibinfo {author} {\bibfnamefont {A.~W.}\ \bibnamefont {Sharpe}}, \ and\ \bibinfo {author} {\bibfnamefont {A.~J.}\ \bibnamefont {Shields}},\ }\bibfield  {title} {\enquote {\bibinfo {title} {A high speed, postprocessing free, quantum random number generator},}\ }\href@noop {} {\bibfield  {journal} {\bibinfo  {journal} {Appl. Phys. Lett.}\ }\textbf {\bibinfo {volume} {93}} (\bibinfo {year} {2008})}\BibitemShut {NoStop}%
\bibitem [{\citenamefont {Wayne}\ \emph {et~al.}(2009)\citenamefont {Wayne}, \citenamefont {Jeffrey}, \citenamefont {Akselrod},\ and\ \citenamefont {Kwiat}}]{wayne_JourModOpt_2009}%
  \BibitemOpen
  \bibfield  {author} {\bibinfo {author} {\bibfnamefont {M.~A.}\ \bibnamefont {Wayne}}, \bibinfo {author} {\bibfnamefont {E.~R.}\ \bibnamefont {Jeffrey}}, \bibinfo {author} {\bibfnamefont {G.~M.}\ \bibnamefont {Akselrod}}, \ and\ \bibinfo {author} {\bibfnamefont {P.~G.}\ \bibnamefont {Kwiat}},\ }\bibfield  {title} {\enquote {\bibinfo {title} {Photon arrival time quantum random number generation},}\ }\href@noop {} {\bibfield  {journal} {\bibinfo  {journal} {J. Mod. Opt.}\ }\textbf {\bibinfo {volume} {56}},\ \bibinfo {pages} {516--522} (\bibinfo {year} {2009})}\BibitemShut {NoStop}%
\bibitem [{\citenamefont {Gabriel}\ \emph {et~al.}(2010)\citenamefont {Gabriel}, \citenamefont {Wittmann}, \citenamefont {Sych}, \citenamefont {Dong}, \citenamefont {Mauerer} \emph {et~al.}}]{gabriel_NatPhot_2010}%
  \BibitemOpen
  \bibfield  {author} {\bibinfo {author} {\bibfnamefont {C.}~\bibnamefont {Gabriel}}, \bibinfo {author} {\bibfnamefont {C.}~\bibnamefont {Wittmann}}, \bibinfo {author} {\bibfnamefont {D.}~\bibnamefont {Sych}}, \bibinfo {author} {\bibfnamefont {R.}~\bibnamefont {Dong}}, \bibinfo {author} {\bibfnamefont {W.}~\bibnamefont {Mauerer}},  \emph {et~al.},\ }\bibfield  {title} {\enquote {\bibinfo {title} {A generator for unique quantum random numbers based on vacuum states},}\ }\href@noop {} {\bibfield  {journal} {\bibinfo  {journal} {Nat. Photonics}\ }\textbf {\bibinfo {volume} {4}},\ \bibinfo {pages} {711--715} (\bibinfo {year} {2010})}\BibitemShut {NoStop}%
\bibitem [{\citenamefont {Nie}\ \emph {et~al.}(2016)\citenamefont {Nie}, \citenamefont {Guan}, \citenamefont {Zhou}, \citenamefont {Zhang}, \citenamefont {Ma} \emph {et~al.}}]{nie_PhysRevA_2016}%
  \BibitemOpen
  \bibfield  {author} {\bibinfo {author} {\bibfnamefont {Y.-Q.}\ \bibnamefont {Nie}}, \bibinfo {author} {\bibfnamefont {J.-Y.}\ \bibnamefont {Guan}}, \bibinfo {author} {\bibfnamefont {H.}~\bibnamefont {Zhou}}, \bibinfo {author} {\bibfnamefont {Q.}~\bibnamefont {Zhang}}, \bibinfo {author} {\bibfnamefont {X.}~\bibnamefont {Ma}},  \emph {et~al.},\ }\bibfield  {title} {\enquote {\bibinfo {title} {Experimental measurement-device-independent quantum random-number generation},}\ }\href@noop {} {\bibfield  {journal} {\bibinfo  {journal} {Phys. Rev. A}\ }\textbf {\bibinfo {volume} {94}},\ \bibinfo {pages} {060301} (\bibinfo {year} {2016})}\BibitemShut {NoStop}%
\bibitem [{\citenamefont {Cao}, \citenamefont {Zhou},\ and\ \citenamefont {Ma}(2015)}]{cao_NewJourPhys_2015}%
  \BibitemOpen
  \bibfield  {author} {\bibinfo {author} {\bibfnamefont {Z.}~\bibnamefont {Cao}}, \bibinfo {author} {\bibfnamefont {H.}~\bibnamefont {Zhou}}, \ and\ \bibinfo {author} {\bibfnamefont {X.}~\bibnamefont {Ma}},\ }\bibfield  {title} {\enquote {\bibinfo {title} {Loss-tolerant measurement-device-independent quantum random number generation},}\ }\href@noop {} {\bibfield  {journal} {\bibinfo  {journal} {New J. Phys.}\ }\textbf {\bibinfo {volume} {17}},\ \bibinfo {pages} {125011} (\bibinfo {year} {2015})}\BibitemShut {NoStop}%
\bibitem [{\citenamefont {Avesani}\ \emph {et~al.}(2018)\citenamefont {Avesani}, \citenamefont {Marangon}, \citenamefont {Vallone},\ and\ \citenamefont {Villoresi}}]{avesani_NatComm_2018}%
  \BibitemOpen
  \bibfield  {author} {\bibinfo {author} {\bibfnamefont {M.}~\bibnamefont {Avesani}}, \bibinfo {author} {\bibfnamefont {D.~G.}\ \bibnamefont {Marangon}}, \bibinfo {author} {\bibfnamefont {G.}~\bibnamefont {Vallone}}, \ and\ \bibinfo {author} {\bibfnamefont {P.}~\bibnamefont {Villoresi}},\ }\bibfield  {title} {\enquote {\bibinfo {title} {Source-device-independent heterodyne-based quantum random number generator at 17 gbps},}\ }\href@noop {} {\bibfield  {journal} {\bibinfo  {journal} {Nat. Commun.}\ }\textbf {\bibinfo {volume} {9}},\ \bibinfo {pages} {5365} (\bibinfo {year} {2018})}\BibitemShut {NoStop}%
\bibitem [{\citenamefont {Marangon}, \citenamefont {Vallone},\ and\ \citenamefont {Villoresi}(2017)}]{marangon_PhysRevLett_2017}%
  \BibitemOpen
  \bibfield  {author} {\bibinfo {author} {\bibfnamefont {D.~G.}\ \bibnamefont {Marangon}}, \bibinfo {author} {\bibfnamefont {G.}~\bibnamefont {Vallone}}, \ and\ \bibinfo {author} {\bibfnamefont {P.}~\bibnamefont {Villoresi}},\ }\bibfield  {title} {\enquote {\bibinfo {title} {Source-device-independent ultrafast quantum random number generation},}\ }\href@noop {} {\bibfield  {journal} {\bibinfo  {journal} {Phys. Rev. Lett.}\ }\textbf {\bibinfo {volume} {118}},\ \bibinfo {pages} {060503} (\bibinfo {year} {2017})}\BibitemShut {NoStop}%
\bibitem [{\citenamefont {Cao}\ \emph {et~al.}(2016)\citenamefont {Cao}, \citenamefont {Zhou}, \citenamefont {Yuan},\ and\ \citenamefont {Ma}}]{cao_PhysRevX_2016}%
  \BibitemOpen
  \bibfield  {author} {\bibinfo {author} {\bibfnamefont {Z.}~\bibnamefont {Cao}}, \bibinfo {author} {\bibfnamefont {H.}~\bibnamefont {Zhou}}, \bibinfo {author} {\bibfnamefont {X.}~\bibnamefont {Yuan}}, \ and\ \bibinfo {author} {\bibfnamefont {X.}~\bibnamefont {Ma}},\ }\bibfield  {title} {\enquote {\bibinfo {title} {Source-independent quantum random number generation},}\ }\href@noop {} {\bibfield  {journal} {\bibinfo  {journal} {Phys. Rev. X}\ }\textbf {\bibinfo {volume} {6}},\ \bibinfo {pages} {011020} (\bibinfo {year} {2016})}\BibitemShut {NoStop}%
\bibitem [{\citenamefont {Brask}\ \emph {et~al.}(2017)\citenamefont {Brask}, \citenamefont {Martin}, \citenamefont {Esposito}, \citenamefont {Houlmann}, \citenamefont {Bowles} \emph {et~al.}}]{brask_PhysRevAppl_2017}%
  \BibitemOpen
  \bibfield  {author} {\bibinfo {author} {\bibfnamefont {J.~B.}\ \bibnamefont {Brask}}, \bibinfo {author} {\bibfnamefont {A.}~\bibnamefont {Martin}}, \bibinfo {author} {\bibfnamefont {W.}~\bibnamefont {Esposito}}, \bibinfo {author} {\bibfnamefont {R.}~\bibnamefont {Houlmann}}, \bibinfo {author} {\bibfnamefont {J.}~\bibnamefont {Bowles}},  \emph {et~al.},\ }\bibfield  {title} {\enquote {\bibinfo {title} {Megahertz-rate semi-device-independent quantum random number generators based on unambiguous state discrimination},}\ }\href@noop {} {\bibfield  {journal} {\bibinfo  {journal} {Phys. Rev. Appl.}\ }\textbf {\bibinfo {volume} {7}},\ \bibinfo {pages} {054018} (\bibinfo {year} {2017})}\BibitemShut {NoStop}%
\bibitem [{\citenamefont {Van~Himbeeck}\ \emph {et~al.}(2017)\citenamefont {Van~Himbeeck}, \citenamefont {Woodhead}, \citenamefont {Cerf}, \citenamefont {Garc{\'\i}a-Patr{\'o}n},\ and\ \citenamefont {Pironio}}]{van_Quantum_2017}%
  \BibitemOpen
  \bibfield  {author} {\bibinfo {author} {\bibfnamefont {T.}~\bibnamefont {Van~Himbeeck}}, \bibinfo {author} {\bibfnamefont {E.}~\bibnamefont {Woodhead}}, \bibinfo {author} {\bibfnamefont {N.~J.}\ \bibnamefont {Cerf}}, \bibinfo {author} {\bibfnamefont {R.}~\bibnamefont {Garc{\'\i}a-Patr{\'o}n}}, \ and\ \bibinfo {author} {\bibfnamefont {S.}~\bibnamefont {Pironio}},\ }\bibfield  {title} {\enquote {\bibinfo {title} {Semi-device-independent framework based on natural physical assumptions},}\ }\href@noop {} {\bibfield  {journal} {\bibinfo  {journal} {Quantum}\ }\textbf {\bibinfo {volume} {1}},\ \bibinfo {pages} {33} (\bibinfo {year} {2017})}\BibitemShut {NoStop}%
\bibitem [{\citenamefont {Xu}\ \emph {et~al.}(2019)\citenamefont {Xu}, \citenamefont {Chen}, \citenamefont {Li}, \citenamefont {Yang}, \citenamefont {Su} \emph {et~al.}}]{xu_QuanSciTech_2019}%
  \BibitemOpen
  \bibfield  {author} {\bibinfo {author} {\bibfnamefont {B.}~\bibnamefont {Xu}}, \bibinfo {author} {\bibfnamefont {Z.}~\bibnamefont {Chen}}, \bibinfo {author} {\bibfnamefont {Z.}~\bibnamefont {Li}}, \bibinfo {author} {\bibfnamefont {J.}~\bibnamefont {Yang}}, \bibinfo {author} {\bibfnamefont {Q.}~\bibnamefont {Su}},  \emph {et~al.},\ }\bibfield  {title} {\enquote {\bibinfo {title} {High speed continuous variable source-independent quantum random number generation},}\ }\href@noop {} {\bibfield  {journal} {\bibinfo  {journal} {Quantum Sci. Technol.}\ }\textbf {\bibinfo {volume} {4}},\ \bibinfo {pages} {025013} (\bibinfo {year} {2019})}\BibitemShut {NoStop}%
\bibitem [{\citenamefont {Tebyanian}\ \emph {et~al.}(2021)\citenamefont {Tebyanian}, \citenamefont {Avesani}, \citenamefont {Vallone},\ and\ \citenamefont {Villoresi}}]{tebyanian_PhysRevA_2021}%
  \BibitemOpen
  \bibfield  {author} {\bibinfo {author} {\bibfnamefont {H.}~\bibnamefont {Tebyanian}}, \bibinfo {author} {\bibfnamefont {M.}~\bibnamefont {Avesani}}, \bibinfo {author} {\bibfnamefont {G.}~\bibnamefont {Vallone}}, \ and\ \bibinfo {author} {\bibfnamefont {P.}~\bibnamefont {Villoresi}},\ }\bibfield  {title} {\enquote {\bibinfo {title} {Semi-device-independent randomness from d-outcome continuous-variable detection},}\ }\href@noop {} {\bibfield  {journal} {\bibinfo  {journal} {Phys. Rev. A}\ }\textbf {\bibinfo {volume} {104}},\ \bibinfo {pages} {062424} (\bibinfo {year} {2021})}\BibitemShut {NoStop}%
\bibitem [{\citenamefont {Avesani}\ \emph {et~al.}(2021)\citenamefont {Avesani}, \citenamefont {Tebyanian}, \citenamefont {Villoresi},\ and\ \citenamefont {Vallone}}]{avesani_PhysRevAppl_2021}%
  \BibitemOpen
  \bibfield  {author} {\bibinfo {author} {\bibfnamefont {M.}~\bibnamefont {Avesani}}, \bibinfo {author} {\bibfnamefont {H.}~\bibnamefont {Tebyanian}}, \bibinfo {author} {\bibfnamefont {P.}~\bibnamefont {Villoresi}}, \ and\ \bibinfo {author} {\bibfnamefont {G.}~\bibnamefont {Vallone}},\ }\bibfield  {title} {\enquote {\bibinfo {title} {Semi-device-independent heterodyne-based quantum random-number generator},}\ }\href@noop {} {\bibfield  {journal} {\bibinfo  {journal} {Phys. Rev. Appl.}\ }\textbf {\bibinfo {volume} {15}},\ \bibinfo {pages} {034034} (\bibinfo {year} {2021})}\BibitemShut {NoStop}%
\bibitem [{\citenamefont {Liu}\ \emph {et~al.}(2018{\natexlab{b}})\citenamefont {Liu}, \citenamefont {Zhao}, \citenamefont {Li}, \citenamefont {Guan}, \citenamefont {Zhang} \emph {et~al.}}]{liu_Nature_2018}%
  \BibitemOpen
  \bibfield  {author} {\bibinfo {author} {\bibfnamefont {Y.}~\bibnamefont {Liu}}, \bibinfo {author} {\bibfnamefont {Q.}~\bibnamefont {Zhao}}, \bibinfo {author} {\bibfnamefont {M.-H.}\ \bibnamefont {Li}}, \bibinfo {author} {\bibfnamefont {J.-Y.}\ \bibnamefont {Guan}}, \bibinfo {author} {\bibfnamefont {Y.}~\bibnamefont {Zhang}},  \emph {et~al.},\ }\bibfield  {title} {\enquote {\bibinfo {title} {Device-independent quantum random-number generation},}\ }\href@noop {} {\bibfield  {journal} {\bibinfo  {journal} {Nature}\ }\textbf {\bibinfo {volume} {562}},\ \bibinfo {pages} {548--551} (\bibinfo {year} {2018}{\natexlab{b}})}\BibitemShut {NoStop}%
\bibitem [{\citenamefont {Liu}\ \emph {et~al.}(2021{\natexlab{b}})\citenamefont {Liu}, \citenamefont {Li}, \citenamefont {Ragy}, \citenamefont {Zhao}, \citenamefont {Bai} \emph {et~al.}}]{liu_NatPhys_2021}%
  \BibitemOpen
  \bibfield  {author} {\bibinfo {author} {\bibfnamefont {W.-Z.}\ \bibnamefont {Liu}}, \bibinfo {author} {\bibfnamefont {M.-H.}\ \bibnamefont {Li}}, \bibinfo {author} {\bibfnamefont {S.}~\bibnamefont {Ragy}}, \bibinfo {author} {\bibfnamefont {S.-R.}\ \bibnamefont {Zhao}}, \bibinfo {author} {\bibfnamefont {B.}~\bibnamefont {Bai}},  \emph {et~al.},\ }\bibfield  {title} {\enquote {\bibinfo {title} {Device-independent randomness expansion against quantum side information},}\ }\href@noop {} {\bibfield  {journal} {\bibinfo  {journal} {Nat. Phys.}\ }\textbf {\bibinfo {volume} {17}},\ \bibinfo {pages} {448--451} (\bibinfo {year} {2021}{\natexlab{b}})}\BibitemShut {NoStop}%
\bibitem [{\citenamefont {Li}\ \emph {et~al.}(2021{\natexlab{a}})\citenamefont {Li}, \citenamefont {Zhang}, \citenamefont {Liu}, \citenamefont {Zhao}, \citenamefont {Bai} \emph {et~al.}}]{li_PhysRevLett_2021}%
  \BibitemOpen
  \bibfield  {author} {\bibinfo {author} {\bibfnamefont {M.-H.}\ \bibnamefont {Li}}, \bibinfo {author} {\bibfnamefont {X.}~\bibnamefont {Zhang}}, \bibinfo {author} {\bibfnamefont {W.-Z.}\ \bibnamefont {Liu}}, \bibinfo {author} {\bibfnamefont {S.-R.}\ \bibnamefont {Zhao}}, \bibinfo {author} {\bibfnamefont {B.}~\bibnamefont {Bai}},  \emph {et~al.},\ }\bibfield  {title} {\enquote {\bibinfo {title} {Experimental realization of device-independent quantum randomness expansion},}\ }\href@noop {} {\bibfield  {journal} {\bibinfo  {journal} {Phys. Rev. Lett.}\ }\textbf {\bibinfo {volume} {126}},\ \bibinfo {pages} {050503} (\bibinfo {year} {2021}{\natexlab{a}})}\BibitemShut {NoStop}%
\bibitem [{\citenamefont {Zhang}\ \emph {et~al.}(2020{\natexlab{d}})\citenamefont {Zhang}, \citenamefont {Shalm}, \citenamefont {Bienfang}, \citenamefont {Stevens}, \citenamefont {Mazurek} \emph {et~al.}}]{zhangyb_PhysRevLett_2020}%
  \BibitemOpen
  \bibfield  {author} {\bibinfo {author} {\bibfnamefont {Y.}~\bibnamefont {Zhang}}, \bibinfo {author} {\bibfnamefont {L.~K.}\ \bibnamefont {Shalm}}, \bibinfo {author} {\bibfnamefont {J.~C.}\ \bibnamefont {Bienfang}}, \bibinfo {author} {\bibfnamefont {M.~J.}\ \bibnamefont {Stevens}}, \bibinfo {author} {\bibfnamefont {M.~D.}\ \bibnamefont {Mazurek}},  \emph {et~al.},\ }\bibfield  {title} {\enquote {\bibinfo {title} {Experimental low-latency device-independent quantum randomness},}\ }\href@noop {} {\bibfield  {journal} {\bibinfo  {journal} {Phys. Rev. Lett.}\ }\textbf {\bibinfo {volume} {124}},\ \bibinfo {pages} {010505} (\bibinfo {year} {2020}{\natexlab{d}})}\BibitemShut {NoStop}%
\bibitem [{\citenamefont {Wei}\ \emph {et~al.}(2011)\citenamefont {Wei}, \citenamefont {Xie}, \citenamefont {Dang},\ and\ \citenamefont {Guo}}]{wei_IEEEPhotTechLett_2011}%
  \BibitemOpen
  \bibfield  {author} {\bibinfo {author} {\bibfnamefont {W.}~\bibnamefont {Wei}}, \bibinfo {author} {\bibfnamefont {G.}~\bibnamefont {Xie}}, \bibinfo {author} {\bibfnamefont {A.}~\bibnamefont {Dang}}, \ and\ \bibinfo {author} {\bibfnamefont {H.}~\bibnamefont {Guo}},\ }\bibfield  {title} {\enquote {\bibinfo {title} {High-speed and bias-free optical random number generator},}\ }\href@noop {} {\bibfield  {journal} {\bibinfo  {journal} {IEEE Photon. Technol. Lett.}\ }\textbf {\bibinfo {volume} {24}},\ \bibinfo {pages} {437--439} (\bibinfo {year} {2011})}\BibitemShut {NoStop}%
\bibitem [{\citenamefont {Argyris}\ \emph {et~al.}(2012)\citenamefont {Argyris}, \citenamefont {Pikasis}, \citenamefont {Deligiannidis},\ and\ \citenamefont {Syvridis}}]{argyris_JourLigTech_2012}%
  \BibitemOpen
  \bibfield  {author} {\bibinfo {author} {\bibfnamefont {A.}~\bibnamefont {Argyris}}, \bibinfo {author} {\bibfnamefont {E.}~\bibnamefont {Pikasis}}, \bibinfo {author} {\bibfnamefont {S.}~\bibnamefont {Deligiannidis}}, \ and\ \bibinfo {author} {\bibfnamefont {D.}~\bibnamefont {Syvridis}},\ }\bibfield  {title} {\enquote {\bibinfo {title} {Sub-tb/s physical random bit generators based on direct detection of amplified spontaneous emission signals},}\ }\href@noop {} {\bibfield  {journal} {\bibinfo  {journal} {J. Light. Technol.}\ }\textbf {\bibinfo {volume} {30}},\ \bibinfo {pages} {1329--1334} (\bibinfo {year} {2012})}\BibitemShut {NoStop}%
\bibitem [{\citenamefont {Li}\ \emph {et~al.}(2014{\natexlab{b}})\citenamefont {Li}, \citenamefont {Wang}, \citenamefont {Li}, \citenamefont {Xu}, \citenamefont {Wang} \emph {et~al.}}]{li_IEEEPhotJour_2014}%
  \BibitemOpen
  \bibfield  {author} {\bibinfo {author} {\bibfnamefont {L.}~\bibnamefont {Li}}, \bibinfo {author} {\bibfnamefont {A.}~\bibnamefont {Wang}}, \bibinfo {author} {\bibfnamefont {P.}~\bibnamefont {Li}}, \bibinfo {author} {\bibfnamefont {H.}~\bibnamefont {Xu}}, \bibinfo {author} {\bibfnamefont {L.}~\bibnamefont {Wang}},  \emph {et~al.},\ }\bibfield  {title} {\enquote {\bibinfo {title} {Random bit generator using delayed self-difference of filtered amplified spontaneous emission},}\ }\href@noop {} {\bibfield  {journal} {\bibinfo  {journal} {IEEE Photon. J.}\ }\textbf {\bibinfo {volume} {6}},\ \bibinfo {pages} {1--9} (\bibinfo {year} {2014}{\natexlab{b}})}\BibitemShut {NoStop}%
\bibitem [{\citenamefont {Zhou}, \citenamefont {Yuan},\ and\ \citenamefont {Ma}(2015)}]{zhou_PhysRevA_2015}%
  \BibitemOpen
  \bibfield  {author} {\bibinfo {author} {\bibfnamefont {H.}~\bibnamefont {Zhou}}, \bibinfo {author} {\bibfnamefont {X.}~\bibnamefont {Yuan}}, \ and\ \bibinfo {author} {\bibfnamefont {X.}~\bibnamefont {Ma}},\ }\bibfield  {title} {\enquote {\bibinfo {title} {Randomness generation based on spontaneous emissions of lasers},}\ }\href@noop {} {\bibfield  {journal} {\bibinfo  {journal} {Phys. Rev. A}\ }\textbf {\bibinfo {volume} {91}},\ \bibinfo {pages} {062316} (\bibinfo {year} {2015})}\BibitemShut {NoStop}%
\bibitem [{\citenamefont {Guo}\ \emph {et~al.}(2010)\citenamefont {Guo}, \citenamefont {Tang}, \citenamefont {Liu},\ and\ \citenamefont {Wei}}]{guo_PhysRevE_2010}%
  \BibitemOpen
  \bibfield  {author} {\bibinfo {author} {\bibfnamefont {H.}~\bibnamefont {Guo}}, \bibinfo {author} {\bibfnamefont {W.}~\bibnamefont {Tang}}, \bibinfo {author} {\bibfnamefont {Y.}~\bibnamefont {Liu}}, \ and\ \bibinfo {author} {\bibfnamefont {W.}~\bibnamefont {Wei}},\ }\bibfield  {title} {\enquote {\bibinfo {title} {Truly random number generation based on measurement of phase noise of a laser},}\ }\href@noop {} {\bibfield  {journal} {\bibinfo  {journal} {Phys. Rev. E}\ }\textbf {\bibinfo {volume} {81}},\ \bibinfo {pages} {051137} (\bibinfo {year} {2010})}\BibitemShut {NoStop}%
\bibitem [{\citenamefont {Xu}\ \emph {et~al.}(2012)\citenamefont {Xu}, \citenamefont {Qi}, \citenamefont {Ma}, \citenamefont {Xu}, \citenamefont {Zheng} \emph {et~al.}}]{xu_OptExp_2012}%
  \BibitemOpen
  \bibfield  {author} {\bibinfo {author} {\bibfnamefont {F.}~\bibnamefont {Xu}}, \bibinfo {author} {\bibfnamefont {B.}~\bibnamefont {Qi}}, \bibinfo {author} {\bibfnamefont {X.}~\bibnamefont {Ma}}, \bibinfo {author} {\bibfnamefont {H.}~\bibnamefont {Xu}}, \bibinfo {author} {\bibfnamefont {H.}~\bibnamefont {Zheng}},  \emph {et~al.},\ }\bibfield  {title} {\enquote {\bibinfo {title} {Ultrafast quantum random number generation based on quantum phase fluctuations},}\ }\href@noop {} {\bibfield  {journal} {\bibinfo  {journal} {Opt. Express}\ }\textbf {\bibinfo {volume} {20}},\ \bibinfo {pages} {12366--12377} (\bibinfo {year} {2012})}\BibitemShut {NoStop}%
\bibitem [{\citenamefont {Abell{\'a}n}\ \emph {et~al.}(2014)\citenamefont {Abell{\'a}n}, \citenamefont {Amaya}, \citenamefont {Jofre}, \citenamefont {Curty}, \citenamefont {Ac{\'\i}n} \emph {et~al.}}]{abellan_OptExp_2014}%
  \BibitemOpen
  \bibfield  {author} {\bibinfo {author} {\bibfnamefont {C.}~\bibnamefont {Abell{\'a}n}}, \bibinfo {author} {\bibfnamefont {W.}~\bibnamefont {Amaya}}, \bibinfo {author} {\bibfnamefont {M.}~\bibnamefont {Jofre}}, \bibinfo {author} {\bibfnamefont {M.}~\bibnamefont {Curty}}, \bibinfo {author} {\bibfnamefont {A.}~\bibnamefont {Ac{\'\i}n}},  \emph {et~al.},\ }\bibfield  {title} {\enquote {\bibinfo {title} {Ultra-fast quantum randomness generation by accelerated phase diffusion in a pulsed laser diode},}\ }\href@noop {} {\bibfield  {journal} {\bibinfo  {journal} {Opt. Express}\ }\textbf {\bibinfo {volume} {22}},\ \bibinfo {pages} {1645--1654} (\bibinfo {year} {2014})}\BibitemShut {NoStop}%
\bibitem [{\citenamefont {Zhang}\ \emph {et~al.}(2016)\citenamefont {Zhang}, \citenamefont {Nie}, \citenamefont {Zhou}, \citenamefont {Liang}, \citenamefont {Ma} \emph {et~al.}}]{zhang_RevSciInstr_2016}%
  \BibitemOpen
  \bibfield  {author} {\bibinfo {author} {\bibfnamefont {X.-G.}\ \bibnamefont {Zhang}}, \bibinfo {author} {\bibfnamefont {Y.-Q.}\ \bibnamefont {Nie}}, \bibinfo {author} {\bibfnamefont {H.}~\bibnamefont {Zhou}}, \bibinfo {author} {\bibfnamefont {H.}~\bibnamefont {Liang}}, \bibinfo {author} {\bibfnamefont {X.}~\bibnamefont {Ma}},  \emph {et~al.},\ }\bibfield  {title} {\enquote {\bibinfo {title} {Fully integrated 3.2 gbps quantum random number generator with real-time extraction},}\ }\href@noop {} {\bibfield  {journal} {\bibinfo  {journal} {Rev. Sci. Instrum.}\ }\textbf {\bibinfo {volume} {87}} (\bibinfo {year} {2016})}\BibitemShut {NoStop}%
\bibitem [{\citenamefont {Abellan}\ \emph {et~al.}(2016)\citenamefont {Abellan}, \citenamefont {Amaya}, \citenamefont {Domenech}, \citenamefont {Mu{\~n}oz}, \citenamefont {Capmany} \emph {et~al.}}]{abellan_Optica_2016}%
  \BibitemOpen
  \bibfield  {author} {\bibinfo {author} {\bibfnamefont {C.}~\bibnamefont {Abellan}}, \bibinfo {author} {\bibfnamefont {W.}~\bibnamefont {Amaya}}, \bibinfo {author} {\bibfnamefont {D.}~\bibnamefont {Domenech}}, \bibinfo {author} {\bibfnamefont {P.}~\bibnamefont {Mu{\~n}oz}}, \bibinfo {author} {\bibfnamefont {J.}~\bibnamefont {Capmany}},  \emph {et~al.},\ }\bibfield  {title} {\enquote {\bibinfo {title} {Quantum entropy source on an inp photonic integrated circuit for random number generation},}\ }\href@noop {} {\bibfield  {journal} {\bibinfo  {journal} {Optica}\ }\textbf {\bibinfo {volume} {3}},\ \bibinfo {pages} {989--994} (\bibinfo {year} {2016})}\BibitemShut {NoStop}%
\bibitem [{\citenamefont {Liu}\ \emph {et~al.}(2016{\natexlab{a}})\citenamefont {Liu}, \citenamefont {Yang}, \citenamefont {Li}, \citenamefont {Su}, \citenamefont {Huang} \emph {et~al.}}]{liu_IEEEPhotTechLett_2016}%
  \BibitemOpen
  \bibfield  {author} {\bibinfo {author} {\bibfnamefont {J.}~\bibnamefont {Liu}}, \bibinfo {author} {\bibfnamefont {J.}~\bibnamefont {Yang}}, \bibinfo {author} {\bibfnamefont {Z.}~\bibnamefont {Li}}, \bibinfo {author} {\bibfnamefont {Q.}~\bibnamefont {Su}}, \bibinfo {author} {\bibfnamefont {W.}~\bibnamefont {Huang}},  \emph {et~al.},\ }\bibfield  {title} {\enquote {\bibinfo {title} {117 gbits/s quantum random number generation with simple structure},}\ }\href@noop {} {\bibfield  {journal} {\bibinfo  {journal} {IEEE Photon. Technol. Lett.}\ }\textbf {\bibinfo {volume} {29}},\ \bibinfo {pages} {283--286} (\bibinfo {year} {2016}{\natexlab{a}})}\BibitemShut {NoStop}%
\bibitem [{\citenamefont {Sun}\ and\ \citenamefont {Xu}(2017)}]{sun_PhysRevA_2017}%
  \BibitemOpen
  \bibfield  {author} {\bibinfo {author} {\bibfnamefont {S.-H.}\ \bibnamefont {Sun}}\ and\ \bibinfo {author} {\bibfnamefont {F.}~\bibnamefont {Xu}},\ }\bibfield  {title} {\enquote {\bibinfo {title} {Experimental study of a quantum random-number generator based on two independent lasers},}\ }\href@noop {} {\bibfield  {journal} {\bibinfo  {journal} {Phys. Rev. A}\ }\textbf {\bibinfo {volume} {96}},\ \bibinfo {pages} {062314} (\bibinfo {year} {2017})}\BibitemShut {NoStop}%
\bibitem [{\citenamefont {{\'A}lvarez}\ \emph {et~al.}(2020)\citenamefont {{\'A}lvarez}, \citenamefont {Sarmiento}, \citenamefont {L{\'a}zaro}, \citenamefont {Gen{\'e}},\ and\ \citenamefont {Torres}}]{alvarez_OptExp_2020}%
  \BibitemOpen
  \bibfield  {author} {\bibinfo {author} {\bibfnamefont {J.-R.}\ \bibnamefont {{\'A}lvarez}}, \bibinfo {author} {\bibfnamefont {S.}~\bibnamefont {Sarmiento}}, \bibinfo {author} {\bibfnamefont {J.}~\bibnamefont {L{\'a}zaro}}, \bibinfo {author} {\bibfnamefont {J.}~\bibnamefont {Gen{\'e}}}, \ and\ \bibinfo {author} {\bibfnamefont {J.}~\bibnamefont {Torres}},\ }\bibfield  {title} {\enquote {\bibinfo {title} {Random number generation by coherent detection of quantum phase noise},}\ }\href@noop {} {\bibfield  {journal} {\bibinfo  {journal} {Opt. Express}\ }\textbf {\bibinfo {volume} {28}},\ \bibinfo {pages} {5538--5547} (\bibinfo {year} {2020})}\BibitemShut {NoStop}%
\bibitem [{\citenamefont {Raffaelli}\ \emph {et~al.}(2018{\natexlab{a}})\citenamefont {Raffaelli}, \citenamefont {Sibson}, \citenamefont {Kennard}, \citenamefont {Mahler}, \citenamefont {Thompson} \emph {et~al.}}]{raffaelli_OptExp_2018}%
  \BibitemOpen
  \bibfield  {author} {\bibinfo {author} {\bibfnamefont {F.}~\bibnamefont {Raffaelli}}, \bibinfo {author} {\bibfnamefont {P.}~\bibnamefont {Sibson}}, \bibinfo {author} {\bibfnamefont {J.~E.}\ \bibnamefont {Kennard}}, \bibinfo {author} {\bibfnamefont {D.~H.}\ \bibnamefont {Mahler}}, \bibinfo {author} {\bibfnamefont {M.~G.}\ \bibnamefont {Thompson}},  \emph {et~al.},\ }\bibfield  {title} {\enquote {\bibinfo {title} {Generation of random numbers by measuring phase fluctuations from a laser diode with a silicon-on-insulator chip},}\ }\href@noop {} {\bibfield  {journal} {\bibinfo  {journal} {Opt. Express}\ }\textbf {\bibinfo {volume} {26}},\ \bibinfo {pages} {19730--19741} (\bibinfo {year} {2018}{\natexlab{a}})}\BibitemShut {NoStop}%
\bibitem [{\citenamefont {Chrysostomidis}\ \emph {et~al.}(2023)\citenamefont {Chrysostomidis}, \citenamefont {Roumpos}, \citenamefont {Outerelo}, \citenamefont {Troncoso-Costas}, \citenamefont {Moskalenko} \emph {et~al.}}]{chrysostomidis_EPJQuanTech_2023}%
  \BibitemOpen
  \bibfield  {author} {\bibinfo {author} {\bibfnamefont {T.}~\bibnamefont {Chrysostomidis}}, \bibinfo {author} {\bibfnamefont {I.}~\bibnamefont {Roumpos}}, \bibinfo {author} {\bibfnamefont {D.~A.}\ \bibnamefont {Outerelo}}, \bibinfo {author} {\bibfnamefont {M.}~\bibnamefont {Troncoso-Costas}}, \bibinfo {author} {\bibfnamefont {V.}~\bibnamefont {Moskalenko}},  \emph {et~al.},\ }\bibfield  {title} {\enquote {\bibinfo {title} {Long term experimental verification of a single chip quantum random number generator fabricated on the inp platform},}\ }\href@noop {} {\bibfield  {journal} {\bibinfo  {journal} {EPJ Quantum Technol.}\ }\textbf {\bibinfo {volume} {10}},\ \bibinfo {pages} {5} (\bibinfo {year} {2023})}\BibitemShut {NoStop}%
\bibitem [{\citenamefont {Jofre}\ \emph {et~al.}(2011)\citenamefont {Jofre}, \citenamefont {Curty}, \citenamefont {Steinlechner}, \citenamefont {Anzolin}, \citenamefont {Torres} \emph {et~al.}}]{jofre_OptExp_2011}%
  \BibitemOpen
  \bibfield  {author} {\bibinfo {author} {\bibfnamefont {M.}~\bibnamefont {Jofre}}, \bibinfo {author} {\bibfnamefont {M.}~\bibnamefont {Curty}}, \bibinfo {author} {\bibfnamefont {F.}~\bibnamefont {Steinlechner}}, \bibinfo {author} {\bibfnamefont {G.}~\bibnamefont {Anzolin}}, \bibinfo {author} {\bibfnamefont {J.}~\bibnamefont {Torres}},  \emph {et~al.},\ }\bibfield  {title} {\enquote {\bibinfo {title} {True random numbers from amplified quantum vacuum},}\ }\href@noop {} {\bibfield  {journal} {\bibinfo  {journal} {Opt. Express}\ }\textbf {\bibinfo {volume} {19}},\ \bibinfo {pages} {20665--20672} (\bibinfo {year} {2011})}\BibitemShut {NoStop}%
\bibitem [{\citenamefont {Haw}\ \emph {et~al.}(2015)\citenamefont {Haw}, \citenamefont {Assad}, \citenamefont {Lance}, \citenamefont {Ng}, \citenamefont {Sharma} \emph {et~al.}}]{haw_PhysRevAppl_2015}%
  \BibitemOpen
  \bibfield  {author} {\bibinfo {author} {\bibfnamefont {J.-Y.}\ \bibnamefont {Haw}}, \bibinfo {author} {\bibfnamefont {S.}~\bibnamefont {Assad}}, \bibinfo {author} {\bibfnamefont {A.}~\bibnamefont {Lance}}, \bibinfo {author} {\bibfnamefont {N.}~\bibnamefont {Ng}}, \bibinfo {author} {\bibfnamefont {V.}~\bibnamefont {Sharma}},  \emph {et~al.},\ }\bibfield  {title} {\enquote {\bibinfo {title} {Maximization of extractable randomness in a quantum random-number generator},}\ }\href@noop {} {\bibfield  {journal} {\bibinfo  {journal} {Phys. Rev. Appl.}\ }\textbf {\bibinfo {volume} {3}},\ \bibinfo {pages} {054004} (\bibinfo {year} {2015})}\BibitemShut {NoStop}%
\bibitem [{\citenamefont {Raffaelli}\ \emph {et~al.}(2018{\natexlab{b}})\citenamefont {Raffaelli}, \citenamefont {Ferranti}, \citenamefont {Mahler}, \citenamefont {Sibson}, \citenamefont {Kennard} \emph {et~al.}}]{raffaelli_QuanSciTech_2018}%
  \BibitemOpen
  \bibfield  {author} {\bibinfo {author} {\bibfnamefont {F.}~\bibnamefont {Raffaelli}}, \bibinfo {author} {\bibfnamefont {G.}~\bibnamefont {Ferranti}}, \bibinfo {author} {\bibfnamefont {D.~H.}\ \bibnamefont {Mahler}}, \bibinfo {author} {\bibfnamefont {P.}~\bibnamefont {Sibson}}, \bibinfo {author} {\bibfnamefont {J.~E.}\ \bibnamefont {Kennard}},  \emph {et~al.},\ }\bibfield  {title} {\enquote {\bibinfo {title} {A homodyne detector integrated onto a photonic chip for measuring quantum states and generating random numbers},}\ }\href@noop {} {\bibfield  {journal} {\bibinfo  {journal} {Quantum Sci. Technol.}\ }\textbf {\bibinfo {volume} {3}},\ \bibinfo {pages} {025003} (\bibinfo {year} {2018}{\natexlab{b}})}\BibitemShut {NoStop}%
\bibitem [{\citenamefont {Bruynsteen}\ \emph {et~al.}(2021{\natexlab{a}})\citenamefont {Bruynsteen}, \citenamefont {Vanhoecke}, \citenamefont {Bauwelinck},\ and\ \citenamefont {Yin}}]{bruynsteen_Optica_2021}%
  \BibitemOpen
  \bibfield  {author} {\bibinfo {author} {\bibfnamefont {C.}~\bibnamefont {Bruynsteen}}, \bibinfo {author} {\bibfnamefont {M.}~\bibnamefont {Vanhoecke}}, \bibinfo {author} {\bibfnamefont {J.}~\bibnamefont {Bauwelinck}}, \ and\ \bibinfo {author} {\bibfnamefont {X.}~\bibnamefont {Yin}},\ }\bibfield  {title} {\enquote {\bibinfo {title} {Integrated balanced homodyne photonic--electronic detector for beyond 20 ghz shot-noise-limited measurements},}\ }\href@noop {} {\bibfield  {journal} {\bibinfo  {journal} {Optica}\ }\textbf {\bibinfo {volume} {8}},\ \bibinfo {pages} {1146--1152} (\bibinfo {year} {2021}{\natexlab{a}})}\BibitemShut {NoStop}%
\bibitem [{\citenamefont {Muller}(1958)}]{muller1958inverse}%
  \BibitemOpen
  \bibfield  {author} {\bibinfo {author} {\bibfnamefont {M.~E.}\ \bibnamefont {Muller}},\ }\bibfield  {title} {\enquote {\bibinfo {title} {An inverse method for the generation of random normal deviates on large-scale computers},}\ }\href@noop {} {\bibfield  {journal} {\bibinfo  {journal} {Math. Comput.}\ }\textbf {\bibinfo {volume} {12}},\ \bibinfo {pages} {167--174} (\bibinfo {year} {1958})}\BibitemShut {NoStop}%
\bibitem [{\citenamefont {Box}\ and\ \citenamefont {Muller}(1958)}]{box1958note}%
  \BibitemOpen
  \bibfield  {author} {\bibinfo {author} {\bibfnamefont {G.~E.}\ \bibnamefont {Box}}\ and\ \bibinfo {author} {\bibfnamefont {M.~E.}\ \bibnamefont {Muller}},\ }\bibfield  {title} {\enquote {\bibinfo {title} {A note on the generation of random normal deviates},}\ }\href@noop {} {\bibfield  {journal} {\bibinfo  {journal} {The annals of mathematical statistics}\ }\textbf {\bibinfo {volume} {29}},\ \bibinfo {pages} {610--611} (\bibinfo {year} {1958})}\BibitemShut {NoStop}%
\bibitem [{\citenamefont {Teichroew}(1953)}]{teichroew1953distribution}%
  \BibitemOpen
  \bibfield  {author} {\bibinfo {author} {\bibfnamefont {D.}~\bibnamefont {Teichroew}},\ }\emph {\bibinfo {title} {Distribution sampling with high speed computers}},\ \href@noop {} {Ph.D. thesis},\ \bibinfo  {school} {North Carolina State College} (\bibinfo {year} {1953})\BibitemShut {NoStop}%
\bibitem [{\citenamefont {Kabal}(2000)}]{kabal2000generating}%
  \BibitemOpen
  \bibfield  {author} {\bibinfo {author} {\bibfnamefont {P.}~\bibnamefont {Kabal}},\ }\bibfield  {title} {\enquote {\bibinfo {title} {Generating gaussian pseudo-random deviates},}\ }\href@noop {} {\bibfield  {journal} {\bibinfo  {journal} {Department of Electrical and Computer Engineering, McGill University, Tech. Rep}\ } (\bibinfo {year} {2000})}\BibitemShut {NoStop}%
\bibitem [{\citenamefont {Knuth}(1981)}]{knuth1981art}%
  \BibitemOpen
  \bibfield  {author} {\bibinfo {author} {\bibfnamefont {D.}~\bibnamefont {Knuth}},\ }\bibfield  {title} {\enquote {\bibinfo {title} {The art of computer programming, 2 (seminumerical algorithms)},}\ }\href@noop {} {\bibfield  {journal} {\bibinfo  {journal} {(No Title)}\ } (\bibinfo {year} {1981})}\BibitemShut {NoStop}%
\bibitem [{\citenamefont {Kumar}, \citenamefont {Qin},\ and\ \citenamefont {All{\'e}aume}(2015)}]{Kumar_NewJPhys_2015}%
  \BibitemOpen
  \bibfield  {author} {\bibinfo {author} {\bibfnamefont {R.}~\bibnamefont {Kumar}}, \bibinfo {author} {\bibfnamefont {H.}~\bibnamefont {Qin}}, \ and\ \bibinfo {author} {\bibfnamefont {R.}~\bibnamefont {All{\'e}aume}},\ }\bibfield  {title} {\enquote {\bibinfo {title} {Coexistence of continuous variable qkd with intense dwdm classical channels},}\ }\href@noop {} {\bibfield  {journal} {\bibinfo  {journal} {New J. Phys.}\ }\textbf {\bibinfo {volume} {17}},\ \bibinfo {pages} {043027} (\bibinfo {year} {2015})}\BibitemShut {NoStop}%
\bibitem [{\citenamefont {Eriksson}\ \emph {et~al.}(2019)\citenamefont {Eriksson}, \citenamefont {Hirano}, \citenamefont {Puttnam}, \citenamefont {Rademacher}, \citenamefont {Lu{\'\i}s} \emph {et~al.}}]{Eriksson_CommunPhys_2019}%
  \BibitemOpen
  \bibfield  {author} {\bibinfo {author} {\bibfnamefont {T.~A.}\ \bibnamefont {Eriksson}}, \bibinfo {author} {\bibfnamefont {T.}~\bibnamefont {Hirano}}, \bibinfo {author} {\bibfnamefont {B.~J.}\ \bibnamefont {Puttnam}}, \bibinfo {author} {\bibfnamefont {G.}~\bibnamefont {Rademacher}}, \bibinfo {author} {\bibfnamefont {R.~S.}\ \bibnamefont {Lu{\'\i}s}},  \emph {et~al.},\ }\bibfield  {title} {\enquote {\bibinfo {title} {Wavelength division multiplexing of continuous variable quantum key distribution and 18.3 tbit/s data channels},}\ }\href@noop {} {\bibfield  {journal} {\bibinfo  {journal} {Commun. Phys.}\ }\textbf {\bibinfo {volume} {2}},\ \bibinfo {pages} {1--8} (\bibinfo {year} {2019})}\BibitemShut {NoStop}%
\bibitem [{\citenamefont {Huang}\ \emph {et~al.}(2015{\natexlab{a}})\citenamefont {Huang}, \citenamefont {Huang}, \citenamefont {Lin}, \citenamefont {Wang},\ and\ \citenamefont {Zeng}}]{Huang_OptLett_2015}%
  \BibitemOpen
  \bibfield  {author} {\bibinfo {author} {\bibfnamefont {D.}~\bibnamefont {Huang}}, \bibinfo {author} {\bibfnamefont {P.}~\bibnamefont {Huang}}, \bibinfo {author} {\bibfnamefont {D.}~\bibnamefont {Lin}}, \bibinfo {author} {\bibfnamefont {C.}~\bibnamefont {Wang}}, \ and\ \bibinfo {author} {\bibfnamefont {G.}~\bibnamefont {Zeng}},\ }\bibfield  {title} {\enquote {\bibinfo {title} {High-speed continuous-variable quantum key distribution without sending a local oscillator},}\ }\href@noop {} {\bibfield  {journal} {\bibinfo  {journal} {Opt. Lett.}\ }\textbf {\bibinfo {volume} {40}},\ \bibinfo {pages} {3695--3698} (\bibinfo {year} {2015}{\natexlab{a}})}\BibitemShut {NoStop}%
\bibitem [{\citenamefont {Aldama}\ \emph {et~al.}(2023{\natexlab{a}})\citenamefont {Aldama}, \citenamefont {Sarmiento}, \citenamefont {Etcheverry}, \citenamefont {Valivarthi}, \citenamefont {Grande} \emph {et~al.}}]{aldama2023small}%
  \BibitemOpen
  \bibfield  {author} {\bibinfo {author} {\bibfnamefont {J.}~\bibnamefont {Aldama}}, \bibinfo {author} {\bibfnamefont {S.}~\bibnamefont {Sarmiento}}, \bibinfo {author} {\bibfnamefont {S.}~\bibnamefont {Etcheverry}}, \bibinfo {author} {\bibfnamefont {R.}~\bibnamefont {Valivarthi}}, \bibinfo {author} {\bibfnamefont {I.~L.}\ \bibnamefont {Grande}},  \emph {et~al.},\ }\bibfield  {title} {\enquote {\bibinfo {title} {Small-form-factor gaussian-modulated coherent-state transmitter for cv-qkd using a gain-switched dfb laser},}\ }\href@noop {} {\bibfield  {journal} {\bibinfo  {journal} {Opt. Express}\ }\textbf {\bibinfo {volume} {31}},\ \bibinfo {pages} {5414--5425} (\bibinfo {year} {2023}{\natexlab{a}})}\BibitemShut {NoStop}%
\bibitem [{\citenamefont {Wang}\ \emph {et~al.}(2015{\natexlab{c}})\citenamefont {Wang}, \citenamefont {Liu}, \citenamefont {Li},\ and\ \citenamefont {Li}}]{Wang_JQuantumElectron_2015}%
  \BibitemOpen
  \bibfield  {author} {\bibinfo {author} {\bibfnamefont {X.}~\bibnamefont {Wang}}, \bibinfo {author} {\bibfnamefont {J.}~\bibnamefont {Liu}}, \bibinfo {author} {\bibfnamefont {X.}~\bibnamefont {Li}}, \ and\ \bibinfo {author} {\bibfnamefont {Y.}~\bibnamefont {Li}},\ }\bibfield  {title} {\enquote {\bibinfo {title} {Generation of stable and high extinction ratio light pulses for continuous variable quantum key distribution},}\ }\href@noop {} {\bibfield  {journal} {\bibinfo  {journal} {IEEE J. Quantum Electron.}\ }\textbf {\bibinfo {volume} {51}},\ \bibinfo {pages} {1--6} (\bibinfo {year} {2015}{\natexlab{c}})}\BibitemShut {NoStop}%
\bibitem [{\citenamefont {Brunner}\ \emph {et~al.}(2017)\citenamefont {Brunner}, \citenamefont {Comandar}, \citenamefont {Karinou}, \citenamefont {Bettelli}, \citenamefont {Hillerkuss} \emph {et~al.}}]{Brunner_ICTON_2017}%
  \BibitemOpen
  \bibfield  {author} {\bibinfo {author} {\bibfnamefont {H.~H.}\ \bibnamefont {Brunner}}, \bibinfo {author} {\bibfnamefont {L.~C.}\ \bibnamefont {Comandar}}, \bibinfo {author} {\bibfnamefont {F.}~\bibnamefont {Karinou}}, \bibinfo {author} {\bibfnamefont {S.}~\bibnamefont {Bettelli}}, \bibinfo {author} {\bibfnamefont {D.}~\bibnamefont {Hillerkuss}},  \emph {et~al.},\ }\bibfield  {title} {\enquote {\bibinfo {title} {A low-complexity heterodyne cv-qkd architecture},}\ }in\ \href@noop {} {\emph {\bibinfo {booktitle} {2017 19th International Conference on Transparent Optical Networks (ICTON)}}}\ (\bibinfo {organization} {IEEE},\ \bibinfo {year} {2017})\ pp.\ \bibinfo {pages} {1--4}\BibitemShut {NoStop}%
\bibitem [{\citenamefont {Kleis}, \citenamefont {Rueckmann},\ and\ \citenamefont {Schaeffer}(2017)}]{Kleis_OptLett_2017}%
  \BibitemOpen
  \bibfield  {author} {\bibinfo {author} {\bibfnamefont {S.}~\bibnamefont {Kleis}}, \bibinfo {author} {\bibfnamefont {M.}~\bibnamefont {Rueckmann}}, \ and\ \bibinfo {author} {\bibfnamefont {C.~G.}\ \bibnamefont {Schaeffer}},\ }\bibfield  {title} {\enquote {\bibinfo {title} {Continuous variable quantum key distribution with a real local oscillator using simultaneous pilot signals},}\ }\href@noop {} {\bibfield  {journal} {\bibinfo  {journal} {Opt. Lett.}\ }\textbf {\bibinfo {volume} {42}},\ \bibinfo {pages} {1588--1591} (\bibinfo {year} {2017})}\BibitemShut {NoStop}%
\bibitem [{\citenamefont {Wang}\ \emph {et~al.}(2020{\natexlab{a}})\citenamefont {Wang}, \citenamefont {Pi}, \citenamefont {Huang}, \citenamefont {Li}, \citenamefont {Shao} \emph {et~al.}}]{Wang_OptExpress_2020}%
  \BibitemOpen
  \bibfield  {author} {\bibinfo {author} {\bibfnamefont {H.}~\bibnamefont {Wang}}, \bibinfo {author} {\bibfnamefont {Y.}~\bibnamefont {Pi}}, \bibinfo {author} {\bibfnamefont {W.}~\bibnamefont {Huang}}, \bibinfo {author} {\bibfnamefont {Y.}~\bibnamefont {Li}}, \bibinfo {author} {\bibfnamefont {Y.}~\bibnamefont {Shao}},  \emph {et~al.},\ }\bibfield  {title} {\enquote {\bibinfo {title} {High-speed gaussian-modulated continuous-variable quantum key distribution with a local local oscillator based on pilot-tone-assisted phase compensation},}\ }\href@noop {} {\bibfield  {journal} {\bibinfo  {journal} {Opt. Express}\ }\textbf {\bibinfo {volume} {28}},\ \bibinfo {pages} {32882--32893} (\bibinfo {year} {2020}{\natexlab{a}})}\BibitemShut {NoStop}%
\bibitem [{\citenamefont {Pan}\ \emph {et~al.}(2022)\citenamefont {Pan}, \citenamefont {Wang}, \citenamefont {Shao}, \citenamefont {Pi}, \citenamefont {Li} \emph {et~al.}}]{Pan2022DM}%
  \BibitemOpen
  \bibfield  {author} {\bibinfo {author} {\bibfnamefont {Y.}~\bibnamefont {Pan}}, \bibinfo {author} {\bibfnamefont {H.}~\bibnamefont {Wang}}, \bibinfo {author} {\bibfnamefont {Y.}~\bibnamefont {Shao}}, \bibinfo {author} {\bibfnamefont {Y.}~\bibnamefont {Pi}}, \bibinfo {author} {\bibfnamefont {Y.}~\bibnamefont {Li}},  \emph {et~al.},\ }\bibfield  {title} {\enquote {\bibinfo {title} {Experimental demonstration of high-rate discrete-modulated continuous-variable quantum key distribution system},}\ }\href {\doibase 10.1364/OL.456978} {\bibfield  {journal} {\bibinfo  {journal} {Opt. Lett.}\ }\textbf {\bibinfo {volume} {47}},\ \bibinfo {pages} {3307--3310} (\bibinfo {year} {2022})}\BibitemShut {NoStop}%
\bibitem [{\citenamefont {Jain}\ \emph {et~al.}(2022)\citenamefont {Jain}, \citenamefont {Chin}, \citenamefont {Mani}, \citenamefont {Lupo}, \citenamefont {Nikolic} \emph {et~al.}}]{jain2022practical}%
  \BibitemOpen
  \bibfield  {author} {\bibinfo {author} {\bibfnamefont {N.}~\bibnamefont {Jain}}, \bibinfo {author} {\bibfnamefont {H.-M.}\ \bibnamefont {Chin}}, \bibinfo {author} {\bibfnamefont {H.}~\bibnamefont {Mani}}, \bibinfo {author} {\bibfnamefont {C.}~\bibnamefont {Lupo}}, \bibinfo {author} {\bibfnamefont {D.~S.}\ \bibnamefont {Nikolic}},  \emph {et~al.},\ }\bibfield  {title} {\enquote {\bibinfo {title} {Practical continuous-variable quantum key distribution with composable security},}\ }\href@noop {} {\bibfield  {journal} {\bibinfo  {journal} {Nat. Commun.}\ }\textbf {\bibinfo {volume} {13}},\ \bibinfo {pages} {4740} (\bibinfo {year} {2022})}\BibitemShut {NoStop}%
\bibitem [{\citenamefont {Huang}\ \emph {et~al.}(2018)\citenamefont {Huang}, \citenamefont {Huang}, \citenamefont {Zhang},\ and\ \citenamefont {Zeng}}]{huang2018quantum}%
  \BibitemOpen
  \bibfield  {author} {\bibinfo {author} {\bibfnamefont {P.}~\bibnamefont {Huang}}, \bibinfo {author} {\bibfnamefont {J.}~\bibnamefont {Huang}}, \bibinfo {author} {\bibfnamefont {Z.}~\bibnamefont {Zhang}}, \ and\ \bibinfo {author} {\bibfnamefont {G.}~\bibnamefont {Zeng}},\ }\bibfield  {title} {\enquote {\bibinfo {title} {Quantum key distribution using basis encoding of gaussian-modulated coherent states},}\ }\href@noop {} {\bibfield  {journal} {\bibinfo  {journal} {Phys. Rev. A}\ }\textbf {\bibinfo {volume} {97}},\ \bibinfo {pages} {042311} (\bibinfo {year} {2018})}\BibitemShut {NoStop}%
\bibitem [{\citenamefont {Jouguet}\ \emph {et~al.}(2012{\natexlab{b}})\citenamefont {Jouguet}, \citenamefont {Kunz-Jacques}, \citenamefont {Diamanti},\ and\ \citenamefont {Leverrier}}]{jouguet2012analysis}%
  \BibitemOpen
  \bibfield  {author} {\bibinfo {author} {\bibfnamefont {P.}~\bibnamefont {Jouguet}}, \bibinfo {author} {\bibfnamefont {S.}~\bibnamefont {Kunz-Jacques}}, \bibinfo {author} {\bibfnamefont {E.}~\bibnamefont {Diamanti}}, \ and\ \bibinfo {author} {\bibfnamefont {A.}~\bibnamefont {Leverrier}},\ }\bibfield  {title} {\enquote {\bibinfo {title} {Analysis of imperfections in practical continuous-variable quantum key distribution},}\ }\href@noop {} {\bibfield  {journal} {\bibinfo  {journal} {Phys. Rev. A}\ }\textbf {\bibinfo {volume} {86}},\ \bibinfo {pages} {032309} (\bibinfo {year} {2012}{\natexlab{b}})}\BibitemShut {NoStop}%
\bibitem [{\citenamefont {Liu}\ \emph {et~al.}(2017{\natexlab{a}})\citenamefont {Liu}, \citenamefont {Wang}, \citenamefont {Wang}, \citenamefont {Du},\ and\ \citenamefont {Li}}]{Liu_PhysRevA_2017}%
  \BibitemOpen
  \bibfield  {author} {\bibinfo {author} {\bibfnamefont {W.}~\bibnamefont {Liu}}, \bibinfo {author} {\bibfnamefont {X.}~\bibnamefont {Wang}}, \bibinfo {author} {\bibfnamefont {N.}~\bibnamefont {Wang}}, \bibinfo {author} {\bibfnamefont {S.}~\bibnamefont {Du}}, \ and\ \bibinfo {author} {\bibfnamefont {Y.}~\bibnamefont {Li}},\ }\bibfield  {title} {\enquote {\bibinfo {title} {Imperfect state preparation in continuous-variable quantum key distribution},}\ }\href@noop {} {\bibfield  {journal} {\bibinfo  {journal} {Phys. Rev. A}\ }\textbf {\bibinfo {volume} {96}},\ \bibinfo {pages} {042312} (\bibinfo {year} {2017}{\natexlab{a}})}\BibitemShut {NoStop}%
\bibitem [{\citenamefont {Qi}\ \emph {et~al.}(2020)\citenamefont {Qi}, \citenamefont {Gunther}, \citenamefont {Evans}, \citenamefont {Williams}, \citenamefont {Camacho} \emph {et~al.}}]{qi2020experimental}%
  \BibitemOpen
  \bibfield  {author} {\bibinfo {author} {\bibfnamefont {B.}~\bibnamefont {Qi}}, \bibinfo {author} {\bibfnamefont {H.}~\bibnamefont {Gunther}}, \bibinfo {author} {\bibfnamefont {P.~G.}\ \bibnamefont {Evans}}, \bibinfo {author} {\bibfnamefont {B.~P.}\ \bibnamefont {Williams}}, \bibinfo {author} {\bibfnamefont {R.~M.}\ \bibnamefont {Camacho}},  \emph {et~al.},\ }\bibfield  {title} {\enquote {\bibinfo {title} {Experimental passive-state preparation for continuous-variable quantum communications},}\ }\href@noop {} {\bibfield  {journal} {\bibinfo  {journal} {Phys. Rev. Appl.}\ }\textbf {\bibinfo {volume} {13}},\ \bibinfo {pages} {054065} (\bibinfo {year} {2020})}\BibitemShut {NoStop}%
\bibitem [{\citenamefont {Huang}\ \emph {et~al.}(2021)\citenamefont {Huang}, \citenamefont {Wang}, \citenamefont {Chen}, \citenamefont {Wang}, \citenamefont {Zhou} \emph {et~al.}}]{huang2021experimental}%
  \BibitemOpen
  \bibfield  {author} {\bibinfo {author} {\bibfnamefont {P.}~\bibnamefont {Huang}}, \bibinfo {author} {\bibfnamefont {T.}~\bibnamefont {Wang}}, \bibinfo {author} {\bibfnamefont {R.}~\bibnamefont {Chen}}, \bibinfo {author} {\bibfnamefont {P.}~\bibnamefont {Wang}}, \bibinfo {author} {\bibfnamefont {Y.}~\bibnamefont {Zhou}},  \emph {et~al.},\ }\bibfield  {title} {\enquote {\bibinfo {title} {Experimental continuous-variable quantum key distribution using a thermal source},}\ }\href@noop {} {\bibfield  {journal} {\bibinfo  {journal} {New J. Phys.}\ }\textbf {\bibinfo {volume} {23}},\ \bibinfo {pages} {113028} (\bibinfo {year} {2021})}\BibitemShut {NoStop}%
\bibitem [{\citenamefont {Qi}\ \emph {et~al.}(2007)\citenamefont {Qi}, \citenamefont {Huang}, \citenamefont {Qian},\ and\ \citenamefont {Lo}}]{Qi_PhysRevA_2007}%
  \BibitemOpen
  \bibfield  {author} {\bibinfo {author} {\bibfnamefont {B.}~\bibnamefont {Qi}}, \bibinfo {author} {\bibfnamefont {L.-L.}\ \bibnamefont {Huang}}, \bibinfo {author} {\bibfnamefont {L.}~\bibnamefont {Qian}}, \ and\ \bibinfo {author} {\bibfnamefont {H.-K.}\ \bibnamefont {Lo}},\ }\bibfield  {title} {\enquote {\bibinfo {title} {Experimental study on the gaussian-modulated coherent-state quantum key distribution over standard telecommunication fibers},}\ }\href@noop {} {\bibfield  {journal} {\bibinfo  {journal} {Phys. Rev. A}\ }\textbf {\bibinfo {volume} {76}},\ \bibinfo {pages} {052323} (\bibinfo {year} {2007})}\BibitemShut {NoStop}%
\bibitem [{\citenamefont {Li}\ \emph {et~al.}(2017{\natexlab{a}})\citenamefont {Li}, \citenamefont {Wang}, \citenamefont {Bai}, \citenamefont {Liu}, \citenamefont {Yang} \emph {et~al.}}]{li2017continuous}%
  \BibitemOpen
  \bibfield  {author} {\bibinfo {author} {\bibfnamefont {Y.-M.}\ \bibnamefont {Li}}, \bibinfo {author} {\bibfnamefont {X.-Y.}\ \bibnamefont {Wang}}, \bibinfo {author} {\bibfnamefont {Z.-L.}\ \bibnamefont {Bai}}, \bibinfo {author} {\bibfnamefont {W.-Y.}\ \bibnamefont {Liu}}, \bibinfo {author} {\bibfnamefont {S.-S.}\ \bibnamefont {Yang}},  \emph {et~al.},\ }\bibfield  {title} {\enquote {\bibinfo {title} {Continuous variable quantum key distribution},}\ }\href@noop {} {\bibfield  {journal} {\bibinfo  {journal} {Chinese Physics B}\ }\textbf {\bibinfo {volume} {26}},\ \bibinfo {pages} {040303} (\bibinfo {year} {2017}{\natexlab{a}})}\BibitemShut {NoStop}%
\bibitem [{\citenamefont {Wang}\ \emph {et~al.}(2018{\natexlab{a}})\citenamefont {Wang}, \citenamefont {Huang}, \citenamefont {Zhou}, \citenamefont {Liu}, \citenamefont {Ma} \emph {et~al.}}]{Wang_OptExpress_2018}%
  \BibitemOpen
  \bibfield  {author} {\bibinfo {author} {\bibfnamefont {T.}~\bibnamefont {Wang}}, \bibinfo {author} {\bibfnamefont {P.}~\bibnamefont {Huang}}, \bibinfo {author} {\bibfnamefont {Y.}~\bibnamefont {Zhou}}, \bibinfo {author} {\bibfnamefont {W.}~\bibnamefont {Liu}}, \bibinfo {author} {\bibfnamefont {H.}~\bibnamefont {Ma}},  \emph {et~al.},\ }\bibfield  {title} {\enquote {\bibinfo {title} {High key rate continuous-variable quantum key distribution with a real local oscillator},}\ }\href@noop {} {\bibfield  {journal} {\bibinfo  {journal} {Opt. Express}\ }\textbf {\bibinfo {volume} {26}},\ \bibinfo {pages} {2794--2806} (\bibinfo {year} {2018}{\natexlab{a}})}\BibitemShut {NoStop}%
\bibitem [{\citenamefont {Laudenbach}\ \emph {et~al.}(2018)\citenamefont {Laudenbach}, \citenamefont {Pacher}, \citenamefont {Fung}, \citenamefont {Poppe}, \citenamefont {Peev} \emph {et~al.}}]{Laudenbach_AdvQuanTech_2018}%
  \BibitemOpen
  \bibfield  {author} {\bibinfo {author} {\bibfnamefont {F.}~\bibnamefont {Laudenbach}}, \bibinfo {author} {\bibfnamefont {C.}~\bibnamefont {Pacher}}, \bibinfo {author} {\bibfnamefont {C.-H.~F.}\ \bibnamefont {Fung}}, \bibinfo {author} {\bibfnamefont {A.}~\bibnamefont {Poppe}}, \bibinfo {author} {\bibfnamefont {M.}~\bibnamefont {Peev}},  \emph {et~al.},\ }\bibfield  {title} {\enquote {\bibinfo {title} {Continuous-variable quantum key distribution with gaussian modulation: The theory of practical implementations},}\ }\href@noop {} {\bibfield  {journal} {\bibinfo  {journal} {Adv. Quantum Technol.}\ }\textbf {\bibinfo {volume} {1}},\ \bibinfo {pages} {1800011} (\bibinfo {year} {2018})}\BibitemShut {NoStop}%
\bibitem [{\citenamefont {Chu}\ \emph {et~al.}(2021{\natexlab{a}})\citenamefont {Chu}, \citenamefont {Zhang}, \citenamefont {Huang}, \citenamefont {Yu}, \citenamefont {Chen} \emph {et~al.}}]{Chu_QuantumSciTech_2021}%
  \BibitemOpen
  \bibfield  {author} {\bibinfo {author} {\bibfnamefont {B.}~\bibnamefont {Chu}}, \bibinfo {author} {\bibfnamefont {Y.}~\bibnamefont {Zhang}}, \bibinfo {author} {\bibfnamefont {Y.}~\bibnamefont {Huang}}, \bibinfo {author} {\bibfnamefont {S.}~\bibnamefont {Yu}}, \bibinfo {author} {\bibfnamefont {Z.}~\bibnamefont {Chen}},  \emph {et~al.},\ }\bibfield  {title} {\enquote {\bibinfo {title} {Practical source monitoring for continuous-variable quantum key distribution},}\ }\href {\doibase 10.1088/2058-9565/abda8f} {\bibfield  {journal} {\bibinfo  {journal} {Quantum Sci. Technol.}\ }\textbf {\bibinfo {volume} {6}},\ \bibinfo {pages} {025012} (\bibinfo {year} {2021}{\natexlab{a}})}\BibitemShut {NoStop}%
\bibitem [{\citenamefont {Weedbrook}\ \emph {et~al.}(2010)\citenamefont {Weedbrook}, \citenamefont {Pirandola}, \citenamefont {Lloyd},\ and\ \citenamefont {Ralph}}]{Weedbrook_PhysRevLett_2010}%
  \BibitemOpen
  \bibfield  {author} {\bibinfo {author} {\bibfnamefont {C.}~\bibnamefont {Weedbrook}}, \bibinfo {author} {\bibfnamefont {S.}~\bibnamefont {Pirandola}}, \bibinfo {author} {\bibfnamefont {S.}~\bibnamefont {Lloyd}}, \ and\ \bibinfo {author} {\bibfnamefont {T.~C.}\ \bibnamefont {Ralph}},\ }\bibfield  {title} {\enquote {\bibinfo {title} {Quantum cryptography approaching the classical limit},}\ }\href {\doibase 10.1103/PhysRevLett.105.110501} {\bibfield  {journal} {\bibinfo  {journal} {Phys. Rev. Lett.}\ }\textbf {\bibinfo {volume} {105}},\ \bibinfo {pages} {110501} (\bibinfo {year} {2010})}\BibitemShut {NoStop}%
\bibitem [{\citenamefont {Weedbrook}, \citenamefont {Pirandola},\ and\ \citenamefont {Ralph}(2012)}]{Weedbrook_PhysRevA_2012}%
  \BibitemOpen
  \bibfield  {author} {\bibinfo {author} {\bibfnamefont {C.}~\bibnamefont {Weedbrook}}, \bibinfo {author} {\bibfnamefont {S.}~\bibnamefont {Pirandola}}, \ and\ \bibinfo {author} {\bibfnamefont {T.~C.}\ \bibnamefont {Ralph}},\ }\bibfield  {title} {\enquote {\bibinfo {title} {Continuous-variable quantum key distribution using thermal states},}\ }\href {\doibase 10.1103/PhysRevA.86.022318} {\bibfield  {journal} {\bibinfo  {journal} {Phys. Rev. A}\ }\textbf {\bibinfo {volume} {86}},\ \bibinfo {pages} {022318} (\bibinfo {year} {2012})}\BibitemShut {NoStop}%
\bibitem [{\citenamefont {Stiller}\ \emph {et~al.}(2015{\natexlab{a}})\citenamefont {Stiller}, \citenamefont {Khan}, \citenamefont {Jain}, \citenamefont {Jouguet}, \citenamefont {Kunz-Jacques} \emph {et~al.}}]{Stiller_CLEO_2015}%
  \BibitemOpen
  \bibfield  {author} {\bibinfo {author} {\bibfnamefont {B.}~\bibnamefont {Stiller}}, \bibinfo {author} {\bibfnamefont {I.}~\bibnamefont {Khan}}, \bibinfo {author} {\bibfnamefont {N.}~\bibnamefont {Jain}}, \bibinfo {author} {\bibfnamefont {P.}~\bibnamefont {Jouguet}}, \bibinfo {author} {\bibfnamefont {S.}~\bibnamefont {Kunz-Jacques}},  \emph {et~al.},\ }\bibfield  {title} {\enquote {\bibinfo {title} {Quantum hacking of continuous-variable quantum key distribution systems: realtime trojan-horse attacks},}\ }in\ \href {\doibase 10.1364/CLEO_QELS.2015.FF1A.7} {\emph {\bibinfo {booktitle} {Conf. on Lasers and Electro-Optics (San Jose)}}}\ (\bibinfo {year} {2015})\ p.\ \bibinfo {pages} {FF1A.7}\BibitemShut {NoStop}%
\bibitem [{\citenamefont {Liu}\ \emph {et~al.}(2017{\natexlab{b}})\citenamefont {Liu}, \citenamefont {Peng}, \citenamefont {Huang}, \citenamefont {Huang},\ and\ \citenamefont {Zeng}}]{Liu_OptExpress_2017}%
  \BibitemOpen
  \bibfield  {author} {\bibinfo {author} {\bibfnamefont {W.}~\bibnamefont {Liu}}, \bibinfo {author} {\bibfnamefont {J.}~\bibnamefont {Peng}}, \bibinfo {author} {\bibfnamefont {P.}~\bibnamefont {Huang}}, \bibinfo {author} {\bibfnamefont {D.}~\bibnamefont {Huang}}, \ and\ \bibinfo {author} {\bibfnamefont {G.}~\bibnamefont {Zeng}},\ }\bibfield  {title} {\enquote {\bibinfo {title} {Monitoring of continuous-variable quantum key distribution system in real environment},}\ }\href@noop {} {\bibfield  {journal} {\bibinfo  {journal} {Opt. Express}\ }\textbf {\bibinfo {volume} {25}},\ \bibinfo {pages} {19429--19443} (\bibinfo {year} {2017}{\natexlab{b}})}\BibitemShut {NoStop}%
\bibitem [{\citenamefont {Wang}\ \emph {et~al.}(2019{\natexlab{b}})\citenamefont {Wang}, \citenamefont {Huang}, \citenamefont {Wang},\ and\ \citenamefont {Zeng}}]{wang_OptExp_2019}%
  \BibitemOpen
  \bibfield  {author} {\bibinfo {author} {\bibfnamefont {T.}~\bibnamefont {Wang}}, \bibinfo {author} {\bibfnamefont {P.}~\bibnamefont {Huang}}, \bibinfo {author} {\bibfnamefont {S.}~\bibnamefont {Wang}}, \ and\ \bibinfo {author} {\bibfnamefont {G.}~\bibnamefont {Zeng}},\ }\bibfield  {title} {\enquote {\bibinfo {title} {Polarization-state tracking based on kalman filter in continuous-variable quantum key distribution},}\ }\href@noop {} {\bibfield  {journal} {\bibinfo  {journal} {Opt. Express}\ }\textbf {\bibinfo {volume} {27}},\ \bibinfo {pages} {26689--26700} (\bibinfo {year} {2019}{\natexlab{b}})}\BibitemShut {NoStop}%
\bibitem [{\citenamefont {Liu}\ \emph {et~al.}(2020)\citenamefont {Liu}, \citenamefont {Cao}, \citenamefont {Wang},\ and\ \citenamefont {Li}}]{liu2020continuous}%
  \BibitemOpen
  \bibfield  {author} {\bibinfo {author} {\bibfnamefont {W.}~\bibnamefont {Liu}}, \bibinfo {author} {\bibfnamefont {Y.}~\bibnamefont {Cao}}, \bibinfo {author} {\bibfnamefont {X.}~\bibnamefont {Wang}}, \ and\ \bibinfo {author} {\bibfnamefont {Y.}~\bibnamefont {Li}},\ }\bibfield  {title} {\enquote {\bibinfo {title} {Continuous-variable quantum key distribution under strong channel polarization disturbance},}\ }\href@noop {} {\bibfield  {journal} {\bibinfo  {journal} {Phys. Rev. A}\ }\textbf {\bibinfo {volume} {102}},\ \bibinfo {pages} {032625} (\bibinfo {year} {2020})}\BibitemShut {NoStop}%
\bibitem [{\citenamefont {Li}\ \emph {et~al.}(2019{\natexlab{a}})\citenamefont {Li}, \citenamefont {Huang}, \citenamefont {Wang}, \citenamefont {Wang}, \citenamefont {Chen} \emph {et~al.}}]{li_OptExp_2019}%
  \BibitemOpen
  \bibfield  {author} {\bibinfo {author} {\bibfnamefont {D.}~\bibnamefont {Li}}, \bibinfo {author} {\bibfnamefont {P.}~\bibnamefont {Huang}}, \bibinfo {author} {\bibfnamefont {T.}~\bibnamefont {Wang}}, \bibinfo {author} {\bibfnamefont {S.}~\bibnamefont {Wang}}, \bibinfo {author} {\bibfnamefont {R.}~\bibnamefont {Chen}},  \emph {et~al.},\ }\bibfield  {title} {\enquote {\bibinfo {title} {Phase compensation based on step-length control in continuous-variable quantum key distribution},}\ }\href@noop {} {\bibfield  {journal} {\bibinfo  {journal} {Opt. Express}\ }\textbf {\bibinfo {volume} {27}},\ \bibinfo {pages} {20670--20687} (\bibinfo {year} {2019}{\natexlab{a}})}\BibitemShut {NoStop}%
\bibitem [{\citenamefont {Wang}\ \emph {et~al.}(2019{\natexlab{c}})\citenamefont {Wang}, \citenamefont {Huang}, \citenamefont {Wang},\ and\ \citenamefont {Zeng}}]{wang_PhysRevA_2019}%
  \BibitemOpen
  \bibfield  {author} {\bibinfo {author} {\bibfnamefont {T.}~\bibnamefont {Wang}}, \bibinfo {author} {\bibfnamefont {P.}~\bibnamefont {Huang}}, \bibinfo {author} {\bibfnamefont {S.}~\bibnamefont {Wang}}, \ and\ \bibinfo {author} {\bibfnamefont {G.}~\bibnamefont {Zeng}},\ }\bibfield  {title} {\enquote {\bibinfo {title} {Carrier-phase estimation for simultaneous quantum key distribution and classical communication using a real local oscillator},}\ }\href@noop {} {\bibfield  {journal} {\bibinfo  {journal} {Phys. Rev. A}\ }\textbf {\bibinfo {volume} {99}},\ \bibinfo {pages} {022318} (\bibinfo {year} {2019}{\natexlab{c}})}\BibitemShut {NoStop}%
\bibitem [{\citenamefont {Xing}\ \emph {et~al.}(2022)\citenamefont {Xing}, \citenamefont {Li}, \citenamefont {Ruan}, \citenamefont {Luo},\ and\ \citenamefont {Zhang}}]{xing_Photonics_2022}%
  \BibitemOpen
  \bibfield  {author} {\bibinfo {author} {\bibfnamefont {Z.}~\bibnamefont {Xing}}, \bibinfo {author} {\bibfnamefont {X.}~\bibnamefont {Li}}, \bibinfo {author} {\bibfnamefont {X.}~\bibnamefont {Ruan}}, \bibinfo {author} {\bibfnamefont {Y.}~\bibnamefont {Luo}}, \ and\ \bibinfo {author} {\bibfnamefont {H.}~\bibnamefont {Zhang}},\ }\bibfield  {title} {\enquote {\bibinfo {title} {Phase compensation for continuous variable quantum key distribution based on convolutional neural network},}\ }in\ \href@noop {} {\emph {\bibinfo {booktitle} {Photonics}}},\ Vol.~\bibinfo {volume} {9}\ (\bibinfo {organization} {MDPI},\ \bibinfo {year} {2022})\ p.\ \bibinfo {pages} {463}\BibitemShut {NoStop}%
\bibitem [{\citenamefont {Chin}\ \emph {et~al.}(2023)\citenamefont {Chin}, \citenamefont {Hajomer}, \citenamefont {Jain}, \citenamefont {Andersen},\ and\ \citenamefont {Gehring}}]{chin_OptFibCommCon_2023}%
  \BibitemOpen
  \bibfield  {author} {\bibinfo {author} {\bibfnamefont {H.-M.}\ \bibnamefont {Chin}}, \bibinfo {author} {\bibfnamefont {A.~A.}\ \bibnamefont {Hajomer}}, \bibinfo {author} {\bibfnamefont {N.}~\bibnamefont {Jain}}, \bibinfo {author} {\bibfnamefont {U.~L.}\ \bibnamefont {Andersen}}, \ and\ \bibinfo {author} {\bibfnamefont {T.}~\bibnamefont {Gehring}},\ }\bibfield  {title} {\enquote {\bibinfo {title} {Machine learning based joint polarization and phase compensation for cv-qkd},}\ }in\ \href@noop {} {\emph {\bibinfo {booktitle} {Optical Fiber Communication Conference}}}\ (\bibinfo {organization} {Optica Publishing Group},\ \bibinfo {year} {2023})\ pp.\ \bibinfo {pages} {Th3J--2}\BibitemShut {NoStop}%
\bibitem [{\citenamefont {Lin}\ \emph {et~al.}(2015)\citenamefont {Lin}, \citenamefont {Huang}, \citenamefont {Huang}, \citenamefont {Wang}, \citenamefont {Peng} \emph {et~al.}}]{lin_OptExp_2015}%
  \BibitemOpen
  \bibfield  {author} {\bibinfo {author} {\bibfnamefont {D.}~\bibnamefont {Lin}}, \bibinfo {author} {\bibfnamefont {P.}~\bibnamefont {Huang}}, \bibinfo {author} {\bibfnamefont {D.}~\bibnamefont {Huang}}, \bibinfo {author} {\bibfnamefont {C.}~\bibnamefont {Wang}}, \bibinfo {author} {\bibfnamefont {J.}~\bibnamefont {Peng}},  \emph {et~al.},\ }\bibfield  {title} {\enquote {\bibinfo {title} {High performance frame synchronization for continuous variable quantum key distribution systems},}\ }\href@noop {} {\bibfield  {journal} {\bibinfo  {journal} {Opt. Express}\ }\textbf {\bibinfo {volume} {23}},\ \bibinfo {pages} {22190--22198} (\bibinfo {year} {2015})}\BibitemShut {NoStop}%
\bibitem [{\citenamefont {Liu}\ \emph {et~al.}(2017{\natexlab{c}})\citenamefont {Liu}, \citenamefont {Zhao}, \citenamefont {Zhang}, \citenamefont {Zheng},\ and\ \citenamefont {Yu}}]{liu_InterConfer_2017}%
  \BibitemOpen
  \bibfield  {author} {\bibinfo {author} {\bibfnamefont {C.}~\bibnamefont {Liu}}, \bibinfo {author} {\bibfnamefont {Y.}~\bibnamefont {Zhao}}, \bibinfo {author} {\bibfnamefont {Y.}~\bibnamefont {Zhang}}, \bibinfo {author} {\bibfnamefont {Z.}~\bibnamefont {Zheng}}, \ and\ \bibinfo {author} {\bibfnamefont {S.}~\bibnamefont {Yu}},\ }\bibfield  {title} {\enquote {\bibinfo {title} {Synchronization schemes for continuous-variable quantum key distribution},}\ }in\ \href@noop {} {\emph {\bibinfo {booktitle} {International Conference on Optoelectronics and Microelectronics Technology and Application}}},\ Vol.\ \bibinfo {volume} {10244}\ (\bibinfo {organization} {SPIE},\ \bibinfo {year} {2017})\ pp.\ \bibinfo {pages} {40--44}\BibitemShut {NoStop}%
\bibitem [{\citenamefont {Chen}\ \emph {et~al.}(2019)\citenamefont {Chen}, \citenamefont {Huang}, \citenamefont {Li}, \citenamefont {Zhu},\ and\ \citenamefont {Zeng}}]{chen_Entropy_2019}%
  \BibitemOpen
  \bibfield  {author} {\bibinfo {author} {\bibfnamefont {R.}~\bibnamefont {Chen}}, \bibinfo {author} {\bibfnamefont {P.}~\bibnamefont {Huang}}, \bibinfo {author} {\bibfnamefont {D.}~\bibnamefont {Li}}, \bibinfo {author} {\bibfnamefont {Y.}~\bibnamefont {Zhu}}, \ and\ \bibinfo {author} {\bibfnamefont {G.}~\bibnamefont {Zeng}},\ }\bibfield  {title} {\enquote {\bibinfo {title} {Robust frame synchronization scheme for continuous-variable quantum key distribution with simple process},}\ }\href@noop {} {\bibfield  {journal} {\bibinfo  {journal} {Entropy}\ }\textbf {\bibinfo {volume} {21}},\ \bibinfo {pages} {1146} (\bibinfo {year} {2019})}\BibitemShut {NoStop}%
\bibitem [{\citenamefont {Dong}\ \emph {et~al.}(2022)\citenamefont {Dong}, \citenamefont {Wang}, \citenamefont {Li}, \citenamefont {Huang},\ and\ \citenamefont {Zeng}}]{dong_PhysRevA_2022}%
  \BibitemOpen
  \bibfield  {author} {\bibinfo {author} {\bibfnamefont {J.}~\bibnamefont {Dong}}, \bibinfo {author} {\bibfnamefont {T.}~\bibnamefont {Wang}}, \bibinfo {author} {\bibfnamefont {L.}~\bibnamefont {Li}}, \bibinfo {author} {\bibfnamefont {P.}~\bibnamefont {Huang}}, \ and\ \bibinfo {author} {\bibfnamefont {G.}~\bibnamefont {Zeng}},\ }\bibfield  {title} {\enquote {\bibinfo {title} {Efficient frame synchronization using a weak coherent state for continuous-variable quantum key distribution},}\ }\href@noop {} {\bibfield  {journal} {\bibinfo  {journal} {Phys. Rev. A}\ }\textbf {\bibinfo {volume} {105}},\ \bibinfo {pages} {052407} (\bibinfo {year} {2022})}\BibitemShut {NoStop}%
\bibitem [{\citenamefont {Li}\ \emph {et~al.}(2022)\citenamefont {Li}, \citenamefont {Song}, \citenamefont {Liu}, \citenamefont {Wen}, \citenamefont {Sun} \emph {et~al.}}]{li_InterConfer_2022}%
  \BibitemOpen
  \bibfield  {author} {\bibinfo {author} {\bibfnamefont {H.}~\bibnamefont {Li}}, \bibinfo {author} {\bibfnamefont {H.}~\bibnamefont {Song}}, \bibinfo {author} {\bibfnamefont {X.}~\bibnamefont {Liu}}, \bibinfo {author} {\bibfnamefont {J.}~\bibnamefont {Wen}}, \bibinfo {author} {\bibfnamefont {S.}~\bibnamefont {Sun}},  \emph {et~al.},\ }\bibfield  {title} {\enquote {\bibinfo {title} {Reliable synchronization technology for continuous variable quantum key distribution system},}\ }in\ \href@noop {} {\emph {\bibinfo {booktitle} {2022 International Conference on 6G Communications and IoT Technologies (6GIoTT)}}}\ (\bibinfo {organization} {IEEE},\ \bibinfo {year} {2022})\ pp.\ \bibinfo {pages} {35--41}\BibitemShut {NoStop}%
\bibitem [{\citenamefont {Pan}\ \emph {et~al.}(2023{\natexlab{a}})\citenamefont {Pan}, \citenamefont {Wang}, \citenamefont {Shao}, \citenamefont {Pi}, \citenamefont {Ye}, \citenamefont {Zhang}, \citenamefont {Li}, \citenamefont {Huang},\ and\ \citenamefont {Xu}}]{pan2023simple}%
  \BibitemOpen
  \bibfield  {author} {\bibinfo {author} {\bibfnamefont {Y.}~\bibnamefont {Pan}}, \bibinfo {author} {\bibfnamefont {H.}~\bibnamefont {Wang}}, \bibinfo {author} {\bibfnamefont {Y.}~\bibnamefont {Shao}}, \bibinfo {author} {\bibfnamefont {Y.}~\bibnamefont {Pi}}, \bibinfo {author} {\bibfnamefont {T.}~\bibnamefont {Ye}}, \bibinfo {author} {\bibfnamefont {S.}~\bibnamefont {Zhang}}, \bibinfo {author} {\bibfnamefont {Y.}~\bibnamefont {Li}}, \bibinfo {author} {\bibfnamefont {W.}~\bibnamefont {Huang}}, \ and\ \bibinfo {author} {\bibfnamefont {B.}~\bibnamefont {Xu}},\ }\bibfield  {title} {\enquote {\bibinfo {title} {Simple and fast polarization tracking algorithm for continuous-variable quantum key distribution system using orthogonal pilot tone},}\ }\href@noop {} {\bibfield  {journal} {\bibinfo  {journal} {J. Light. Technol.}\ } (\bibinfo {year} {2023}{\natexlab{a}})}\BibitemShut {NoStop}%
\bibitem [{\citenamefont {Mo}\ \emph {et~al.}(2005)\citenamefont {Mo}, \citenamefont {Zhu}, \citenamefont {Han}, \citenamefont {Gui} \emph {et~al.}}]{Mo_OptLett_2005}%
  \BibitemOpen
  \bibfield  {author} {\bibinfo {author} {\bibfnamefont {X.-F.}\ \bibnamefont {Mo}}, \bibinfo {author} {\bibfnamefont {B.}~\bibnamefont {Zhu}}, \bibinfo {author} {\bibfnamefont {Z.-F.}\ \bibnamefont {Han}}, \bibinfo {author} {\bibfnamefont {Y.-Z.}\ \bibnamefont {Gui}},  \emph {et~al.},\ }\bibfield  {title} {\enquote {\bibinfo {title} {Faraday--michelson system for quantum cryptography},}\ }\href@noop {} {\bibfield  {journal} {\bibinfo  {journal} {Opt. Lett.}\ }\textbf {\bibinfo {volume} {30}},\ \bibinfo {pages} {2632--2634} (\bibinfo {year} {2005})}\BibitemShut {NoStop}%
\bibitem [{\citenamefont {da~Silva}\ and\ \citenamefont {von~der Weid}(2009)}]{Silva_JMOe_2009}%
  \BibitemOpen
  \bibfield  {author} {\bibinfo {author} {\bibfnamefont {T.~F.}\ \bibnamefont {da~Silva}}\ and\ \bibinfo {author} {\bibfnamefont {J.~P.}\ \bibnamefont {von~der Weid}},\ }\bibfield  {title} {\enquote {\bibinfo {title} {Optical transmission of frequency-coded quantum bits with wdm synchronization},}\ }\href@noop {} {\bibfield  {journal} {\bibinfo  {journal} {J. Microwaves, Optoelectron. Electromagn. Appl.}\ }\textbf {\bibinfo {volume} {8}},\ \bibinfo {pages} {163S--178S} (\bibinfo {year} {2009})}\BibitemShut {NoStop}%
\bibitem [{\citenamefont {Beveratos}\ \emph {et~al.}(2002)\citenamefont {Beveratos}, \citenamefont {Brouri}, \citenamefont {Gacoin}, \citenamefont {Villing}, \citenamefont {Poizat} \emph {et~al.}}]{Beveratos_PhysRevLett_2002}%
  \BibitemOpen
  \bibfield  {author} {\bibinfo {author} {\bibfnamefont {A.}~\bibnamefont {Beveratos}}, \bibinfo {author} {\bibfnamefont {R.}~\bibnamefont {Brouri}}, \bibinfo {author} {\bibfnamefont {T.}~\bibnamefont {Gacoin}}, \bibinfo {author} {\bibfnamefont {A.}~\bibnamefont {Villing}}, \bibinfo {author} {\bibfnamefont {J.-P.}\ \bibnamefont {Poizat}},  \emph {et~al.},\ }\bibfield  {title} {\enquote {\bibinfo {title} {Single photon quantum cryptography},}\ }\href@noop {} {\bibfield  {journal} {\bibinfo  {journal} {Phys. Rev. Lett.}\ }\textbf {\bibinfo {volume} {89}},\ \bibinfo {pages} {187901} (\bibinfo {year} {2002})}\BibitemShut {NoStop}%
\bibitem [{\citenamefont {Wang}\ \emph {et~al.}(2022{\natexlab{c}})\citenamefont {Wang}, \citenamefont {Zuo}, \citenamefont {Li}, \citenamefont {Huang}, \citenamefont {Guo} \emph {et~al.}}]{wang2022continuous}%
  \BibitemOpen
  \bibfield  {author} {\bibinfo {author} {\bibfnamefont {T.}~\bibnamefont {Wang}}, \bibinfo {author} {\bibfnamefont {Z.}~\bibnamefont {Zuo}}, \bibinfo {author} {\bibfnamefont {L.}~\bibnamefont {Li}}, \bibinfo {author} {\bibfnamefont {P.}~\bibnamefont {Huang}}, \bibinfo {author} {\bibfnamefont {Y.}~\bibnamefont {Guo}},  \emph {et~al.},\ }\bibfield  {title} {\enquote {\bibinfo {title} {Continuous-variable quantum key distribution without synchronized clocks},}\ }\href@noop {} {\bibfield  {journal} {\bibinfo  {journal} {Phys. Rev. Appl.}\ }\textbf {\bibinfo {volume} {18}},\ \bibinfo {pages} {014064} (\bibinfo {year} {2022}{\natexlab{c}})}\BibitemShut {NoStop}%
\bibitem [{\citenamefont {Yuen}\ and\ \citenamefont {Shapiro}(1980)}]{yuen_IEEETransInforThe_1980}%
  \BibitemOpen
  \bibfield  {author} {\bibinfo {author} {\bibfnamefont {H.}~\bibnamefont {Yuen}}\ and\ \bibinfo {author} {\bibfnamefont {J.}~\bibnamefont {Shapiro}},\ }\bibfield  {title} {\enquote {\bibinfo {title} {Optical communication with two-photon coherent states--part iii: Quantum measurements realizable with photoemissive detectors},}\ }\href@noop {} {\bibfield  {journal} {\bibinfo  {journal} {IEEE Trans. Inf. Theory}\ }\textbf {\bibinfo {volume} {26}},\ \bibinfo {pages} {78--92} (\bibinfo {year} {1980})}\BibitemShut {NoStop}%
\bibitem [{\citenamefont {Smithey}\ \emph {et~al.}(1993)\citenamefont {Smithey}, \citenamefont {Beck}, \citenamefont {Raymer},\ and\ \citenamefont {Faridani}}]{smithey_PhysRevLett_1993}%
  \BibitemOpen
  \bibfield  {author} {\bibinfo {author} {\bibfnamefont {D.}~\bibnamefont {Smithey}}, \bibinfo {author} {\bibfnamefont {M.}~\bibnamefont {Beck}}, \bibinfo {author} {\bibfnamefont {M.~G.}\ \bibnamefont {Raymer}}, \ and\ \bibinfo {author} {\bibfnamefont {A.}~\bibnamefont {Faridani}},\ }\bibfield  {title} {\enquote {\bibinfo {title} {Measurement of the wigner distribution and the density matrix of a light mode using optical homodyne tomography: Application to squeezed states and the vacuum},}\ }\href@noop {} {\bibfield  {journal} {\bibinfo  {journal} {Phys. Rev. Lett.}\ }\textbf {\bibinfo {volume} {70}},\ \bibinfo {pages} {1244} (\bibinfo {year} {1993})}\BibitemShut {NoStop}%
\bibitem [{\citenamefont {Ou}\ and\ \citenamefont {Kimble}(1995)}]{ou_PhysRevA_1995}%
  \BibitemOpen
  \bibfield  {author} {\bibinfo {author} {\bibfnamefont {Z.}~\bibnamefont {Ou}}\ and\ \bibinfo {author} {\bibfnamefont {H.}~\bibnamefont {Kimble}},\ }\bibfield  {title} {\enquote {\bibinfo {title} {Probability distribution of photoelectric currents in photodetection processes and its connection to the measurement of a quantum state},}\ }\href@noop {} {\bibfield  {journal} {\bibinfo  {journal} {Phys. Rev. A}\ }\textbf {\bibinfo {volume} {52}},\ \bibinfo {pages} {3126} (\bibinfo {year} {1995})}\BibitemShut {NoStop}%
\bibitem [{\citenamefont {Hansen}\ \emph {et~al.}(2001)\citenamefont {Hansen}, \citenamefont {Aichele}, \citenamefont {Hettich}, \citenamefont {Lodahl}, \citenamefont {Lvovsky} \emph {et~al.}}]{Hansen_OptLett_2001}%
  \BibitemOpen
  \bibfield  {author} {\bibinfo {author} {\bibfnamefont {H.}~\bibnamefont {Hansen}}, \bibinfo {author} {\bibfnamefont {T.}~\bibnamefont {Aichele}}, \bibinfo {author} {\bibfnamefont {C.}~\bibnamefont {Hettich}}, \bibinfo {author} {\bibfnamefont {P.}~\bibnamefont {Lodahl}}, \bibinfo {author} {\bibfnamefont {A.}~\bibnamefont {Lvovsky}},  \emph {et~al.},\ }\bibfield  {title} {\enquote {\bibinfo {title} {Ultrasensitive pulsed, balanced homodyne detector: application to time-domain quantum measurements},}\ }\href@noop {} {\bibfield  {journal} {\bibinfo  {journal} {Opt. Lett.}\ }\textbf {\bibinfo {volume} {26}},\ \bibinfo {pages} {1714--1716} (\bibinfo {year} {2001})}\BibitemShut {NoStop}%
\bibitem [{\citenamefont {Haderka}\ \emph {et~al.}(2009)\citenamefont {Haderka}, \citenamefont {Mich{\'a}lek}, \citenamefont {Urb{\'a}{\v{s}}ek},\ and\ \citenamefont {Je{\v{z}}ek}}]{Haderka_ApplOpt_2009}%
  \BibitemOpen
  \bibfield  {author} {\bibinfo {author} {\bibfnamefont {O.}~\bibnamefont {Haderka}}, \bibinfo {author} {\bibfnamefont {V.}~\bibnamefont {Mich{\'a}lek}}, \bibinfo {author} {\bibfnamefont {V.}~\bibnamefont {Urb{\'a}{\v{s}}ek}}, \ and\ \bibinfo {author} {\bibfnamefont {M.}~\bibnamefont {Je{\v{z}}ek}},\ }\bibfield  {title} {\enquote {\bibinfo {title} {Fast time-domain balanced homodyne detection of light},}\ }\href@noop {} {\bibfield  {journal} {\bibinfo  {journal} {Appl. Opt.}\ }\textbf {\bibinfo {volume} {48}},\ \bibinfo {pages} {2884--2889} (\bibinfo {year} {2009})}\BibitemShut {NoStop}%
\bibitem [{\citenamefont {Chi}\ \emph {et~al.}(2011)\citenamefont {Chi}, \citenamefont {Qi}, \citenamefont {Zhu}, \citenamefont {Qian}, \citenamefont {Lo} \emph {et~al.}}]{chi_NewJourPhys_2011}%
  \BibitemOpen
  \bibfield  {author} {\bibinfo {author} {\bibfnamefont {Y.-M.}\ \bibnamefont {Chi}}, \bibinfo {author} {\bibfnamefont {B.}~\bibnamefont {Qi}}, \bibinfo {author} {\bibfnamefont {W.}~\bibnamefont {Zhu}}, \bibinfo {author} {\bibfnamefont {L.}~\bibnamefont {Qian}}, \bibinfo {author} {\bibfnamefont {H.-K.}\ \bibnamefont {Lo}},  \emph {et~al.},\ }\bibfield  {title} {\enquote {\bibinfo {title} {A balanced homodyne detector for high-rate gaussian-modulated coherent-state quantum key distribution},}\ }\href@noop {} {\bibfield  {journal} {\bibinfo  {journal} {New J. Phys.}\ }\textbf {\bibinfo {volume} {13}},\ \bibinfo {pages} {013003} (\bibinfo {year} {2011})}\BibitemShut {NoStop}%
\bibitem [{\citenamefont {Kumar}\ \emph {et~al.}(2012)\citenamefont {Kumar}, \citenamefont {Barrios}, \citenamefont {MacRae}, \citenamefont {Cairns}, \citenamefont {Huntington} \emph {et~al.}}]{Kumar_OptComm_2012}%
  \BibitemOpen
  \bibfield  {author} {\bibinfo {author} {\bibfnamefont {R.}~\bibnamefont {Kumar}}, \bibinfo {author} {\bibfnamefont {E.}~\bibnamefont {Barrios}}, \bibinfo {author} {\bibfnamefont {A.}~\bibnamefont {MacRae}}, \bibinfo {author} {\bibfnamefont {E.}~\bibnamefont {Cairns}}, \bibinfo {author} {\bibfnamefont {E.}~\bibnamefont {Huntington}},  \emph {et~al.},\ }\bibfield  {title} {\enquote {\bibinfo {title} {Versatile wideband balanced detector for quantum optical homodyne tomography},}\ }\href@noop {} {\bibfield  {journal} {\bibinfo  {journal} {Opt. Commun.}\ }\textbf {\bibinfo {volume} {285}},\ \bibinfo {pages} {5259--5267} (\bibinfo {year} {2012})}\BibitemShut {NoStop}%
\bibitem [{\citenamefont {Wang}\ \emph {et~al.}(2012)\citenamefont {Wang}, \citenamefont {liang Bai}, \citenamefont {Du}, \citenamefont {Li},\ and\ \citenamefont {Peng}}]{Wang2012UltrastableFT}%
  \BibitemOpen
  \bibfield  {author} {\bibinfo {author} {\bibfnamefont {X.}~\bibnamefont {Wang}}, \bibinfo {author} {\bibfnamefont {Z.}~\bibnamefont {liang Bai}}, \bibinfo {author} {\bibfnamefont {P.}~\bibnamefont {Du}}, \bibinfo {author} {\bibfnamefont {Y.}~\bibnamefont {Li}}, \ and\ \bibinfo {author} {\bibfnamefont {K.}~\bibnamefont {Peng}},\ }\bibfield  {title} {\enquote {\bibinfo {title} {Ultrastable fiber-based time-domain balanced homodyne detector for quantum communication},}\ }\href {https://api.semanticscholar.org/CorpusID:124891505} {\bibfield  {journal} {\bibinfo  {journal} {Chin. Phys. Lett.}\ }\textbf {\bibinfo {volume} {29}},\ \bibinfo {pages} {124202--124202} (\bibinfo {year} {2012})}\BibitemShut {NoStop}%
\bibitem [{\citenamefont {Duan}\ \emph {et~al.}(2013)\citenamefont {Duan}, \citenamefont {Jian}, \citenamefont {Chao}, \citenamefont {Peng},\ and\ \citenamefont {Gui-Hua}}]{Huang_ChinPhysLett_2013}%
  \BibitemOpen
  \bibfield  {author} {\bibinfo {author} {\bibfnamefont {H.}~\bibnamefont {Duan}}, \bibinfo {author} {\bibfnamefont {F.}~\bibnamefont {Jian}}, \bibinfo {author} {\bibfnamefont {W.}~\bibnamefont {Chao}}, \bibinfo {author} {\bibfnamefont {H.}~\bibnamefont {Peng}}, \ and\ \bibinfo {author} {\bibfnamefont {Z.}~\bibnamefont {Gui-Hua}},\ }\bibfield  {title} {\enquote {\bibinfo {title} {A 300-mhz bandwidth balanced homodyne detector for continuous variable quantum key distribution},}\ }\href@noop {} {\bibfield  {journal} {\bibinfo  {journal} {Chinese Phys. Lett.}\ }\textbf {\bibinfo {volume} {30}},\ \bibinfo {pages} {114209} (\bibinfo {year} {2013})}\BibitemShut {NoStop}%
\bibitem [{\citenamefont {Cooper}, \citenamefont {S{\"o}ller},\ and\ \citenamefont {Smith}(2013)}]{Cooper_JModOpt_2013}%
  \BibitemOpen
  \bibfield  {author} {\bibinfo {author} {\bibfnamefont {M.}~\bibnamefont {Cooper}}, \bibinfo {author} {\bibfnamefont {C.}~\bibnamefont {S{\"o}ller}}, \ and\ \bibinfo {author} {\bibfnamefont {B.~J.}\ \bibnamefont {Smith}},\ }\bibfield  {title} {\enquote {\bibinfo {title} {High-stability time-domain balanced homodyne detector for ultrafast optical pulse applications},}\ }\href@noop {} {\bibfield  {journal} {\bibinfo  {journal} {J. Mod. Opt.}\ }\textbf {\bibinfo {volume} {60}},\ \bibinfo {pages} {611--616} (\bibinfo {year} {2013})}\BibitemShut {NoStop}%
\bibitem [{\citenamefont {Zhang}\ \emph {et~al.}(2018)\citenamefont {Zhang}, \citenamefont {Zhang}, \citenamefont {Li}, \citenamefont {Yu},\ and\ \citenamefont {Guo}}]{Zhang_IEEEPhotonJ_2018}%
  \BibitemOpen
  \bibfield  {author} {\bibinfo {author} {\bibfnamefont {X.}~\bibnamefont {Zhang}}, \bibinfo {author} {\bibfnamefont {Y.}~\bibnamefont {Zhang}}, \bibinfo {author} {\bibfnamefont {Z.}~\bibnamefont {Li}}, \bibinfo {author} {\bibfnamefont {S.}~\bibnamefont {Yu}}, \ and\ \bibinfo {author} {\bibfnamefont {H.}~\bibnamefont {Guo}},\ }\bibfield  {title} {\enquote {\bibinfo {title} {1.2-ghz balanced homodyne detector for continuous-variable quantum information technology},}\ }\href@noop {} {\bibfield  {journal} {\bibinfo  {journal} {IEEE Photon. J.}\ }\textbf {\bibinfo {volume} {10}},\ \bibinfo {pages} {1--10} (\bibinfo {year} {2018})}\BibitemShut {NoStop}%
\bibitem [{\citenamefont {Liu}\ \emph {et~al.}(2022)\citenamefont {Liu}, \citenamefont {Cao}, \citenamefont {Wang}, \citenamefont {Liu}, \citenamefont {Lu} \emph {et~al.}}]{liu_OptExp_2022}%
  \BibitemOpen
  \bibfield  {author} {\bibinfo {author} {\bibfnamefont {J.}~\bibnamefont {Liu}}, \bibinfo {author} {\bibfnamefont {Y.}~\bibnamefont {Cao}}, \bibinfo {author} {\bibfnamefont {P.}~\bibnamefont {Wang}}, \bibinfo {author} {\bibfnamefont {S.}~\bibnamefont {Liu}}, \bibinfo {author} {\bibfnamefont {Z.}~\bibnamefont {Lu}},  \emph {et~al.},\ }\bibfield  {title} {\enquote {\bibinfo {title} {Impact of homodyne receiver bandwidth and signal modulation patterns on the continuous-variable quantum key distribution},}\ }\href@noop {} {\bibfield  {journal} {\bibinfo  {journal} {Opt. Express}\ }\textbf {\bibinfo {volume} {30}},\ \bibinfo {pages} {27912--27925} (\bibinfo {year} {2022})}\BibitemShut {NoStop}%
\bibitem [{\citenamefont {Jin}\ \emph {et~al.}(2015)\citenamefont {Jin}, \citenamefont {Su}, \citenamefont {Zheng}, \citenamefont {Chen}, \citenamefont {Wang} \emph {et~al.}}]{jin_OptExp_2015}%
  \BibitemOpen
  \bibfield  {author} {\bibinfo {author} {\bibfnamefont {X.}~\bibnamefont {Jin}}, \bibinfo {author} {\bibfnamefont {J.}~\bibnamefont {Su}}, \bibinfo {author} {\bibfnamefont {Y.}~\bibnamefont {Zheng}}, \bibinfo {author} {\bibfnamefont {C.}~\bibnamefont {Chen}}, \bibinfo {author} {\bibfnamefont {W.}~\bibnamefont {Wang}},  \emph {et~al.},\ }\bibfield  {title} {\enquote {\bibinfo {title} {Balanced homodyne detection with high common mode rejection ratio based on parameter compensation of two arbitrary photodiodes},}\ }\href@noop {} {\bibfield  {journal} {\bibinfo  {journal} {Opt. Express}\ }\textbf {\bibinfo {volume} {23}},\ \bibinfo {pages} {23859--23866} (\bibinfo {year} {2015})}\BibitemShut {NoStop}%
\bibitem [{\citenamefont {Du}\ \emph {et~al.}(2018)\citenamefont {Du}, \citenamefont {Li}, \citenamefont {Liu}, \citenamefont {Wang},\ and\ \citenamefont {Li}}]{du_JOSAB_2018}%
  \BibitemOpen
  \bibfield  {author} {\bibinfo {author} {\bibfnamefont {S.}~\bibnamefont {Du}}, \bibinfo {author} {\bibfnamefont {Z.}~\bibnamefont {Li}}, \bibinfo {author} {\bibfnamefont {W.}~\bibnamefont {Liu}}, \bibinfo {author} {\bibfnamefont {X.}~\bibnamefont {Wang}}, \ and\ \bibinfo {author} {\bibfnamefont {Y.}~\bibnamefont {Li}},\ }\bibfield  {title} {\enquote {\bibinfo {title} {High-speed time-domain balanced homodyne detector for nanosecond optical field applications},}\ }\href@noop {} {\bibfield  {journal} {\bibinfo  {journal} {J. Opt. Soc. Am. B}\ }\textbf {\bibinfo {volume} {35}},\ \bibinfo {pages} {481--486} (\bibinfo {year} {2018})}\BibitemShut {NoStop}%
\bibitem [{\citenamefont {Milovan{\v{c}}ev}\ \emph {et~al.}(2021)\citenamefont {Milovan{\v{c}}ev}, \citenamefont {Honz}, \citenamefont {Voki{\'c}}, \citenamefont {Laudenbach}, \citenamefont {H{\"u}bel} \emph {et~al.}}]{milovanvcev2021ultra}%
  \BibitemOpen
  \bibfield  {author} {\bibinfo {author} {\bibfnamefont {D.}~\bibnamefont {Milovan{\v{c}}ev}}, \bibinfo {author} {\bibfnamefont {F.}~\bibnamefont {Honz}}, \bibinfo {author} {\bibfnamefont {N.}~\bibnamefont {Voki{\'c}}}, \bibinfo {author} {\bibfnamefont {F.}~\bibnamefont {Laudenbach}}, \bibinfo {author} {\bibfnamefont {H.}~\bibnamefont {H{\"u}bel}},  \emph {et~al.},\ }\bibfield  {title} {\enquote {\bibinfo {title} {Ultra-low noise balanced receiver with> 20 db quantum-to-classical noise clearance at 1 ghz},}\ }in\ \href@noop {} {\emph {\bibinfo {booktitle} {2021 European Conference on Optical Communication (ECOC)}}}\ (\bibinfo {organization} {IEEE},\ \bibinfo {year} {2021})\ pp.\ \bibinfo {pages} {1--4}\BibitemShut {NoStop}%
\bibitem [{\citenamefont {Honz}\ \emph {et~al.}(2021)\citenamefont {Honz}, \citenamefont {Milovan{\v{c}}ev}, \citenamefont {Voki{\'c}}, \citenamefont {Pacher},\ and\ \citenamefont {Schrenk}}]{honz2021broadband}%
  \BibitemOpen
  \bibfield  {author} {\bibinfo {author} {\bibfnamefont {F.}~\bibnamefont {Honz}}, \bibinfo {author} {\bibfnamefont {D.}~\bibnamefont {Milovan{\v{c}}ev}}, \bibinfo {author} {\bibfnamefont {N.}~\bibnamefont {Voki{\'c}}}, \bibinfo {author} {\bibfnamefont {C.}~\bibnamefont {Pacher}}, \ and\ \bibinfo {author} {\bibfnamefont {B.}~\bibnamefont {Schrenk}},\ }\bibfield  {title} {\enquote {\bibinfo {title} {Broadband balanced homodyne detector for high-rate (> 10 gb/s) vacuum-noise quantum random number generation},}\ }in\ \href@noop {} {\emph {\bibinfo {booktitle} {2021 European Conference on Optical Communication (ECOC)}}}\ (\bibinfo {organization} {IEEE},\ \bibinfo {year} {2021})\ pp.\ \bibinfo {pages} {1--4}\BibitemShut {NoStop}%
\bibitem [{\citenamefont {Bruynsteen}\ \emph {et~al.}(2021{\natexlab{b}})\citenamefont {Bruynsteen}, \citenamefont {Vanhoecke}, \citenamefont {Bauwelinck},\ and\ \citenamefont {Yin}}]{bruynsteen2021integrated}%
  \BibitemOpen
  \bibfield  {author} {\bibinfo {author} {\bibfnamefont {C.}~\bibnamefont {Bruynsteen}}, \bibinfo {author} {\bibfnamefont {M.}~\bibnamefont {Vanhoecke}}, \bibinfo {author} {\bibfnamefont {J.}~\bibnamefont {Bauwelinck}}, \ and\ \bibinfo {author} {\bibfnamefont {X.}~\bibnamefont {Yin}},\ }\bibfield  {title} {\enquote {\bibinfo {title} {Integrated balanced homodyne photonic--electronic detector for beyond 20 ghz shot-noise-limited measurements},}\ }\href@noop {} {\bibfield  {journal} {\bibinfo  {journal} {Optica}\ }\textbf {\bibinfo {volume} {8}},\ \bibinfo {pages} {1146--1152} (\bibinfo {year} {2021}{\natexlab{b}})}\BibitemShut {NoStop}%
\bibitem [{\citenamefont {Tasker}\ \emph {et~al.}(2021)\citenamefont {Tasker}, \citenamefont {Frazer}, \citenamefont {Ferranti}, \citenamefont {Allen}, \citenamefont {Brunel} \emph {et~al.}}]{tasker2021silicon}%
  \BibitemOpen
  \bibfield  {author} {\bibinfo {author} {\bibfnamefont {J.~F.}\ \bibnamefont {Tasker}}, \bibinfo {author} {\bibfnamefont {J.}~\bibnamefont {Frazer}}, \bibinfo {author} {\bibfnamefont {G.}~\bibnamefont {Ferranti}}, \bibinfo {author} {\bibfnamefont {E.~J.}\ \bibnamefont {Allen}}, \bibinfo {author} {\bibfnamefont {L.~F.}\ \bibnamefont {Brunel}},  \emph {et~al.},\ }\bibfield  {title} {\enquote {\bibinfo {title} {Silicon photonics interfaced with integrated electronics for 9 ghz measurement of squeezed light},}\ }\href@noop {} {\bibfield  {journal} {\bibinfo  {journal} {Nat. Photonics}\ }\textbf {\bibinfo {volume} {15}},\ \bibinfo {pages} {11--15} (\bibinfo {year} {2021})}\BibitemShut {NoStop}%
\bibitem [{\citenamefont {Jia}\ \emph {et~al.}(2023)\citenamefont {Jia}, \citenamefont {Wang}, \citenamefont {Hu}, \citenamefont {Hua}, \citenamefont {Zhang}, \citenamefont {Guo}, \citenamefont {Zhang}, \citenamefont {Xiao}, \citenamefont {Yu}, \citenamefont {Zou} \emph {et~al.}}]{jia2023silicon}%
  \BibitemOpen
  \bibfield  {author} {\bibinfo {author} {\bibfnamefont {Y.}~\bibnamefont {Jia}}, \bibinfo {author} {\bibfnamefont {X.}~\bibnamefont {Wang}}, \bibinfo {author} {\bibfnamefont {X.}~\bibnamefont {Hu}}, \bibinfo {author} {\bibfnamefont {X.}~\bibnamefont {Hua}}, \bibinfo {author} {\bibfnamefont {Y.}~\bibnamefont {Zhang}}, \bibinfo {author} {\bibfnamefont {X.}~\bibnamefont {Guo}}, \bibinfo {author} {\bibfnamefont {S.}~\bibnamefont {Zhang}}, \bibinfo {author} {\bibfnamefont {X.}~\bibnamefont {Xiao}}, \bibinfo {author} {\bibfnamefont {S.}~\bibnamefont {Yu}}, \bibinfo {author} {\bibfnamefont {J.}~\bibnamefont {Zou}},  \emph {et~al.},\ }\bibfield  {title} {\enquote {\bibinfo {title} {Silicon photonics-integrated time-domain balanced homodyne detector for quantum tomography and quantum key distribution},}\ }\href@noop {} {\bibfield  {journal} {\bibinfo  {journal} {New Journal of Physics}\ }\textbf {\bibinfo {volume} {25}},\ \bibinfo {pages} {103030} (\bibinfo {year} {2023})}\BibitemShut {NoStop}%
\bibitem [{\citenamefont {Wang}\ \emph {et~al.}(2023{\natexlab{b}})\citenamefont {Wang}, \citenamefont {Guo}, \citenamefont {Jia}, \citenamefont {Zhang}, \citenamefont {Lu} \emph {et~al.}}]{wang_JourLigTech_2023}%
  \BibitemOpen
  \bibfield  {author} {\bibinfo {author} {\bibfnamefont {X.}~\bibnamefont {Wang}}, \bibinfo {author} {\bibfnamefont {X.}~\bibnamefont {Guo}}, \bibinfo {author} {\bibfnamefont {Y.}~\bibnamefont {Jia}}, \bibinfo {author} {\bibfnamefont {Y.}~\bibnamefont {Zhang}}, \bibinfo {author} {\bibfnamefont {Z.}~\bibnamefont {Lu}},  \emph {et~al.},\ }\bibfield  {title} {\enquote {\bibinfo {title} {Accurate shot-noise-limited calibration of a time-domain balanced homodyne detector for continuous-variable quantum key distribution},}\ }\href@noop {} {\bibfield  {journal} {\bibinfo  {journal} {J. Light. Technol.}\ } (\bibinfo {year} {2023}{\natexlab{b}})}\BibitemShut {NoStop}%
\bibitem [{\citenamefont {Liu}\ \emph {et~al.}(2016{\natexlab{b}})\citenamefont {Liu}, \citenamefont {Wang}, \citenamefont {Bai},\ and\ \citenamefont {Li}}]{jian_ActaPhysSin_2016}%
  \BibitemOpen
  \bibfield  {author} {\bibinfo {author} {\bibfnamefont {J.}~\bibnamefont {Liu}}, \bibinfo {author} {\bibfnamefont {X.}~\bibnamefont {Wang}}, \bibinfo {author} {\bibfnamefont {Z.}~\bibnamefont {Bai}}, \ and\ \bibinfo {author} {\bibfnamefont {Y.}~\bibnamefont {Li}},\ }\bibfield  {title} {\enquote {\bibinfo {title} {High precision auto-balance of the time-domain pulsed homodyne detector},}\ }\href@noop {} {\bibfield  {journal} {\bibinfo  {journal} {Acta Phys. Sin.}\ }\textbf {\bibinfo {volume} {65}} (\bibinfo {year} {2016}{\natexlab{b}})}\BibitemShut {NoStop}%
\bibitem [{\citenamefont {Tang}\ \emph {et~al.}(2020)\citenamefont {Tang}, \citenamefont {Kumar}, \citenamefont {Ren}, \citenamefont {Wonfor}, \citenamefont {Penty} \emph {et~al.}}]{tang_OptComm_2020}%
  \BibitemOpen
  \bibfield  {author} {\bibinfo {author} {\bibfnamefont {X.}~\bibnamefont {Tang}}, \bibinfo {author} {\bibfnamefont {R.}~\bibnamefont {Kumar}}, \bibinfo {author} {\bibfnamefont {S.}~\bibnamefont {Ren}}, \bibinfo {author} {\bibfnamefont {A.}~\bibnamefont {Wonfor}}, \bibinfo {author} {\bibfnamefont {R.~V.}\ \bibnamefont {Penty}},  \emph {et~al.},\ }\bibfield  {title} {\enquote {\bibinfo {title} {Performance of continuous variable quantum key distribution system at different detector bandwidth},}\ }\href@noop {} {\bibfield  {journal} {\bibinfo  {journal} {Opt. Commun.}\ }\textbf {\bibinfo {volume} {471}},\ \bibinfo {pages} {126034} (\bibinfo {year} {2020})}\BibitemShut {NoStop}%
\bibitem [{\citenamefont {Pereira}\ \emph {et~al.}(2021)\citenamefont {Pereira}, \citenamefont {Almeida}, \citenamefont {Fac{\~a}o}, \citenamefont {Pinto},\ and\ \citenamefont {Silva}}]{pereira_EPJQuanTech_2021}%
  \BibitemOpen
  \bibfield  {author} {\bibinfo {author} {\bibfnamefont {D.}~\bibnamefont {Pereira}}, \bibinfo {author} {\bibfnamefont {M.}~\bibnamefont {Almeida}}, \bibinfo {author} {\bibfnamefont {M.}~\bibnamefont {Fac{\~a}o}}, \bibinfo {author} {\bibfnamefont {A.~N.}\ \bibnamefont {Pinto}}, \ and\ \bibinfo {author} {\bibfnamefont {N.~A.}\ \bibnamefont {Silva}},\ }\bibfield  {title} {\enquote {\bibinfo {title} {Impact of receiver imbalances on the security of continuous variables quantum key distribution},}\ }\href@noop {} {\bibfield  {journal} {\bibinfo  {journal} {EPJ Quantum Technol.}\ }\textbf {\bibinfo {volume} {8}},\ \bibinfo {pages} {1--12} (\bibinfo {year} {2021})}\BibitemShut {NoStop}%
\bibitem [{\citenamefont {Fossier}\ \emph {et~al.}(2009{\natexlab{b}})\citenamefont {Fossier}, \citenamefont {Diamanti}, \citenamefont {Debuisschert}, \citenamefont {Villing}, \citenamefont {Tualle-Brouri} \emph {et~al.}}]{Fossier_NewJPhys_2009}%
  \BibitemOpen
  \bibfield  {author} {\bibinfo {author} {\bibfnamefont {S.}~\bibnamefont {Fossier}}, \bibinfo {author} {\bibfnamefont {E.}~\bibnamefont {Diamanti}}, \bibinfo {author} {\bibfnamefont {T.}~\bibnamefont {Debuisschert}}, \bibinfo {author} {\bibfnamefont {A.}~\bibnamefont {Villing}}, \bibinfo {author} {\bibfnamefont {R.}~\bibnamefont {Tualle-Brouri}},  \emph {et~al.},\ }\bibfield  {title} {\enquote {\bibinfo {title} {Field test of a continuous-variable quantum key distribution prototype},}\ }\href@noop {} {\bibfield  {journal} {\bibinfo  {journal} {New J. Phys.}\ }\textbf {\bibinfo {volume} {11}},\ \bibinfo {pages} {045023} (\bibinfo {year} {2009}{\natexlab{b}})}\BibitemShut {NoStop}%
\bibitem [{\citenamefont {Zhang}\ \emph {et~al.}(2020{\natexlab{e}})\citenamefont {Zhang}, \citenamefont {Huang}, \citenamefont {Chen}, \citenamefont {Li}, \citenamefont {Yu} \emph {et~al.}}]{Zhang_PhysRevApplied_2020}%
  \BibitemOpen
  \bibfield  {author} {\bibinfo {author} {\bibfnamefont {Y.}~\bibnamefont {Zhang}}, \bibinfo {author} {\bibfnamefont {Y.}~\bibnamefont {Huang}}, \bibinfo {author} {\bibfnamefont {Z.}~\bibnamefont {Chen}}, \bibinfo {author} {\bibfnamefont {Z.}~\bibnamefont {Li}}, \bibinfo {author} {\bibfnamefont {S.}~\bibnamefont {Yu}},  \emph {et~al.},\ }\bibfield  {title} {\enquote {\bibinfo {title} {One-time shot-noise unit calibration method for continuous-variable quantum key distribution},}\ }\href@noop {} {\bibfield  {journal} {\bibinfo  {journal} {Phys. Rev. Appl.}\ }\textbf {\bibinfo {volume} {13}},\ \bibinfo {pages} {024058} (\bibinfo {year} {2020}{\natexlab{e}})}\BibitemShut {NoStop}%
\bibitem [{\citenamefont {Chin}\ \emph {et~al.}(2021)\citenamefont {Chin}, \citenamefont {Jain}, \citenamefont {Zibar}, \citenamefont {Andersen},\ and\ \citenamefont {Gehring}}]{Chin2021Machine}%
  \BibitemOpen
  \bibfield  {author} {\bibinfo {author} {\bibfnamefont {H.-M.}\ \bibnamefont {Chin}}, \bibinfo {author} {\bibfnamefont {N.}~\bibnamefont {Jain}}, \bibinfo {author} {\bibfnamefont {D.}~\bibnamefont {Zibar}}, \bibinfo {author} {\bibfnamefont {U.~L.}\ \bibnamefont {Andersen}}, \ and\ \bibinfo {author} {\bibfnamefont {T.}~\bibnamefont {Gehring}},\ }\bibfield  {title} {\enquote {\bibinfo {title} {Machine learning aided carrier recovery in continuous-variable quantum key distribution},}\ }\href@noop {} {\bibfield  {journal} {\bibinfo  {journal} {npj Quantum Information}\ }\textbf {\bibinfo {volume} {7}},\ \bibinfo {pages} {20} (\bibinfo {year} {2021})}\BibitemShut {NoStop}%
\bibitem [{\citenamefont {Hajomer}\ \emph {et~al.}(2023{\natexlab{b}})\citenamefont {Hajomer}, \citenamefont {Derkach}, \citenamefont {Jain}, \citenamefont {Chin}, \citenamefont {Andersen} \emph {et~al.}}]{hajomer2023longdistance}%
  \BibitemOpen
  \bibfield  {author} {\bibinfo {author} {\bibfnamefont {A.~A.~E.}\ \bibnamefont {Hajomer}}, \bibinfo {author} {\bibfnamefont {I.}~\bibnamefont {Derkach}}, \bibinfo {author} {\bibfnamefont {N.}~\bibnamefont {Jain}}, \bibinfo {author} {\bibfnamefont {H.-M.}\ \bibnamefont {Chin}}, \bibinfo {author} {\bibfnamefont {U.~L.}\ \bibnamefont {Andersen}},  \emph {et~al.},\ }\href@noop {} {\enquote {\bibinfo {title} {Long-distance continuous-variable quantum key distribution over 100 km fiber with local local oscillator},}\ } (\bibinfo {year} {2023}{\natexlab{b}}),\ \Eprint {http://arxiv.org/abs/2305.08156} {arXiv:2305.08156 [quant-ph]} \BibitemShut {NoStop}%
\bibitem [{\citenamefont {Huang}\ \emph {et~al.}(2020{\natexlab{a}})\citenamefont {Huang}, \citenamefont {Zhang}, \citenamefont {Xu}, \citenamefont {Huang},\ and\ \citenamefont {Yu}}]{Huang_JPhysB_2020}%
  \BibitemOpen
  \bibfield  {author} {\bibinfo {author} {\bibfnamefont {Y.}~\bibnamefont {Huang}}, \bibinfo {author} {\bibfnamefont {Y.}~\bibnamefont {Zhang}}, \bibinfo {author} {\bibfnamefont {B.}~\bibnamefont {Xu}}, \bibinfo {author} {\bibfnamefont {L.}~\bibnamefont {Huang}}, \ and\ \bibinfo {author} {\bibfnamefont {S.}~\bibnamefont {Yu}},\ }\bibfield  {title} {\enquote {\bibinfo {title} {A modified practical homodyne detector model for continuous-variable quantum key distribution: detailed security analysis and improvement by the phase-sensitive amplifier},}\ }\href@noop {} {\bibfield  {journal} {\bibinfo  {journal} {J. Phys. B: At., Mol. Opt. Phys.}\ } (\bibinfo {year} {2020}{\natexlab{a}})}\BibitemShut {NoStop}%
\bibitem [{\citenamefont {Huang}, \citenamefont {Zhang},\ and\ \citenamefont {Yu}(2021)}]{Huang2021ContinuousVariableMQ}%
  \BibitemOpen
  \bibfield  {author} {\bibinfo {author} {\bibfnamefont {L.}~\bibnamefont {Huang}}, \bibinfo {author} {\bibfnamefont {Y.}~\bibnamefont {Zhang}}, \ and\ \bibinfo {author} {\bibfnamefont {S.}~\bibnamefont {Yu}},\ }\bibfield  {title} {\enquote {\bibinfo {title} {Continuous-variable measurement-device-independent quantum key distribution with one-time shot-noise unit calibration},}\ }\href {https://api.semanticscholar.org/CorpusID:235292460} {\bibfield  {journal} {\bibinfo  {journal} {Chinese Phys. Lett.}\ }\textbf {\bibinfo {volume} {38}} (\bibinfo {year} {2021})}\BibitemShut {NoStop}%
\bibitem [{\citenamefont {Chu}\ \emph {et~al.}(2021{\natexlab{b}})\citenamefont {Chu}, \citenamefont {Zhang}, \citenamefont {Huang}, \citenamefont {Yu}, \citenamefont {Chen} \emph {et~al.}}]{Chu2021PracticalSM}%
  \BibitemOpen
  \bibfield  {author} {\bibinfo {author} {\bibfnamefont {B.}~\bibnamefont {Chu}}, \bibinfo {author} {\bibfnamefont {Y.}~\bibnamefont {Zhang}}, \bibinfo {author} {\bibfnamefont {Y.}~\bibnamefont {Huang}}, \bibinfo {author} {\bibfnamefont {S.}~\bibnamefont {Yu}}, \bibinfo {author} {\bibfnamefont {Z.}~\bibnamefont {Chen}},  \emph {et~al.},\ }\bibfield  {title} {\enquote {\bibinfo {title} {Practical source monitoring for continuous-variable quantum key distribution},}\ }\href {https://api.semanticscholar.org/CorpusID:233885425} {\bibfield  {journal} {\bibinfo  {journal} {Quantum Sci. Technol.}\ }\textbf {\bibinfo {volume} {6}} (\bibinfo {year} {2021}{\natexlab{b}})}\BibitemShut {NoStop}%
\bibitem [{\citenamefont {Cubukcu}(2012)}]{Cubukcu2012RootRC}%
  \BibitemOpen
  \bibfield  {author} {\bibinfo {author} {\bibfnamefont {E.}~\bibnamefont {Cubukcu}},\ }\bibfield  {title} {\enquote {\bibinfo {title} {Root raised cosine (rrc) filters and pulse shaping in communication systems},}\ }in\ \href@noop {} {\emph {\bibinfo {booktitle} {AIAA Conference}}},\ \bibinfo {series and number} {\bibinfo {number} {JSC-CN-26387}}\ (\bibinfo {year} {2012})\BibitemShut {NoStop}%
\bibitem [{\citenamefont {Marie}\ and\ \citenamefont {All{\'e}aume}(2017)}]{Marie_PhysRevA_2017}%
  \BibitemOpen
  \bibfield  {author} {\bibinfo {author} {\bibfnamefont {A.}~\bibnamefont {Marie}}\ and\ \bibinfo {author} {\bibfnamefont {R.}~\bibnamefont {All{\'e}aume}},\ }\bibfield  {title} {\enquote {\bibinfo {title} {Self-coherent phase reference sharing for continuous-variable quantum key distribution},}\ }\href@noop {} {\bibfield  {journal} {\bibinfo  {journal} {Phys. Rev. A}\ }\textbf {\bibinfo {volume} {95}},\ \bibinfo {pages} {012316} (\bibinfo {year} {2017})}\BibitemShut {NoStop}%
\bibitem [{\citenamefont {Corvaja}(2017)}]{Corvaja_PhysRevA_2017}%
  \BibitemOpen
  \bibfield  {author} {\bibinfo {author} {\bibfnamefont {R.}~\bibnamefont {Corvaja}},\ }\bibfield  {title} {\enquote {\bibinfo {title} {Phase-noise limitations in continuous-variable quantum key distribution with homodyne detection},}\ }\href@noop {} {\bibfield  {journal} {\bibinfo  {journal} {Phys. Rev. A}\ }\textbf {\bibinfo {volume} {95}},\ \bibinfo {pages} {022315} (\bibinfo {year} {2017})}\BibitemShut {NoStop}%
\bibitem [{\citenamefont {Qi}\ and\ \citenamefont {Lim}(2018)}]{Qi_PhysRevApplied_2018}%
  \BibitemOpen
  \bibfield  {author} {\bibinfo {author} {\bibfnamefont {B.}~\bibnamefont {Qi}}\ and\ \bibinfo {author} {\bibfnamefont {C.~C.~W.}\ \bibnamefont {Lim}},\ }\bibfield  {title} {\enquote {\bibinfo {title} {Noise analysis of simultaneous quantum key distribution and classical communication scheme using a true local oscillator},}\ }\href@noop {} {\bibfield  {journal} {\bibinfo  {journal} {Phys. Rev. Appl.}\ }\textbf {\bibinfo {volume} {9}},\ \bibinfo {pages} {054008} (\bibinfo {year} {2018})}\BibitemShut {NoStop}%
\bibitem [{\citenamefont {Zou}, \citenamefont {Mao},\ and\ \citenamefont {Chen}(2019)}]{Zou_JApplPhys_2019}%
  \BibitemOpen
  \bibfield  {author} {\bibinfo {author} {\bibfnamefont {M.}~\bibnamefont {Zou}}, \bibinfo {author} {\bibfnamefont {Y.}~\bibnamefont {Mao}}, \ and\ \bibinfo {author} {\bibfnamefont {T.-Y.}\ \bibnamefont {Chen}},\ }\bibfield  {title} {\enquote {\bibinfo {title} {Phase estimation using homodyne detection for continuous variable quantum key distribution},}\ }\href@noop {} {\bibfield  {journal} {\bibinfo  {journal} {J. Appl. Phys.}\ }\textbf {\bibinfo {volume} {126}},\ \bibinfo {pages} {063105} (\bibinfo {year} {2019})}\BibitemShut {NoStop}%
\bibitem [{\citenamefont {Van~Assche}, \citenamefont {Cardinal},\ and\ \citenamefont {Cerf}(2004)}]{van_IEEETransInforTheory_2004}%
  \BibitemOpen
  \bibfield  {author} {\bibinfo {author} {\bibfnamefont {G.}~\bibnamefont {Van~Assche}}, \bibinfo {author} {\bibfnamefont {J.}~\bibnamefont {Cardinal}}, \ and\ \bibinfo {author} {\bibfnamefont {N.~J.}\ \bibnamefont {Cerf}},\ }\bibfield  {title} {\enquote {\bibinfo {title} {Reconciliation of a quantum-distributed gaussian key},}\ }\href@noop {} {\bibfield  {journal} {\bibinfo  {journal} {IEEE Trans. Inf. Theory}\ }\textbf {\bibinfo {volume} {50}},\ \bibinfo {pages} {394--400} (\bibinfo {year} {2004})}\BibitemShut {NoStop}%
\bibitem [{\citenamefont {Mountogiannakis}, \citenamefont {Papanastasiou},\ and\ \citenamefont {Pirandola}(2022)}]{mountogiannakis_PhysRevA_2022}%
  \BibitemOpen
  \bibfield  {author} {\bibinfo {author} {\bibfnamefont {A.~G.}\ \bibnamefont {Mountogiannakis}}, \bibinfo {author} {\bibfnamefont {P.}~\bibnamefont {Papanastasiou}}, \ and\ \bibinfo {author} {\bibfnamefont {S.}~\bibnamefont {Pirandola}},\ }\bibfield  {title} {\enquote {\bibinfo {title} {Data postprocessing for the one-way heterodyne protocol under composable finite-size security},}\ }\href@noop {} {\bibfield  {journal} {\bibinfo  {journal} {Phys. Rev. A}\ }\textbf {\bibinfo {volume} {106}},\ \bibinfo {pages} {042606} (\bibinfo {year} {2022})}\BibitemShut {NoStop}%
\bibitem [{\citenamefont {Yang}\ \emph {et~al.}(2023{\natexlab{b}})\citenamefont {Yang}, \citenamefont {Yan}, \citenamefont {Yang}, \citenamefont {Lu}, \citenamefont {Lu} \emph {et~al.}}]{yang2023information}%
  \BibitemOpen
  \bibfield  {author} {\bibinfo {author} {\bibfnamefont {S.}~\bibnamefont {Yang}}, \bibinfo {author} {\bibfnamefont {Z.}~\bibnamefont {Yan}}, \bibinfo {author} {\bibfnamefont {H.}~\bibnamefont {Yang}}, \bibinfo {author} {\bibfnamefont {Q.}~\bibnamefont {Lu}}, \bibinfo {author} {\bibfnamefont {Z.}~\bibnamefont {Lu}},  \emph {et~al.},\ }\bibfield  {title} {\enquote {\bibinfo {title} {Information reconciliation of continuous-variables quantum key distribution: principles, implementations and applications},}\ }\href@noop {} {\bibfield  {journal} {\bibinfo  {journal} {EPJ Quantum Technol.}\ }\textbf {\bibinfo {volume} {10}},\ \bibinfo {pages} {40} (\bibinfo {year} {2023}{\natexlab{b}})}\BibitemShut {NoStop}%
\bibitem [{\citenamefont {Ma}\ \emph {et~al.}(2011)\citenamefont {Ma}, \citenamefont {Fung}, \citenamefont {Boileau},\ and\ \citenamefont {Chau}}]{ma2011universally}%
  \BibitemOpen
  \bibfield  {author} {\bibinfo {author} {\bibfnamefont {X.}~\bibnamefont {Ma}}, \bibinfo {author} {\bibfnamefont {C.-H.~F.}\ \bibnamefont {Fung}}, \bibinfo {author} {\bibfnamefont {J.-C.}\ \bibnamefont {Boileau}}, \ and\ \bibinfo {author} {\bibfnamefont {H.}~\bibnamefont {Chau}},\ }\bibfield  {title} {\enquote {\bibinfo {title} {Universally composable and customizable post-processing for practical quantum key distribution},}\ }\href@noop {} {\bibfield  {journal} {\bibinfo  {journal} {Comput. Secur.}\ }\textbf {\bibinfo {volume} {30}},\ \bibinfo {pages} {172--177} (\bibinfo {year} {2011})}\BibitemShut {NoStop}%
\bibitem [{\citenamefont {Wen}\ \emph {et~al.}(2021)\citenamefont {Wen}, \citenamefont {Li}, \citenamefont {Mao}, \citenamefont {Wen},\ and\ \citenamefont {Chen}}]{wen2021improved}%
  \BibitemOpen
  \bibfield  {author} {\bibinfo {author} {\bibfnamefont {X.}~\bibnamefont {Wen}}, \bibinfo {author} {\bibfnamefont {Q.}~\bibnamefont {Li}}, \bibinfo {author} {\bibfnamefont {H.}~\bibnamefont {Mao}}, \bibinfo {author} {\bibfnamefont {X.}~\bibnamefont {Wen}}, \ and\ \bibinfo {author} {\bibfnamefont {N.}~\bibnamefont {Chen}},\ }\bibfield  {title} {\enquote {\bibinfo {title} {An improved slice reconciliation protocol for continuous-variable quantum key distribution},}\ }\href@noop {} {\bibfield  {journal} {\bibinfo  {journal} {Entropy}\ }\textbf {\bibinfo {volume} {23}},\ \bibinfo {pages} {1317} (\bibinfo {year} {2021})}\BibitemShut {NoStop}%
\bibitem [{\citenamefont {Leverrier}, \citenamefont {Grosshans},\ and\ \citenamefont {Grangier}(2010)}]{Leverrier_PRA_Finite}%
  \BibitemOpen
  \bibfield  {author} {\bibinfo {author} {\bibfnamefont {A.}~\bibnamefont {Leverrier}}, \bibinfo {author} {\bibfnamefont {F.}~\bibnamefont {Grosshans}}, \ and\ \bibinfo {author} {\bibfnamefont {P.}~\bibnamefont {Grangier}},\ }\bibfield  {title} {\enquote {\bibinfo {title} {Finite-size analysis of a continuous-variable quantum key distribution},}\ }\href {\doibase 10.1103/PhysRevA.81.062343} {\bibfield  {journal} {\bibinfo  {journal} {Phys. Rev. A}\ }\textbf {\bibinfo {volume} {81}},\ \bibinfo {pages} {062343} (\bibinfo {year} {2010})}\BibitemShut {NoStop}%
\bibitem [{\citenamefont {Ma}\ \emph {et~al.}(2023)\citenamefont {Ma}, \citenamefont {Yang}, \citenamefont {Zhang}, \citenamefont {Shao}, \citenamefont {Liu} \emph {et~al.}}]{ma_SciChinaInforSci_2023}%
  \BibitemOpen
  \bibfield  {author} {\bibinfo {author} {\bibfnamefont {L.}~\bibnamefont {Ma}}, \bibinfo {author} {\bibfnamefont {J.}~\bibnamefont {Yang}}, \bibinfo {author} {\bibfnamefont {T.}~\bibnamefont {Zhang}}, \bibinfo {author} {\bibfnamefont {Y.}~\bibnamefont {Shao}}, \bibinfo {author} {\bibfnamefont {J.}~\bibnamefont {Liu}},  \emph {et~al.},\ }\bibfield  {title} {\enquote {\bibinfo {title} {Practical continuous-variable quantum key distribution with feasible optimization parameters},}\ }\href@noop {} {\bibfield  {journal} {\bibinfo  {journal} {Sci. China Inf. Sci.}\ }\textbf {\bibinfo {volume} {66}},\ \bibinfo {pages} {1--12} (\bibinfo {year} {2023})}\BibitemShut {NoStop}%
\bibitem [{\citenamefont {Jouguet}, \citenamefont {Kunz-Jacques},\ and\ \citenamefont {Leverrier}(2011)}]{jouguet_PhysRevA_2011}%
  \BibitemOpen
  \bibfield  {author} {\bibinfo {author} {\bibfnamefont {P.}~\bibnamefont {Jouguet}}, \bibinfo {author} {\bibfnamefont {S.}~\bibnamefont {Kunz-Jacques}}, \ and\ \bibinfo {author} {\bibfnamefont {A.}~\bibnamefont {Leverrier}},\ }\bibfield  {title} {\enquote {\bibinfo {title} {Long-distance continuous-variable quantum key distribution with a gaussian modulation},}\ }\href@noop {} {\bibfield  {journal} {\bibinfo  {journal} {Phys. Rev. A}\ }\textbf {\bibinfo {volume} {84}},\ \bibinfo {pages} {062317} (\bibinfo {year} {2011})}\BibitemShut {NoStop}%
\bibitem [{\citenamefont {Jouguet}, \citenamefont {Elkouss},\ and\ \citenamefont {Kunz-Jacques}(2014)}]{jouguet_PhysRevA_2014}%
  \BibitemOpen
  \bibfield  {author} {\bibinfo {author} {\bibfnamefont {P.}~\bibnamefont {Jouguet}}, \bibinfo {author} {\bibfnamefont {D.}~\bibnamefont {Elkouss}}, \ and\ \bibinfo {author} {\bibfnamefont {S.}~\bibnamefont {Kunz-Jacques}},\ }\bibfield  {title} {\enquote {\bibinfo {title} {High-bit-rate continuous-variable quantum key distribution},}\ }\href@noop {} {\bibfield  {journal} {\bibinfo  {journal} {Phys. Rev. A}\ }\textbf {\bibinfo {volume} {90}},\ \bibinfo {pages} {042329} (\bibinfo {year} {2014})}\BibitemShut {NoStop}%
\bibitem [{\citenamefont {liang Bai}, \citenamefont {Yang},\ and\ \citenamefont {Li}(2017)}]{Bai2017HighefficiencyRF}%
  \BibitemOpen
  \bibfield  {author} {\bibinfo {author} {\bibfnamefont {Z.}~\bibnamefont {liang Bai}}, \bibinfo {author} {\bibfnamefont {S.}~\bibnamefont {Yang}}, \ and\ \bibinfo {author} {\bibfnamefont {Y.}~\bibnamefont {Li}},\ }\bibfield  {title} {\enquote {\bibinfo {title} {High-efficiency reconciliation for continuous variable quantum key distribution},}\ }\href {https://api.semanticscholar.org/CorpusID:126250767} {\bibfield  {journal} {\bibinfo  {journal} {Jpn. J. Appl. Phys.}\ }\textbf {\bibinfo {volume} {56}} (\bibinfo {year} {2017})}\BibitemShut {NoStop}%
\bibitem [{\citenamefont {Wang}\ \emph {et~al.}(2018{\natexlab{b}})\citenamefont {Wang}, \citenamefont {Zhang}, \citenamefont {Yu},\ and\ \citenamefont {Guo}}]{wang_SciRep_2018}%
  \BibitemOpen
  \bibfield  {author} {\bibinfo {author} {\bibfnamefont {X.}~\bibnamefont {Wang}}, \bibinfo {author} {\bibfnamefont {Y.}~\bibnamefont {Zhang}}, \bibinfo {author} {\bibfnamefont {S.}~\bibnamefont {Yu}}, \ and\ \bibinfo {author} {\bibfnamefont {H.}~\bibnamefont {Guo}},\ }\bibfield  {title} {\enquote {\bibinfo {title} {High speed error correction for continuous-variable quantum key distribution with multi-edge type ldpc code},}\ }\href@noop {} {\bibfield  {journal} {\bibinfo  {journal} {Sci. Rep.}\ }\textbf {\bibinfo {volume} {8}},\ \bibinfo {pages} {10543} (\bibinfo {year} {2018}{\natexlab{b}})}\BibitemShut {NoStop}%
\bibitem [{\citenamefont {Milicevic}\ \emph {et~al.}(2018)\citenamefont {Milicevic}, \citenamefont {Feng}, \citenamefont {Zhang},\ and\ \citenamefont {Gulak}}]{milicevic_npjQuanInfor_2018}%
  \BibitemOpen
  \bibfield  {author} {\bibinfo {author} {\bibfnamefont {M.}~\bibnamefont {Milicevic}}, \bibinfo {author} {\bibfnamefont {C.}~\bibnamefont {Feng}}, \bibinfo {author} {\bibfnamefont {L.~M.}\ \bibnamefont {Zhang}}, \ and\ \bibinfo {author} {\bibfnamefont {P.~G.}\ \bibnamefont {Gulak}},\ }\bibfield  {title} {\enquote {\bibinfo {title} {Quasi-cyclic multi-edge ldpc codes for long-distance quantum cryptography},}\ }\href@noop {} {\bibfield  {journal} {\bibinfo  {journal} {npj Quantum Inf.}\ }\textbf {\bibinfo {volume} {4}},\ \bibinfo {pages} {21} (\bibinfo {year} {2018})}\BibitemShut {NoStop}%
\bibitem [{\citenamefont {Zhao}\ \emph {et~al.}(2018{\natexlab{b}})\citenamefont {Zhao}, \citenamefont {Shen}, \citenamefont {Xiao},\ and\ \citenamefont {Wang}}]{zhao_SciChinaPhys_2018}%
  \BibitemOpen
  \bibfield  {author} {\bibinfo {author} {\bibfnamefont {S.}~\bibnamefont {Zhao}}, \bibinfo {author} {\bibfnamefont {Z.}~\bibnamefont {Shen}}, \bibinfo {author} {\bibfnamefont {H.}~\bibnamefont {Xiao}}, \ and\ \bibinfo {author} {\bibfnamefont {L.}~\bibnamefont {Wang}},\ }\bibfield  {title} {\enquote {\bibinfo {title} {Multidimensional reconciliation protocol for continuous-variable quantum key agreement with polar coding},}\ }\href@noop {} {\bibfield  {journal} {\bibinfo  {journal} {Sci. China: Phys., Mech. Astron.}\ }\textbf {\bibinfo {volume} {61}},\ \bibinfo {pages} {1--4} (\bibinfo {year} {2018}{\natexlab{b}})}\BibitemShut {NoStop}%
\bibitem [{\citenamefont {Zhou}\ \emph {et~al.}(2019{\natexlab{a}})\citenamefont {Zhou}, \citenamefont {Wang}, \citenamefont {Zhang}, \citenamefont {Zhang}, \citenamefont {Yu} \emph {et~al.}}]{zhou_PhysRevAppl_2019}%
  \BibitemOpen
  \bibfield  {author} {\bibinfo {author} {\bibfnamefont {C.}~\bibnamefont {Zhou}}, \bibinfo {author} {\bibfnamefont {X.}~\bibnamefont {Wang}}, \bibinfo {author} {\bibfnamefont {Y.}~\bibnamefont {Zhang}}, \bibinfo {author} {\bibfnamefont {Z.}~\bibnamefont {Zhang}}, \bibinfo {author} {\bibfnamefont {S.}~\bibnamefont {Yu}},  \emph {et~al.},\ }\bibfield  {title} {\enquote {\bibinfo {title} {Continuous-variable quantum key distribution with rateless reconciliation protocol},}\ }\href@noop {} {\bibfield  {journal} {\bibinfo  {journal} {Phys. Rev. Appl.}\ }\textbf {\bibinfo {volume} {12}},\ \bibinfo {pages} {054013} (\bibinfo {year} {2019}{\natexlab{a}})}\BibitemShut {NoStop}%
\bibitem [{\citenamefont {Li}\ \emph {et~al.}(2020)\citenamefont {Li}, \citenamefont {Zhang}, \citenamefont {Li}, \citenamefont {Xu}, \citenamefont {Ma} \emph {et~al.}}]{li_SciRep_2020}%
  \BibitemOpen
  \bibfield  {author} {\bibinfo {author} {\bibfnamefont {Y.}~\bibnamefont {Li}}, \bibinfo {author} {\bibfnamefont {X.}~\bibnamefont {Zhang}}, \bibinfo {author} {\bibfnamefont {Y.}~\bibnamefont {Li}}, \bibinfo {author} {\bibfnamefont {B.}~\bibnamefont {Xu}}, \bibinfo {author} {\bibfnamefont {L.}~\bibnamefont {Ma}},  \emph {et~al.},\ }\bibfield  {title} {\enquote {\bibinfo {title} {High-throughput gpu layered decoder of quasi-cyclic multi-edge type low density parity check codes in continuous-variable quantum key distribution systems},}\ }\href@noop {} {\bibfield  {journal} {\bibinfo  {journal} {Sci. Rep.}\ }\textbf {\bibinfo {volume} {10}},\ \bibinfo {pages} {14561} (\bibinfo {year} {2020})}\BibitemShut {NoStop}%
\bibitem [{\citenamefont {Yang}, \citenamefont {Lu},\ and\ \citenamefont {Li}(2020)}]{yang_jourLightTech_2020}%
  \BibitemOpen
  \bibfield  {author} {\bibinfo {author} {\bibfnamefont {S.}~\bibnamefont {Yang}}, \bibinfo {author} {\bibfnamefont {Z.}~\bibnamefont {Lu}}, \ and\ \bibinfo {author} {\bibfnamefont {Y.}~\bibnamefont {Li}},\ }\bibfield  {title} {\enquote {\bibinfo {title} {High-speed post-processing in continuous-variable quantum key distribution based on fpga implementation},}\ }\href@noop {} {\bibfield  {journal} {\bibinfo  {journal} {J. Light. Technol.}\ }\textbf {\bibinfo {volume} {38}},\ \bibinfo {pages} {3935--3941} (\bibinfo {year} {2020})}\BibitemShut {NoStop}%
\bibitem [{\citenamefont {Mani}\ \emph {et~al.}(2021)\citenamefont {Mani}, \citenamefont {Gehring}, \citenamefont {Grabenweger}, \citenamefont {{\"O}mer}, \citenamefont {Pacher} \emph {et~al.}}]{mani_PhysRevA_2021}%
  \BibitemOpen
  \bibfield  {author} {\bibinfo {author} {\bibfnamefont {H.}~\bibnamefont {Mani}}, \bibinfo {author} {\bibfnamefont {T.}~\bibnamefont {Gehring}}, \bibinfo {author} {\bibfnamefont {P.}~\bibnamefont {Grabenweger}}, \bibinfo {author} {\bibfnamefont {B.}~\bibnamefont {{\"O}mer}}, \bibinfo {author} {\bibfnamefont {C.}~\bibnamefont {Pacher}},  \emph {et~al.},\ }\bibfield  {title} {\enquote {\bibinfo {title} {Multiedge-type low-density parity-check codes for continuous-variable quantum key distribution},}\ }\href@noop {} {\bibfield  {journal} {\bibinfo  {journal} {Phys. Rev. A}\ }\textbf {\bibinfo {volume} {103}},\ \bibinfo {pages} {062419} (\bibinfo {year} {2021})}\BibitemShut {NoStop}%
\bibitem [{\citenamefont {Jeong}, \citenamefont {Jung},\ and\ \citenamefont {Ha}(2022)}]{jeong_npjQuanInfor_2022}%
  \BibitemOpen
  \bibfield  {author} {\bibinfo {author} {\bibfnamefont {S.}~\bibnamefont {Jeong}}, \bibinfo {author} {\bibfnamefont {H.}~\bibnamefont {Jung}}, \ and\ \bibinfo {author} {\bibfnamefont {J.}~\bibnamefont {Ha}},\ }\bibfield  {title} {\enquote {\bibinfo {title} {Rate-compatible multi-edge type low-density parity-check code ensembles for continuous-variable quantum key distribution systems},}\ }\href@noop {} {\bibfield  {journal} {\bibinfo  {journal} {npj Quantum Inf.}\ }\textbf {\bibinfo {volume} {8}},\ \bibinfo {pages} {6} (\bibinfo {year} {2022})}\BibitemShut {NoStop}%
\bibitem [{\citenamefont {Li}\ \emph {et~al.}(2019{\natexlab{b}})\citenamefont {Li}, \citenamefont {Wen}, \citenamefont {Mao},\ and\ \citenamefont {Wen}}]{li_QuanInforProc_2019}%
  \BibitemOpen
  \bibfield  {author} {\bibinfo {author} {\bibfnamefont {Q.}~\bibnamefont {Li}}, \bibinfo {author} {\bibfnamefont {X.}~\bibnamefont {Wen}}, \bibinfo {author} {\bibfnamefont {H.}~\bibnamefont {Mao}}, \ and\ \bibinfo {author} {\bibfnamefont {X.}~\bibnamefont {Wen}},\ }\bibfield  {title} {\enquote {\bibinfo {title} {An improved multidimensional reconciliation algorithm for continuous-variable quantum key distribution},}\ }\href@noop {} {\bibfield  {journal} {\bibinfo  {journal} {Quantum Inf. Process.}\ }\textbf {\bibinfo {volume} {18}},\ \bibinfo {pages} {1--20} (\bibinfo {year} {2019}{\natexlab{b}})}\BibitemShut {NoStop}%
\bibitem [{\citenamefont {Zhou}\ \emph {et~al.}(2019{\natexlab{b}})\citenamefont {Zhou}, \citenamefont {Wang}, \citenamefont {Zhang}, \citenamefont {Zhang} \emph {et~al.}}]{Zhou2019ContinuousVariableQK}%
  \BibitemOpen
  \bibfield  {author} {\bibinfo {author} {\bibfnamefont {C.}~\bibnamefont {Zhou}}, \bibinfo {author} {\bibfnamefont {X.}~\bibnamefont {Wang}}, \bibinfo {author} {\bibfnamefont {Y.}~\bibnamefont {Zhang}}, \bibinfo {author} {\bibfnamefont {Z.}~\bibnamefont {Zhang}},  \emph {et~al.},\ }\bibfield  {title} {\enquote {\bibinfo {title} {Continuous-variable quantum key distribution with rateless reconciliation protocol},}\ }\href {https://api.semanticscholar.org/CorpusID:199552277} {\bibfield  {journal} {\bibinfo  {journal} {Phys. Rev. Appl.}\ } (\bibinfo {year} {2019}{\natexlab{b}})}\BibitemShut {NoStop}%
\bibitem [{\citenamefont {Jiang}\ \emph {et~al.}(2018)\citenamefont {Jiang}, \citenamefont {Yang}, \citenamefont {Huang},\ and\ \citenamefont {Zeng}}]{jiang_IEEEPhotJour_2018}%
  \BibitemOpen
  \bibfield  {author} {\bibinfo {author} {\bibfnamefont {X.-Q.}\ \bibnamefont {Jiang}}, \bibinfo {author} {\bibfnamefont {S.}~\bibnamefont {Yang}}, \bibinfo {author} {\bibfnamefont {P.}~\bibnamefont {Huang}}, \ and\ \bibinfo {author} {\bibfnamefont {G.}~\bibnamefont {Zeng}},\ }\bibfield  {title} {\enquote {\bibinfo {title} {High-speed reconciliation for cvqkd based on spatially coupled ldpc codes},}\ }\href@noop {} {\bibfield  {journal} {\bibinfo  {journal} {IEEE Photon. J.}\ }\textbf {\bibinfo {volume} {10}},\ \bibinfo {pages} {1--10} (\bibinfo {year} {2018})}\BibitemShut {NoStop}%
\bibitem [{\citenamefont {Zhang}\ \emph {et~al.}(2020{\natexlab{f}})\citenamefont {Zhang}, \citenamefont {Jiang}, \citenamefont {Feng}, \citenamefont {Qiu},\ and\ \citenamefont {Bai}}]{zhang_Entropy_2020}%
  \BibitemOpen
  \bibfield  {author} {\bibinfo {author} {\bibfnamefont {K.}~\bibnamefont {Zhang}}, \bibinfo {author} {\bibfnamefont {X.-Q.}\ \bibnamefont {Jiang}}, \bibinfo {author} {\bibfnamefont {Y.}~\bibnamefont {Feng}}, \bibinfo {author} {\bibfnamefont {R.}~\bibnamefont {Qiu}}, \ and\ \bibinfo {author} {\bibfnamefont {E.}~\bibnamefont {Bai}},\ }\bibfield  {title} {\enquote {\bibinfo {title} {High efficiency continuous-variable quantum key distribution based on atsc 3.0 ldpc codes},}\ }\href@noop {} {\bibfield  {journal} {\bibinfo  {journal} {Entropy}\ }\textbf {\bibinfo {volume} {22}},\ \bibinfo {pages} {1087} (\bibinfo {year} {2020}{\natexlab{f}})}\BibitemShut {NoStop}%
\bibitem [{\citenamefont {Guo}\ \emph {et~al.}(2020)\citenamefont {Guo}, \citenamefont {He}, \citenamefont {Guo}, \citenamefont {Xue}, \citenamefont {Feng} \emph {et~al.}}]{guo_QuanInforProc_2020}%
  \BibitemOpen
  \bibfield  {author} {\bibinfo {author} {\bibfnamefont {D.}~\bibnamefont {Guo}}, \bibinfo {author} {\bibfnamefont {C.}~\bibnamefont {He}}, \bibinfo {author} {\bibfnamefont {T.}~\bibnamefont {Guo}}, \bibinfo {author} {\bibfnamefont {Z.}~\bibnamefont {Xue}}, \bibinfo {author} {\bibfnamefont {Q.}~\bibnamefont {Feng}},  \emph {et~al.},\ }\bibfield  {title} {\enquote {\bibinfo {title} {Comprehensive high-speed reconciliation for continuous-variable quantum key distribution},}\ }\href@noop {} {\bibfield  {journal} {\bibinfo  {journal} {Quantum Inf. Process.}\ }\textbf {\bibinfo {volume} {19}},\ \bibinfo {pages} {320} (\bibinfo {year} {2020})}\BibitemShut {NoStop}%
\bibitem [{\citenamefont {Yang}\ \emph {et~al.}(2021)\citenamefont {Yang}, \citenamefont {Liu}, \citenamefont {Lu}, \citenamefont {Bai}, \citenamefont {Wang} \emph {et~al.}}]{yang_IEEEAcc_2021}%
  \BibitemOpen
  \bibfield  {author} {\bibinfo {author} {\bibfnamefont {S.-S.}\ \bibnamefont {Yang}}, \bibinfo {author} {\bibfnamefont {J.-Q.}\ \bibnamefont {Liu}}, \bibinfo {author} {\bibfnamefont {Z.-G.}\ \bibnamefont {Lu}}, \bibinfo {author} {\bibfnamefont {Z.-L.}\ \bibnamefont {Bai}}, \bibinfo {author} {\bibfnamefont {X.-Y.}\ \bibnamefont {Wang}},  \emph {et~al.},\ }\bibfield  {title} {\enquote {\bibinfo {title} {An fpga-based ldpc decoder with ultra-long codes for continuous-variable quantum key distribution},}\ }\href@noop {} {\bibfield  {journal} {\bibinfo  {journal} {IEEE Access}\ }\textbf {\bibinfo {volume} {9}},\ \bibinfo {pages} {47687--47697} (\bibinfo {year} {2021})}\BibitemShut {NoStop}%
\bibitem [{\citenamefont {Xie}\ \emph {et~al.}(2022)\citenamefont {Xie}, \citenamefont {Zhang}, \citenamefont {Wang},\ and\ \citenamefont {Huang}}]{xie_Photonics_2022}%
  \BibitemOpen
  \bibfield  {author} {\bibinfo {author} {\bibfnamefont {J.}~\bibnamefont {Xie}}, \bibinfo {author} {\bibfnamefont {L.}~\bibnamefont {Zhang}}, \bibinfo {author} {\bibfnamefont {Y.}~\bibnamefont {Wang}}, \ and\ \bibinfo {author} {\bibfnamefont {D.}~\bibnamefont {Huang}},\ }\bibfield  {title} {\enquote {\bibinfo {title} {Deep neural network based reconciliation for cv-qkd},}\ }in\ \href@noop {} {\emph {\bibinfo {booktitle} {Photonics}}},\ Vol.~\bibinfo {volume} {9}\ (\bibinfo {organization} {MDPI},\ \bibinfo {year} {2022})\ p.\ \bibinfo {pages} {110}\BibitemShut {NoStop}%
\bibitem [{\citenamefont {Sun}\ and\ \citenamefont {Liang}(2023)}]{sun_OptEng_2023}%
  \BibitemOpen
  \bibfield  {author} {\bibinfo {author} {\bibfnamefont {X.}~\bibnamefont {Sun}}\ and\ \bibinfo {author} {\bibfnamefont {H.}~\bibnamefont {Liang}},\ }\bibfield  {title} {\enquote {\bibinfo {title} {Implementation of encoder and decoder for low-density parity-check codes in continuous-variable quantum key distribution on a field programmable gate array},}\ }\href@noop {} {\bibfield  {journal} {\bibinfo  {journal} {Opt. Eng.}\ }\textbf {\bibinfo {volume} {62}},\ \bibinfo {pages} {014105--014105} (\bibinfo {year} {2023})}\BibitemShut {NoStop}%
\bibitem [{\citenamefont {Zhou}\ \emph {et~al.}(2023)\citenamefont {Zhou}, \citenamefont {Li}, \citenamefont {Ma}, \citenamefont {Luo}, \citenamefont {Yang} \emph {et~al.}}]{zhou2023high}%
  \BibitemOpen
  \bibfield  {author} {\bibinfo {author} {\bibfnamefont {C.}~\bibnamefont {Zhou}}, \bibinfo {author} {\bibfnamefont {Y.}~\bibnamefont {Li}}, \bibinfo {author} {\bibfnamefont {L.}~\bibnamefont {Ma}}, \bibinfo {author} {\bibfnamefont {Y.}~\bibnamefont {Luo}}, \bibinfo {author} {\bibfnamefont {J.}~\bibnamefont {Yang}},  \emph {et~al.},\ }\bibfield  {title} {\enquote {\bibinfo {title} {A high-throughput and fpga-based ldpc decoder for continuous-variable quantum key distribution system},}\ }in\ \href@noop {} {\emph {\bibinfo {booktitle} {Quantum and Nonlinear Optics X}}},\ Vol.\ \bibinfo {volume} {12775}\ (\bibinfo {organization} {SPIE},\ \bibinfo {year} {2023})\ pp.\ \bibinfo {pages} {62--66}\BibitemShut {NoStop}%
\bibitem [{\citenamefont {Zhou}\ \emph {et~al.}(2022)\citenamefont {Zhou}, \citenamefont {Li}, \citenamefont {Ma}, \citenamefont {Yang}, \citenamefont {Huang} \emph {et~al.}}]{zhou2022high}%
  \BibitemOpen
  \bibfield  {author} {\bibinfo {author} {\bibfnamefont {C.}~\bibnamefont {Zhou}}, \bibinfo {author} {\bibfnamefont {Y.}~\bibnamefont {Li}}, \bibinfo {author} {\bibfnamefont {L.}~\bibnamefont {Ma}}, \bibinfo {author} {\bibfnamefont {J.}~\bibnamefont {Yang}}, \bibinfo {author} {\bibfnamefont {W.}~\bibnamefont {Huang}},  \emph {et~al.},\ }\bibfield  {title} {\enquote {\bibinfo {title} {High-throughput decoder of quasi-cyclic ldpc codes with limited precision for continuous-variable quantum key distribution systems},}\ }\href@noop {} {\bibfield  {journal} {\bibinfo  {journal} {arXiv preprint arXiv:2207.01860}\ } (\bibinfo {year} {2022})}\BibitemShut {NoStop}%
\bibitem [{\citenamefont {Jiang}\ \emph {et~al.}(2017)\citenamefont {Jiang}, \citenamefont {Huang}, \citenamefont {Huang}, \citenamefont {Lin},\ and\ \citenamefont {Zeng}}]{jiang_PhysRevA_2017}%
  \BibitemOpen
  \bibfield  {author} {\bibinfo {author} {\bibfnamefont {X.-Q.}\ \bibnamefont {Jiang}}, \bibinfo {author} {\bibfnamefont {P.}~\bibnamefont {Huang}}, \bibinfo {author} {\bibfnamefont {D.}~\bibnamefont {Huang}}, \bibinfo {author} {\bibfnamefont {D.}~\bibnamefont {Lin}}, \ and\ \bibinfo {author} {\bibfnamefont {G.}~\bibnamefont {Zeng}},\ }\bibfield  {title} {\enquote {\bibinfo {title} {Secret information reconciliation based on punctured low-density parity-check codes for continuous-variable quantum key distribution},}\ }\href@noop {} {\bibfield  {journal} {\bibinfo  {journal} {Phys. Rev. A}\ }\textbf {\bibinfo {volume} {95}},\ \bibinfo {pages} {022318} (\bibinfo {year} {2017})}\BibitemShut {NoStop}%
\bibitem [{\citenamefont {Zhou}\ \emph {et~al.}(2021)\citenamefont {Zhou}, \citenamefont {Wang}, \citenamefont {Zhang}, \citenamefont {Yu}, \citenamefont {Chen} \emph {et~al.}}]{zhou_SciChinaPhys_2021}%
  \BibitemOpen
  \bibfield  {author} {\bibinfo {author} {\bibfnamefont {C.}~\bibnamefont {Zhou}}, \bibinfo {author} {\bibfnamefont {X.}~\bibnamefont {Wang}}, \bibinfo {author} {\bibfnamefont {Z.}~\bibnamefont {Zhang}}, \bibinfo {author} {\bibfnamefont {S.}~\bibnamefont {Yu}}, \bibinfo {author} {\bibfnamefont {Z.}~\bibnamefont {Chen}},  \emph {et~al.},\ }\bibfield  {title} {\enquote {\bibinfo {title} {Rate compatible reconciliation for continuous-variable quantum key distribution using raptor-like ldpc codes},}\ }\href@noop {} {\bibfield  {journal} {\bibinfo  {journal} {Sci. China: Phys., Mech. Astron.}\ }\textbf {\bibinfo {volume} {64}},\ \bibinfo {pages} {260311} (\bibinfo {year} {2021})}\BibitemShut {NoStop}%
\bibitem [{\citenamefont {Cao}\ \emph {et~al.}(2023)\citenamefont {Cao}, \citenamefont {Chen}, \citenamefont {Chai}, \citenamefont {Liang},\ and\ \citenamefont {Yuan}}]{cao_PhysRevAppl_2023}%
  \BibitemOpen
  \bibfield  {author} {\bibinfo {author} {\bibfnamefont {Z.}~\bibnamefont {Cao}}, \bibinfo {author} {\bibfnamefont {X.}~\bibnamefont {Chen}}, \bibinfo {author} {\bibfnamefont {G.}~\bibnamefont {Chai}}, \bibinfo {author} {\bibfnamefont {K.}~\bibnamefont {Liang}}, \ and\ \bibinfo {author} {\bibfnamefont {Y.}~\bibnamefont {Yuan}},\ }\bibfield  {title} {\enquote {\bibinfo {title} {Rate-adaptive polar-coding-based reconciliation for continuous-variable quantum key distribution at low signal-to-noise ratio},}\ }\href@noop {} {\bibfield  {journal} {\bibinfo  {journal} {Phys. Rev. Appl.}\ }\textbf {\bibinfo {volume} {19}},\ \bibinfo {pages} {044023} (\bibinfo {year} {2023})}\BibitemShut {NoStop}%
\bibitem [{\citenamefont {Yang}\ \emph {et~al.}(2024)\citenamefont {Yang}, \citenamefont {Liu}, \citenamefont {Yang}, \citenamefont {Lu}, \citenamefont {Li},\ and\ \citenamefont {Li}}]{yang2024high}%
  \BibitemOpen
  \bibfield  {author} {\bibinfo {author} {\bibfnamefont {H.}~\bibnamefont {Yang}}, \bibinfo {author} {\bibfnamefont {S.}~\bibnamefont {Liu}}, \bibinfo {author} {\bibfnamefont {S.}~\bibnamefont {Yang}}, \bibinfo {author} {\bibfnamefont {Z.}~\bibnamefont {Lu}}, \bibinfo {author} {\bibfnamefont {Y.}~\bibnamefont {Li}}, \ and\ \bibinfo {author} {\bibfnamefont {Y.}~\bibnamefont {Li}},\ }\bibfield  {title} {\enquote {\bibinfo {title} {High-efficiency rate-adaptive reconciliation in continuous-variable quantum key distribution},}\ }\href@noop {} {\bibfield  {journal} {\bibinfo  {journal} {Physical Review A}\ }\textbf {\bibinfo {volume} {109}},\ \bibinfo {pages} {012604} (\bibinfo {year} {2024})}\BibitemShut {NoStop}%
\bibitem [{\citenamefont {Bennett}\ \emph {et~al.}(1994)\citenamefont {Bennett}, \citenamefont {Brassard}, \citenamefont {Cr{\'e}peau},\ and\ \citenamefont {Maurer}}]{Bennett1994GeneralizedPA}%
  \BibitemOpen
  \bibfield  {author} {\bibinfo {author} {\bibfnamefont {C.~H.}\ \bibnamefont {Bennett}}, \bibinfo {author} {\bibfnamefont {G.}~\bibnamefont {Brassard}}, \bibinfo {author} {\bibfnamefont {C.}~\bibnamefont {Cr{\'e}peau}}, \ and\ \bibinfo {author} {\bibfnamefont {U.}~\bibnamefont {Maurer}},\ }\bibfield  {title} {\enquote {\bibinfo {title} {Generalized privacy amplification},}\ }\href {https://api.semanticscholar.org/CorpusID:10242750} {\bibfield  {journal} {\bibinfo  {journal} {Proceedings of 1994 IEEE International Symposium on Information Theory}\ ,\ \bibinfo {pages} {350--}} (\bibinfo {year} {1994})}\BibitemShut {NoStop}%
\bibitem [{\citenamefont {Tomamichel}\ \emph {et~al.}(2010)\citenamefont {Tomamichel}, \citenamefont {Renner}, \citenamefont {Schaffner},\ and\ \citenamefont {Smith}}]{Tomamichel2010LeftoverHA}%
  \BibitemOpen
  \bibfield  {author} {\bibinfo {author} {\bibfnamefont {M.}~\bibnamefont {Tomamichel}}, \bibinfo {author} {\bibfnamefont {R.}~\bibnamefont {Renner}}, \bibinfo {author} {\bibfnamefont {C.}~\bibnamefont {Schaffner}}, \ and\ \bibinfo {author} {\bibfnamefont {A.~D.}\ \bibnamefont {Smith}},\ }\bibfield  {title} {\enquote {\bibinfo {title} {Leftover hashing against quantum side information},}\ }\href {https://api.semanticscholar.org/CorpusID:1263804} {\bibfield  {journal} {\bibinfo  {journal} {2010 IEEE International Symposium on Information Theory}\ ,\ \bibinfo {pages} {2703--2707}} (\bibinfo {year} {2010})}\BibitemShut {NoStop}%
\bibitem [{\citenamefont {Wang}\ \emph {et~al.}(2018{\natexlab{c}})\citenamefont {Wang}, \citenamefont {Zhang}, \citenamefont {Yu},\ and\ \citenamefont {Guo}}]{wang_IEEEPhotJour_2018}%
  \BibitemOpen
  \bibfield  {author} {\bibinfo {author} {\bibfnamefont {X.}~\bibnamefont {Wang}}, \bibinfo {author} {\bibfnamefont {Y.}~\bibnamefont {Zhang}}, \bibinfo {author} {\bibfnamefont {S.}~\bibnamefont {Yu}}, \ and\ \bibinfo {author} {\bibfnamefont {H.}~\bibnamefont {Guo}},\ }\bibfield  {title} {\enquote {\bibinfo {title} {High-speed implementation of length-compatible privacy amplification in continuous-variable quantum key distribution},}\ }\href@noop {} {\bibfield  {journal} {\bibinfo  {journal} {IEEE Photon. J.}\ }\textbf {\bibinfo {volume} {10}},\ \bibinfo {pages} {1--9} (\bibinfo {year} {2018}{\natexlab{c}})}\BibitemShut {NoStop}%
\bibitem [{\citenamefont {Yan}\ \emph {et~al.}(2022)\citenamefont {Yan}, \citenamefont {Li}, \citenamefont {Mao}, \citenamefont {Xu},\ and\ \citenamefont {Abd El-Latif}}]{yan_JourLightTech_2022}%
  \BibitemOpen
  \bibfield  {author} {\bibinfo {author} {\bibfnamefont {B.-Z.}\ \bibnamefont {Yan}}, \bibinfo {author} {\bibfnamefont {Q.}~\bibnamefont {Li}}, \bibinfo {author} {\bibfnamefont {H.-K.}\ \bibnamefont {Mao}}, \bibinfo {author} {\bibfnamefont {H.-W.}\ \bibnamefont {Xu}}, \ and\ \bibinfo {author} {\bibfnamefont {A.~A.}\ \bibnamefont {Abd El-Latif}},\ }\bibfield  {title} {\enquote {\bibinfo {title} {Large-scale and high-speed fpga-based privacy amplification for quantum key distribution},}\ }\href@noop {} {\bibfield  {journal} {\bibinfo  {journal} {J. Light. Technol.}\ }\textbf {\bibinfo {volume} {41}},\ \bibinfo {pages} {169--175} (\bibinfo {year} {2022})}\BibitemShut {NoStop}%
\bibitem [{\citenamefont {Huang}\ \emph {et~al.}(2015{\natexlab{b}})\citenamefont {Huang}, \citenamefont {Lin}, \citenamefont {Wang}, \citenamefont {Liu}, \citenamefont {Fang} \emph {et~al.}}]{Huang_OptExpress_2015}%
  \BibitemOpen
  \bibfield  {author} {\bibinfo {author} {\bibfnamefont {D.}~\bibnamefont {Huang}}, \bibinfo {author} {\bibfnamefont {D.}~\bibnamefont {Lin}}, \bibinfo {author} {\bibfnamefont {C.}~\bibnamefont {Wang}}, \bibinfo {author} {\bibfnamefont {W.}~\bibnamefont {Liu}}, \bibinfo {author} {\bibfnamefont {S.}~\bibnamefont {Fang}},  \emph {et~al.},\ }\bibfield  {title} {\enquote {\bibinfo {title} {Continuous-variable quantum key distribution with 1 mbps secure key rate},}\ }\href@noop {} {\bibfield  {journal} {\bibinfo  {journal} {Opt. Express}\ }\textbf {\bibinfo {volume} {23}},\ \bibinfo {pages} {17511--17519} (\bibinfo {year} {2015}{\natexlab{b}})}\BibitemShut {NoStop}%
\bibitem [{\citenamefont {Huang}\ \emph {et~al.}(2016{\natexlab{c}})\citenamefont {Huang}, \citenamefont {Huang}, \citenamefont {Wang}, \citenamefont {Li}, \citenamefont {Zhou} \emph {et~al.}}]{Huang_PhysRevA_2016}%
  \BibitemOpen
  \bibfield  {author} {\bibinfo {author} {\bibfnamefont {D.}~\bibnamefont {Huang}}, \bibinfo {author} {\bibfnamefont {P.}~\bibnamefont {Huang}}, \bibinfo {author} {\bibfnamefont {T.}~\bibnamefont {Wang}}, \bibinfo {author} {\bibfnamefont {H.}~\bibnamefont {Li}}, \bibinfo {author} {\bibfnamefont {Y.}~\bibnamefont {Zhou}},  \emph {et~al.},\ }\bibfield  {title} {\enquote {\bibinfo {title} {Continuous-variable quantum key distribution based on a plug-and-play dual-phase-modulated coherent-states protocol},}\ }\href@noop {} {\bibfield  {journal} {\bibinfo  {journal} {Phys. Rev. A}\ }\textbf {\bibinfo {volume} {94}},\ \bibinfo {pages} {032305} (\bibinfo {year} {2016}{\natexlab{c}})}\BibitemShut {NoStop}%
\bibitem [{\citenamefont {Li}\ \emph {et~al.}(2017{\natexlab{b}})\citenamefont {Li}, \citenamefont {Wang}, \citenamefont {Bai}, \citenamefont {Liu}, \citenamefont {Yang} \emph {et~al.}}]{Li_ChinPhysB_2017}%
  \BibitemOpen
  \bibfield  {author} {\bibinfo {author} {\bibfnamefont {Y.-M.}\ \bibnamefont {Li}}, \bibinfo {author} {\bibfnamefont {X.-Y.}\ \bibnamefont {Wang}}, \bibinfo {author} {\bibfnamefont {Z.-L.}\ \bibnamefont {Bai}}, \bibinfo {author} {\bibfnamefont {W.-Y.}\ \bibnamefont {Liu}}, \bibinfo {author} {\bibfnamefont {S.-S.}\ \bibnamefont {Yang}},  \emph {et~al.},\ }\bibfield  {title} {\enquote {\bibinfo {title} {Continuous variable quantum key distribution},}\ }\href@noop {} {\bibfield  {journal} {\bibinfo  {journal} {Chin. Phys. B}\ }\textbf {\bibinfo {volume} {26}},\ \bibinfo {pages} {040303} (\bibinfo {year} {2017}{\natexlab{b}})}\BibitemShut {NoStop}%
\bibitem [{\citenamefont {Wang}\ \emph {et~al.}(2019{\natexlab{d}})\citenamefont {Wang}, \citenamefont {Guo}, \citenamefont {Wang}, \citenamefont {Liu},\ and\ \citenamefont {Li}}]{Wang_OptExpress_2019}%
  \BibitemOpen
  \bibfield  {author} {\bibinfo {author} {\bibfnamefont {X.}~\bibnamefont {Wang}}, \bibinfo {author} {\bibfnamefont {S.}~\bibnamefont {Guo}}, \bibinfo {author} {\bibfnamefont {P.}~\bibnamefont {Wang}}, \bibinfo {author} {\bibfnamefont {W.}~\bibnamefont {Liu}}, \ and\ \bibinfo {author} {\bibfnamefont {Y.}~\bibnamefont {Li}},\ }\bibfield  {title} {\enquote {\bibinfo {title} {Realistic rate--distance limit of continuous-variable quantum key distribution},}\ }\href@noop {} {\bibfield  {journal} {\bibinfo  {journal} {Opt. Express}\ }\textbf {\bibinfo {volume} {27}},\ \bibinfo {pages} {13372--13386} (\bibinfo {year} {2019}{\natexlab{d}})}\BibitemShut {NoStop}%
\bibitem [{\citenamefont {Li}\ \emph {et~al.}(2016{\natexlab{d}})\citenamefont {Li}, \citenamefont {Wang}, \citenamefont {Huang}, \citenamefont {Huang}, \citenamefont {Wang} \emph {et~al.}}]{Li_OptExpress_2015}%
  \BibitemOpen
  \bibfield  {author} {\bibinfo {author} {\bibfnamefont {H.}~\bibnamefont {Li}}, \bibinfo {author} {\bibfnamefont {C.}~\bibnamefont {Wang}}, \bibinfo {author} {\bibfnamefont {P.}~\bibnamefont {Huang}}, \bibinfo {author} {\bibfnamefont {D.}~\bibnamefont {Huang}}, \bibinfo {author} {\bibfnamefont {T.}~\bibnamefont {Wang}},  \emph {et~al.},\ }\bibfield  {title} {\enquote {\bibinfo {title} {Practical continuous-variable quantum key distribution without finite sampling bandwidth effects},}\ }\href@noop {} {\bibfield  {journal} {\bibinfo  {journal} {Opt. Express}\ }\textbf {\bibinfo {volume} {24}},\ \bibinfo {pages} {20481--20493} (\bibinfo {year} {2016}{\natexlab{d}})}\BibitemShut {NoStop}%
\bibitem [{\citenamefont {Wang}\ \emph {et~al.}(2015{\natexlab{d}})\citenamefont {Wang}, \citenamefont {Huang}, \citenamefont {Huang}, \citenamefont {Lin}, \citenamefont {Peng} \emph {et~al.}}]{Wang_SciRep_2015}%
  \BibitemOpen
  \bibfield  {author} {\bibinfo {author} {\bibfnamefont {C.}~\bibnamefont {Wang}}, \bibinfo {author} {\bibfnamefont {D.}~\bibnamefont {Huang}}, \bibinfo {author} {\bibfnamefont {P.}~\bibnamefont {Huang}}, \bibinfo {author} {\bibfnamefont {D.}~\bibnamefont {Lin}}, \bibinfo {author} {\bibfnamefont {J.}~\bibnamefont {Peng}},  \emph {et~al.},\ }\bibfield  {title} {\enquote {\bibinfo {title} {25 mhz clock continuous-variable quantum key distribution system over 50 km fiber channel},}\ }\href@noop {} {\bibfield  {journal} {\bibinfo  {journal} {Sci. Rep.}\ }\textbf {\bibinfo {volume} {5}},\ \bibinfo {pages} {1--8} (\bibinfo {year} {2015}{\natexlab{d}})}\BibitemShut {NoStop}%
\bibitem [{\citenamefont {Shen}\ \emph {et~al.}(2014)\citenamefont {Shen}, \citenamefont {Chen}, \citenamefont {Zou},\ and\ \citenamefont {Yuan}}]{Shen_SciRep_2015}%
  \BibitemOpen
  \bibfield  {author} {\bibinfo {author} {\bibfnamefont {Y.}~\bibnamefont {Shen}}, \bibinfo {author} {\bibfnamefont {Y.}~\bibnamefont {Chen}}, \bibinfo {author} {\bibfnamefont {H.}~\bibnamefont {Zou}}, \ and\ \bibinfo {author} {\bibfnamefont {J.}~\bibnamefont {Yuan}},\ }\bibfield  {title} {\enquote {\bibinfo {title} {A fiber-based quasi-continuous-wave quantum key distribution system},}\ }\href@noop {} {\bibfield  {journal} {\bibinfo  {journal} {Sci. Rep.}\ }\textbf {\bibinfo {volume} {4}},\ \bibinfo {pages} {1--5} (\bibinfo {year} {2014})}\BibitemShut {NoStop}%
\bibitem [{\citenamefont {Li}\ \emph {et~al.}(2019{\natexlab{c}})\citenamefont {Li}, \citenamefont {Huang}, \citenamefont {Wang}, \citenamefont {Wang}, \citenamefont {Chen} \emph {et~al.}}]{Li_OptExpress_2019}%
  \BibitemOpen
  \bibfield  {author} {\bibinfo {author} {\bibfnamefont {D.}~\bibnamefont {Li}}, \bibinfo {author} {\bibfnamefont {P.}~\bibnamefont {Huang}}, \bibinfo {author} {\bibfnamefont {T.}~\bibnamefont {Wang}}, \bibinfo {author} {\bibfnamefont {S.}~\bibnamefont {Wang}}, \bibinfo {author} {\bibfnamefont {R.}~\bibnamefont {Chen}},  \emph {et~al.},\ }\bibfield  {title} {\enquote {\bibinfo {title} {Phase compensation based on step-length control in continuous-variable quantum key distribution},}\ }\href@noop {} {\bibfield  {journal} {\bibinfo  {journal} {Opt. Express}\ }\textbf {\bibinfo {volume} {27}},\ \bibinfo {pages} {20670--20687} (\bibinfo {year} {2019}{\natexlab{c}})}\BibitemShut {NoStop}%
\bibitem [{\citenamefont {Ziebell}\ \emph {et~al.}(2015)\citenamefont {Ziebell}, \citenamefont {Persechino}, \citenamefont {Harris}, \citenamefont {Galland}, \citenamefont {Marris-Morini} \emph {et~al.}}]{ziebell2015towards}%
  \BibitemOpen
  \bibfield  {author} {\bibinfo {author} {\bibfnamefont {M.}~\bibnamefont {Ziebell}}, \bibinfo {author} {\bibfnamefont {M.}~\bibnamefont {Persechino}}, \bibinfo {author} {\bibfnamefont {N.}~\bibnamefont {Harris}}, \bibinfo {author} {\bibfnamefont {C.}~\bibnamefont {Galland}}, \bibinfo {author} {\bibfnamefont {D.}~\bibnamefont {Marris-Morini}},  \emph {et~al.},\ }\bibfield  {title} {\enquote {\bibinfo {title} {Towards on-chip continuous-variable quantum key distribution},}\ }in\ \href@noop {} {\emph {\bibinfo {booktitle} {The European Conference on Lasers and Electro-Optics}}}\ (\bibinfo {organization} {Optica Publishing Group},\ \bibinfo {year} {2015})\ p.\ \bibinfo {pages} {JSV\_4\_2}\BibitemShut {NoStop}%
\bibitem [{\citenamefont {Wang}\ \emph {et~al.}(2013{\natexlab{b}})\citenamefont {Wang}, \citenamefont {Bai}, \citenamefont {Wang}, \citenamefont {Li},\ and\ \citenamefont {Peng}}]{Wang_ChinesePhysLett_2013}%
  \BibitemOpen
  \bibfield  {author} {\bibinfo {author} {\bibfnamefont {X.}~\bibnamefont {Wang}}, \bibinfo {author} {\bibfnamefont {Z.}~\bibnamefont {Bai}}, \bibinfo {author} {\bibfnamefont {S.}~\bibnamefont {Wang}}, \bibinfo {author} {\bibfnamefont {Y.-M.}\ \bibnamefont {Li}}, \ and\ \bibinfo {author} {\bibfnamefont {K.}~\bibnamefont {Peng}},\ }\bibfield  {title} {\enquote {\bibinfo {title} {Four-state modulation continuous variable quantum key distribution over a 30-km fiber and analysis of excess noise},}\ }\href@noop {} {\bibfield  {journal} {\bibinfo  {journal} {Chinese Phys. Lett.}\ }\textbf {\bibinfo {volume} {30}},\ \bibinfo {pages} {010305} (\bibinfo {year} {2013}{\natexlab{b}})}\BibitemShut {NoStop}%
\bibitem [{\citenamefont {Hirano}\ \emph {et~al.}(2017)\citenamefont {Hirano}, \citenamefont {Ichikawa}, \citenamefont {Matsubara}, \citenamefont {Ono}, \citenamefont {Oguri} \emph {et~al.}}]{hirano2017implementation}%
  \BibitemOpen
  \bibfield  {author} {\bibinfo {author} {\bibfnamefont {T.}~\bibnamefont {Hirano}}, \bibinfo {author} {\bibfnamefont {T.}~\bibnamefont {Ichikawa}}, \bibinfo {author} {\bibfnamefont {T.}~\bibnamefont {Matsubara}}, \bibinfo {author} {\bibfnamefont {M.}~\bibnamefont {Ono}}, \bibinfo {author} {\bibfnamefont {Y.}~\bibnamefont {Oguri}},  \emph {et~al.},\ }\bibfield  {title} {\enquote {\bibinfo {title} {Implementation of continuous-variable quantum key distribution with discrete modulation},}\ }\href@noop {} {\bibfield  {journal} {\bibinfo  {journal} {Quantum Sci. Technol.}\ }\textbf {\bibinfo {volume} {2}},\ \bibinfo {pages} {024010} (\bibinfo {year} {2017})}\BibitemShut {NoStop}%
\bibitem [{\citenamefont {Wang}\ \emph {et~al.}(2017{\natexlab{b}})\citenamefont {Wang}, \citenamefont {Liu}, \citenamefont {Wang},\ and\ \citenamefont {Li}}]{Wang_PhysRevA_2017}%
  \BibitemOpen
  \bibfield  {author} {\bibinfo {author} {\bibfnamefont {X.}~\bibnamefont {Wang}}, \bibinfo {author} {\bibfnamefont {W.}~\bibnamefont {Liu}}, \bibinfo {author} {\bibfnamefont {P.}~\bibnamefont {Wang}}, \ and\ \bibinfo {author} {\bibfnamefont {Y.}~\bibnamefont {Li}},\ }\bibfield  {title} {\enquote {\bibinfo {title} {Experimental study on all-fiber-based unidimensional continuous-variable quantum key distribution},}\ }\href@noop {} {\bibfield  {journal} {\bibinfo  {journal} {Phys. Rev. A}\ }\textbf {\bibinfo {volume} {95}},\ \bibinfo {pages} {062330} (\bibinfo {year} {2017}{\natexlab{b}})}\BibitemShut {NoStop}%
\bibitem [{\citenamefont {Zhao}\ \emph {et~al.}(2022)\citenamefont {Zhao}, \citenamefont {Li}, \citenamefont {Xu}, \citenamefont {Huang}, \citenamefont {Wang} \emph {et~al.}}]{zhao2022simple}%
  \BibitemOpen
  \bibfield  {author} {\bibinfo {author} {\bibfnamefont {H.}~\bibnamefont {Zhao}}, \bibinfo {author} {\bibfnamefont {H.}~\bibnamefont {Li}}, \bibinfo {author} {\bibfnamefont {Y.}~\bibnamefont {Xu}}, \bibinfo {author} {\bibfnamefont {P.}~\bibnamefont {Huang}}, \bibinfo {author} {\bibfnamefont {T.}~\bibnamefont {Wang}},  \emph {et~al.},\ }\bibfield  {title} {\enquote {\bibinfo {title} {Simple continuous-variable quantum key distribution scheme using a sagnac-based gaussian modulator},}\ }\href@noop {} {\bibfield  {journal} {\bibinfo  {journal} {Opt. Lett.}\ }\textbf {\bibinfo {volume} {47}},\ \bibinfo {pages} {2939--2942} (\bibinfo {year} {2022})}\BibitemShut {NoStop}%
\bibitem [{\citenamefont {Tian}\ \emph {et~al.}(2023)\citenamefont {Tian}, \citenamefont {Zhang}, \citenamefont {Liu}, \citenamefont {Wang}, \citenamefont {Lu} \emph {et~al.}}]{Tian2023DM}%
  \BibitemOpen
  \bibfield  {author} {\bibinfo {author} {\bibfnamefont {Y.}~\bibnamefont {Tian}}, \bibinfo {author} {\bibfnamefont {Y.}~\bibnamefont {Zhang}}, \bibinfo {author} {\bibfnamefont {S.}~\bibnamefont {Liu}}, \bibinfo {author} {\bibfnamefont {P.}~\bibnamefont {Wang}}, \bibinfo {author} {\bibfnamefont {Z.}~\bibnamefont {Lu}},  \emph {et~al.},\ }\bibfield  {title} {\enquote {\bibinfo {title} {High-performance long-distance discrete-modulation continuous-variable quantum key distribution},}\ }\href {\doibase 10.1364/OL.492082} {\bibfield  {journal} {\bibinfo  {journal} {Opt. Lett.}\ }\textbf {\bibinfo {volume} {48}},\ \bibinfo {pages} {2953--2956} (\bibinfo {year} {2023})}\BibitemShut {NoStop}%
\bibitem [{\citenamefont {Pi}\ \emph {et~al.}(2023)\citenamefont {Pi}, \citenamefont {Wang}, \citenamefont {Pan}, \citenamefont {Shao}, \citenamefont {Li} \emph {et~al.}}]{Pi2023SubMbps}%
  \BibitemOpen
  \bibfield  {author} {\bibinfo {author} {\bibfnamefont {Y.}~\bibnamefont {Pi}}, \bibinfo {author} {\bibfnamefont {H.}~\bibnamefont {Wang}}, \bibinfo {author} {\bibfnamefont {Y.}~\bibnamefont {Pan}}, \bibinfo {author} {\bibfnamefont {Y.}~\bibnamefont {Shao}}, \bibinfo {author} {\bibfnamefont {Y.}~\bibnamefont {Li}},  \emph {et~al.},\ }\bibfield  {title} {\enquote {\bibinfo {title} {Sub-mbps key-rate continuous-variable quantum key distribution with local local oscillator over 100-km fiber},}\ }\href {\doibase 10.1364/OL.485913} {\bibfield  {journal} {\bibinfo  {journal} {Opt. Lett.}\ }\textbf {\bibinfo {volume} {48}},\ \bibinfo {pages} {1766--1769} (\bibinfo {year} {2023})}\BibitemShut {NoStop}%
\bibitem [{\citenamefont {Hajomer}\ \emph {et~al.}(2024)\citenamefont {Hajomer}, \citenamefont {Derkach}, \citenamefont {Jain}, \citenamefont {Chin}, \citenamefont {Andersen} \emph {et~al.}}]{hajomer2024long}%
  \BibitemOpen
  \bibfield  {author} {\bibinfo {author} {\bibfnamefont {A.~A.}\ \bibnamefont {Hajomer}}, \bibinfo {author} {\bibfnamefont {I.}~\bibnamefont {Derkach}}, \bibinfo {author} {\bibfnamefont {N.}~\bibnamefont {Jain}}, \bibinfo {author} {\bibfnamefont {H.-M.}\ \bibnamefont {Chin}}, \bibinfo {author} {\bibfnamefont {U.~L.}\ \bibnamefont {Andersen}},  \emph {et~al.},\ }\bibfield  {title} {\enquote {\bibinfo {title} {Long-distance continuous-variable quantum key distribution over 100-km fiber with local local oscillator},}\ }\href@noop {} {\bibfield  {journal} {\bibinfo  {journal} {Science Advances}\ }\textbf {\bibinfo {volume} {10}},\ \bibinfo {pages} {eadi9474} (\bibinfo {year} {2024})}\BibitemShut {NoStop}%
\bibitem [{\citenamefont {Brunner}\ \emph {et~al.}(2023)\citenamefont {Brunner}, \citenamefont {Fung}, \citenamefont {Peev}, \citenamefont {M{\'e}ndez}, \citenamefont {Ort{\'i}z} \emph {et~al.}}]{Brunner2023DemonstrationOA}%
  \BibitemOpen
  \bibfield  {author} {\bibinfo {author} {\bibfnamefont {H.~H.}\ \bibnamefont {Brunner}}, \bibinfo {author} {\bibfnamefont {C.-H.~F.}\ \bibnamefont {Fung}}, \bibinfo {author} {\bibfnamefont {M.}~\bibnamefont {Peev}}, \bibinfo {author} {\bibfnamefont {R.~B.}\ \bibnamefont {M{\'e}ndez}}, \bibinfo {author} {\bibfnamefont {L.}~\bibnamefont {Ort{\'i}z}},  \emph {et~al.},\ }\bibfield  {title} {\enquote {\bibinfo {title} {Demonstration of a switched cv-qkd network},}\ }\href {https://api.semanticscholar.org/CorpusID:262074294} {\bibfield  {journal} {\bibinfo  {journal} {EPJ Quantum Technol.}\ }\textbf {\bibinfo {volume} {10}} (\bibinfo {year} {2023})}\BibitemShut {NoStop}%
\bibitem [{\citenamefont {Williams}\ \emph {et~al.}(2023)\citenamefont {Williams}, \citenamefont {Qi}, \citenamefont {Alshowkan}, \citenamefont {Evans},\ and\ \citenamefont {Peters}}]{williams2023continuousvariable}%
  \BibitemOpen
  \bibfield  {author} {\bibinfo {author} {\bibfnamefont {B.~P.}\ \bibnamefont {Williams}}, \bibinfo {author} {\bibfnamefont {B.}~\bibnamefont {Qi}}, \bibinfo {author} {\bibfnamefont {M.}~\bibnamefont {Alshowkan}}, \bibinfo {author} {\bibfnamefont {P.~G.}\ \bibnamefont {Evans}}, \ and\ \bibinfo {author} {\bibfnamefont {N.~A.}\ \bibnamefont {Peters}},\ }\href@noop {} {\enquote {\bibinfo {title} {Continuous-variable quantum key distribution field-test with true local oscillator},}\ } (\bibinfo {year} {2023}),\ \Eprint {http://arxiv.org/abs/2309.03959} {arXiv:2309.03959 [quant-ph]} \BibitemShut {NoStop}%
\bibitem [{\citenamefont {Pi{\'e}tri}\ \emph {et~al.}(2023)\citenamefont {Pi{\'e}tri}, \citenamefont {Vidarte}, \citenamefont {Schiavon}, \citenamefont {Grangier}, \citenamefont {Rhouni} \emph {et~al.}}]{pietri2023cv}%
  \BibitemOpen
  \bibfield  {author} {\bibinfo {author} {\bibfnamefont {Y.}~\bibnamefont {Pi{\'e}tri}}, \bibinfo {author} {\bibfnamefont {L.~T.}\ \bibnamefont {Vidarte}}, \bibinfo {author} {\bibfnamefont {M.}~\bibnamefont {Schiavon}}, \bibinfo {author} {\bibfnamefont {P.}~\bibnamefont {Grangier}}, \bibinfo {author} {\bibfnamefont {A.}~\bibnamefont {Rhouni}},  \emph {et~al.},\ }\bibfield  {title} {\enquote {\bibinfo {title} {Cv-qkd receiver platform based on a silicon photonic integrated circuit},}\ }in\ \href@noop {} {\emph {\bibinfo {booktitle} {2023 Optical Fiber Communications Conference and Exhibition (OFC)}}}\ (\bibinfo {organization} {IEEE},\ \bibinfo {year} {2023})\ pp.\ \bibinfo {pages} {1--3}\BibitemShut {NoStop}%
\bibitem [{\citenamefont {Aldama}\ \emph {et~al.}(2023{\natexlab{b}})\citenamefont {Aldama}, \citenamefont {Sarmiento}, \citenamefont {Etcheverry}, \citenamefont {Grande}, \citenamefont {Vidarte} \emph {et~al.}}]{aldama2023inp}%
  \BibitemOpen
  \bibfield  {author} {\bibinfo {author} {\bibfnamefont {J.}~\bibnamefont {Aldama}}, \bibinfo {author} {\bibfnamefont {S.}~\bibnamefont {Sarmiento}}, \bibinfo {author} {\bibfnamefont {S.}~\bibnamefont {Etcheverry}}, \bibinfo {author} {\bibfnamefont {I.~L.}\ \bibnamefont {Grande}}, \bibinfo {author} {\bibfnamefont {L.~T.}\ \bibnamefont {Vidarte}},  \emph {et~al.},\ }\bibfield  {title} {\enquote {\bibinfo {title} {Inp-based cv-qkd pic transmitter},}\ }in\ \href@noop {} {\emph {\bibinfo {booktitle} {Optical Fiber Communication Conference}}}\ (\bibinfo {organization} {Optica Publishing Group},\ \bibinfo {year} {2023})\ pp.\ \bibinfo {pages} {M1I--3}\BibitemShut {NoStop}%
\bibitem [{\citenamefont {Li}\ \emph {et~al.}(2023{\natexlab{b}})\citenamefont {Li}, \citenamefont {Wang}, \citenamefont {Li}, \citenamefont {Huang}, \citenamefont {Guo} \emph {et~al.}}]{li2023continuous}%
  \BibitemOpen
  \bibfield  {author} {\bibinfo {author} {\bibfnamefont {L.}~\bibnamefont {Li}}, \bibinfo {author} {\bibfnamefont {T.}~\bibnamefont {Wang}}, \bibinfo {author} {\bibfnamefont {X.}~\bibnamefont {Li}}, \bibinfo {author} {\bibfnamefont {P.}~\bibnamefont {Huang}}, \bibinfo {author} {\bibfnamefont {Y.}~\bibnamefont {Guo}},  \emph {et~al.},\ }\bibfield  {title} {\enquote {\bibinfo {title} {Continuous-variable quantum key distribution with on-chip light sources},}\ }\href@noop {} {\bibfield  {journal} {\bibinfo  {journal} {Photonics Res.}\ }\textbf {\bibinfo {volume} {11}},\ \bibinfo {pages} {504--516} (\bibinfo {year} {2023}{\natexlab{b}})}\BibitemShut {NoStop}%
\bibitem [{\citenamefont {Ma}\ \emph {et~al.}(2013{\natexlab{a}})\citenamefont {Ma}, \citenamefont {Sun}, \citenamefont {Jiang},\ and\ \citenamefont {Liang}}]{ma2013local}%
  \BibitemOpen
  \bibfield  {author} {\bibinfo {author} {\bibfnamefont {X.-C.}\ \bibnamefont {Ma}}, \bibinfo {author} {\bibfnamefont {S.-H.}\ \bibnamefont {Sun}}, \bibinfo {author} {\bibfnamefont {M.-S.}\ \bibnamefont {Jiang}}, \ and\ \bibinfo {author} {\bibfnamefont {L.-M.}\ \bibnamefont {Liang}},\ }\bibfield  {title} {\enquote {\bibinfo {title} {Local oscillator fluctuation opens a loophole for eve in practical continuous-variable quantum-key-distribution systems},}\ }\href@noop {} {\bibfield  {journal} {\bibinfo  {journal} {Phys. Rev. A}\ }\textbf {\bibinfo {volume} {88}},\ \bibinfo {pages} {022339} (\bibinfo {year} {2013}{\natexlab{a}})}\BibitemShut {NoStop}%
\bibitem [{\citenamefont {Ma}\ \emph {et~al.}(2013{\natexlab{b}})\citenamefont {Ma}, \citenamefont {Sun}, \citenamefont {Jiang},\ and\ \citenamefont {Liang}}]{ma2013wavelength}%
  \BibitemOpen
  \bibfield  {author} {\bibinfo {author} {\bibfnamefont {X.-C.}\ \bibnamefont {Ma}}, \bibinfo {author} {\bibfnamefont {S.-H.}\ \bibnamefont {Sun}}, \bibinfo {author} {\bibfnamefont {M.-S.}\ \bibnamefont {Jiang}}, \ and\ \bibinfo {author} {\bibfnamefont {L.-M.}\ \bibnamefont {Liang}},\ }\bibfield  {title} {\enquote {\bibinfo {title} {Wavelength attack on practical continuous-variable quantum-key-distribution system with a heterodyne protocol},}\ }\href@noop {} {\bibfield  {journal} {\bibinfo  {journal} {Phys. Rev. A}\ }\textbf {\bibinfo {volume} {87}},\ \bibinfo {pages} {052309} (\bibinfo {year} {2013}{\natexlab{b}})}\BibitemShut {NoStop}%
\bibitem [{\citenamefont {Huang}\ \emph {et~al.}(2013)\citenamefont {Huang}, \citenamefont {Weedbrook}, \citenamefont {Yin}, \citenamefont {Wang}, \citenamefont {Li} \emph {et~al.}}]{huang2013quantum}%
  \BibitemOpen
  \bibfield  {author} {\bibinfo {author} {\bibfnamefont {J.-Z.}\ \bibnamefont {Huang}}, \bibinfo {author} {\bibfnamefont {C.}~\bibnamefont {Weedbrook}}, \bibinfo {author} {\bibfnamefont {Z.-Q.}\ \bibnamefont {Yin}}, \bibinfo {author} {\bibfnamefont {S.}~\bibnamefont {Wang}}, \bibinfo {author} {\bibfnamefont {H.-W.}\ \bibnamefont {Li}},  \emph {et~al.},\ }\bibfield  {title} {\enquote {\bibinfo {title} {Quantum hacking of a continuous-variable quantum-key-distribution system using a wavelength attack},}\ }\href@noop {} {\bibfield  {journal} {\bibinfo  {journal} {Phys. Rev. A}\ }\textbf {\bibinfo {volume} {87}},\ \bibinfo {pages} {062329} (\bibinfo {year} {2013})}\BibitemShut {NoStop}%
\bibitem [{\citenamefont {Jouguet}, \citenamefont {Kunz-Jacques},\ and\ \citenamefont {Diamanti}(2013)}]{jouguet2013preventing}%
  \BibitemOpen
  \bibfield  {author} {\bibinfo {author} {\bibfnamefont {P.}~\bibnamefont {Jouguet}}, \bibinfo {author} {\bibfnamefont {S.}~\bibnamefont {Kunz-Jacques}}, \ and\ \bibinfo {author} {\bibfnamefont {E.}~\bibnamefont {Diamanti}},\ }\bibfield  {title} {\enquote {\bibinfo {title} {Preventing calibration attacks on the local oscillator in continuous-variable quantum key distribution},}\ }\href@noop {} {\bibfield  {journal} {\bibinfo  {journal} {Phys. Rev. A}\ }\textbf {\bibinfo {volume} {87}},\ \bibinfo {pages} {062313} (\bibinfo {year} {2013})}\BibitemShut {NoStop}%
\bibitem [{\citenamefont {Grande}\ \emph {et~al.}(2021)\citenamefont {Grande}, \citenamefont {Etcheverry}, \citenamefont {Aldama}, \citenamefont {Ghasemi}, \citenamefont {Nolan} \emph {et~al.}}]{grande2021adaptable}%
  \BibitemOpen
  \bibfield  {author} {\bibinfo {author} {\bibfnamefont {I.~L.}\ \bibnamefont {Grande}}, \bibinfo {author} {\bibfnamefont {S.}~\bibnamefont {Etcheverry}}, \bibinfo {author} {\bibfnamefont {J.}~\bibnamefont {Aldama}}, \bibinfo {author} {\bibfnamefont {S.}~\bibnamefont {Ghasemi}}, \bibinfo {author} {\bibfnamefont {D.}~\bibnamefont {Nolan}},  \emph {et~al.},\ }\bibfield  {title} {\enquote {\bibinfo {title} {Adaptable transmitter for discrete and continuous variable quantum key distribution},}\ }\href@noop {} {\bibfield  {journal} {\bibinfo  {journal} {Opt. Express}\ }\textbf {\bibinfo {volume} {29}},\ \bibinfo {pages} {14815--14827} (\bibinfo {year} {2021})}\BibitemShut {NoStop}%
\bibitem [{\citenamefont {Sarmiento}\ \emph {et~al.}(2022)\citenamefont {Sarmiento}, \citenamefont {Etcheverry}, \citenamefont {Aldama}, \citenamefont {Lopez}, \citenamefont {Vidarte} \emph {et~al.}}]{sarmiento2022continuous}%
  \BibitemOpen
  \bibfield  {author} {\bibinfo {author} {\bibfnamefont {S.}~\bibnamefont {Sarmiento}}, \bibinfo {author} {\bibfnamefont {S.}~\bibnamefont {Etcheverry}}, \bibinfo {author} {\bibfnamefont {J.}~\bibnamefont {Aldama}}, \bibinfo {author} {\bibfnamefont {I.}~\bibnamefont {Lopez}}, \bibinfo {author} {\bibfnamefont {L.}~\bibnamefont {Vidarte}},  \emph {et~al.},\ }\bibfield  {title} {\enquote {\bibinfo {title} {Continuous-variable quantum key distribution over a 15 km multi-core fiber},}\ }\href@noop {} {\bibfield  {journal} {\bibinfo  {journal} {New J. Phys.}\ }\textbf {\bibinfo {volume} {24}},\ \bibinfo {pages} {063011} (\bibinfo {year} {2022})}\BibitemShut {NoStop}%
\bibitem [{\citenamefont {Aldama}\ \emph {et~al.}(2022)\citenamefont {Aldama}, \citenamefont {Sarmiento}, \citenamefont {Grande}, \citenamefont {Signorini}, \citenamefont {Vidarte} \emph {et~al.}}]{aldama2022integrated}%
  \BibitemOpen
  \bibfield  {author} {\bibinfo {author} {\bibfnamefont {J.}~\bibnamefont {Aldama}}, \bibinfo {author} {\bibfnamefont {S.}~\bibnamefont {Sarmiento}}, \bibinfo {author} {\bibfnamefont {I.~H.~L.}\ \bibnamefont {Grande}}, \bibinfo {author} {\bibfnamefont {S.}~\bibnamefont {Signorini}}, \bibinfo {author} {\bibfnamefont {L.~T.}\ \bibnamefont {Vidarte}},  \emph {et~al.},\ }\bibfield  {title} {\enquote {\bibinfo {title} {Integrated qkd and qrng photonic technologies},}\ }\href@noop {} {\bibfield  {journal} {\bibinfo  {journal} {J. Lightwave Technol.}\ }\textbf {\bibinfo {volume} {40}},\ \bibinfo {pages} {7498--7517} (\bibinfo {year} {2022})}\BibitemShut {NoStop}%
\bibitem [{\citenamefont {Wang}\ \emph {et~al.}(2018{\natexlab{d}})\citenamefont {Wang}, \citenamefont {Huang}, \citenamefont {Zhou}, \citenamefont {Liu},\ and\ \citenamefont {Zeng}}]{Wang_PhysRevA_2018}%
  \BibitemOpen
  \bibfield  {author} {\bibinfo {author} {\bibfnamefont {T.}~\bibnamefont {Wang}}, \bibinfo {author} {\bibfnamefont {P.}~\bibnamefont {Huang}}, \bibinfo {author} {\bibfnamefont {Y.}~\bibnamefont {Zhou}}, \bibinfo {author} {\bibfnamefont {W.}~\bibnamefont {Liu}}, \ and\ \bibinfo {author} {\bibfnamefont {G.}~\bibnamefont {Zeng}},\ }\bibfield  {title} {\enquote {\bibinfo {title} {Pilot-multiplexed continuous-variable quantum key distribution with a real local oscillator},}\ }\href@noop {} {\bibfield  {journal} {\bibinfo  {journal} {Phys. Rev. A}\ }\textbf {\bibinfo {volume} {97}},\ \bibinfo {pages} {012310} (\bibinfo {year} {2018}{\natexlab{d}})}\BibitemShut {NoStop}%
\bibitem [{\citenamefont {Laudenbach}\ \emph {et~al.}(2019)\citenamefont {Laudenbach}, \citenamefont {Schrenk}, \citenamefont {Pacher}, \citenamefont {Hentschel}, \citenamefont {Fung} \emph {et~al.}}]{laudenbach2019pilot}%
  \BibitemOpen
  \bibfield  {author} {\bibinfo {author} {\bibfnamefont {F.}~\bibnamefont {Laudenbach}}, \bibinfo {author} {\bibfnamefont {B.}~\bibnamefont {Schrenk}}, \bibinfo {author} {\bibfnamefont {C.}~\bibnamefont {Pacher}}, \bibinfo {author} {\bibfnamefont {M.}~\bibnamefont {Hentschel}}, \bibinfo {author} {\bibfnamefont {C.-H.~F.}\ \bibnamefont {Fung}},  \emph {et~al.},\ }\bibfield  {title} {\enquote {\bibinfo {title} {Pilot-assisted intradyne reception for high-speed continuous-variable quantum key distribution with true local oscillator},}\ }\href@noop {} {\bibfield  {journal} {\bibinfo  {journal} {Quantum}\ }\textbf {\bibinfo {volume} {3}},\ \bibinfo {pages} {193} (\bibinfo {year} {2019})}\BibitemShut {NoStop}%
\bibitem [{\citenamefont {Karinou}\ \emph {et~al.}(2018)\citenamefont {Karinou}, \citenamefont {Brunner}, \citenamefont {Fung}, \citenamefont {Comandar}, \citenamefont {Bettelli} \emph {et~al.}}]{Karinou_IEEE_2018}%
  \BibitemOpen
  \bibfield  {author} {\bibinfo {author} {\bibfnamefont {F.}~\bibnamefont {Karinou}}, \bibinfo {author} {\bibfnamefont {H.~H.}\ \bibnamefont {Brunner}}, \bibinfo {author} {\bibfnamefont {C.-H.~F.}\ \bibnamefont {Fung}}, \bibinfo {author} {\bibfnamefont {L.~C.}\ \bibnamefont {Comandar}}, \bibinfo {author} {\bibfnamefont {S.}~\bibnamefont {Bettelli}},  \emph {et~al.},\ }\bibfield  {title} {\enquote {\bibinfo {title} {Toward the integration of cv quantum key distribution in deployed optical networks},}\ }\href {\doibase 10.1109/LPT.2018.2810334} {\bibfield  {journal} {\bibinfo  {journal} {IEEE Photon. Technol. Lett.}\ }\textbf {\bibinfo {volume} {30}},\ \bibinfo {pages} {650--653} (\bibinfo {year} {2018})}\BibitemShut {NoStop}%
\bibitem [{\citenamefont {G{\"u}nthner}\ \emph {et~al.}(2017)\citenamefont {G{\"u}nthner}, \citenamefont {Khan}, \citenamefont {Elser}, \citenamefont {Stiller}, \citenamefont {Bayraktar} \emph {et~al.}}]{gunthner2017quantum}%
  \BibitemOpen
  \bibfield  {author} {\bibinfo {author} {\bibfnamefont {K.}~\bibnamefont {G{\"u}nthner}}, \bibinfo {author} {\bibfnamefont {I.}~\bibnamefont {Khan}}, \bibinfo {author} {\bibfnamefont {D.}~\bibnamefont {Elser}}, \bibinfo {author} {\bibfnamefont {B.}~\bibnamefont {Stiller}}, \bibinfo {author} {\bibfnamefont {{\"O}.}~\bibnamefont {Bayraktar}},  \emph {et~al.},\ }\bibfield  {title} {\enquote {\bibinfo {title} {Quantum-limited measurements of optical signals from a geostationary satellite},}\ }\href@noop {} {\bibfield  {journal} {\bibinfo  {journal} {Optica}\ }\textbf {\bibinfo {volume} {4}},\ \bibinfo {pages} {611--616} (\bibinfo {year} {2017})}\BibitemShut {NoStop}%
\bibitem [{\citenamefont {Wang}\ \emph {et~al.}(2020{\natexlab{b}})\citenamefont {Wang}, \citenamefont {Huang}, \citenamefont {Liu}, \citenamefont {Wang}, \citenamefont {Wang} \emph {et~al.}}]{wang2020phase}%
  \BibitemOpen
  \bibfield  {author} {\bibinfo {author} {\bibfnamefont {S.}~\bibnamefont {Wang}}, \bibinfo {author} {\bibfnamefont {P.}~\bibnamefont {Huang}}, \bibinfo {author} {\bibfnamefont {M.}~\bibnamefont {Liu}}, \bibinfo {author} {\bibfnamefont {T.}~\bibnamefont {Wang}}, \bibinfo {author} {\bibfnamefont {P.}~\bibnamefont {Wang}},  \emph {et~al.},\ }\bibfield  {title} {\enquote {\bibinfo {title} {Phase compensation for free-space continuous-variable quantum key distribution},}\ }\href@noop {} {\bibfield  {journal} {\bibinfo  {journal} {Opt. Express}\ }\textbf {\bibinfo {volume} {28}},\ \bibinfo {pages} {10737--10745} (\bibinfo {year} {2020}{\natexlab{b}})}\BibitemShut {NoStop}%
\bibitem [{\citenamefont {Zhang}\ \emph {et~al.}(2023)\citenamefont {Zhang}, \citenamefont {Huang}, \citenamefont {Wang}, \citenamefont {Wei},\ and\ \citenamefont {Zeng}}]{zhang2023experimental}%
  \BibitemOpen
  \bibfield  {author} {\bibinfo {author} {\bibfnamefont {M.}~\bibnamefont {Zhang}}, \bibinfo {author} {\bibfnamefont {P.}~\bibnamefont {Huang}}, \bibinfo {author} {\bibfnamefont {P.}~\bibnamefont {Wang}}, \bibinfo {author} {\bibfnamefont {S.}~\bibnamefont {Wei}}, \ and\ \bibinfo {author} {\bibfnamefont {G.}~\bibnamefont {Zeng}},\ }\bibfield  {title} {\enquote {\bibinfo {title} {Experimental free-space continuous-variable quantum key distribution with thermal source},}\ }\href@noop {} {\bibfield  {journal} {\bibinfo  {journal} {Opt. Lett.}\ }\textbf {\bibinfo {volume} {48}},\ \bibinfo {pages} {1184--1187} (\bibinfo {year} {2023})}\BibitemShut {NoStop}%
\bibitem [{\citenamefont {Qi}\ \emph {et~al.}(2010{\natexlab{b}})\citenamefont {Qi}, \citenamefont {Zhu}, \citenamefont {Qian},\ and\ \citenamefont {Lo}}]{Qi_NewJPhys_2010}%
  \BibitemOpen
  \bibfield  {author} {\bibinfo {author} {\bibfnamefont {B.}~\bibnamefont {Qi}}, \bibinfo {author} {\bibfnamefont {W.}~\bibnamefont {Zhu}}, \bibinfo {author} {\bibfnamefont {L.}~\bibnamefont {Qian}}, \ and\ \bibinfo {author} {\bibfnamefont {H.-K.}\ \bibnamefont {Lo}},\ }\bibfield  {title} {\enquote {\bibinfo {title} {Feasibility of quantum key distribution through a dense wavelength division multiplexing network},}\ }\href@noop {} {\bibfield  {journal} {\bibinfo  {journal} {New J. Phys.}\ }\textbf {\bibinfo {volume} {12}},\ \bibinfo {pages} {103042} (\bibinfo {year} {2010}{\natexlab{b}})}\BibitemShut {NoStop}%
\bibitem [{\citenamefont {Li}\ \emph {et~al.}(2014{\natexlab{c}})\citenamefont {Li}, \citenamefont {Wang}, \citenamefont {Wang},\ and\ \citenamefont {Bai}}]{li2014influence}%
  \BibitemOpen
  \bibfield  {author} {\bibinfo {author} {\bibfnamefont {Y.}~\bibnamefont {Li}}, \bibinfo {author} {\bibfnamefont {N.}~\bibnamefont {Wang}}, \bibinfo {author} {\bibfnamefont {X.}~\bibnamefont {Wang}}, \ and\ \bibinfo {author} {\bibfnamefont {Z.}~\bibnamefont {Bai}},\ }\bibfield  {title} {\enquote {\bibinfo {title} {Influence of guided acoustic wave brillouin scattering on excess noise in fiber-based continuous variable quantum key distribution},}\ }\href@noop {} {\bibfield  {journal} {\bibinfo  {journal} {JOSA B}\ }\textbf {\bibinfo {volume} {31}},\ \bibinfo {pages} {2379--2383} (\bibinfo {year} {2014}{\natexlab{c}})}\BibitemShut {NoStop}%
\bibitem [{\citenamefont {Eriksson}\ \emph {et~al.}(2020)\citenamefont {Eriksson}, \citenamefont {Lu{\'\i}s}, \citenamefont {Puttnam}, \citenamefont {Rademacher}, \citenamefont {Fujiwara} \emph {et~al.}}]{Eriksson_JLightwaveTechnol_2020}%
  \BibitemOpen
  \bibfield  {author} {\bibinfo {author} {\bibfnamefont {T.~A.}\ \bibnamefont {Eriksson}}, \bibinfo {author} {\bibfnamefont {R.~S.}\ \bibnamefont {Lu{\'\i}s}}, \bibinfo {author} {\bibfnamefont {B.~J.}\ \bibnamefont {Puttnam}}, \bibinfo {author} {\bibfnamefont {G.}~\bibnamefont {Rademacher}}, \bibinfo {author} {\bibfnamefont {M.}~\bibnamefont {Fujiwara}},  \emph {et~al.},\ }\bibfield  {title} {\enquote {\bibinfo {title} {Wavelength division multiplexing of 194 continuous variable quantum key distribution channels},}\ }\href@noop {} {\bibfield  {journal} {\bibinfo  {journal} {J. Light. Technol.}\ }\textbf {\bibinfo {volume} {38}},\ \bibinfo {pages} {2214--2218} (\bibinfo {year} {2020})}\BibitemShut {NoStop}%
\bibitem [{\citenamefont {Du}, \citenamefont {Tian},\ and\ \citenamefont {Li}(2020)}]{Du_PhysRevApplied_2020}%
  \BibitemOpen
  \bibfield  {author} {\bibinfo {author} {\bibfnamefont {S.}~\bibnamefont {Du}}, \bibinfo {author} {\bibfnamefont {Y.}~\bibnamefont {Tian}}, \ and\ \bibinfo {author} {\bibfnamefont {Y.}~\bibnamefont {Li}},\ }\bibfield  {title} {\enquote {\bibinfo {title} {Impact of four-wave-mixing noise from dense wavelength-division-multiplexing systems on entangled-state continuous-variable quantum key distribution},}\ }\href@noop {} {\bibfield  {journal} {\bibinfo  {journal} {Phys. Rev. Appl.}\ }\textbf {\bibinfo {volume} {14}},\ \bibinfo {pages} {024013} (\bibinfo {year} {2020})}\BibitemShut {NoStop}%
\bibitem [{\citenamefont {Milovančev}\ \emph {et~al.}(2021)\citenamefont {Milovančev}, \citenamefont {Vokić}, \citenamefont {Laudenbach}, \citenamefont {Pacher}, \citenamefont {Hübel} \emph {et~al.}}]{Milovančev_2021_High}%
  \BibitemOpen
  \bibfield  {author} {\bibinfo {author} {\bibfnamefont {D.}~\bibnamefont {Milovančev}}, \bibinfo {author} {\bibfnamefont {N.}~\bibnamefont {Vokić}}, \bibinfo {author} {\bibfnamefont {F.}~\bibnamefont {Laudenbach}}, \bibinfo {author} {\bibfnamefont {C.}~\bibnamefont {Pacher}}, \bibinfo {author} {\bibfnamefont {H.}~\bibnamefont {Hübel}},  \emph {et~al.},\ }\bibfield  {title} {\enquote {\bibinfo {title} {High rate cv-qkd secured mobile wdm fronthaul for dense 5g radio networks},}\ }\href {\doibase 10.1109/JLT.2021.3068963} {\bibfield  {journal} {\bibinfo  {journal} {J. Light. Technol.}\ }\textbf {\bibinfo {volume} {39}},\ \bibinfo {pages} {3445--3457} (\bibinfo {year} {2021})}\BibitemShut {NoStop}%
\bibitem [{\citenamefont {Kleis}\ \emph {et~al.}(2019)\citenamefont {Kleis}, \citenamefont {Steinmayer}, \citenamefont {Derksen},\ and\ \citenamefont {Schaeffer}}]{Kleis_OE_2019}%
  \BibitemOpen
  \bibfield  {author} {\bibinfo {author} {\bibfnamefont {S.}~\bibnamefont {Kleis}}, \bibinfo {author} {\bibfnamefont {J.}~\bibnamefont {Steinmayer}}, \bibinfo {author} {\bibfnamefont {R.~H.}\ \bibnamefont {Derksen}}, \ and\ \bibinfo {author} {\bibfnamefont {C.~G.}\ \bibnamefont {Schaeffer}},\ }\bibfield  {title} {\enquote {\bibinfo {title} {Experimental investigation of heterodyne quantum key distribution in the s-band or l-band embedded in a commercial c-band dwdm system},}\ }\href {\doibase 10.1364/OE.27.016540} {\bibfield  {journal} {\bibinfo  {journal} {Opt. Express}\ }\textbf {\bibinfo {volume} {27}},\ \bibinfo {pages} {16540--16549} (\bibinfo {year} {2019})}\BibitemShut {NoStop}%
\bibitem [{\citenamefont {Chu}\ \emph {et~al.}(2020)\citenamefont {Chu}, \citenamefont {Zhang}, \citenamefont {Zhao}, \citenamefont {Xu}, \citenamefont {Chen} \emph {et~al.}}]{Chu_2020}%
  \BibitemOpen
  \bibfield  {author} {\bibinfo {author} {\bibfnamefont {B.}~\bibnamefont {Chu}}, \bibinfo {author} {\bibfnamefont {Y.}~\bibnamefont {Zhang}}, \bibinfo {author} {\bibfnamefont {Y.}~\bibnamefont {Zhao}}, \bibinfo {author} {\bibfnamefont {Y.}~\bibnamefont {Xu}}, \bibinfo {author} {\bibfnamefont {X.}~\bibnamefont {Chen}},  \emph {et~al.},\ }\bibfield  {title} {\enquote {\bibinfo {title} {Crosstalk-induced impact of coexisting dwdm network on continuous-variable qkd},}\ }in\ \href {\doibase 10.1109/DRCN48652.2020.1570604589} {\emph {\bibinfo {booktitle} {2020 16th International Conference on the Design of Reliable Communication Networks DRCN 2020}}}\ (\bibinfo {year} {2020})\ pp.\ \bibinfo {pages} {1--5}\BibitemShut {NoStop}%
\bibitem [{\citenamefont {Wang}\ \emph {et~al.}(2018{\natexlab{e}})\citenamefont {Wang}, \citenamefont {Huang}, \citenamefont {Wang},\ and\ \citenamefont {Zeng}}]{wang2018atmospheric}%
  \BibitemOpen
  \bibfield  {author} {\bibinfo {author} {\bibfnamefont {S.}~\bibnamefont {Wang}}, \bibinfo {author} {\bibfnamefont {P.}~\bibnamefont {Huang}}, \bibinfo {author} {\bibfnamefont {T.}~\bibnamefont {Wang}}, \ and\ \bibinfo {author} {\bibfnamefont {G.}~\bibnamefont {Zeng}},\ }\bibfield  {title} {\enquote {\bibinfo {title} {Atmospheric effects on continuous-variable quantum key distribution},}\ }\href@noop {} {\bibfield  {journal} {\bibinfo  {journal} {New J. Phys.}\ }\textbf {\bibinfo {volume} {20}},\ \bibinfo {pages} {083037} (\bibinfo {year} {2018}{\natexlab{e}})}\BibitemShut {NoStop}%
\bibitem [{\citenamefont {Papanastasiou}, \citenamefont {Weedbrook},\ and\ \citenamefont {Pirandola}(2018)}]{papanastasiou2018continuous}%
  \BibitemOpen
  \bibfield  {author} {\bibinfo {author} {\bibfnamefont {P.}~\bibnamefont {Papanastasiou}}, \bibinfo {author} {\bibfnamefont {C.}~\bibnamefont {Weedbrook}}, \ and\ \bibinfo {author} {\bibfnamefont {S.}~\bibnamefont {Pirandola}},\ }\bibfield  {title} {\enquote {\bibinfo {title} {Continuous-variable quantum key distribution in uniform fast-fading channels},}\ }\href@noop {} {\bibfield  {journal} {\bibinfo  {journal} {Phys. Rev. A}\ }\textbf {\bibinfo {volume} {97}},\ \bibinfo {pages} {032311} (\bibinfo {year} {2018})}\BibitemShut {NoStop}%
\bibitem [{\citenamefont {Ghalaii}\ and\ \citenamefont {Pirandola}(2023)}]{ghalaii2023continuous}%
  \BibitemOpen
  \bibfield  {author} {\bibinfo {author} {\bibfnamefont {M.}~\bibnamefont {Ghalaii}}\ and\ \bibinfo {author} {\bibfnamefont {S.}~\bibnamefont {Pirandola}},\ }\bibfield  {title} {\enquote {\bibinfo {title} {Continuous-variable measurement-device-independent quantum key distribution in free-space channels},}\ }\href@noop {} {\bibfield  {journal} {\bibinfo  {journal} {Phys. Rev. A}\ }\textbf {\bibinfo {volume} {108}},\ \bibinfo {pages} {042621} (\bibinfo {year} {2023})}\BibitemShut {NoStop}%
\bibitem [{\citenamefont {Heim}\ \emph {et~al.}(2014)\citenamefont {Heim}, \citenamefont {Peuntinger}, \citenamefont {Killoran}, \citenamefont {Khan}, \citenamefont {Wittmann} \emph {et~al.}}]{heim2014atmospheric}%
  \BibitemOpen
  \bibfield  {author} {\bibinfo {author} {\bibfnamefont {B.}~\bibnamefont {Heim}}, \bibinfo {author} {\bibfnamefont {C.}~\bibnamefont {Peuntinger}}, \bibinfo {author} {\bibfnamefont {N.}~\bibnamefont {Killoran}}, \bibinfo {author} {\bibfnamefont {I.}~\bibnamefont {Khan}}, \bibinfo {author} {\bibfnamefont {C.}~\bibnamefont {Wittmann}},  \emph {et~al.},\ }\bibfield  {title} {\enquote {\bibinfo {title} {Atmospheric continuous-variable quantum communication},}\ }\href@noop {} {\bibfield  {journal} {\bibinfo  {journal} {New J. Phys.}\ }\textbf {\bibinfo {volume} {16}},\ \bibinfo {pages} {113018} (\bibinfo {year} {2014})}\BibitemShut {NoStop}%
\bibitem [{\citenamefont {Dequal}\ \emph {et~al.}(2021)\citenamefont {Dequal}, \citenamefont {Trigo~Vidarte}, \citenamefont {Roman~Rodriguez}, \citenamefont {Vallone}, \citenamefont {Villoresi} \emph {et~al.}}]{dequal2021feasibility}%
  \BibitemOpen
  \bibfield  {author} {\bibinfo {author} {\bibfnamefont {D.}~\bibnamefont {Dequal}}, \bibinfo {author} {\bibfnamefont {L.}~\bibnamefont {Trigo~Vidarte}}, \bibinfo {author} {\bibfnamefont {V.}~\bibnamefont {Roman~Rodriguez}}, \bibinfo {author} {\bibfnamefont {G.}~\bibnamefont {Vallone}}, \bibinfo {author} {\bibfnamefont {P.}~\bibnamefont {Villoresi}},  \emph {et~al.},\ }\bibfield  {title} {\enquote {\bibinfo {title} {Feasibility of satellite-to-ground continuous-variable quantum key distribution},}\ }\href@noop {} {\bibfield  {journal} {\bibinfo  {journal} {npj Quantum Inf.}\ }\textbf {\bibinfo {volume} {7}},\ \bibinfo {pages} {3} (\bibinfo {year} {2021})}\BibitemShut {NoStop}%
\bibitem [{\citenamefont {Wang}\ \emph {et~al.}(2019{\natexlab{e}})\citenamefont {Wang}, \citenamefont {Huang}, \citenamefont {Wang},\ and\ \citenamefont {Zeng}}]{wang2019feasibility}%
  \BibitemOpen
  \bibfield  {author} {\bibinfo {author} {\bibfnamefont {S.}~\bibnamefont {Wang}}, \bibinfo {author} {\bibfnamefont {P.}~\bibnamefont {Huang}}, \bibinfo {author} {\bibfnamefont {T.}~\bibnamefont {Wang}}, \ and\ \bibinfo {author} {\bibfnamefont {G.}~\bibnamefont {Zeng}},\ }\bibfield  {title} {\enquote {\bibinfo {title} {Feasibility of all-day quantum communication with coherent detection},}\ }\href@noop {} {\bibfield  {journal} {\bibinfo  {journal} {Phys. Rev. Appl.}\ }\textbf {\bibinfo {volume} {12}},\ \bibinfo {pages} {024041} (\bibinfo {year} {2019}{\natexlab{e}})}\BibitemShut {NoStop}%
\bibitem [{\citenamefont {Wang}\ \emph {et~al.}(2020{\natexlab{c}})\citenamefont {Wang}, \citenamefont {Huang}, \citenamefont {Wang},\ and\ \citenamefont {Zeng}}]{wang2020dynamic}%
  \BibitemOpen
  \bibfield  {author} {\bibinfo {author} {\bibfnamefont {S.}~\bibnamefont {Wang}}, \bibinfo {author} {\bibfnamefont {P.}~\bibnamefont {Huang}}, \bibinfo {author} {\bibfnamefont {T.}~\bibnamefont {Wang}}, \ and\ \bibinfo {author} {\bibfnamefont {G.}~\bibnamefont {Zeng}},\ }\bibfield  {title} {\enquote {\bibinfo {title} {Dynamic polarization control for free-space continuous-variable quantum key distribution},}\ }\href@noop {} {\bibfield  {journal} {\bibinfo  {journal} {Opt. Lett.}\ }\textbf {\bibinfo {volume} {45}},\ \bibinfo {pages} {5921--5924} (\bibinfo {year} {2020}{\natexlab{c}})}\BibitemShut {NoStop}%
\bibitem [{\citenamefont {Wang}\ \emph {et~al.}(2021{\natexlab{a}})\citenamefont {Wang}, \citenamefont {Huang}, \citenamefont {Chen},\ and\ \citenamefont {Zeng}}]{wang2021robust}%
  \BibitemOpen
  \bibfield  {author} {\bibinfo {author} {\bibfnamefont {P.}~\bibnamefont {Wang}}, \bibinfo {author} {\bibfnamefont {P.}~\bibnamefont {Huang}}, \bibinfo {author} {\bibfnamefont {R.}~\bibnamefont {Chen}}, \ and\ \bibinfo {author} {\bibfnamefont {G.}~\bibnamefont {Zeng}},\ }\bibfield  {title} {\enquote {\bibinfo {title} {Robust frame synchronization for free-space continuous-variable quantum key distribution},}\ }\href@noop {} {\bibfield  {journal} {\bibinfo  {journal} {Opt. Express}\ }\textbf {\bibinfo {volume} {29}},\ \bibinfo {pages} {25048--25063} (\bibinfo {year} {2021}{\natexlab{a}})}\BibitemShut {NoStop}%
\bibitem [{\citenamefont {Wei}\ \emph {et~al.}(2023)\citenamefont {Wei}, \citenamefont {Huang}, \citenamefont {Wang}, \citenamefont {Wang},\ and\ \citenamefont {Zeng}}]{wei2023high}%
  \BibitemOpen
  \bibfield  {author} {\bibinfo {author} {\bibfnamefont {S.}~\bibnamefont {Wei}}, \bibinfo {author} {\bibfnamefont {P.}~\bibnamefont {Huang}}, \bibinfo {author} {\bibfnamefont {S.}~\bibnamefont {Wang}}, \bibinfo {author} {\bibfnamefont {T.}~\bibnamefont {Wang}}, \ and\ \bibinfo {author} {\bibfnamefont {G.}~\bibnamefont {Zeng}},\ }\bibfield  {title} {\enquote {\bibinfo {title} {High-precision data acquisition for free-space continuous-variable quantum key distribution},}\ }\href@noop {} {\bibfield  {journal} {\bibinfo  {journal} {Opt. Express}\ }\textbf {\bibinfo {volume} {31}},\ \bibinfo {pages} {7383--7397} (\bibinfo {year} {2023})}\BibitemShut {NoStop}%
\bibitem [{\citenamefont {Wang}\ \emph {et~al.}(2021{\natexlab{b}})\citenamefont {Wang}, \citenamefont {Huang}, \citenamefont {Wang},\ and\ \citenamefont {Zeng}}]{wang2021feasibility}%
  \BibitemOpen
  \bibfield  {author} {\bibinfo {author} {\bibfnamefont {S.}~\bibnamefont {Wang}}, \bibinfo {author} {\bibfnamefont {P.}~\bibnamefont {Huang}}, \bibinfo {author} {\bibfnamefont {T.}~\bibnamefont {Wang}}, \ and\ \bibinfo {author} {\bibfnamefont {G.}~\bibnamefont {Zeng}},\ }\bibfield  {title} {\enquote {\bibinfo {title} {Feasibility of continuous-variable quantum key distribution through fog},}\ }\href@noop {} {\bibfield  {journal} {\bibinfo  {journal} {Opt. Lett.}\ }\textbf {\bibinfo {volume} {46}},\ \bibinfo {pages} {5858--5861} (\bibinfo {year} {2021}{\natexlab{b}})}\BibitemShut {NoStop}%
\bibitem [{\citenamefont {Su}\ \emph {et~al.}(2009)\citenamefont {Su}, \citenamefont {Wang}, \citenamefont {Wang}, \citenamefont {Jia}, \citenamefont {Xie} \emph {et~al.}}]{su_EurophLett_2009}%
  \BibitemOpen
  \bibfield  {author} {\bibinfo {author} {\bibfnamefont {X.}~\bibnamefont {Su}}, \bibinfo {author} {\bibfnamefont {W.}~\bibnamefont {Wang}}, \bibinfo {author} {\bibfnamefont {Y.}~\bibnamefont {Wang}}, \bibinfo {author} {\bibfnamefont {X.}~\bibnamefont {Jia}}, \bibinfo {author} {\bibfnamefont {C.}~\bibnamefont {Xie}},  \emph {et~al.},\ }\bibfield  {title} {\enquote {\bibinfo {title} {Continuous variable quantum key distribution based on optical entangled states without signal modulation},}\ }\href@noop {} {\bibfield  {journal} {\bibinfo  {journal} {Europhys. Lett.}\ }\textbf {\bibinfo {volume} {87}},\ \bibinfo {pages} {20005} (\bibinfo {year} {2009})}\BibitemShut {NoStop}%
\bibitem [{\citenamefont {Madsen}\ \emph {et~al.}(2012{\natexlab{b}})\citenamefont {Madsen}, \citenamefont {Usenko}, \citenamefont {Lassen}, \citenamefont {Filip},\ and\ \citenamefont {Andersen}}]{madsen_NatComm_2012}%
  \BibitemOpen
  \bibfield  {author} {\bibinfo {author} {\bibfnamefont {L.~S.}\ \bibnamefont {Madsen}}, \bibinfo {author} {\bibfnamefont {V.~C.}\ \bibnamefont {Usenko}}, \bibinfo {author} {\bibfnamefont {M.}~\bibnamefont {Lassen}}, \bibinfo {author} {\bibfnamefont {R.}~\bibnamefont {Filip}}, \ and\ \bibinfo {author} {\bibfnamefont {U.~L.}\ \bibnamefont {Andersen}},\ }\bibfield  {title} {\enquote {\bibinfo {title} {Continuous variable quantum key distribution with modulated entangled states},}\ }\href@noop {} {\bibfield  {journal} {\bibinfo  {journal} {Nat. Commun.}\ }\textbf {\bibinfo {volume} {3}},\ \bibinfo {pages} {1083} (\bibinfo {year} {2012}{\natexlab{b}})}\BibitemShut {NoStop}%
\bibitem [{\citenamefont {Gehring}\ \emph {et~al.}(2015)\citenamefont {Gehring}, \citenamefont {H{\"a}ndchen}, \citenamefont {Duhme}, \citenamefont {Furrer}, \citenamefont {Franz} \emph {et~al.}}]{Gehring_NatCommun_2015}%
  \BibitemOpen
  \bibfield  {author} {\bibinfo {author} {\bibfnamefont {T.}~\bibnamefont {Gehring}}, \bibinfo {author} {\bibfnamefont {V.}~\bibnamefont {H{\"a}ndchen}}, \bibinfo {author} {\bibfnamefont {J.}~\bibnamefont {Duhme}}, \bibinfo {author} {\bibfnamefont {F.}~\bibnamefont {Furrer}}, \bibinfo {author} {\bibfnamefont {T.}~\bibnamefont {Franz}},  \emph {et~al.},\ }\bibfield  {title} {\enquote {\bibinfo {title} {Implementation of continuous-variable quantum key distribution with composable and one-sided-device-independent security against coherent attacks},}\ }\href@noop {} {\bibfield  {journal} {\bibinfo  {journal} {Nat. Commun.}\ }\textbf {\bibinfo {volume} {6}},\ \bibinfo {pages} {1--7} (\bibinfo {year} {2015})}\BibitemShut {NoStop}%
\bibitem [{\citenamefont {Wang}\ \emph {et~al.}(2018{\natexlab{f}})\citenamefont {Wang}, \citenamefont {Du}, \citenamefont {Liu}, \citenamefont {Wang}, \citenamefont {Li} \emph {et~al.}}]{wang2018long}%
  \BibitemOpen
  \bibfield  {author} {\bibinfo {author} {\bibfnamefont {N.}~\bibnamefont {Wang}}, \bibinfo {author} {\bibfnamefont {S.}~\bibnamefont {Du}}, \bibinfo {author} {\bibfnamefont {W.}~\bibnamefont {Liu}}, \bibinfo {author} {\bibfnamefont {X.}~\bibnamefont {Wang}}, \bibinfo {author} {\bibfnamefont {Y.}~\bibnamefont {Li}},  \emph {et~al.},\ }\bibfield  {title} {\enquote {\bibinfo {title} {Long-distance continuous-variable quantum key distribution with entangled states},}\ }\href@noop {} {\bibfield  {journal} {\bibinfo  {journal} {Phys. Rev. Appl.}\ }\textbf {\bibinfo {volume} {10}},\ \bibinfo {pages} {064028} (\bibinfo {year} {2018}{\natexlab{f}})}\BibitemShut {NoStop}%
\bibitem [{\citenamefont {Ren}, \citenamefont {Wang},\ and\ \citenamefont {Su}(2022)}]{ren_SciChinaInforSci_2022}%
  \BibitemOpen
  \bibfield  {author} {\bibinfo {author} {\bibfnamefont {S.}~\bibnamefont {Ren}}, \bibinfo {author} {\bibfnamefont {Y.}~\bibnamefont {Wang}}, \ and\ \bibinfo {author} {\bibfnamefont {X.}~\bibnamefont {Su}},\ }\bibfield  {title} {\enquote {\bibinfo {title} {Hybrid quantum key distribution network},}\ }\href@noop {} {\bibfield  {journal} {\bibinfo  {journal} {Sci. China Inf. Sci.}\ }\textbf {\bibinfo {volume} {65}},\ \bibinfo {pages} {200502} (\bibinfo {year} {2022})}\BibitemShut {NoStop}%
\bibitem [{\citenamefont {Feng}\ \emph {et~al.}(2017)\citenamefont {Feng}, \citenamefont {Wan}, \citenamefont {Li},\ and\ \citenamefont {Zhang}}]{feng_OptLett_2017}%
  \BibitemOpen
  \bibfield  {author} {\bibinfo {author} {\bibfnamefont {J.}~\bibnamefont {Feng}}, \bibinfo {author} {\bibfnamefont {Z.}~\bibnamefont {Wan}}, \bibinfo {author} {\bibfnamefont {Y.}~\bibnamefont {Li}}, \ and\ \bibinfo {author} {\bibfnamefont {K.}~\bibnamefont {Zhang}},\ }\bibfield  {title} {\enquote {\bibinfo {title} {Distribution of continuous variable quantum entanglement at a telecommunication wavelength over 20 km of optical fiber},}\ }\href@noop {} {\bibfield  {journal} {\bibinfo  {journal} {Opt. Lett.}\ }\textbf {\bibinfo {volume} {42}},\ \bibinfo {pages} {3399--3402} (\bibinfo {year} {2017})}\BibitemShut {NoStop}%
\bibitem [{\citenamefont {Du}\ \emph {et~al.}(2023)\citenamefont {Du}, \citenamefont {Wang}, \citenamefont {Liu}, \citenamefont {Tian},\ and\ \citenamefont {Li}}]{du2023continuous}%
  \BibitemOpen
  \bibfield  {author} {\bibinfo {author} {\bibfnamefont {S.}~\bibnamefont {Du}}, \bibinfo {author} {\bibfnamefont {P.}~\bibnamefont {Wang}}, \bibinfo {author} {\bibfnamefont {J.}~\bibnamefont {Liu}}, \bibinfo {author} {\bibfnamefont {Y.}~\bibnamefont {Tian}}, \ and\ \bibinfo {author} {\bibfnamefont {Y.}~\bibnamefont {Li}},\ }\bibfield  {title} {\enquote {\bibinfo {title} {Continuous variable quantum key distribution with a shared partially characterized entangled source},}\ }\href@noop {} {\bibfield  {journal} {\bibinfo  {journal} {Photonics Res.}\ }\textbf {\bibinfo {volume} {11}},\ \bibinfo {pages} {463--475} (\bibinfo {year} {2023})}\BibitemShut {NoStop}%
\bibitem [{\citenamefont {Chen}\ \emph {et~al.}(2023)\citenamefont {Chen}, \citenamefont {Wang}, \citenamefont {Yu}, \citenamefont {Li},\ and\ \citenamefont {Guo}}]{chen2023continuous}%
  \BibitemOpen
  \bibfield  {author} {\bibinfo {author} {\bibfnamefont {Z.}~\bibnamefont {Chen}}, \bibinfo {author} {\bibfnamefont {X.}~\bibnamefont {Wang}}, \bibinfo {author} {\bibfnamefont {S.}~\bibnamefont {Yu}}, \bibinfo {author} {\bibfnamefont {Z.}~\bibnamefont {Li}}, \ and\ \bibinfo {author} {\bibfnamefont {H.}~\bibnamefont {Guo}},\ }\bibfield  {title} {\enquote {\bibinfo {title} {Continuous-mode quantum key distribution with digital signal processing},}\ }\href@noop {} {\bibfield  {journal} {\bibinfo  {journal} {npj Quantum Inf.}\ }\textbf {\bibinfo {volume} {9}},\ \bibinfo {pages} {28} (\bibinfo {year} {2023})}\BibitemShut {NoStop}%
\bibitem [{\citenamefont {Bian}\ \emph {et~al.}(2024)\citenamefont {Bian} \emph {et~al.}}]{bian2023chip}%
  \BibitemOpen
  \bibfield  {author} {\bibinfo {author} {\bibfnamefont {Y.}~\bibnamefont {Bian}} \emph {et~al.},\ }\href@noop {} {\bibfield  {journal} {\bibinfo  {journal} {In prepration}\ } (\bibinfo {year} {2024})}\BibitemShut {NoStop}%
\bibitem [{\citenamefont {Wang}\ \emph {et~al.}(2020{\natexlab{d}})\citenamefont {Wang}, \citenamefont {Sciarrino}, \citenamefont {Laing},\ and\ \citenamefont {Thompson}}]{wang2020integrated}%
  \BibitemOpen
  \bibfield  {author} {\bibinfo {author} {\bibfnamefont {J.}~\bibnamefont {Wang}}, \bibinfo {author} {\bibfnamefont {F.}~\bibnamefont {Sciarrino}}, \bibinfo {author} {\bibfnamefont {A.}~\bibnamefont {Laing}}, \ and\ \bibinfo {author} {\bibfnamefont {M.~G.}\ \bibnamefont {Thompson}},\ }\bibfield  {title} {\enquote {\bibinfo {title} {Integrated photonic quantum technologies},}\ }\href@noop {} {\bibfield  {journal} {\bibinfo  {journal} {Nature Photonics}\ }\textbf {\bibinfo {volume} {14}},\ \bibinfo {pages} {273--284} (\bibinfo {year} {2020}{\natexlab{d}})}\BibitemShut {NoStop}%
\bibitem [{\citenamefont {Li}\ \emph {et~al.}(2021{\natexlab{b}})\citenamefont {Li}, \citenamefont {Huang}, \citenamefont {Wang},\ and\ \citenamefont {Zeng}}]{li2021practical}%
  \BibitemOpen
  \bibfield  {author} {\bibinfo {author} {\bibfnamefont {L.}~\bibnamefont {Li}}, \bibinfo {author} {\bibfnamefont {P.}~\bibnamefont {Huang}}, \bibinfo {author} {\bibfnamefont {T.}~\bibnamefont {Wang}}, \ and\ \bibinfo {author} {\bibfnamefont {G.}~\bibnamefont {Zeng}},\ }\bibfield  {title} {\enquote {\bibinfo {title} {Practical security of a chip-based continuous-variable quantum-key-distribution system},}\ }\href@noop {} {\bibfield  {journal} {\bibinfo  {journal} {Phys. Rev. A}\ }\textbf {\bibinfo {volume} {103}},\ \bibinfo {pages} {032611} (\bibinfo {year} {2021}{\natexlab{b}})}\BibitemShut {NoStop}%
\bibitem [{\citenamefont {Wang}\ \emph {et~al.}(2022{\natexlab{d}})\citenamefont {Wang}, \citenamefont {Jia}, \citenamefont {Guo}, \citenamefont {Liu}, \citenamefont {Wang} \emph {et~al.}}]{wang2022silicon}%
  \BibitemOpen
  \bibfield  {author} {\bibinfo {author} {\bibfnamefont {X.}~\bibnamefont {Wang}}, \bibinfo {author} {\bibfnamefont {Y.}~\bibnamefont {Jia}}, \bibinfo {author} {\bibfnamefont {X.}~\bibnamefont {Guo}}, \bibinfo {author} {\bibfnamefont {J.}~\bibnamefont {Liu}}, \bibinfo {author} {\bibfnamefont {S.}~\bibnamefont {Wang}},  \emph {et~al.},\ }\bibfield  {title} {\enquote {\bibinfo {title} {Silicon photonics integrated dynamic polarization controller},}\ }\href@noop {} {\bibfield  {journal} {\bibinfo  {journal} {Chinese Opt. Lett.}\ }\textbf {\bibinfo {volume} {20}},\ \bibinfo {pages} {041301} (\bibinfo {year} {2022}{\natexlab{d}})}\BibitemShut {NoStop}%
\bibitem [{\citenamefont {Luo}\ \emph {et~al.}(2023)\citenamefont {Luo}, \citenamefont {Cao}, \citenamefont {Shi}, \citenamefont {Wan}, \citenamefont {Zhang}, \citenamefont {Li}, \citenamefont {Chen}, \citenamefont {Li}, \citenamefont {Li}, \citenamefont {Wang} \emph {et~al.}}]{luo2023recent}%
  \BibitemOpen
  \bibfield  {author} {\bibinfo {author} {\bibfnamefont {W.}~\bibnamefont {Luo}}, \bibinfo {author} {\bibfnamefont {L.}~\bibnamefont {Cao}}, \bibinfo {author} {\bibfnamefont {Y.}~\bibnamefont {Shi}}, \bibinfo {author} {\bibfnamefont {L.}~\bibnamefont {Wan}}, \bibinfo {author} {\bibfnamefont {H.}~\bibnamefont {Zhang}}, \bibinfo {author} {\bibfnamefont {S.}~\bibnamefont {Li}}, \bibinfo {author} {\bibfnamefont {G.}~\bibnamefont {Chen}}, \bibinfo {author} {\bibfnamefont {Y.}~\bibnamefont {Li}}, \bibinfo {author} {\bibfnamefont {S.}~\bibnamefont {Li}}, \bibinfo {author} {\bibfnamefont {Y.}~\bibnamefont {Wang}},  \emph {et~al.},\ }\bibfield  {title} {\enquote {\bibinfo {title} {Recent progress in quantum photonic chips for quantum communication and internet},}\ }\href@noop {} {\bibfield  {journal} {\bibinfo  {journal} {Light: Science \& Applications}\ }\textbf {\bibinfo {volume} {12}},\ \bibinfo {pages} {175} (\bibinfo {year} {2023})}\BibitemShut {NoStop}%
\bibitem [{\citenamefont {Fröhlich}\ \emph {et~al.}(2013)\citenamefont {Fröhlich}, \citenamefont {Dynes}, \citenamefont {Lucamarini} \emph {et~al.}}]{QNetNature2013}%
  \BibitemOpen
  \bibfield  {author} {\bibinfo {author} {\bibfnamefont {B.}~\bibnamefont {Fröhlich}}, \bibinfo {author} {\bibfnamefont {J.}~\bibnamefont {Dynes}}, \bibinfo {author} {\bibfnamefont {M.}~\bibnamefont {Lucamarini}},  \emph {et~al.},\ }\bibfield  {title} {\enquote {\bibinfo {title} {A quantum access network},}\ }\href {\doibase 10.1038/nature12493} {\bibfield  {journal} {\bibinfo  {journal} {Nature}\ }\textbf {\bibinfo {volume} {501}},\ \bibinfo {pages} {69--72} (\bibinfo {year} {2013})}\BibitemShut {NoStop}%
\bibitem [{\citenamefont {Huang}\ \emph {et~al.}(2020{\natexlab{b}})\citenamefont {Huang}, \citenamefont {Zhang}, \citenamefont {Shen}, \citenamefont {Huang},\ and\ \citenamefont {Yu}}]{huang2020experimental}%
  \BibitemOpen
  \bibfield  {author} {\bibinfo {author} {\bibfnamefont {Y.}~\bibnamefont {Huang}}, \bibinfo {author} {\bibfnamefont {Y.}~\bibnamefont {Zhang}}, \bibinfo {author} {\bibfnamefont {T.}~\bibnamefont {Shen}}, \bibinfo {author} {\bibfnamefont {G.}~\bibnamefont {Huang}}, \ and\ \bibinfo {author} {\bibfnamefont {S.}~\bibnamefont {Yu}},\ }\bibfield  {title} {\enquote {\bibinfo {title} {Experimental demonstration of upstream continuous-variable qkd access network},}\ }in\ \href@noop {} {\emph {\bibinfo {booktitle} {CLEO: QELS\_Fundamental Science}}}\ (\bibinfo {organization} {Optica Publishing Group},\ \bibinfo {year} {2020})\ pp.\ \bibinfo {pages} {JTu2A--24}\BibitemShut {NoStop}%
\bibitem [{\citenamefont {Xu}\ \emph {et~al.}(2023)\citenamefont {Xu}, \citenamefont {Wang}, \citenamefont {Zhao}, \citenamefont {Huang},\ and\ \citenamefont {Zeng}}]{xu2023round}%
  \BibitemOpen
  \bibfield  {author} {\bibinfo {author} {\bibfnamefont {Y.}~\bibnamefont {Xu}}, \bibinfo {author} {\bibfnamefont {T.}~\bibnamefont {Wang}}, \bibinfo {author} {\bibfnamefont {H.}~\bibnamefont {Zhao}}, \bibinfo {author} {\bibfnamefont {P.}~\bibnamefont {Huang}}, \ and\ \bibinfo {author} {\bibfnamefont {G.}~\bibnamefont {Zeng}},\ }\bibfield  {title} {\enquote {\bibinfo {title} {Round-trip multi-band quantum access network},}\ }\href {\doibase 10.1364/PRJ.492448} {\bibfield  {journal} {\bibinfo  {journal} {Photon. Res.}\ }\textbf {\bibinfo {volume} {11}},\ \bibinfo {pages} {1449--1464} (\bibinfo {year} {2023})}\BibitemShut {NoStop}%
\bibitem [{\citenamefont {Bian}\ \emph {et~al.}(2023{\natexlab{b}})\citenamefont {Bian}, \citenamefont {Pan}, \citenamefont {Ma}, \citenamefont {Wang}, \citenamefont {Dou} \emph {et~al.}}]{bian2023first}%
  \BibitemOpen
  \bibfield  {author} {\bibinfo {author} {\bibfnamefont {Y.}~\bibnamefont {Bian}}, \bibinfo {author} {\bibfnamefont {Y.}~\bibnamefont {Pan}}, \bibinfo {author} {\bibfnamefont {L.}~\bibnamefont {Ma}}, \bibinfo {author} {\bibfnamefont {H.}~\bibnamefont {Wang}}, \bibinfo {author} {\bibfnamefont {J.}~\bibnamefont {Dou}},  \emph {et~al.},\ }\bibfield  {title} {\enquote {\bibinfo {title} {First demonstration of an 8-node mbps quantum access network based on passive optical distribution network facilities},}\ }in\ \href@noop {} {\emph {\bibinfo {booktitle} {Laser Science}}}\ (\bibinfo {organization} {Optica Publishing Group},\ \bibinfo {year} {2023})\ pp.\ \bibinfo {pages} {JTu7A--4}\BibitemShut {NoStop}%
\bibitem [{\citenamefont {Huang}, \citenamefont {He},\ and\ \citenamefont {Zeng}(2013)}]{huang2013bound}%
  \BibitemOpen
  \bibfield  {author} {\bibinfo {author} {\bibfnamefont {P.}~\bibnamefont {Huang}}, \bibinfo {author} {\bibfnamefont {G.-Q.}\ \bibnamefont {He}}, \ and\ \bibinfo {author} {\bibfnamefont {G.-H.}\ \bibnamefont {Zeng}},\ }\bibfield  {title} {\enquote {\bibinfo {title} {Bound on noise of coherent source for secure continuous-variable quantum key distribution},}\ }\href@noop {} {\bibfield  {journal} {\bibinfo  {journal} {Int. J. Theor. Phys.}\ }\textbf {\bibinfo {volume} {52}},\ \bibinfo {pages} {1572--1582} (\bibinfo {year} {2013})}\BibitemShut {NoStop}%
\bibitem [{\citenamefont {Stiller}\ \emph {et~al.}(2015{\natexlab{b}})\citenamefont {Stiller}, \citenamefont {Khan}, \citenamefont {Jain}, \citenamefont {Jouguet}, \citenamefont {Kunz-Jacques} \emph {et~al.}}]{Stiller:15}%
  \BibitemOpen
  \bibfield  {author} {\bibinfo {author} {\bibfnamefont {B.}~\bibnamefont {Stiller}}, \bibinfo {author} {\bibfnamefont {I.}~\bibnamefont {Khan}}, \bibinfo {author} {\bibfnamefont {N.}~\bibnamefont {Jain}}, \bibinfo {author} {\bibfnamefont {P.}~\bibnamefont {Jouguet}}, \bibinfo {author} {\bibfnamefont {S.}~\bibnamefont {Kunz-Jacques}},  \emph {et~al.},\ }\bibfield  {title} {\enquote {\bibinfo {title} {Quantum hacking of continuous-variable quantum key distribution systems: realtime trojan-horse attacks},}\ }in\ \href {\doibase 10.1364/CLEO_QELS.2015.FF1A.7} {\emph {\bibinfo {booktitle} {CLEO: 2015}}}\ (\bibinfo  {publisher} {Optica Publishing Group},\ \bibinfo {year} {2015})\ p.\ \bibinfo {pages} {FF1A.7}\BibitemShut {NoStop}%
\bibitem [{\citenamefont {Qin}, \citenamefont {Kumar},\ and\ \citenamefont {All\'eaume}(2016)}]{Qin2016QuantumHacking}%
  \BibitemOpen
  \bibfield  {author} {\bibinfo {author} {\bibfnamefont {H.}~\bibnamefont {Qin}}, \bibinfo {author} {\bibfnamefont {R.}~\bibnamefont {Kumar}}, \ and\ \bibinfo {author} {\bibfnamefont {R.}~\bibnamefont {All\'eaume}},\ }\bibfield  {title} {\enquote {\bibinfo {title} {Quantum hacking: Saturation attack on practical continuous-variable quantum key distribution},}\ }\href {\doibase 10.1103/PhysRevA.94.012325} {\bibfield  {journal} {\bibinfo  {journal} {Phys. Rev. A}\ }\textbf {\bibinfo {volume} {94}},\ \bibinfo {pages} {012325} (\bibinfo {year} {2016})}\BibitemShut {NoStop}%
\bibitem [{\citenamefont {Zhao}\ \emph {et~al.}(2018{\natexlab{c}})\citenamefont {Zhao}, \citenamefont {Zhang}, \citenamefont {Huang}, \citenamefont {Xu}, \citenamefont {Yu} \emph {et~al.}}]{zhao2018polarization}%
  \BibitemOpen
  \bibfield  {author} {\bibinfo {author} {\bibfnamefont {Y.}~\bibnamefont {Zhao}}, \bibinfo {author} {\bibfnamefont {Y.}~\bibnamefont {Zhang}}, \bibinfo {author} {\bibfnamefont {Y.}~\bibnamefont {Huang}}, \bibinfo {author} {\bibfnamefont {B.}~\bibnamefont {Xu}}, \bibinfo {author} {\bibfnamefont {S.}~\bibnamefont {Yu}},  \emph {et~al.},\ }\bibfield  {title} {\enquote {\bibinfo {title} {Polarization attack on continuous-variable quantum key distribution},}\ }\href@noop {} {\bibfield  {journal} {\bibinfo  {journal} {J. Phys. B: At., Mol. Opt. Phys.}\ }\textbf {\bibinfo {volume} {52}},\ \bibinfo {pages} {015501} (\bibinfo {year} {2018}{\natexlab{c}})}\BibitemShut {NoStop}%
\bibitem [{\citenamefont {Qin}\ \emph {et~al.}(2018)\citenamefont {Qin}, \citenamefont {Kumar}, \citenamefont {Makarov},\ and\ \citenamefont {All\'eaume}}]{Qin2018Homodyne}%
  \BibitemOpen
  \bibfield  {author} {\bibinfo {author} {\bibfnamefont {H.}~\bibnamefont {Qin}}, \bibinfo {author} {\bibfnamefont {R.}~\bibnamefont {Kumar}}, \bibinfo {author} {\bibfnamefont {V.}~\bibnamefont {Makarov}}, \ and\ \bibinfo {author} {\bibfnamefont {R.}~\bibnamefont {All\'eaume}},\ }\bibfield  {title} {\enquote {\bibinfo {title} {Homodyne-detector-blinding attack in continuous-variable quantum key distribution},}\ }\href {\doibase 10.1103/PhysRevA.98.012312} {\bibfield  {journal} {\bibinfo  {journal} {Phys. Rev. A}\ }\textbf {\bibinfo {volume} {98}},\ \bibinfo {pages} {012312} (\bibinfo {year} {2018})}\BibitemShut {NoStop}%
\bibitem [{\citenamefont {Zheng}\ \emph {et~al.}(2019{\natexlab{b}})\citenamefont {Zheng}, \citenamefont {Huang}, \citenamefont {Huang}, \citenamefont {Peng},\ and\ \citenamefont {Zeng}}]{zheng2019security}%
  \BibitemOpen
  \bibfield  {author} {\bibinfo {author} {\bibfnamefont {Y.}~\bibnamefont {Zheng}}, \bibinfo {author} {\bibfnamefont {P.}~\bibnamefont {Huang}}, \bibinfo {author} {\bibfnamefont {A.}~\bibnamefont {Huang}}, \bibinfo {author} {\bibfnamefont {J.}~\bibnamefont {Peng}}, \ and\ \bibinfo {author} {\bibfnamefont {G.}~\bibnamefont {Zeng}},\ }\bibfield  {title} {\enquote {\bibinfo {title} {Security analysis of practical continuous-variable quantum key distribution systems under laser seeding attack},}\ }\href@noop {} {\bibfield  {journal} {\bibinfo  {journal} {Opt. Express}\ }\textbf {\bibinfo {volume} {27}},\ \bibinfo {pages} {27369--27384} (\bibinfo {year} {2019}{\natexlab{b}})}\BibitemShut {NoStop}%
\bibitem [{\citenamefont {Zheng}\ \emph {et~al.}(2019{\natexlab{c}})\citenamefont {Zheng}, \citenamefont {Huang}, \citenamefont {Huang}, \citenamefont {Peng},\ and\ \citenamefont {Zeng}}]{Zheng_PhysRevA_2019}%
  \BibitemOpen
  \bibfield  {author} {\bibinfo {author} {\bibfnamefont {Y.}~\bibnamefont {Zheng}}, \bibinfo {author} {\bibfnamefont {P.}~\bibnamefont {Huang}}, \bibinfo {author} {\bibfnamefont {A.}~\bibnamefont {Huang}}, \bibinfo {author} {\bibfnamefont {J.}~\bibnamefont {Peng}}, \ and\ \bibinfo {author} {\bibfnamefont {G.}~\bibnamefont {Zeng}},\ }\bibfield  {title} {\enquote {\bibinfo {title} {Practical security of continuous-variable quantum key distribution with reduced optical attenuation},}\ }\href {\doibase 10.1103/PhysRevA.100.012313} {\bibfield  {journal} {\bibinfo  {journal} {Phys. Rev. A}\ }\textbf {\bibinfo {volume} {100}},\ \bibinfo {pages} {012313} (\bibinfo {year} {2019}{\natexlab{c}})}\BibitemShut {NoStop}%
\bibitem [{\citenamefont {Ren}\ \emph {et~al.}(2019)\citenamefont {Ren}, \citenamefont {Kumar}, \citenamefont {Wonfor}, \citenamefont {Tang}, \citenamefont {Penty} \emph {et~al.}}]{ren2019reference}%
  \BibitemOpen
  \bibfield  {author} {\bibinfo {author} {\bibfnamefont {S.}~\bibnamefont {Ren}}, \bibinfo {author} {\bibfnamefont {R.}~\bibnamefont {Kumar}}, \bibinfo {author} {\bibfnamefont {A.}~\bibnamefont {Wonfor}}, \bibinfo {author} {\bibfnamefont {X.}~\bibnamefont {Tang}}, \bibinfo {author} {\bibfnamefont {R.}~\bibnamefont {Penty}},  \emph {et~al.},\ }\bibfield  {title} {\enquote {\bibinfo {title} {Reference pulse attack on continuous variable quantum key distribution with local local oscillator under trusted phase noise},}\ }\href@noop {} {\bibfield  {journal} {\bibinfo  {journal} {J. Opt. Soc. Am. B}\ }\textbf {\bibinfo {volume} {36}},\ \bibinfo {pages} {B7--B15} (\bibinfo {year} {2019})}\BibitemShut {NoStop}%
\bibitem [{\citenamefont {Shao}\ \emph {et~al.}(2022)\citenamefont {Shao}, \citenamefont {Li}, \citenamefont {Wang}, \citenamefont {Pan}, \citenamefont {Pi} \emph {et~al.}}]{Shao2022Phase}%
  \BibitemOpen
  \bibfield  {author} {\bibinfo {author} {\bibfnamefont {Y.}~\bibnamefont {Shao}}, \bibinfo {author} {\bibfnamefont {Y.}~\bibnamefont {Li}}, \bibinfo {author} {\bibfnamefont {H.}~\bibnamefont {Wang}}, \bibinfo {author} {\bibfnamefont {Y.}~\bibnamefont {Pan}}, \bibinfo {author} {\bibfnamefont {Y.}~\bibnamefont {Pi}},  \emph {et~al.},\ }\bibfield  {title} {\enquote {\bibinfo {title} {Phase-reference-intensity attack on continuous-variable quantum key distribution with a local local oscillator},}\ }\href {\doibase 10.1103/PhysRevA.105.032601} {\bibfield  {journal} {\bibinfo  {journal} {Phys. Rev. A}\ }\textbf {\bibinfo {volume} {105}},\ \bibinfo {pages} {032601} (\bibinfo {year} {2022})}\BibitemShut {NoStop}%
\bibitem [{\citenamefont {Huang}, \citenamefont {Huang},\ and\ \citenamefont {Peng}(2019)}]{Huang_OptExpress_2019}%
  \BibitemOpen
  \bibfield  {author} {\bibinfo {author} {\bibfnamefont {B.}~\bibnamefont {Huang}}, \bibinfo {author} {\bibfnamefont {Y.}~\bibnamefont {Huang}}, \ and\ \bibinfo {author} {\bibfnamefont {Z.}~\bibnamefont {Peng}},\ }\bibfield  {title} {\enquote {\bibinfo {title} {Practical security of the continuous-variable quantum key distribution with real local oscillators under phase attack},}\ }\href@noop {} {\bibfield  {journal} {\bibinfo  {journal} {Opt. Express}\ }\textbf {\bibinfo {volume} {27}},\ \bibinfo {pages} {20621--20631} (\bibinfo {year} {2019})}\BibitemShut {NoStop}%
\bibitem [{\citenamefont {Fan}\ \emph {et~al.}(2023)\citenamefont {Fan}, \citenamefont {Bian}, \citenamefont {Wu}, \citenamefont {Zhang},\ and\ \citenamefont {Yu}}]{Fan2023Quantum}%
  \BibitemOpen
  \bibfield  {author} {\bibinfo {author} {\bibfnamefont {L.}~\bibnamefont {Fan}}, \bibinfo {author} {\bibfnamefont {Y.}~\bibnamefont {Bian}}, \bibinfo {author} {\bibfnamefont {M.}~\bibnamefont {Wu}}, \bibinfo {author} {\bibfnamefont {Y.}~\bibnamefont {Zhang}}, \ and\ \bibinfo {author} {\bibfnamefont {S.}~\bibnamefont {Yu}},\ }\bibfield  {title} {\enquote {\bibinfo {title} {Quantum hacking against discrete-modulated continuous-variable quantum key distribution using modified local oscillator intensity attack with random fluctuations},}\ }\href {\doibase 10.1103/PhysRevApplied.20.024073} {\bibfield  {journal} {\bibinfo  {journal} {Phys. Rev. Appl.}\ }\textbf {\bibinfo {volume} {20}},\ \bibinfo {pages} {024073} (\bibinfo {year} {2023})}\BibitemShut {NoStop}%
\bibitem [{\citenamefont {Pan}\ \emph {et~al.}(2023{\natexlab{b}})\citenamefont {Pan}, \citenamefont {Bian}, \citenamefont {Wang}, \citenamefont {Dou}, \citenamefont {Shao} \emph {et~al.}}]{PTMPAccessNetwork}%
  \BibitemOpen
  \bibfield  {author} {\bibinfo {author} {\bibfnamefont {Y.}~\bibnamefont {Pan}}, \bibinfo {author} {\bibfnamefont {Y.}~\bibnamefont {Bian}}, \bibinfo {author} {\bibfnamefont {H.}~\bibnamefont {Wang}}, \bibinfo {author} {\bibfnamefont {J.}~\bibnamefont {Dou}}, \bibinfo {author} {\bibfnamefont {Y.}~\bibnamefont {Shao}},  \emph {et~al.},\ }\bibfield  {title} {\enquote {\bibinfo {title} {Experimental demonstration of 4-user quantum access network based on passive optical network},}\ }in\ \href@noop {} {\emph {\bibinfo {booktitle} {European Conference on Optical Communications (ECOC 2023)}}}\ (\bibinfo  {publisher} {IEEE},\ \bibinfo {year} {2023})\ p.\ \bibinfo {pages} {P51}\BibitemShut {NoStop}%
\bibitem [{\citenamefont {Pan}\ \emph {et~al.}(2024)\citenamefont {Pan}, \citenamefont {Bian}, \citenamefont {Wang}, \citenamefont {Dou}, \citenamefont {Shao} \emph {et~al.}}]{PanHigh2024}%
  \BibitemOpen
  \bibfield  {author} {\bibinfo {author} {\bibfnamefont {Y.}~\bibnamefont {Pan}}, \bibinfo {author} {\bibfnamefont {Y.}~\bibnamefont {Bian}}, \bibinfo {author} {\bibfnamefont {H.}~\bibnamefont {Wang}}, \bibinfo {author} {\bibfnamefont {J.}~\bibnamefont {Dou}}, \bibinfo {author} {\bibfnamefont {Y.}~\bibnamefont {Shao}},  \emph {et~al.},\ }\bibfield  {title} {\enquote {\bibinfo {title} {High-performance continuous-variable quantum key distribution access network},}\ }in\ \href@noop {} {\emph {\bibinfo {booktitle} {Optical Fiber Communication Conference and Exposition (OFC 2024)}}}\ (\bibinfo  {publisher} {Optica Publishing Group},\ \bibinfo {year} {2024})\ p.\ \bibinfo {pages} {Th1C}\BibitemShut {NoStop}%
\bibitem [{\citenamefont {Shao}\ \emph {et~al.}(2021)\citenamefont {Shao}, \citenamefont {Wang}, \citenamefont {Pi}, \citenamefont {Huang}, \citenamefont {Li} \emph {et~al.}}]{shao2021phase}%
  \BibitemOpen
  \bibfield  {author} {\bibinfo {author} {\bibfnamefont {Y.}~\bibnamefont {Shao}}, \bibinfo {author} {\bibfnamefont {H.}~\bibnamefont {Wang}}, \bibinfo {author} {\bibfnamefont {Y.}~\bibnamefont {Pi}}, \bibinfo {author} {\bibfnamefont {W.}~\bibnamefont {Huang}}, \bibinfo {author} {\bibfnamefont {Y.}~\bibnamefont {Li}},  \emph {et~al.},\ }\bibfield  {title} {\enquote {\bibinfo {title} {Phase noise model for continuous-variable quantum key distribution using a local local oscillator},}\ }\href@noop {} {\bibfield  {journal} {\bibinfo  {journal} {Physical Review A}\ }\textbf {\bibinfo {volume} {104}},\ \bibinfo {pages} {032608} (\bibinfo {year} {2021})}\BibitemShut {NoStop}%
\bibitem [{\citenamefont {Liao}\ \emph {et~al.}(2022)\citenamefont {Liao}, \citenamefont {Wang}, \citenamefont {Liu}, \citenamefont {Mao},\ and\ \citenamefont {Fu}}]{liao2022detecting}%
  \BibitemOpen
  \bibfield  {author} {\bibinfo {author} {\bibfnamefont {Q.}~\bibnamefont {Liao}}, \bibinfo {author} {\bibfnamefont {Z.}~\bibnamefont {Wang}}, \bibinfo {author} {\bibfnamefont {H.}~\bibnamefont {Liu}}, \bibinfo {author} {\bibfnamefont {Y.}~\bibnamefont {Mao}}, \ and\ \bibinfo {author} {\bibfnamefont {X.}~\bibnamefont {Fu}},\ }\bibfield  {title} {\enquote {\bibinfo {title} {Detecting practical quantum attacks for continuous-variable quantum key distribution using density-based spatial clustering of applications with noise},}\ }\href@noop {} {\bibfield  {journal} {\bibinfo  {journal} {Phys. Rev. A}\ }\textbf {\bibinfo {volume} {106}},\ \bibinfo {pages} {022607} (\bibinfo {year} {2022})}\BibitemShut {NoStop}%
\bibitem [{\citenamefont {Sajeed}\ \emph {et~al.}(2021)\citenamefont {Sajeed}, \citenamefont {Chaiwongkhot}, \citenamefont {Huang}, \citenamefont {Qin}, \citenamefont {Egorov} \emph {et~al.}}]{sajeed2021approach}%
  \BibitemOpen
  \bibfield  {author} {\bibinfo {author} {\bibfnamefont {S.}~\bibnamefont {Sajeed}}, \bibinfo {author} {\bibfnamefont {P.}~\bibnamefont {Chaiwongkhot}}, \bibinfo {author} {\bibfnamefont {A.}~\bibnamefont {Huang}}, \bibinfo {author} {\bibfnamefont {H.}~\bibnamefont {Qin}}, \bibinfo {author} {\bibfnamefont {V.}~\bibnamefont {Egorov}},  \emph {et~al.},\ }\bibfield  {title} {\enquote {\bibinfo {title} {An approach for security evaluation and certification of a complete quantum communication system},}\ }\href@noop {} {\bibfield  {journal} {\bibinfo  {journal} {Sci. Rep.}\ }\textbf {\bibinfo {volume} {11}},\ \bibinfo {pages} {5110} (\bibinfo {year} {2021})}\BibitemShut {NoStop}%
\bibitem [{\citenamefont {Tian}\ \emph {et~al.}(2022)\citenamefont {Tian}, \citenamefont {Wang}, \citenamefont {Liu}, \citenamefont {Du}, \citenamefont {Liu} \emph {et~al.}}]{tian2022experimental}%
  \BibitemOpen
  \bibfield  {author} {\bibinfo {author} {\bibfnamefont {Y.}~\bibnamefont {Tian}}, \bibinfo {author} {\bibfnamefont {P.}~\bibnamefont {Wang}}, \bibinfo {author} {\bibfnamefont {J.}~\bibnamefont {Liu}}, \bibinfo {author} {\bibfnamefont {S.}~\bibnamefont {Du}}, \bibinfo {author} {\bibfnamefont {W.}~\bibnamefont {Liu}},  \emph {et~al.},\ }\bibfield  {title} {\enquote {\bibinfo {title} {Experimental demonstration of continuous-variable measurement-device-independent quantum key distribution over optical fiber},}\ }\href@noop {} {\bibfield  {journal} {\bibinfo  {journal} {Optica}\ }\textbf {\bibinfo {volume} {9}},\ \bibinfo {pages} {492--500} (\bibinfo {year} {2022})}\BibitemShut {NoStop}%
\bibitem [{\citenamefont {Hajomer}\ \emph {et~al.}(2023{\natexlab{c}})\citenamefont {Hajomer}, \citenamefont {Nguyen}, \citenamefont {Andersen},\ and\ \citenamefont {Gehring}}]{hajomer2023high}%
  \BibitemOpen
  \bibfield  {author} {\bibinfo {author} {\bibfnamefont {A.~A.}\ \bibnamefont {Hajomer}}, \bibinfo {author} {\bibfnamefont {H.~Q.}\ \bibnamefont {Nguyen}}, \bibinfo {author} {\bibfnamefont {U.~L.}\ \bibnamefont {Andersen}}, \ and\ \bibinfo {author} {\bibfnamefont {T.}~\bibnamefont {Gehring}},\ }\bibfield  {title} {\enquote {\bibinfo {title} {High-rate continuous-variable measurement-device-independent quantum key distribution},}\ }in\ \href@noop {} {\emph {\bibinfo {booktitle} {Optical Fiber Communication Conference}}}\ (\bibinfo {organization} {Optica Publishing Group},\ \bibinfo {year} {2023})\ pp.\ \bibinfo {pages} {M2I--2}\BibitemShut {NoStop}%
\bibitem [{\citenamefont {Hajomer}, \citenamefont {Andersen},\ and\ \citenamefont {Gehring}(2023)}]{hajomer2023real}%
  \BibitemOpen
  \bibfield  {author} {\bibinfo {author} {\bibfnamefont {A.~A.}\ \bibnamefont {Hajomer}}, \bibinfo {author} {\bibfnamefont {U.~L.}\ \bibnamefont {Andersen}}, \ and\ \bibinfo {author} {\bibfnamefont {T.}~\bibnamefont {Gehring}},\ }\bibfield  {title} {\enquote {\bibinfo {title} {Real-world data encryption with continuous-variable measurement device-independent quantum key distribution},}\ }\href@noop {} {\bibfield  {journal} {\bibinfo  {journal} {arXiv preprint arXiv:2303.01611}\ } (\bibinfo {year} {2023})}\BibitemShut {NoStop}%
\bibitem [{\citenamefont {Guo}\ \emph {et~al.}(2021)\citenamefont {Guo}, \citenamefont {Li}, \citenamefont {Yu},\ and\ \citenamefont {Zhang}}]{guo_FunRes_2021}%
  \BibitemOpen
  \bibfield  {author} {\bibinfo {author} {\bibfnamefont {H.}~\bibnamefont {Guo}}, \bibinfo {author} {\bibfnamefont {Z.}~\bibnamefont {Li}}, \bibinfo {author} {\bibfnamefont {S.}~\bibnamefont {Yu}}, \ and\ \bibinfo {author} {\bibfnamefont {Y.}~\bibnamefont {Zhang}},\ }\bibfield  {title} {\enquote {\bibinfo {title} {Toward practical quantum key distribution using telecom components},}\ }\href@noop {} {\bibfield  {journal} {\bibinfo  {journal} {Fundam. Res.}\ }\textbf {\bibinfo {volume} {1}},\ \bibinfo {pages} {96--98} (\bibinfo {year} {2021})}\BibitemShut {NoStop}%
\end{thebibliography}%

\end{document}